\pdfoutput=1
\documentclass[11pt]{article}
\usepackage{amsfonts, amsmath, amssymb, amsthm}
\usepackage[english]{babel}
\usepackage{booktabs}
\usepackage{bm}
\usepackage{eucal}
\usepackage{enumerate}
\usepackage[shortlabels]{enumitem}
\usepackage{float}
\usepackage{dsfont}
\usepackage{epsfig}
\usepackage {fontenc}
\usepackage[hmargin=1.6cm,vmargin=3cm]{geometry}
\usepackage{lscape}
\usepackage{mathrsfs}
\usepackage{natbib}
\usepackage{rotating}
\usepackage{verbatim}
\usepackage[table]{xcolor}
\usepackage{dirtytalk}
\usepackage{multirow}
\usepackage{mdframed}
\usepackage{subfigure}

\date{}

\renewcommand{\thefootnote}{$\dagger$}

\begin{document}
\title{Robust and flexible inference for the covariate-specific ROC curve}
\author{\textsc{Vanda In\'acio}, \textsc{Vanda M. Louren\c co}, \textsc{Miguel de Carvalho},\\ \textsc{Richard A. Parker}, and \textsc{Vincent Gnanapragasam}}
\date{}
\maketitle 

\begin{abstract}
\footnotesize{Diagnostic tests are of critical importance in health care and medical research. Motivated by the impact that atypical and outlying test outcomes might have on the assessment of the discriminatory ability of a diagnostic test, we develop a flexible and robust model for conducting inference about the covariate-specific receiver operating characteristic (ROC) curve that safeguards against outlying test results while also accommodating for possible nonlinear effects of the covariates. Specifically, we postulate a location-scale additive regression model for the test outcomes in both the diseased and nondiseased populations, combining additive cubic B-splines and M-estimation for the regression function, while the residuals are estimated via a weighted empirical distribution function. The results of the simulation study show that our approach successfully recovers the true covariate-specific ROC curve and corresponding area under the curve on a variety of conceivable test outcomes contamination scenarios. Our method is applied to a dataset derived from a prostate cancer study where we seek to assess the ability of the Prostate Health Index to discriminate between men with and without Gleason 7 or above prostate cancer, and if and how such discriminatory capacity changes with age.}
\end{abstract}
\let\thefootnote\relax\footnotetext{Vanda In\'acio, School of Mathematics, University of Edinburgh, Scotland, UK (\textit{vanda.inacio@ed.ac.uk}). Vanda M. Louren\c co, Faculdade de Ci\^encias e Tecnologia, Universidade Nova de Lisboa, Portugal (\textit{vmml@fct.unl.pt}). Miguel de Carvalho, School of Mathematics, University of Edinburgh, Scotland, UK (\textit{Miguel.deCarvalho@ed.ac.uk}), Richard A. Parker, Edinburgh Clinical Trials Unit, Usher Institute, University of Edinburgh, Scotland, UK  (\textit{Richard.Parker@ed.ac.uk}), Vincent Gnanapragasam, Cambridge Urology Translational Research and Clinical Trials Office, Cambridge University Hospitals NHS Foundation Trust, and Academic Urology Group, Department of Surgery, University of Cambridge and Department of Urology, Cambridge University Hospitals Trust, UK (\textit{vjg29@cam.ac.uk}).}

\textsc{key words:} Additive model; Covariate-adjustment; Diagnostic test; Outliers; Receiver operating characteristic curve; M-estimation

\section{\large{\textsf{INTRODUCTION}}}
The evaluation of the performance of a medical test for screening and diagnosing disease is an important step towards advancing health in individuals and communities. The major goal of a diagnostic test is to distinguish diseased from nondiseased individuals or, more generally, to distinguish between different disease stages. Before the widespread use of a test, its ability to discriminate between the different states must be rigorously vetted. Note that here we use the term `diagnostic test' to broadly encompass any continuous classifier, which may include a single biological marker or a composite score resulting from the combination of multiple biomarkers. We further note that we will be assuming the existence of a so-called gold standard test, i.e., a perfect test that correctly classifies all individuals as being diseased or nondiseased. Compared to the diagnosis made by the gold standard test, the goal is to assess how well the candidate test, which is possibly less invasive and/or costly, performs. The receiver operating characteristic (ROC) curve is the most popular graphical tool used for evaluating the discriminatory ability of continuous-outcome tests. The ROC curve is a plot of the false positive fraction (probability that a nondiseased subject tests positive) against the true positive fraction (probability that a diseased subject tests positive) for all possible threshold values that can be used to convert continuous test outcomes into binary ones. Further background on ROC curves is provided in Section 2.

It has been recognised that the performance of a test may be affected by covariates, such as age and/or gender and, in such situations, ignoring covariate information might result in erroneous inferences about a test's accuracy. The full understanding of how covariates impact a test's performance is thus of paramount importance in order to determine the optimal and suboptimal populations, as defined by the covariate values, in which to perform the tests. The covariate-specific or conditional ROC curve, which is an ROC curve that conditions on a specific covariate value, arises as the natural tool to use in this context. For a recent overview of available ROC regression methods we refer to \cite{Inacio2020}.

Motivated by the fact that atypical/outlying test outcomes (due, for instance, to experimental, biological, or coding errors) may put at risk the reliability of the inferences about the test's accuracy \citep[e.g.,][]{Walach2017}, we develop a robust additive (based on regression splines) modelling framework for conducting inference about the covariate-specific ROC curve that mitigates the impact that outliers can have on inferences, while simultaneously allowing for nonlinear effects of the covariates. Here and below, by an outlier or atypical test outcome we mean an outcome that is clearly separated from the majority or bulk of the test outcomes, or that in some way deviates from the general patterns present in the test results \citep[][p.~124]{Racine2019}. Our estimation method for the covariate-specific ROC curve is similar in spirit to those developed by \cite{Pepe1998}, \cite{Gonzalez2011}, \cite{Rodriguez2011SC}, and \cite{Rodriguez2014}, which postulates a location-scale regression model for the test outcomes in both the diseased and nondiseased populations (termed in the literature as `induced' approach). Yet, unlike previous approaches: (i) our specification for the regression function relies on an additive cubic B-splines formulation, with M-estimation used for the regression coefficients, hence safeguarding against outlying test outcomes, and (ii) the distribution of the regression errors is modelled via a weighted empirical distribution function of the standardised residuals, therefore downweighting the influence of outliers when estimating the covariate-specific ROC curve and its associated summary indices. These features result in a widely applicable approach that can be used for many populations and for a large number of diseases and continuous diagnostic tests. In addition, from a computational perspective, our method is extremely fast and can be easily implemented in any software package. We acknowledge that the approaches of 
 \cite{Gonzalez2011}, \cite{Rodriguez2011SC}, and \cite{Rodriguez2014} also allow for nonlinear effects of the covariates on the mean (and also on the variance) function but, unlike our proposed approach, they do it through the use of kernel methods (the former two approaches) and Gaussian processes (the latter approach).
 
The remainder of this paper is organised as follows. In the next section, we introduce our modelling approach to conduct inference about the covariate-specific ROC curve. The performance of our method is validated in Section 3 using simulated data under different test results' contamination scenarios. In Section 4 our approach is applied to assess the age-specific accuracy of the Prostate Health Index as a biomarker for prostate cancer. Concluding remarks are offered in Section 5.

\section{\large{\textsf{ ROBUST AND FLEXIBLE INFERENCE FOR THE COVARIATE-SPECIFIC ROC CURVE}}}
\subsection{Preliminaries}
 We start with some background on ROC curves. Let $Y$ be the continuous random variable denoting the outcome of the diagnostic test and $D$ the binary variable indicating the presence $(D=1)$ or absence ($D=0$) of disease. Throughout, we use the subscripts $D$ and $\bar{D}$ to denote quantities conditional on $D=1$ and $D=0$, respectively. For example, $Y_{D}$ and $Y_{\bar{D}}$ denote the test outcomes in the diseased and nondiseased populations, with cumulative distribution functions given by $F_D$ and $F_{\bar{D}}$, respectively. Further, let $c$ be the threshold value used for defining a positive test result. Without loss of generality, we proceed with the assumption that larger values of $Y$ are more indicative of disease; that is, a subject is diagnosed as diseased when his/her test outcome is equal or greater than $c$, $Y\geq c$, and he or she is diagnosed as nondiseased when the outcome is below $c$, $Y<c$. Hence, for each possible threshold $c$, the true positive fraction (TPF) and false positive fraction (FPF) corresponding to such decision criterion are
\begin{align*}
\text{TPF}(c)&=\Pr(Y\geq c\mid D=1)=\Pr(Y_{D}\geq c) = 1- F_{D}(c),\\
\text{FPF}(c)&=\Pr(Y\geq c\mid D=0)=\Pr(Y_{\bar{D}}\geq c) = 1- F_{\bar{D}}(c).
\end{align*}
The ROC curve is defined as the set of points $\{(\text{FPF}(c), \text{TPF}(c)): c\in\mathbb{R}\}$ and, as it is clear from this definition, it lies in the unit square.  Letting $t=\text{FPF}(c)$, the ROC curve can be alternatively expressed as $\{(t,\text{ROC}(t)): t\in[0,1]\}$, with
\begin{equation*}
\text{ROC}(t)= 1-F_{D}\{F_{\bar{D}}^{-1}(1-t)\}.
\end{equation*}
ROC curves measure how separated the test outcomes in the diseased and nondiseased populations are (see Figure 1 of the Supplementary Materials). When the test outcomes in the two populations completely overlap, the ROC curve is the diagonal line of the unit square, that is, $\text{FPF}(c)=\text{TPF}(c)$ for all $c$, thus indicating a noninformative test. Conversely, the more separated the distributions of the test outcomes are, the closer the ROC curve is to the point $(0,1)$ in the unit square and the better the diagnostic accuracy. A curve that reaches the point $(0,1)$ has $\text{FPF}(c)=0$ and $\text{TPF}(c)=1$, for some threshold $c$ and, hence, corresponds to a test that perfectly determines the true disease status.

It is common to summarise the information of the ROC curve into a single summary index and, undeniably, the most popular one is the area under the ROC curve (AUC), given by
\begin{equation*}
\text{AUC}=\int_0^1\text{ROC}(t)\text{d}t.
\end{equation*}
For a useless test that classifies individuals as diseased or nondiseased no better than chance, $\text{AUC}=0.5$, whereas for a perfect test, $\text{AUC}=1$. In addition to its geometric definition, the AUC has also a probabilistic interpretation \citep[][p.~78]{Pepe2003},
\begin{equation*}
\text{AUC}=\Pr(Y_D\geq Y_{\bar{D}}),
\end{equation*}
that is, the AUC is the probability that the test outcome for a randomly chosen diseased subject exceeds the one exhibited by a randomly selected nondiseased individual.

\subsection{Modelling Framework for the Covariate-Specific ROC curve}
Let $\mathbf{X}$ denotes the covariate vector and, for ease of notation, we will be assuming that the covariate vectors $\mathbf{X}_{\bar{D}}$ and $\mathbf{X}_{D}$ are the same in both populations. However, this is not necessarily always the case as, for instance, disease stage, which is a diseased-specific covariate, might be of interest. The key object of our modelling framework is the covariate-specific ROC curve, which for a given covariate value $\mathbf{x}$, is defined as
\begin{equation*}
\text{ROC}(t\mid\mathbf{x})=1-F_{D}\{F_{\bar{D}}^{-1}(1-t\mid\mathbf{x})\mid \mathbf{x}\},\qquad 0\leq t \leq 1,\\
\end{equation*}
where $F_{D}(y\mid\mathbf{x})=\Pr(Y_D\leq y \mid \mathbf{X}_D=\mathbf{x})$ is the conditional cumulative distribution function in the diseased population, with $F_{\bar{D}}(y\mid\mathbf{x})$ being analogously defined. The covariate-specific counterpart of the AUC is given by
\begin{equation}\label{cauc}
\text{AUC}(\mathbf{x})=\int_{0}^{1}\text{ROC}(t\mid\mathbf{x})\text{d}t.
\end{equation}
Note that in this setting, for each possible value $\mathbf{x}$, we might obtain a different ROC curve/AUC and, therefore, also a possible different accuracy.

We follow an induced approach and we further assume that the relationship between covariates and test outcomes in each population is given by a location-scale regression model, i.e.,
\begin{equation}\label{locscale}
Y_D=\mu_{D}(\mathbf{x})+\sigma_D\varepsilon_D,\qquad Y_{\bar{D}}=\mu_{\bar{D}}(\mathbf{x})+\sigma_{\bar{D}}\varepsilon_{\bar{D}},
\end{equation}
where $\mu_{D}(\mathbf{x})=E(Y_D\mid \mathbf{X}_D=\mathbf{x})$ and $\sigma_D$ are the conditional mean function and scale parameter, respectively, in the diseased population, with $\mu_{\bar{D}}(\mathbf{x})$ and $\sigma_{\bar{D}}$ similarly defined. The errors $\varepsilon_D$ and $\varepsilon_{\bar{D}}$ are independent of each other and independent of the covariates $\mathbf{X}_D$ and $\mathbf{X}_{\bar{D}}$, with mean zero, unit variance, and cumulative distribution functions given by $F_{\varepsilon_{D}}$ and $F_{\varepsilon_{\bar{D}}}$, respectively. The independence between the error and the covariates in the location-scale regression model, allows one to rewrite the conditional cumulative distribution function of the test outcomes in terms of the cumulative distribution function of the regression errors
\begin{equation}\label{cdfls}
F_{D}(y\mid\mathbf{x})=F_{\varepsilon_{D}}\left(\frac{y-\mu_{D}(\mathbf{x})}{\sigma_D}\right),\qquad F_{\bar{D}}(y\mid\mathbf{x})=F_{\varepsilon_{\bar{D}}}\left(\frac{y-\mu_{\bar{D}}(\mathbf{x})}{\sigma_{\bar{D}}}\right).
\end{equation}
An analogous relationship holds between the conditional quantile function and the quantile function of the error terms, namely
\begin{equation*}
F_{D}^{-1}(t\mid\mathbf{x})=\mu_{D}(\mathbf{x})+\sigma_{D}F_{\varepsilon_{D}}^{-1}(t),\qquad F_{\bar{D}}^{-1}(t\mid\mathbf{x})=\mu_{\bar{D}}(\mathbf{x})+\sigma_{\bar{D}}F_{\varepsilon_{\bar{D}}}^{-1}(t).
\end{equation*}
The covariate-specific ROC curve can therefore be expressed as
\begin{equation*}
\text{ROC}(t\mid\mathbf{x})=1-F_{\varepsilon_{D}}\left\{\frac{\mu_{\bar{D}}(\mathbf{x})-\mu_{D}(\mathbf{x})}{\sigma_D}+\frac{\sigma_{\bar{D}}}{\sigma_D}F_{\varepsilon_{\bar{D}}}^{-1}(1-t)\right\}, \qquad 0\leq t \leq 1. 
\end{equation*}
An advantage of this formulation is that the distribution and quantile functions of the regression errors are not conditional, thus alleviating the computational burden. Note that under this approach the effect of covariates on the ROC curve is expressed in terms of their effects on the mean functions of each population. 

\subsection{Proposed Robust and Flexible Estimator and its Implementation}
Let $\{(\mathbf{x}_{\bar{D}i},y_{\bar{D}i})\}_{i=1}^{n_{\bar{D}}}$ and $\{(\mathbf{x}_{Dj},y_{Dj})\}_{j=1}^{n_D}$ be two independent random samples of covariates and test outcomes from the nondiseased and diseased populations of size $n_{\bar{D}}$ and $n_D$, respectively. Further, for all $i=1,\ldots,n_{\bar{D}}$ and $j=1,\ldots,n_D$, let $\mathbf{x}_{\bar{D}i}=(x_{\bar{D}i,1},\ldots,x_{\bar{D}i,p})^{\prime}$ and $\mathbf{x}_{Dj}=(x_{Dj,1},\ldots,x_{Dj,p})^{\prime}$  be $p$-dimensional vectors of covariates. 

\subsubsection{Modelling the Mean Function}
From the location-scale models in \eqref{locscale}, what needs to be specified is the regression function in each population. We will describe our modelling approach for the diseased population, but everything follows similarly for the nondiseased population. Since nonlinear relationships between test outcomes and continuous covariates often occur, we assume a flexible additive formulation for the mean function, namely
\begin{equation*}
\mu_{D}(\mathbf{x}_{Dj}) = \beta_{D0} + f_{D1}(x_{Dj,1})+\cdots+ f_{Dp}(x_{Dj,p}), \qquad j=1,\ldots,n_D,
\end{equation*}
where $f_{Dh}(\cdot)$, $h=1,\ldots,p$, are smooth functions, each approximated by a linear combination of cubic B-splines basis functions defined over a sequence of knots $\xi_{Dh0}<\xi_{Dh1}<\cdots<\xi_{DhK_{Dh}}<\xi_{Dh,K_{Dh}+1}$. The knots $\xi_{Dh0}$ and $\xi_{Dh,K_{Dh}+1}$ are boundary knots, while the remaining ones are interior knots. We then write
\begin{align*}
f_{Dh}(x_{Dj,h})=\sum_{k=1}^{K_{Dh}+3}B_{hk}(x_{Dj,h})\beta_{Dhk}=\mathbf{B}_{\xi_{Dh}}^{\prime}(x_{Dj,h})\boldsymbol{\beta}_{Dh},\quad j=1,\ldots,n_D,\quad h = 1,\ldots,p,
\end{align*}
where $B_{k}(x)$ denotes the $k$th cubic B-spline basis function \citep[][Chapter 9]{deboor1978} evaluated at $x$, $\mathbf{B}_{\xi_{Dh}}(x_{Dj,h})=(B_{h1}(x_{Dj,h}),\ldots, B_{h,K_{Dh}+3}(x_{Dj,h}))^{\prime}$, $\boldsymbol{\beta}_{Dh}=(\beta_{Dh1},\ldots,\beta_{Dh,K_{Dh}+3})^{\prime}$. The mean function is thus expressed as
\begin{align}\label{meanf}
\mu_{D}(\mathbf{x}_{Dj})&=\beta_{D0} + \mathbf{B}^{\prime}_{\mathbf{\xi}_{D1}}(x_{Dj,1})\boldsymbol{\beta}_{D1}+\cdots + \mathbf{B}^{\prime}_{\mathbf{\xi}_{Dp}}(x_{Dj,p})\boldsymbol{\beta}_{Dp}\nonumber\\
&=\mathbf{z}_{Dj}^{\prime}\boldsymbol{\beta}_D,
\end{align}
where $\mathbf{z}_{Dj}=(1,\mathbf{B}^{\prime}_{\mathbf{\xi}_{D1}}(x_{Dj,1}),\ldots, \mathbf{B}^{\prime}_{\mathbf{\xi}_{Dp}}(x_{Dj,p}))^{\prime}$ and $\boldsymbol{\beta}_{D}=(\beta_{D0},\boldsymbol{\beta}_{D1},\ldots,\boldsymbol{\beta}_{Dp})^{\prime}$.
It is well-known that both the number and location of knots characterising the B-splines basis functions are key choices that have the potential to impact the inferences, more so the former than the latter. As noted in \cite{Durrleman1989}, usually, only a few number of knots, say a maximum of three or four, are needed to adequately describe most of the phenomena likely to be observed in medical statistics. In this paper, the selection of the number of knots is assisted by a robust version of the Akaike information criterion (see Section 2.4). Regarding the location of the $K_{Dh}$ interior knots, we follow \cite{Rosenberg1995} and $\xi_{Dhk}$ is set equal to the $k/(K_{Dh}+1)$ quantile of $\mathbf{x}_{D,h}=(x_{D1,h},\ldots,x_{Dn_D,h})$, for $k=1,\ldots,K_{Dh}$ and $h=1,\ldots,p$, thus assuring an approximate equal number of observations at each interval defined by the knots. The boundary knots $\xi_{Dh0}$ and $\xi_{Dh,K_{Dh}+1}$ are set equal to the minimum and maximum of $\mathbf{x}_{D,h}$, respectively. For the ease of presentation, we have assumed that all $p$ covariates are continuous, but we can also easily deal with categorical covariates, as well as, interactions between categorical covariates and interactions between a (smooth) continuous covariate and a categorical one.

\subsubsection{Robust Estimation}
The representation in \eqref{meanf} reduces the estimation of $\mu_{D}(\mathbf{x}_{Dj})$ to the estimation of the coefficients $\boldsymbol{\beta}_D$. Moreover, this expression is linear in the coefficient vector  $\boldsymbol{\beta}_D$ and therefore the estimation of $\mu_{D}(\mathbf{x}_{Dj})$ can be viewed as an optimisation problem that is linear in the transformed variables $\mathbf{z}_{Dj}$, therefore allowing the use of well-established estimation techniques for multiple regression models. Estimation by ordinary least squares would be the most natural option. However, least squares type of approaches, because they rely on (minimising) a quadratic loss function, are extremely sensitive to outliers. Even a single atypical test outcome can drastically affect the estimated regression coefficients. Moreover, the scale parameter $\sigma_D$ is traditionally estimated by the square root of $\widehat{\sigma}^{2}_D=(n_D-Q_D)^{-1}\sum_{j=1}^{n_D}(y_{Dj}-\mathbf{z}_{Dj}^{\prime}\widehat{\boldsymbol{\beta}}_D^{\text{OLS}})^2$, which is not robust either. Note that here $Q_D$ is the dimension of the vector $\mathbf{z}_{Dj}$ and $\widehat{\boldsymbol{\beta}}_D^{\text{OLS}}$ is the least squares estimate of $\boldsymbol{\beta}_D$. It could be tempting to remove the outlying test outcomes using, for instance, graphical or residual analysis, and then obtaining the least squares estimates of the regression coefficients based on the `clean' sample. However, this strategy, might be not only impractical, but might also lead to inferences that are neither valid nor robust \citep{Welsh2002}, not to mention the reduction in sample size.
 One way to circumvent this problem is to minimise a less rapidly increasing function than the squared one, so that the influence of test outcomes with large residuals is reduced. For instance, least absolute deviation regression, which minimises the absolute value loss function, $\sum_{j=1}^{n_D}\mathopen|y_{Dj}-\mathbf{z}_{Dj}^{\prime}\boldsymbol{\beta}_D\mathclose|$, leads to estimators that are highly resistant to outliers (in the response variable). However, the drawback is that such estimators are relatively inefficient. An elegant compromise between the squared and absolute value loss functions was proposed by \cite{Huber1964}, who suggested to estimate $\boldsymbol{\beta}_D$ as
 \begin{equation}\label{rhohuber}
 \widehat{\boldsymbol{\beta}}_D=\underset{\boldsymbol{\beta}_D}{\arg\min}\sum_{j=1}^{n_D}\rho\left(\frac{y_{Dj}-\mathbf{z}_{Dj}^{\prime}\boldsymbol{\beta}_D}{\widehat{\sigma}_D}\right),\qquad \rho(u)= \begin{cases} \frac{u^2}{2},\quad  \mathopen|u\mathclose|\leq b, \\  b\mathopen|u\mathclose|  -
\frac{b^2}{2},\quad  \mathopen|u\mathclose| > b,\end{cases}
 \end{equation}
where $b$ is a tuning constant and $\widehat{\sigma}_D$ is a robust estimate of scale. Huber's loss function is quadratic for small values of the standardised residuals but grows linearly for large absolute values of the standardised residuals. The tuning constant $b$ describes where the transition from a quadratic to a linear loss function takes place, thus allowing for different compromises between robustness and efficiency, and acting like a threshold such that observations with standardised residuals larger, in absolute value, than $b$ have a reduced effect in the estimation. For larger values of $b$, Huber's loss function becomes more similar to the least squares loss function, whereas for small values of $b$, it is more similar to the absolute value loss function. The typical choice of $b$ is $1.345$, for which \cite{Huber1964} showed that the resulting estimator is, asymptotically, $95\%$ as efficient as the least squares estimator when the true distribution of the errors is normal. Although we do not make any distributional assumption about $\varepsilon_D$, we shall use $b=1.345$. In \eqref{rhohuber}, the robust estimate of the scale $\widehat{\sigma}_D$, needed to ensure that the resulting estimate of $\boldsymbol{\beta}_D$ is scale equivariant, is in our case the re-scaled median absolute deviation of the residuals
\begin{equation}\label{mad}
\widehat{\sigma}_D=1.4826\,\underset{j=1,\ldots,n_D}{\text{median}}\mathopen|y_{Dj}-\mathbf{z}_{Dj}^{\prime}\widehat{\boldsymbol{\beta}}_D\mathclose|,
\end{equation}
with the constant $1.4826$ based on a normality assumption. Huber's estimator falls under the general category of M-estimators \citep[e.g.][Chapters 2--5]{Maronna2019}. The M-estimator minimises \eqref{rhohuber} or, equivalently, solves the system of estimating equations
\begin{equation}\label{psihuber}
\sum_{j=1}^{n_D}\psi\left(\frac{y_{Dj}-\mathbf{z}_{Dj}^{\prime}\boldsymbol{\beta}_D}{\widehat{\sigma}_D}\right)\mathbf{z}_{Dj}=\mathbf{0}_{Q_D},\qquad \psi(u)=\frac{\text{d}}{\text{d} u}\rho(u)=\begin{cases}  u,\quad  \mathopen|u\mathclose|\leq b, \\  b\,\text{sign}(u),\quad  \mathopen|u\mathclose| > b,\ \end{cases}
\end{equation}
where $\text{sign}(u)=I(u>0)-I(u<0)$ and with $\text{sign}(0)=0$ and $\mathbf{0}_{Q_D}$ denotes a vector of zeros of length $Q_D$. Defining the weight function $\omega(u)$ by
\begin{equation*}
\omega(u)=\frac{\psi(u)}{u}=\begin{cases}  1,\quad  \mathopen|u\mathclose|\leq b, \\  \frac{b}{\mathopen|u\mathclose| },\quad  \mathopen|u\mathclose| > b,\ \end{cases}
\end{equation*}
allows us to rewrite Equation \eqref{psihuber} as
\begin{equation}\label{huberweights}
\sum_{j=1}^{n_D}\omega_{Dj}\left(y_{Dj}-\mathbf{z}_{Dj}^{\prime}\boldsymbol{\beta}_D\right)\mathbf{z}_{Dj}=\mathbf{0}_{Q_D},\qquad \omega_{Dj}=\omega\left(\frac{y_{Dj}-\mathbf{z}_{Dj}^{\prime}\boldsymbol{\beta}_D}{\widehat{\sigma}_D}\right).
\end{equation}
In Figure 2 of the Supplementary Materials we present a comparison between Huber's $\rho$, $\psi$, and $\omega$ functions and the corresponding least squares and least absolute deviation counterparts for a better understanding of their behaviour. Note that, for instance, least squares assigns equal weight to all observations, whereas Huber's based weight function assigns decreasing weights for observations with large, in absolute value, standardised residuals. The system of equations in \eqref{huberweights} can be written in matrix form as
\begin{equation*}
\mathbf{Z}_D^{\prime}\boldsymbol{\Omega}_D\mathbf{Z}_D\boldsymbol{\beta}_D=\mathbf{Z}_D^{\prime}\boldsymbol{\Omega}_D\mathbf{y}_D,
\end{equation*}
 where $\mathbf{Z}_D$ is a matrix with $\mathbf{z}_{Dj}^{\prime}$ as its $j$th row, $\boldsymbol{\Omega}_D$ is a diagonal matrix with entries given by $\omega_{Dj}$, for $j=1,\ldots,n_D$, and $\mathbf{y}_D=(y_{D1},\ldots,y_{Dn_D})^{\prime}$, and therefore can be regarded as a weighted least squares problem whose solution is given by $\widehat{\boldsymbol{\beta}}_D=(\mathbf{Z}_D^{\prime}\boldsymbol{\Omega}_D\mathbf{Z}_D)^{-1}\mathbf{Z}_D^{\prime}\boldsymbol{\Omega}_D\mathbf{y}_D$. Because the weights depend upon the estimated regression coefficients and scale parameters and, in turn, these depend upon the weights, the iteratively reweighted least squares procedure is employed. The algorithm can be briefly summarised by the following two steps.
\begin{enumerate}[start=1,label={\upshape\bfseries Step~\arabic*:},wide = 0pt, leftmargin = 3em]
\item Obtain an initial estimate $\widehat{\boldsymbol{\beta}}_{D}^{(0)}$, which can be based, for instance, on a least squares fit. Use $\widehat{\boldsymbol{\beta}}_{D}^{(0)}$ to obtain $\widehat{\sigma}_D^{(0)}$ using the re-scaled median absolute deviation of the residuals as in \eqref{mad}. Compute an initial estimate of $\boldsymbol{\Omega}^{(0)}$ using $\widehat{\boldsymbol{\beta}}_{D}^{(0)}$ and $\widehat{\sigma}_D^{(0)}$.
\item At iteration $k=1,2,\ldots$, solve for the new weighted least squares estimate \linebreak  $\widehat{\boldsymbol{\beta}}_{D}^{(k)}=(\mathbf{Z}_{D}^{\prime}\boldsymbol{\Omega}_D^{(k-1)}\mathbf{Z}_{D})^{-1}\mathbf{Z}_{D}^{\prime}\boldsymbol{\Omega}_D^{(k-1)}\mathbf{y}_D$. This estimate will be used to obtain $\widehat{\sigma}_D^{(k)}$ and to compute $\boldsymbol{\Omega}_{D}^{(k)}$ which, in turn, will form the basis of $\widehat{\boldsymbol{\beta}}_{D}^{(k+1)}$. The iterative procedure is run until some convergence criterion is met.
\end{enumerate}
The converged estimate $\widehat{\boldsymbol{\beta}}_{D}$ is taken as our final robust estimate of $\boldsymbol{\beta}_{D}$ and used to obtain the final estimate $\widehat{\sigma}_D$ of $\sigma_D$. We note here that $\widehat{\boldsymbol{\beta}}_{D}$ based on Huber's loss function is not robust against outliers in the covariates.

Once estimates $\widehat{\boldsymbol{\beta}}_D$ and $\widehat{\sigma}_D$ have been obtained, the distribution function of the error $\varepsilon_D$ is estimated on the basis of a weighted empirical distribution function of the standardised residuals,
 \begin{align}\label{wecdf}
 \widehat{F}_{\varepsilon_{D}}(y)=\frac{1}{\sum_{l=1}^{n_D}\omega_{Dl}^{*}}\sum_{j=1}^{n_D}\omega_{Dj}^{*}I\left(\widehat{\varepsilon}_{Dj}\leq y\right),\quad \widehat{\varepsilon}_{Dj}=\frac{y_{Dj}-\widehat{\mu}_{D}(\mathbf{x}_{Dj})}{\widehat{\sigma}_D},\quad \widehat{\mu}_{D}(\mathbf{x}_{Dj})=\mathbf{z}^{T}_{Dj}\widehat{\boldsymbol{\beta}}_D,\quad \omega_{Dj}^{*} = \begin{cases}
 1,\quad  \mathopen|\widehat{\varepsilon}_{Dj}\mathclose|\leq v, \\ \omega_{Dj}, \quad  \mathopen|\widehat{\varepsilon}_{Dj}\mathclose|> v,
 \end{cases}
 \end{align}
 where $v$ is a tuning constant and, using the normal distribution as a benchmark, values between 2 and 3 are deemed as reasonable. We set $v=3$ for the results reported in this paper. The purpose of using a weighted version of the empirical distribution function is to downweight the influence of outliers on its estimation and, consequently, on the estimation of the covariate-specific ROC curve and associated AUC. The empirical distribution function is recovered when $\omega_{Dj}^{*}=1$, for all $j=1,\ldots,n_D$. 
 
 Finally, the ROC curve estimate can be written as 
 \begin{equation}\label{rocls}
 \widehat{\text{ROC}}(t\mid\mathbf{x})=1-\widehat{F}_{\varepsilon_{D}}\left\{\frac{\widehat{\mu}_{\bar{D}}(\mathbf{x})-\widehat{\mu}_{D}(\mathbf{x})}{\widehat{\sigma}_D}+\frac{\widehat{\sigma}_{\bar{D}}}{\widehat{\sigma}_D}\widehat{F}_{\varepsilon_{\bar{D}}}^{-1}(1-t)\right\},
 \end{equation}
and the corresponding AUC admits the following closed-form expression, derived in the Appendix, and which can be regarded as a weighted robust covariate-specific Mann--Whitney type of statistic
 \begin{equation}\label{rmws}
\widehat{\text{AUC}}(\mathbf{x})=\frac{1}{\sum_{l=1}^{n_D}\omega_{Dl}^{*}\sum_{l=1}^{n_{\bar{D}}}\omega_{\bar{D}l}^{*}}\sum_{j=1}^{n_D}\sum_{i=1}^{n_{\bar{D}}}\omega_{Dj}^{*}\omega_{\bar{D}i}^{*}I\{\widehat{\mu}_{\bar{D}}(\mathbf{x})+\widehat{\sigma}_{\bar{D}}\widehat{\varepsilon}_{\bar{D}i}\leq \widehat{\mu}_{D}(\mathbf{x})+\widehat{\sigma}_{D}\widehat{\varepsilon}_{Dj}\}.
 \end{equation}
 
 \subsubsection{Implementation}
 Some final comments on implementation are in order. Our procedure is easily implemented in \texttt{R} \citep{R20} using the \texttt{bs} function from the package \texttt{splines} (to create the cubic B-splines basis expansions) in combination with the \texttt{rlm} routine from the \texttt{MASS} package \citep{MASS_package}, which performs the robust estimation procedure described above to obtain $\widehat{\boldsymbol{\beta}}_D$ and $\widehat{\sigma}_D$. Of course, M-estimation for generalised additive models based on a smoothing parameter/penalty (as, e.g., in \citealt{Wong2014}) would be an alternative route, but these tend to involve intricate and computationally expensive algorithms. Under our approach, regularisation is achieved through selecting the number of interior knots, which we do with the aid of a robust Akaike information criterion, as we explain in the next section. This results in a simple and fast, yet effective, estimation procedure. The \texttt{R} code implementing our approach is publicly available at (github link upon acceptance). 
 
 \subsection{Robust Akaike Information Criterion}
The issue of selecting the number of interior knots for each smooth function of a continuous covariate can be regarded as a model selection problem. Here, and because the classical Akaike information criterion (AIC) is sensitive to outlying observations, such choice is assisted through the use of a robust version of the AIC, denoted by rAIC, that is suited for M-estimation and which was proposed by \cite{Tharmaratnam2013}. Specifically, the authors suggest to use
\begin{equation}\label{rAIC}
\text{rAIC}_{D}=2\,n_D\log\widehat{\sigma}_{D}+4\,\text{trace}(J^{-1}_{D,n_{D}}U_{D,n_{D}}),
\end{equation}
where the empirical information matrices in the trace term (the penalty term) are calculated as follows
\begin{equation*}
J_{D,n_D}=\frac{1}{n_D}\sum_{j=1}^{n_D}\psi^{\prime}\left(\frac{y_{Dj}-\mathbf{z}_{Dj}^{\prime}\widehat{\boldsymbol{\beta}}_D}{\widehat{\sigma}_{D}}\right)\frac{\mathbf{z}_{Dj}\mathbf{z}_{Dj}^{\prime}}{\widehat{\sigma}_{D}^2},\qquad
U_{D,n_D}=\frac{1}{n_D}\sum_{j=1}^{n_D}\psi^{2}\left(\frac{y_{Dj}-\mathbf{z}_{Dj}^{\prime}\widehat{\boldsymbol{\beta}}_D}{\widehat{\sigma}_{D}}\right)\frac{\mathbf{z}_{Dj}\mathbf{z}_{Dj}^{\prime}}{\widehat{\sigma}_{D}^2}.
\end{equation*}
Models with a varying number of interior knots will be fitted, $\boldsymbol{\beta}_D$ and $\sigma_D$ are re-estimated in each model and the corresponding rAIC is computed, and the model with the smallest rAIC will be selected. When several continuous covariates are involved, our strategy involves exploring the set of all possible models. This is viable because not only our fitting procedure is extremely fast, but also because in medical diagnostic studies the number of continuous covariates available is often reduced and, as mentioned before, usually a modest number of knots suffices to describe the relationship between covariates and test outcomes. On a related task, the rAIC can also be used to select between a linear or smooth effect of a given (continuous) covariate. It is important to remark that the penalty term needs to be changed to $2\,\text{trace}(J^{-1}_{D,n_{D}}U_{D,n_{D}})$ if instead of using the $\psi$ function in \eqref{psihuber}, one uses $2\,\psi(u)$ (as, e.g., in \citealt{Tharmaratnam2013}).

\subsection{Robust Bootstrap-based Inference for the Robust  and Flexible Covariate-Specific ROC Curve}
Confidence intervals for the covariate-specific ROC curve and corresponding AUC can be obtained through the bootstrap. Some bootstrap samples may have a proportion of outliers much higher than in the original one, thus placing at risk the contamination level tolerated by Huber's M-estimator and consequently severely affecting the recomputed quantities (regression coefficients, standard deviations, etc). Hence,
we use a bootstrap of the residuals to resample the (robust) regression model in each population. Specifically, each standardised residual $\widehat{\varepsilon}_{Dj}$ ($\widehat{\varepsilon}_{\bar{D}i}$) is sampled with probability proportional to $\omega_{Dj}^{*}$ ($\omega_{\bar{D}i}^{*}$), as defined in \eqref{wecdf}, for $j=1,\ldots, n_D$ ($i=1,\ldots,n_{\bar{D}}$). 
The details of our bootstrap scheme, for $b=1,\ldots,B$, are as follows:
\begin{enumerate}[start=1,label={\upshape\bfseries Step~\arabic*:},wide = 0pt, leftmargin = 3em]
\item Sample with replacement from the estimated standardised residuals $\{\widehat{\varepsilon}_{\bar{D}i}\}_{i=1}^{n_{\bar{D}}}$ and $\{\widehat{\varepsilon}_{Dj}\}_{j=1}^{n_D}$, with probabilities $\{\omega_{\bar{D}i}^{*}/\sum_{l=1}^{n_{\bar{D}}}\omega_{\bar{D}l}^{*}\}_{i=1}^{n_{\bar{D}}}$ and  $\{\omega_{Dj}^{*}/\sum_{l=1}^{n_D}\omega_{Dl}^{*}\}_{j=1}^{n_D}$, 
to form bootstrap sets $\{\widehat{\varepsilon}_{\bar{D}i}^{(b)}\}_{i=1}^{n_{\bar{D}}}$ and $\{\widehat{\varepsilon}_{Dj}^{(b)}\}_{j=1}^{n_D}$.
\item Use the mean function and variance estimates from the observed data to construct bootstrap samples $\{(\mathbf{x}_{\bar{D}i},y_{\bar{D}i}^{(b)})\}_{i=1}^{n_{\bar{D}}}$ and $\{(\mathbf{x}_{Dj},y_{Dj}^{(b)})\}_{j=1}^{n_{D}}$, where
\begin{equation*}
y_{\bar{D}i}^{(b)}=\widehat{\mu}(\mathbf{x}_{\bar{D}i})+\widehat{\sigma}_{\bar{D}}\widehat{\varepsilon}_{\bar{D}i}^{(b)},\qquad
y_{Dj}^{(b)}=\widehat{\mu}(\mathbf{x}_{Dj})+\widehat{\sigma}_{D}\widehat{\varepsilon}_{Dj}^{(b)}.
\end{equation*}
\item Repeat the estimation process with the $b$th bootstrap sample, thus obtaining $\widehat{\text{ROC}}^{(b)}(p\mid\mathbf{x})$ and $\widehat{\text{AUC}}^{(b)}(\mathbf{x})$.
\end{enumerate}
Once this process has been completed, and according to the percentile method, a bootstrap confidence interval for, e.g.,  $\text{AUC}(\mathbf{x})$, of confidence level $1-\alpha$ is given by
\begin{equation*}
\left(\widehat{\text{AUC}}^{\alpha/2}(\mathbf{x}),\widehat{\text{AUC}}^{1-\alpha/2}(\mathbf{x})\right),
\end{equation*}
where $\widehat{\text{AUC}}^{\tau}(\mathbf{x})$ represents the $\tau$th percentile of the ensemble of estimates $\{\widehat{\text{AUC}}^{(b)}(\mathbf{x})\}_{b=1}^{B}$. 

\section{\large{\textsf{SIMULATION STUDY}}}
To evaluate the empirical performance of our robust and flexible approach for conducting inference about the covariate-specific ROC curve and corresponding AUC, we analysed simulated data under four different scenarios (described in the next section). For each scenario, $1000$ data sets were generated using sample sizes of $( n_{\bar{D}}, n_{D})=(100, 100)$,  $(n_{\bar{D}}, n_{D})=(200, 100)$, and $(n_{\bar{D}}, n_{D})=(200, 200)$. The following percentages of test outcomes contamination, in each population, were considered: $2\%$, $5\%$, and $10\%$. The case of no contamination (original simulated datasets) was also considered in order to ascertain the performance of our method when a robust approach is not needed at all. 

\subsection{Simulation Scenarios}
In Scenario I, we consider different homoscedastic linear mean regression models for the
nondiseased and diseased populations, namely
\begin{equation*}
y_{\bar{D}i}\mid x_{\bar{D}i,1}\overset{\text{ind.}}\sim\text{N}\left(0.5+x_{\bar{D}i,1},1.5^2\right),\quad y_{Dj}\mid x_{Dj,1}\overset{\text{ind.}}\sim\text{N}\left(2+4x_{Dj,1}, 2^2\right),\quad i=1,\ldots,n_{\bar{D}},\quad j=1,\ldots,n_D.
\end{equation*}
The primary purpose of including this scenario is to allow us assessing the impact of using a cubic B-splines basis formulation for the mean function of each population when the underlying true effect is, in fact, linear. Data for Scenario II are governed by the following nonlinear mean regression models
\begin{equation*}
y_{\bar{D}i}\mid x_{\bar{D}i,1}\overset{\text{ind.}}\sim\text{N}\left(\sin\{\pi x_{\bar{D}i,1}\}, 0.5^2\right),\qquad y_{Dj}\mid x_{Dj,1}\overset{\text{ind.}}\sim\text{N}\left(1+x_{Dj,1}^2, 1^2\right).
\end{equation*}
Scenario III involves heteroscedastic nonlinear mean regression models for the diseased and nondiseased populations
\begin{equation*}
y_{\bar{D}i}\mid x_{\bar{D}i,1}\overset{\text{ind.}}\sim\text{N}\left(0.5\sin\{2\pi x_{\bar{D}i,1}\}, (1 + 0.75x_{\bar{D}i,1})^2\right),\quad y_{Dj}\mid x_{Dj,1}\overset{\text{ind.}}\sim\text{N}\left(0.5+\sin\{\pi x_{Dj,1}\},(1 + x_{Djx,1})^2\right).
\end{equation*}
Note that our model is actually misspecified in this case as it does not allow the variance to change with the covariates and the goal of including this scenario is exactly to assess the performance of our approach when the assumption of constant variance does not hold. Finally, in Scenario IV we have considered the case where two continuous covariates affect the test outcomes
\begin{equation*}
y_{\bar{D}i}\mid x_{\bar{D}i,1},x_{\bar{D}i,2}\overset{\text{ind.}}\sim\text{N}\left(0.5 + x_{\bar{D}i,1} + x_{\bar{D}i,2}^2, 1.5^2\right),\qquad y_{Dj}\mid x_{Dj,1},x_{Dj,2}\overset{\text{ind.}}\sim\text{N}\left(2 + 4 x_{Dj,1}^3 + 1.5 x_{Dj,2}, 2^2\right).
\end{equation*}
In all cases, the continuous covariates $x_1$ and $x_2$, are independently generated from uniform distributions, namely
\begin{equation*}
x_{\bar{D}i,1}\overset{\text{i.i.d.}}\sim\text{U}(0,1),\quad x_{\bar{D}i,2}\overset{\text{i.i.d.}}\sim\text{U}(0,2),\quad x_{Dj,1}\overset{\text{i.i.d.}}\sim\text{U}(0,1),\quad x_{Dj,2}\overset{\text{i.i.d.}}\sim\text{U}(0,2).
\end{equation*}
Further, in all scenarios, the contaminated data were generated by randomly selecting a given percentage of test outcomes and replacing them by $\text{N}\{\mu_{\bar{D}}(\mathbf{x}_{\bar{D}})+\kappa_{\bar{D}}\sigma_{\bar{D}}(\mathbf{x}_{\bar{D}}),\sigma_{\bar{D}}^{2}(\mathbf{x}_{\bar{D}})\}$ and $\text{N}\{\mu_{D}(\mathbf{x}_{D})+\kappa_D\sigma_{D}(\mathbf{x}_{D}),\sigma_{D}^{2}(\mathbf{x}_{D})\}$ (shift in the location outliers) in the nondiseased and diseased populations, respectively. Note that for all scenarios but the third we have $\sigma_{\bar{D}}(\mathbf{x}_{\bar{D}})\equiv \sigma_{\bar{D}}$ and $\sigma_{D}(\mathbf{x}_{D})\equiv \sigma_{D}$. Additionally, we have considered $ \kappa_{\bar{D}}=15$ and $\kappa_D=20$, which at a first glance might seems excessive but it is indeed in line with what we observe in our data application in Section 4 (see also the left panel of Figure \ref{exploratory}). The impact of the magnitude of those values on the estimates will be discussed in the Results section.

\subsection{Models}
For each simulated dataset we fit our robust and flexible approach considering no interior knots for each continuous covariate in each population (i.e., $K_{\bar{D}1}=K_{\bar{D}2}=K_{D1}=K_{D2}=0$). A further inspection to this choice is discussed in the next section. Our model is compared to the semiparametric approach of \cite{Pepe1998}, which is based on a location-scale regression model for the test outcomes in each population  that relies on a linear formulation for the mean function and with the regression coefficients estimated, for instance, by least squares. In addition to the original approach proposed by 
\cite{Pepe1998}, we have also considered an extension of this method by using a cubic B-splines trend, also with no interior knots, so that direct comparisons to our approach are easier and fairer. The only difference between ours and this approach is the objective function (least squares versus Huber's $\rho$ function). In addition, our method is also compared to the nonparametric approach of \cite{Rodriguez2011SC}, which relies on kernel-based estimators for the mean and variance functions of the location-scale model. The main difference to the method of \cite{Gonzalez2011} is the order of the local polynomial smoothers used for estimating the regression function; while \cite{Gonzalez2011} employed a local constant fit (order $0$), \cite{Rodriguez2011SC} considered a linear fit (order $1$).  Because local constant regression suffers from boundary--bias problems, we only considered the latter approach. All competing methods were implemented using the \texttt{ROCnReg} package \citep{Rodriguez2020} which, in turn, relies on the \texttt{np} package \citep{Hayfield2008} for kernel estimation. Still on the kernel method, it is important to remark that the bandwidth parameters involved in the estimation process were selected using least-squares cross-validation and that this approach, as it stands now, can only deal with one continuous covariate.

\subsection{Results}
\begin{figure}[H]
\begin{center}
\subfigure{
\includegraphics[width = 4.65cm]{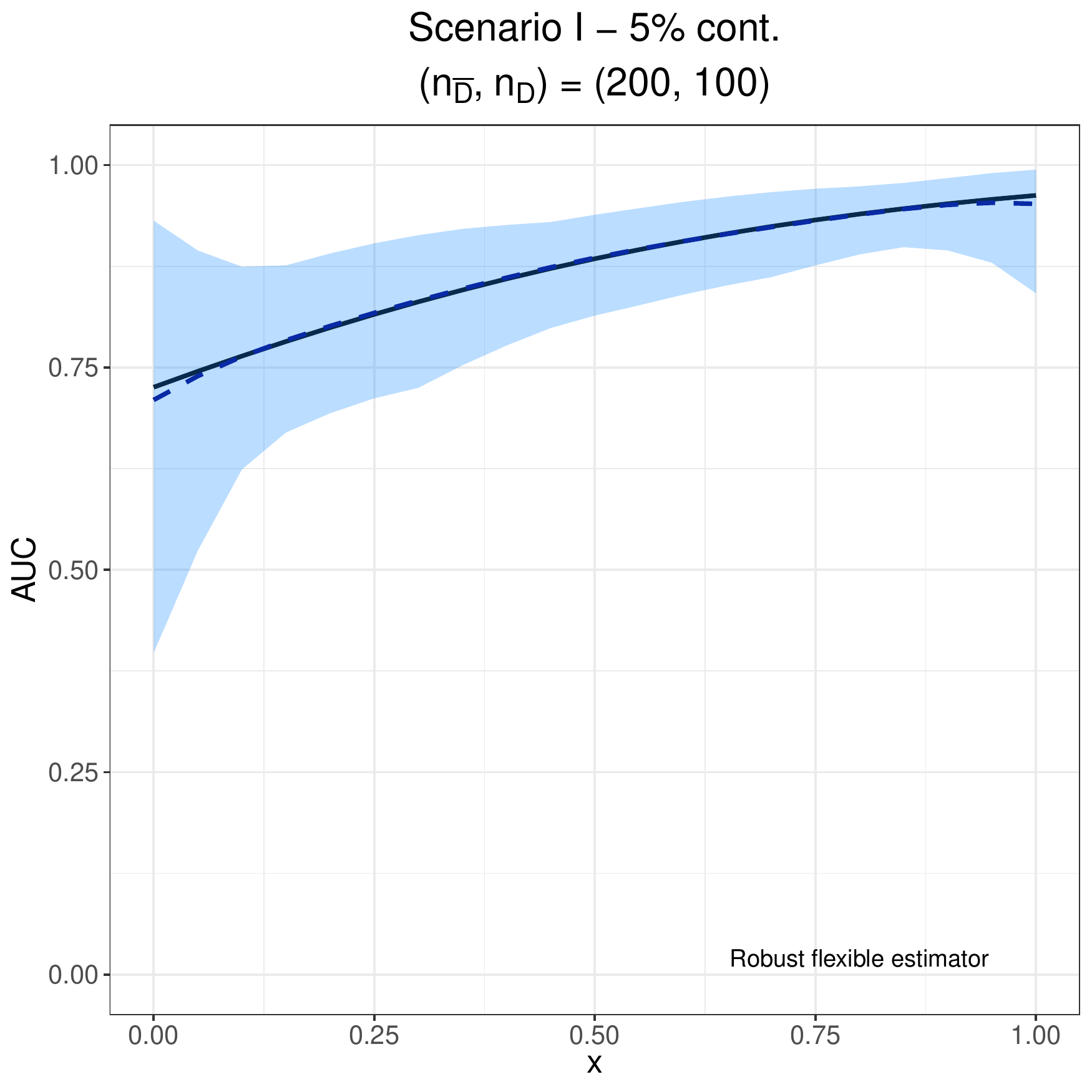}
\includegraphics[width = 4.65cm]{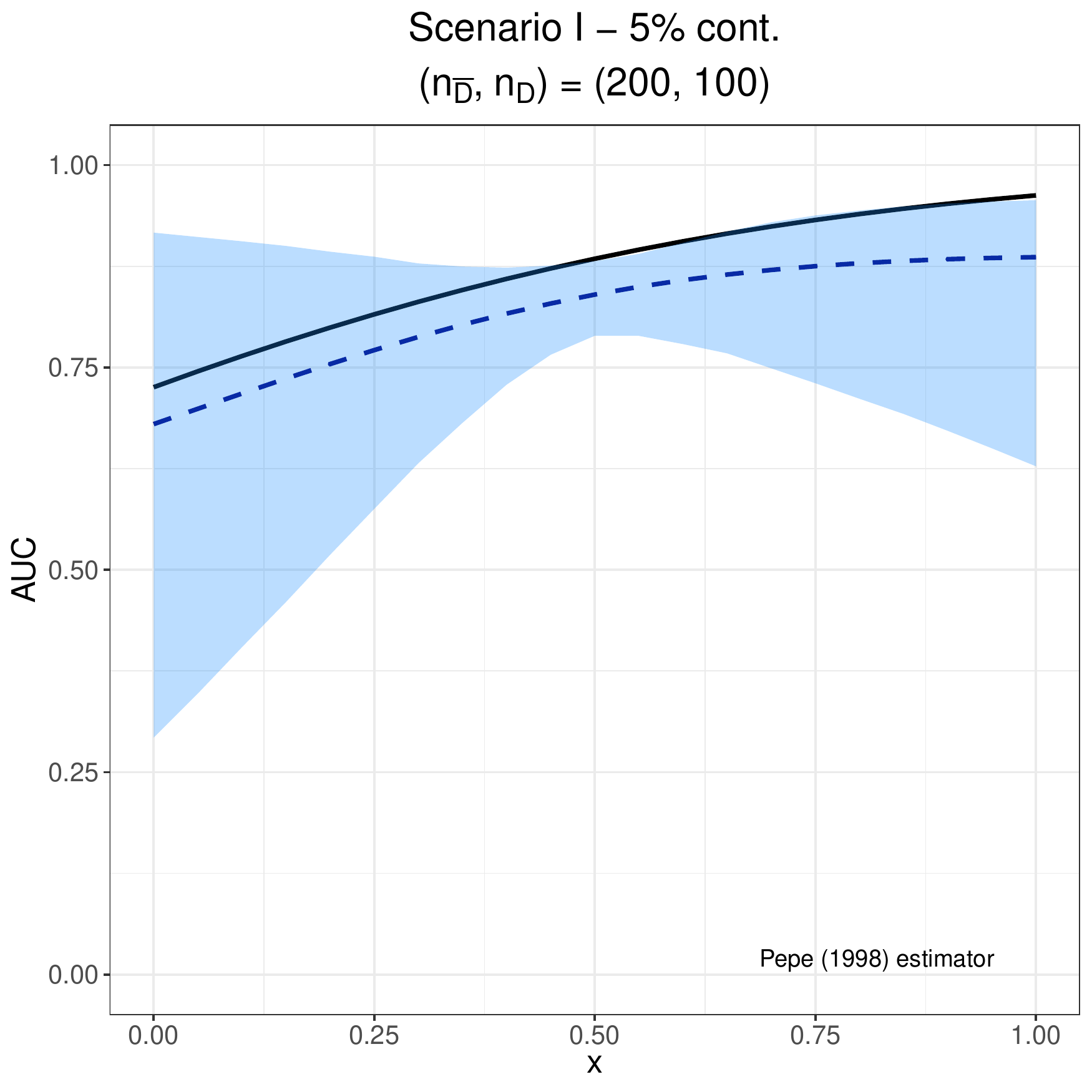}
\includegraphics[width = 4.65cm]{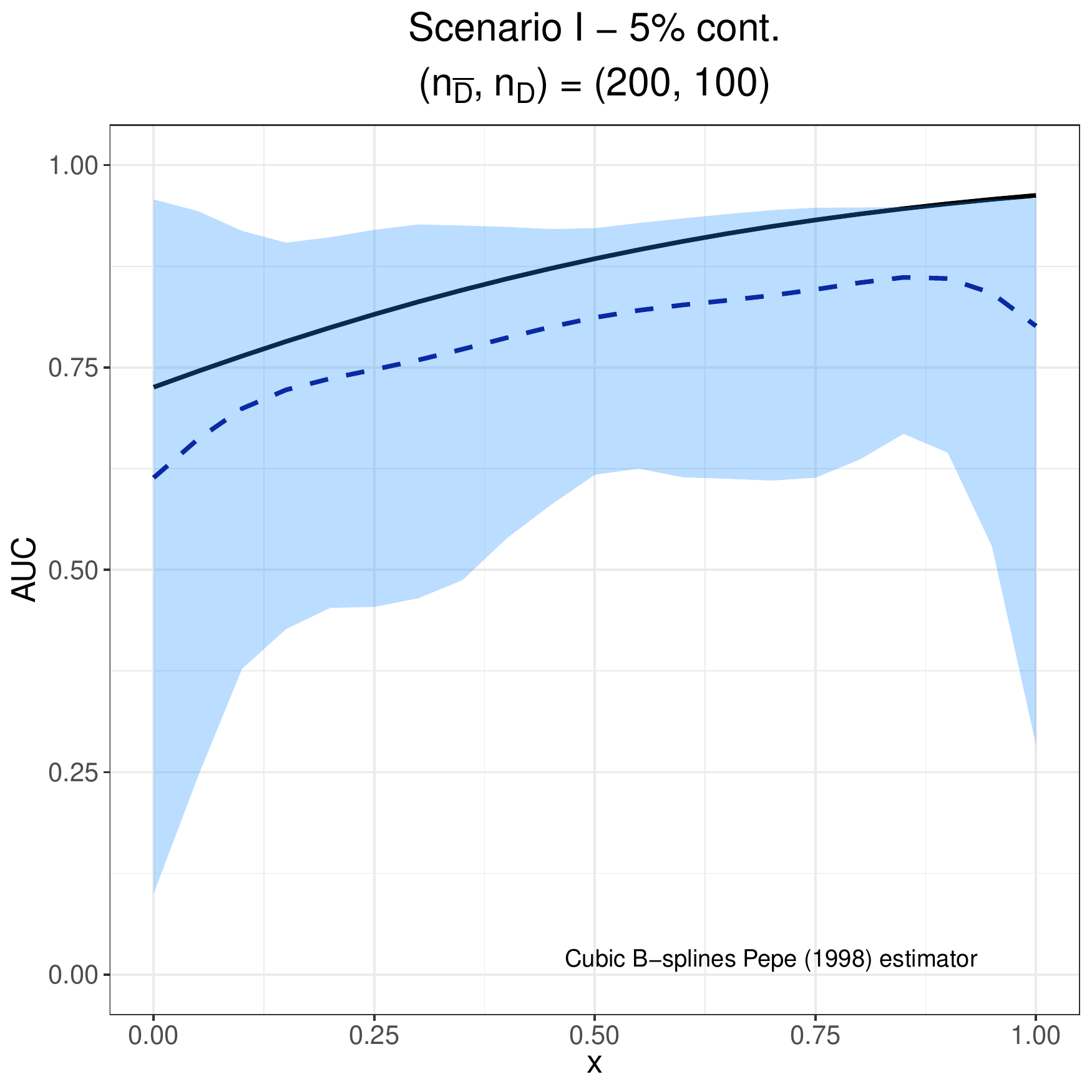}
\includegraphics[width = 4.65cm]{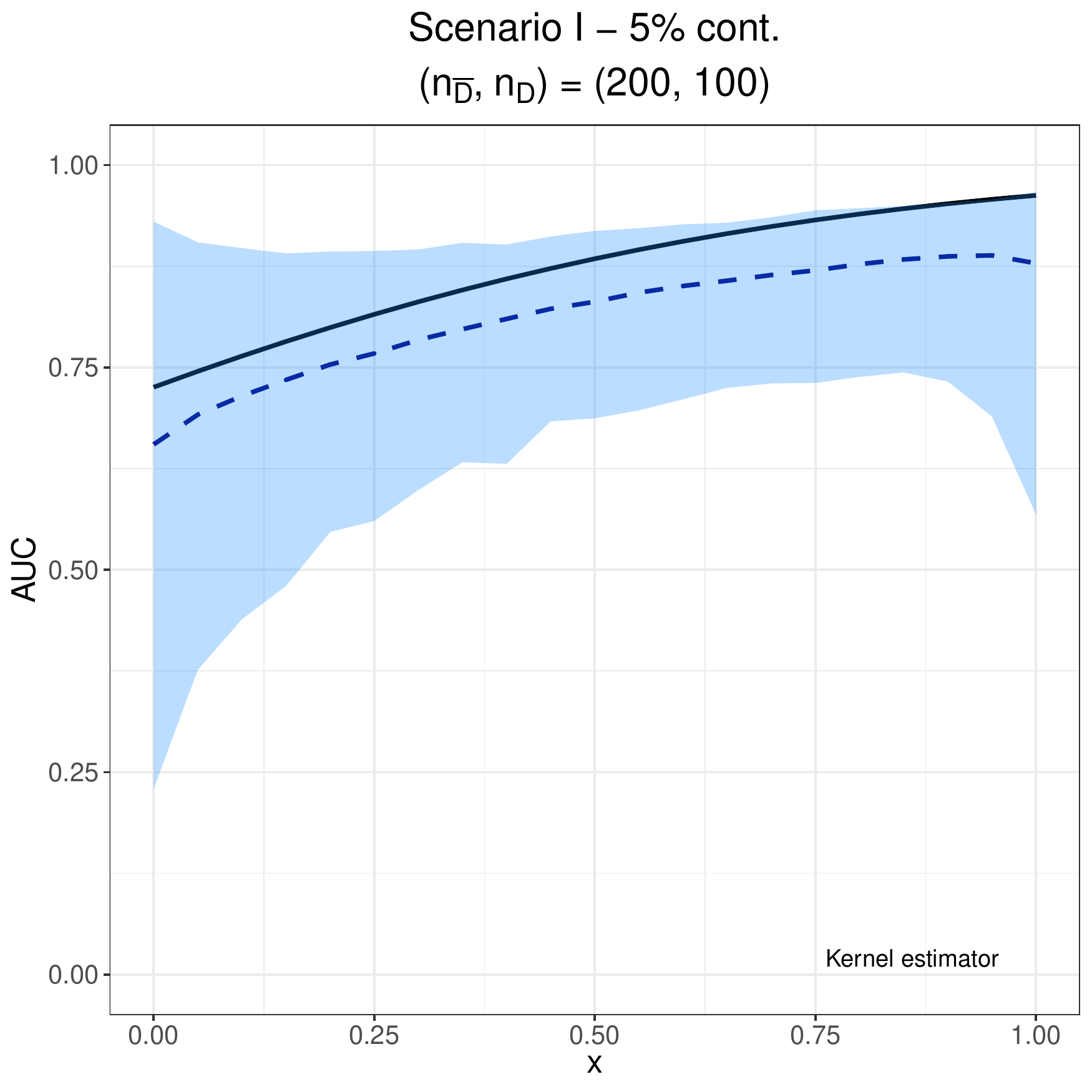}
}
\vspace{0.3cm}
\subfigure{
\includegraphics[width = 4.65cm]{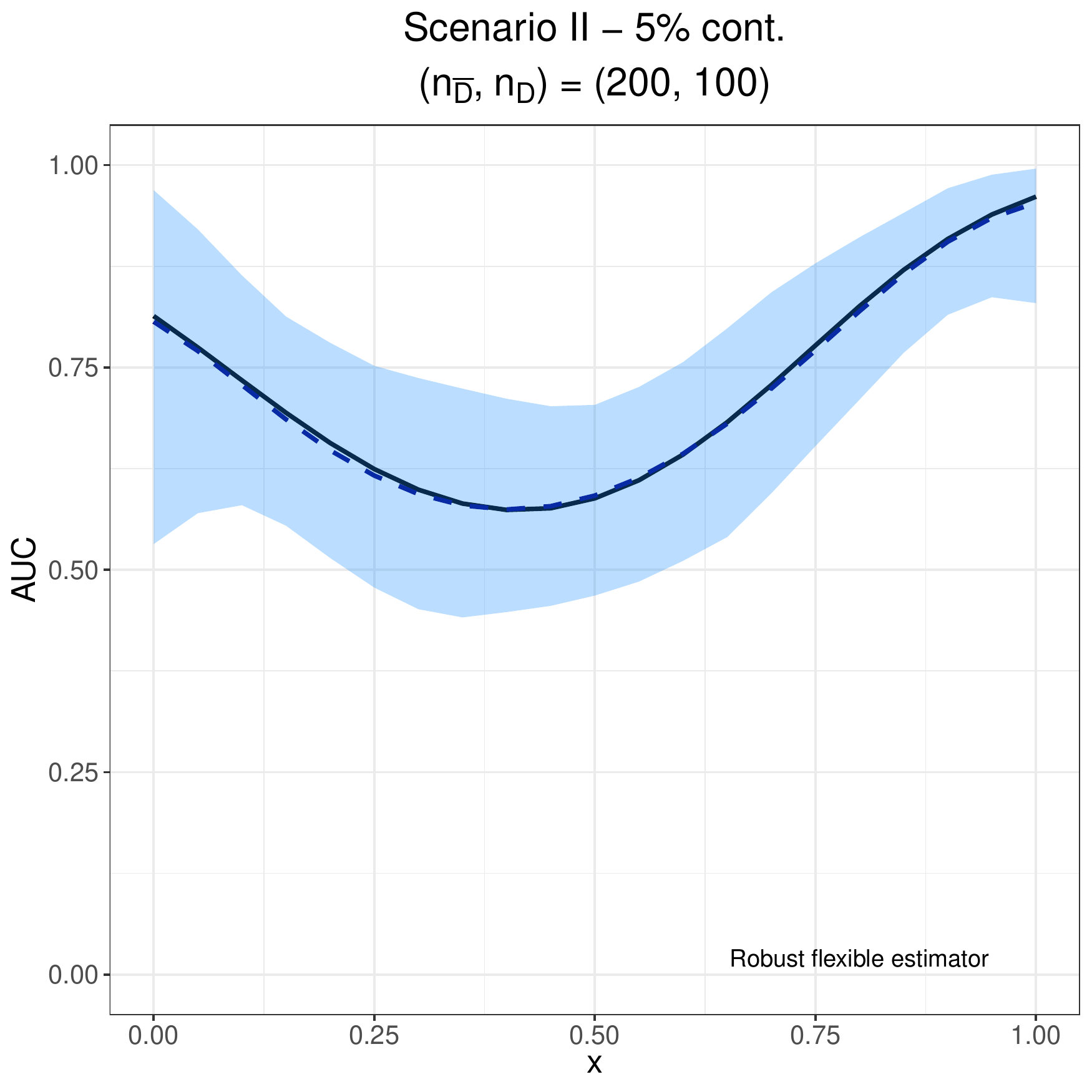}
\includegraphics[width = 4.65cm]{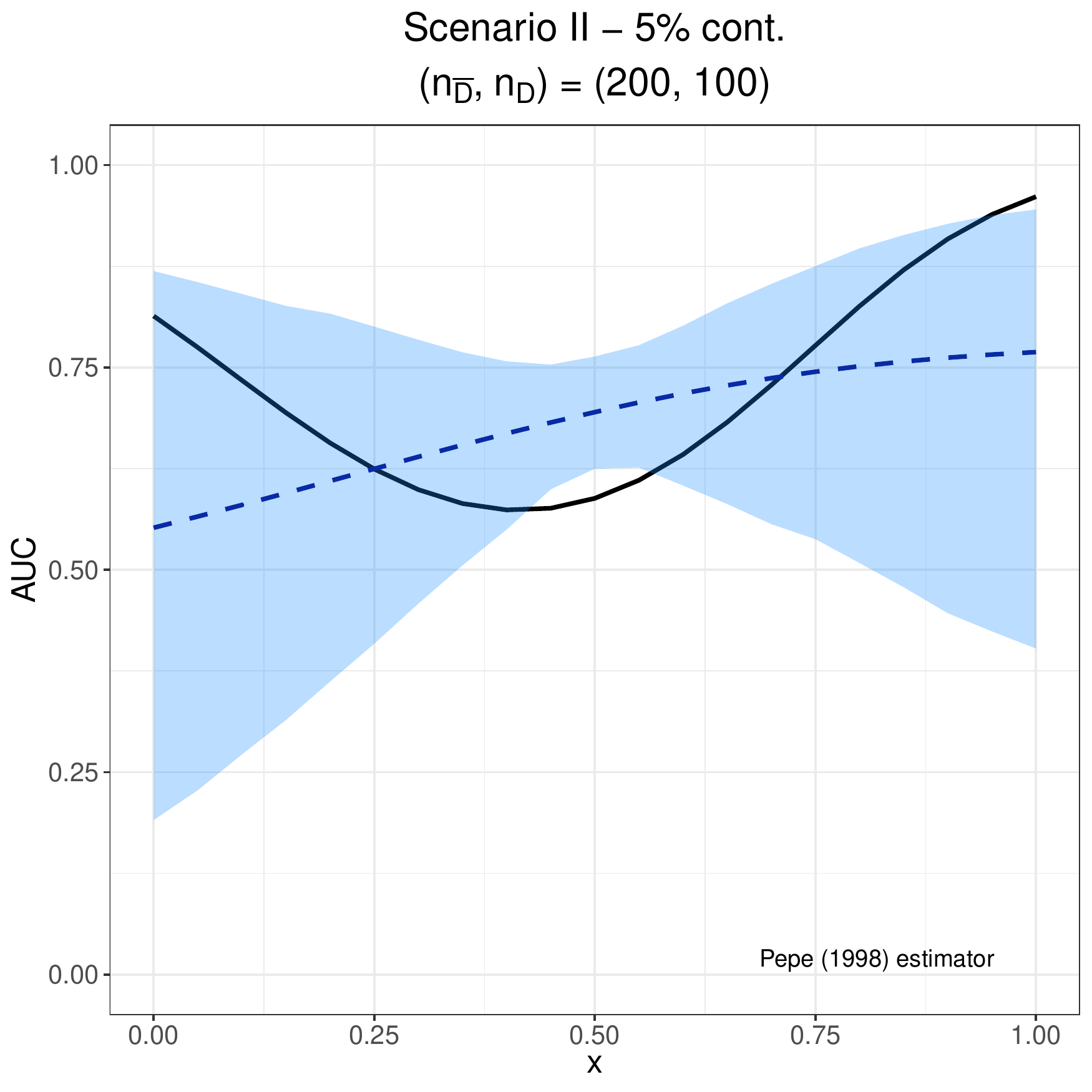}
\includegraphics[width = 4.65cm]{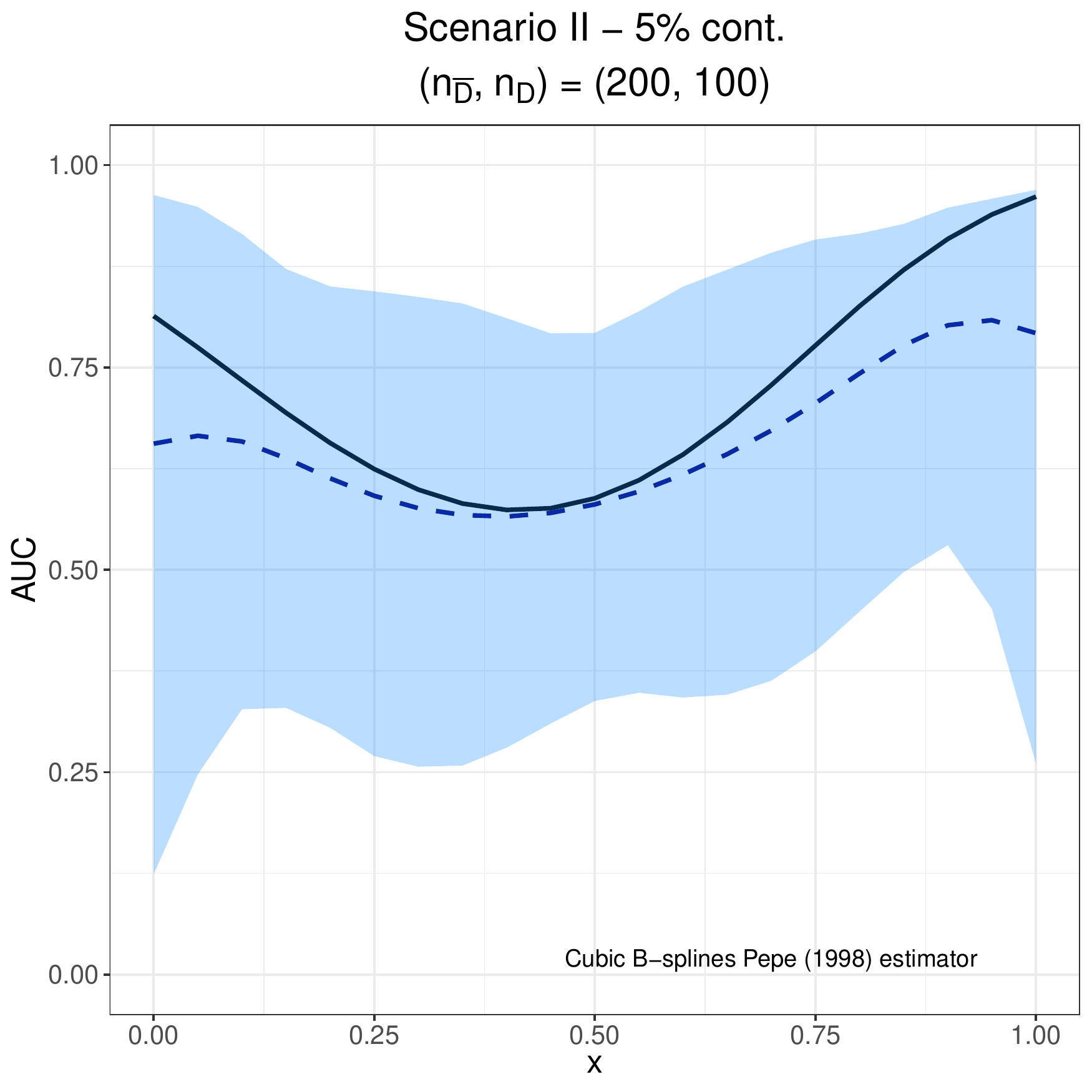}
\includegraphics[width = 4.65cm]{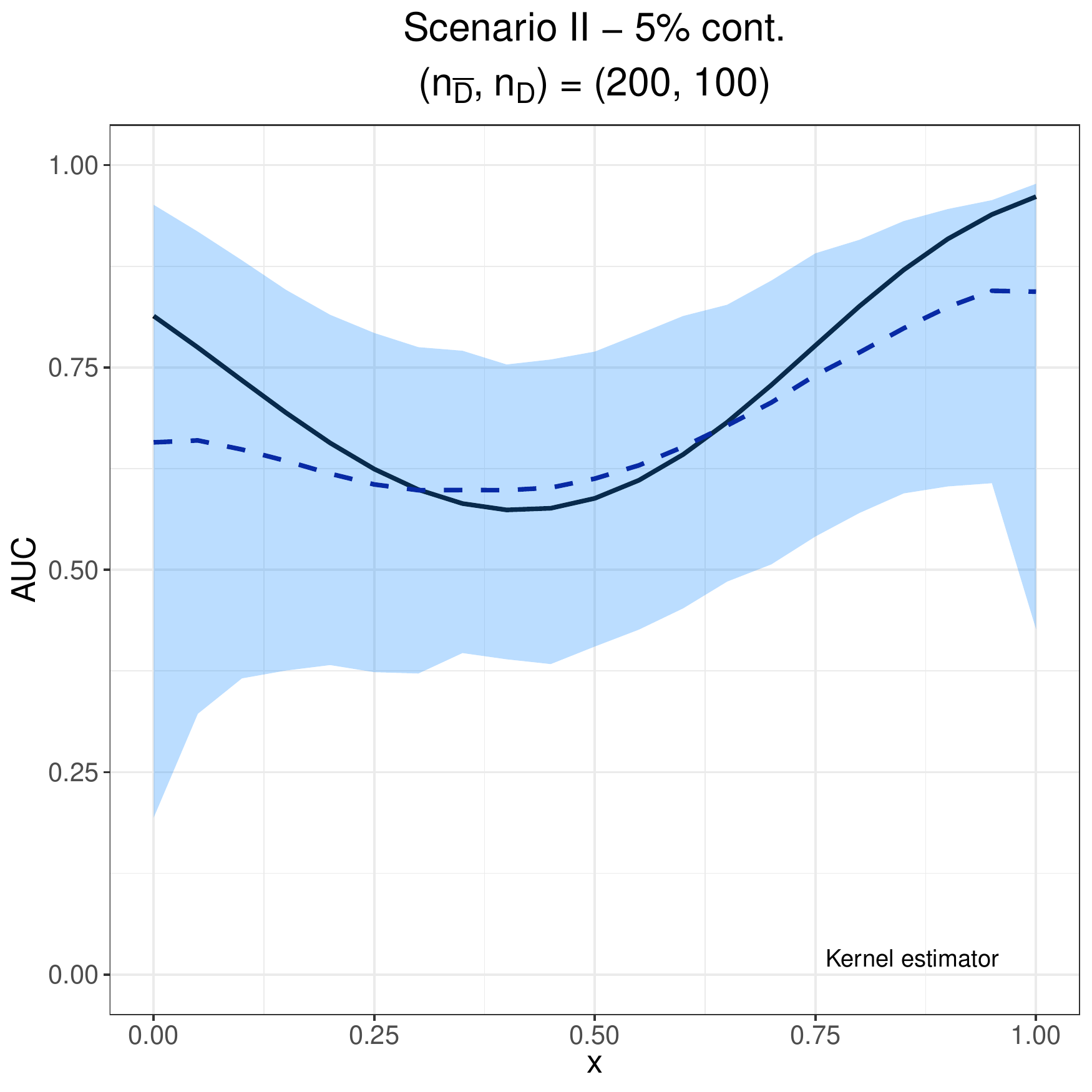}
}
\vspace{0.3cm}
\subfigure{
\includegraphics[width = 4.65cm]{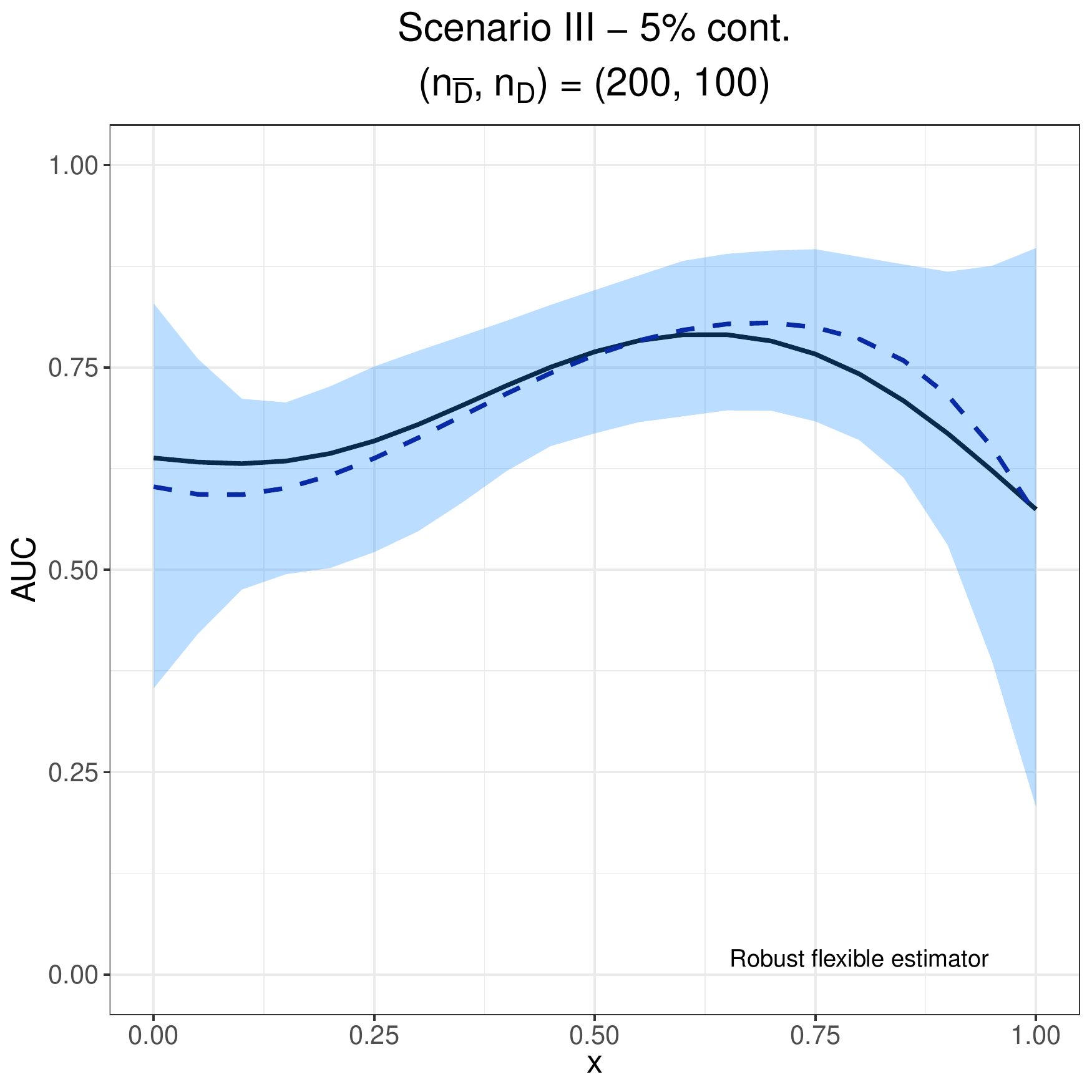}
\includegraphics[width = 4.65cm]{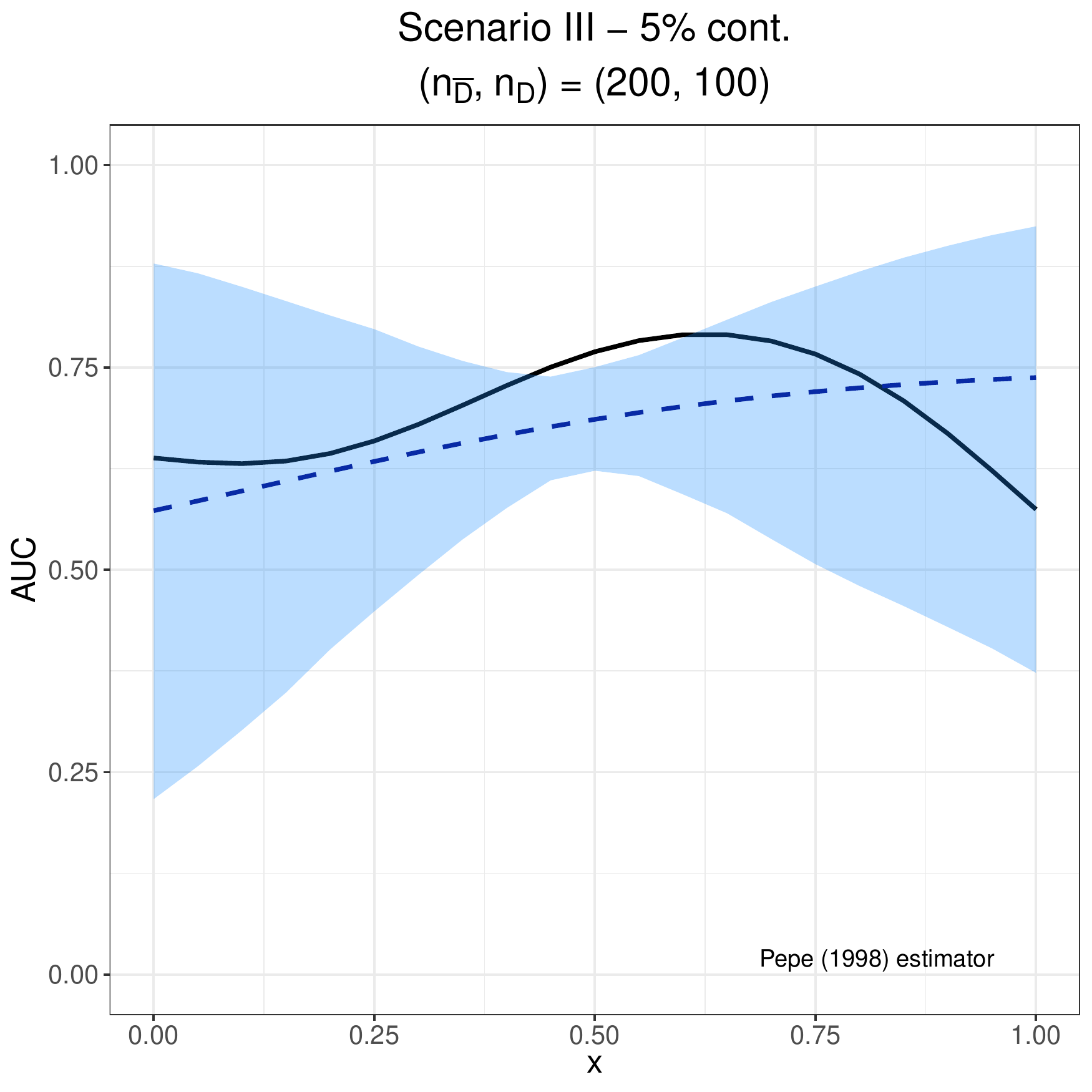}
\includegraphics[width = 4.65cm]{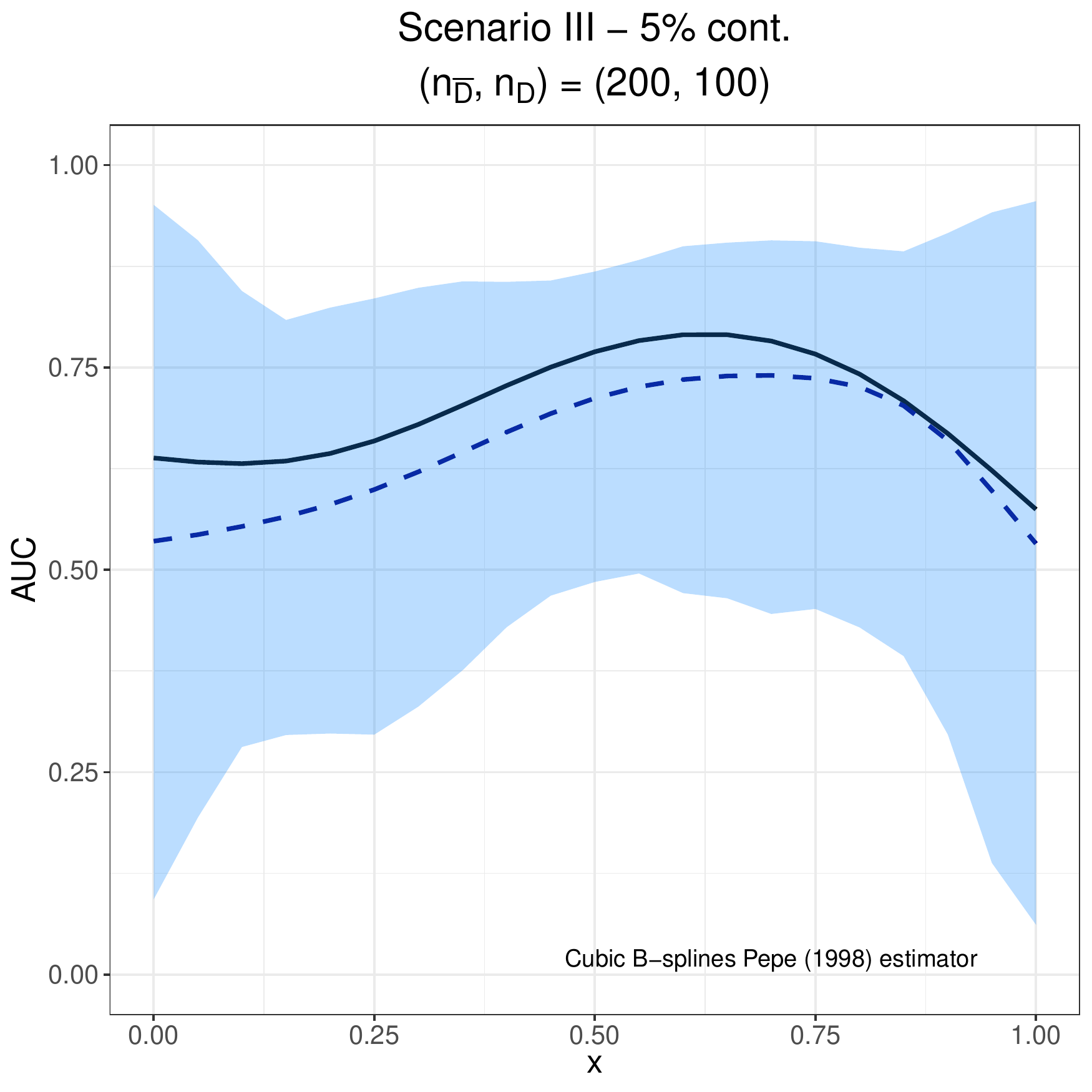}
\includegraphics[width = 4.65cm]{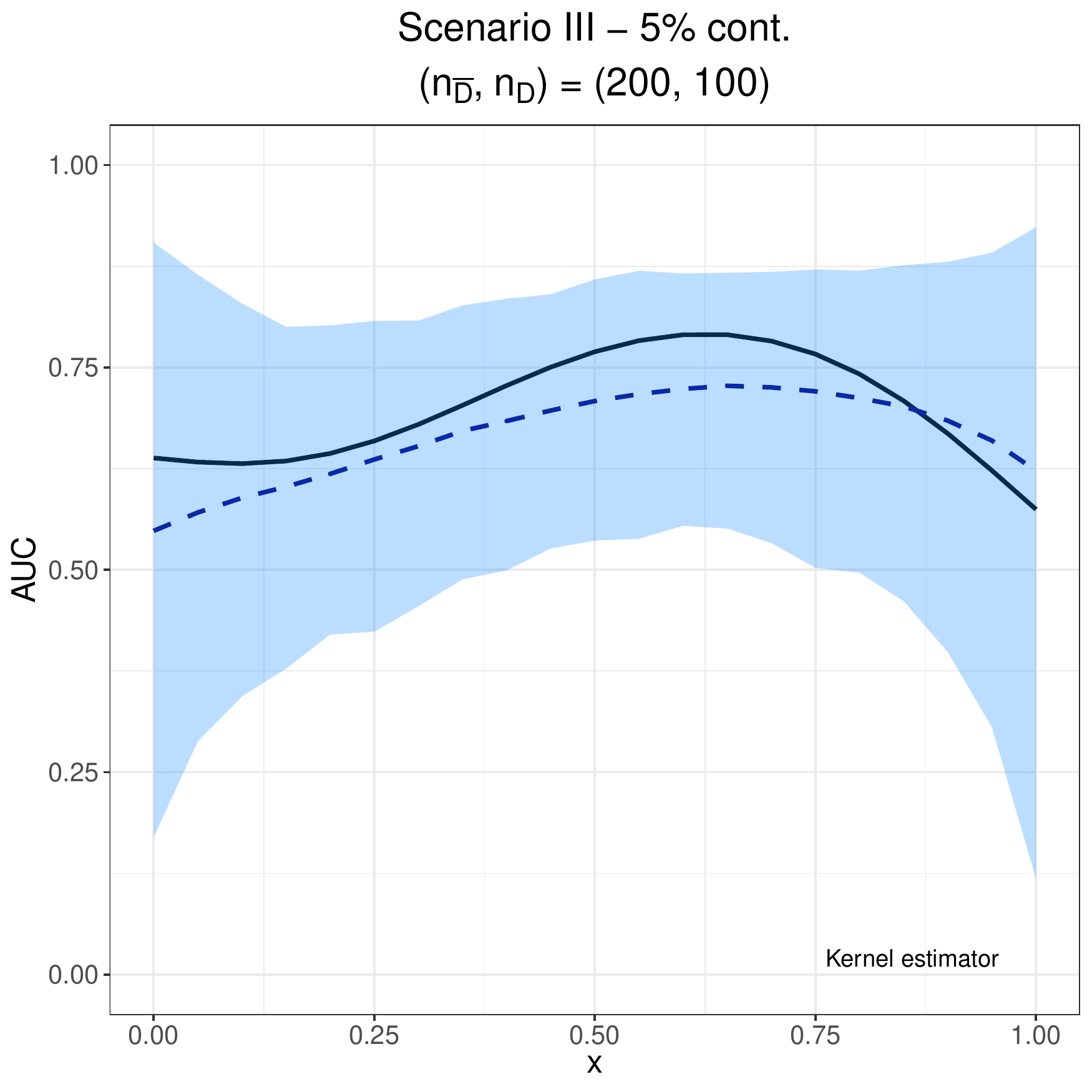}
}
\end{center}
\caption{\footnotesize{True covariate-specific AUC (solid line) versus the mean of the Monte Carlo estimates (dashed line) along with the $2.5\%$ and $97.5\%$ simulation quantiles (shaded area) for the case of $5\%$ contamination. The first row displays the results for Scenario I, the second row for Scenario II, and the third row for Scenario III. The first column corresponds to our flexible and robust estimator, the second column to the estimator proposed by \cite{Pepe1998}, the third one to the cubic B-splines extension of \cite{Pepe1998}, and the fourth column to the kernel estimator. For all scenarios $(n_{\bar{D}}, n_D)=(200,100)$.}}
\label{simresults}
\end{figure}

The case $(n_{\bar{D}},n_{D})=(200,100)$, which is similar to the prostate cancer application in Section 4, is shown here and we first analyse Scenarios I--III. The estimated (mean of the $1000$ Monte Carlo estimates) covariate-specific AUC along with the $2.5\%$ and $97.5\%$ simulation quantiles in Figure \ref{simresults} illustrate the ability of our model to accurately and precisely capture complex functional forms in a case where the contamination in each population is $5\%$. As can be observed in Figure \ref{simresults}, the three non-robust estimators have a very poor performance, showing some bias and wide simulation quantiles bands. Further, and obviously, the original estimator proposed by \cite{Pepe1998} is inadequate for scenarios involving nonlinear trends. Also, note that in Scenario III, where the underlying regression models in the two populations are heteroscedastic, our estimator still has a very decent performance, although we expect it to deteriorate for more substantial changes in the variance along with the covariate. We further note that in this scenario involving heteroscedasticity and when there is no contamination of the test outcomes, the kernel approach, because it models the variance as a function of covariates, it is the one showing less bias (Figure 11 in the Supplementary Materials). 

The remaining sample sizes and percentages of contamination are shown in Figures 3 to 14 in the Supplementary Materials and although similar conclusions were found, some comments are in order. First, even in the case of no contamination, Figures 3, 7, and 11 in the Supplementary Materials, corresponding, respectively, to Scenario I, II, and III, the performance of our robust and flexible estimator is basically on par with that of the non-robust estimators (with the exception of \cite{Pepe1998} for Scenarios II and III). Second, in the case of a $2\%$ contamination (see Figures 4, 8, and 12 in the Supplementary Materials), the non-robust estimators already show some bias and an increase in the width of the simulation bands. This is, of course, much more marked for the case of $10\%$ contamination. In turn, the performance of our robust estimator, although it starts showing some tiny amount of bias, remains quite good.

For Scenario IV, which involves two continuous covariates, only our estimator was considered. We regard this scenario mainly as a proof of concept when there are multiple continuous covariates and the results obtained from fitting the competing approaches were similar to those reported for Scenarios I--III. Nonetheless, for the three sample sizes and different percentages of contamination considered, our approach performs very well and is able to recover the different profiles of the true covariate-specific surface (Figure \ref{simresultssc4} and Figures 15--18 in the Supplementary Materials).

We shall remark that although the covariate-specific AUC admits the closed-form expression in \eqref{rmws}, its calculation can be very time-consuming, especially for large datasets. As a consequence, here and in the Application section, the integral in \eqref{cauc} was approximated using Simpson's rule. In our experience Simpson's rule provides almost identical results to the ones obtained using the closed-form expression.

Because we rely on the robust AIC to assist in the selection of the number of knots needed in the regression function, we have investigated the behaviour of this criterion when performing such a task. Specifically, over the $1000$ simulated datasets, for each scenario considered, for the different sample sizes in each population ($100$ and $200$) and for the different contamination percentages, we computed the number of times the robust AIC favoured the model with no interior knots over a model with three interior knots. For this latter model, following the rule discussed in Section 2, the knots are located at the $0.25$, $0.5$, and $0.75$ quantiles of the covariates. Note that for Scenario IV, as a slight simplification, we have assumed the same number of knots for both continuous covariates (i.e., $(K_{D1}, K_{D2})=(0,0)$ and $(K_{D1}, K_{D2})=(3,3)$, with the same applying in the nondiseased population). Results are displayed in Tables 1--4 in the Supplementary Materials and show that, most of the time, the robust AIC favoured the simpler model with no interior knots over the more complex model with three interior knots. For instance, in Scenario I, where the regression function assumes a linear form in both populations, our intuition would dictate that the model with no interior knots should be selected for a large number of the simulated datasets and Table 1 (Supplementary Materials) confirms exactly this. Also, in Scenario 4, the model with no interior knots for the two covariates (and that involves seven regression parameters) is favoured most of the time over the model that uses three interior knots for each of the covariates (and that involves thirteen regression parameters).

We conclude this section with some extra important remarks. Although we have assumed that both populations were subject to contamination, it may happen that only test outcomes from one of the populations are contaminated. Simulation results (not shown) indicate that in such cases the robust estimator still outperforms the non-robust competitors. However, and interestingly, even when assuming balanced sample sizes, contamination in the nondiseased population seems to impact much more the ability of the non-robust estimators to recover the true functional form of the AUC than contamination in the diseased population. Our intuitive explanation, bearing in mind Equation \eqref{rocls}, is that estimation of the quantile function of the standardised residuals is more impacted by outliers than the estimation of the cumulative distribution function (of the standardised residuals). Further, a shift of  $15\sigma_{\bar{D}}(\mathbf{x}_{\bar{D}})$ and $20\sigma_{D}(\mathbf{x}_D)$ in the location of the distribution of the test outcomes in the nondiseased and diseased populations, respectively, was considered. Our computational experiments show that the performance of the non-robust estimators is affected by the magnitude  of those shifts and, as expected, the larger the shift, the worse the performance. For instance, for a very large shift, which we acknowledge to be unlikely in practice but only to make our point, even the case of a contamination of $2\%$ would be enough to strongly impact their performance. To make this point concrete, Figure 19 in the Supplementary Materials shows the results, under Scenario I, $2\%$ of contamination and $(n_{\bar{D}},n_D)=(200,200)$, of considering $\kappa_D=\kappa_{\bar{D}}=50$. As can be noticed, and especially when compared to row 3 of Figure 4, there is a substantial increase in the bias and width of the $95\%$ simulation bands. On the other hand, as can be observed in Figure 19 (Supplementary Materials), the performance of our robust estimator is unchanged. We have, however, noticed that if the outliers are too small, in magnitude (e.g, by considering a shift of 5 times the standard deviation), they might pass unnoticed when computing the weighted empirical distribution function of the standardised residuals (see \eqref{wecdf}), and this causes some bias for contaminations of $10\%$ and onwards. We should also mention that having also simulated contaminated samples considering radial outliers, which arise by multiplying the scale of the distribution of test outcomes in each group by a given factor, results remained basically the same  and therefore are not shown here. Finally, our computational experiments also revealed that with contamination percentages of about $15\%$ and onwards (in each population), the performance of our estimator starts deteriorating.

\section{\large{\textsf{APPLICATION}}}
\subsection{Motivation and Exploratory Analysis}
Prostate cancer (PCa) is the second most frequent cancer diagnosed in men, only after lung cancer, and amounts to the fifth highest cause of death worldwide \citep{Rawla2019}. Gleason histological scoring system is the most reliable system used for the grading of prostate cancer, but it requires invasive tissue biopsies. This, and the rising incidence of prostate cancer worldwide, have led to the search of less invasive biomarkers that can accurately predict the presence of PCa. The Prostate Health Index (PHI), that combines three prostate specific antigen subforms into a single score using a mathematical formula, has been introduced \citep{Le2010} and since then several studies have shown that it significantly improves prediction of a positive biopsy when compared to the prostate specific antigen across different ranges \citep[e.g.,][]{Stephan2013,Wang2014,delacalle2015}. The PHI is now approved by the US Food and Drug Administration and it has also been adopted into the US National Cancer Network guidelines.  We apply our methods to data from a study designed to assess the added value of the PHI to multi-parametric magnetic resonance imaging in detecting significant prostate cancers  (Gleason $\geq$ 7) in a repeat biopsy population \citep{Gnanapragasam2016}. Here our goal is slightly distinct and we seek to assess, if and how, the ability of the PHI to discriminate between men with benign or Gleason 6 PCa (which throughout we refer as the nondiseased goup and for which $n_{\bar{D}}=185$) and men with Gleason 7 or above PCa (which we term as the diseased group and for which $n_D=94$), changes with age. To the best of our knowledge, this is the first attempt to study the possible age effect on the accuracy of the PHI to distinguish between those two PCa groups. In Figure \ref{exploratory} (left panel) we show the histograms of the PHI levels in the two populations and it can be observed that, as expected, men belonging to the group defined by Gleason $\geq 7$ tend to have higher PHI values than those with a benign lesion or with a Gleason of 6. We can also notice that although the majority of PHI values lie below $100$ in the nondiseased group and below $150$ in the diseased group, there are two PHI scores, one from each group, above $200$. 

\subsection{Unconditional and Age-Specific ROC Analysis}
We start our analysis by calculating the AUC when ignoring the potential age effect and we have computed it in a robust way (so that it is more easily comparable to the covariate-specific AUCs we will present later) as
\begin{equation*}
\widehat{\text{AUC}}=\frac{1}{\sum_{l=1}^{n_D}\omega_{Dl}^{*}\sum_{l=1}^{n_{\bar{D}}}\omega_{\bar{D}l}^{*}}\sum_{j=1}^{n_D}\sum_{i=1}^{n_{\bar{D}}}\omega_{Dj}^{*}\omega_{\bar{D}i}^{*}\left\{I(y_{\bar{Di}}<y_{Dj}) + \frac{1}{2}I(y_{\bar{Di}}=y_{Dj})\right\},
\end{equation*}
where the weights $\omega_{\bar{D}i}^{*}$ and $\omega_{Dj}^{*}$ are defined similarly as in \eqref{wecdf} and arise from fitting, in each group, a robust regression model with the PHI scores as the responses and with only an intercept term. Although PHI outcomes are defined on a continuous scale, in practice ties can occur, and so the extra term $(1/2)\times I(y_{\bar{Di}}=y_{Dj})$ corrects for such possible ties. The resulting AUC estimate (95\% bootstrap confidence interval based on $1000$ resamples) is $0.74$ $(0.68, 0.81)$, revealing a reasonably good capacity of the PHI levels to discriminate between men with a Gleason of 6 or a benign lesion and men with Gleason $\geq 7$.

We now turn our attention to the inclusion of age in the analysis. In Figure \ref{exploratory} (middle and right panels) are depicted the scatter plots of the data in each group along with the estimated regression functions; the robust AIC in \eqref{rAIC} led to $K_{D1}=K_{\bar{D}1}=0$ (no interior knots), with these selected from the set $\{0,1,2,3,4\}$. Firstly, both scatter plots do not indicate any departure from the homoscedasticity assumption. Note that the higher PHI outcomes are properly weighted under our robust scheme; for instance, the PHI scores above $200$ in the nondiseased and diseased groups receive a weight of $0.09$ and $0.19$, respectively. Secondly, as a result of the weighting scheme, such high PHI values do not push the regression functions towards them as much as the analogous least squares counterparts (shown in Figure 20 of the Supplementary Materials). Note that for a fairer comparison we have also included, in Figure 20 of the Supplementary Materials, an approach that models the mean function through a cubic B-splines basis expansion with no interior knots. Thirdly, while in the nondiseased group the PHI does not show any noticeable dynamic along age, in the diseased group there seems to be evidence that older ages are associated with higher PHI outcomes. In Figure \ref{rocsauc} (left and middle panels), we present two different age-specific ROC curves, namely, for ages of 57 and 73 years old, with the corresponding AUCs being $0.71$ $(0.55, 0.87)$ and $0.78$ $(0.68, 0.89)$, respectively. As can be seen, the ROC curves are somewhat jagged, which is due to the fact of them being based on the (weighted) empirical distribution function of the standardised residuals. To inspect the age effect further, Figure \ref{rocsauc} (right panel) shows a plot of the age-specific AUC for ages between 55 and 75 years old and we can observe that the  capacity of the PHI levels to distinguish between men with benign or Gleason 6 PCa and men with Gleason $\geq 7$ PCa slightly increases with age, ranging from $0.70$ $(0.52, 0.87)$ for a men of 55 years old to $0.83$ $(0.71, 0.93)$ for a men of 75 years old. The AUC estimate obtained when ignoring the age effect was $0.74$ and so, roughly, for individuals younger than 70 years we would be slightly overestimating the accuracy of the PHI scores and for individuals older than 70 years old such accuracy would be slightly underestimated. Nonetheless, note that the unconditional AUC estimate and corresponding $95\%$ confidence interval are contained in the $95\%$ bootstrap confidence band for all ages considered and so it is difficult to draw firm conclusions about the age effect. We remark that AUC predictions were only considered for ages in the the interval $(55, 75)$ as this corresponds to the range where both groups had a reasonable number of observations. We further remark that when computing the $95\%$ bootstrap confidence bands, the number of internal knots selected for the observed data (in this case this was $0$ for both groups) was used when re-computing the estimates for the generated bootstrap samples. Also we highlight that it took less than one minute to run our model (including the $1000$ bootstrap resamples) on a MacBook Pro with 2.3GHz Intel i5 processor and 8GB RAM.
Finally, in Figure 21 of the Supplementary Materials we present the age-specific AUC estimates obtained when considering the three non-robust estimators detailed in the Simulation Study section, and as can be observed they are not markedly different from the point estimate provided by our approach. This should come as no surprise as the estimated mean functions were also not too distinct, which makes sense as there are only two PHI outcomes, one in each group, that lie well above the remaining scores. Also, all approaches agree that the accuracy of the PHI scores to distinguish between the two groups of PCa slightly increases with age.

\section{\large{\textsf{CONCLUDING REMARKS}}}
In this work we have developed a flexible and robust modelling framework for estimating the covariate-specific ROC curve and corresponding AUC that assumes a location-scale regression model in both the diseased and nondiseased populations and that combines an additive cubic B-splines formulation for the mean function with M-estimation. Our approach is thus able to simultaneously accommodate nonlinear effects of the covariates and outlying test outcomes. The proposed methodology has the additional appealing features of being simple and computationally inexpensive. The simulation study conducted illustrated the ability of our method to recover the true shape of the covariate-specific ROC curve and AUC in a variety of complex scenarios involving different test outcome distributions and contamination percentages. Our investigation into the potential of the Prostate Health Index to distinguish between men with a benign lesion or a Gleason 6 prostate cancer and men with aggressive prostate cancer (Gleason 7 or above) found that its accuracy slightly increases with age. Although in this particular case the overall message of our analysis agrees with that provided by the non-robust estimators, our approach enabled us to identify one outlying test outcome in each population. 

Our method can be trivially adapted to also estimate the covariate-specific Youden index and its corresponding optimal threshold. In particular, since
\begin{equation}\label{YI}
\text{YI}(\mathbf{x})=\max_c \{F_{\bar{D}}(c\mid\mathbf{x})-F_{D}(c\mid\mathbf{x})\},
\end{equation}
one can make use of the result in \eqref{cdfls} and estimate the cumulative distribution function of the standardised residuals using \eqref{wecdf}. The covariate-specific optimal threshold is the one maximising \eqref{YI}.

Finally, throughout we have assumed that only the test outcomes were prone to outliers. However, if covariates are also contaminated, our approach can be easily extended to cope with this case by considering MM-estimation techniques instead of the M-estimation method used here.

\section*{Acknowledgments}
We are grateful to Gerda Claeskens for the help and insight about the robust AIC used in this paper. The work of VI, VML, and MdC was partially supported by FCT (Funda\c c\~ao para a Ci\^encia e a Tecnologia, Portugal) through the projects PTDC/MAT-STA/28649/2017 and UID/MAT/00006/2020 (VI and MdC) and UIDB/00297/2020 and the sabbatical grant SFRH/BSAB/142919/2018 (VML). VML further acknowledges mobility Erasmus+ funding via contracts 29191/002/2017/STT and 29191/036/2018/STT.

\section*{\large{\textsf{APPENDIX}}}
Here we deduce the representation of our weighted robust covariate-specific AUC in the form of \eqref{rmws}.
The derivation is based on simple calculus and its main steps are outlined below. We start by noting that
\begin{align*}
\widehat{\text{AUC}(\mathbf{x})}&=\int_{0}^{1}\widehat{\text{ROC}}(t\mid\mathbf{x})\text{d}t \\
&=\int_{0}^{1}\left[1-\widehat{F}_{\varepsilon_{D}}\left\{\frac{\widehat{\mu}_{\bar{D}}(\mathbf{x})-\widehat{\mu}_{D}(\mathbf{x})}{\widehat{\sigma}_D}+\frac{\widehat{\sigma}_{\bar{D}}}{\widehat{\sigma}_D}\widehat{F}_{\varepsilon_{\bar{D}}}^{-1}(1-t)\right\}\right]\text{d}t\\
&=\int_{0}^{1}\sum_{j=1}^{n_D}\frac{\omega_{Dj}^{*}}{\sum_{l=1}^{n_D}\omega_{Dl}^{*}}I\left\{\widehat{\varepsilon}_{Dj}\geq\frac{\widehat{\mu}_{\bar{D}}(\mathbf{x})-\widehat{\mu}_{D}(\mathbf{x})}{\widehat{\sigma}_D}+\frac{\widehat{\sigma}_{\bar{D}}}{\widehat{\sigma}_D}\widehat{F}_{\varepsilon_{\bar{D}}}^{-1}(1-t)\right\}\text{d}t,
\end{align*}
which implies that
\begin{align*}
\widehat{\text{AUC}(\mathbf{x})} &= \frac{1}{\sum_{l=1}^{n_D}\omega_{Dl}^{*}}\sum_{j=1}^{n_D}\omega_{Dj}^{*}\int_{0}^{1}I\left\{t\geq 1 - \widehat{F}_{\varepsilon_{\bar{D}}}\left(\frac{\widehat{\mu}_{D}(\mathbf{x})-\widehat{\mu}_{\bar{D}}(\mathbf{x})}{\widehat{\sigma}_{\bar{D}}}+\frac{\widehat{\sigma}_D}{\widehat{\sigma}_{\bar{D}}}\widehat{\varepsilon}_{Dj}\right)\right\}\text{d}t\\
&=\frac{1}{\sum_{l=1}^{n_D}\omega_{Dl}^{*}}\sum_{j=1}^{n_D}\omega_{Dj}^{*}\int_{1 - \widehat{F}_{\varepsilon_{\bar{D}}}\left(\frac{\widehat{\mu}_{D}(\mathbf{x})-\widehat{\mu}_{\bar{D}}(\mathbf{x})}{\widehat{\sigma}_{\bar{D}}}+\frac{\widehat{\sigma}_D}{\widehat{\sigma}_{\bar{D}}}\widehat{\varepsilon}_{Dj}\right)}^{1}\text{d}t\\
&=\frac{1}{\sum_{l=1}^{n_D}\omega_{Dl}^{*}}\sum_{j=1}^{n_D}\omega_{Dj}^{*}\sum_{i=1}^{n_{\bar{D}}}\frac{\omega_{\bar{D}i}^{*}}{\sum_{l=1}^{n_{\bar{D}}}\omega_{\bar{D}l}^{*}}I\left\{\widehat{\varepsilon}_{\bar{D}i}\leq\frac{\widehat{\mu}_{D}(\mathbf{x})-\widehat{\mu}_{\bar{D}}(\mathbf{x})}{\widehat{\sigma}_{\bar{D}}}+\frac{\widehat{\sigma}_D}{\widehat{\sigma}_{\bar{D}}}\widehat{\varepsilon}_{Dj}\right\}\\
&=\frac{1}{\sum_{l=1}^{n_D}\omega_{Dl}^{*}\sum_{l=1}^{n_{\bar{D}}}\omega_{\bar{D}l}^{*}}\sum_{j=1}^{n_D}\sum_{i=1}^{n_{\bar{D}}}\omega_{Dj}^{*}\omega_{\bar{D}i}^{*}I\{\widehat{\mu}_{\bar{D}}(\mathbf{x})+\widehat{\sigma}_{\bar{D}}\widehat{\varepsilon}_{\bar{D}i}\leq \widehat{\mu}_{D}(\mathbf{x})+\widehat{\sigma}_{D}\widehat{\varepsilon}_{Dj}\}.
\end{align*}

\bibliographystyle{hapalike}
\bibliography{references}

\begin{thebibliography}{}

\bibitem[de~Boor, 1978]{deboor1978}
de~Boor, C. (1978).
\newblock {\em A {P}ractical {G}uide to {S}plines}.
\newblock Springer-Verlag, New York.

\bibitem[De~La~Calle et~al., 2015]{delacalle2015}
De~La~Calle, C., Patil, D., Wei, J.~T., Scherr, D.~S., Sokoll, L., Chan, D.~W.,
  Siddiqui, J., Mosquera, J.~M., Rubin, M.~A., and Sanda, M.~G. (2015).
\newblock Multicenter evaluation of the prostate health index to detect
  aggressive prostate cancer in biopsy naive men.
\newblock {\em The Journal of Urology}, 194(1):65--72.

\bibitem[Durrleman and Simon, 1989]{Durrleman1989}
Durrleman, S. and Simon, R. (1989).
\newblock Flexible regression models with cubic splines.
\newblock {\em Statistics in Medicine}, 8(5):551--561.

\bibitem[Gnanapragasam et~al., 2016]{Gnanapragasam2016}
Gnanapragasam, V., Burling, K., George, A., Stearn, S., Warren, A., Barrett,
  T., Koo, B., Gallagher, F., Doble, A., Kastner, C., and Parker, R. (2016).
\newblock The prostate health index adds predictive value to multi-parametric
  {MRI} in detecting significant prostate cancers in a repeat biopsy
  population.
\newblock {\em Scientific Reports}, 6(1):1--8.

\bibitem[Gonz{\'a}lez-Manteiga et~al., 2011]{Gonzalez2011}
Gonz{\'a}lez-Manteiga, W., Pardo-Fern{\'a}ndez, J.~C., and Keilegom, I.~v.
  (2011).
\newblock {ROC} curves in non-parametric location-scale regression models.
\newblock {\em Scandinavian Journal of Statistics}, 38(1):169--184.

\bibitem[Hayfield and Racine, 2008]{Hayfield2008}
Hayfield, T. and Racine, J.~S. (2008).
\newblock Nonparametric econometrics: The np package.
\newblock {\em Journal of Statistical Software}, 27(5):1--32.

\bibitem[Huber, 1964]{Huber1964}
Huber, P.~J. (1964).
\newblock Robust estimation of a location parameter.
\newblock {\em The Annals of Mathematical Statistics}, 45(1):73--101.

\bibitem[In\'acio and Rodr{\'\i}guez-{\'A}lvarez, 2021]{Inacio2020}
In\'acio, V. and Rodr{\'\i}guez-{\'A}lvarez, M.~X. (2021).
\newblock Statistical evaluation of medical tests.
\newblock {\em Annual Review of Statistics and Its Application (accepted,
  arXiv:2007.07687)}.

\bibitem[Le et~al., 2010]{Le2010}
Le, B.~V., Griffin, C.~R., Loeb, S., Carvalhal, G.~F., Kan, D., Baumann, N.~A.,
  and Catalona, W.~J. (2010).
\newblock [-2] {P}roenzyme prostate specific antigen is more accurate than
  total and free prostate specific antigen in differentiating prostate cancer
  from benign disease in a prospective prostate cancer screening study.
\newblock {\em The Journal of Urology}, 183(4):1355--1359.

\bibitem[Maronna et~al., 2019]{Maronna2019}
Maronna, R.~A., Martin, R.~D., Yohai, V.~J., and Salibi{\'a}n-Barrera, M.
  (2019).
\newblock {\em Robust Statistics: Theory and Methods (with {R})}.
\newblock John Wiley \& Sons.

\bibitem[Pepe, 1998]{Pepe1998}
Pepe, M.~S. (1998).
\newblock Three approaches to regression analysis of receiver operating
  characteristic curves for continuous test results.
\newblock {\em Biometrics}, 54(1):124--135.

\bibitem[Pepe, 2003]{Pepe2003}
Pepe, M.~S. (2003).
\newblock {\em The Statistical Evaluation of Medical Tests for Classification
  and Prediction}.
\newblock Oxford University Press.

\bibitem[Racine, 2019]{Racine2019}
Racine, J.~S. (2019).
\newblock {\em Reproducible Econometrics Using {R}}.
\newblock Oxford University Press.

\bibitem[Rawla, 2019]{Rawla2019}
Rawla, P. (2019).
\newblock Epidemiology of prostate cancer.
\newblock {\em World Journal of Oncology}, 10(2):63.

\bibitem[{\texttt{R} Core Team}, 2020]{R20}
{\texttt{R} Core Team} (2020).
\newblock {\em \texttt{R}: {A} Language and Environment for Statistical
  Computing}.
\newblock \texttt{R} Foundation for Statistical Computing, Vienna, Austria.

\bibitem[Rodr{\'\i}guez and Mart{\'\i}nez, 2014]{Rodriguez2014}
Rodr{\'\i}guez, A. and Mart{\'\i}nez, J.~C. (2014).
\newblock Bayesian semiparametric estimation of covariate-dependent {ROC}
  curves.
\newblock {\em Biostatistics}, 15(2):353--369.

\bibitem[Rodr{\'\i}guez-{\'A}lvarez and In\'acio, 2020]{Rodriguez2020}
Rodr{\'\i}guez-{\'A}lvarez, M.~X. and In\'acio, V. (2020).
\newblock {ROCnReg}: An {R} package for receiver operating characteristic curve
  inference with and without covariate information.
\newblock {\em arXiv:2003.13111}.

\bibitem[Rodr{\'\i}guez-{\'A}lvarez et~al., 2011]{Rodriguez2011SC}
Rodr{\'\i}guez-{\'A}lvarez, M.~X., Roca-Pardi{\~n}as, J., and
  Cadarso-Su{\'a}rez, C. (2011).
\newblock {ROC} curve and covariates: extending induced methodology to the
  non-parametric framework.
\newblock {\em Statistics and Computing}, 21(4):483--499.

\bibitem[Rosenberg, 1995]{Rosenberg1995}
Rosenberg, P.~S. (1995).
\newblock Hazard function estimation using {B}-splines.
\newblock {\em Biometrics}, 51(3):874--887.

\bibitem[Stephan et~al., 2013]{Stephan2013}
Stephan, C., Vincendeau, S., Houlgatte, A., Cammann, H., Jung, K., and
  Semjonow, A. (2013).
\newblock Multicenter evaluation of [- 2] proprostate-specific antigen and the
  prostate health index for detecting prostate cancer.
\newblock {\em Clinical Chemistry}, 59(1):306--314.

\bibitem[Tharmaratnam and Claeskens, 2013]{Tharmaratnam2013}
Tharmaratnam, K. and Claeskens, G. (2013).
\newblock A comparison of robust versions of the {AIC} based on {M}-, {S}-and
  {MM}-estimators.
\newblock {\em Statistics}, 47(1):216--235.

\bibitem[Venables and Ripley, 2002]{MASS_package}
Venables, W.~N. and Ripley, B.~D. (2002).
\newblock {\em Modern Applied Statistics with S}.
\newblock Springer, New York, fourth edition.

\bibitem[Walach et~al., 2017]{Walach2017}
Walach, J., Filzmoser, P., Hron, K., Walczak, B., and Najdekr, L. (2017).
\newblock Robust biomarker identification in a two-class problem based on
  pairwise log-ratios.
\newblock {\em Chemometrics and Intelligent Laboratory Systems}, 171:277--285.

\bibitem[Wang et~al., 2014]{Wang2014}
Wang, W., Wang, M., Wang, L., Adams, T.~S., Tian, Y., and Xu, J. (2014).
\newblock Diagnostic ability of\% p2psa and prostate health index for
  aggressive prostate cancer: a meta-analysis.
\newblock {\em Scientific Reports}, 4:5012.

\bibitem[Welsh and Ronchetti, 2002]{Welsh2002}
Welsh, A.~H. and Ronchetti, E. (2002).
\newblock A journey in single steps: robust one-step {M}-estimation in linear
  regression.
\newblock {\em Journal of Statistical Planning and Inference},
  103(1-2):287--310.

\bibitem[Wong et~al., 2014]{Wong2014}
Wong, R.~K., Yao, F., and Lee, T.~C. (2014).
\newblock Robust estimation for generalized additive models.
\newblock {\em Journal of Computational and Graphical Statistics},
  23(1):270--289.

\end{thebibliography}
\newpage

\begin{figure}[H]
\begin{center}
\subfigure{
\includegraphics[width = 5.35cm]{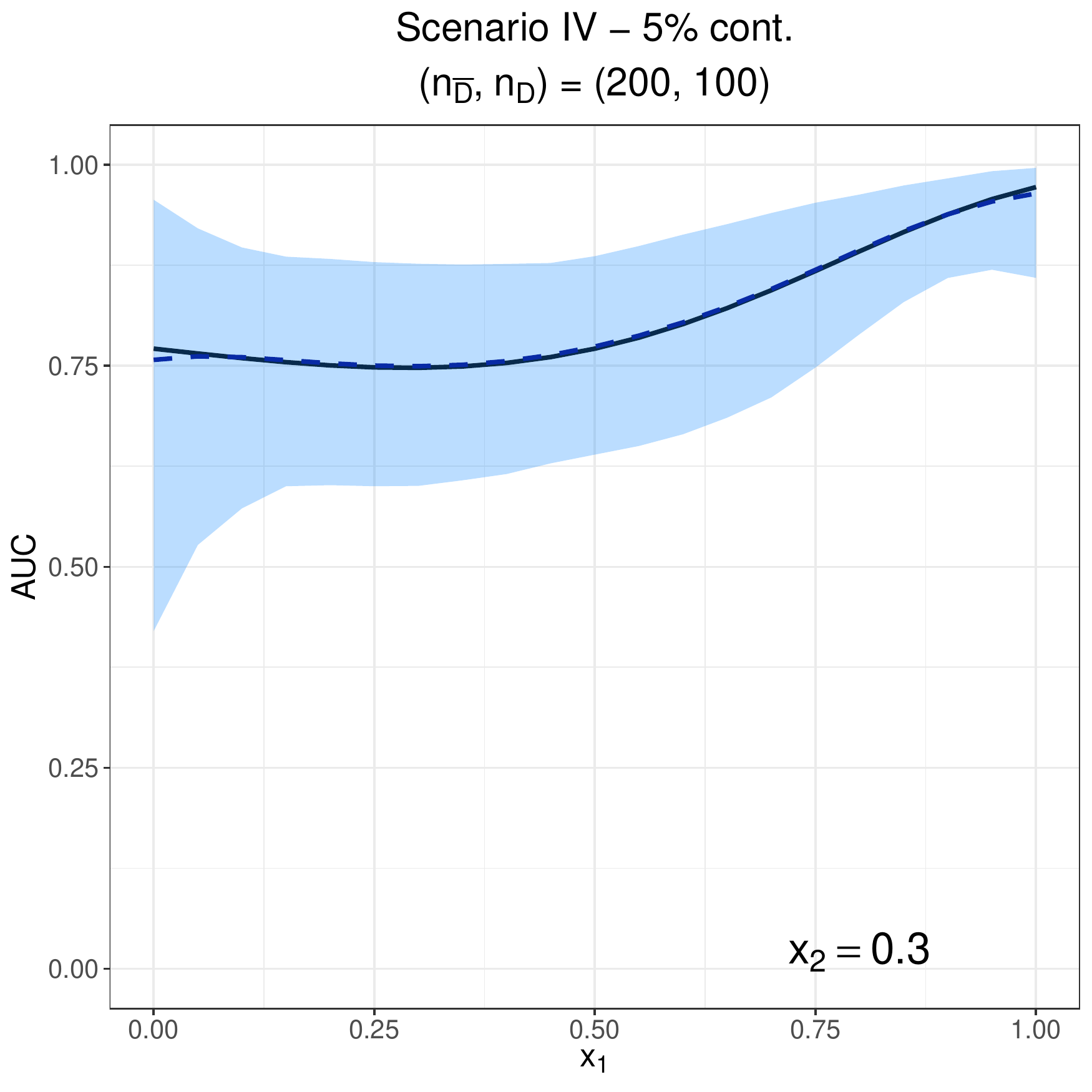}
\includegraphics[width = 5.35cm]{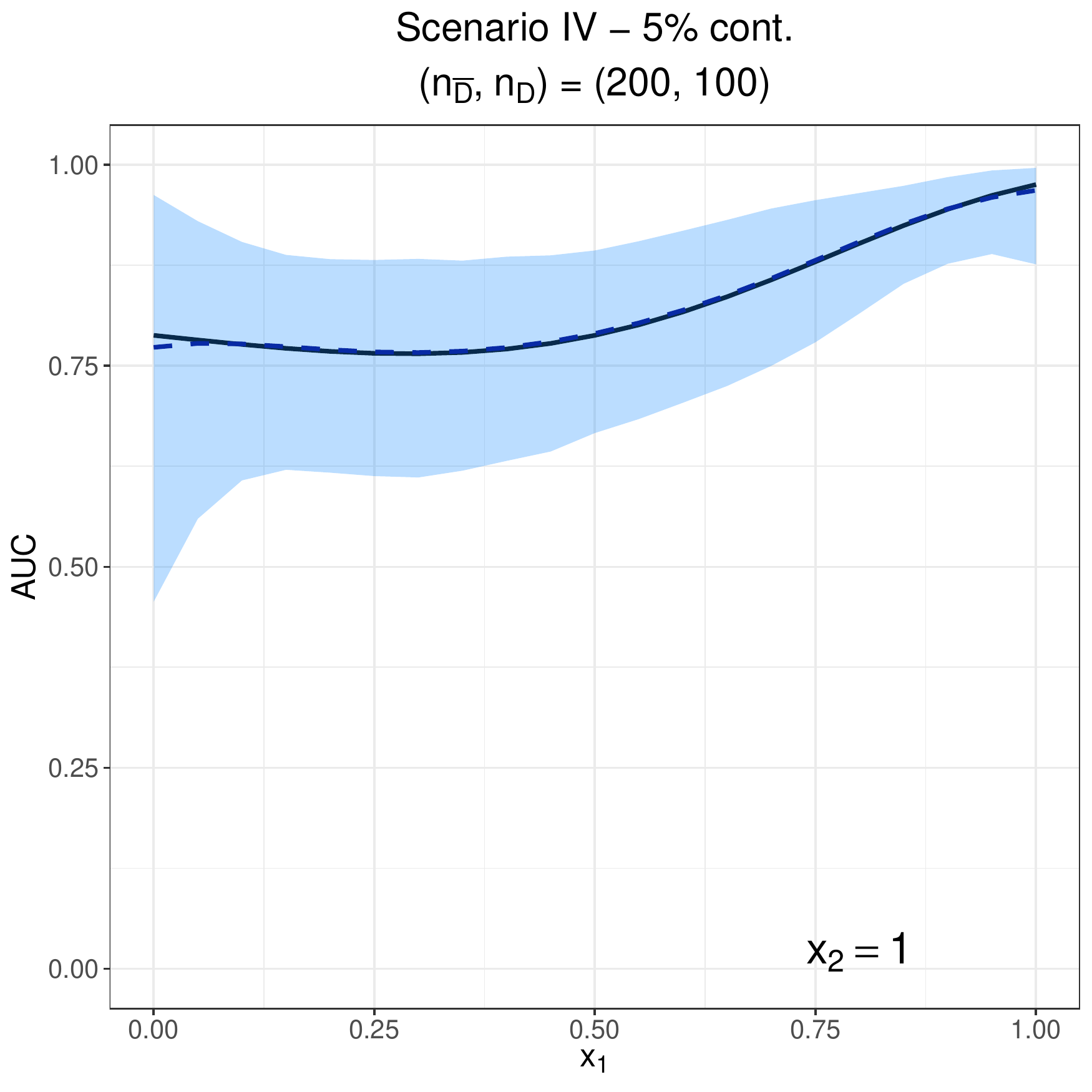}
\includegraphics[width = 5.35cm]{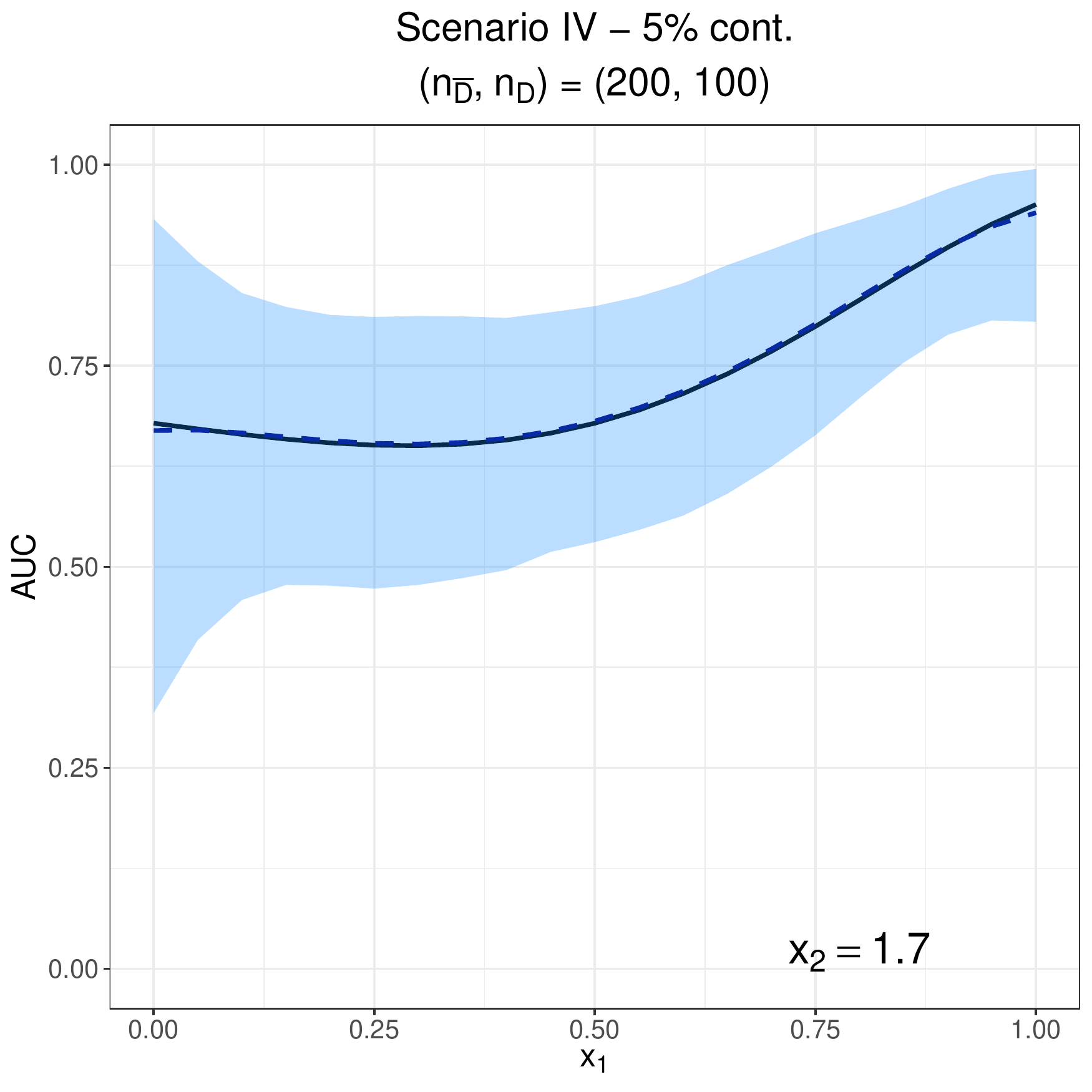}
}
\vspace{0.3cm}
\subfigure{
\includegraphics[width = 5.35cm]{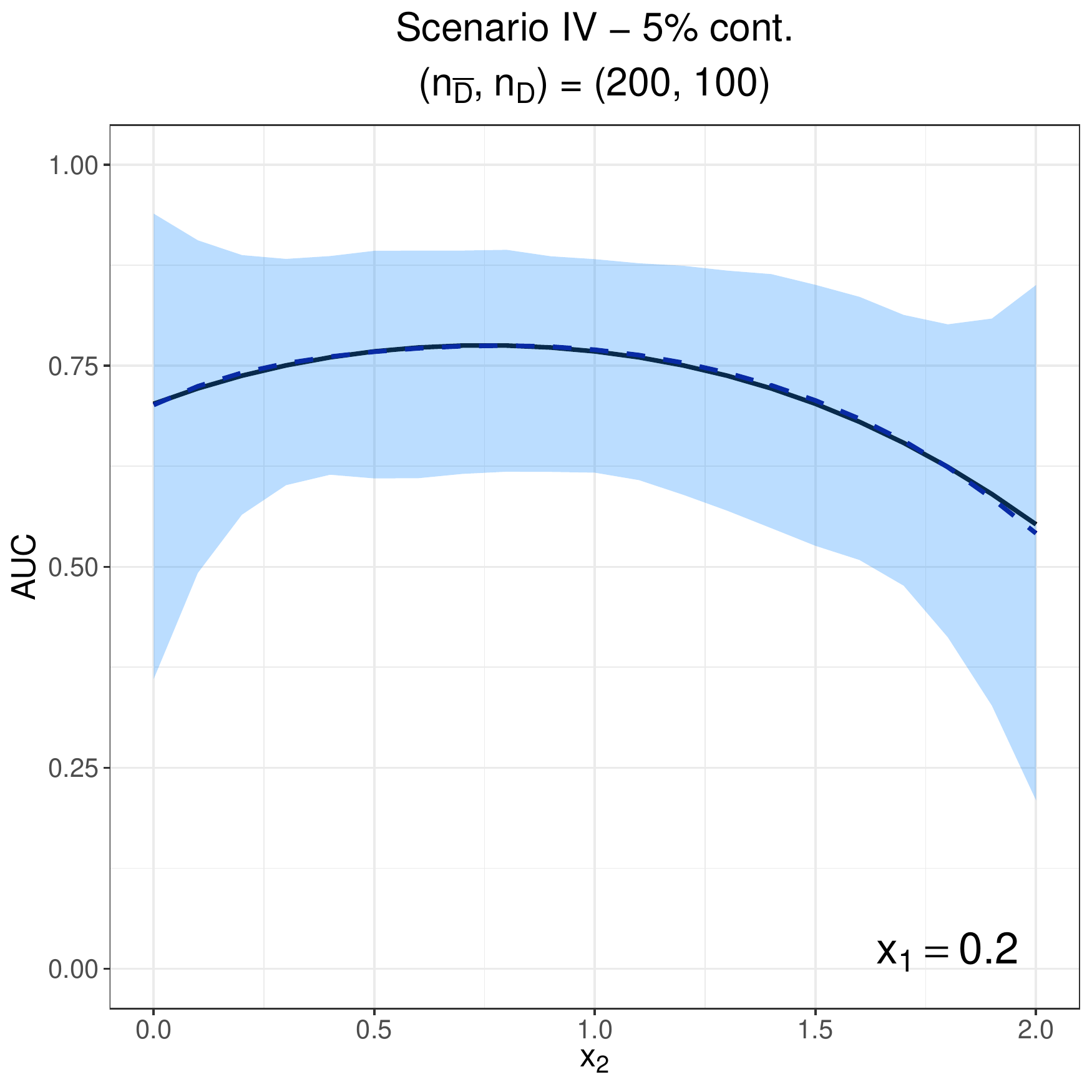}
\includegraphics[width = 5.35cm]{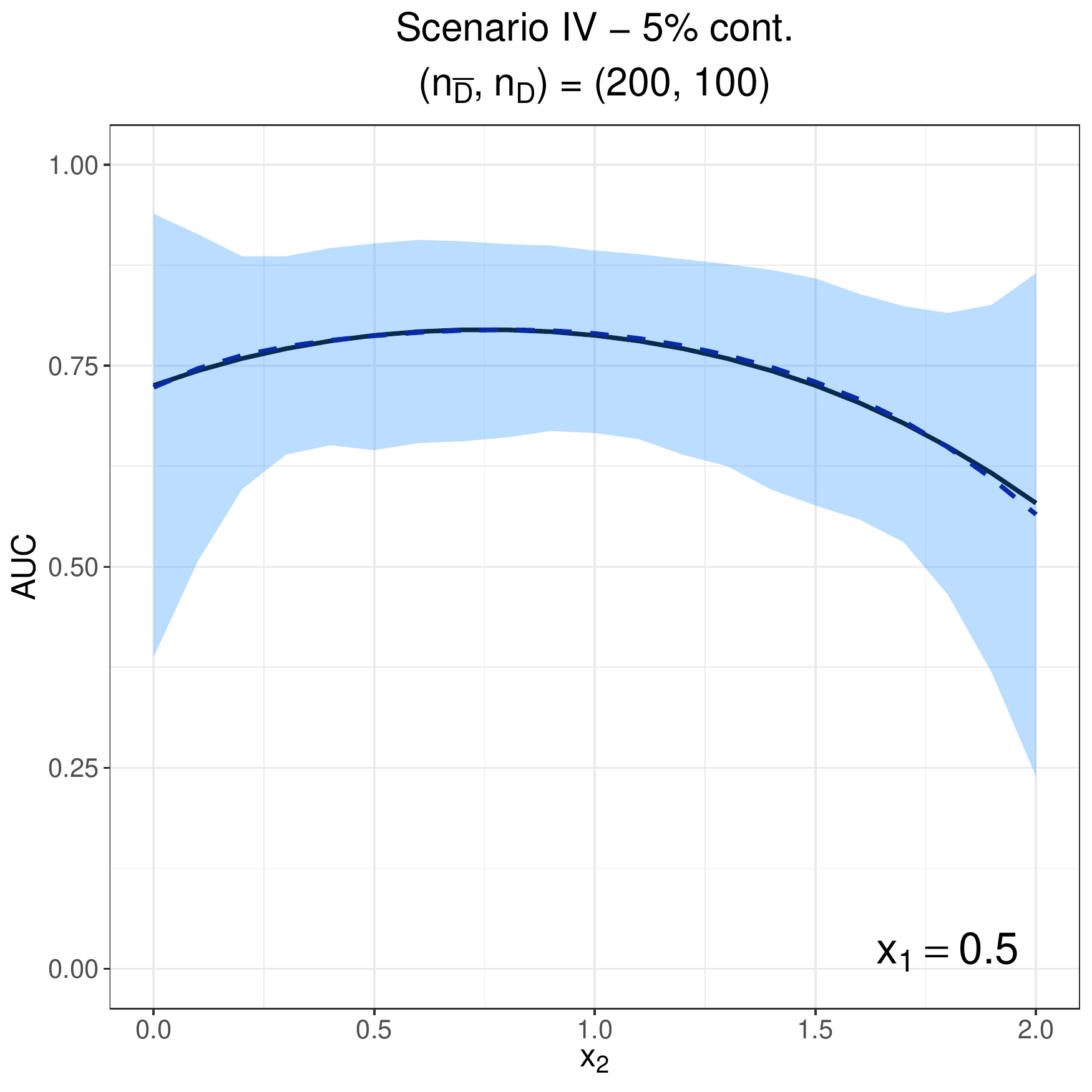}
\includegraphics[width = 5.35cm]{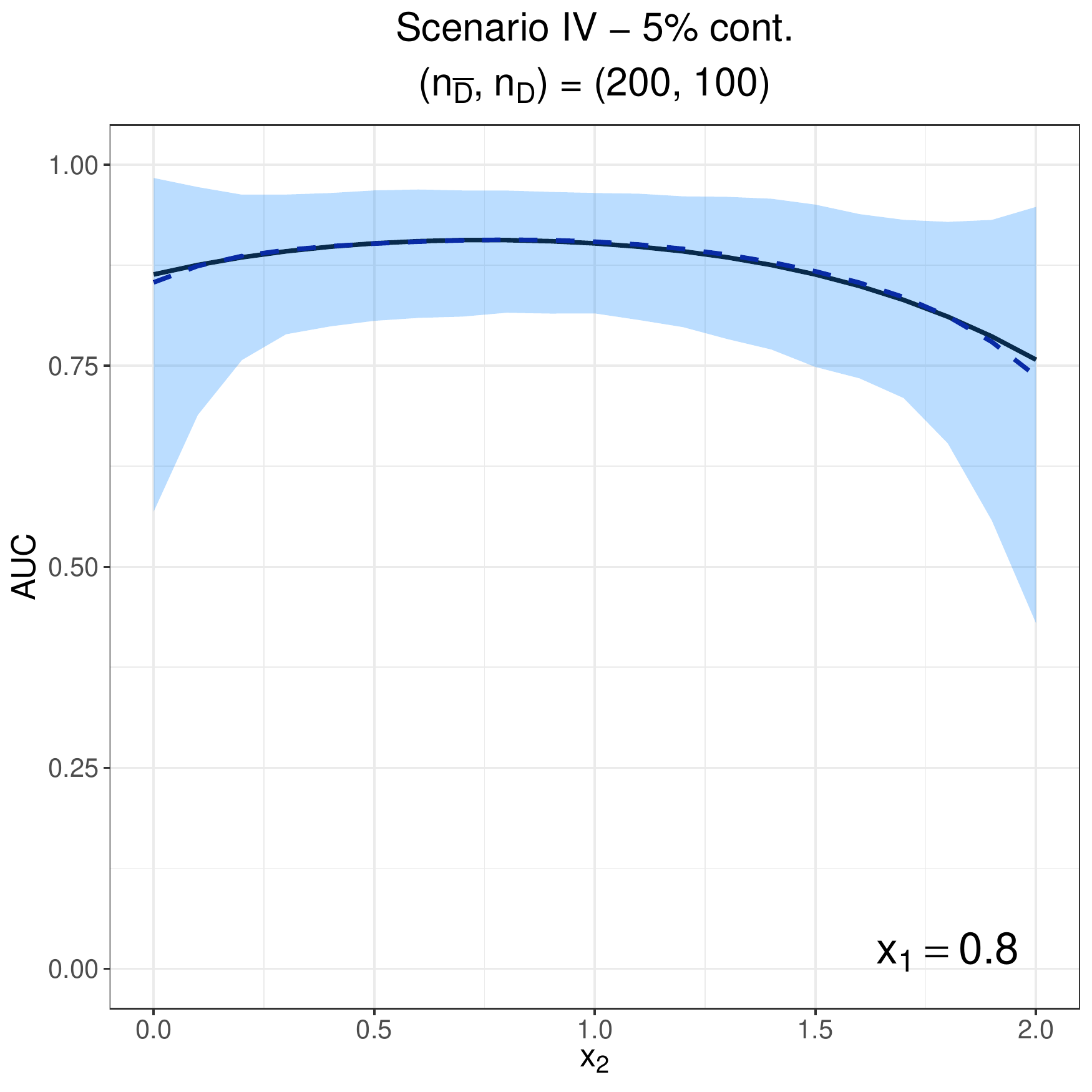}
}
\end{center}
\caption{\footnotesize{Scenario IV. Multiple profiles of the true covariate-specific AUC (solid line) versus the mean of the Monte Carlo estimates (dashed line) along with the $2.5\%$ and $97.5\%$ simulation quantiles (shaded area) for the case of $5\%$ contamination and for $(n_{\bar{D}}, n_D)=(200,100)$.}}
\label{simresultssc4}
\end{figure}

\begin{figure}[H]
\begin{center}
\subfigure{
\includegraphics[height = 7.25cm]{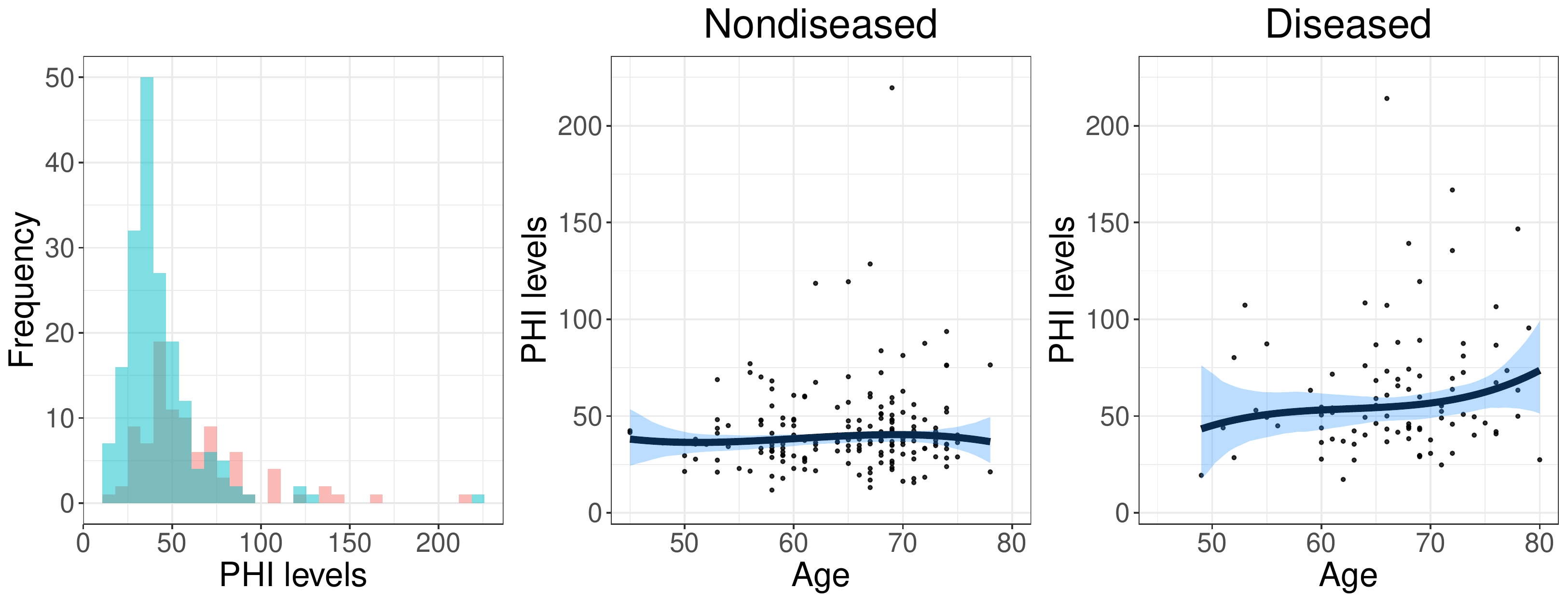}
}
\end{center}
\caption{\footnotesize{Left panel: histogram of the PHI scores from the nondiseased (blue) and diseased (red) populations. Middle and right panels: regression functions resulting from fitting our approach. The solid line is the point estimate, while the shaded areas represent the $95\%$ bootstrap confidence bands (based on $1000$ resamples).}}
\label{exploratory}
\end{figure}

\begin{figure}[H]
\begin{center}
\subfigure{
\includegraphics[height = 7.25cm]{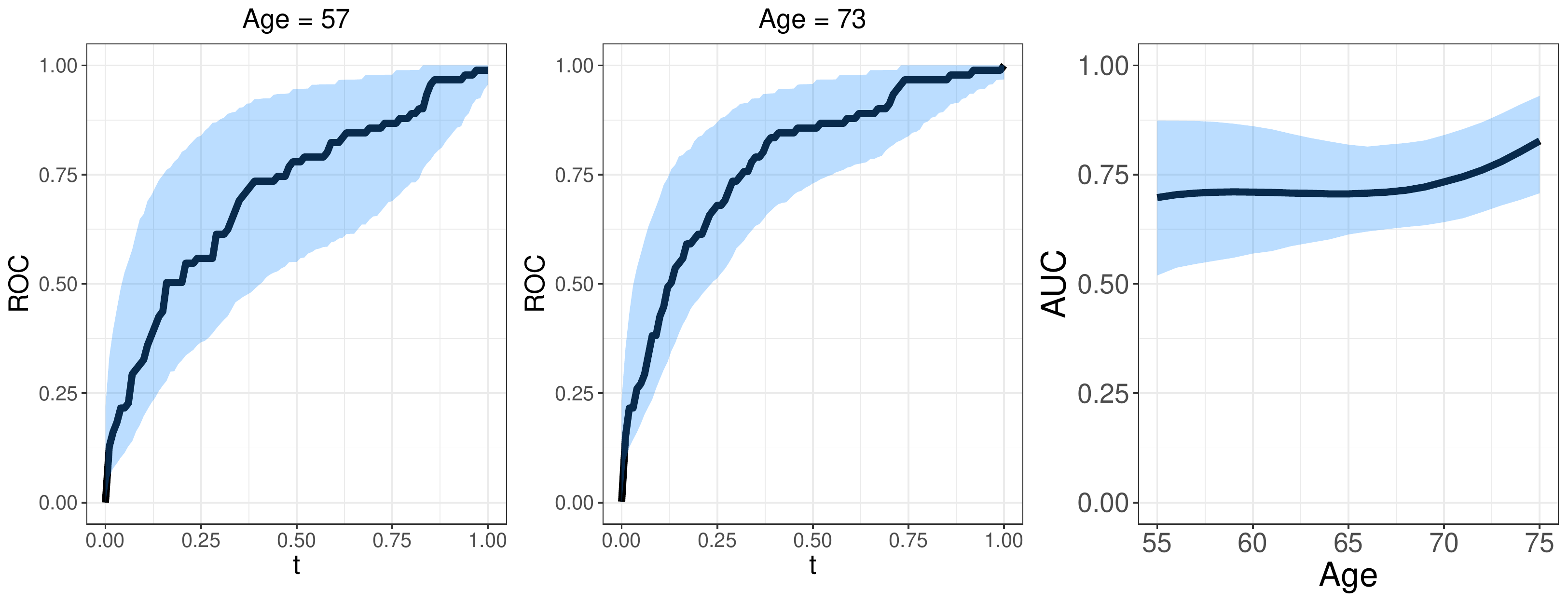}
}
\end{center}
\caption{\footnotesize{Left and middle panels: Two age-specific ROC curves. Right panel: Age-specific AUC. The solid line is the point estimate, while the shaded areas represent the $95\%$ bootstrap confidence bands (based on $1000$ resamples).}}
\label{rocsauc}
\end{figure}

\newpage

\section*{\large{\textsf{SUPPLEMENTARY MATERIALS}}}
\setcounter{figure}{0}  
In this supplementary file we provide additional figures and tables for the Simulation Study and Application sections in the main paper.

\begin{figure}[H]
	\begin{center}
		\subfigure{
			\includegraphics[width = 4.45cm]{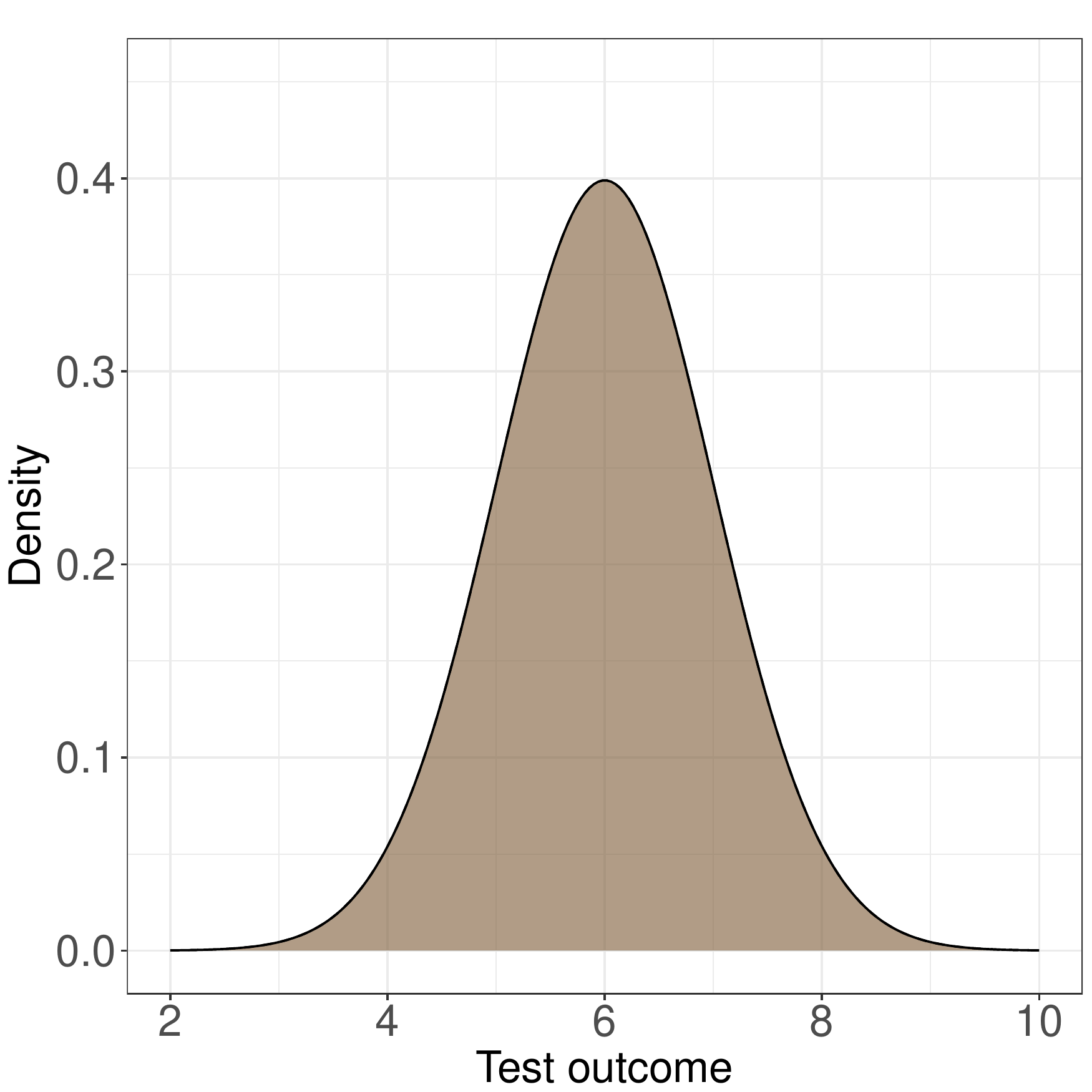}
			\includegraphics[width = 4.45cm]{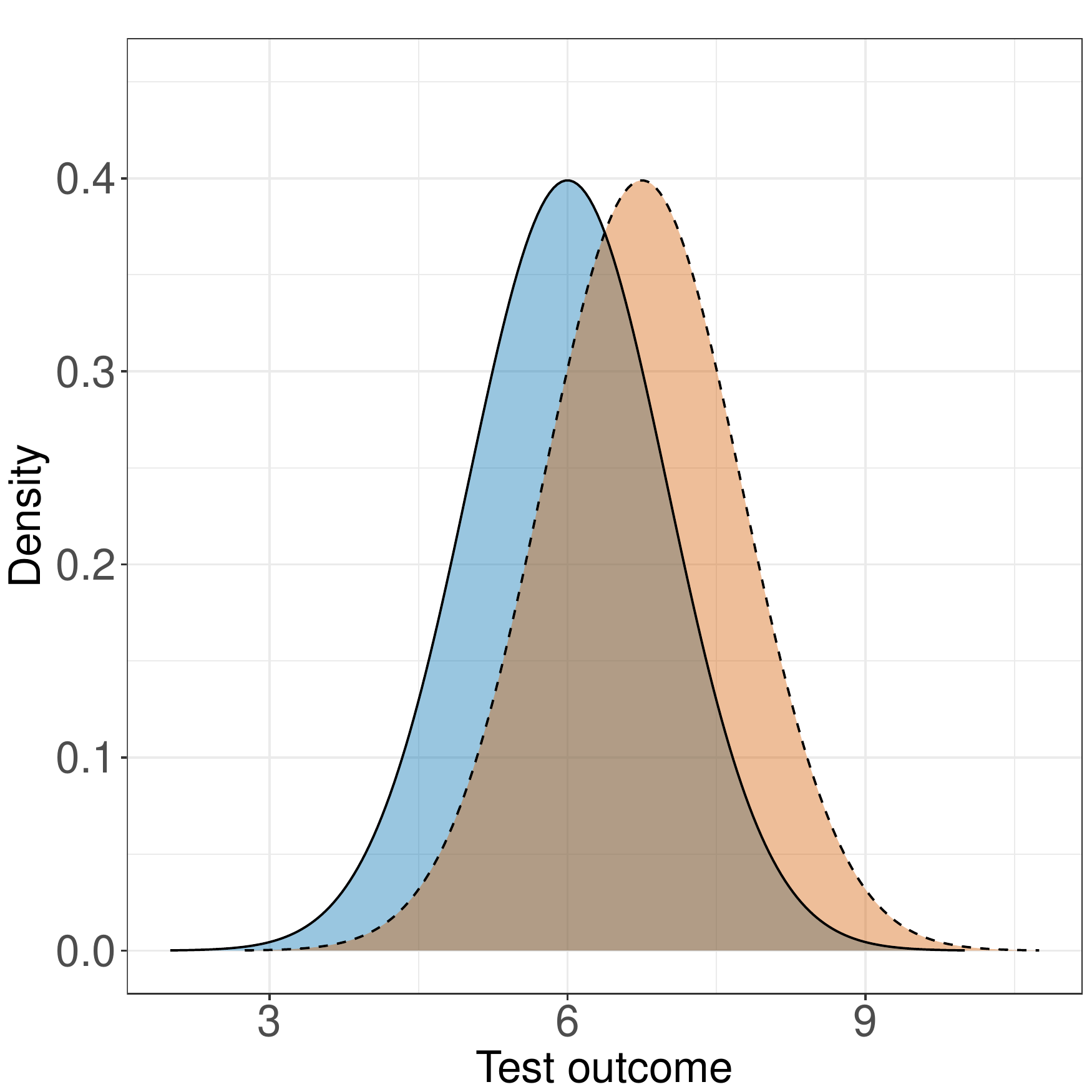}
			\includegraphics[width = 4.45cm]{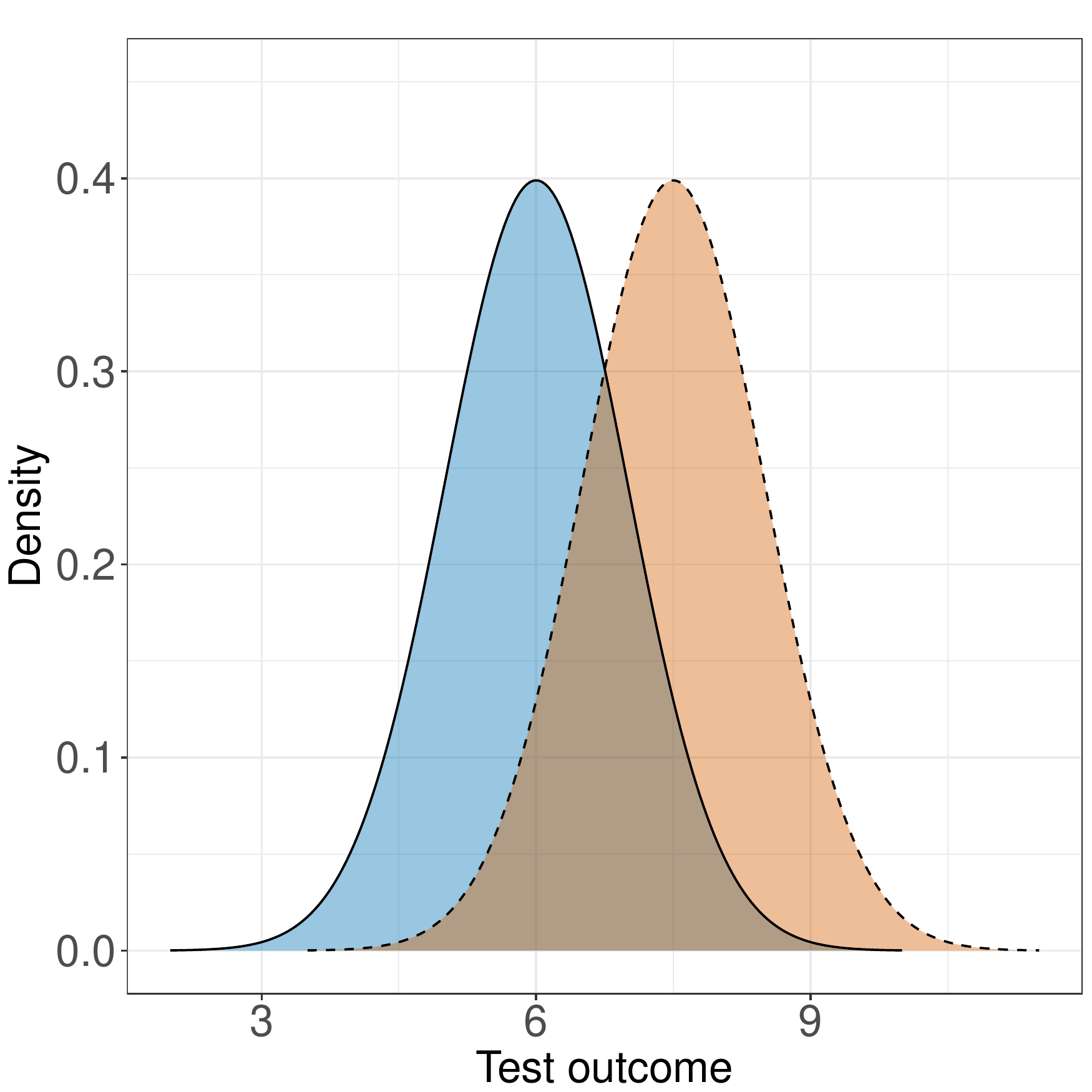}
			\includegraphics[width = 4.45cm]{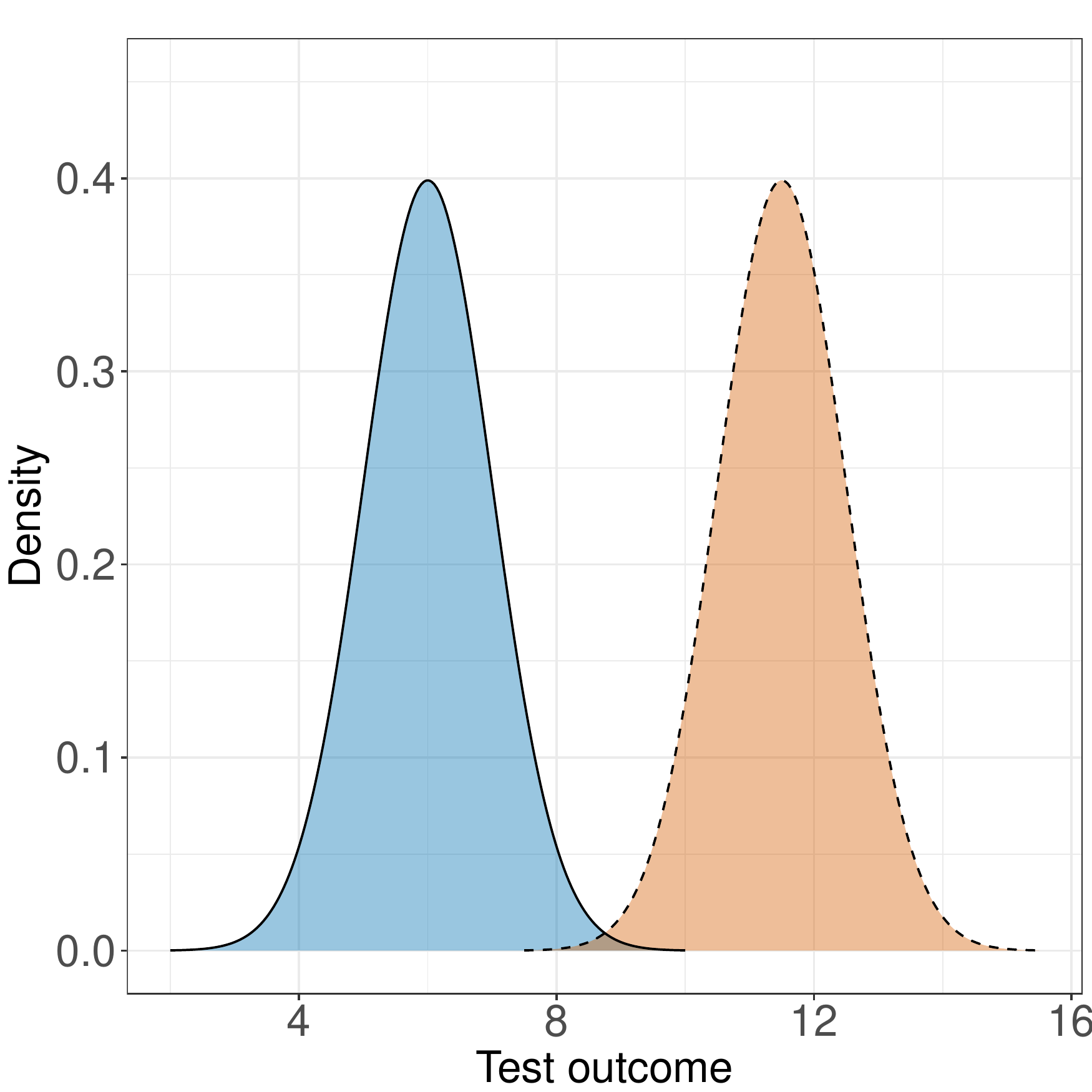}
		}
		\vspace{0.3cm}
		\subfigure{
			\includegraphics[height = 4.45cm]{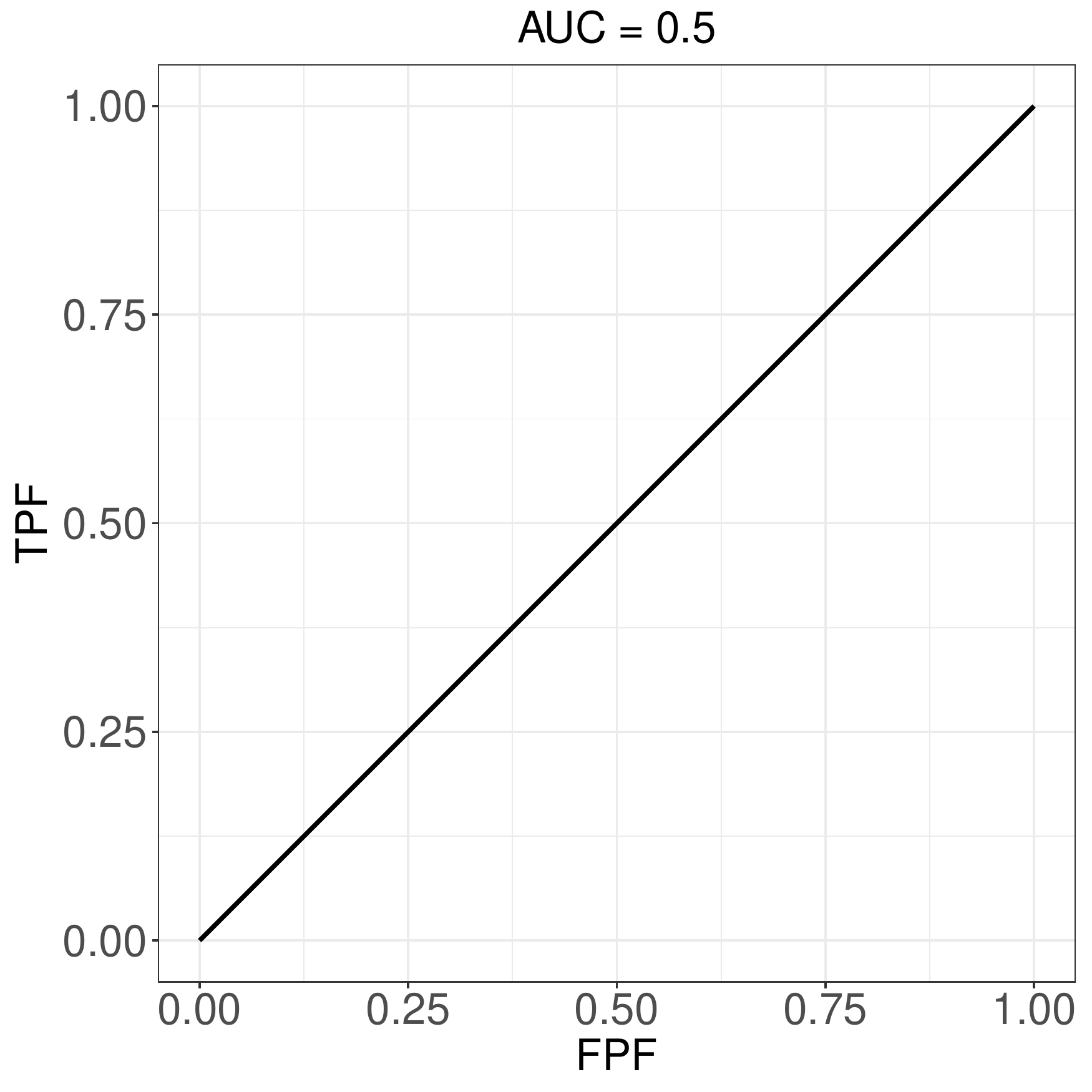}
			\includegraphics[height = 4.45cm]{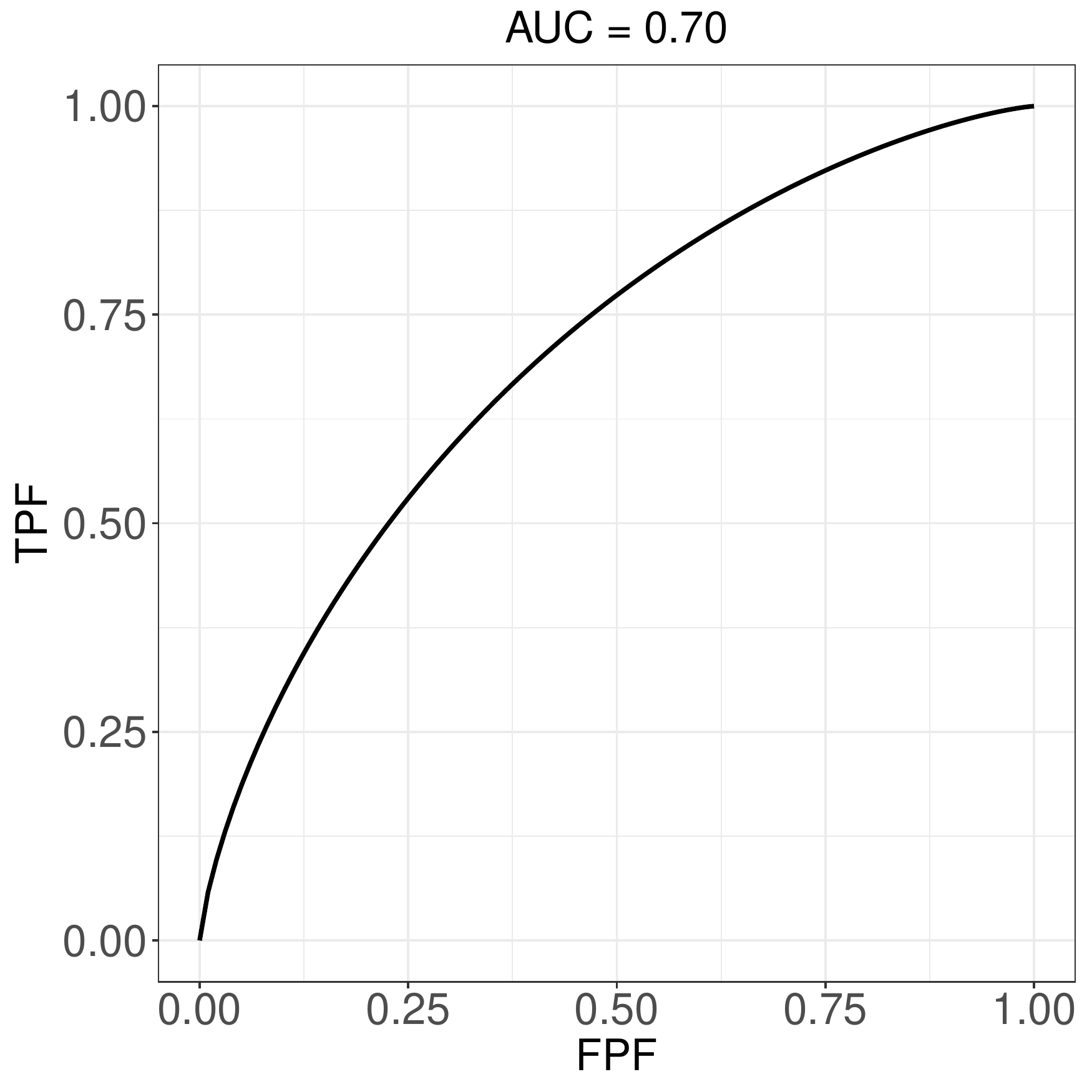}
			\includegraphics[height = 4.45cm]{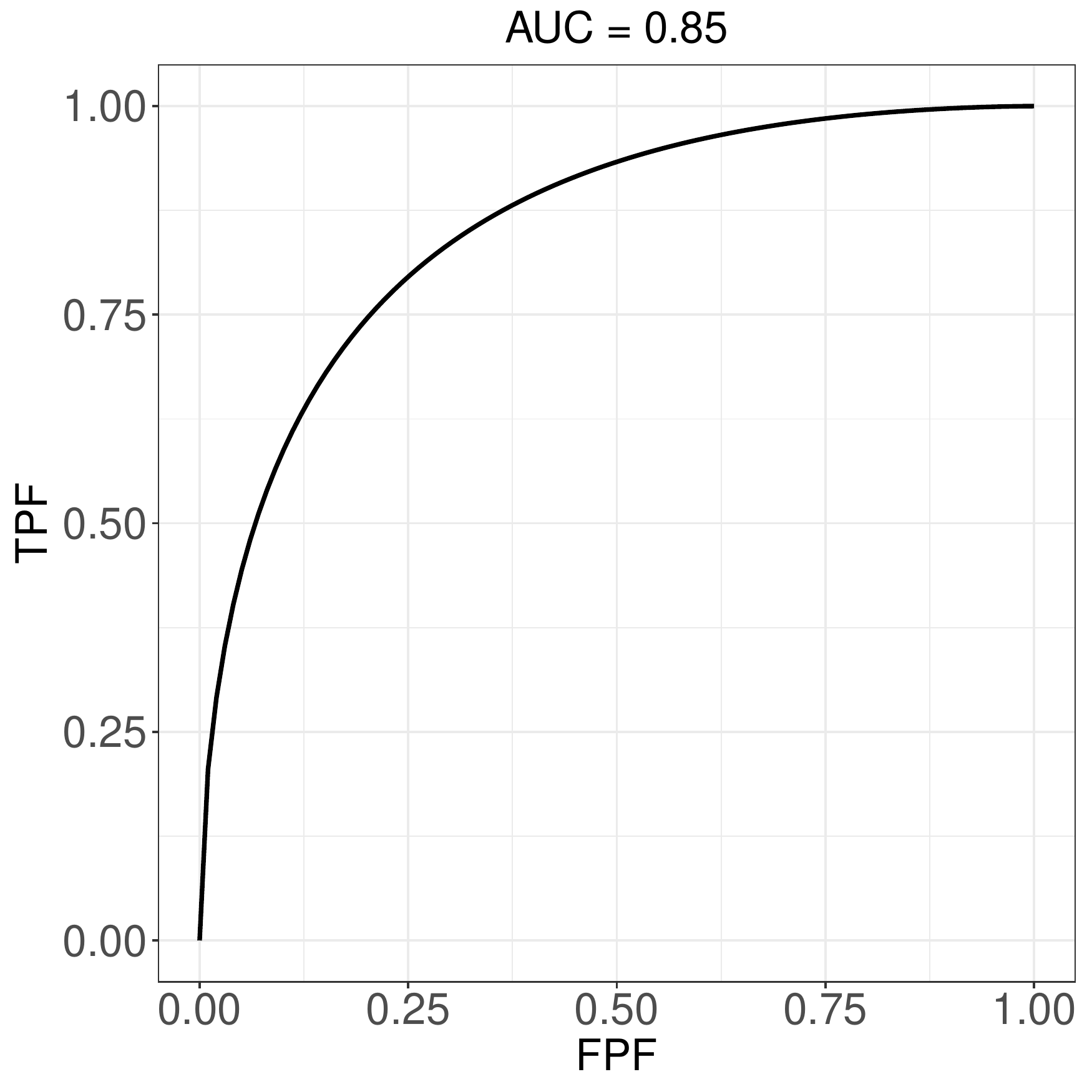}
			\includegraphics[height = 4.45cm]{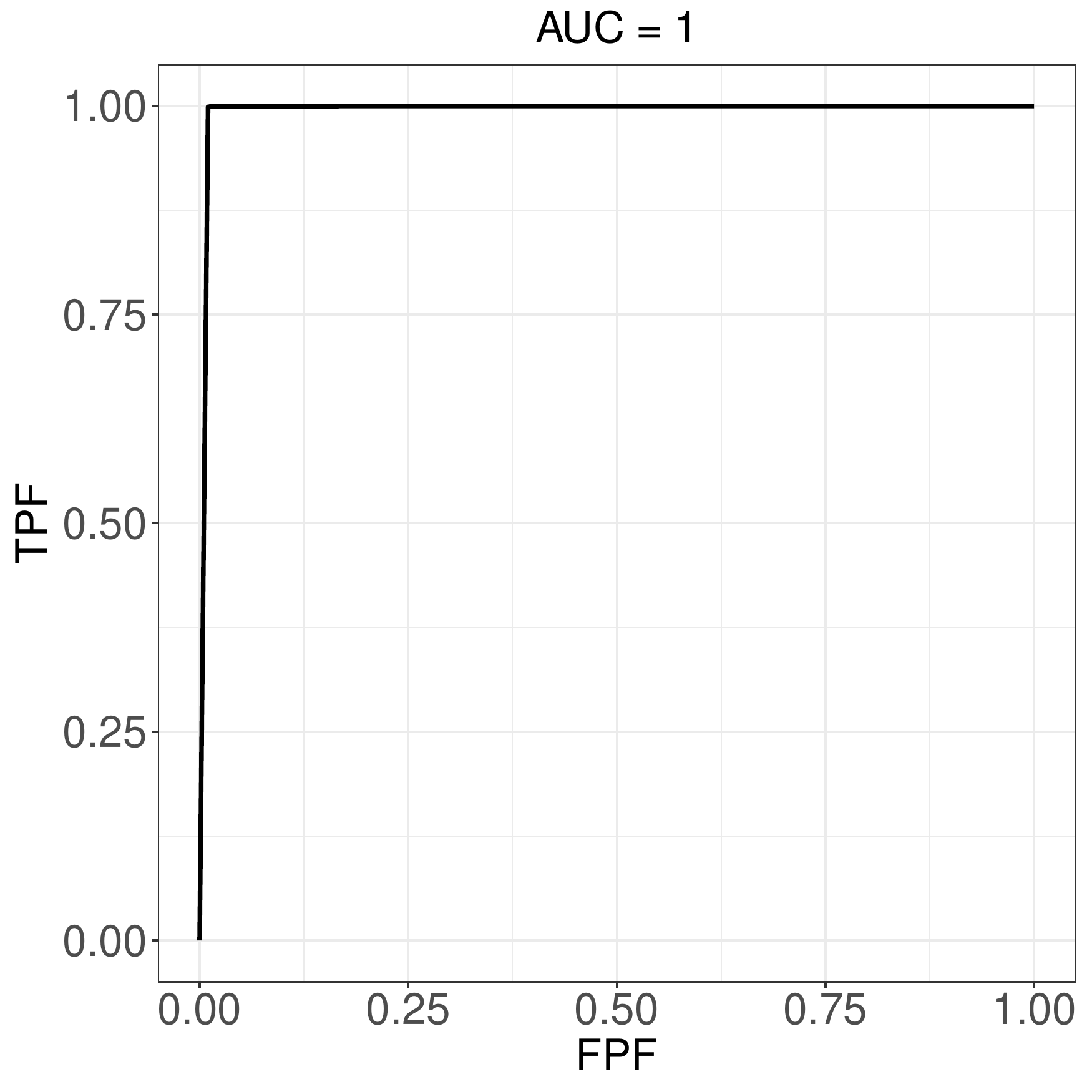}
		}
	\end{center}
	\caption{\footnotesize{Hypothetical densities of test outcomes in the diseased (dotted line, orange) and nondiseased (solid line, blue) populations (top) along with the corresponding ROC curves (bottom).}}
\end{figure}

\begin{figure}[H]
	\begin{center}
		\subfigure{
			\includegraphics[height = 5.35cm]{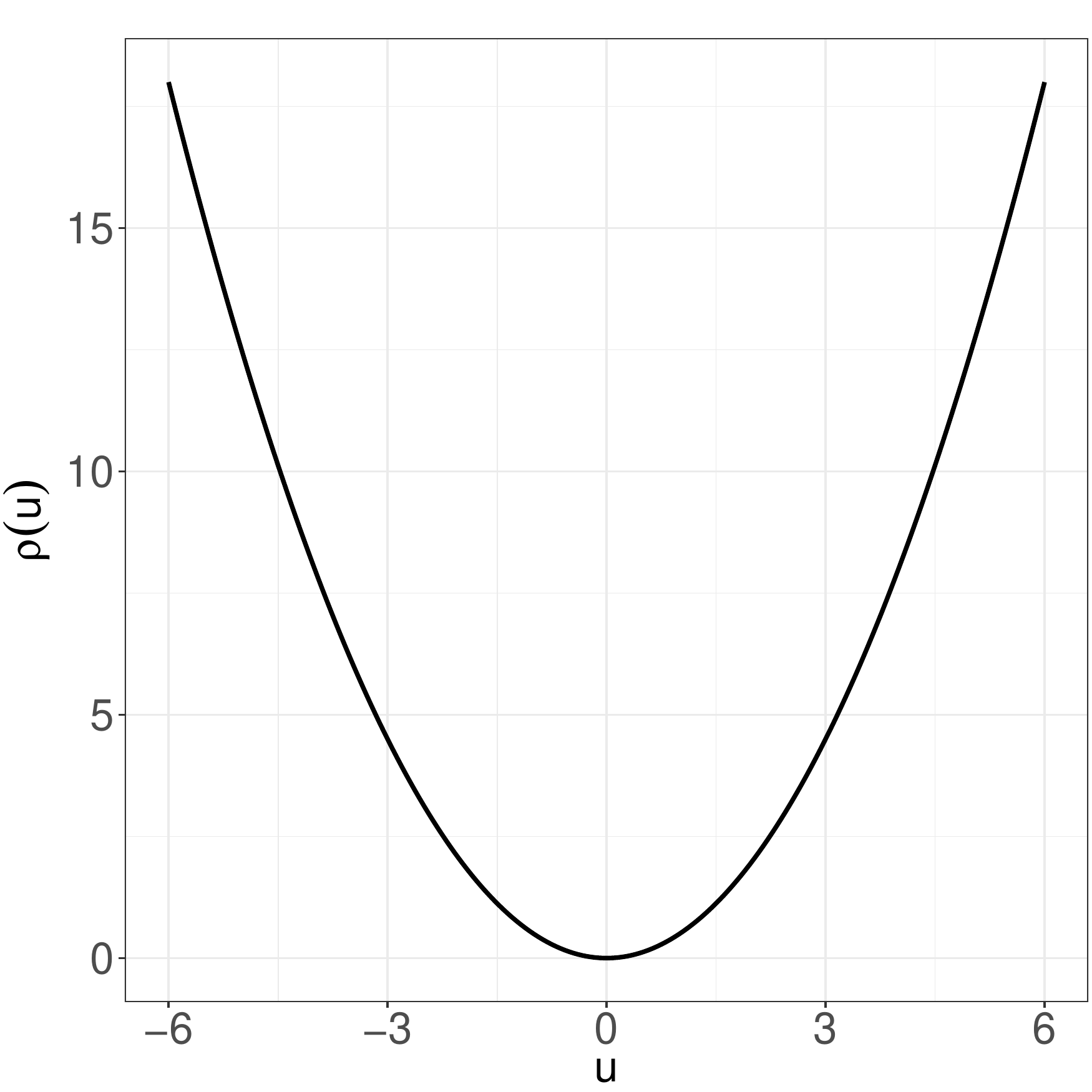}
			\includegraphics[height = 5.35cm]{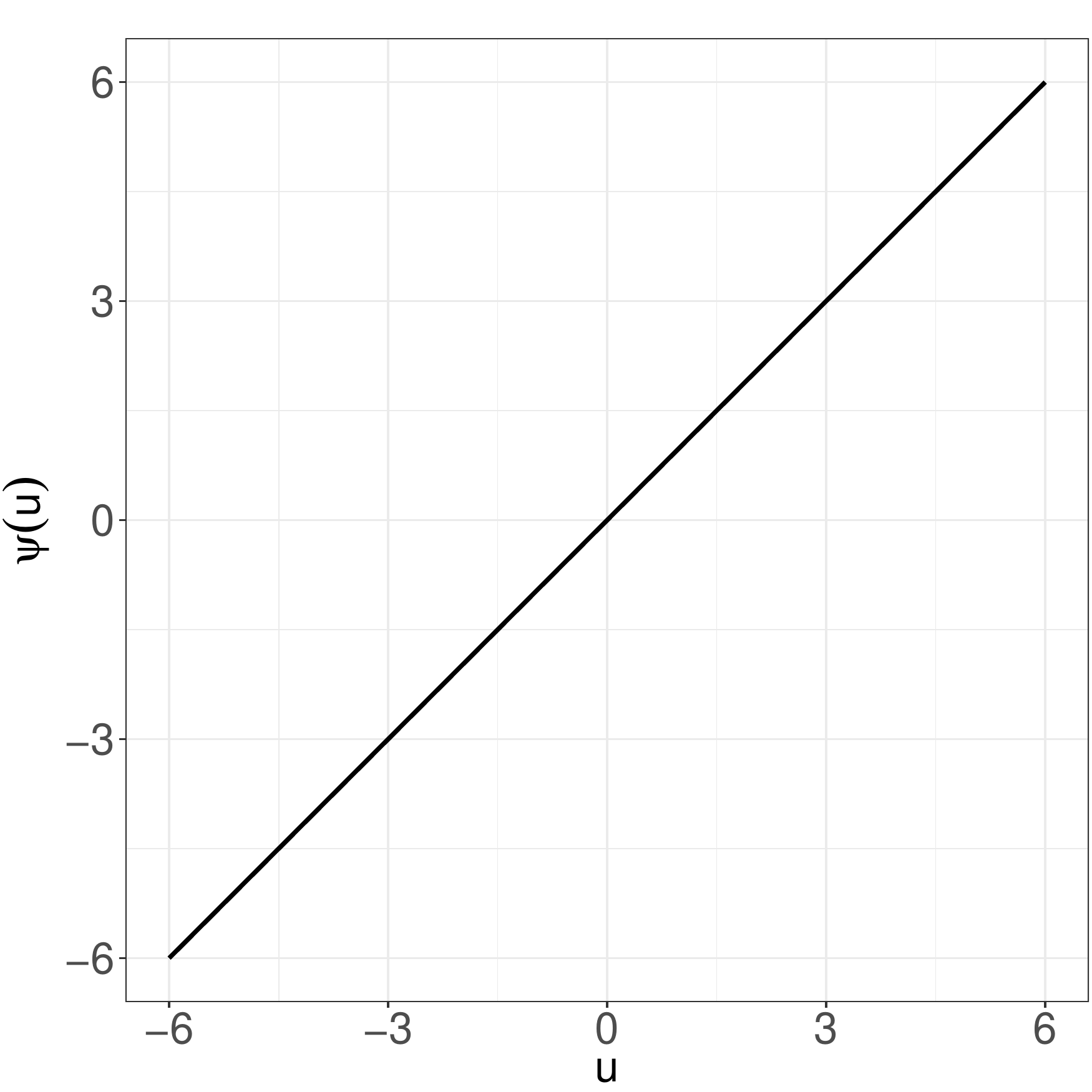}
			\includegraphics[height = 5.35cm]{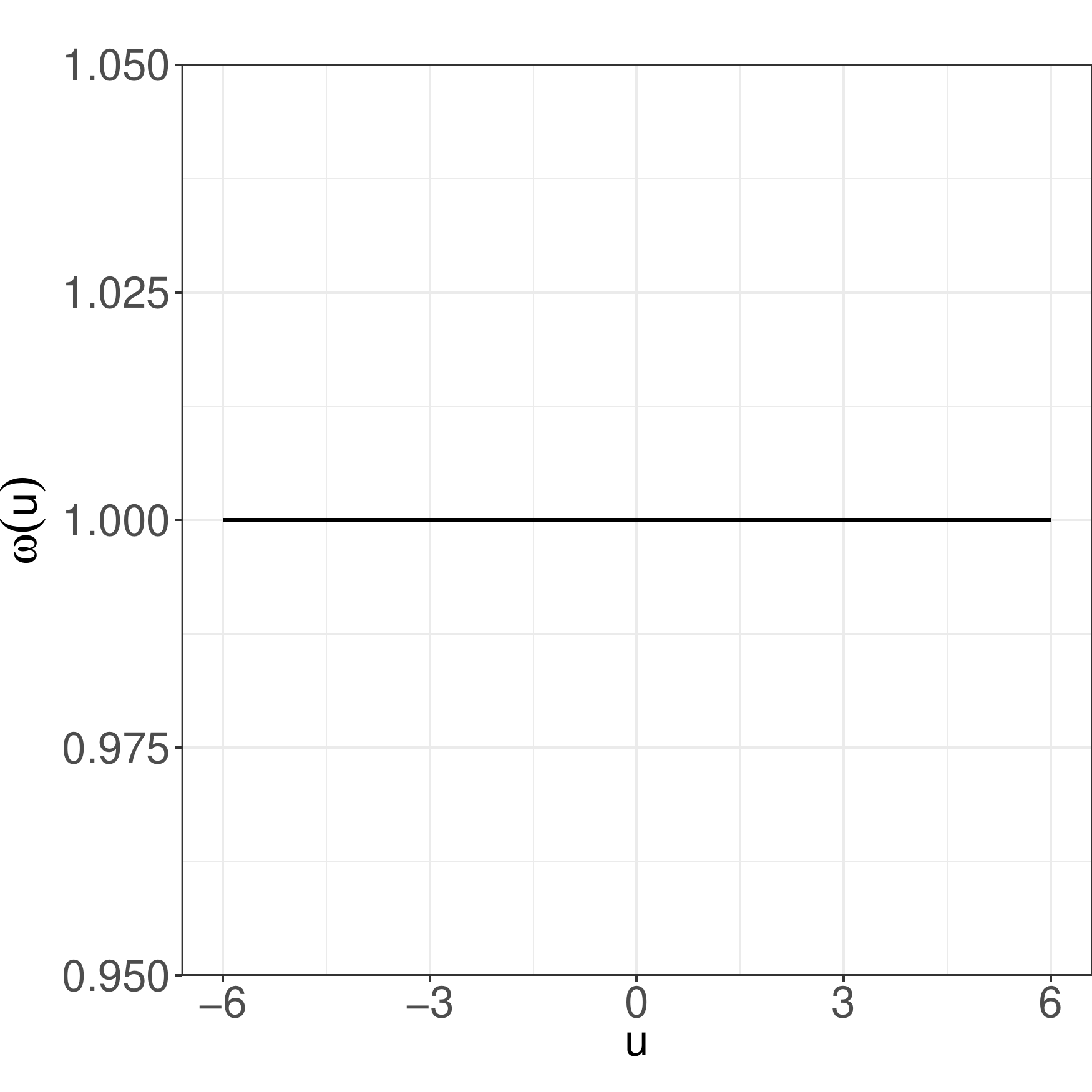}
		}
		\vspace{0.3cm}				
		\subfigure{
			\includegraphics[width = 5.35cm]{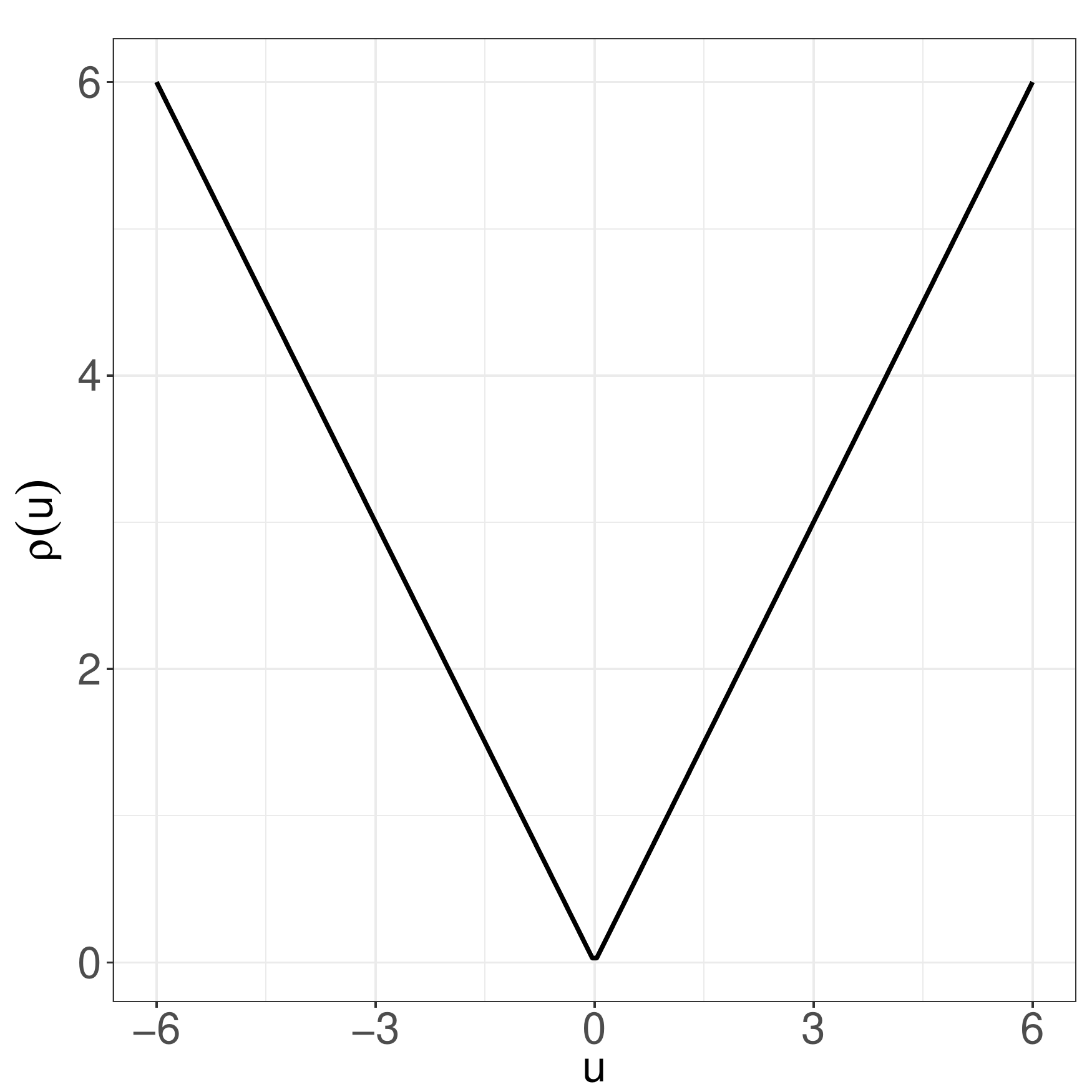}
			\includegraphics[width = 5.35cm]{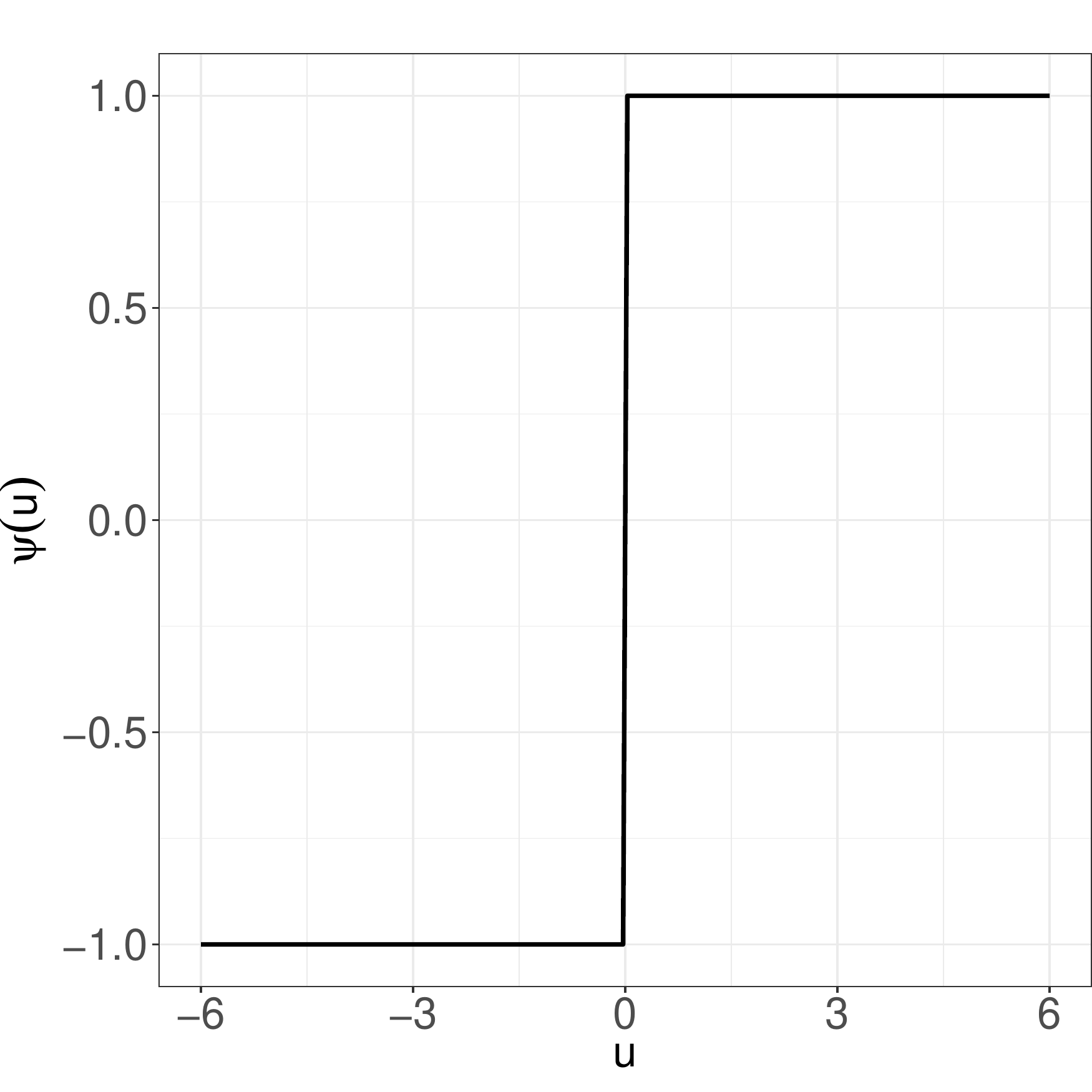}
			\includegraphics[width = 5.35cm]{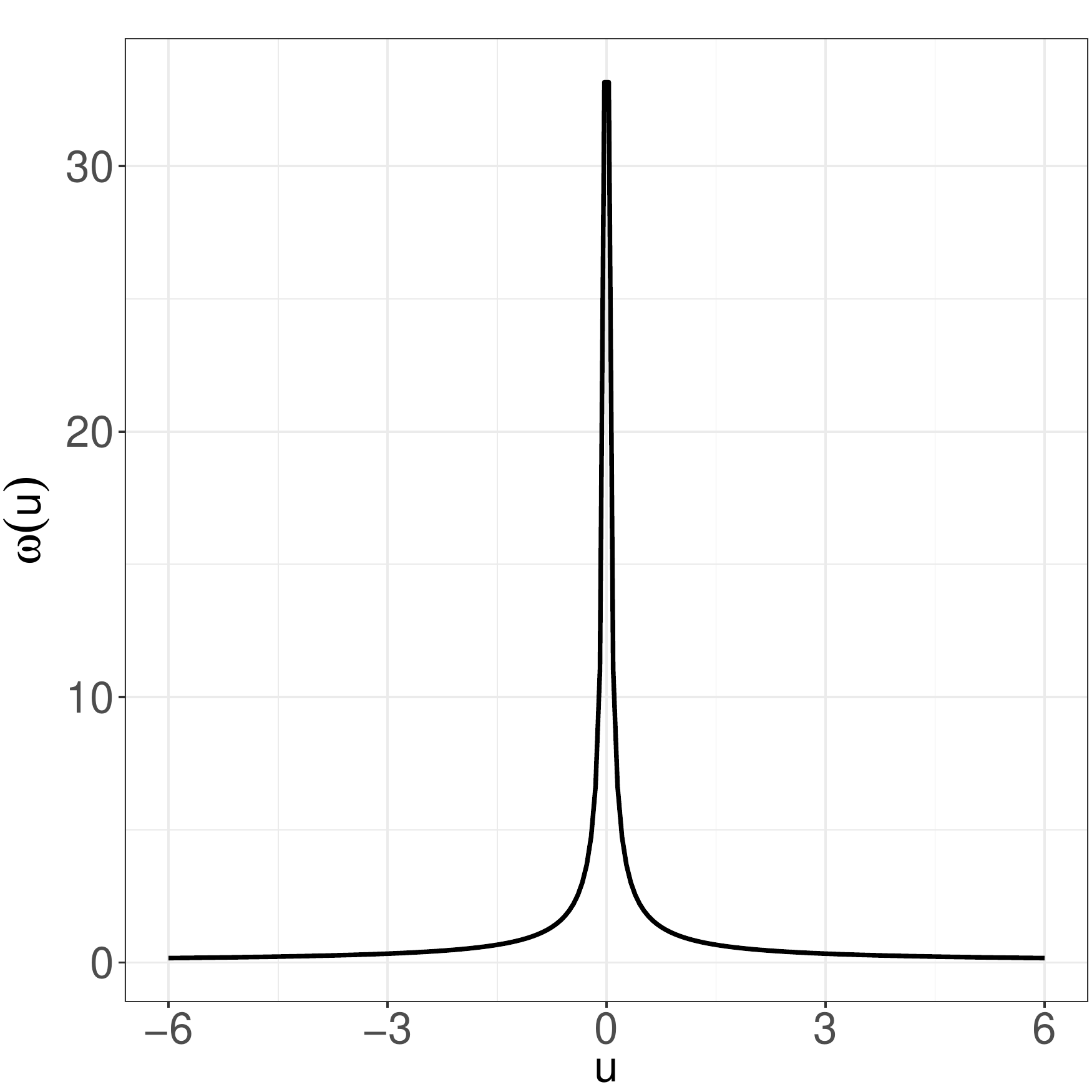}
		}
		\vspace{0.3cm}		
		\subfigure{
			\includegraphics[width = 5.35cm]{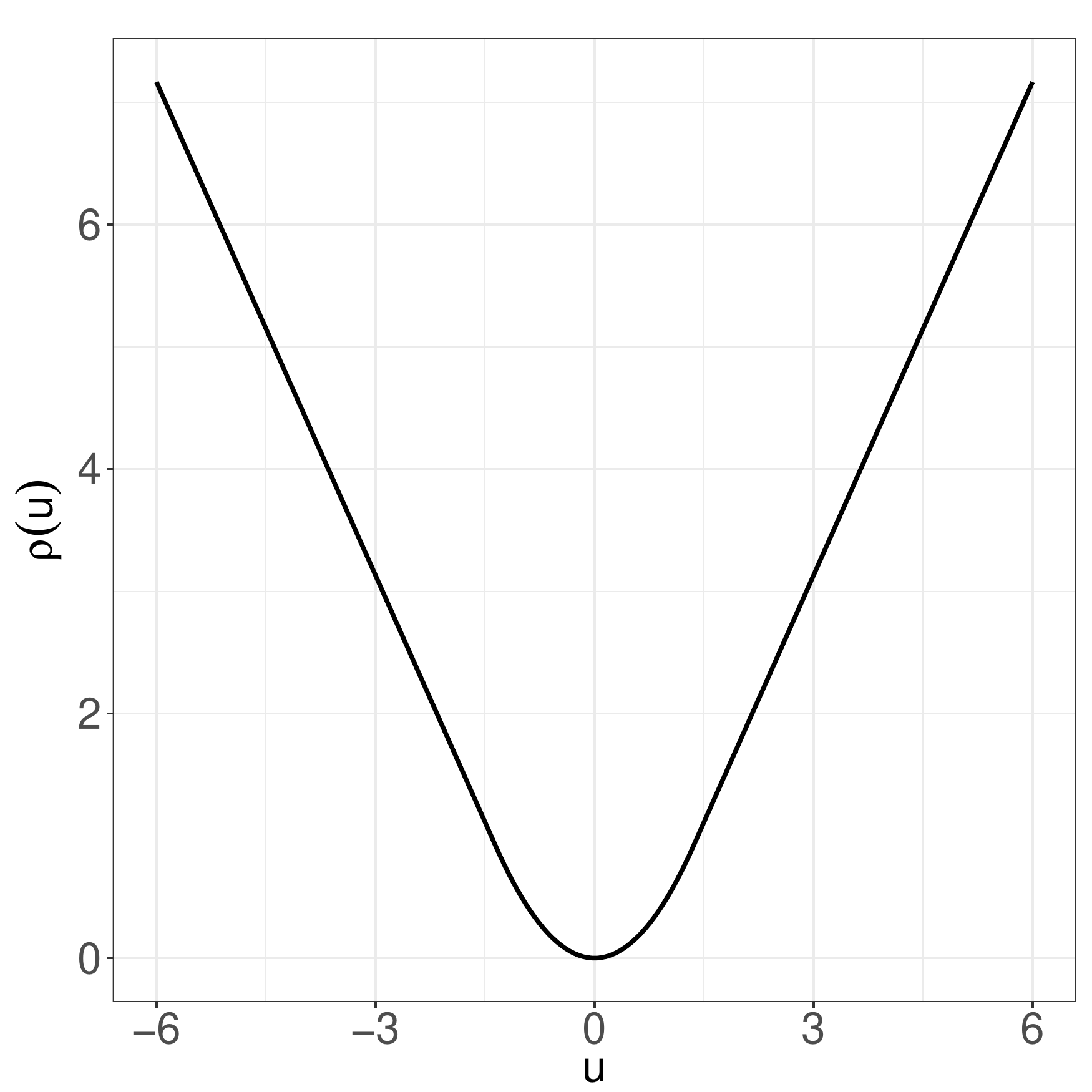}
			\includegraphics[width = 5.35cm]{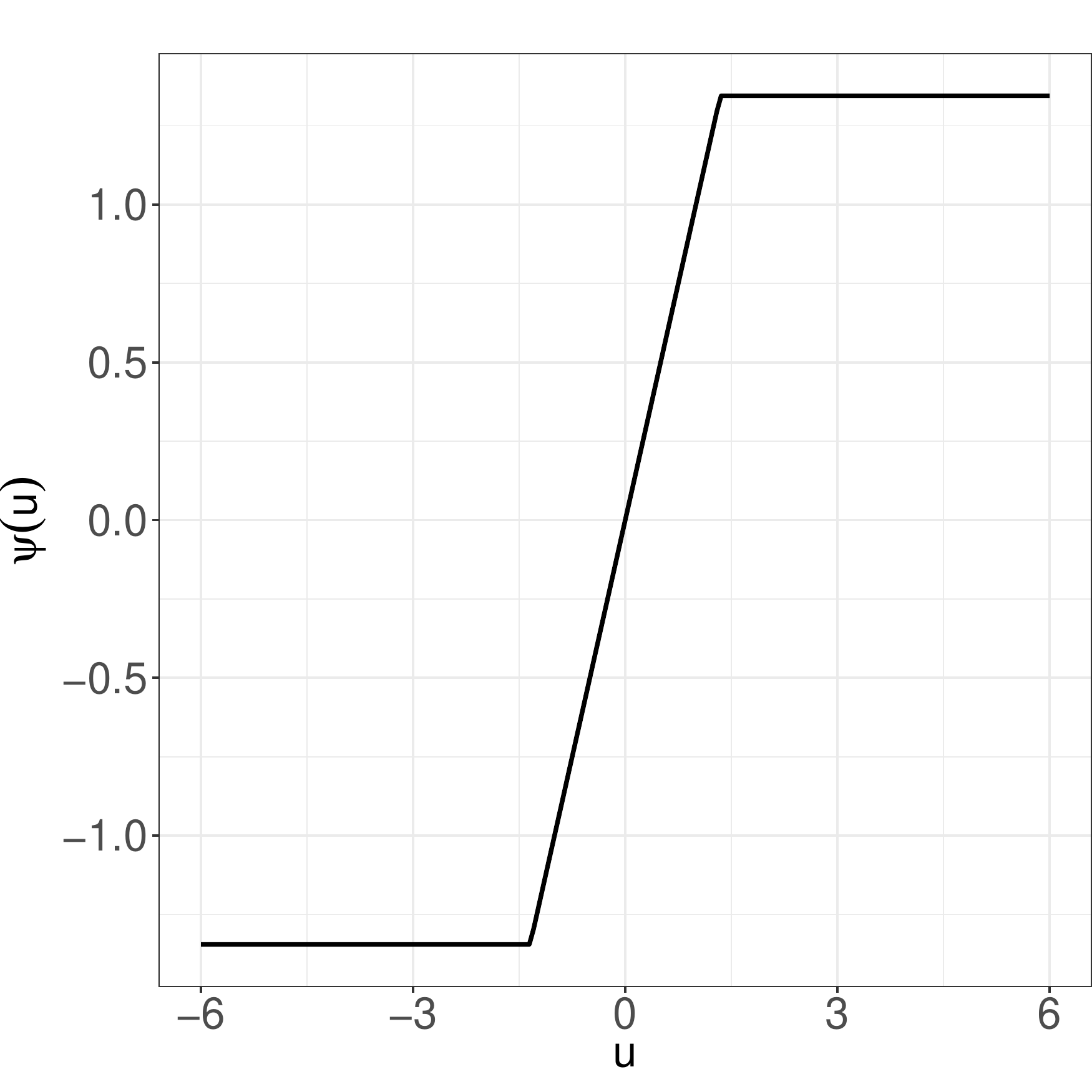}
			\includegraphics[width = 5.35cm]{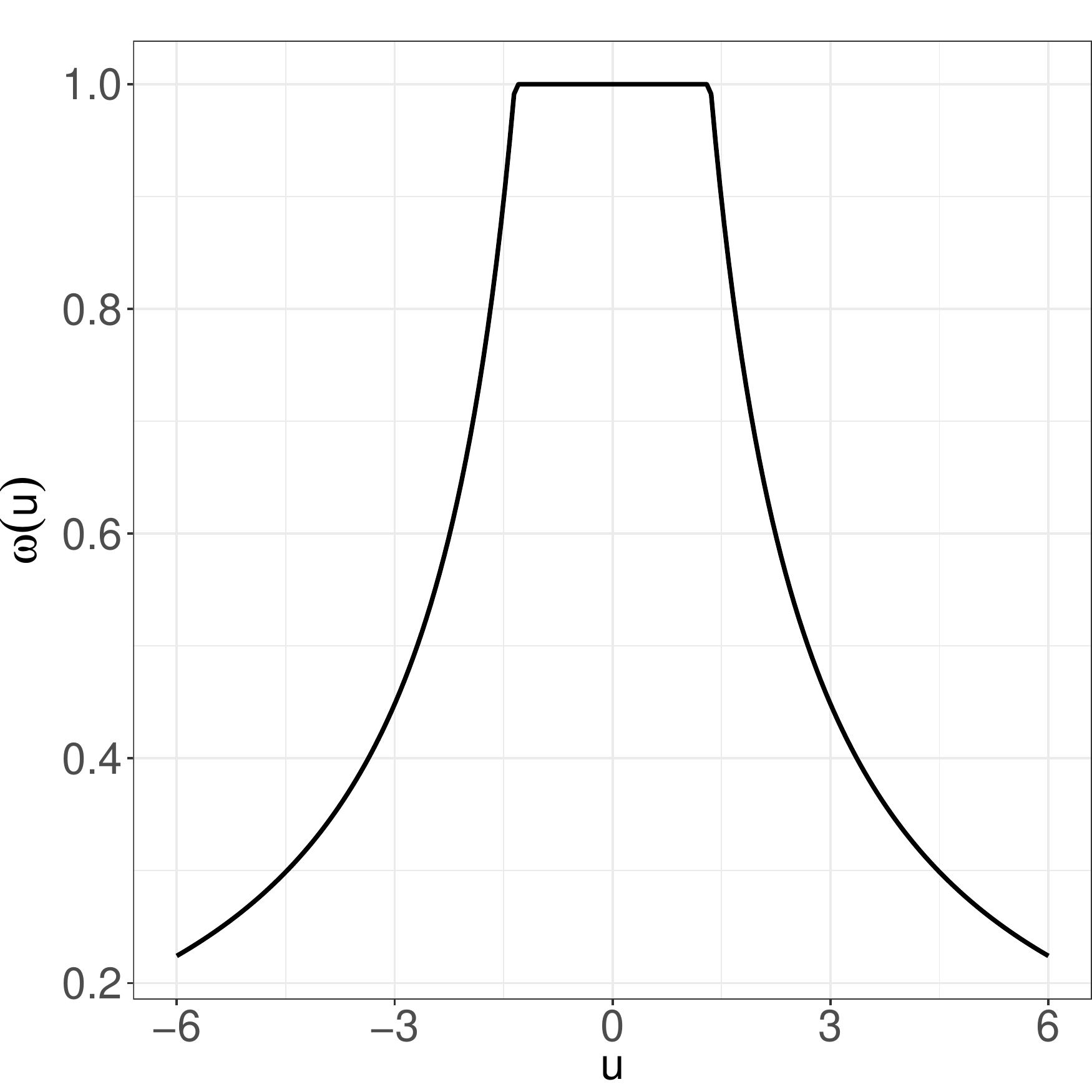}
		}
	\end{center}
	\caption{\footnotesize{$\rho$, $\psi$, and $\omega$ functions for the least-squares (first row), least absolute deviations (second row) and Huber (third row) estimators.}}
\end{figure}

\begin{figure}[H]
	\begin{center}
		\subfigure{
			\includegraphics[width = 4.65cm]{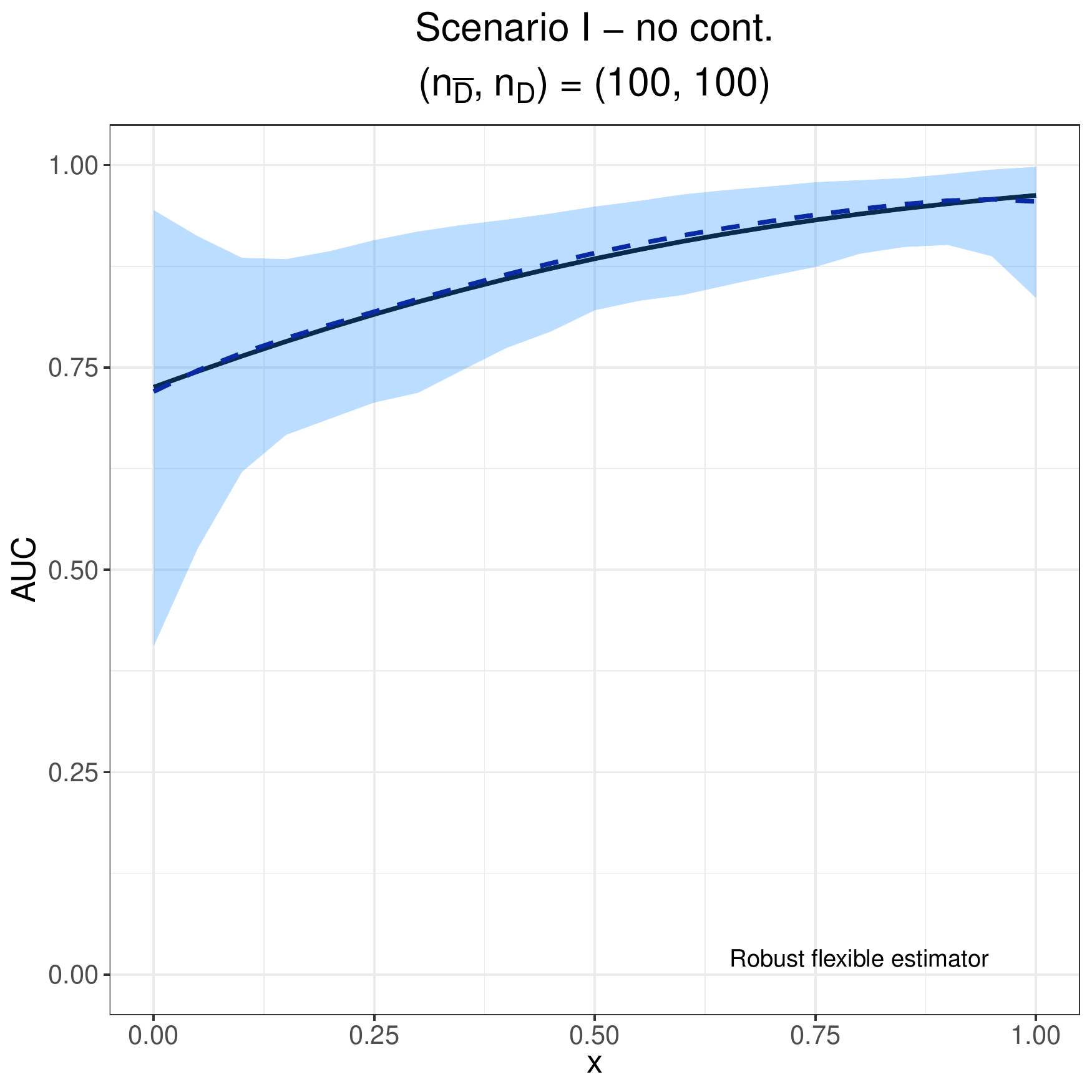}
			\includegraphics[width = 4.65cm]{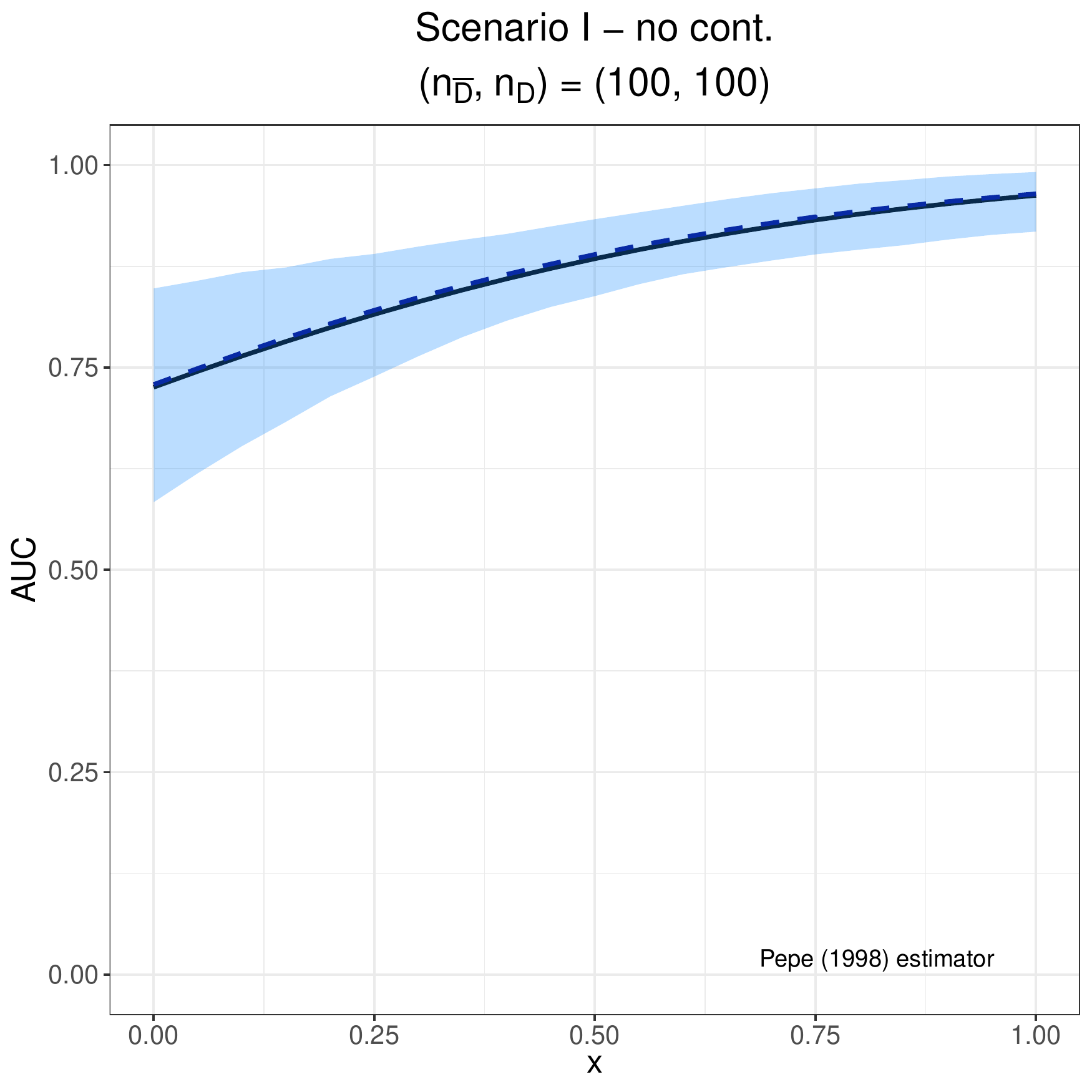}
			\includegraphics[width = 4.65cm]{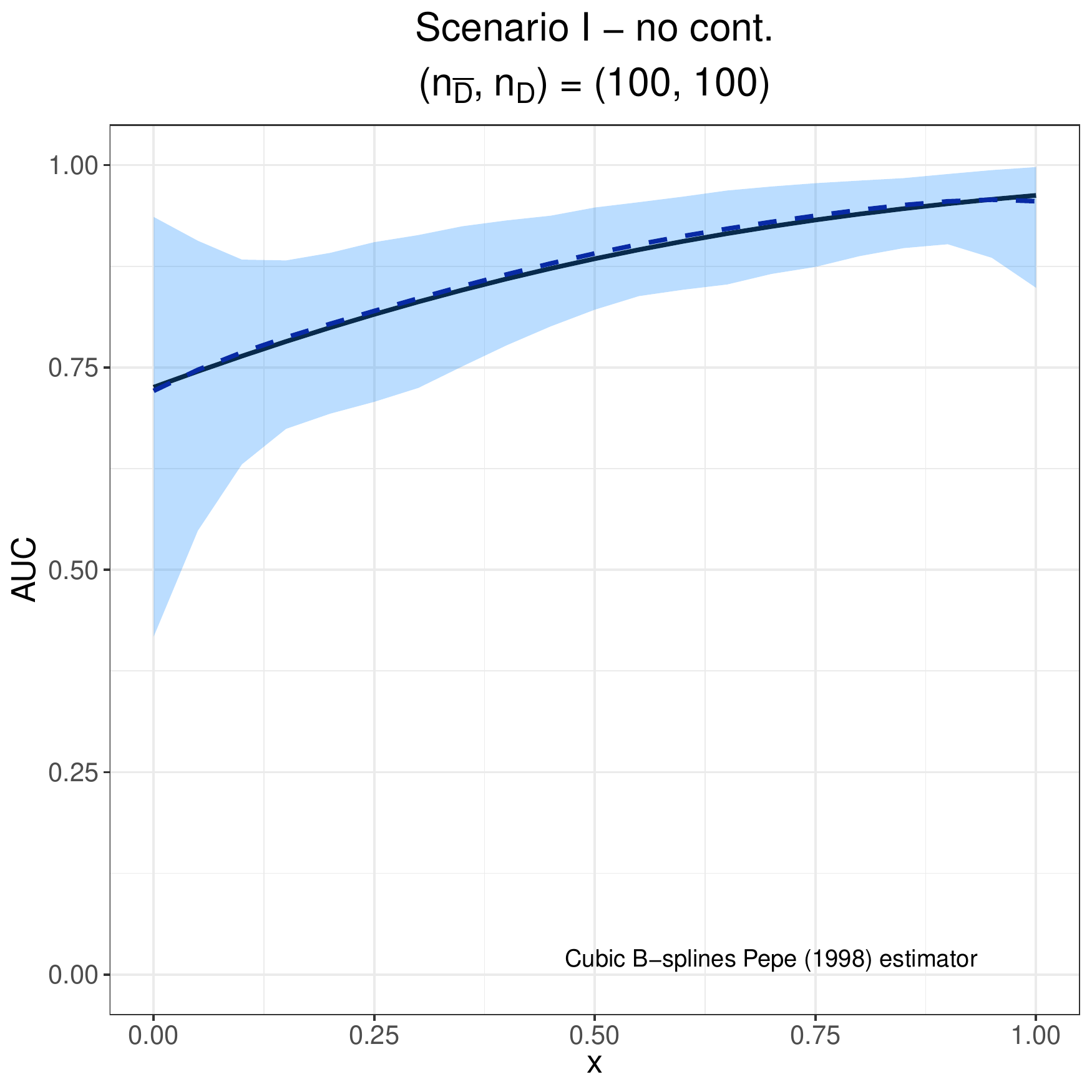}
			\includegraphics[width = 4.65cm]{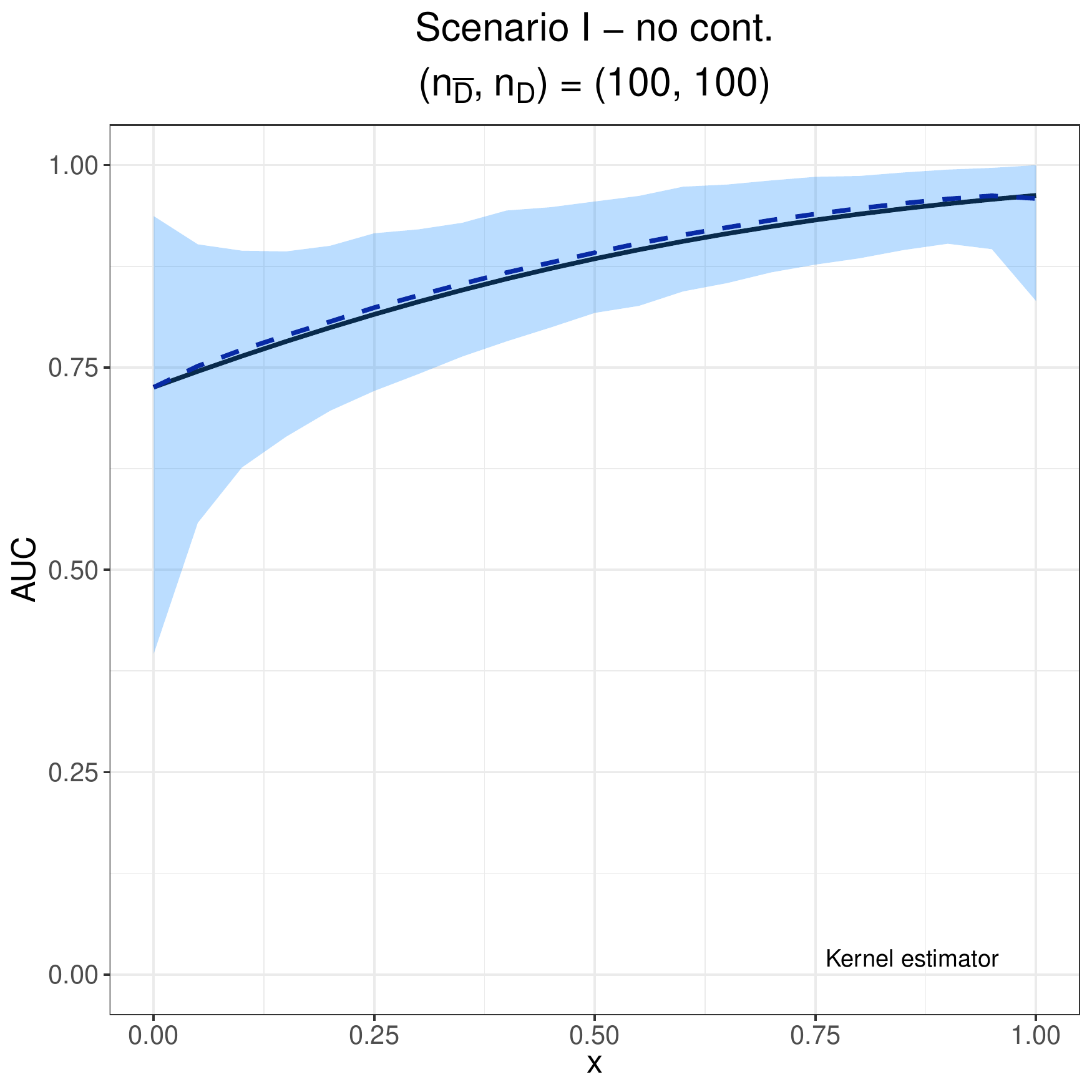}
		}
		\vspace{0.3cm}
		\subfigure{
			\includegraphics[width = 4.65cm]{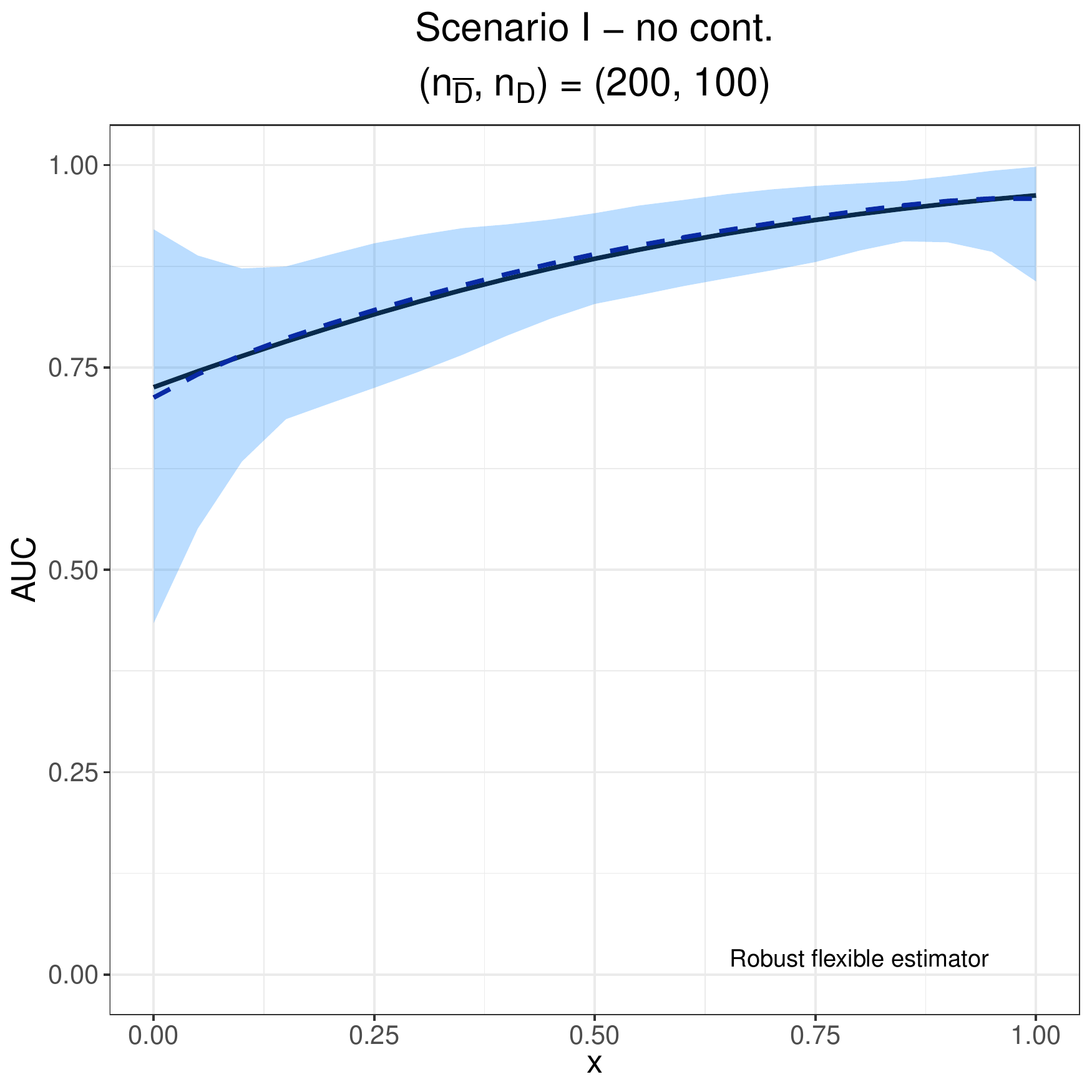}
			\includegraphics[width = 4.65cm]{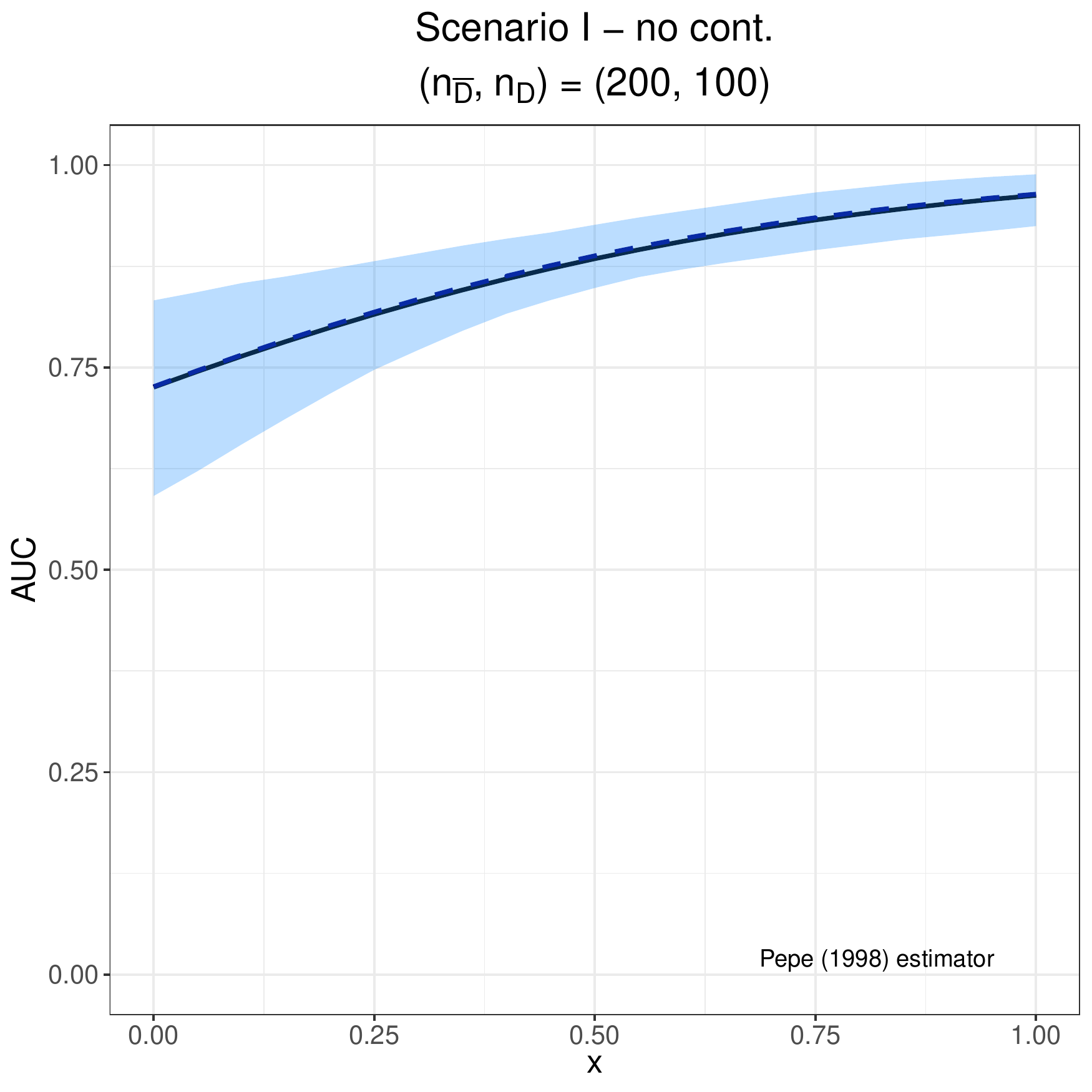}
			\includegraphics[width = 4.65cm]{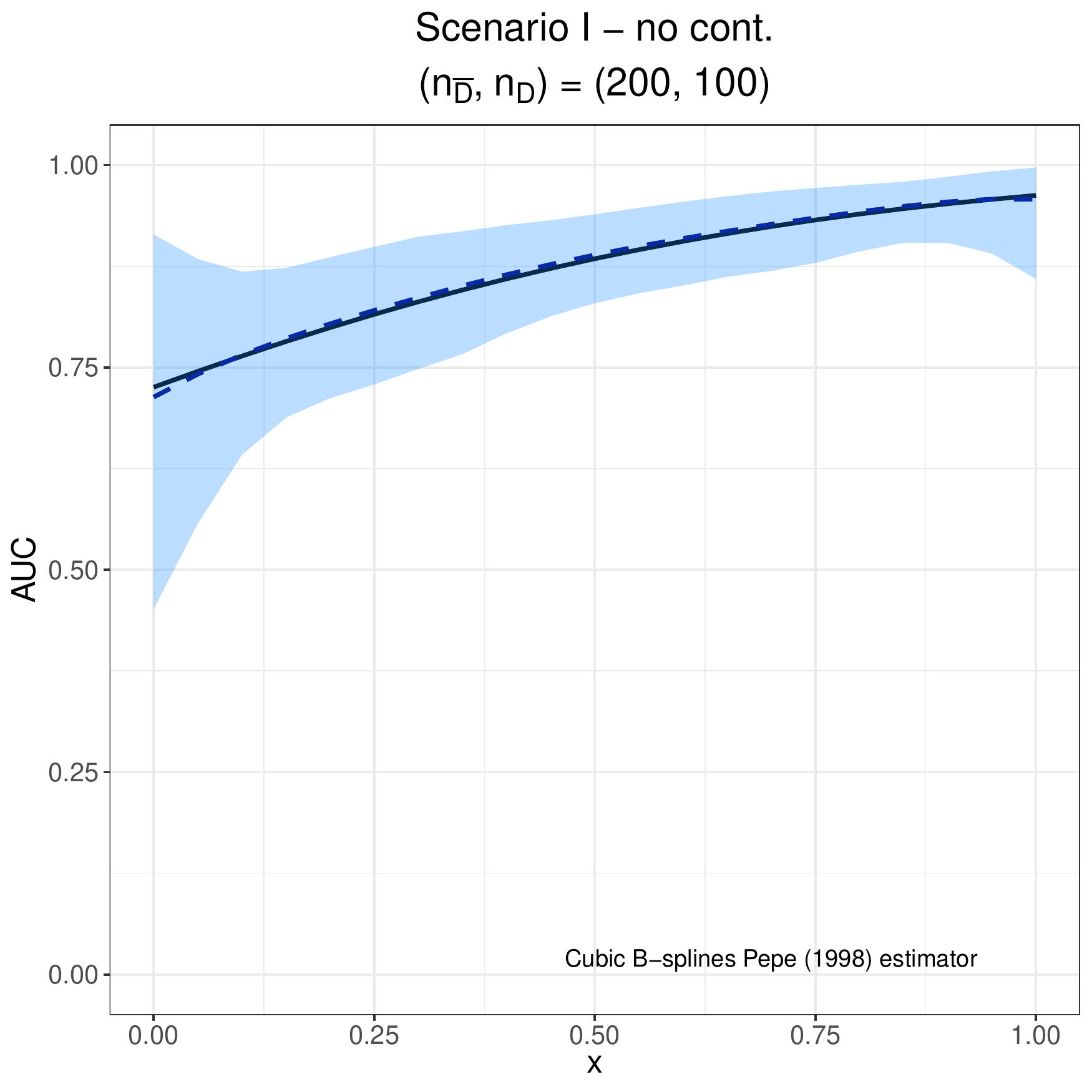}
			\includegraphics[width = 4.65cm]{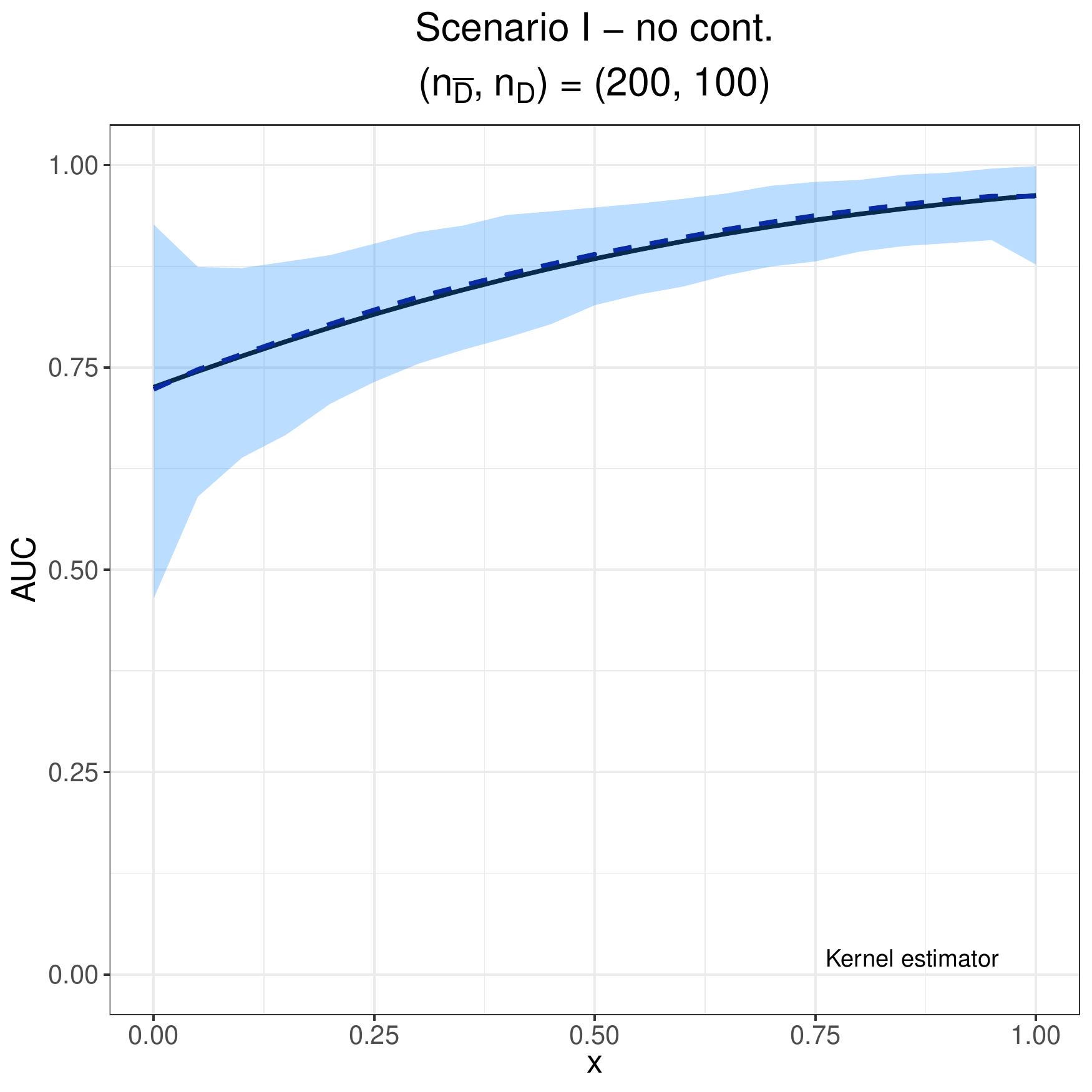}
		}
		\vspace{0.3cm}
		\subfigure{
			\includegraphics[width = 4.65cm]{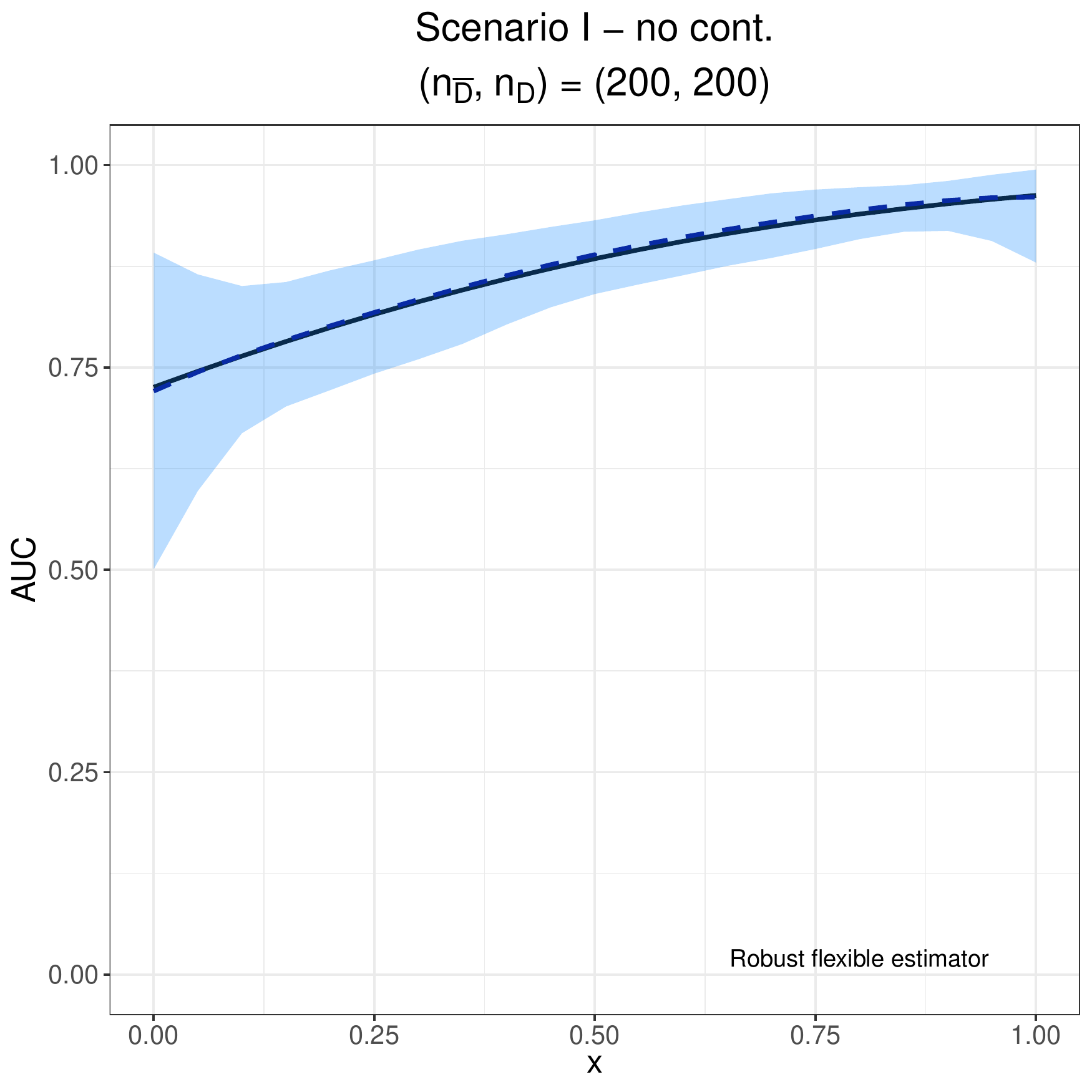}
			\includegraphics[width = 4.65cm]{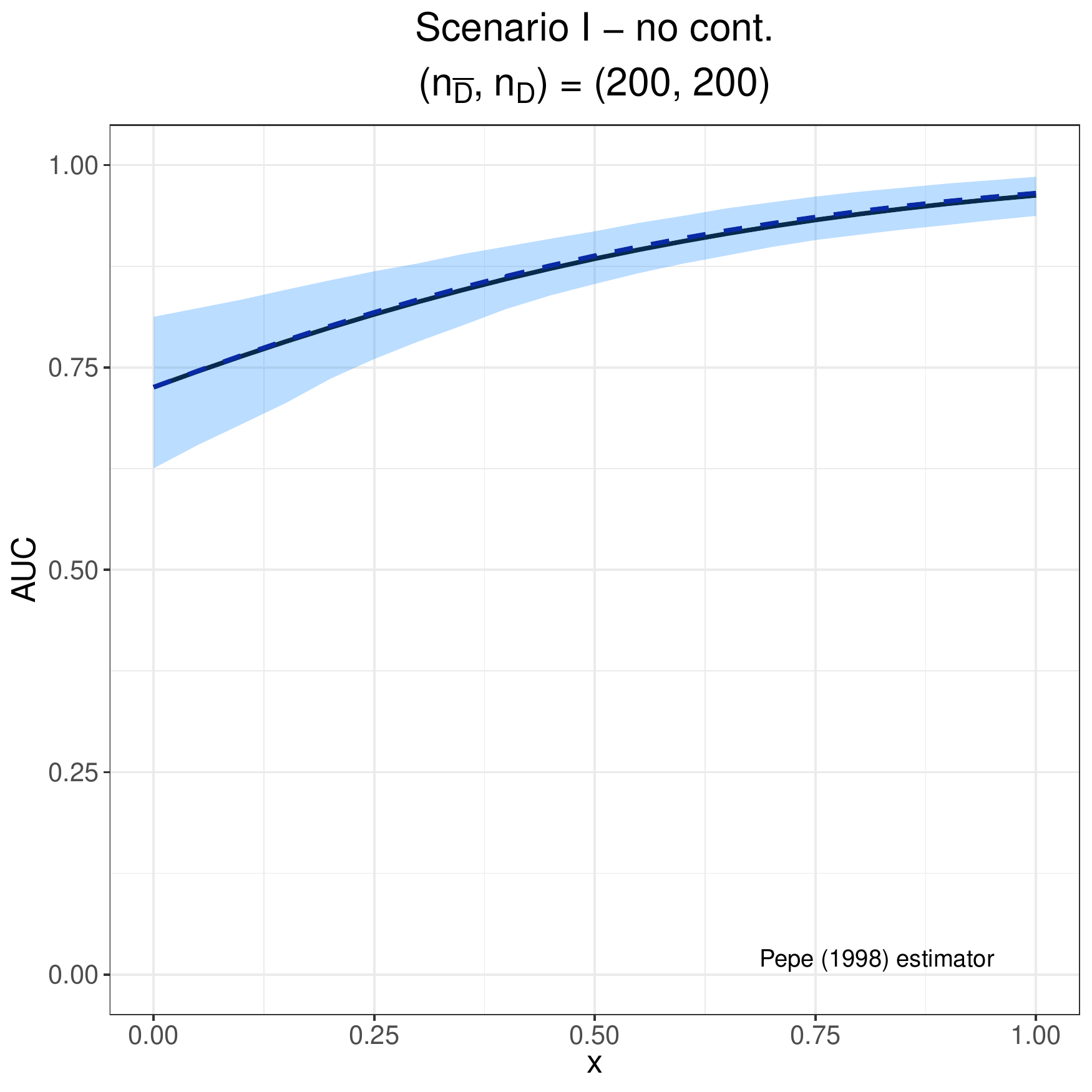}
			\includegraphics[width = 4.65cm]{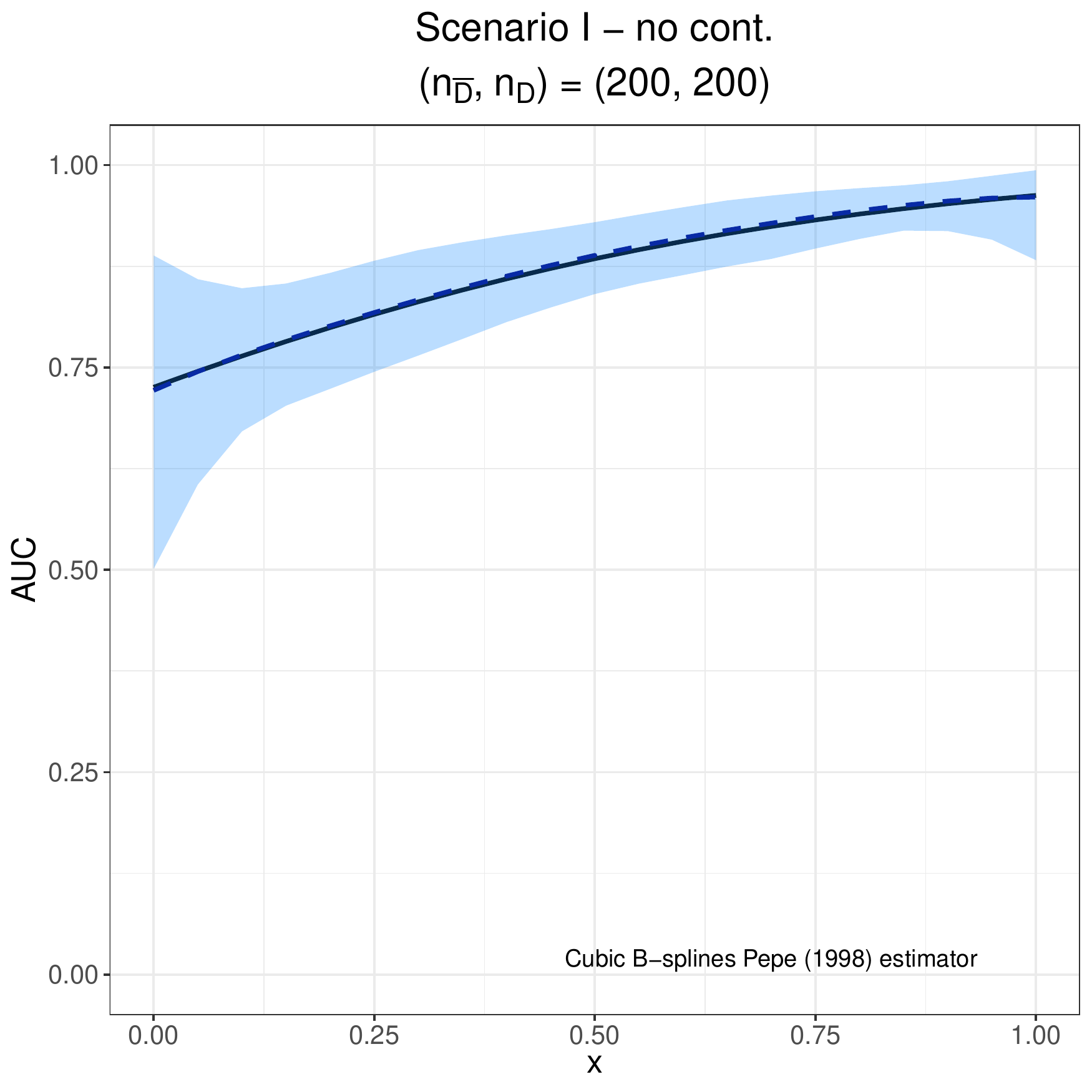}
			\includegraphics[width = 4.65cm]{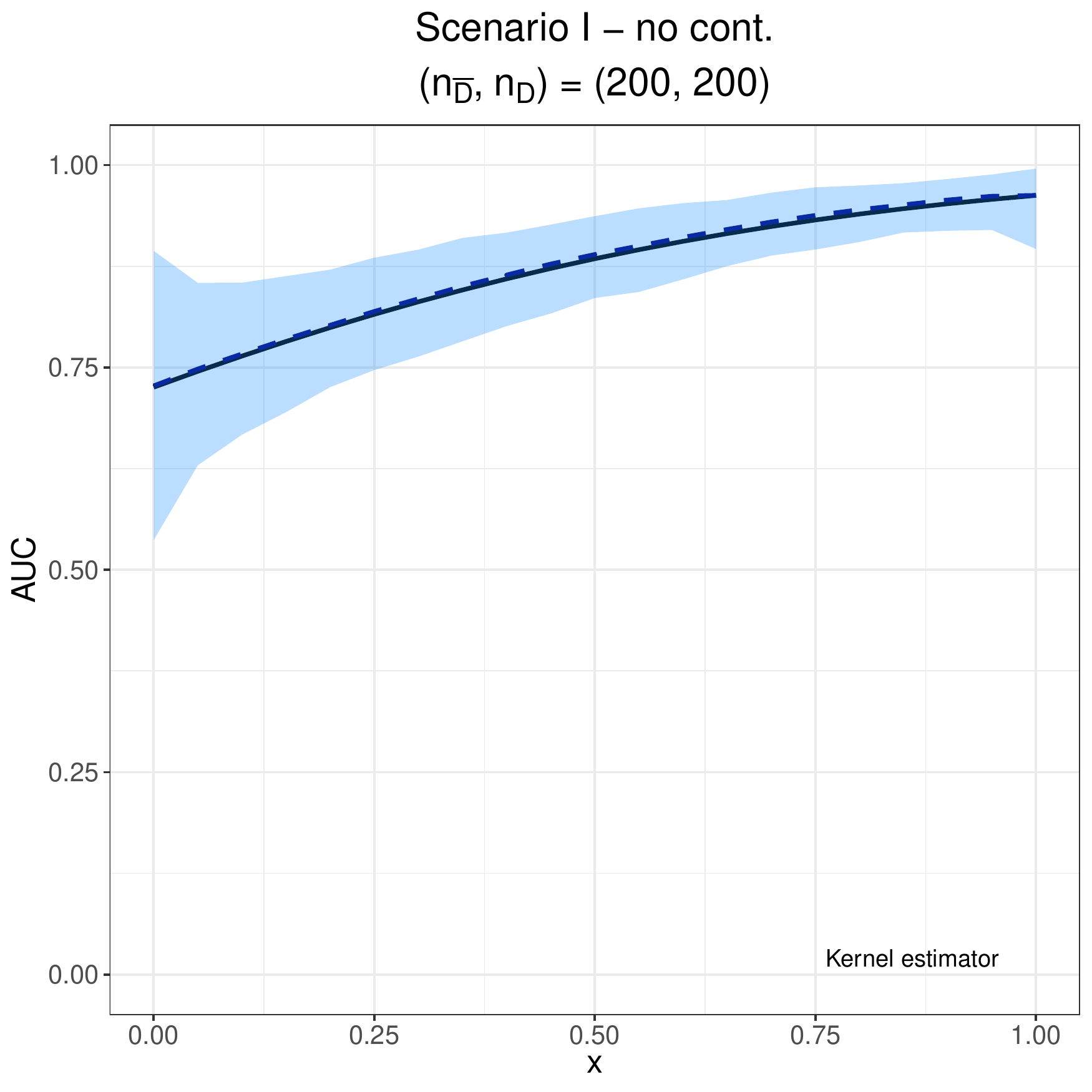}
		}
	\end{center}
	\caption{\footnotesize{Scenario I. True covariate-specific AUC (solid line) versus the mean of the Monte Carlo estimates (dashed line) along with the $2.5\%$ and $97.5\%$ simulation quantiles (shaded area) for the case of no contamination. The first row displays the results for $(n_{\bar{D}}, n_D)=(100,100)$, the second row for $(n_{\bar{D}}, n_D)=(200,100)$, and the third row for $(n_{\bar{D}}, n_D)=(200,200)$. The first column corresponds to our flexible and robust estimator, the second column to the estimator proposed by Pepe (1998), the third one to the cubic B-splines extension of Pepe (1998), and the fourth column to the kernel estimator.}}
\end{figure}

\begin{figure}[H]
	\begin{center}
		\subfigure{
			\includegraphics[width = 4.65cm]{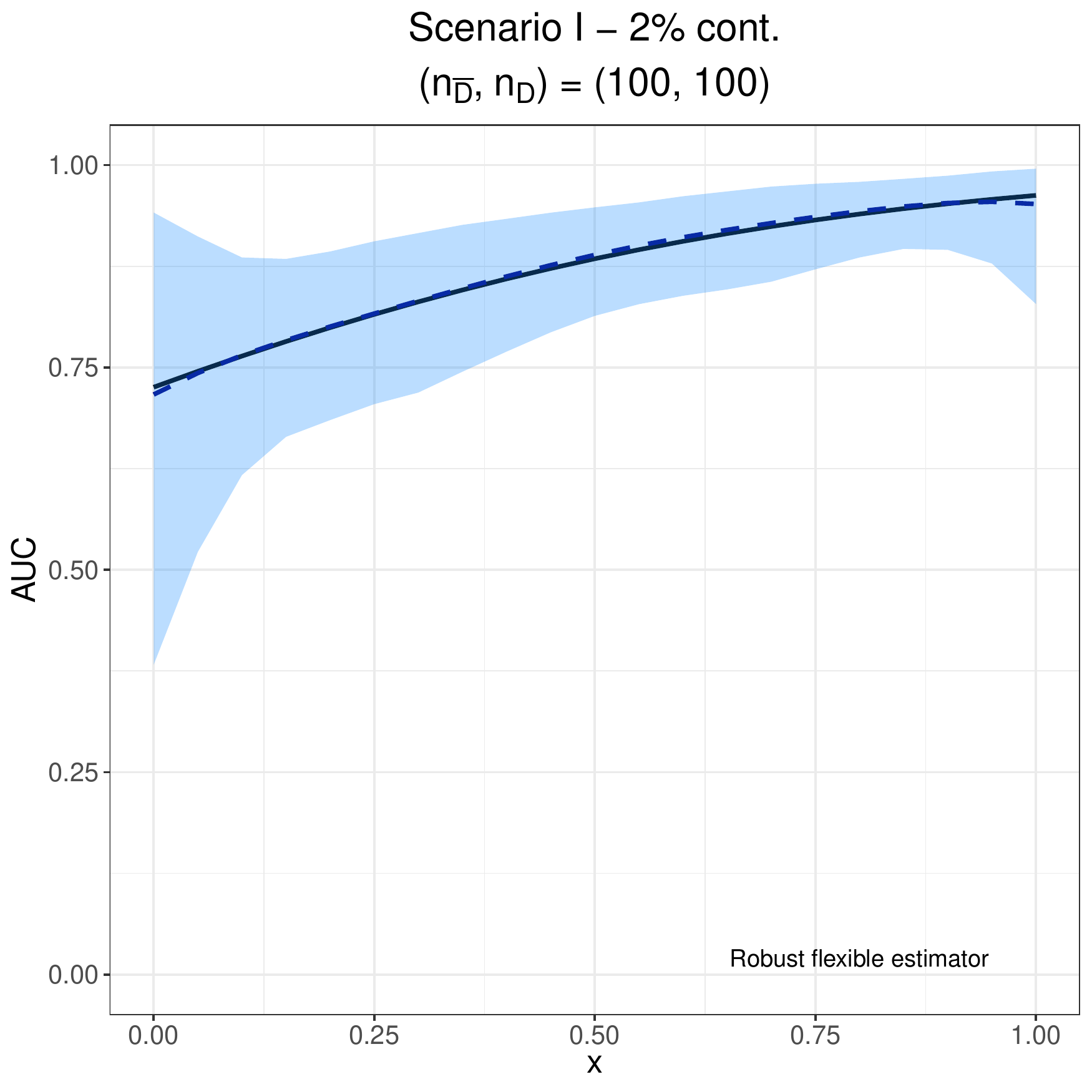}
			\includegraphics[width = 4.65cm]{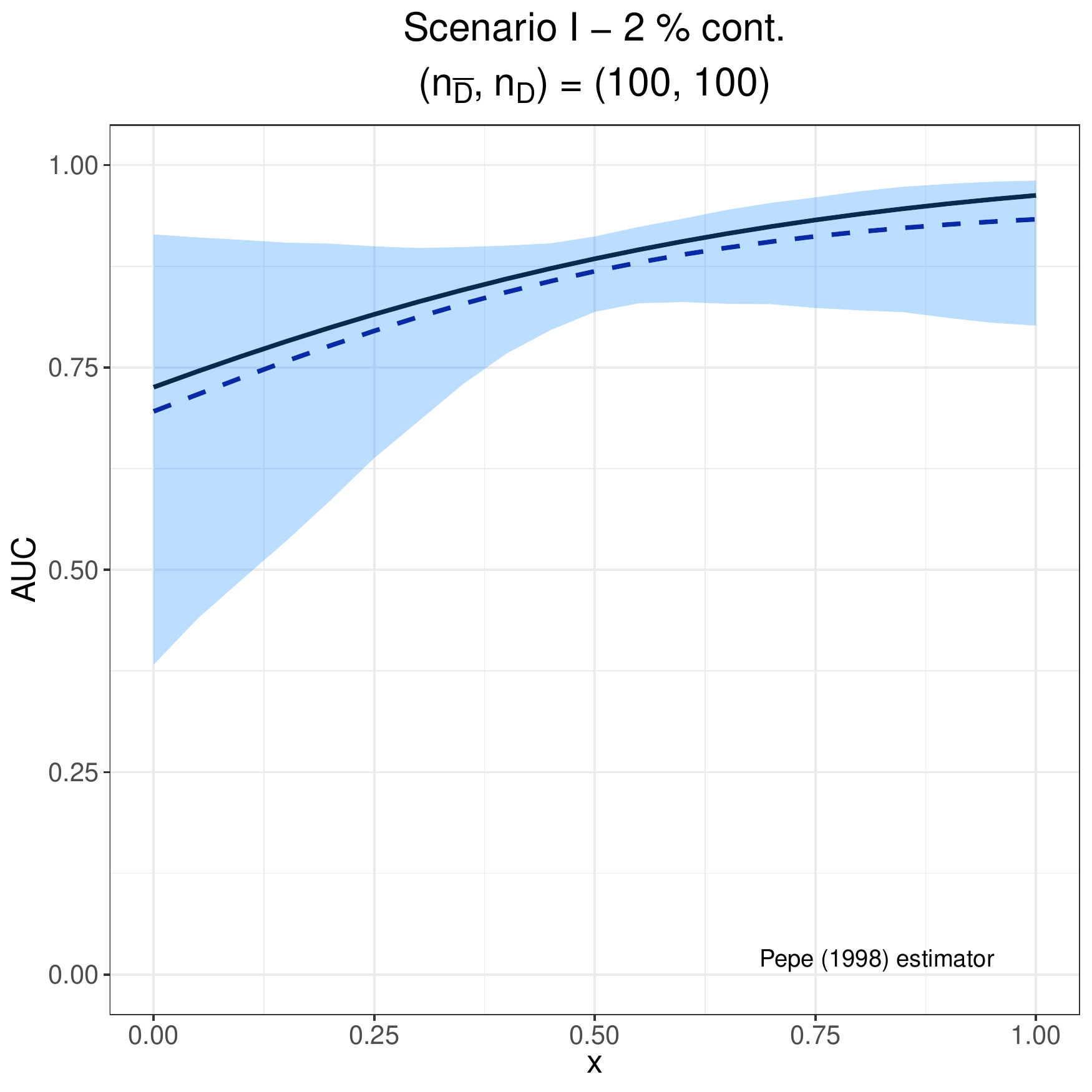}
			\includegraphics[width = 4.65cm]{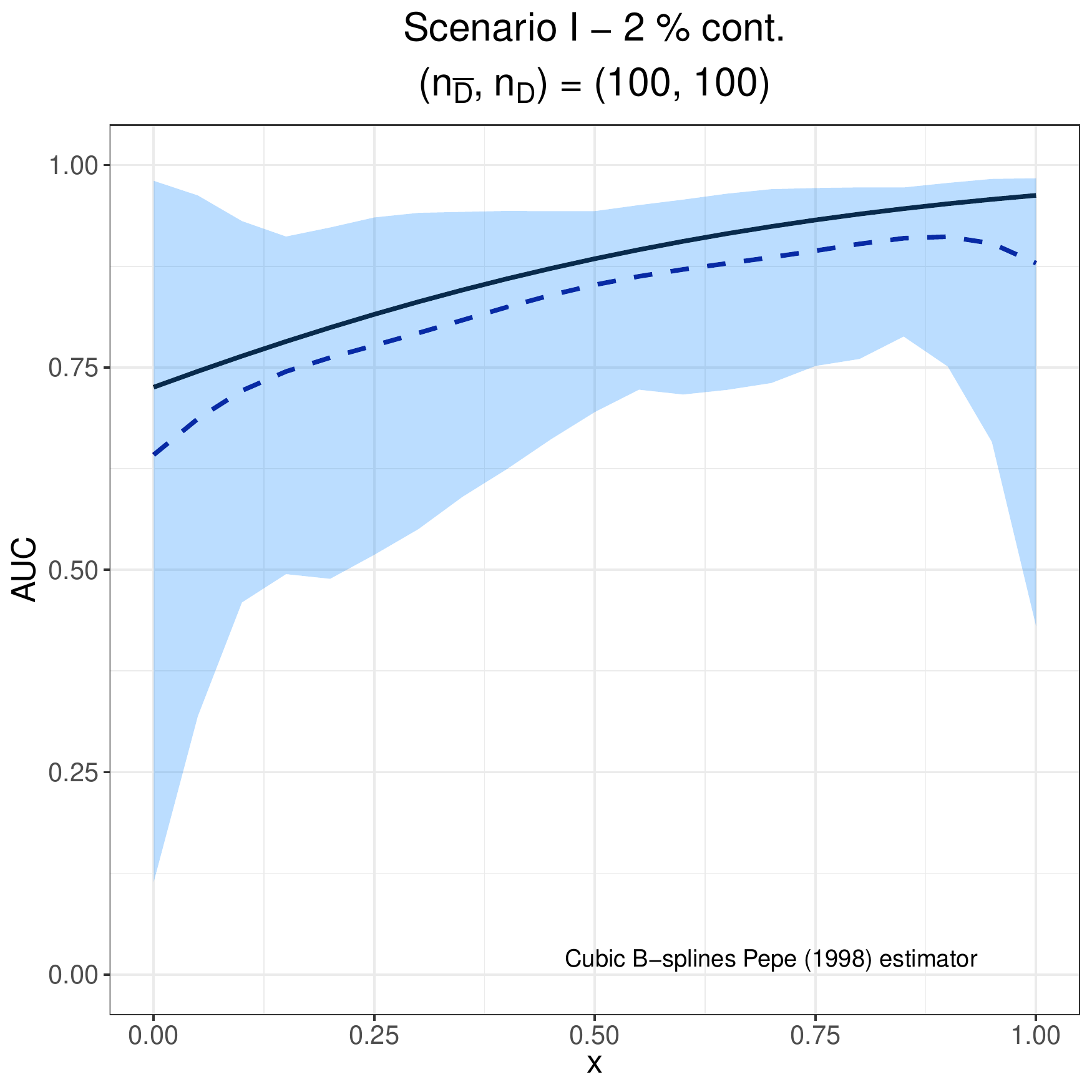}
			\includegraphics[width = 4.65cm]{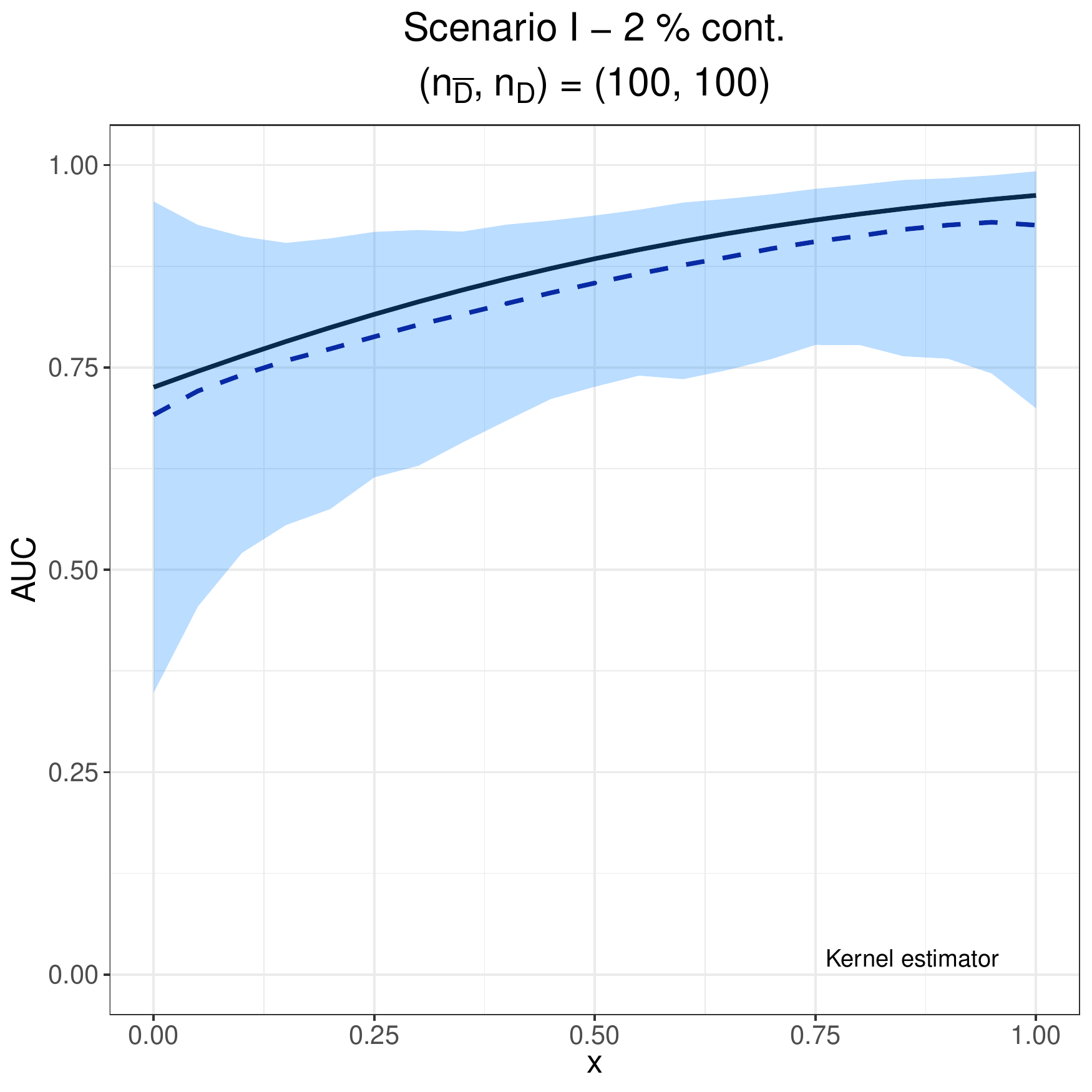}
		}
		\vspace{0.3cm}
		\subfigure{
			\includegraphics[width = 4.65cm]{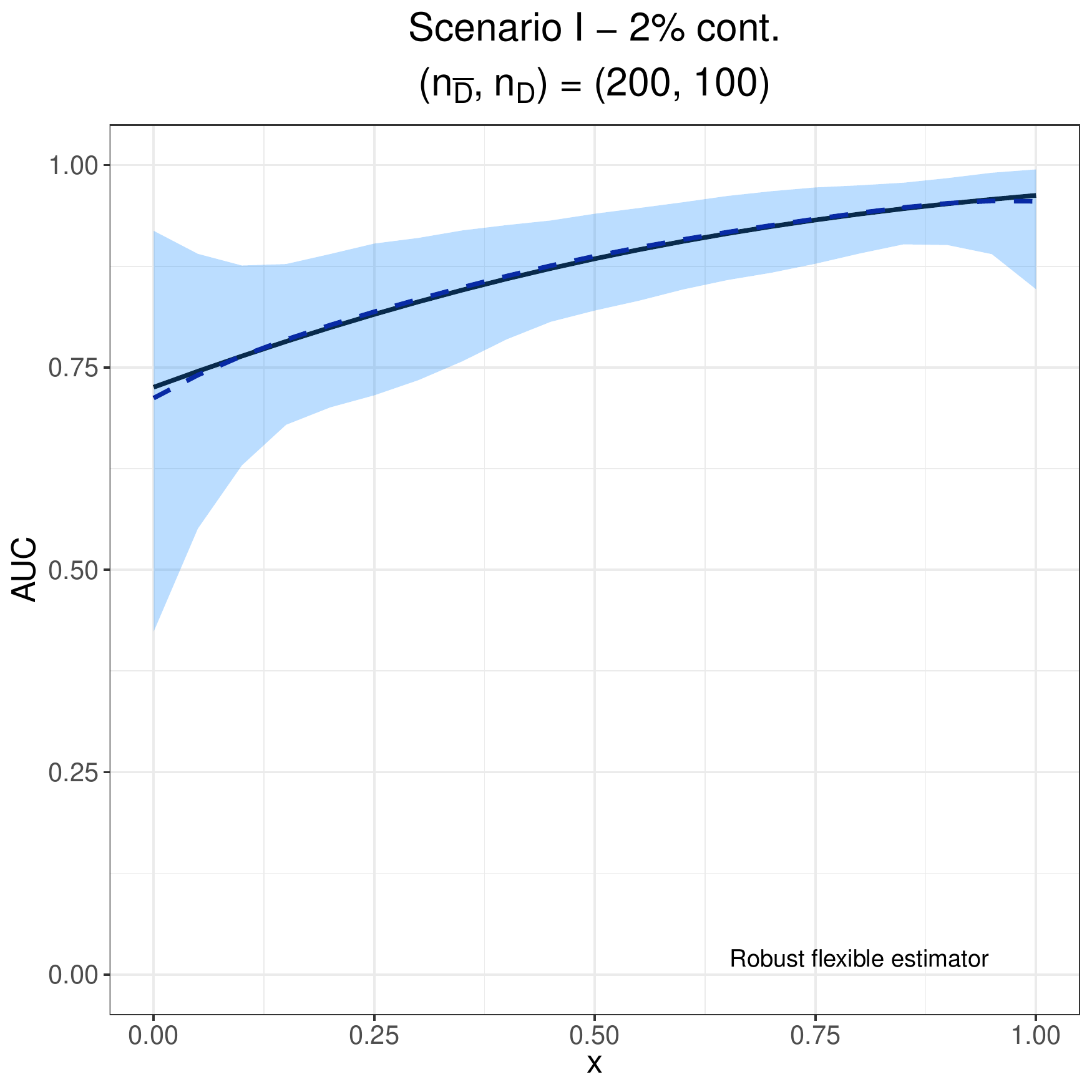}
			\includegraphics[width = 4.65cm]{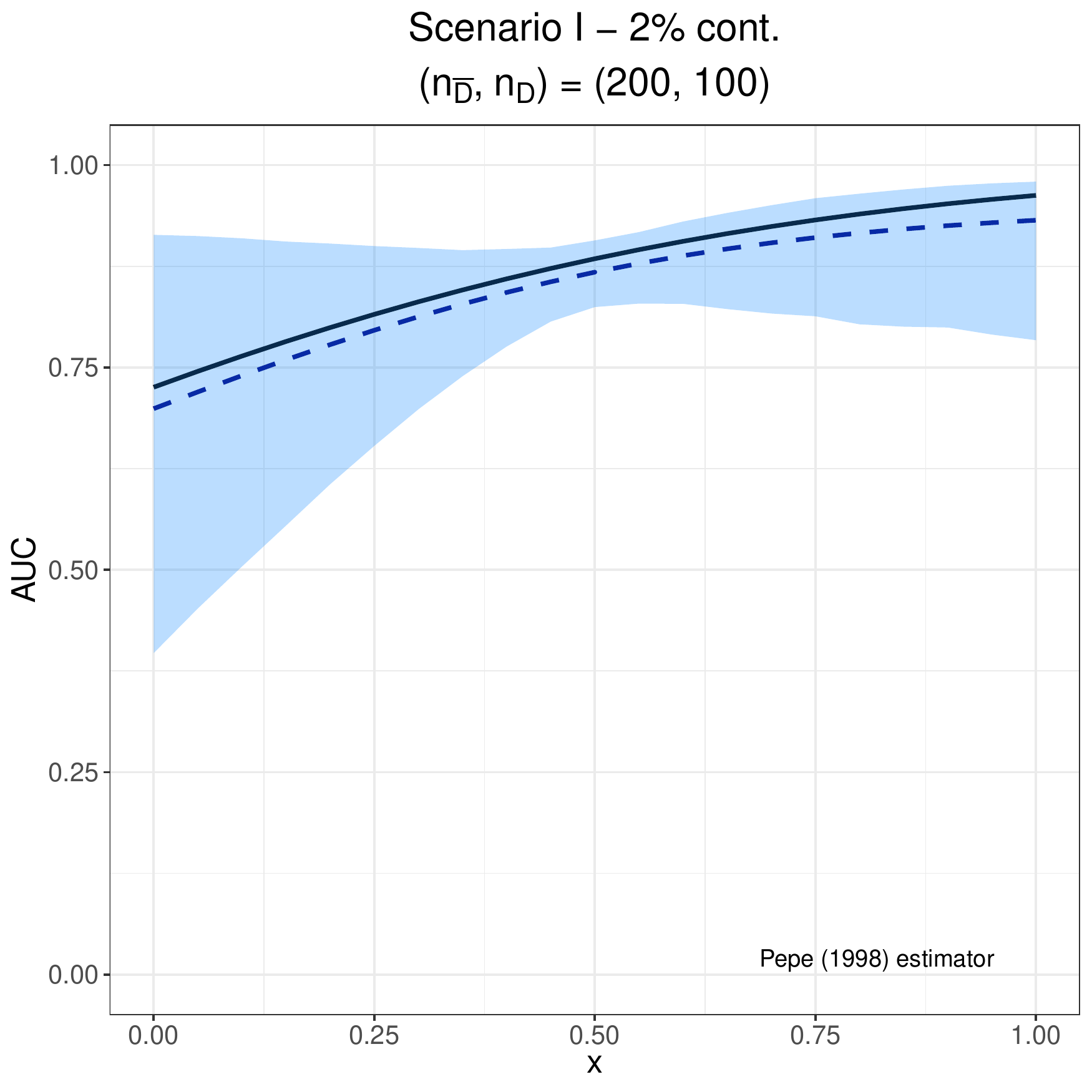}
			\includegraphics[width = 4.65cm]{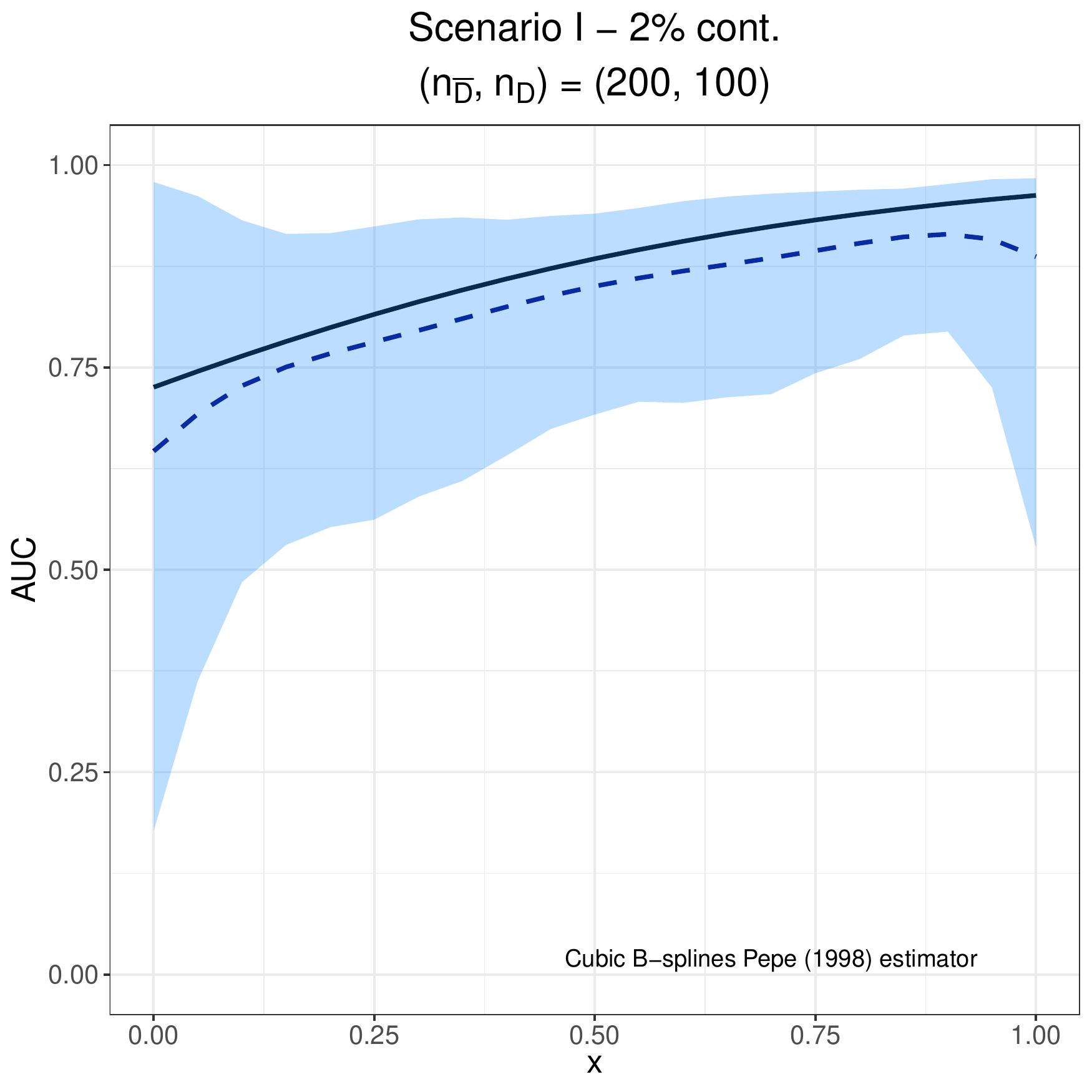}
			\includegraphics[width = 4.65cm]{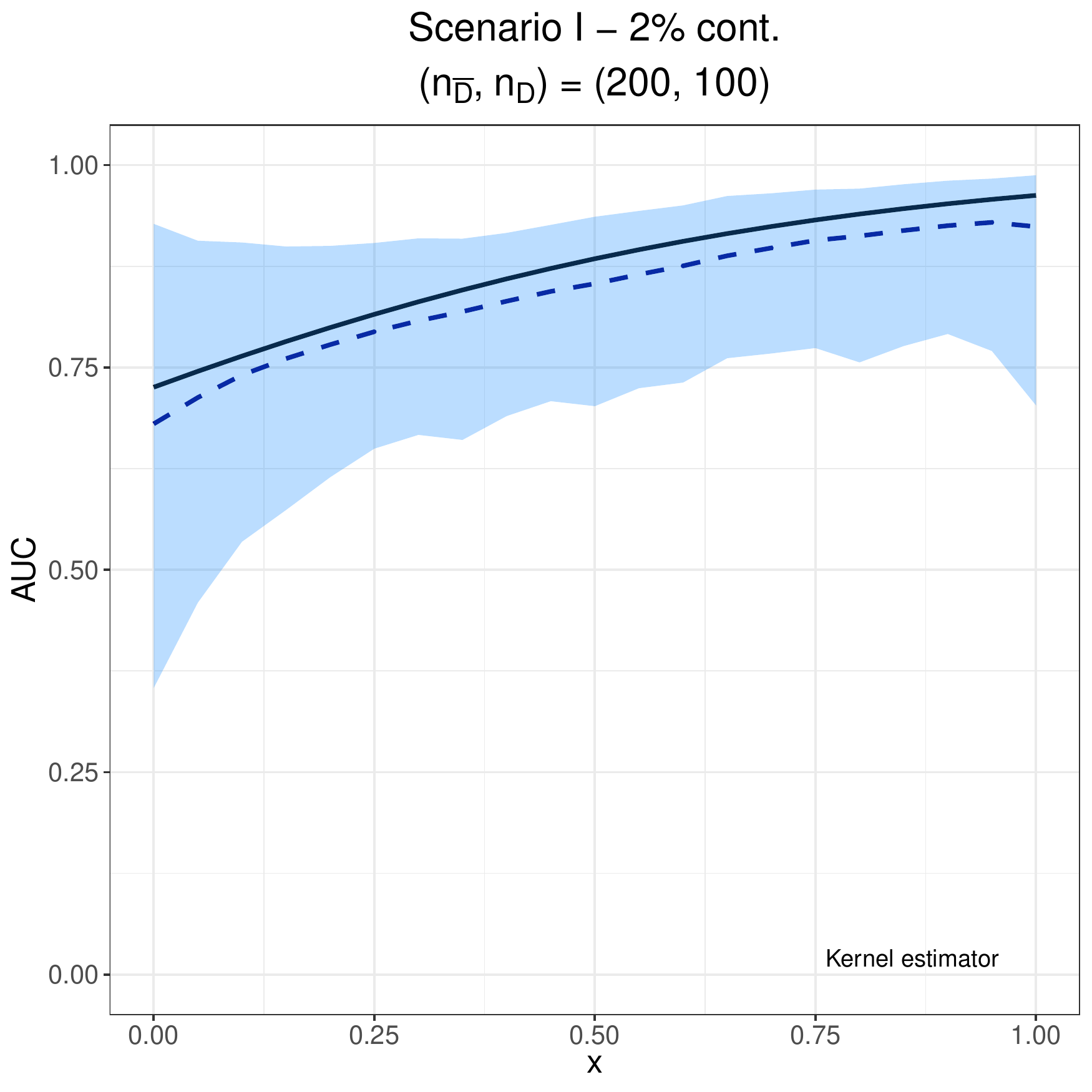}
		}
		\vspace{0.3cm}
		\subfigure{
			\includegraphics[width = 4.65cm]{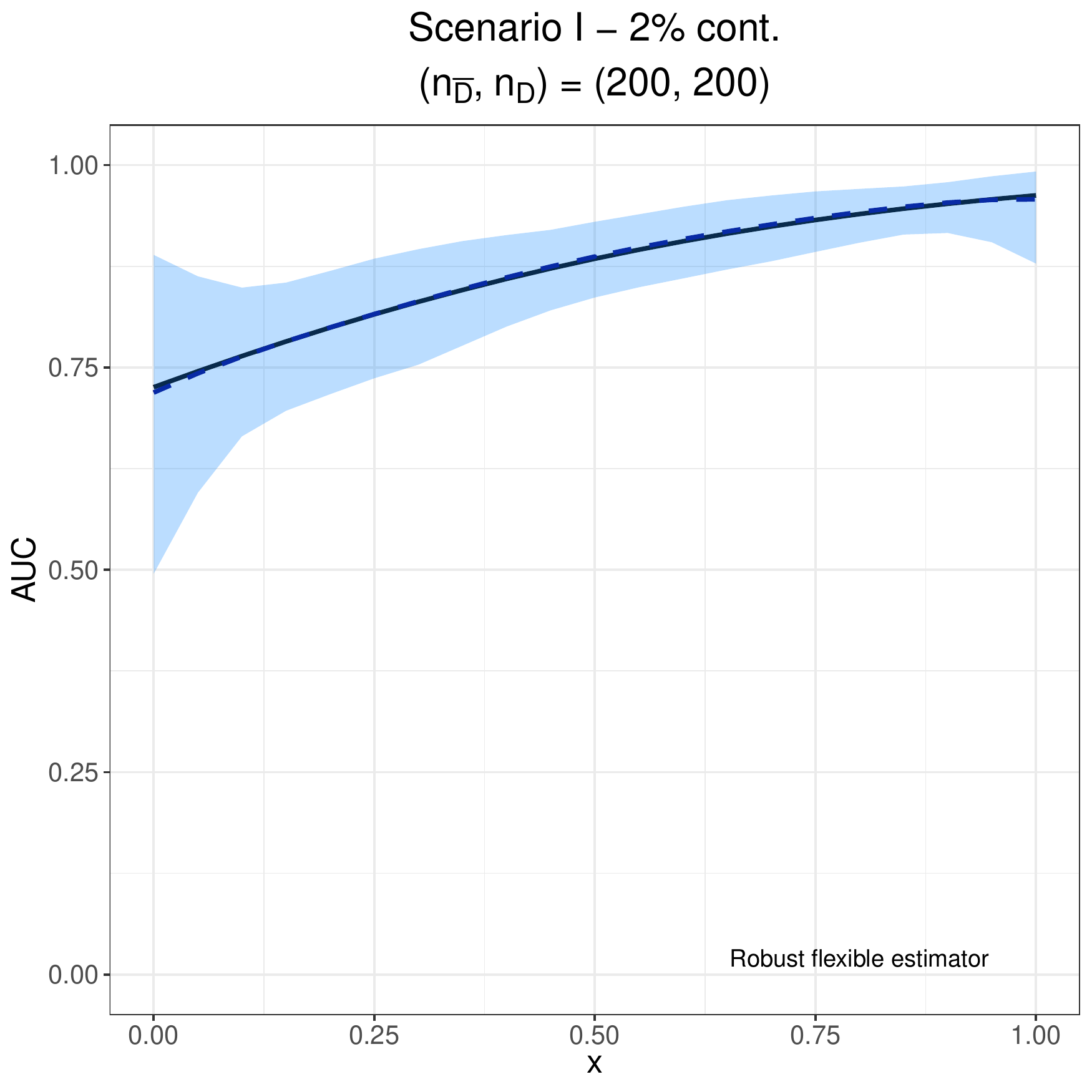}
			\includegraphics[width = 4.65cm]{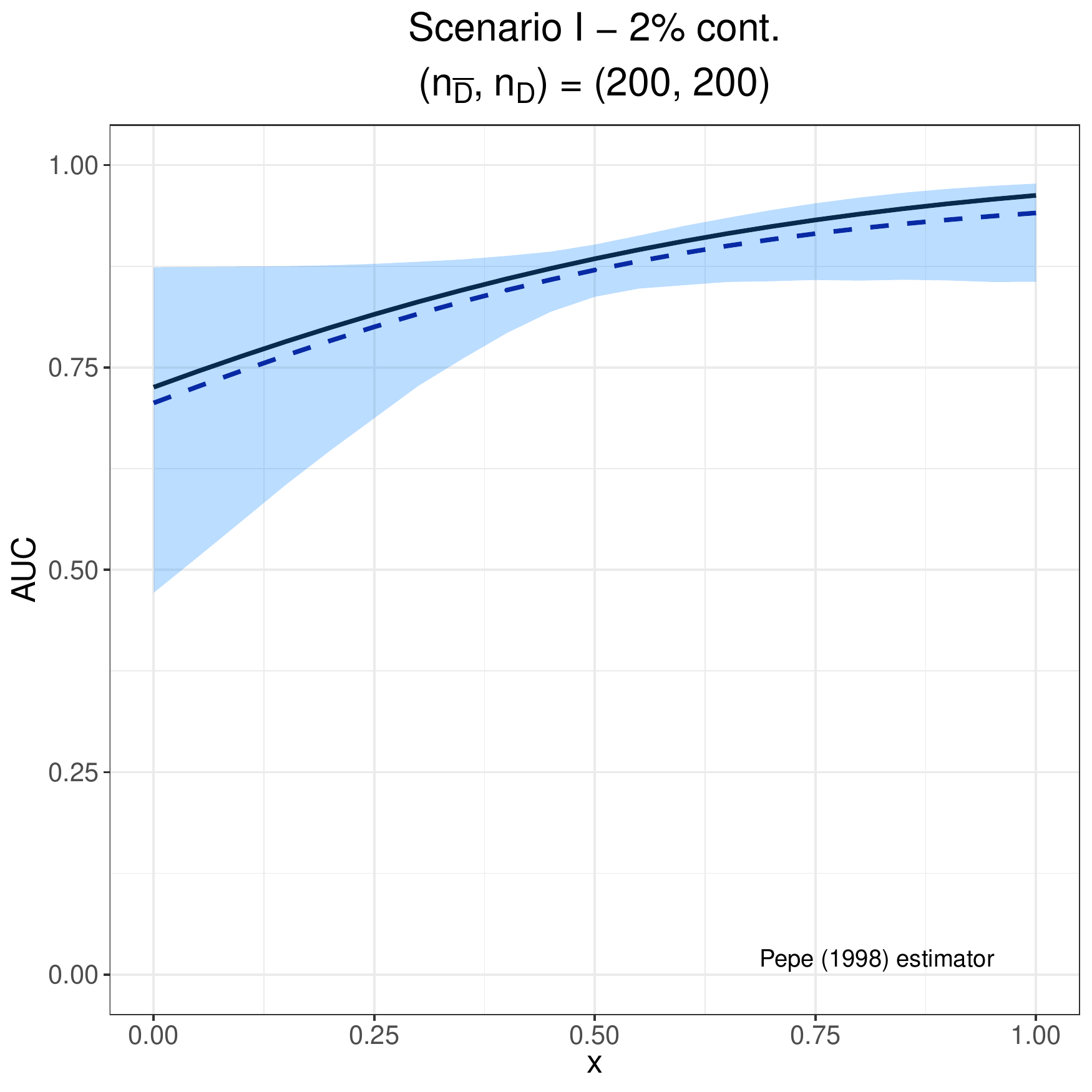}
			\includegraphics[width = 4.65cm]{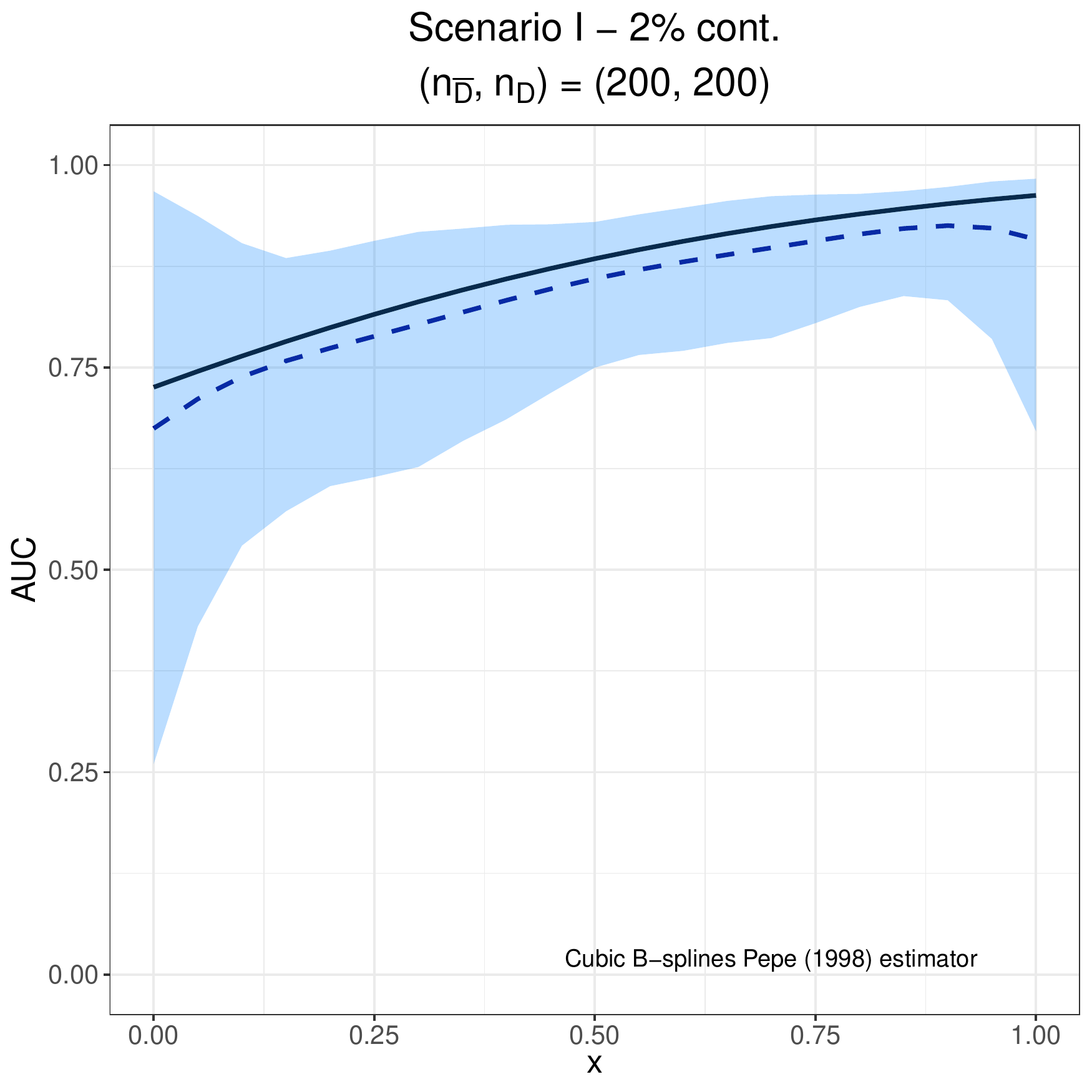}
			\includegraphics[width = 4.65cm]{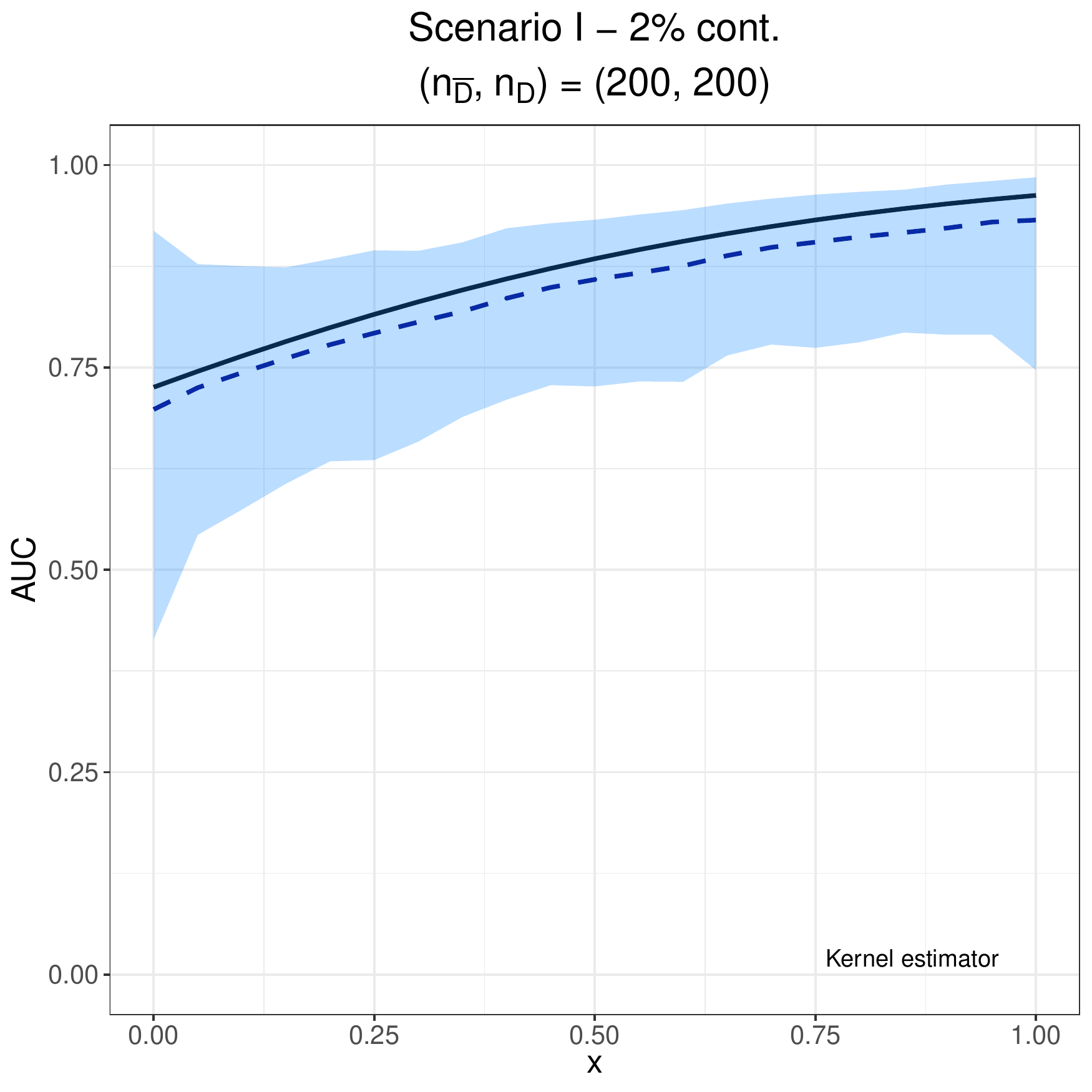}
		}
	\end{center}
	\caption{\footnotesize{Scenario I. True covariate-specific AUC (solid line) versus the mean of the Monte Carlo estimates (dashed line) along with the $2.5\%$ and $97.5\%$ simulation quantiles (shaded area) for the case of $2\%$ of contamination. The first row displays the results for $(n_{\bar{D}}, n_D)=(100,100)$, the second row for $(n_{\bar{D}}, n_D)=(200,100)$, and the third row for $(n_{\bar{D}}, n_D)=(200,200)$. The first column corresponds to our flexible and robust estimator, the second column to the estimator proposed by Pepe (1998), the third one to the cubic B-splines extension of Pepe (1998), and the fourth column to the kernel estimator.}}
\end{figure}

\begin{figure}[H]
	\begin{center}
		\subfigure{
			\includegraphics[width = 4.65cm]{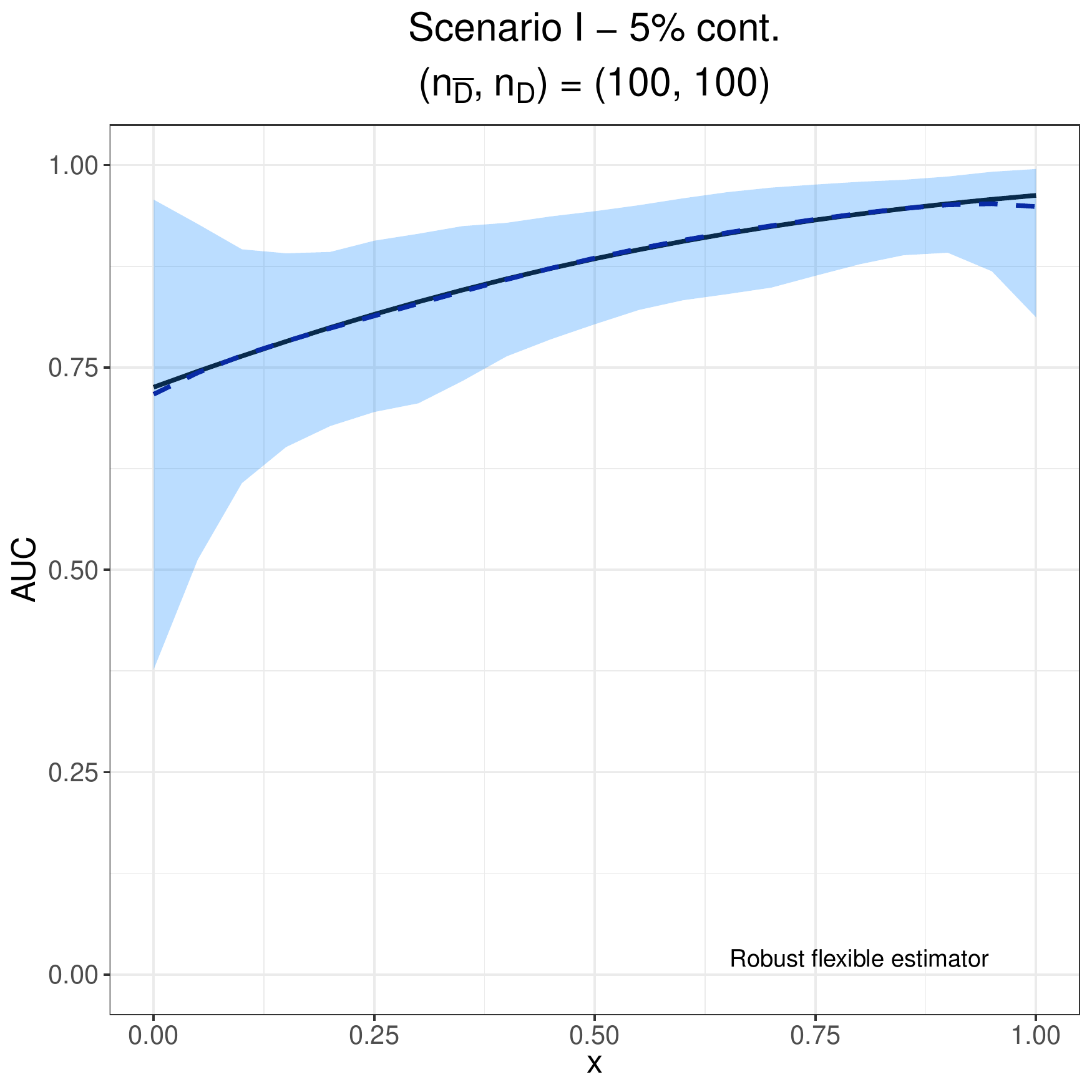}
			\includegraphics[width = 4.65cm]{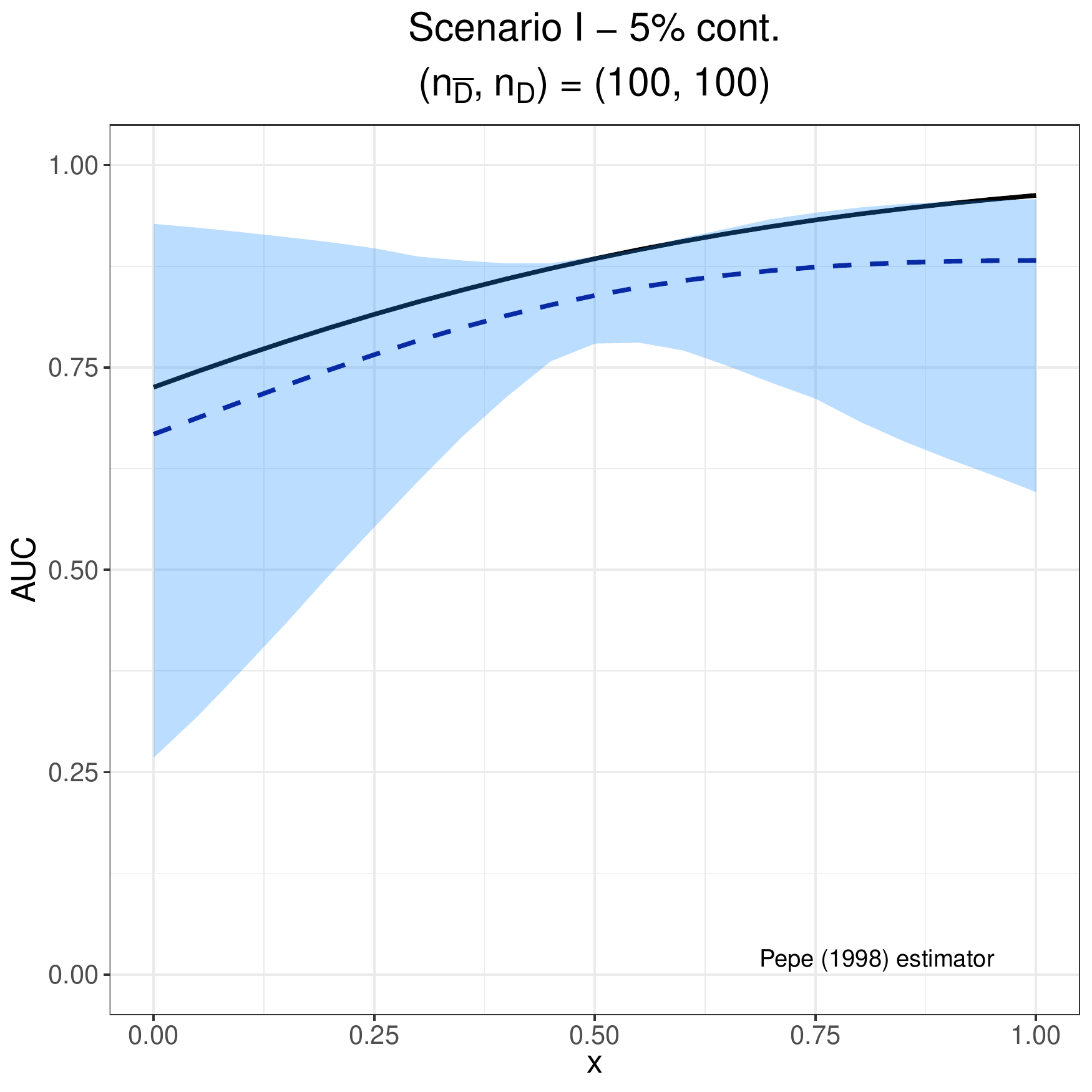}
			\includegraphics[width = 4.65cm]{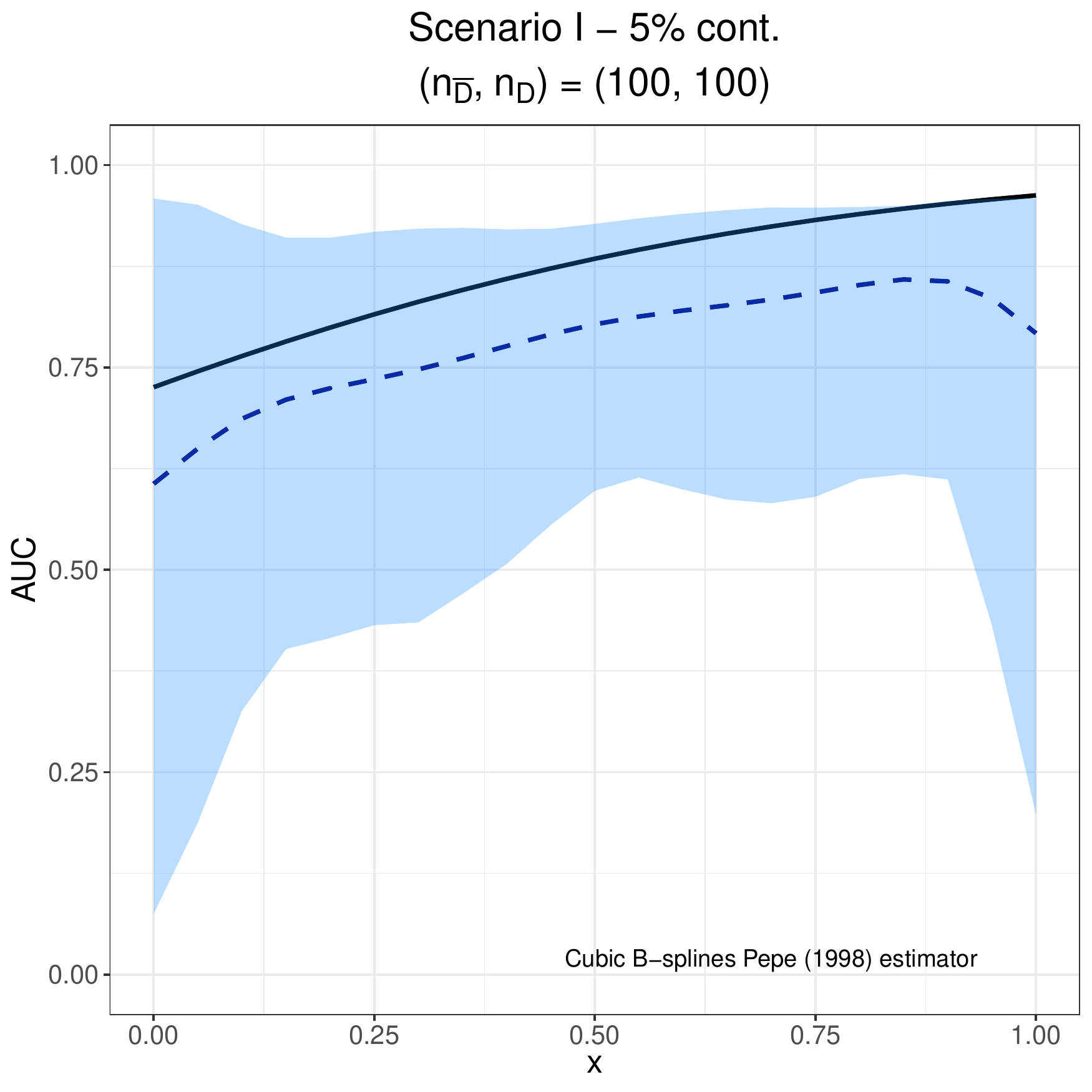}
			\includegraphics[width = 4.65cm]{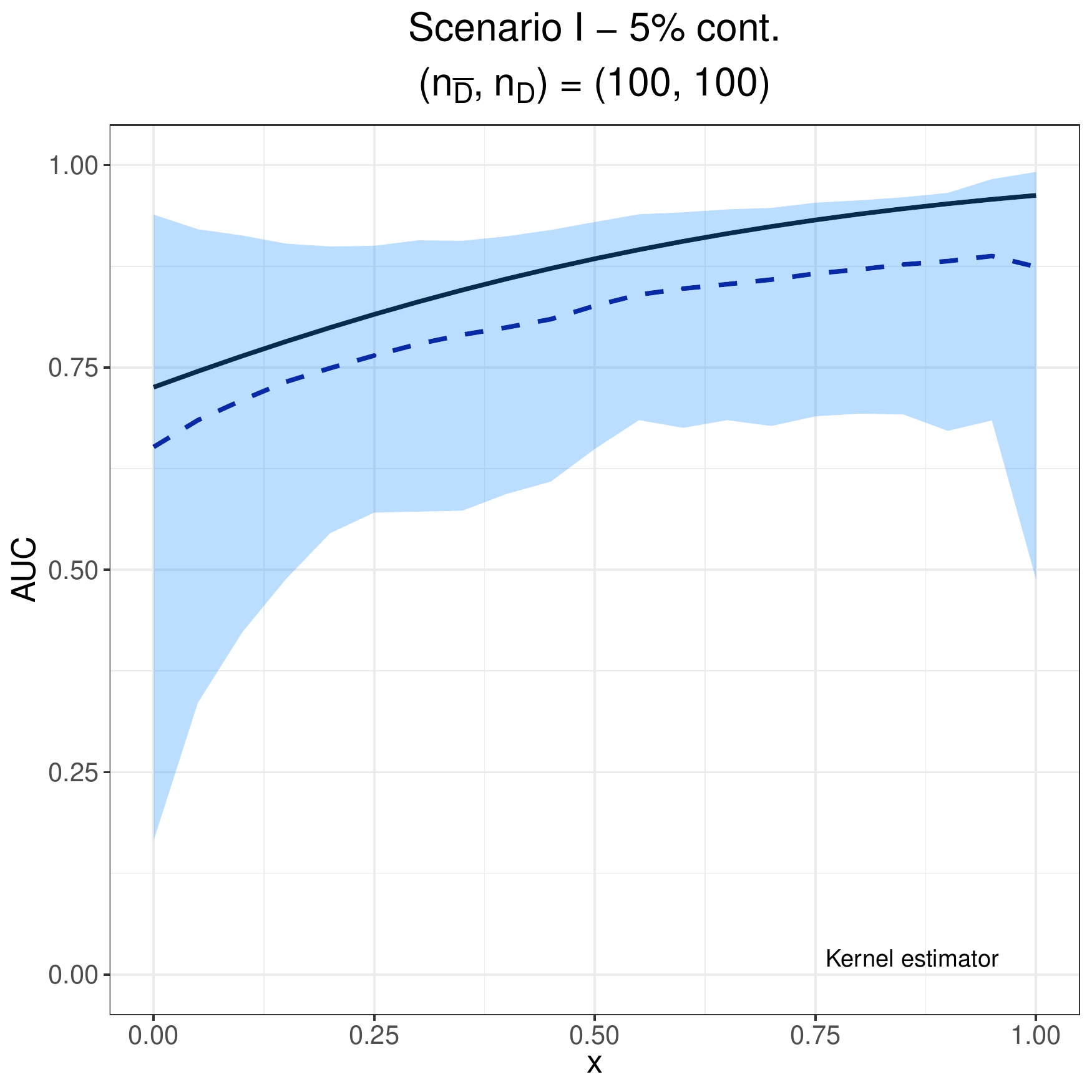}
		}
		\vspace{0.3cm}
		\subfigure{
			\includegraphics[width = 4.65cm]{aucrcont5sampsize2sc1.pdf}
			\includegraphics[width = 4.65cm]{aucspcont5sampsize2sc1.pdf}
			\includegraphics[width = 4.65cm]{aucspbscont5sampsize2sc1.pdf}
			\includegraphics[width = 4.65cm]{auckercont5sampsize2sc1.pdf}
		}
		\vspace{0.3cm}
		\subfigure{
			\includegraphics[width = 4.65cm]{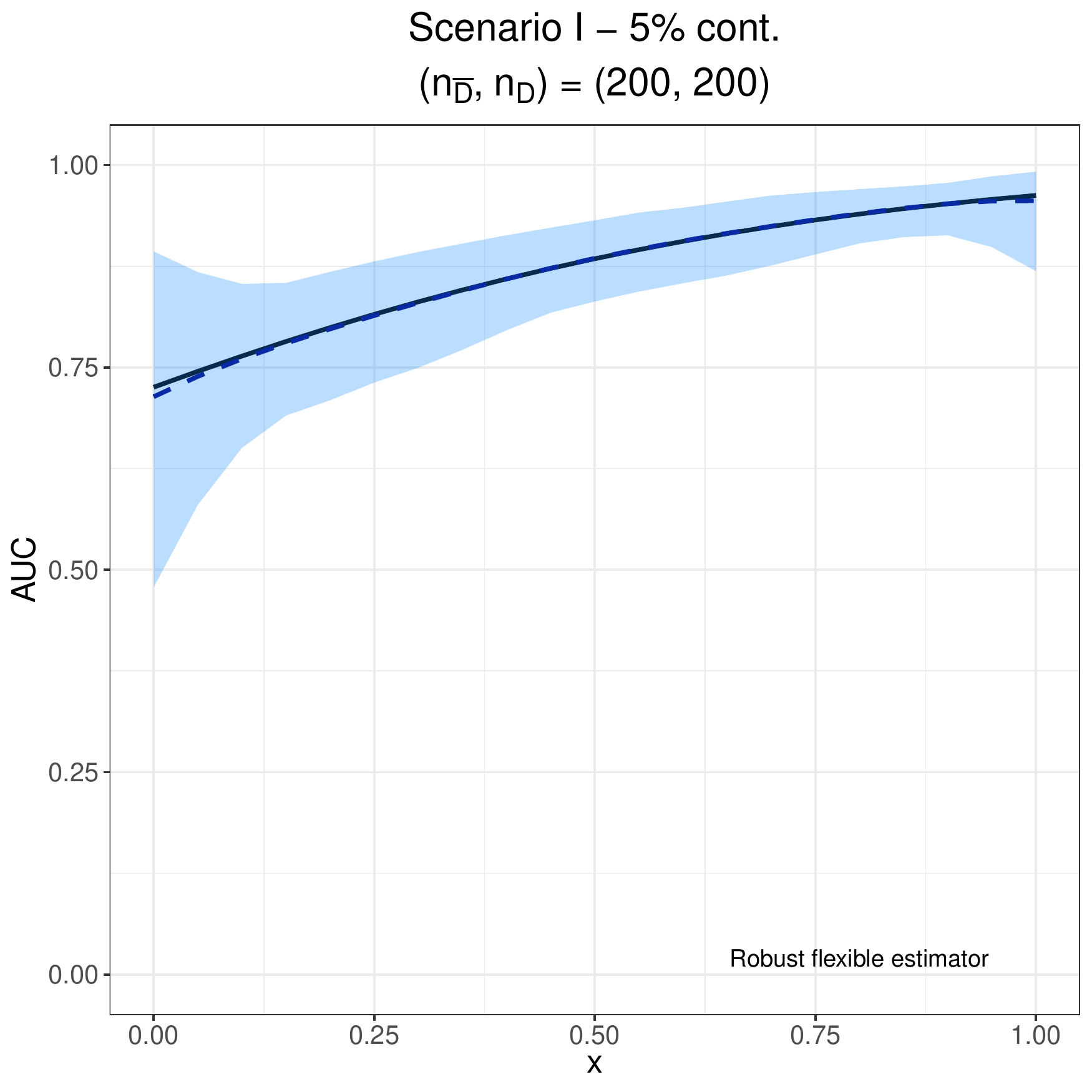}
			\includegraphics[width = 4.65cm]{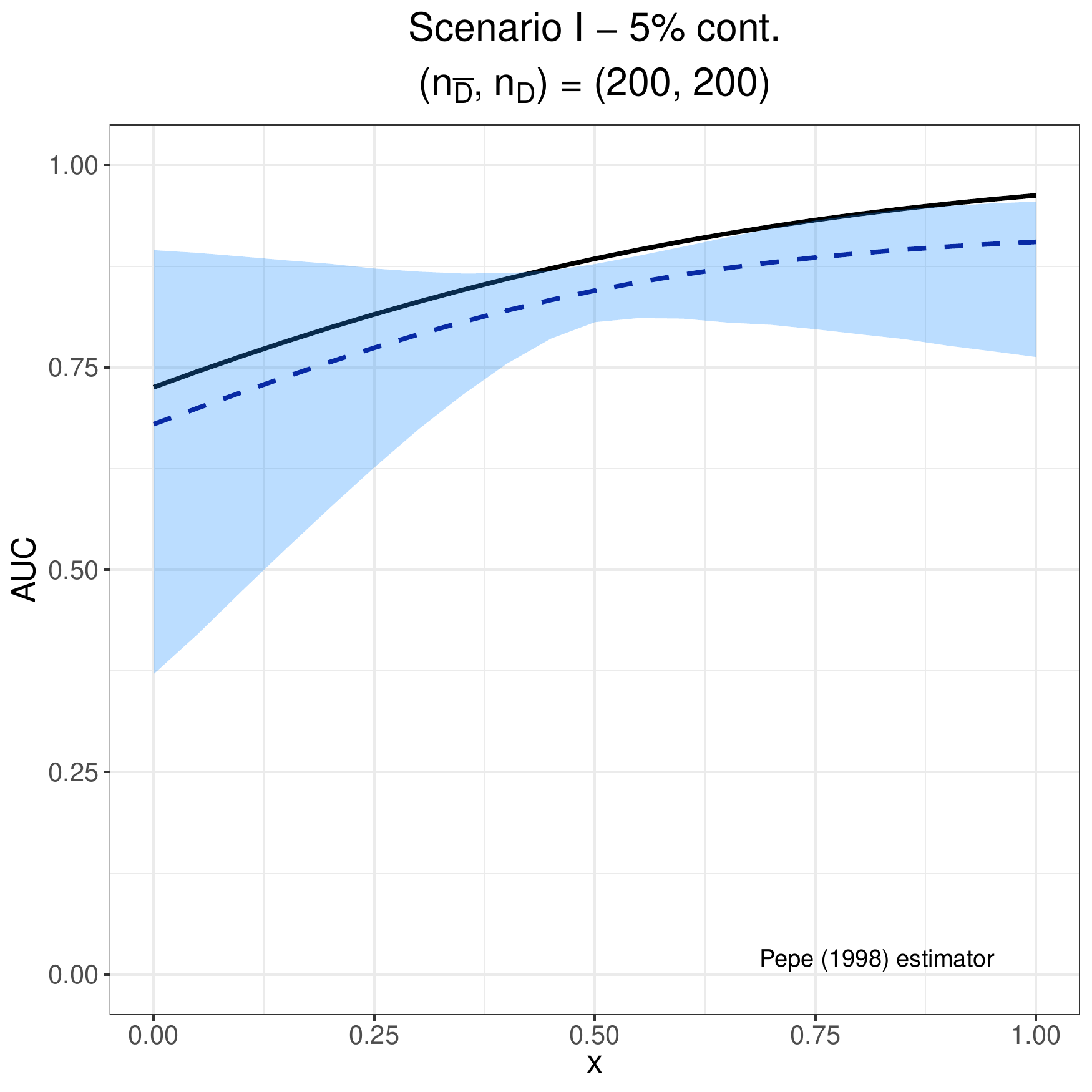}
			\includegraphics[width = 4.65cm]{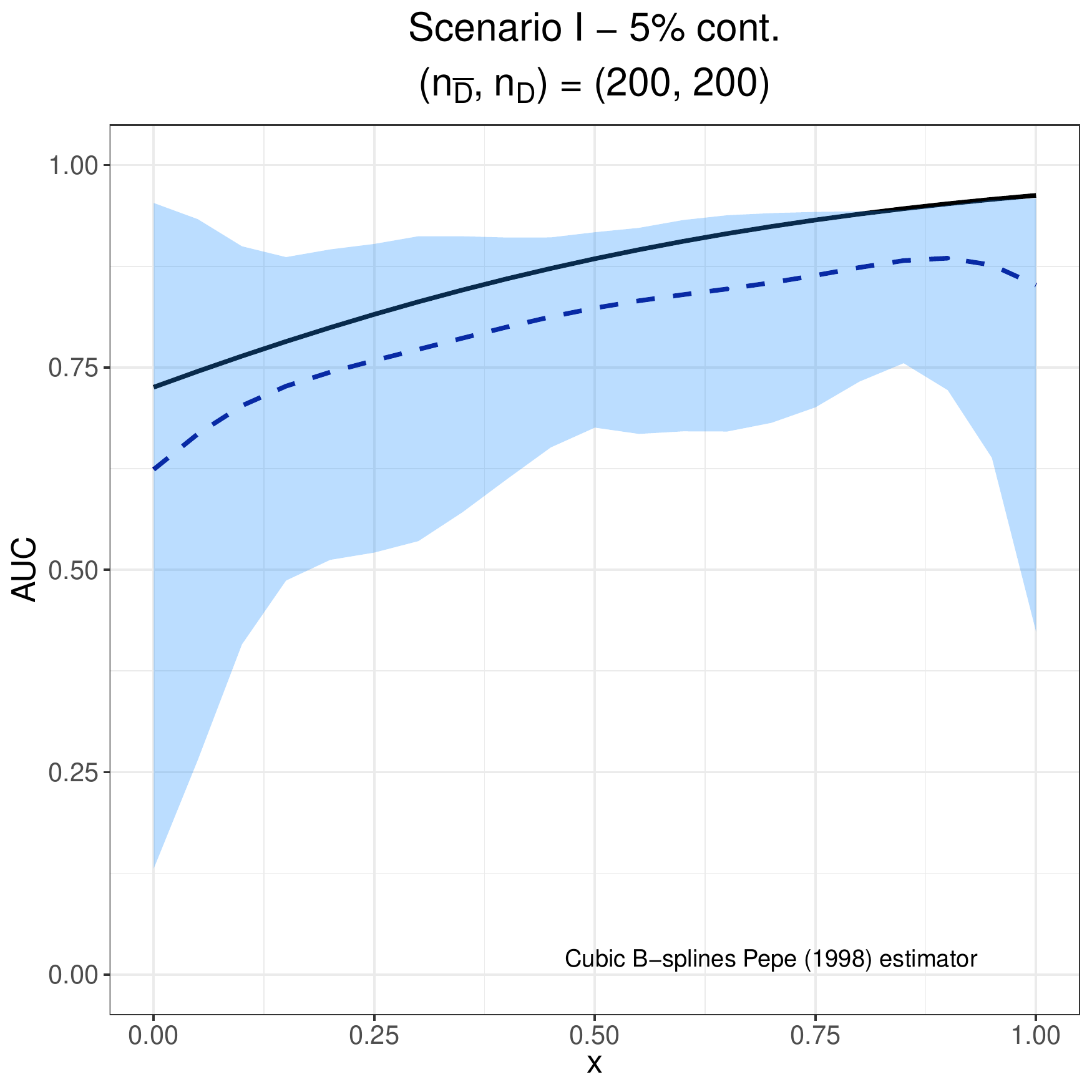}
			\includegraphics[width = 4.65cm]{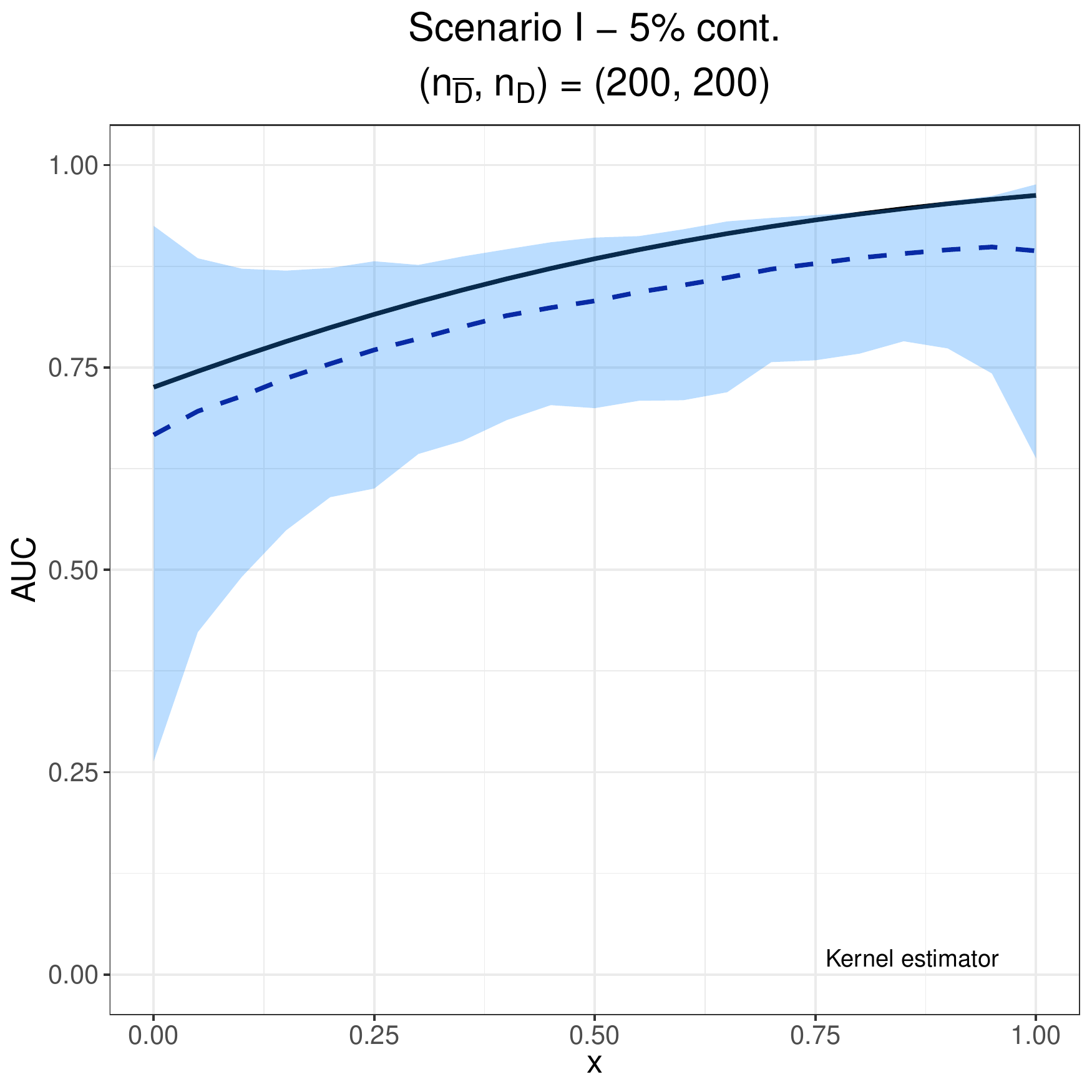}
		}
	\end{center}
	\caption{\footnotesize{Scenario I. True covariate-specific AUC (solid line) versus the mean of the Monte Carlo estimates (dashed line) along with the $2.5\%$ and $97.5\%$ simulation quantiles (shaded area) for the case of $5\%$ of contamination. The first row displays the results for $(n_{\bar{D}}, n_D)=(100,100)$, the second row for $(n_{\bar{D}}, n_D)=(200,100)$, and the third row for $(n_{\bar{D}}, n_D)=(200,200)$. The first column corresponds to our flexible and robust estimator, the second column to the estimator proposed by Pepe (1998), the third one to the cubic B-splines extension of Pepe (1998), and the fourth column to the kernel estimator.}}
\end{figure}

\begin{figure}[H]
	\begin{center}
		\subfigure{
			\includegraphics[width = 4.65cm]{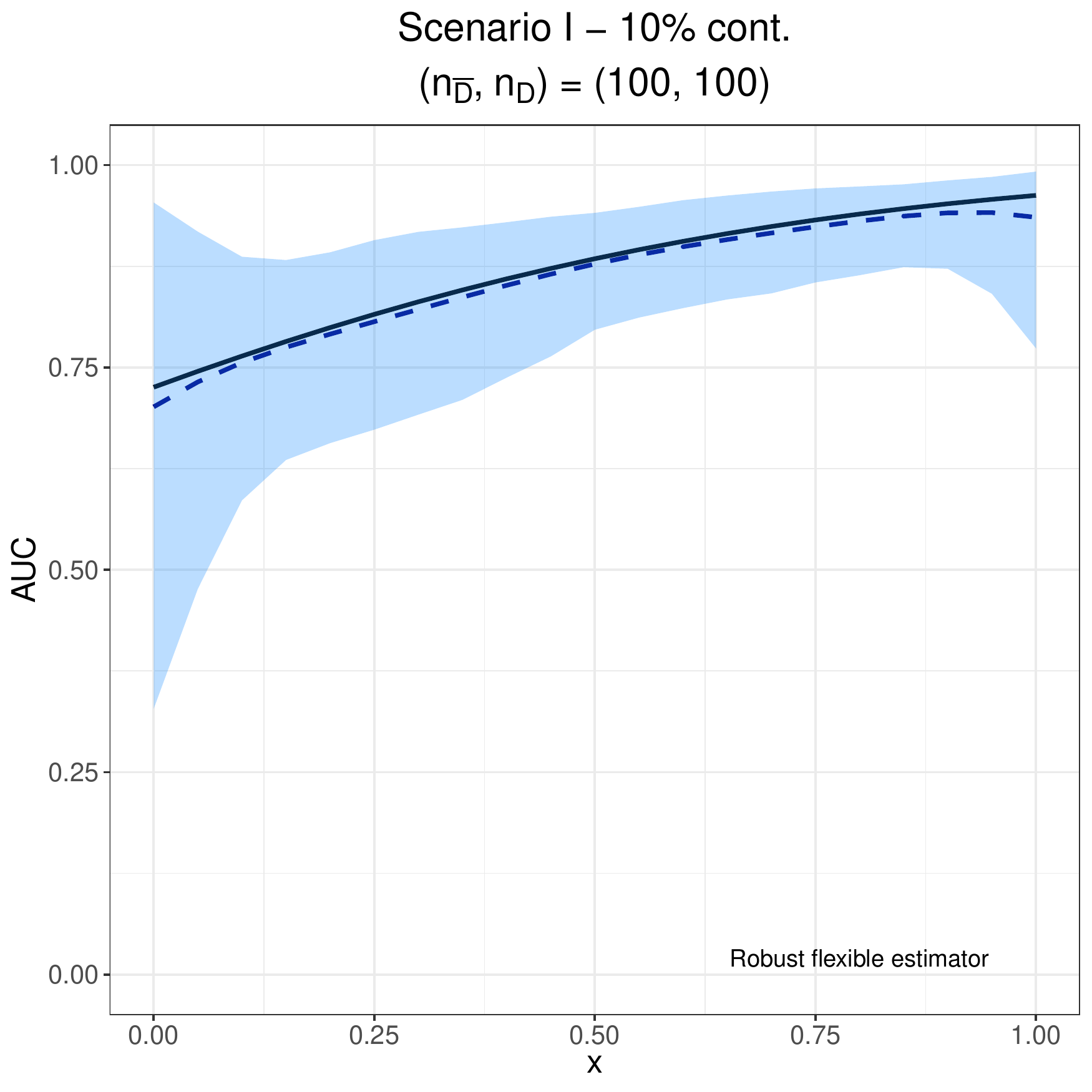}
			\includegraphics[width = 4.65cm]{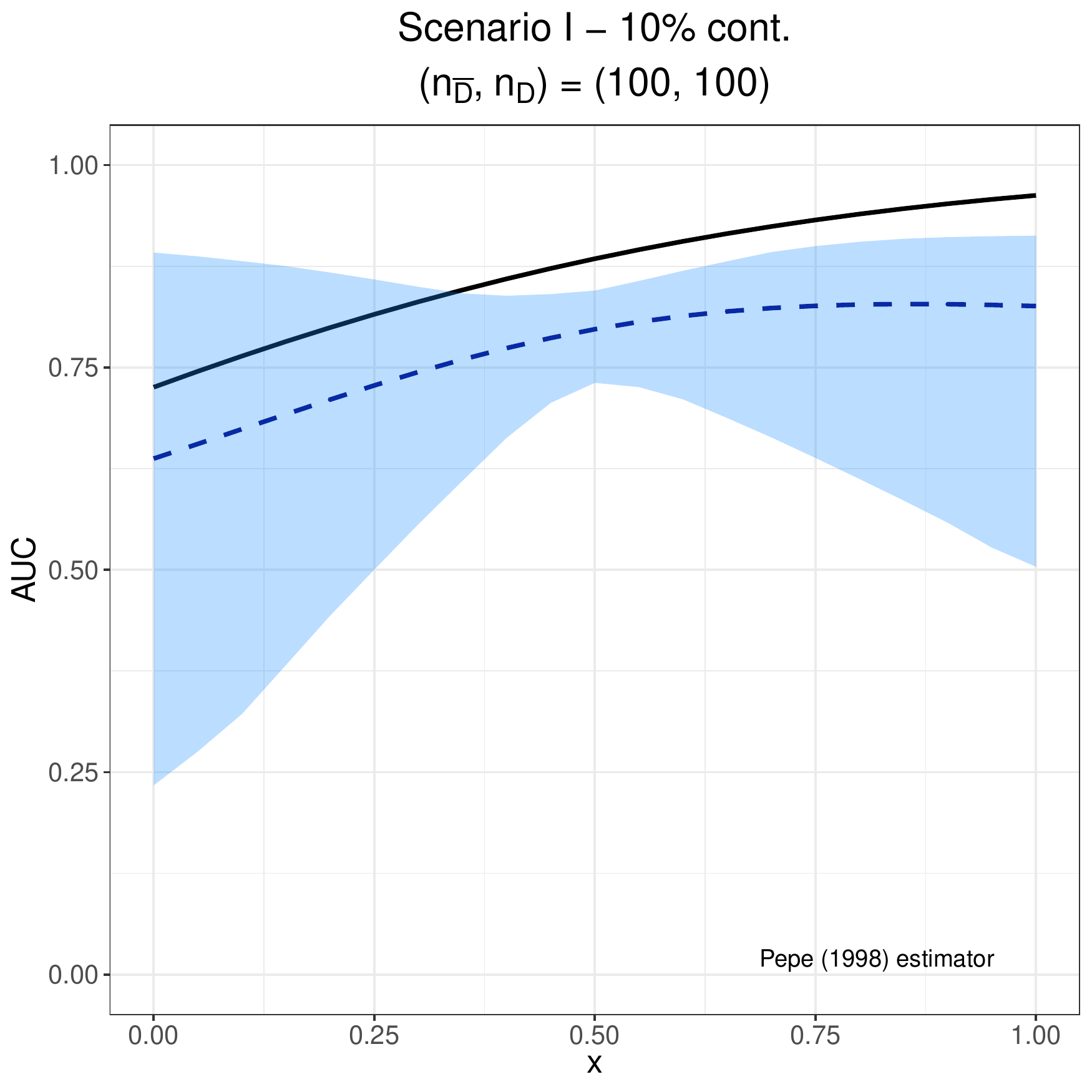}
			\includegraphics[width = 4.65cm]{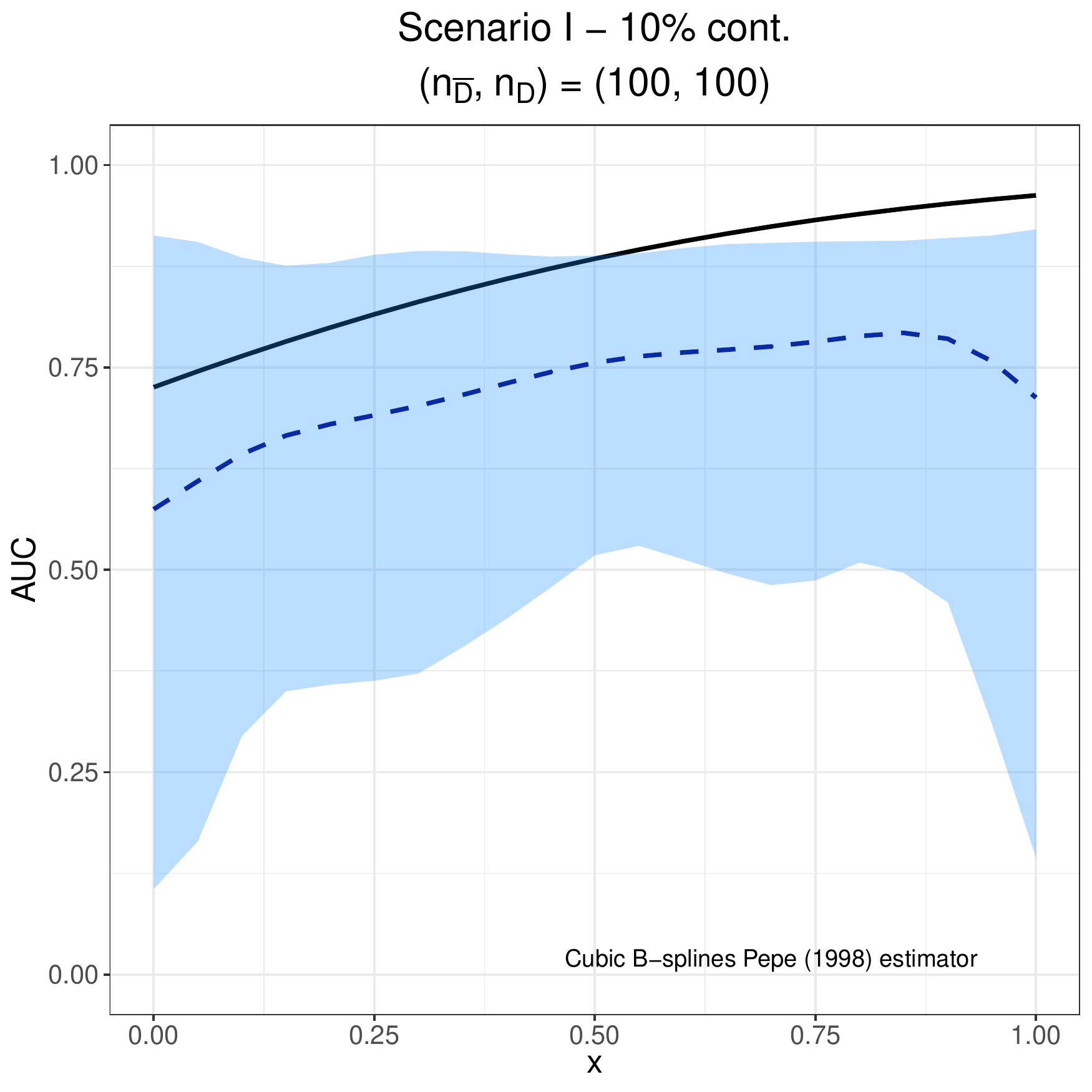}
			\includegraphics[width = 4.65cm]{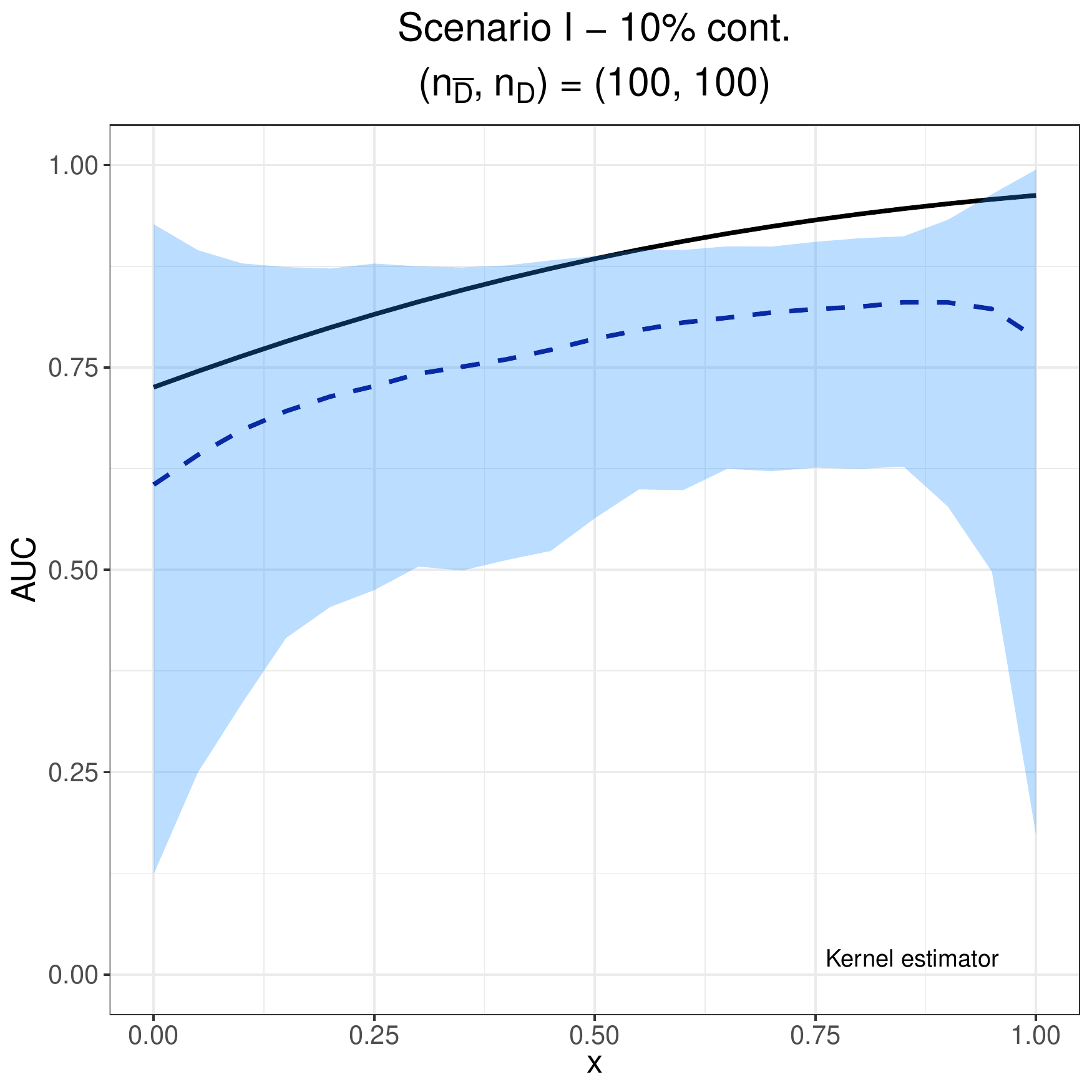}
		}
		\vspace{0.3cm}
		\subfigure{
			\includegraphics[width = 4.65cm]{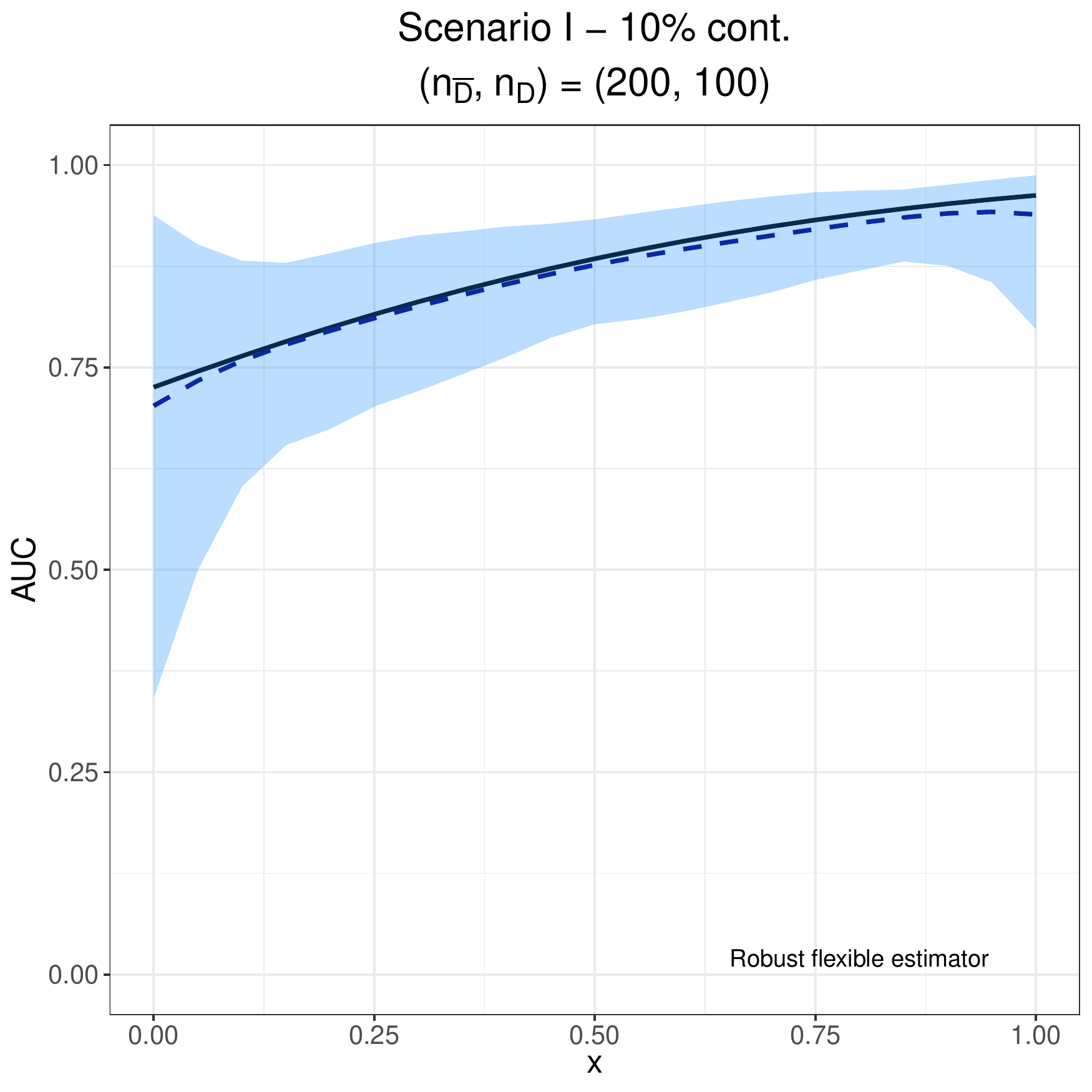}
			\includegraphics[width = 4.65cm]{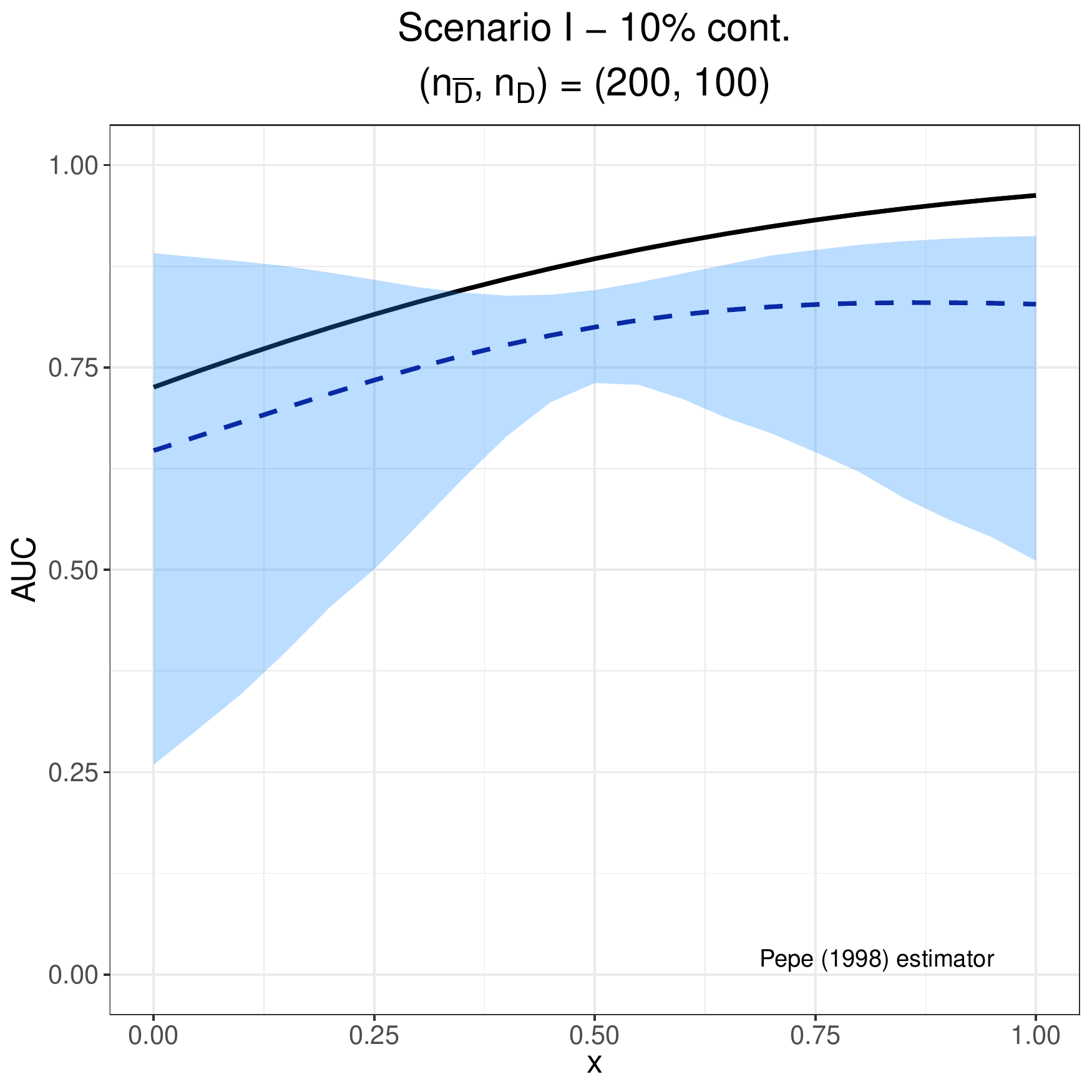}
			\includegraphics[width = 4.65cm]{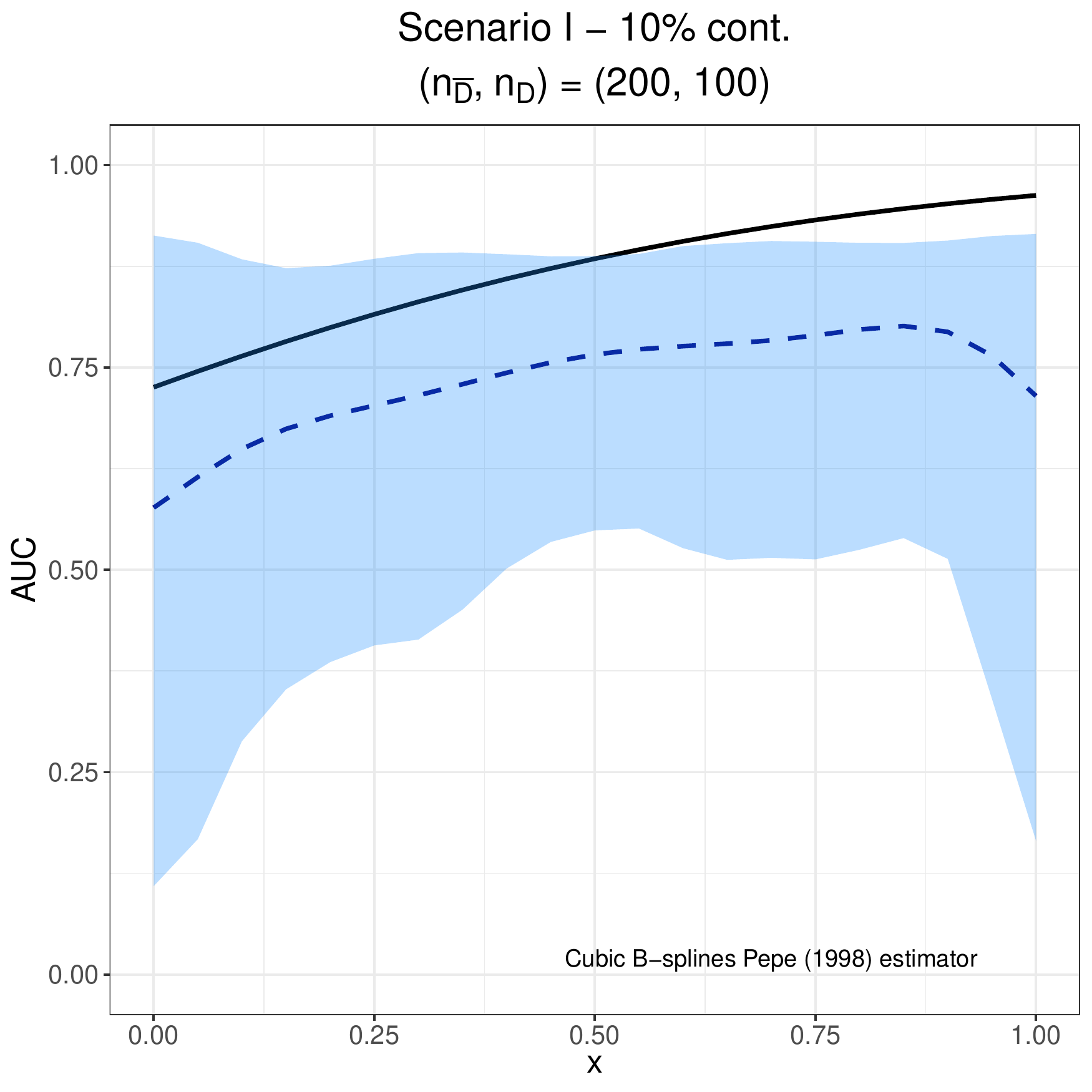}
			\includegraphics[width = 4.65cm]{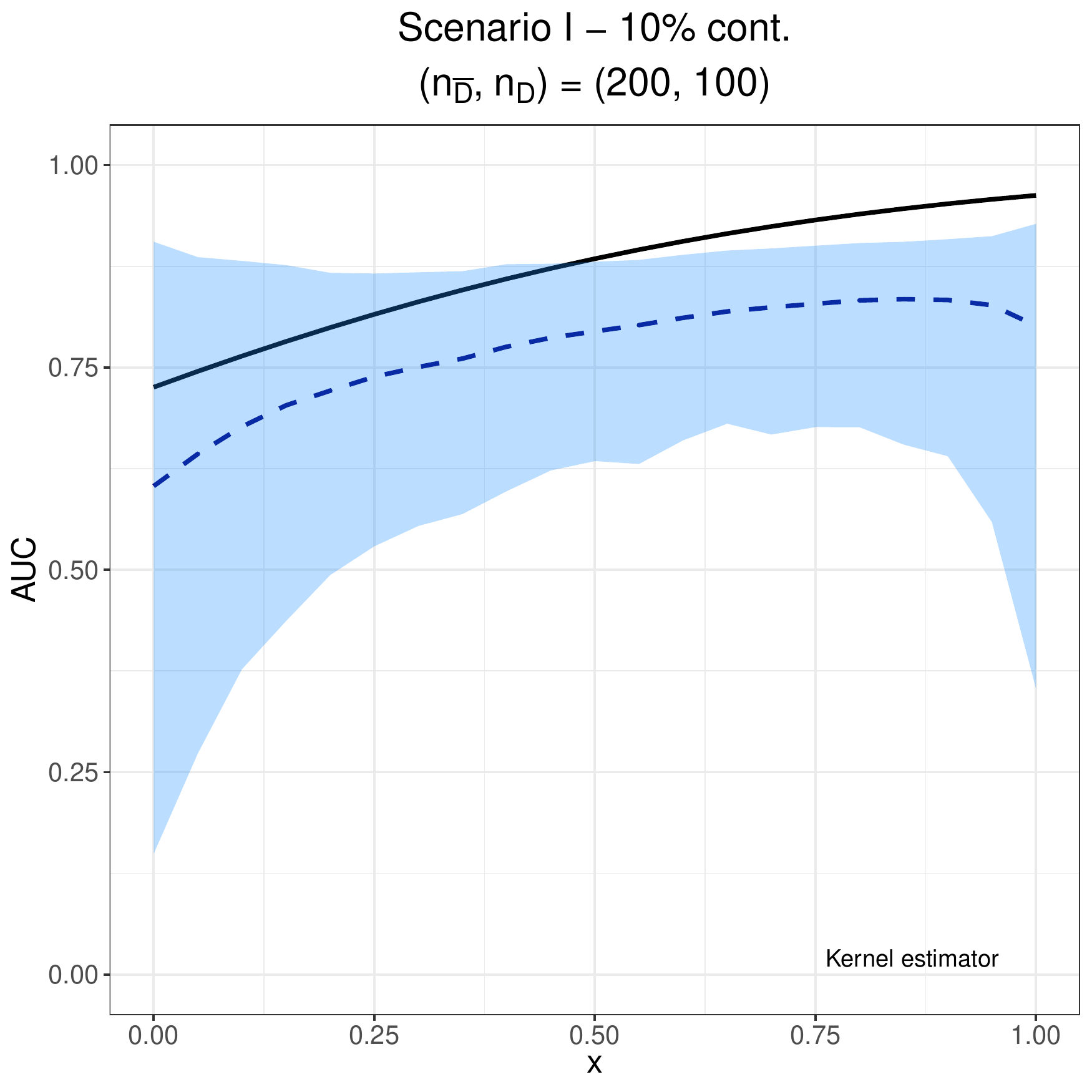}
		}
		\vspace{0.3cm}
		\subfigure{
			\includegraphics[width = 4.65cm]{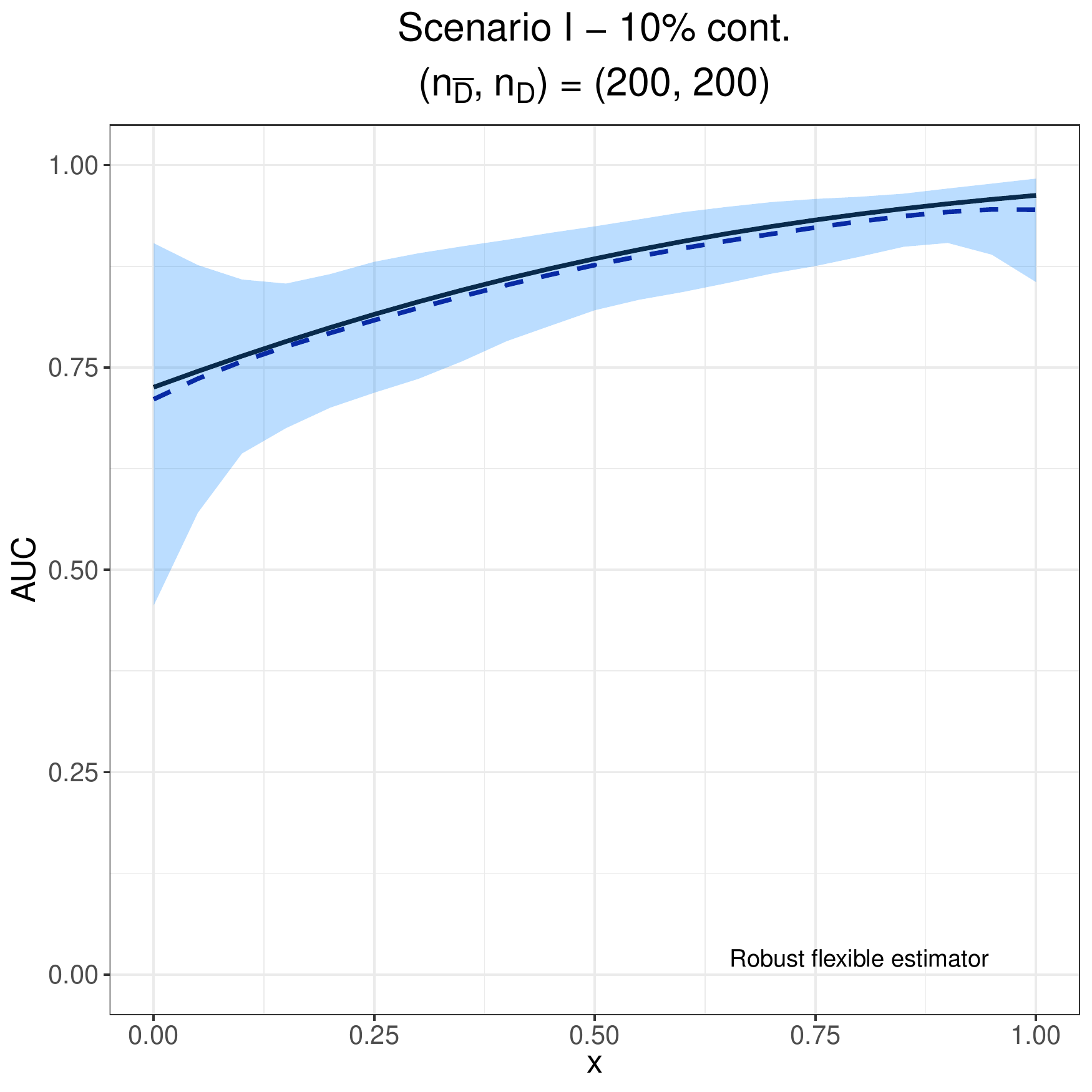}
			\includegraphics[width = 4.65cm]{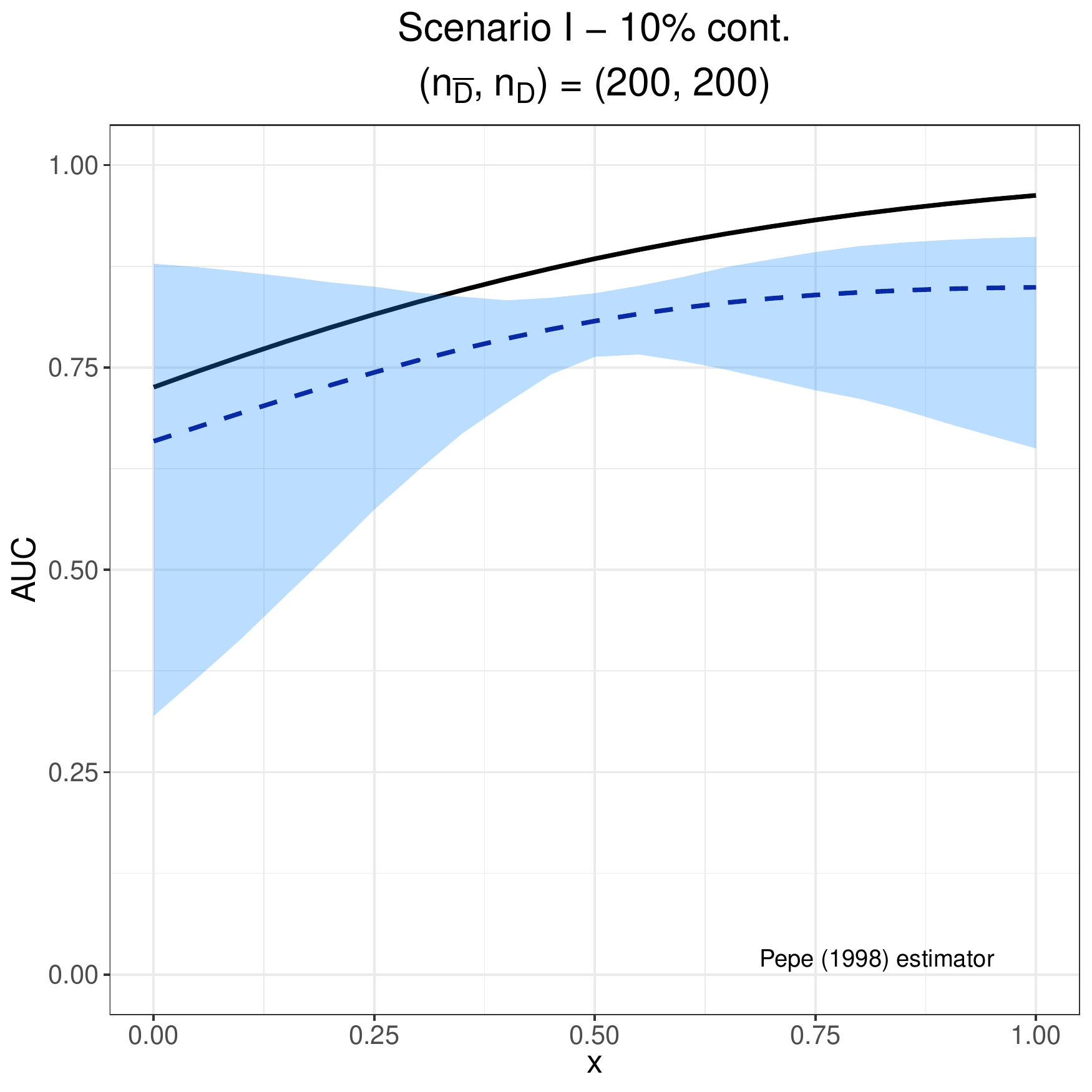}
			\includegraphics[width = 4.65cm]{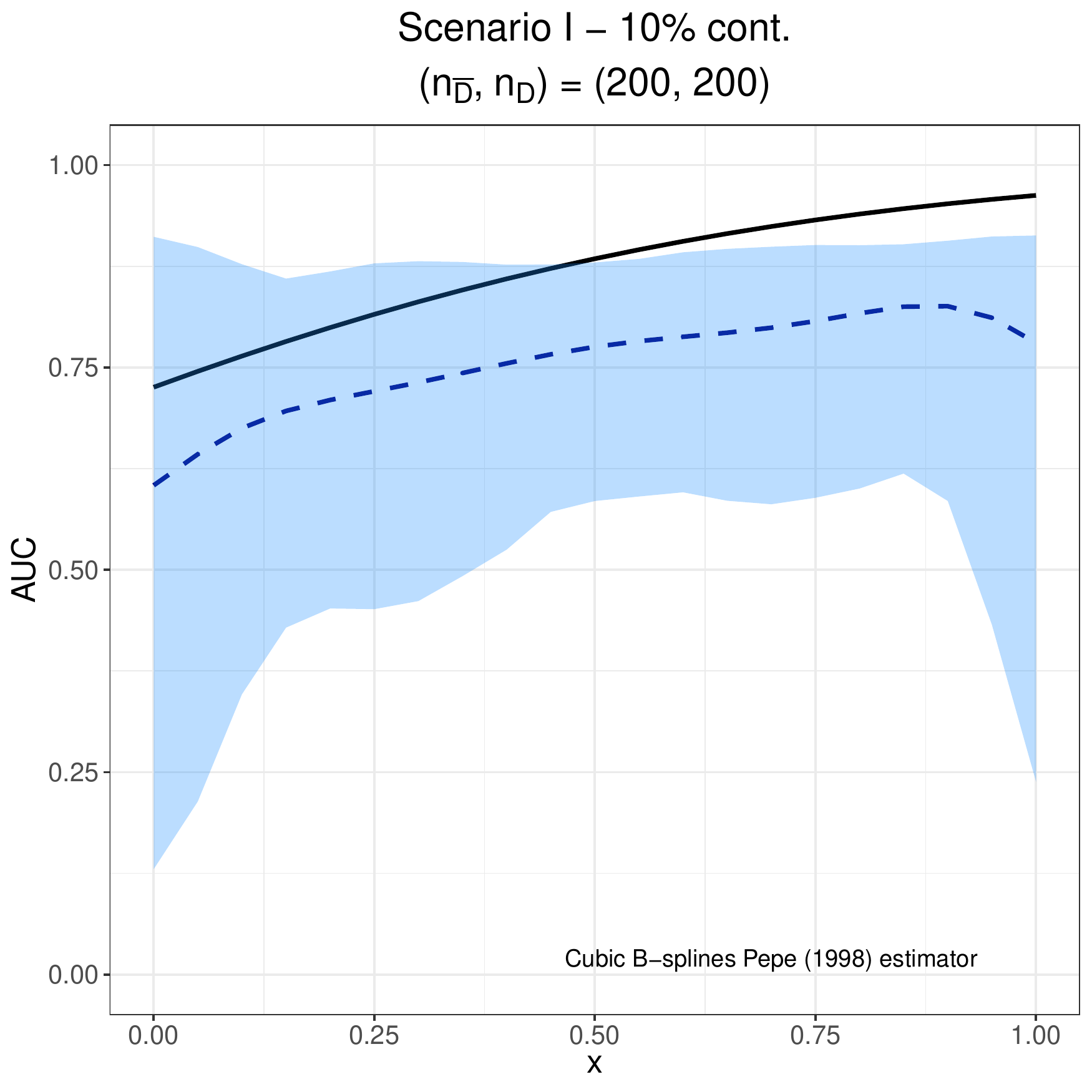}
			\includegraphics[width = 4.65cm]{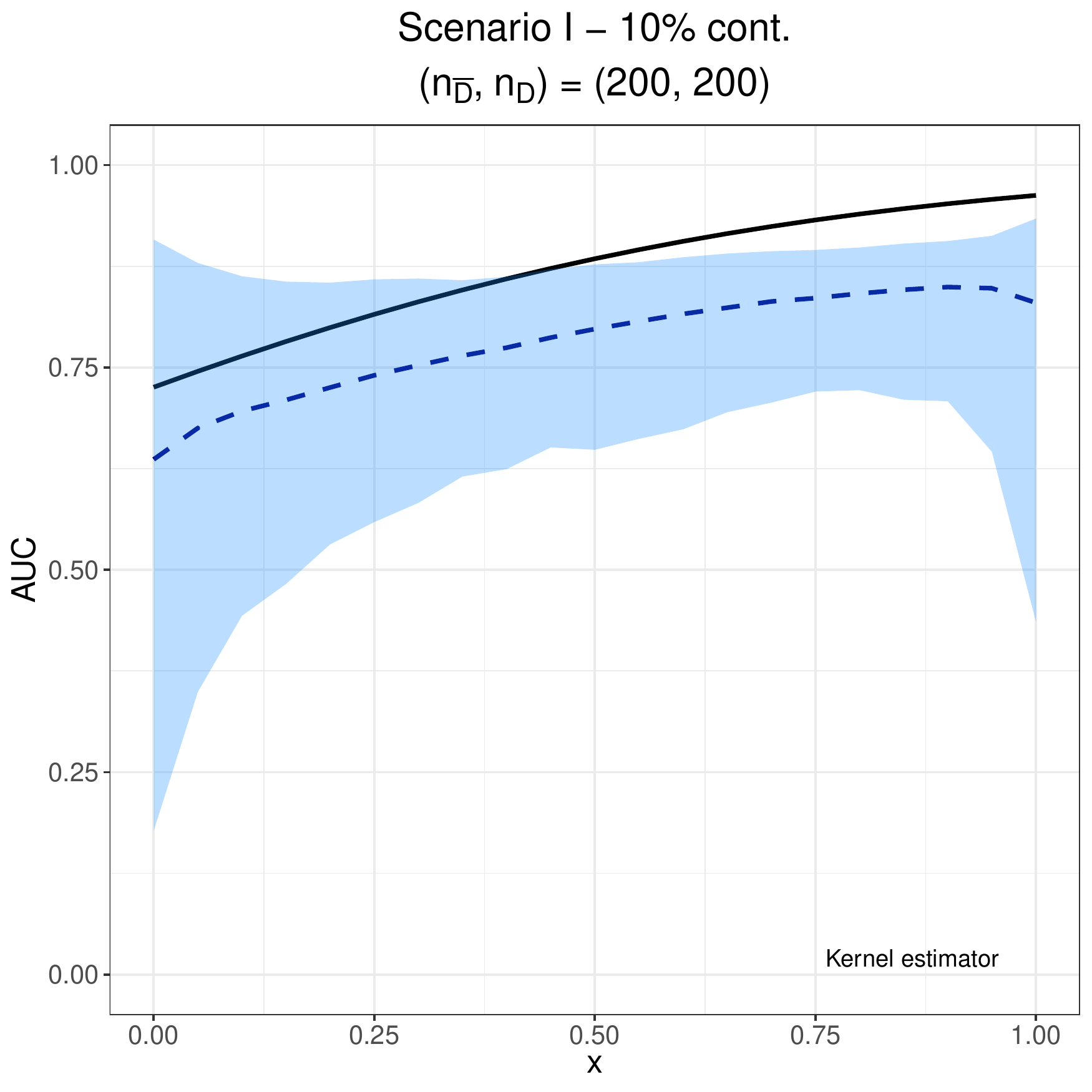}
		}
	\end{center}
	\caption{\footnotesize{Scenario I. True covariate-specific AUC (solid line) versus the mean of the Monte Carlo estimates (dashed line) along with the $2.5\%$ and $97.5\%$ simulation quantiles (shaded area) for the case of $10\%$ of contamination. The first row displays the results for $(n_{\bar{D}}, n_D)=(100,100)$, the second row for $(n_{\bar{D}}, n_D)=(200,100)$, and the third row for $(n_{\bar{D}}, n_D)=(200,200)$. The first column corresponds to our flexible and robust estimator, the second column to the estimator proposed by Pepe (1998), the third one to the cubic B-splines extension of Pepe (1998), and the fourth column to the kernel estimator.}}
\end{figure}

\begin{figure}[H]
	\begin{center}
		\subfigure{
			\includegraphics[width = 4.65cm]{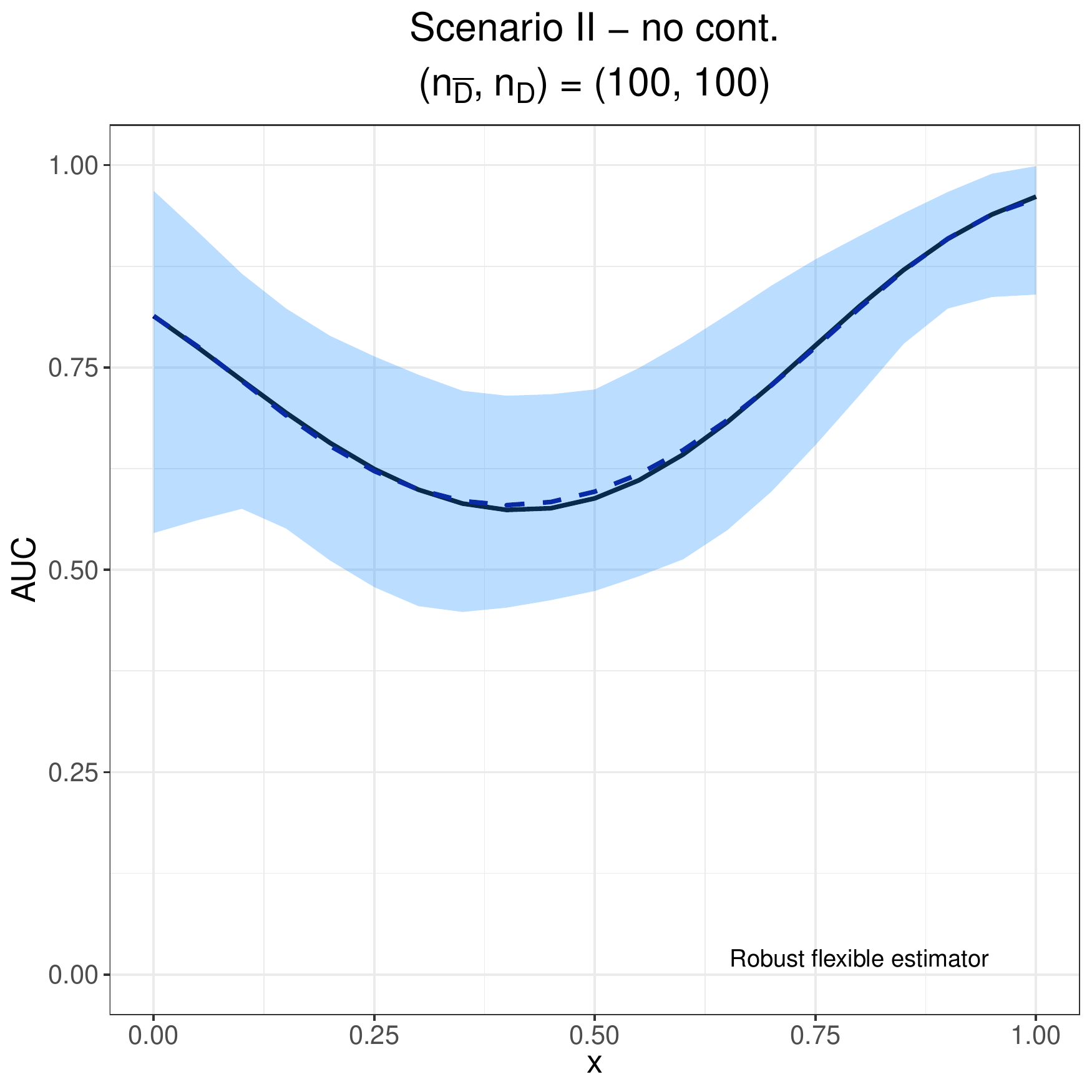}
			\includegraphics[width = 4.65cm]{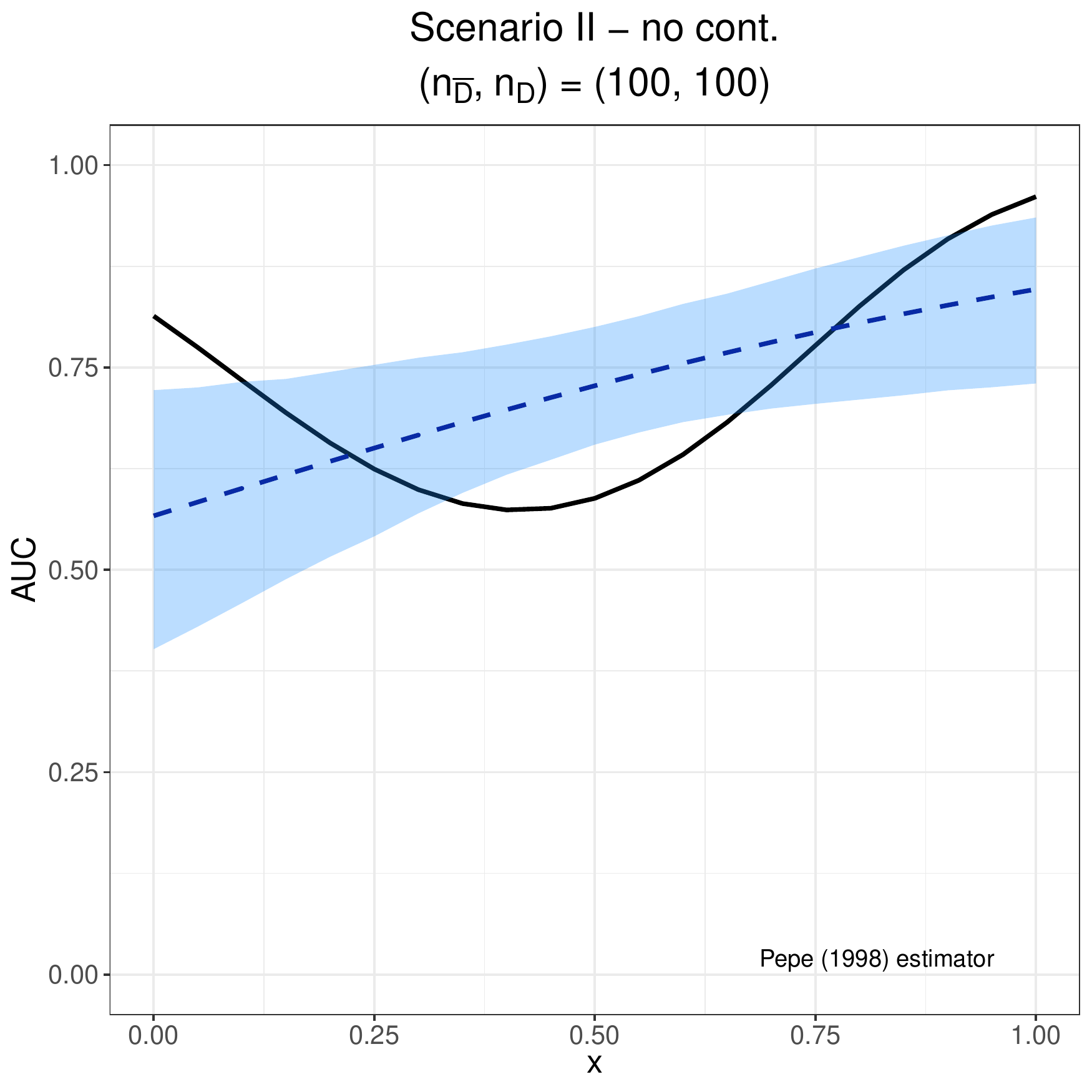}
			\includegraphics[width = 4.65cm]{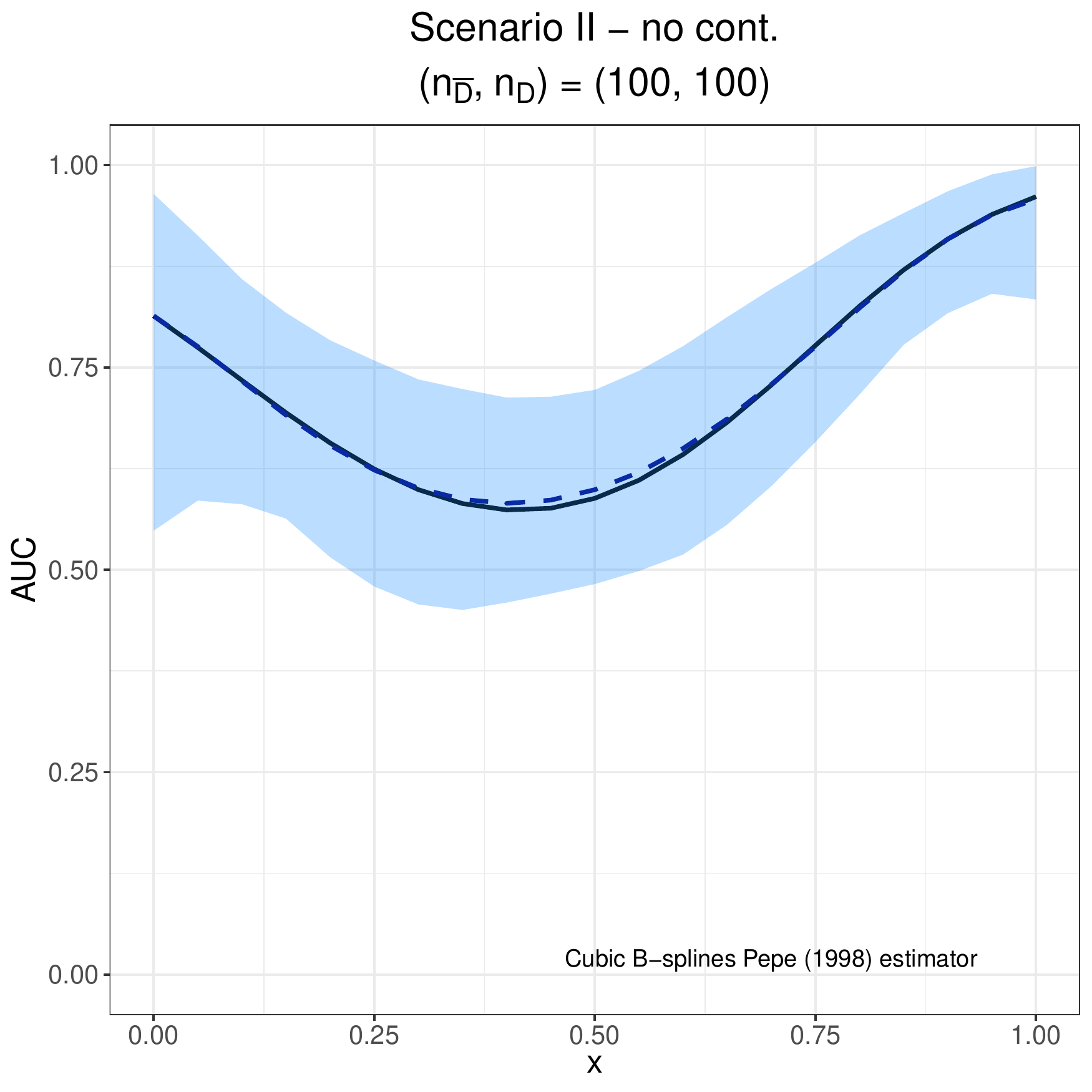}
			\includegraphics[width = 4.65cm]{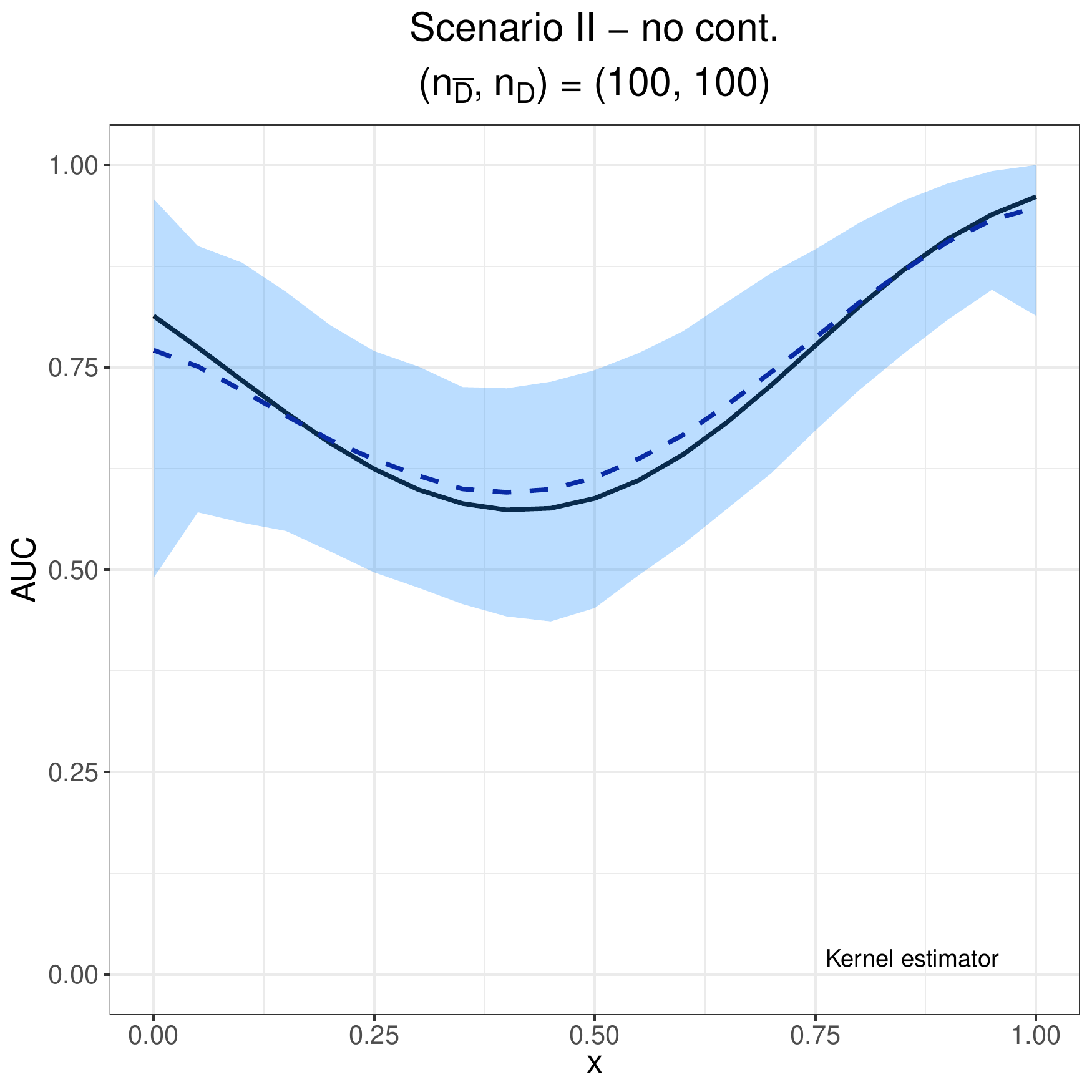}
		}
		\vspace{0.3cm}
		\subfigure{
			\includegraphics[width = 4.65cm]{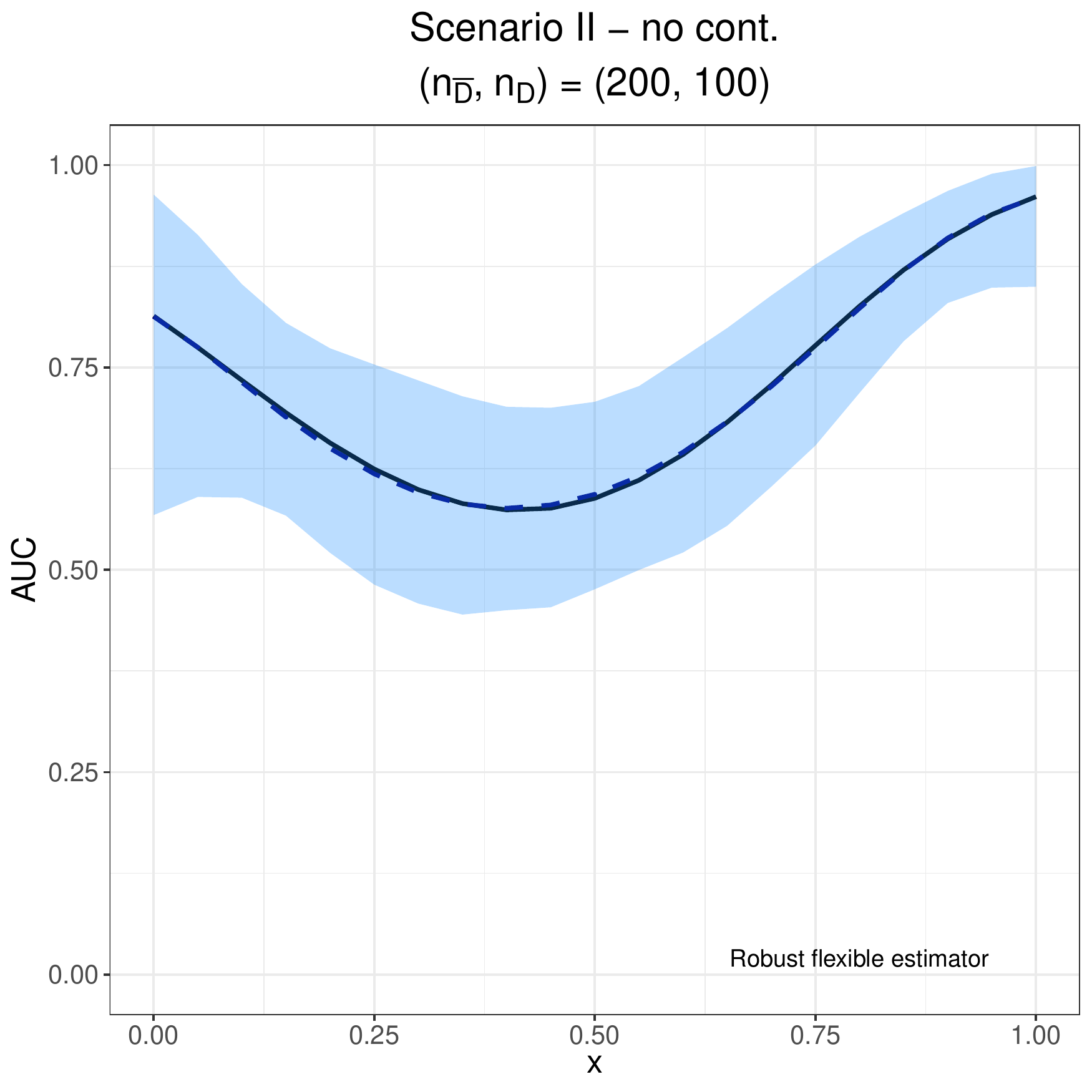}
			\includegraphics[width = 4.65cm]{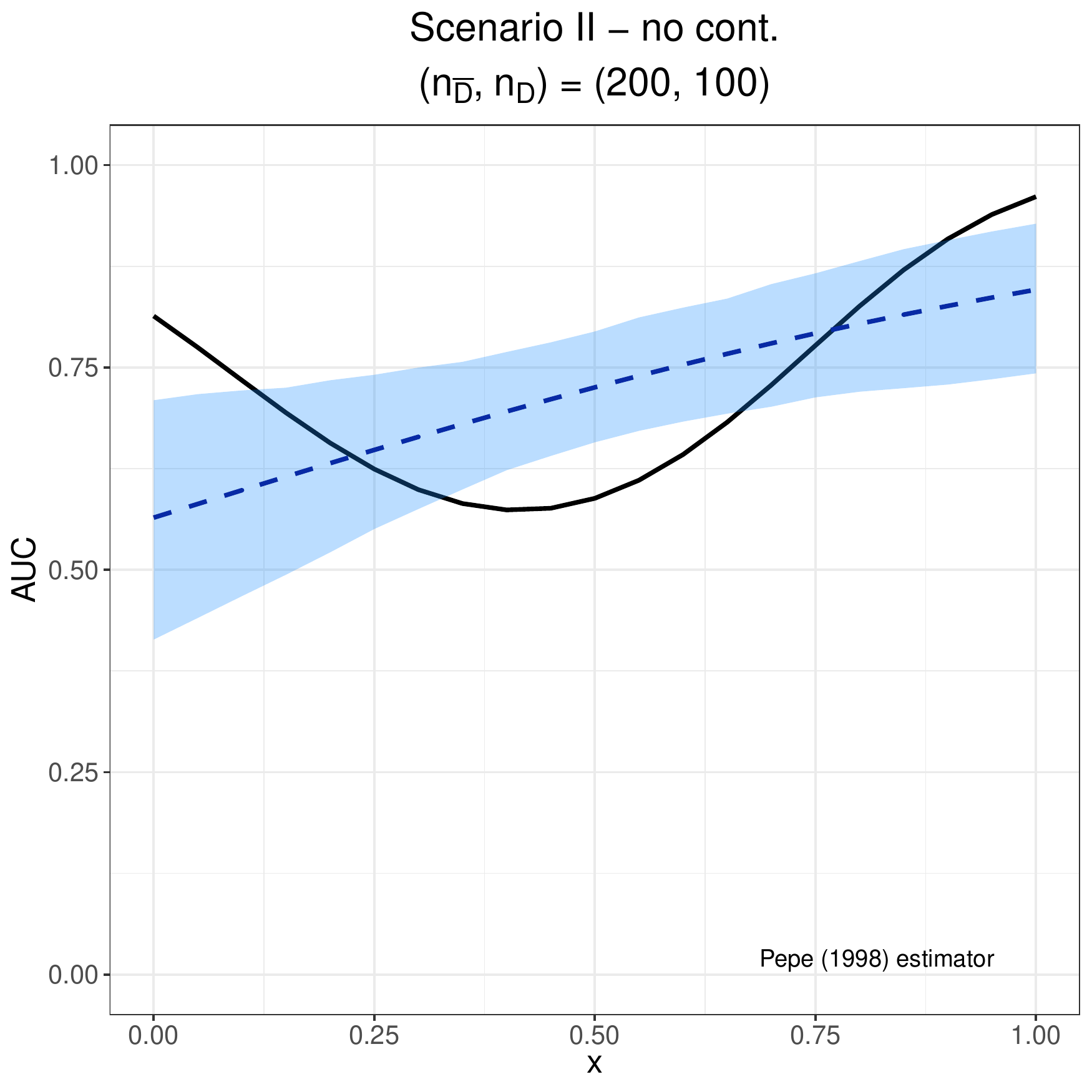}
			\includegraphics[width = 4.65cm]{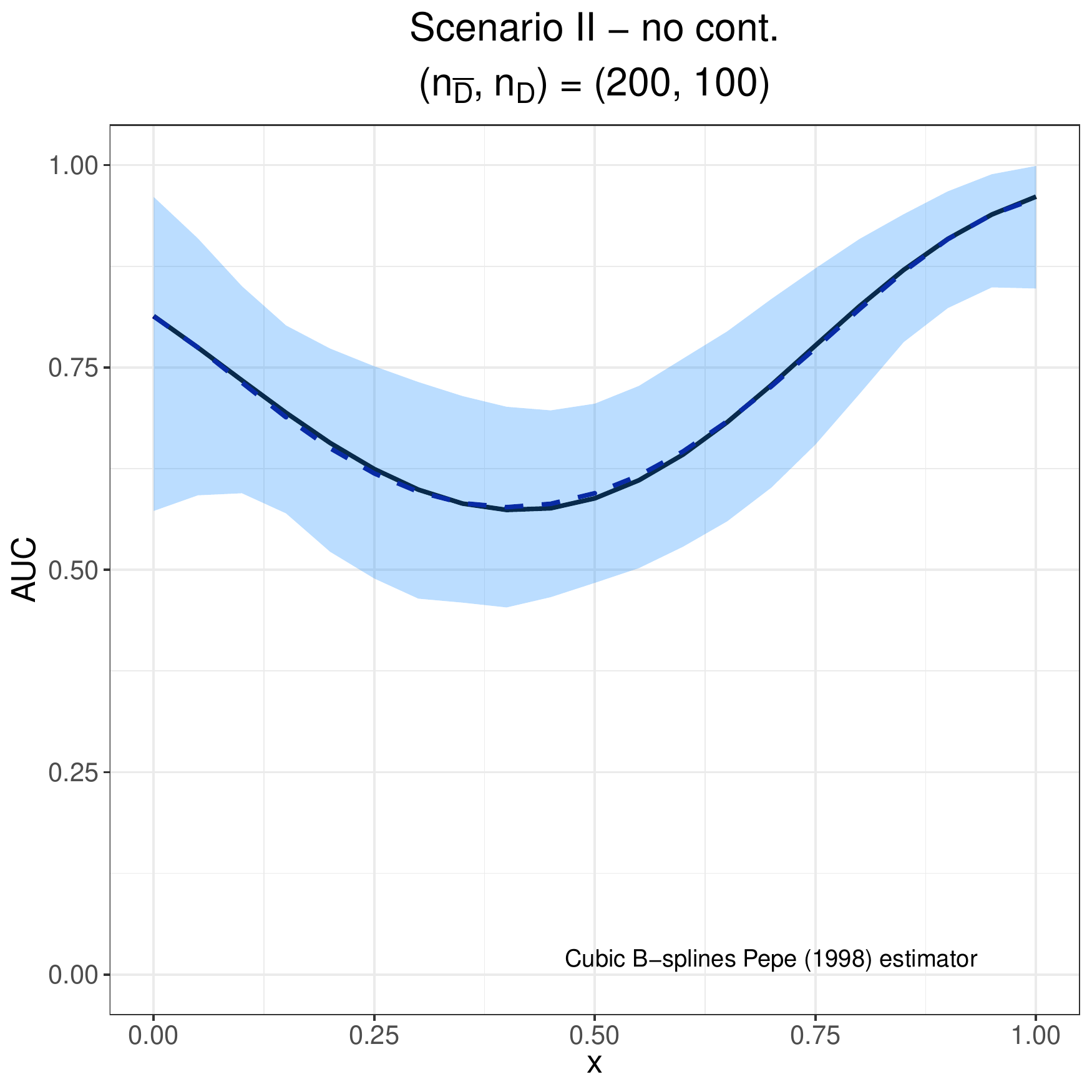}
			\includegraphics[width = 4.65cm]{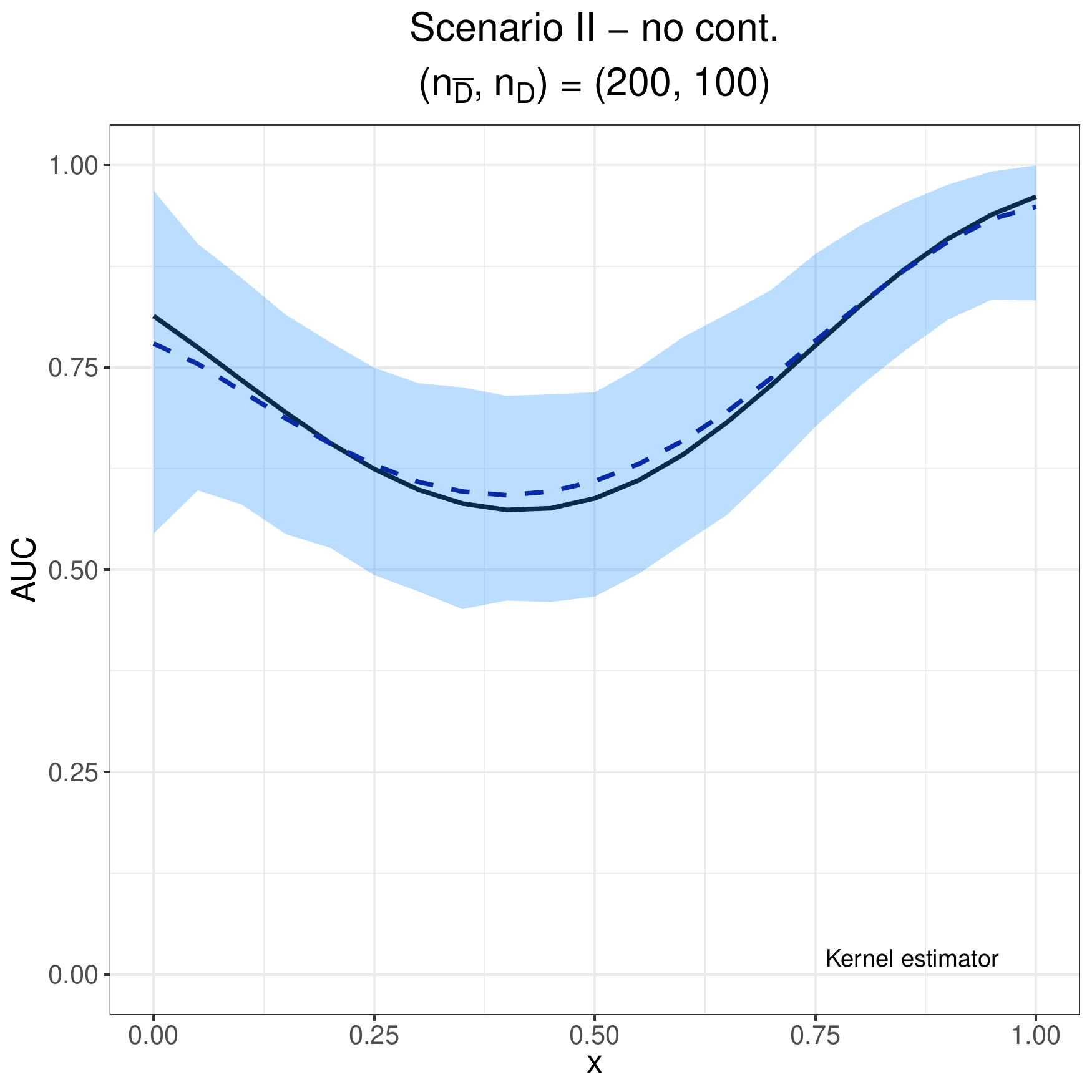}
		}
		\vspace{0.3cm}
		\subfigure{
			\includegraphics[width = 4.65cm]{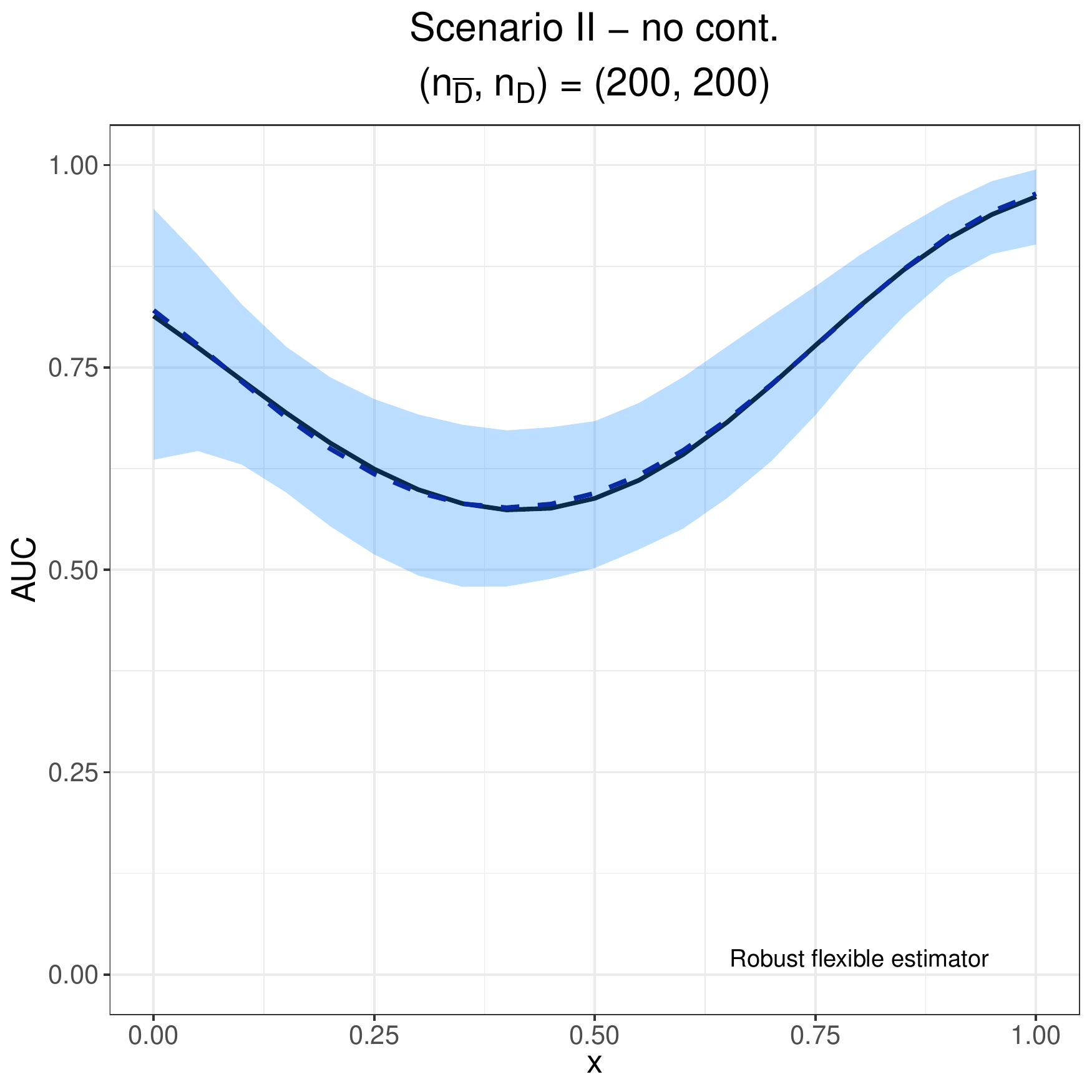}
			\includegraphics[width = 4.65cm]{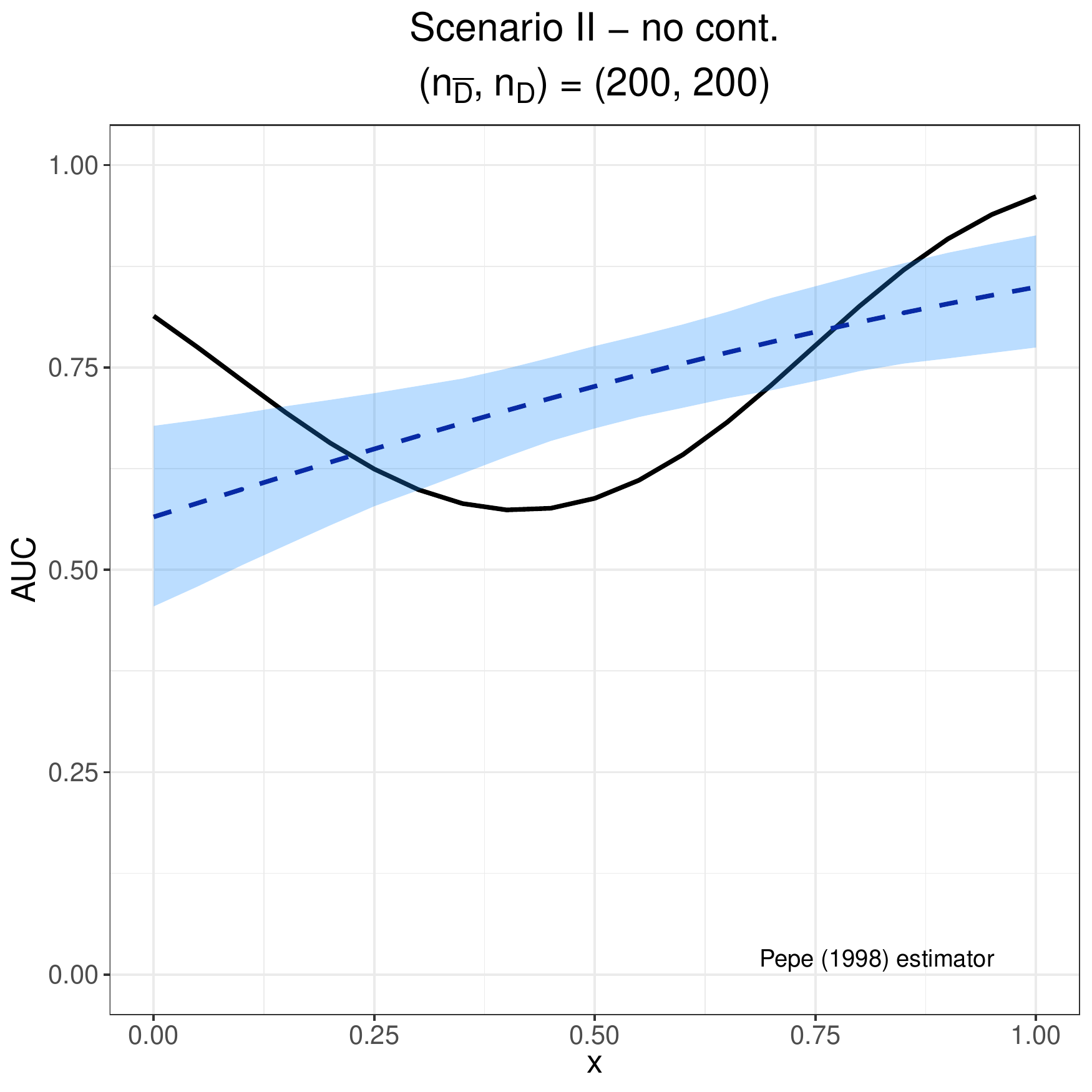}
			\includegraphics[width = 4.65cm]{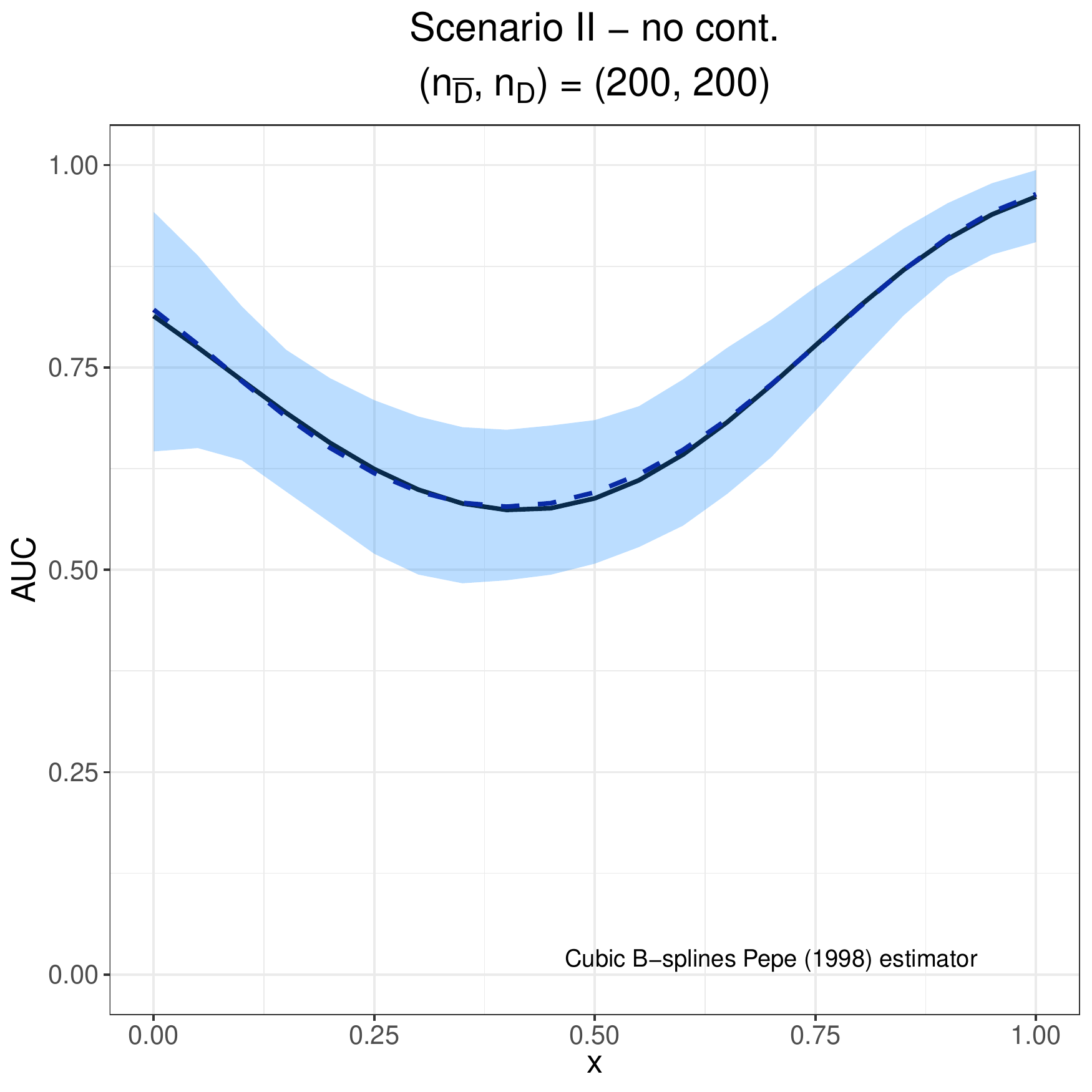}
			\includegraphics[width = 4.65cm]{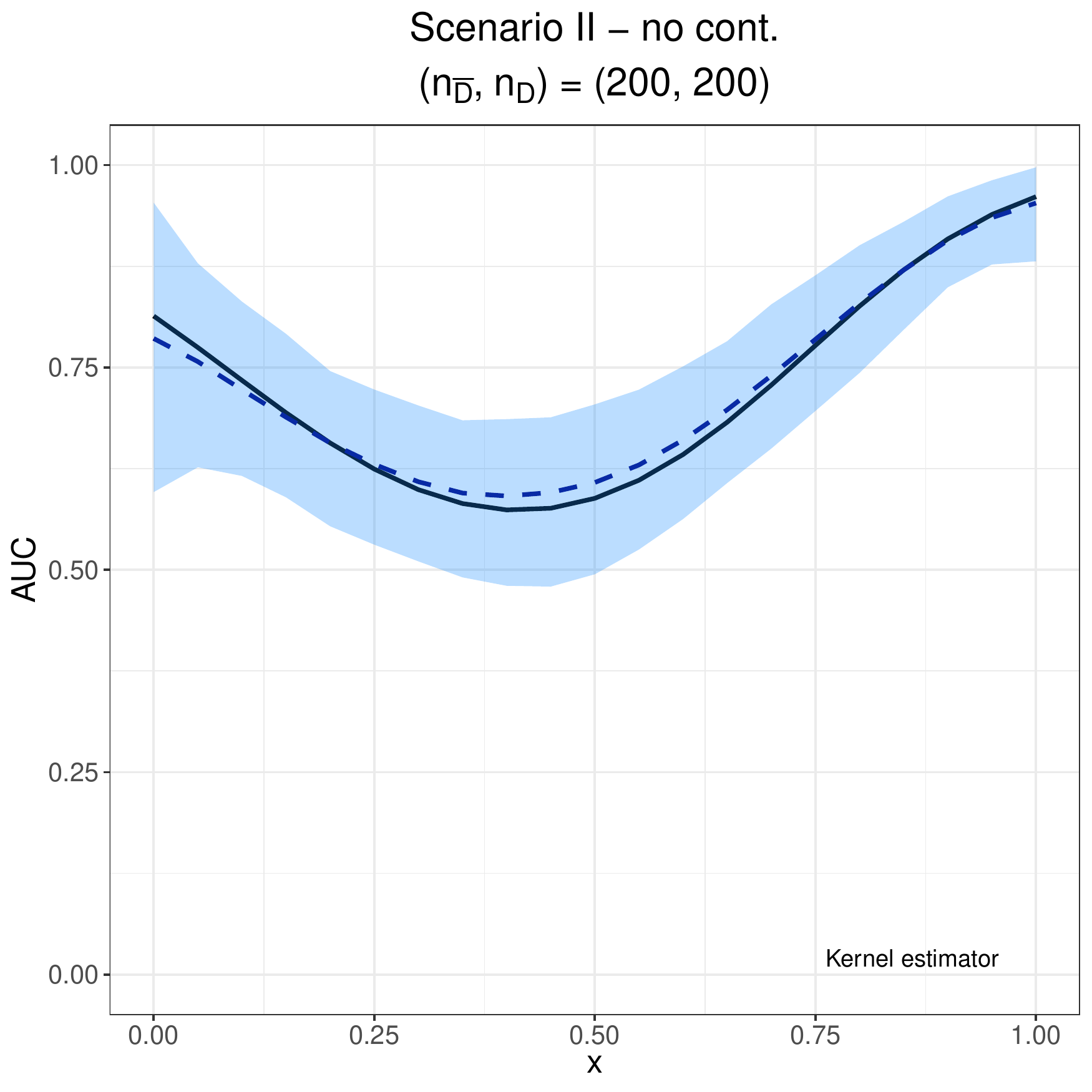}
		}
	\end{center}
	\caption{\footnotesize{Scenario II. True covariate-specific AUC (solid line) versus the mean of the Monte Carlo estimates (dashed line) along with the $2.5\%$ and $97.5\%$ simulation quantiles (shaded area) for the case of no contamination. The first row displays the results for $(n_{\bar{D}}, n_D)=(100,100)$, the second row for $(n_{\bar{D}}, n_D)=(200,100)$, and the third row for $(n_{\bar{D}}, n_D)=(200,200)$. The first column corresponds to our flexible and robust estimator, the second column to the estimator proposed by Pepe (1998), the third one to the cubic B-splines extension of Pepe (1998), and the fourth column to the kernel estimator.}}
\end{figure}

\begin{figure}[H]
	\begin{center}
		\subfigure{
			\includegraphics[width = 4.65cm]{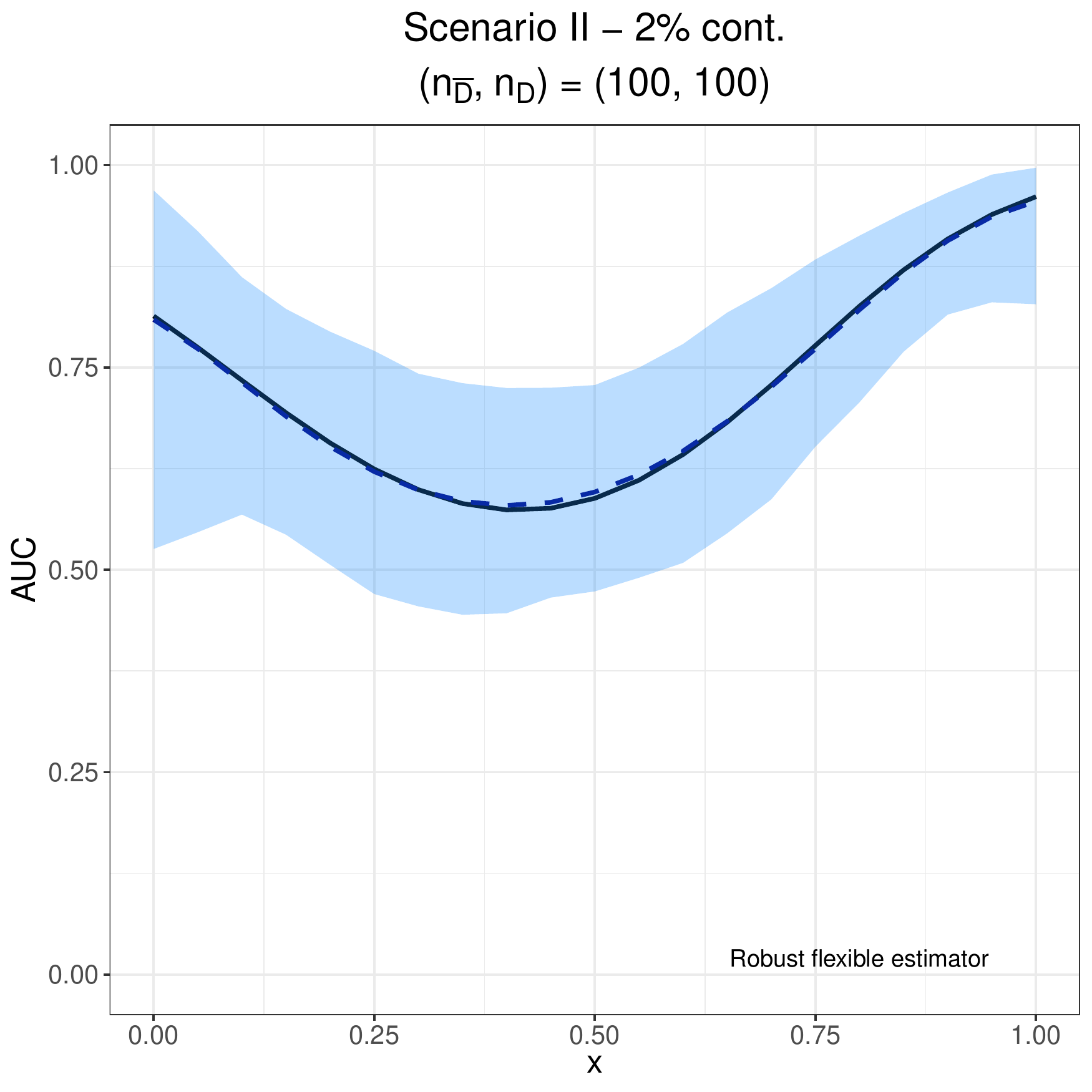}
			\includegraphics[width = 4.65cm]{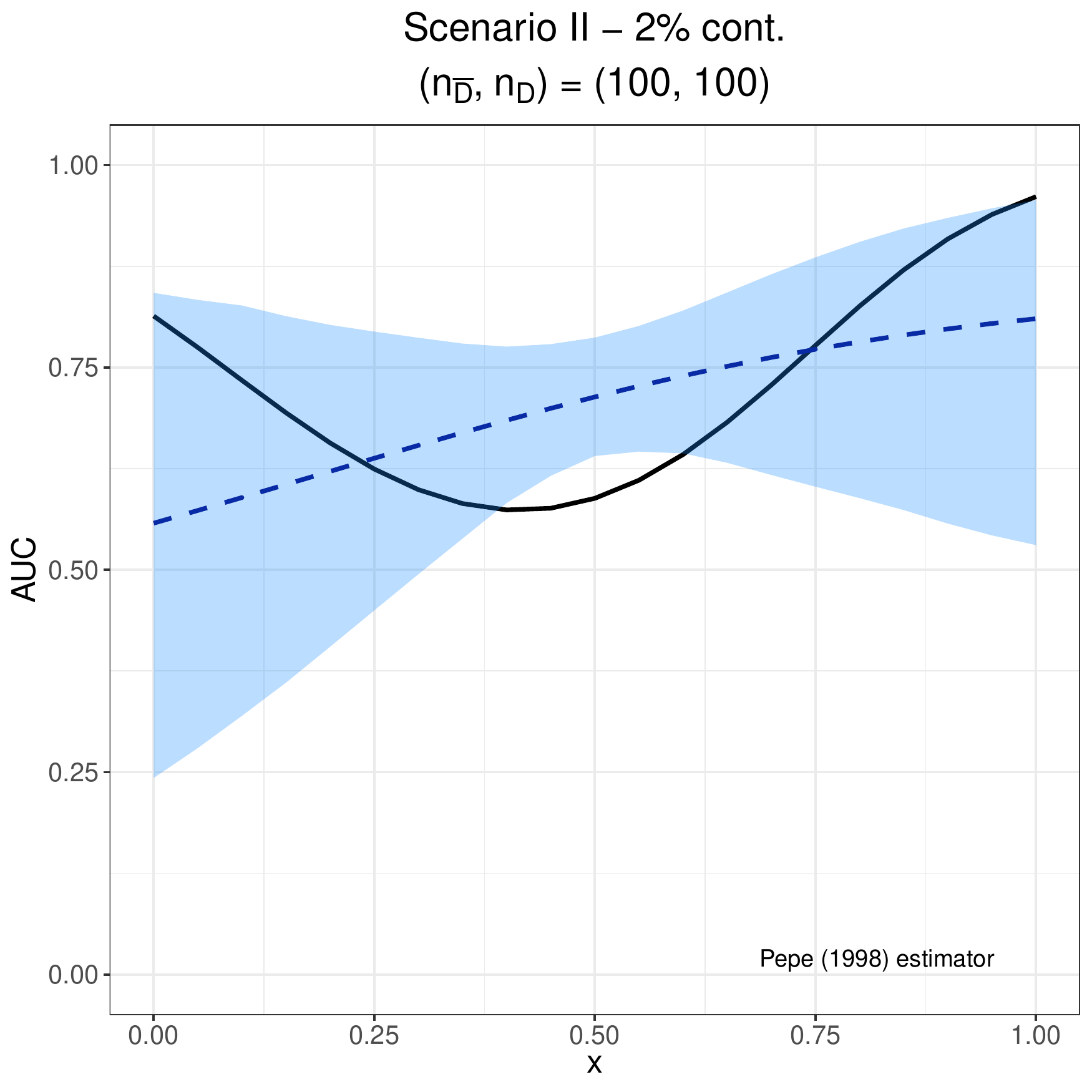}
			\includegraphics[width = 4.65cm]{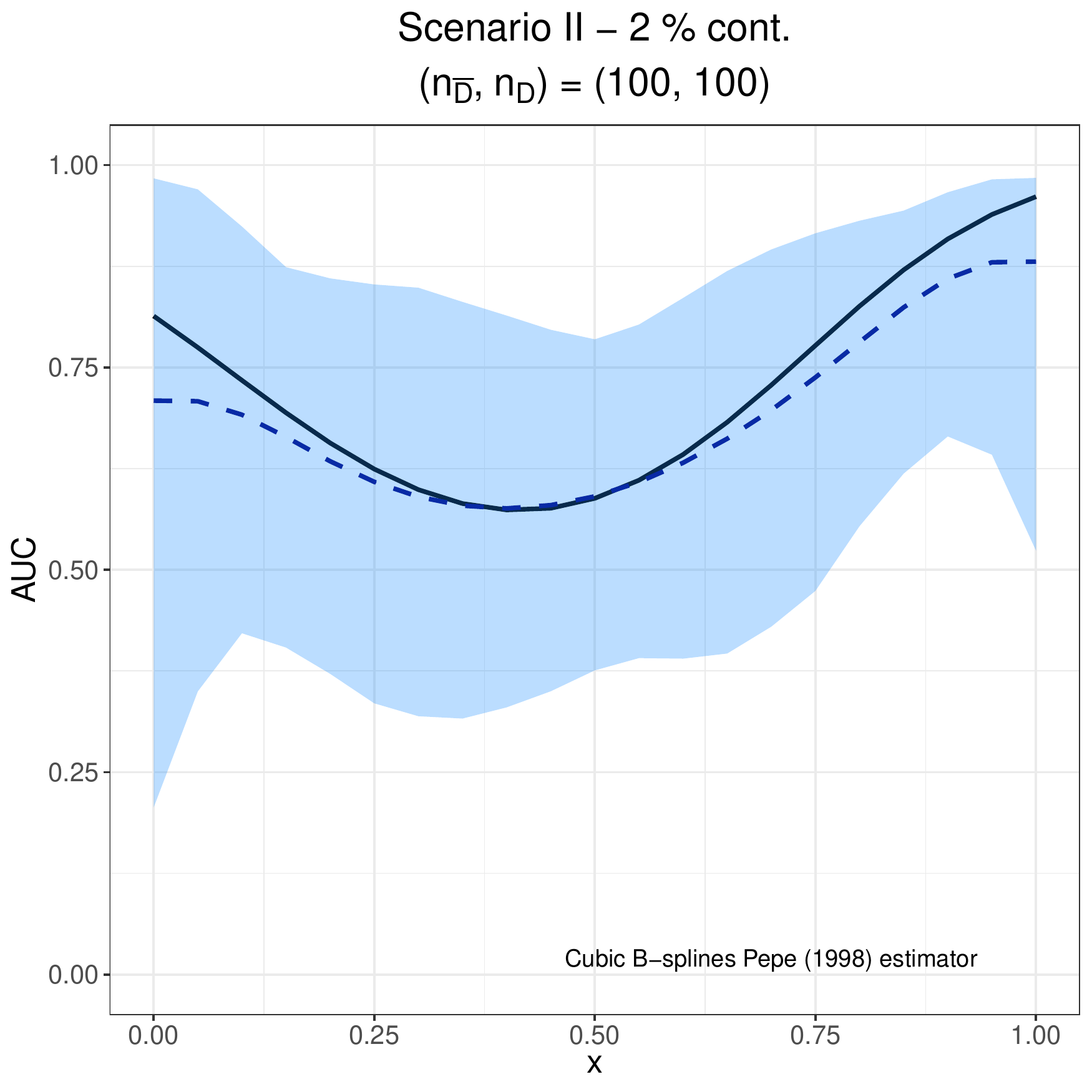}
			\includegraphics[width = 4.65cm]{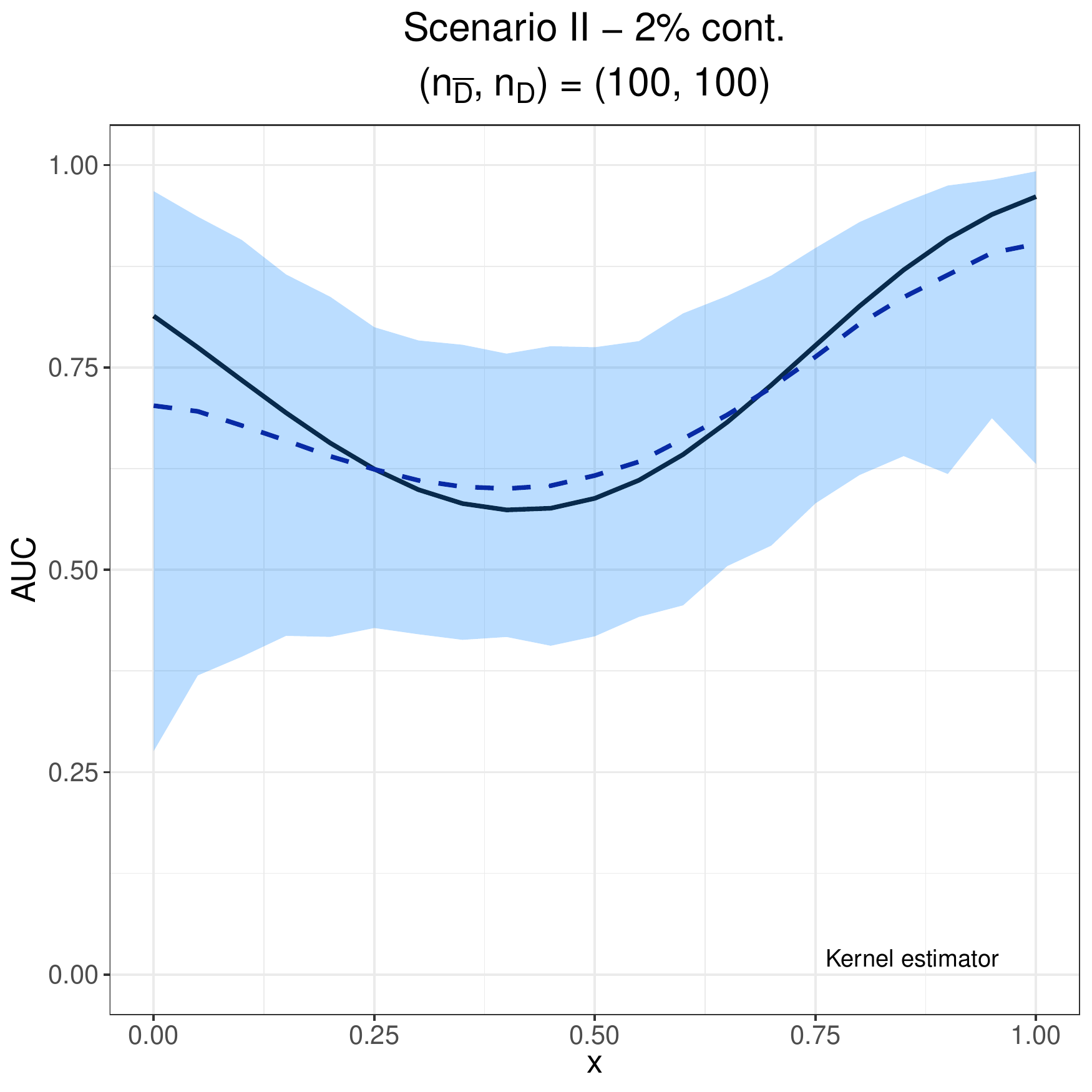}
		}
		\vspace{0.3cm}
		\subfigure{
			\includegraphics[width = 4.65cm]{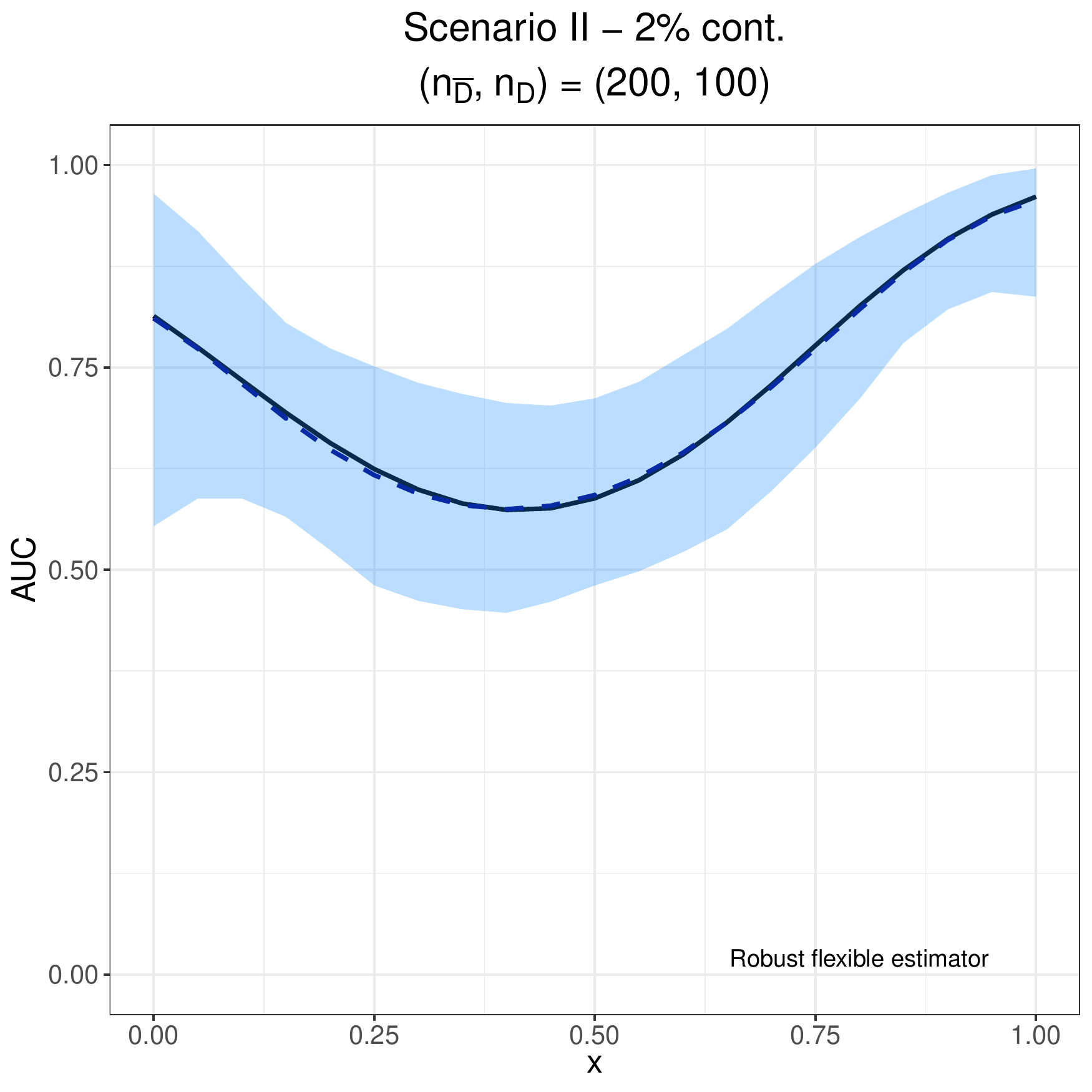}
			\includegraphics[width = 4.65cm]{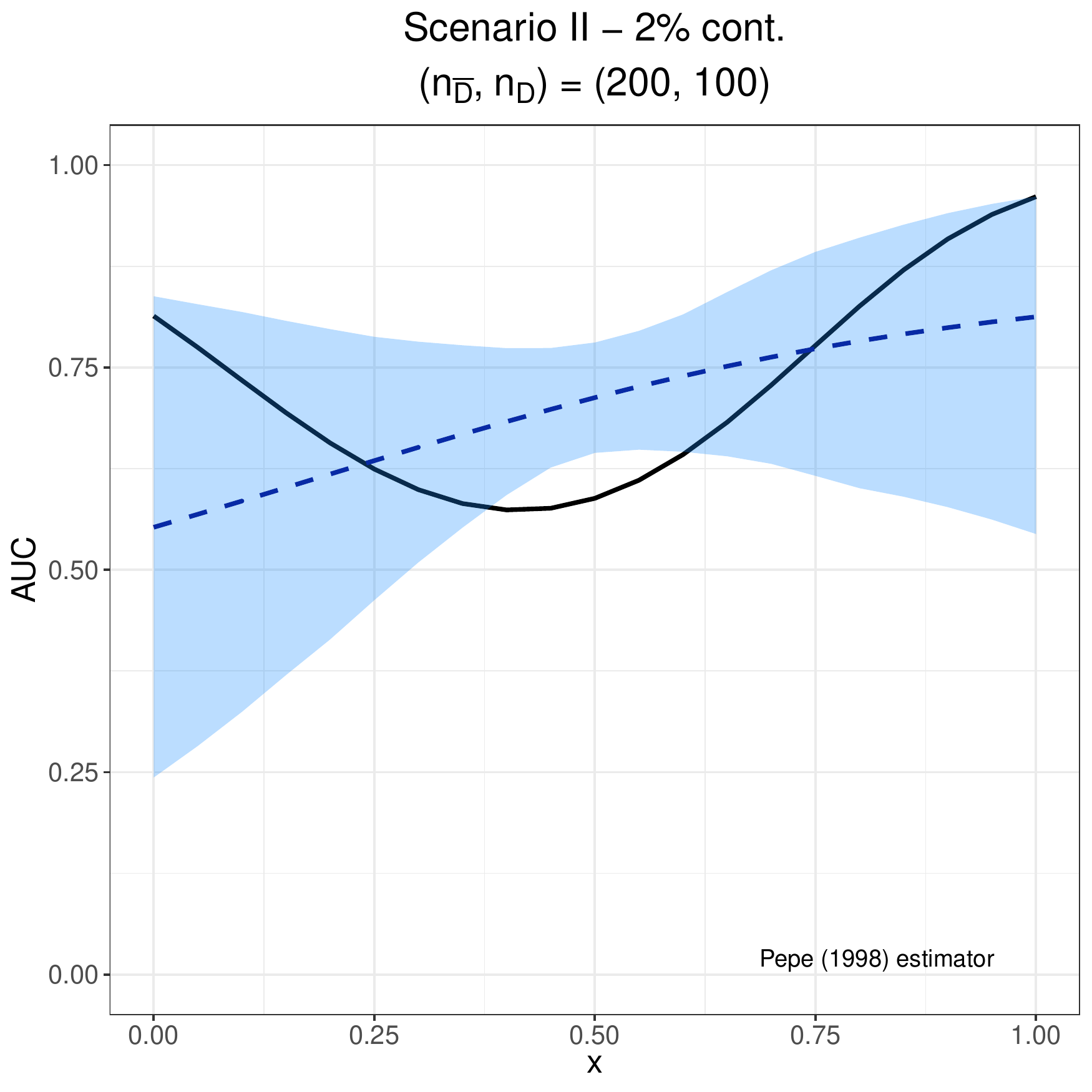}
			\includegraphics[width = 4.65cm]{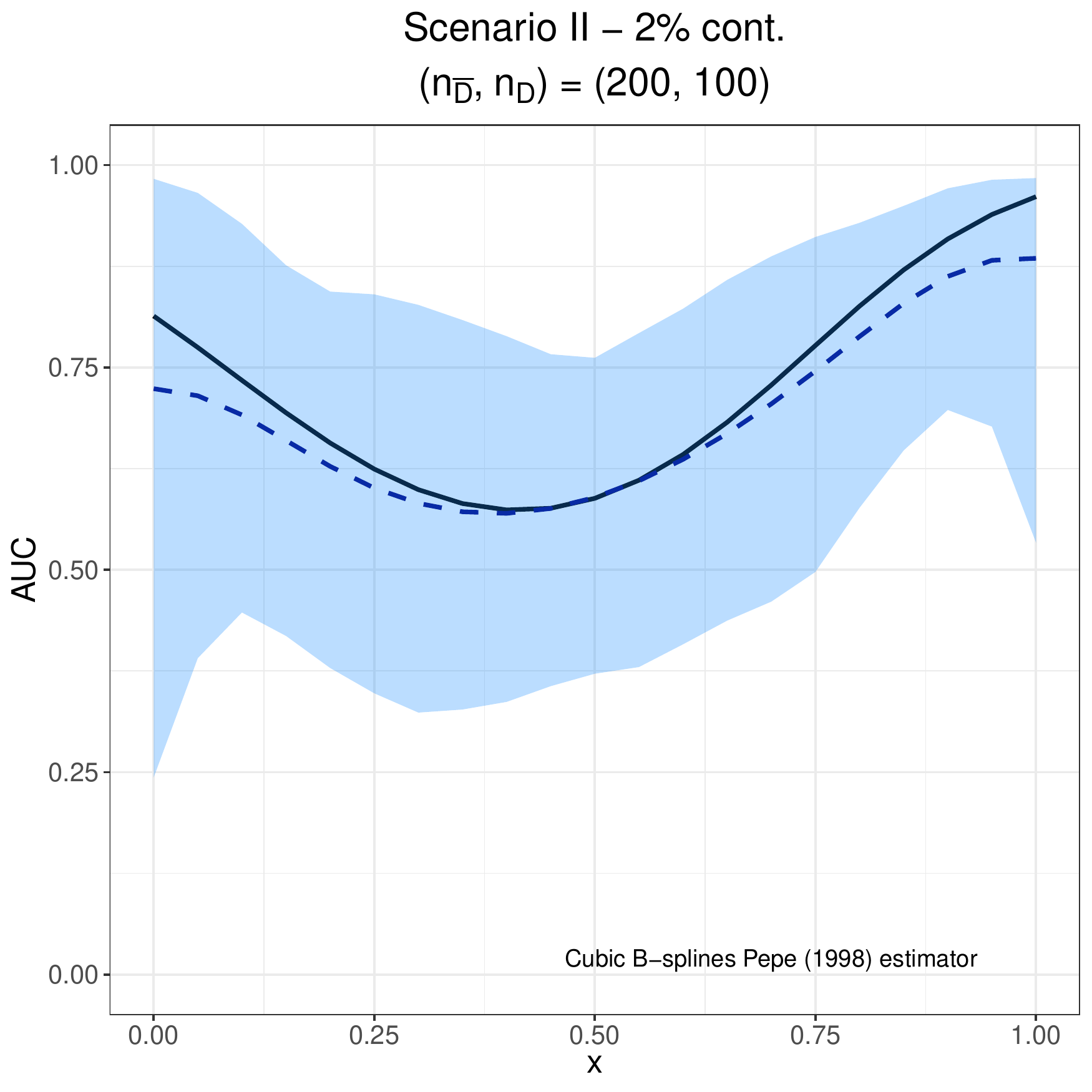}
			\includegraphics[width = 4.65cm]{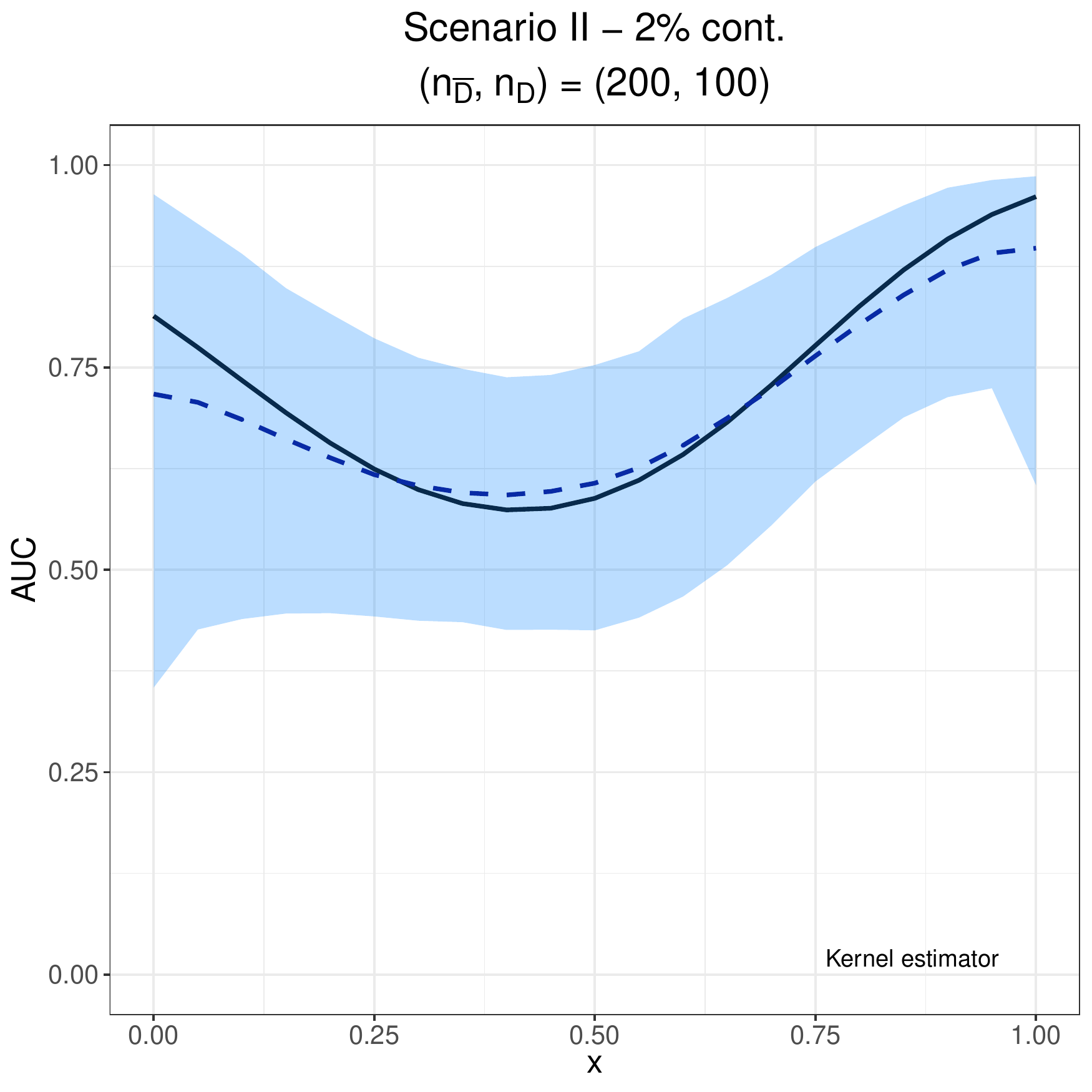}
		}
		\vspace{0.3cm}
		\subfigure{
			\includegraphics[width = 4.65cm]{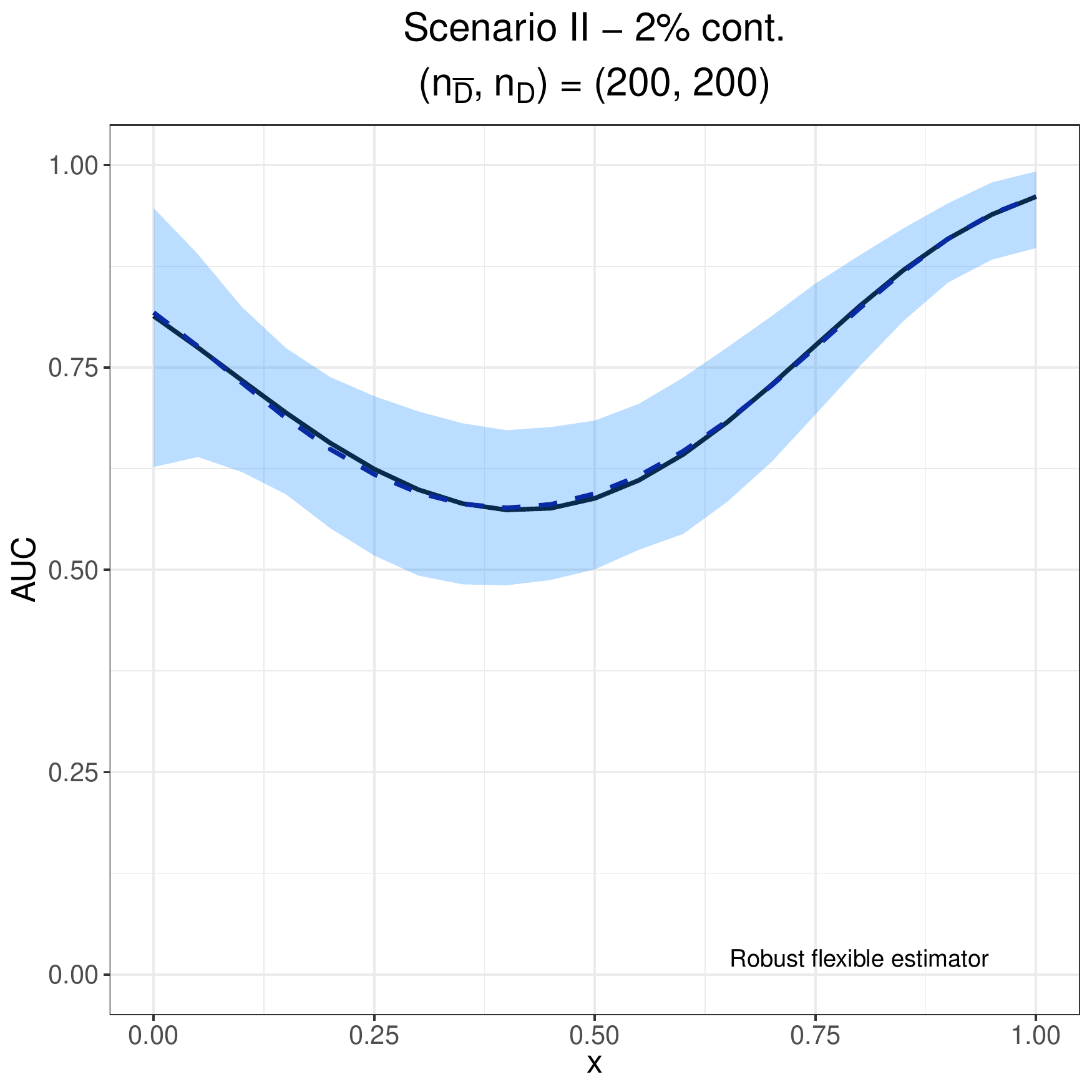}
			\includegraphics[width = 4.65cm]{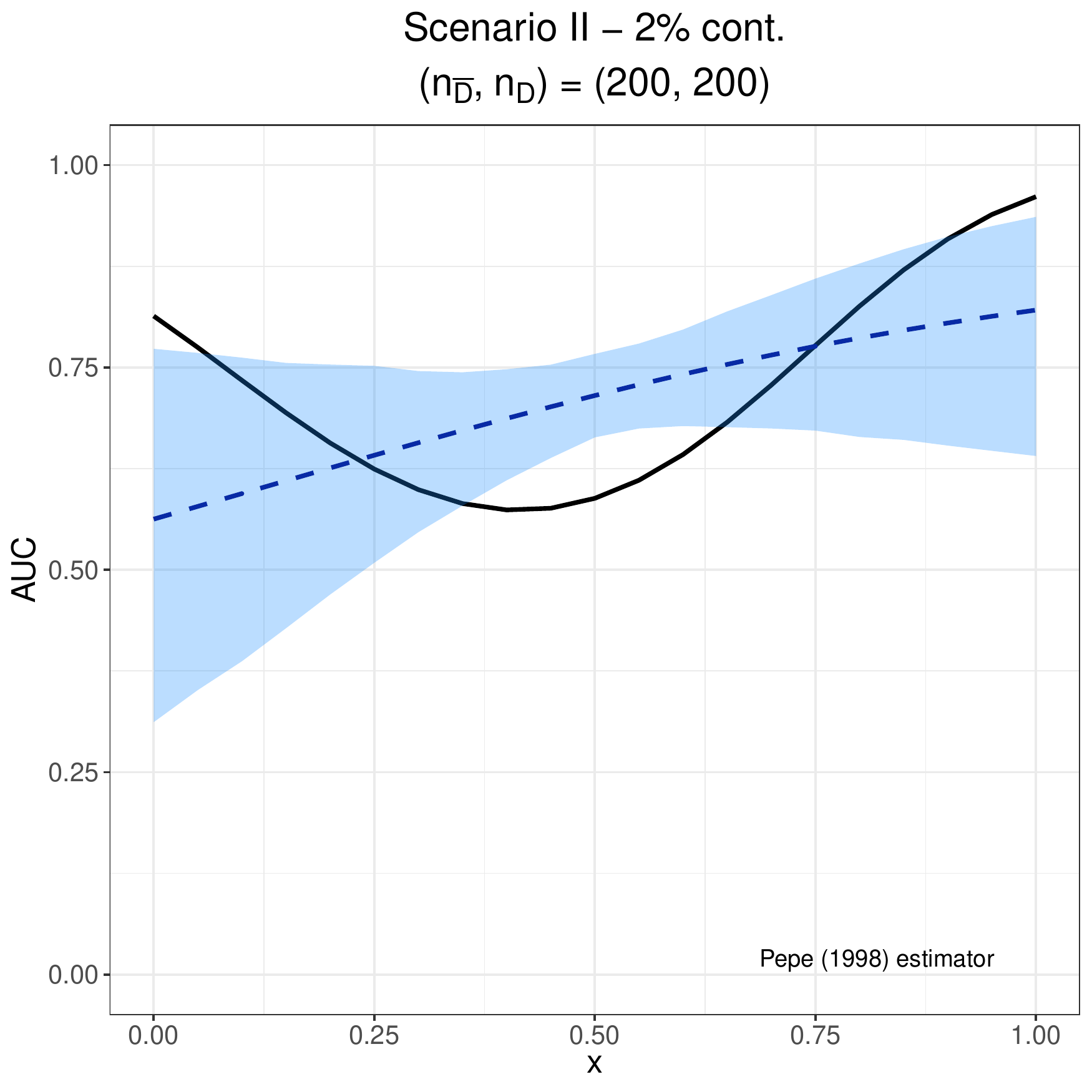}
			\includegraphics[width = 4.65cm]{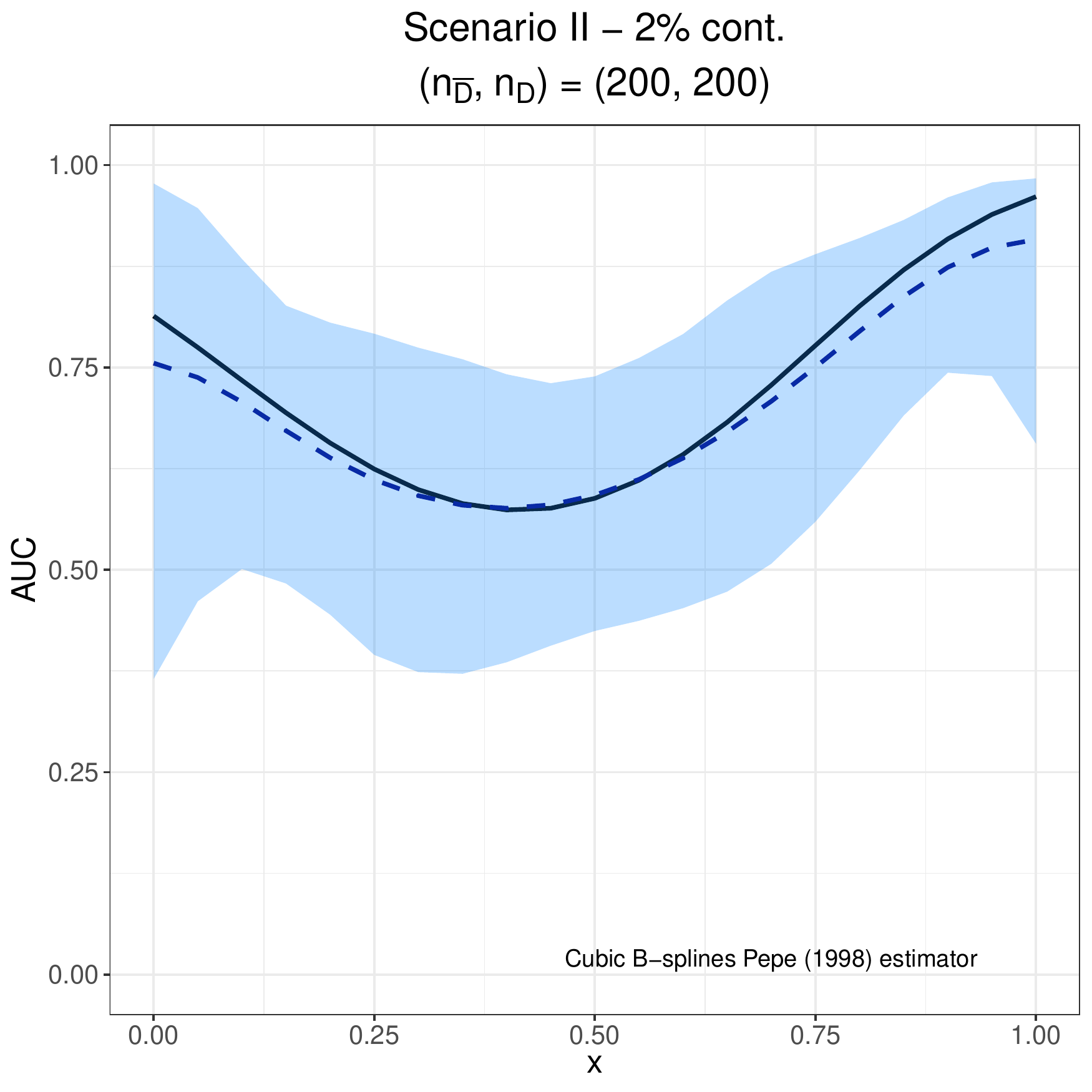}
			\includegraphics[width = 4.65cm]{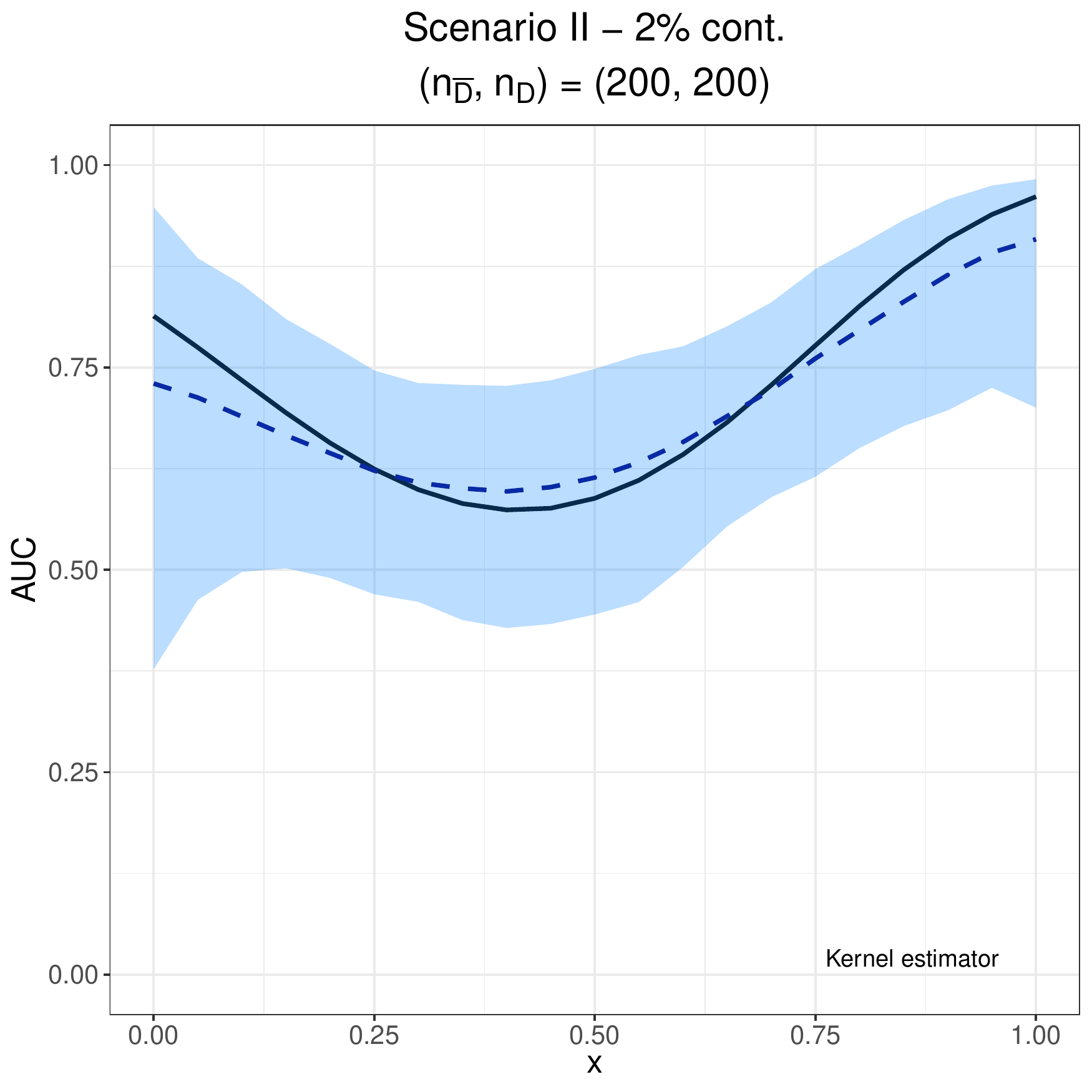}
		}
	\end{center}
	\caption{\footnotesize{Scenario II. True covariate-specific AUC (solid line) versus the mean of the Monte Carlo estimates (dashed line) along with the $2.5\%$ and $97.5\%$ simulation quantiles (shaded area) for the case of $2\%$ of contamination. The first row displays the results for $(n_{\bar{D}}, n_D)=(100,100)$, the second row for $(n_{\bar{D}}, n_D)=(200,100)$, and the third row for $(n_{\bar{D}}, n_D)=(200,200)$. The first column corresponds to our flexible and robust estimator, the second column to the estimator proposed by Pepe (1998), the third one to the cubic B-splines extension of Pepe (1998), and the fourth column to the kernel estimator.}}
\end{figure}

\begin{figure}[H]
	\begin{center}
		\subfigure{
			\includegraphics[width = 4.65cm]{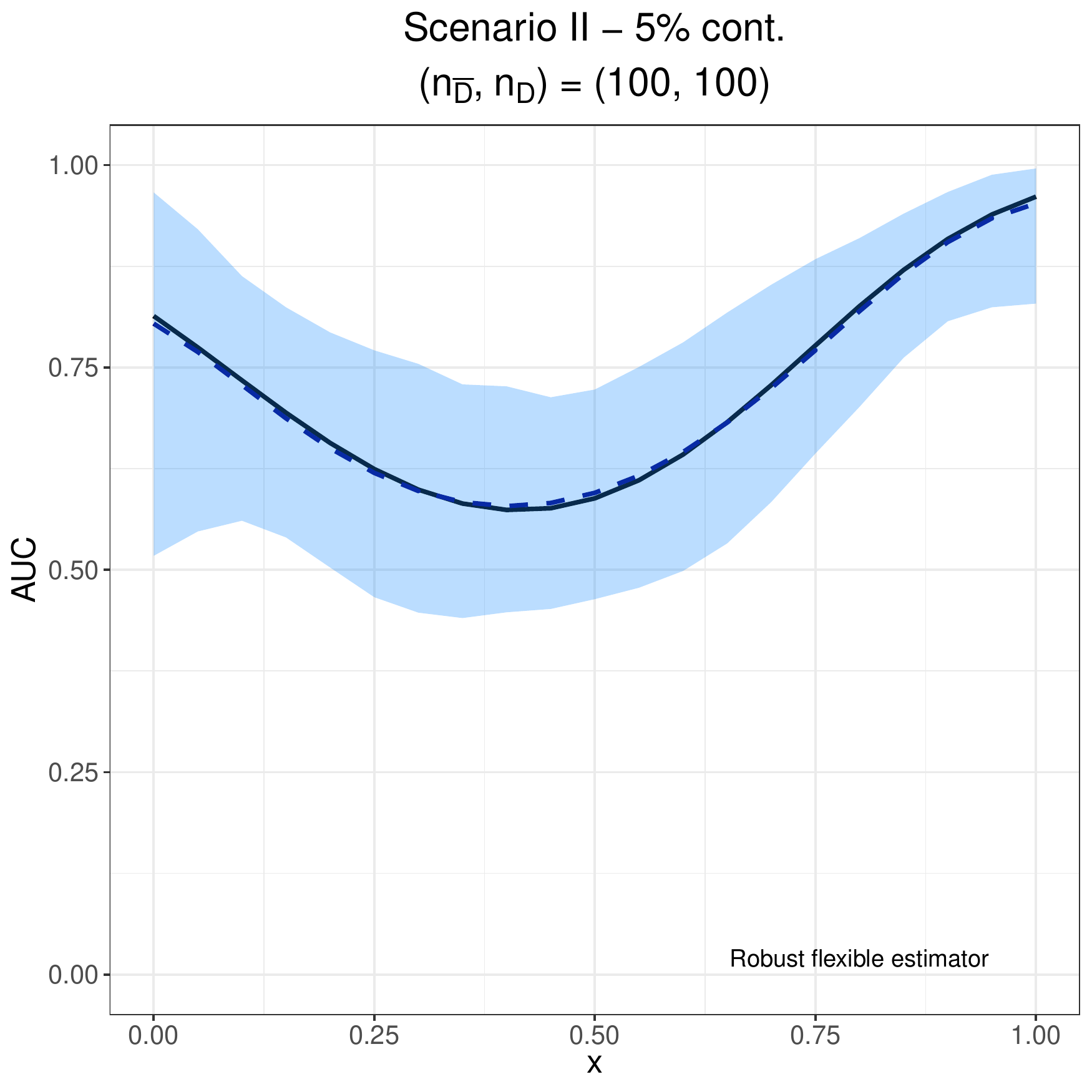}
			\includegraphics[width = 4.65cm]{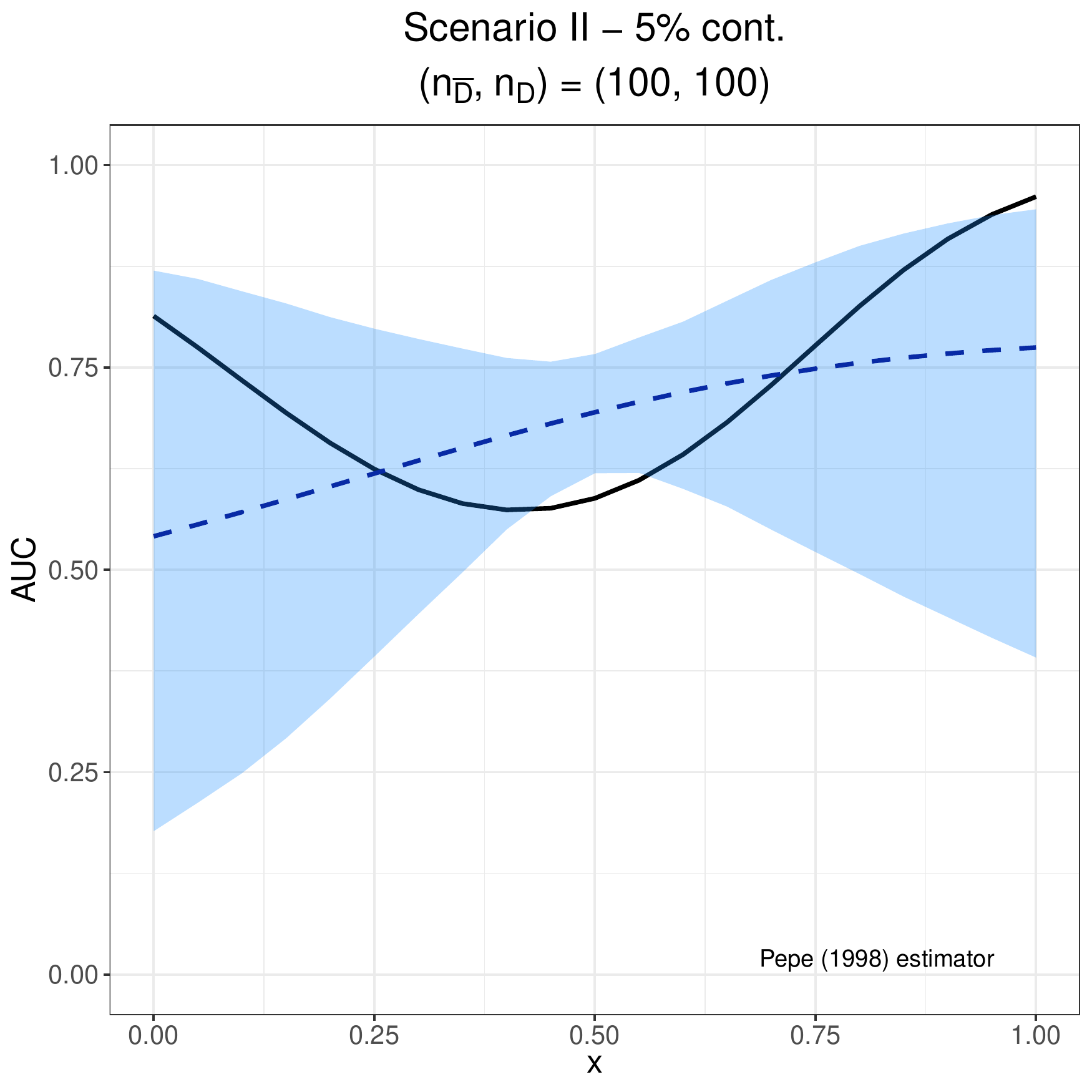}
			\includegraphics[width = 4.65cm]{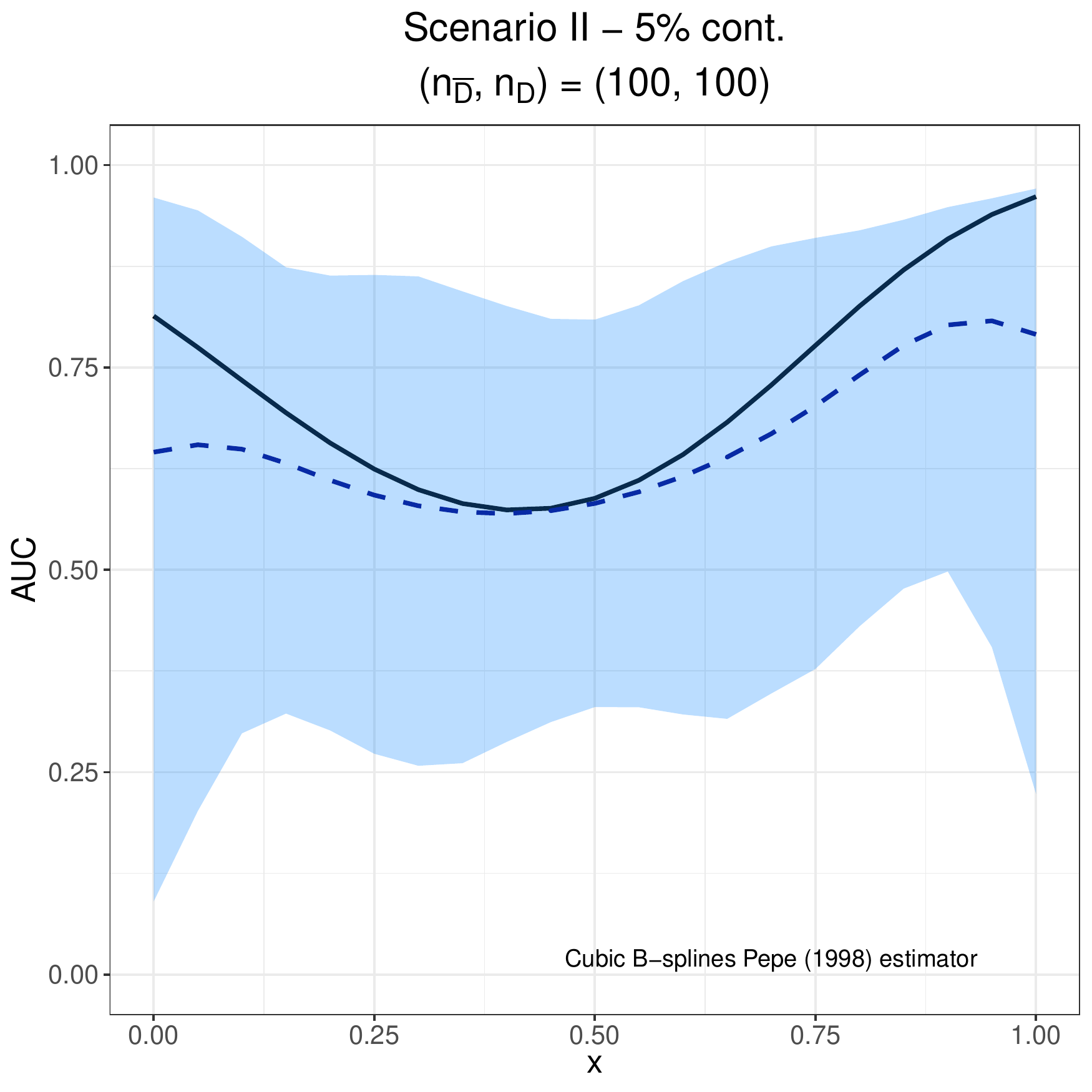}
			\includegraphics[width = 4.65cm]{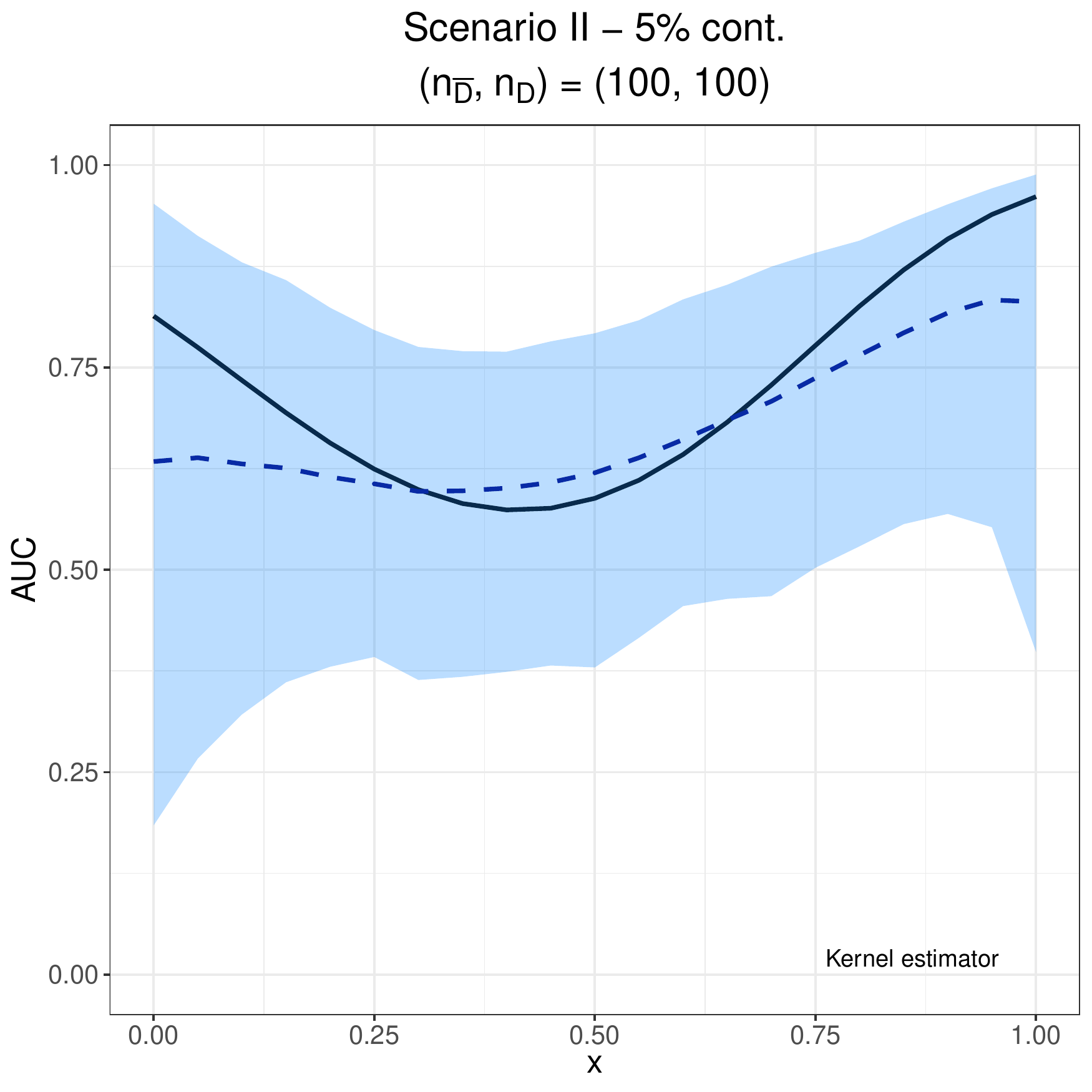}
		}
		\vspace{0.3cm}
		\subfigure{
			\includegraphics[width = 4.65cm]{aucrcont5sampsize2sc2.pdf}
			\includegraphics[width = 4.65cm]{aucspcont5sampsize2sc2.pdf}
			\includegraphics[width = 4.65cm]{aucspbscont5sampsize2sc2.pdf}
			\includegraphics[width = 4.65cm]{auckercont5sampsize2sc2.pdf}
		}
		\vspace{0.3cm}
		\subfigure{
			\includegraphics[width = 4.65cm]{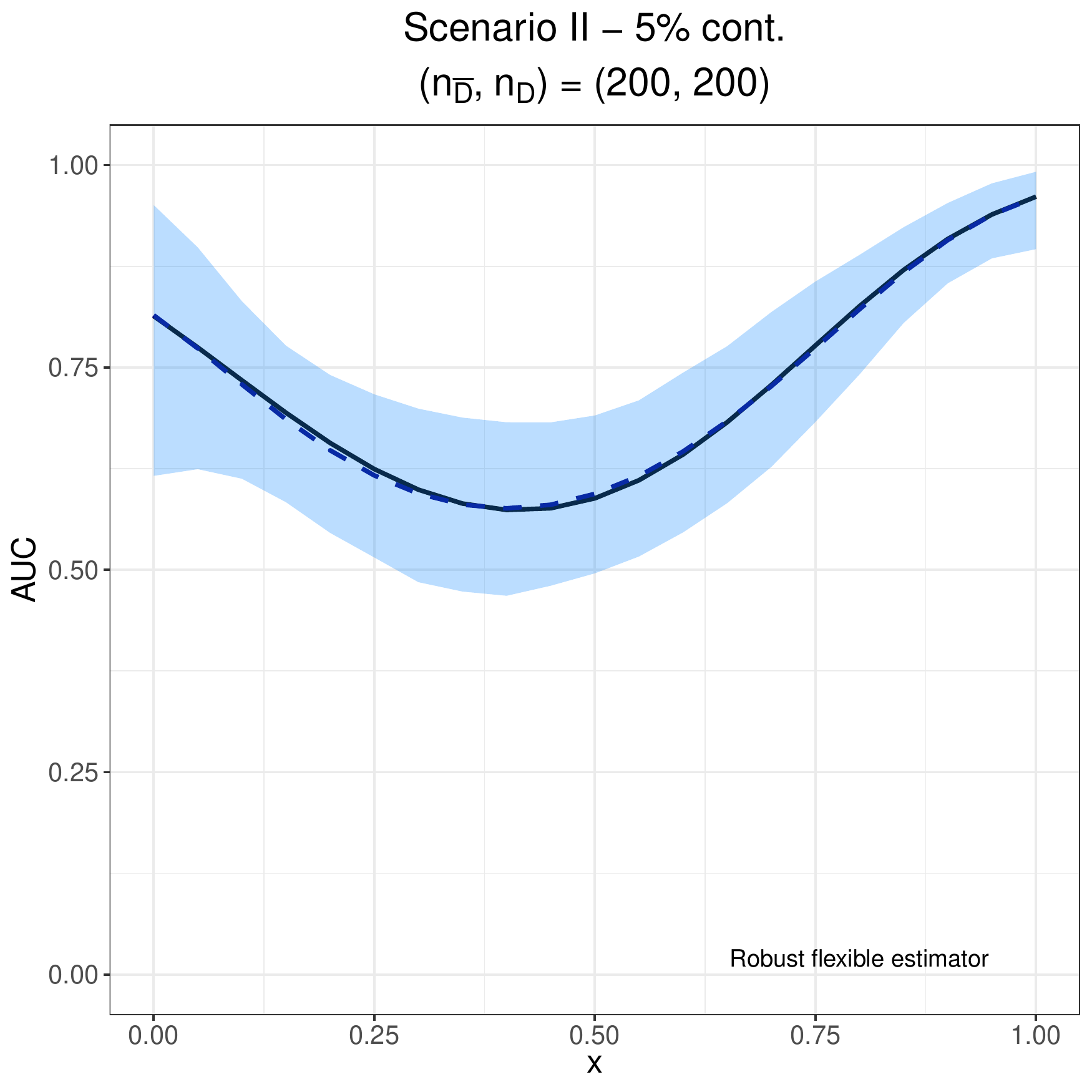}
			\includegraphics[width = 4.65cm]{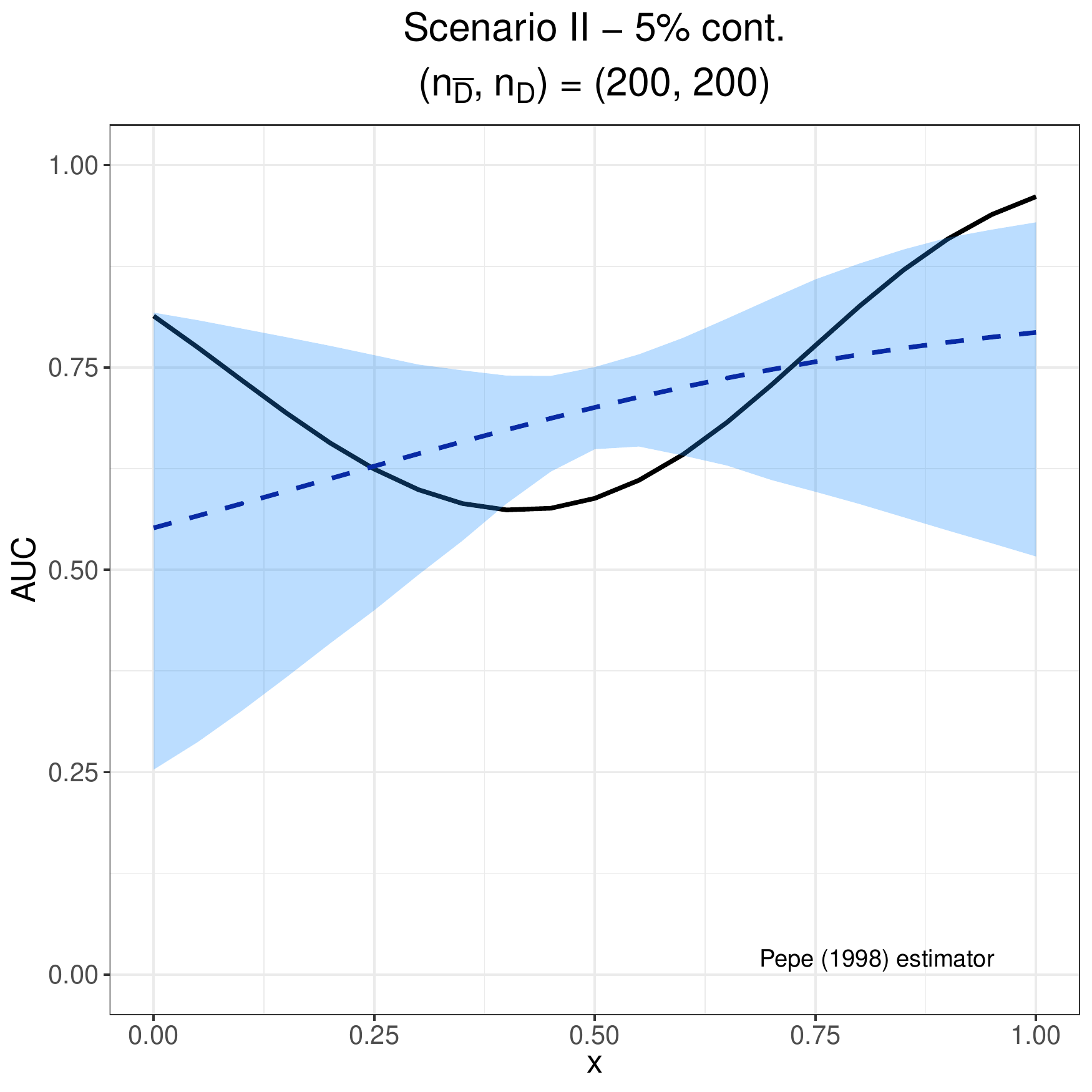}
			\includegraphics[width = 4.65cm]{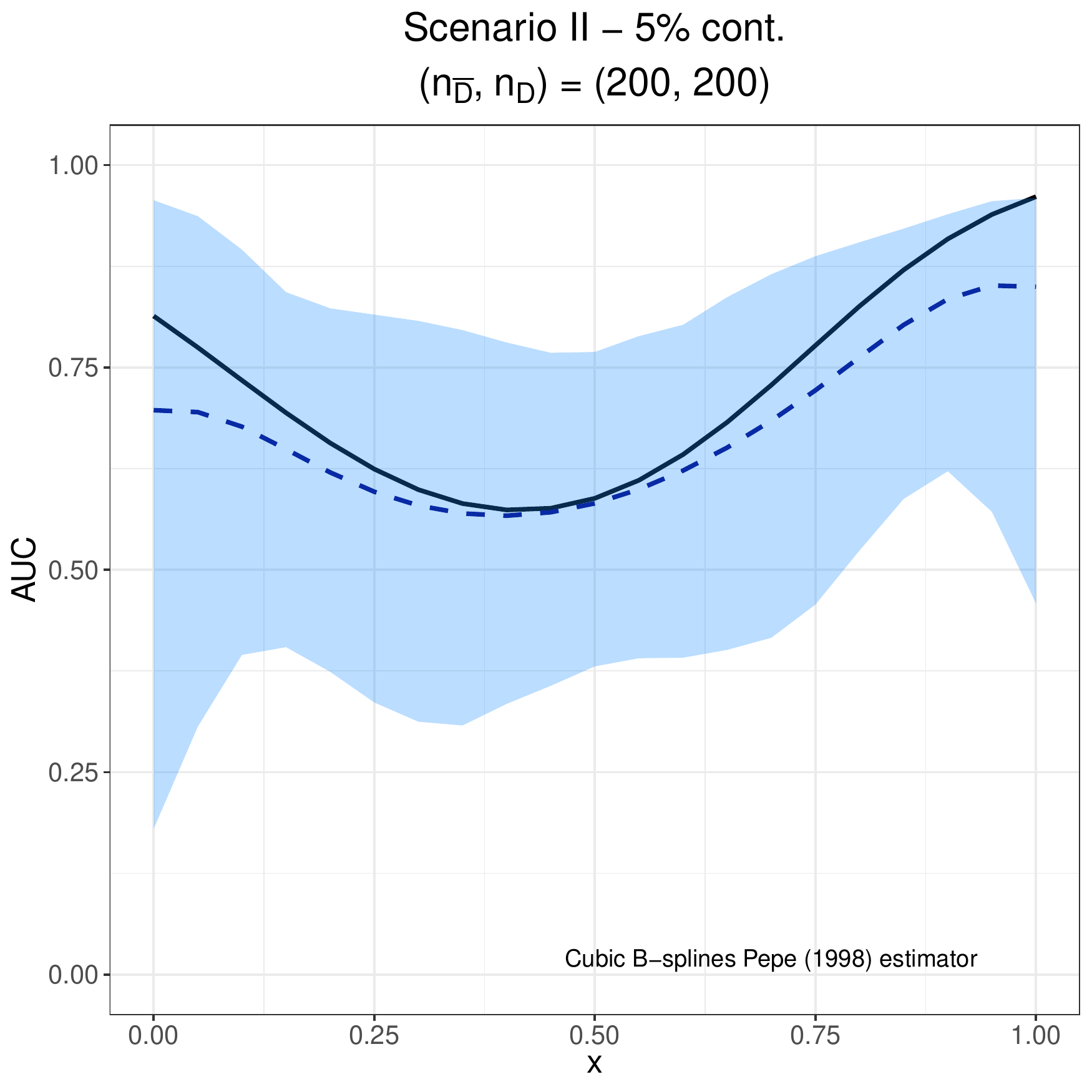}
			\includegraphics[width = 4.65cm]{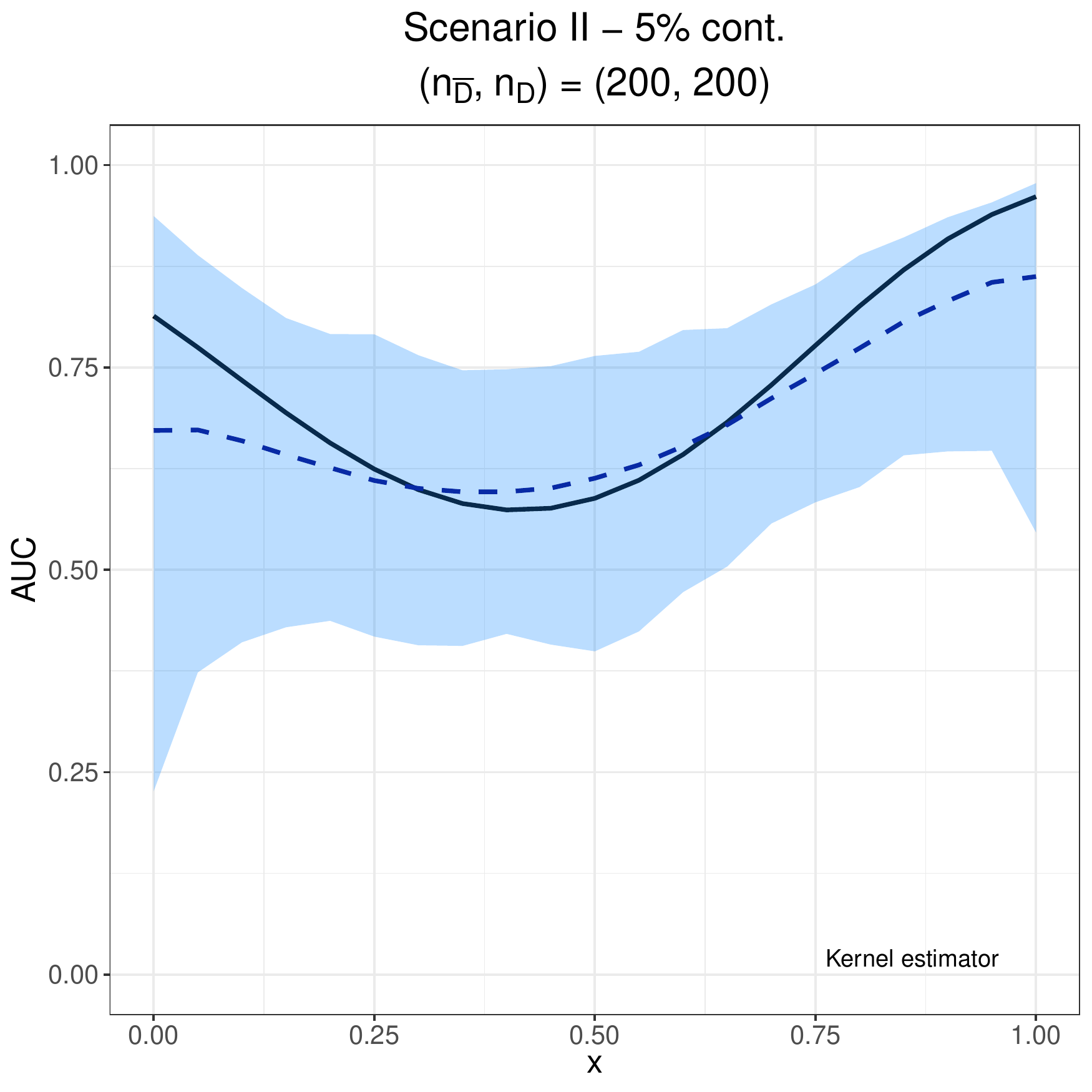}
		}
	\end{center}
	\caption{\footnotesize{Scenario II. True covariate-specific AUC (solid line) versus the mean of the Monte Carlo estimates (dashed line) along with the $2.5\%$ and $97.5\%$ simulation quantiles (shaded area) for the case of $5\%$ of contamination. The first row displays the results for $(n_{\bar{D}}, n_D)=(100,100)$, the second row for $(n_{\bar{D}}, n_D)=(200,100)$, and the third row for $(n_{\bar{D}}, n_D)=(200,200)$. The first column corresponds to our flexible and robust estimator, the second column to the estimator proposed by Pepe (1998), the third one to the cubic B-splines extension of Pepe (1998), and the fourth column to the kernel estimator.}}
\end{figure}

\begin{figure}[H]
	\begin{center}
		\subfigure{
			\includegraphics[width = 4.65cm]{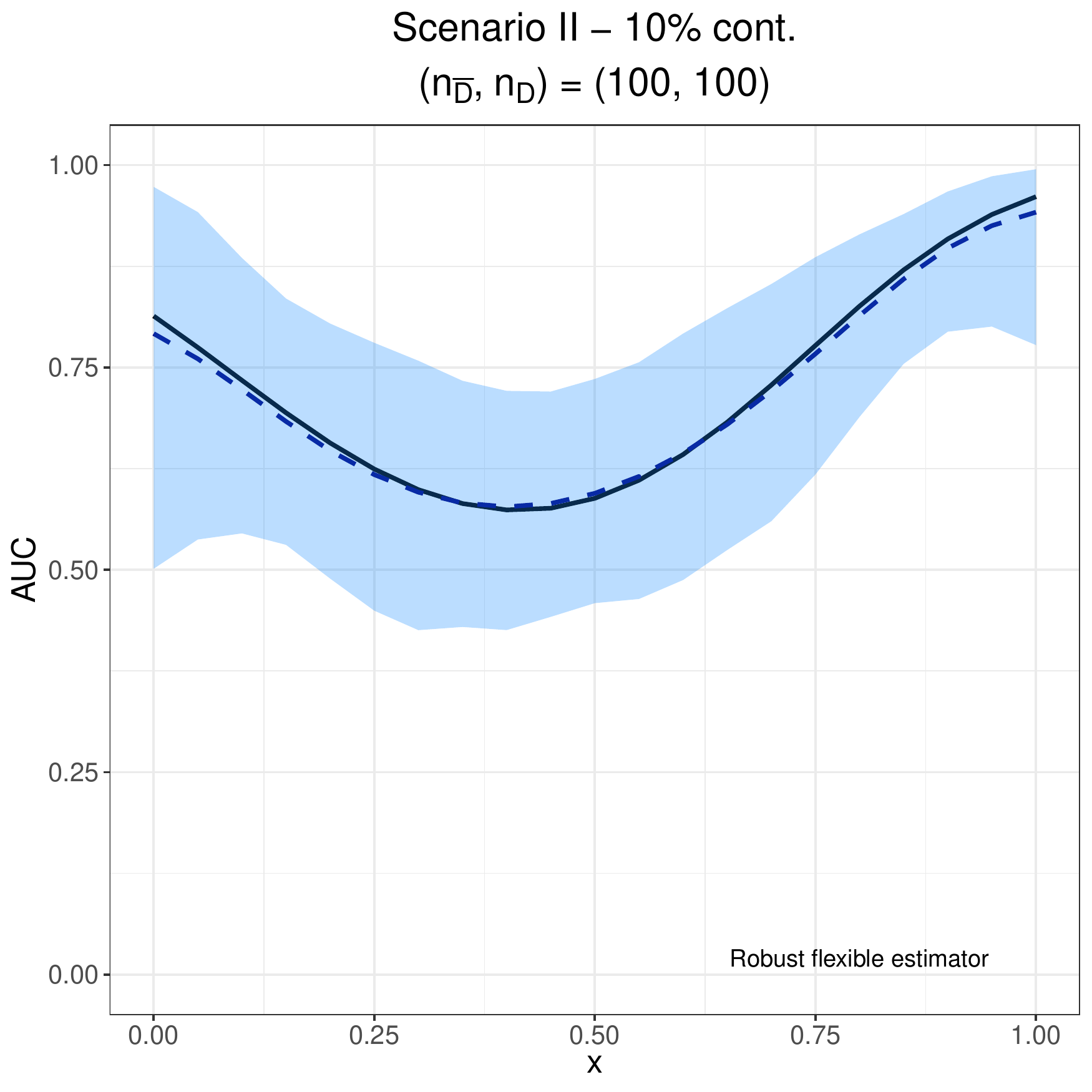}
			\includegraphics[width = 4.65cm]{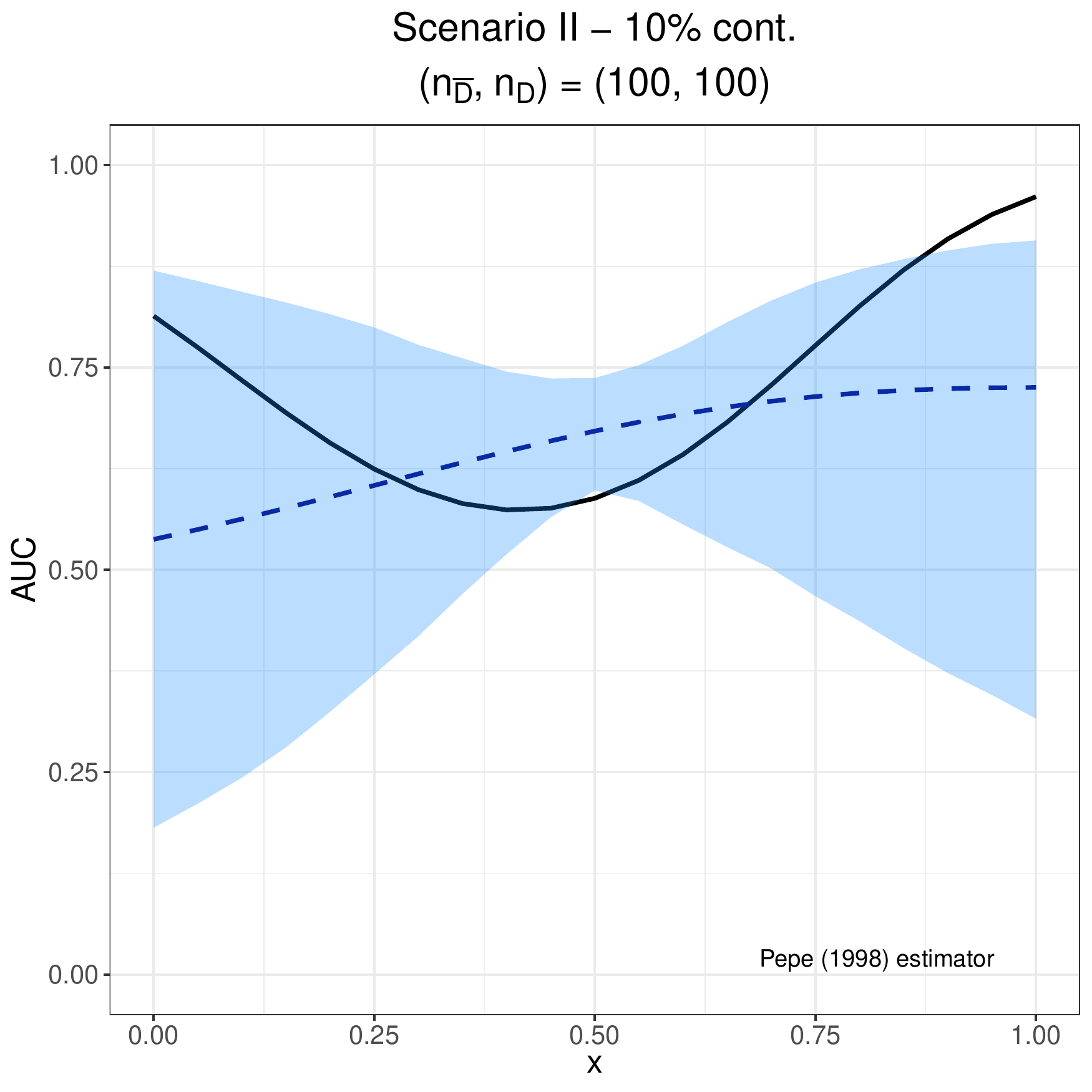}
			\includegraphics[width = 4.65cm]{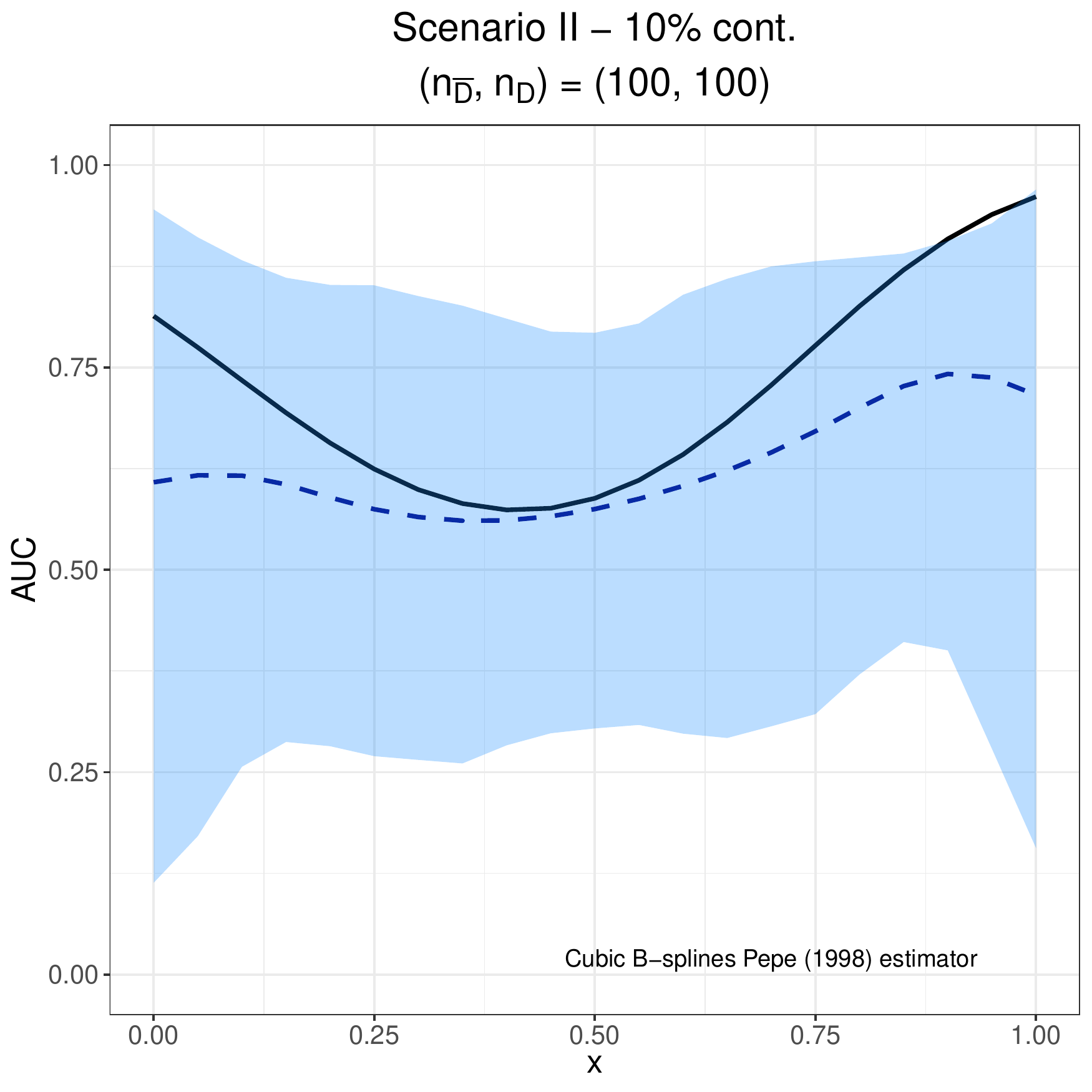}
			\includegraphics[width = 4.65cm]{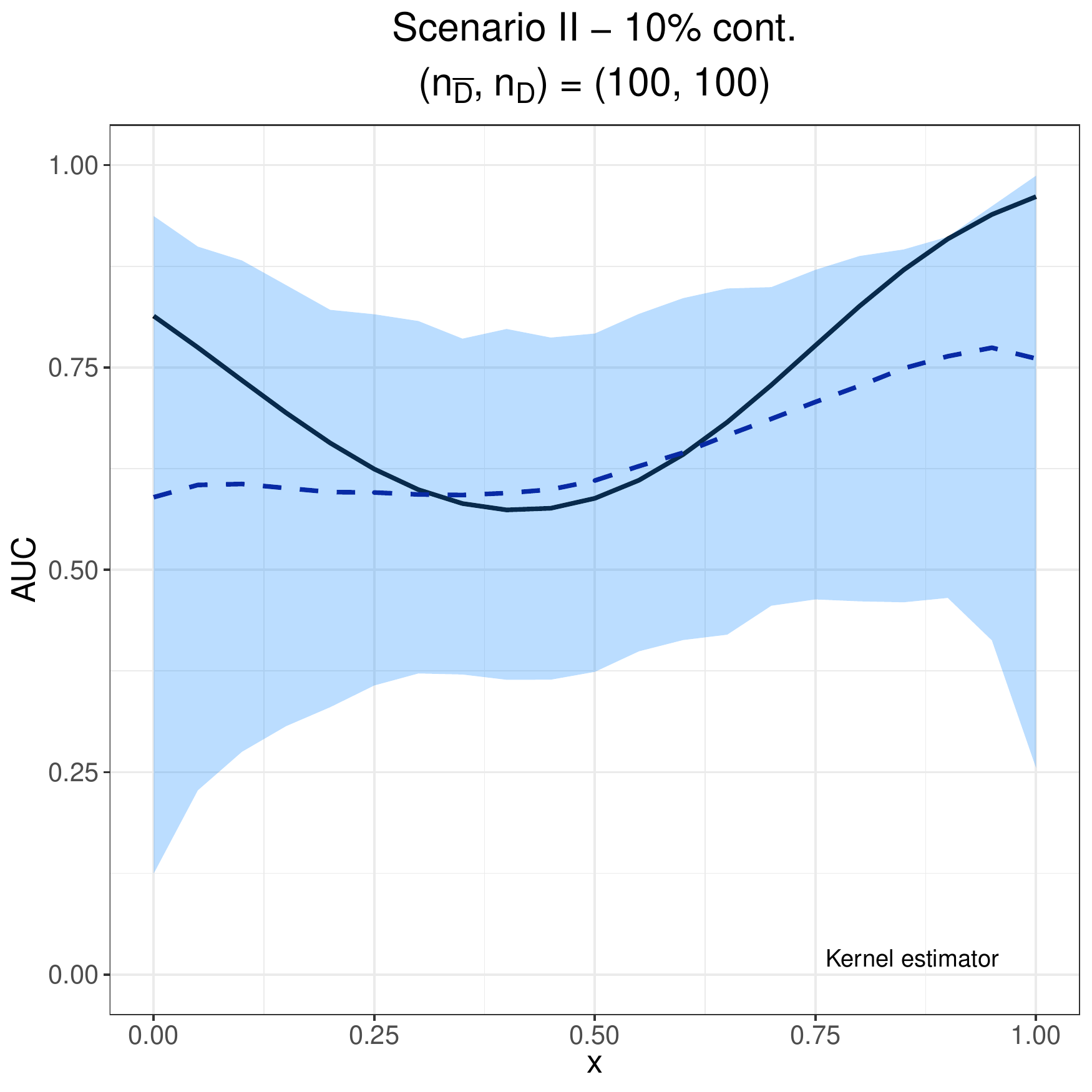}
		}
		\vspace{0.3cm}
		\subfigure{
			\includegraphics[width = 4.65cm]{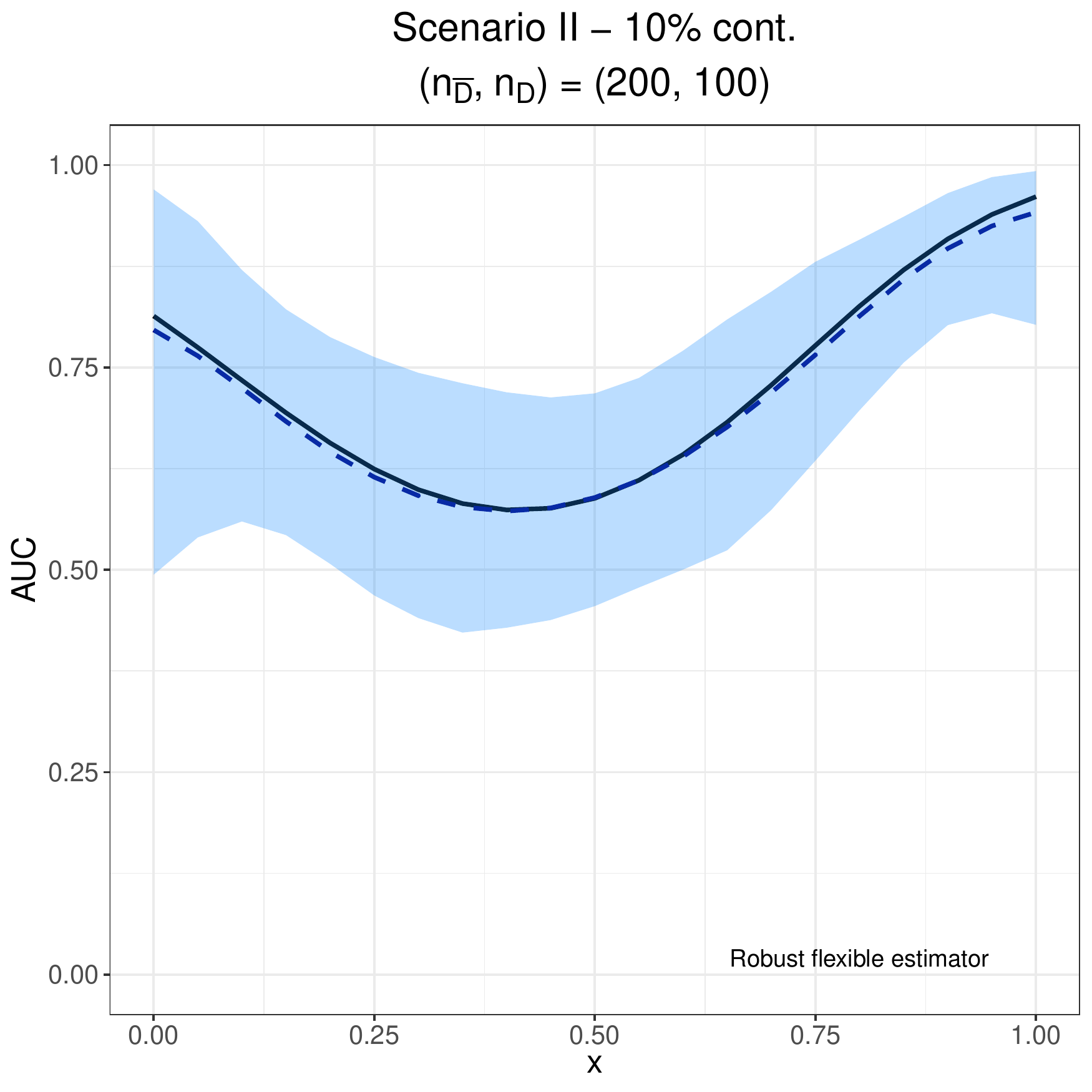}
			\includegraphics[width = 4.65cm]{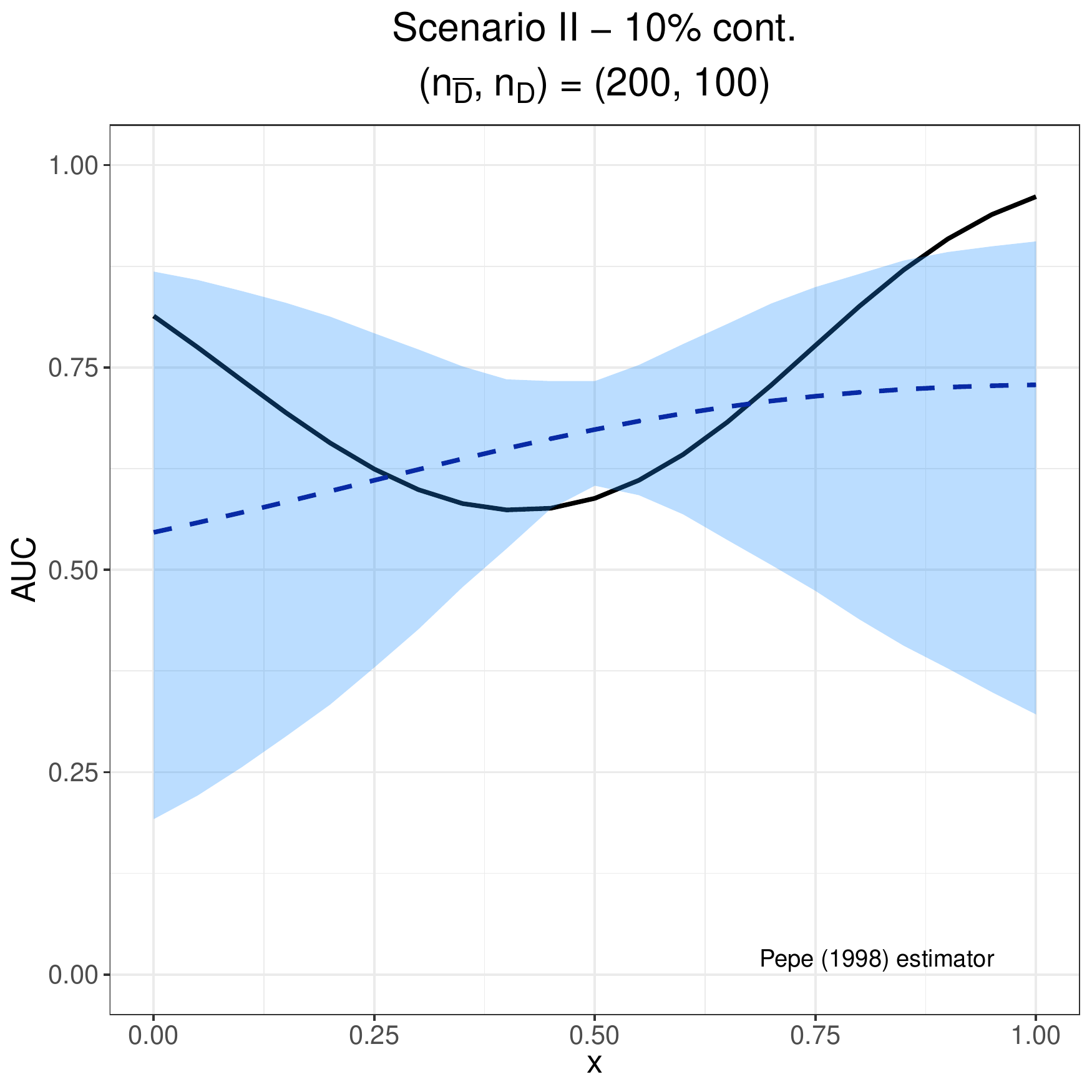}
			\includegraphics[width = 4.65cm]{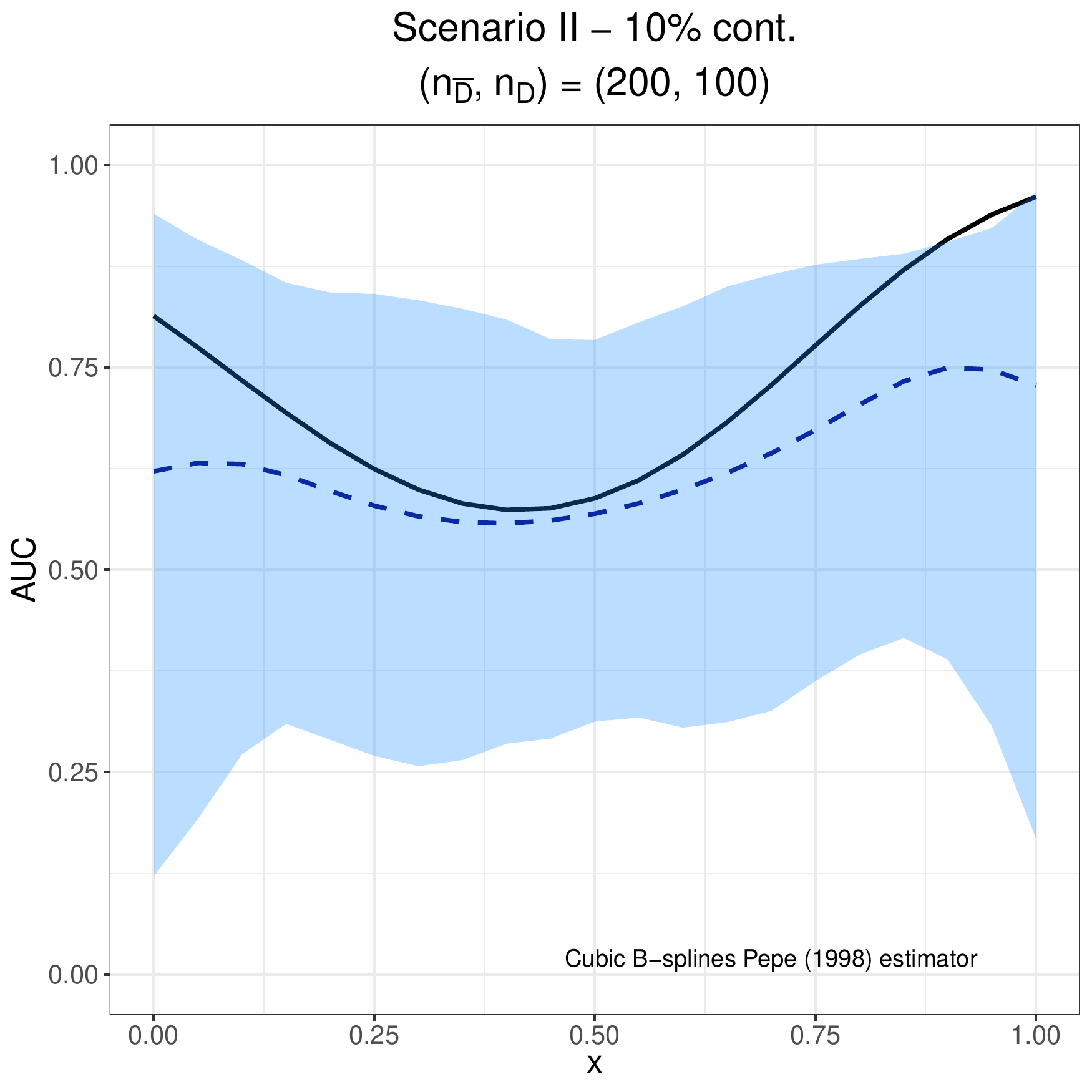}
			\includegraphics[width = 4.65cm]{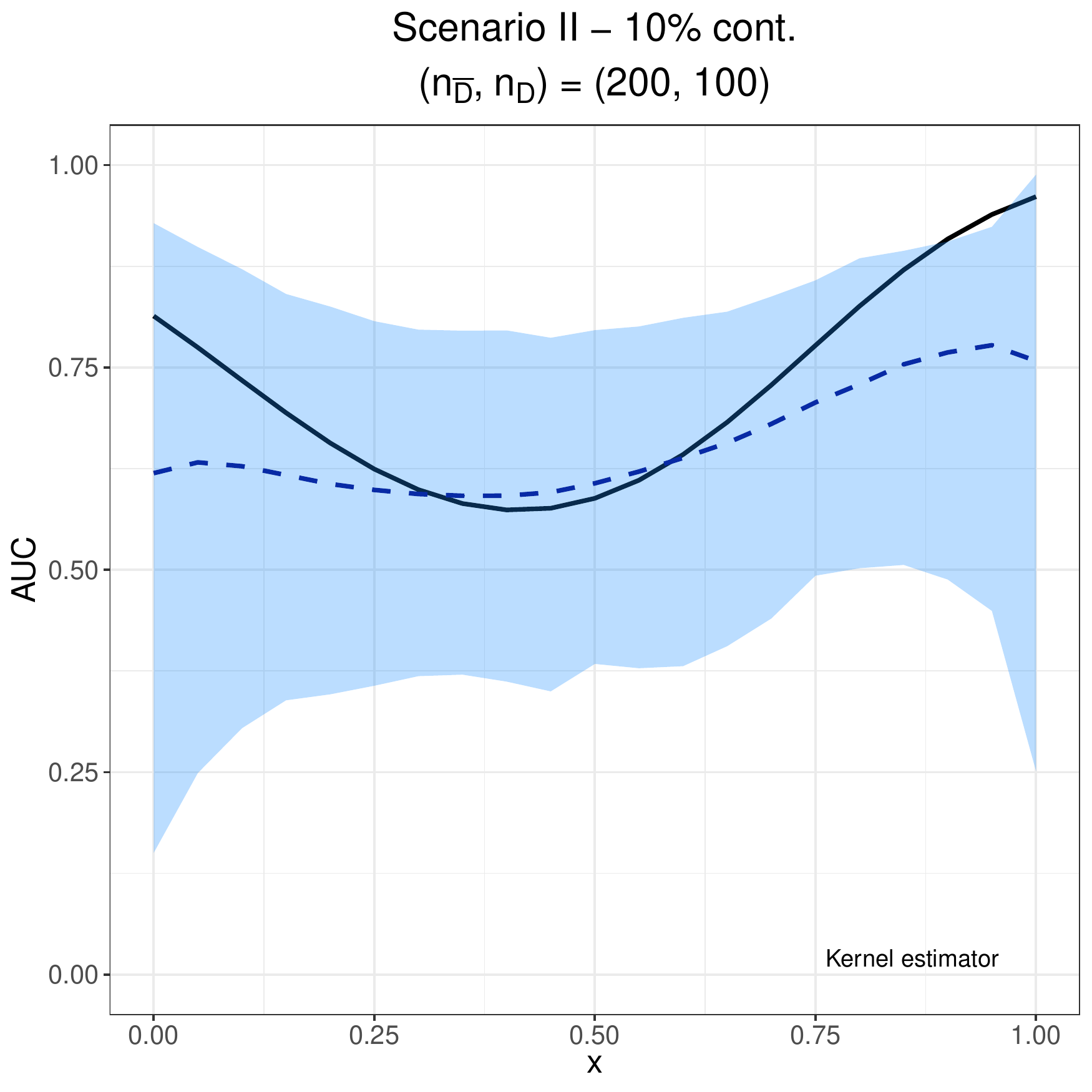}
		}
		\vspace{0.3cm}
		\subfigure{
			\includegraphics[width = 4.65cm]{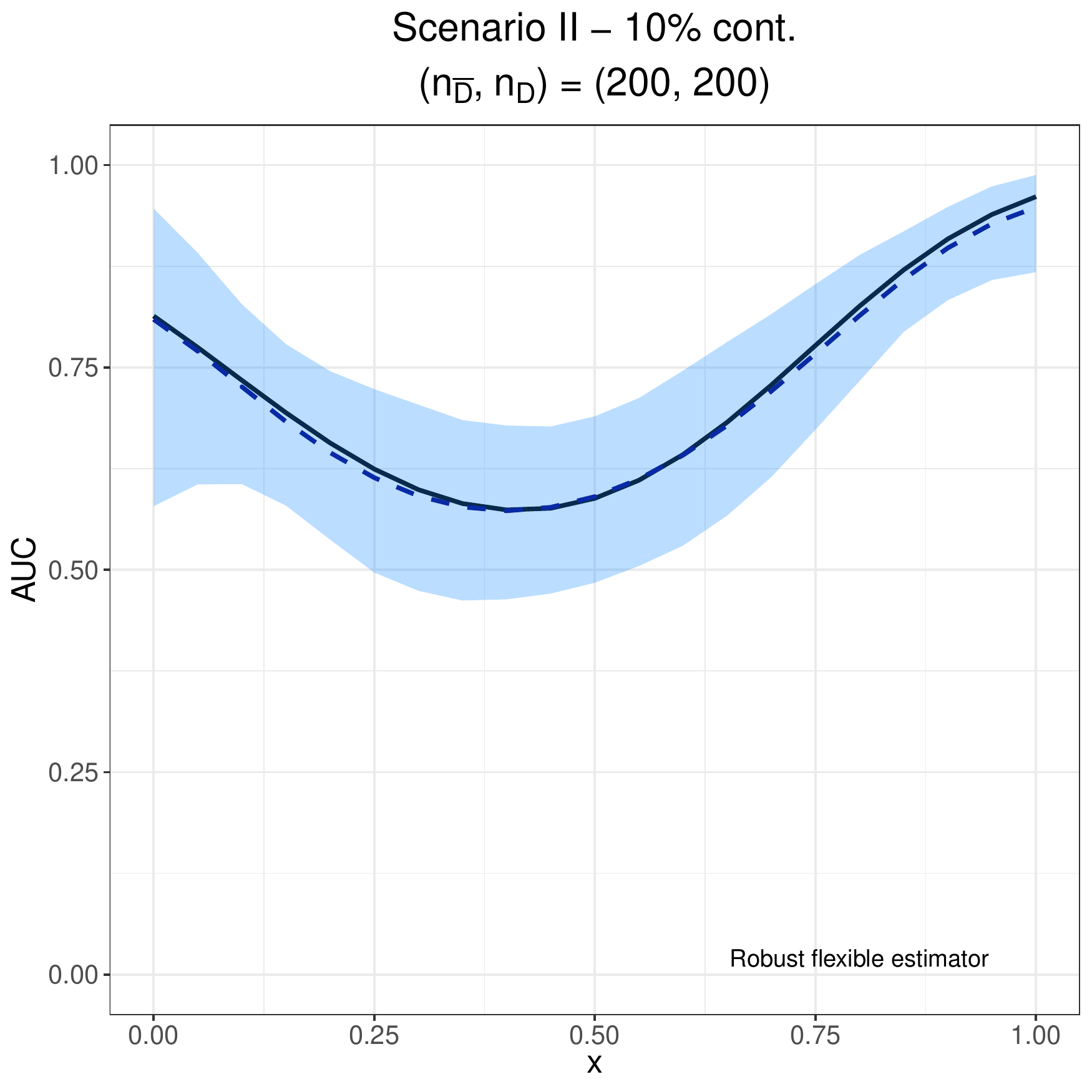}
			\includegraphics[width = 4.65cm]{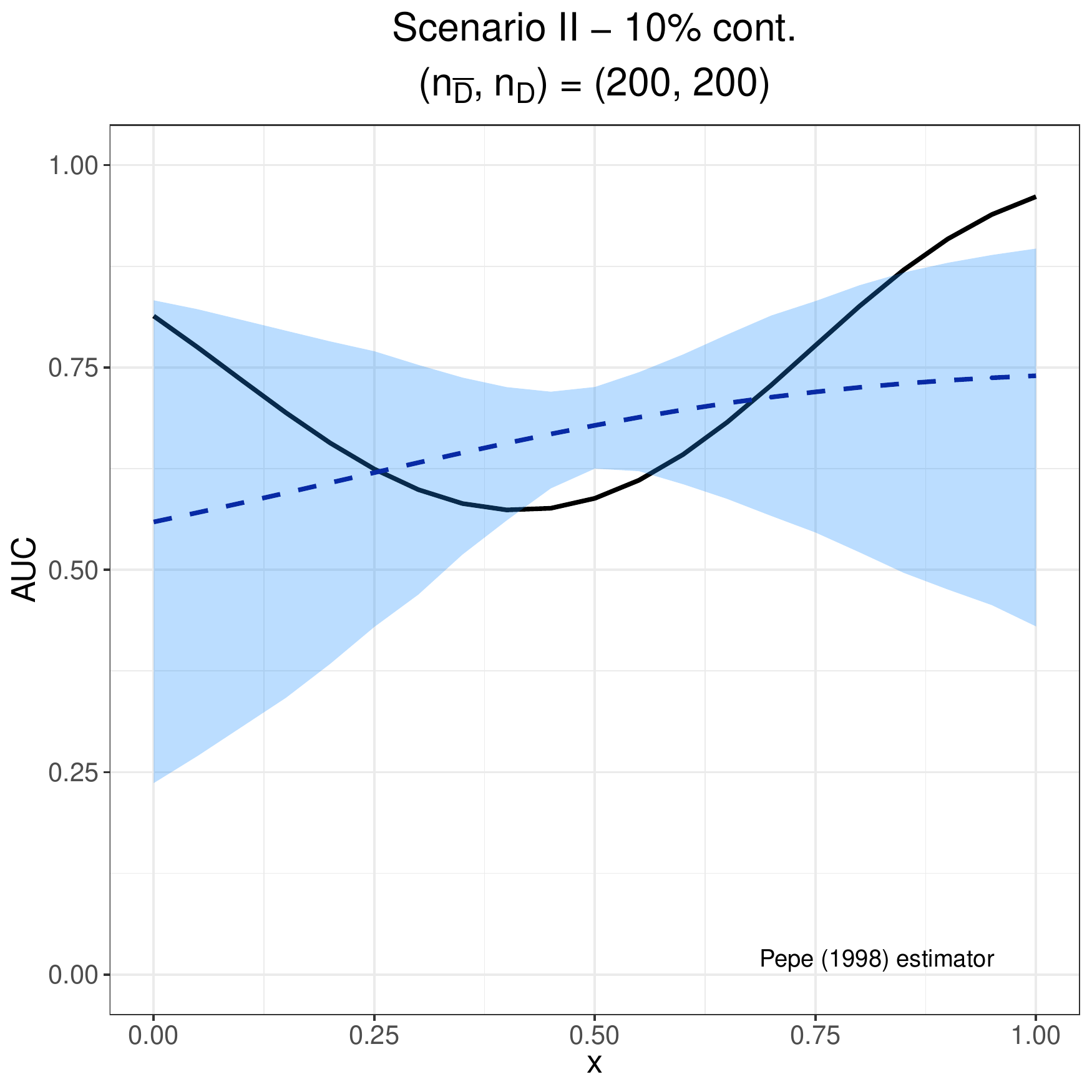}
			\includegraphics[width = 4.65cm]{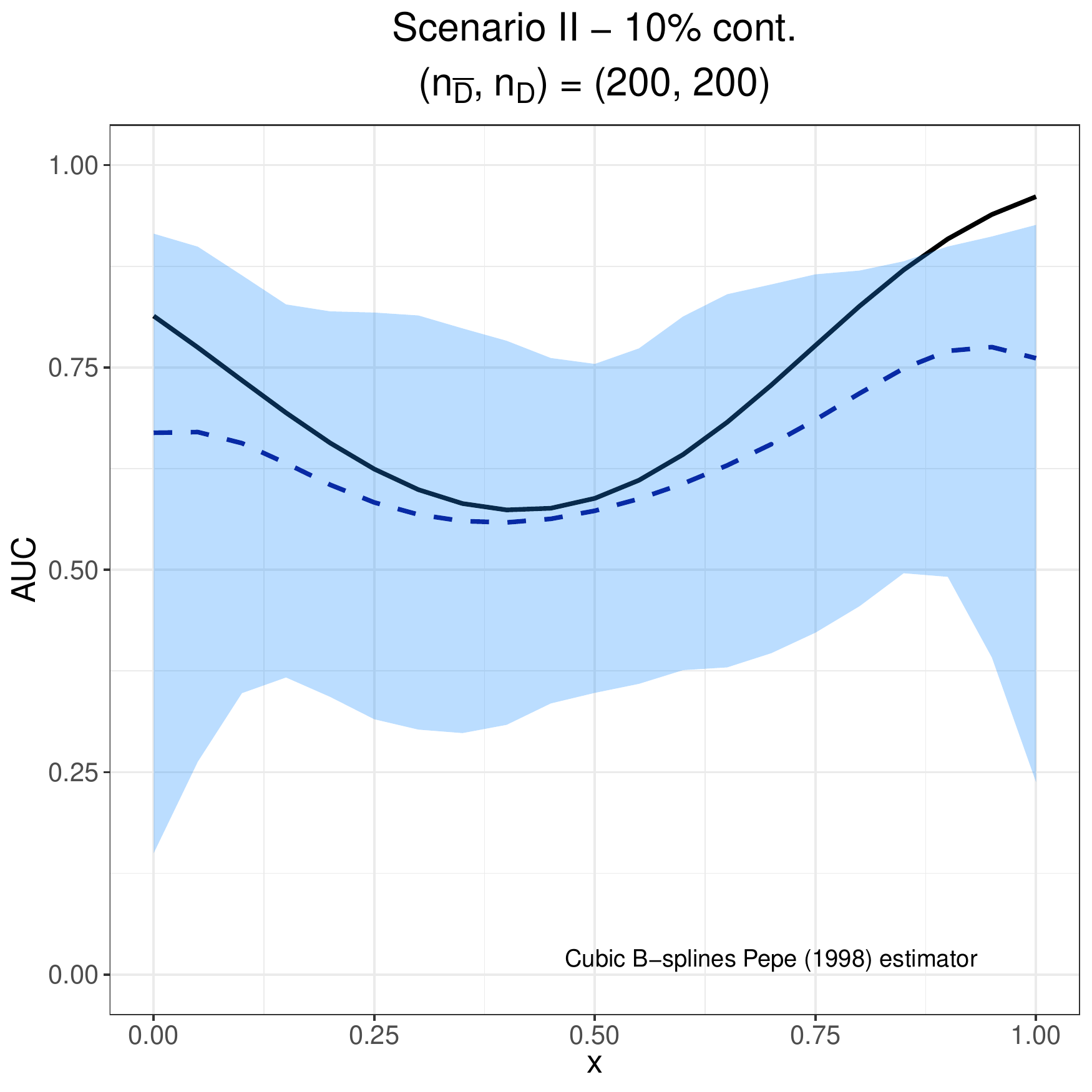}
			\includegraphics[width = 4.65cm]{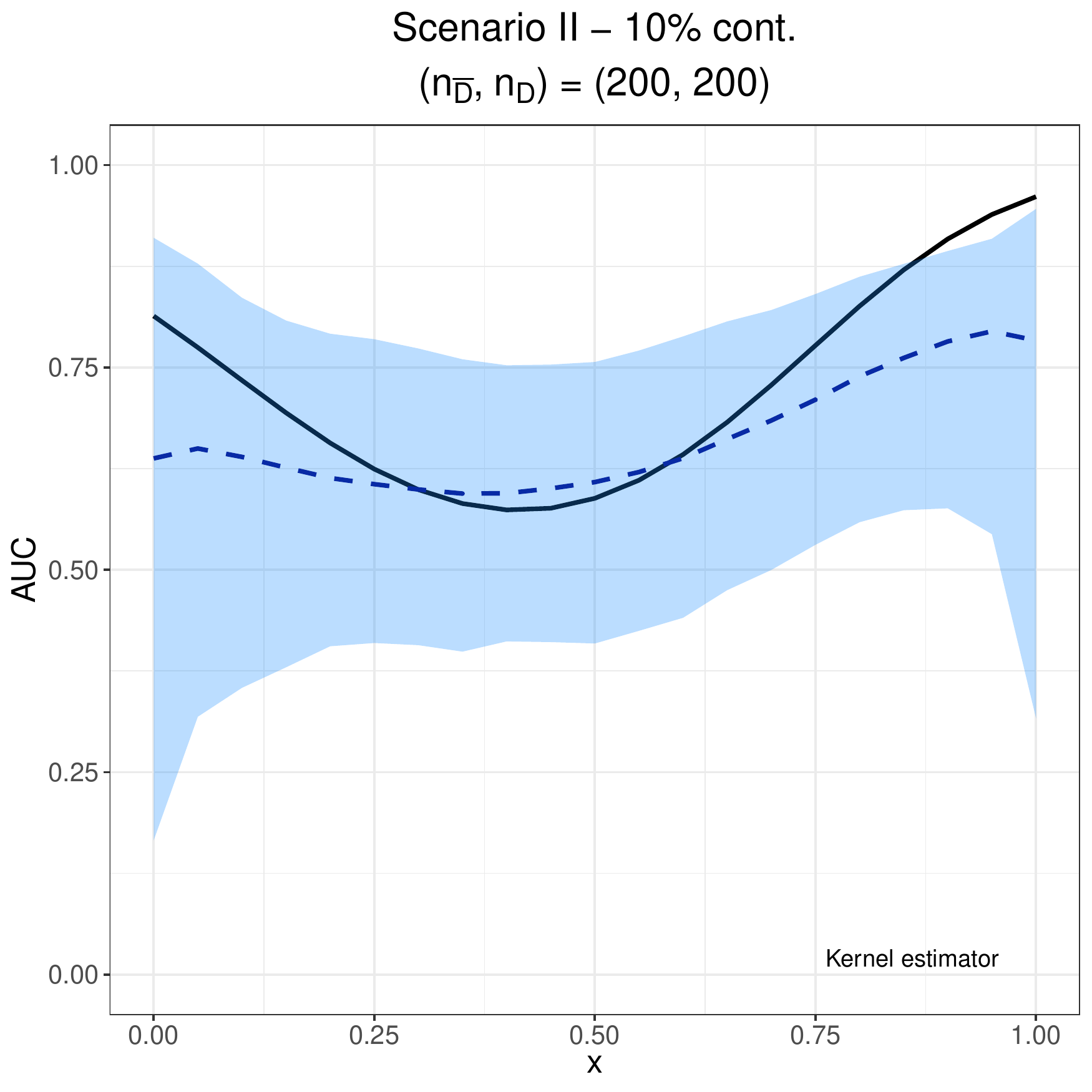}
		}
	\end{center}
	\caption{\footnotesize{Scenario II. True covariate-specific AUC (solid line) versus the mean of the Monte Carlo estimates (dashed line) along with the $2.5\%$ and $97.5\%$ simulation quantiles (shaded area) for the case of $10\%$ of contamination. The first row displays the results for $(n_{\bar{D}}, n_D)=(100,100)$, the second row for $(n_{\bar{D}}, n_D)=(200,100)$, and the third row for $(n_{\bar{D}}, n_D)=(200,200)$. The first column corresponds to our flexible and robust estimator, the second column to the estimator proposed by Pepe (1998), the third one to the cubic B-splines extension of Pepe (1998), and the fourth column to the kernel estimator.}}
\end{figure}

\begin{figure}[H]
	\begin{center}
		\subfigure{
			\includegraphics[width = 4.65cm]{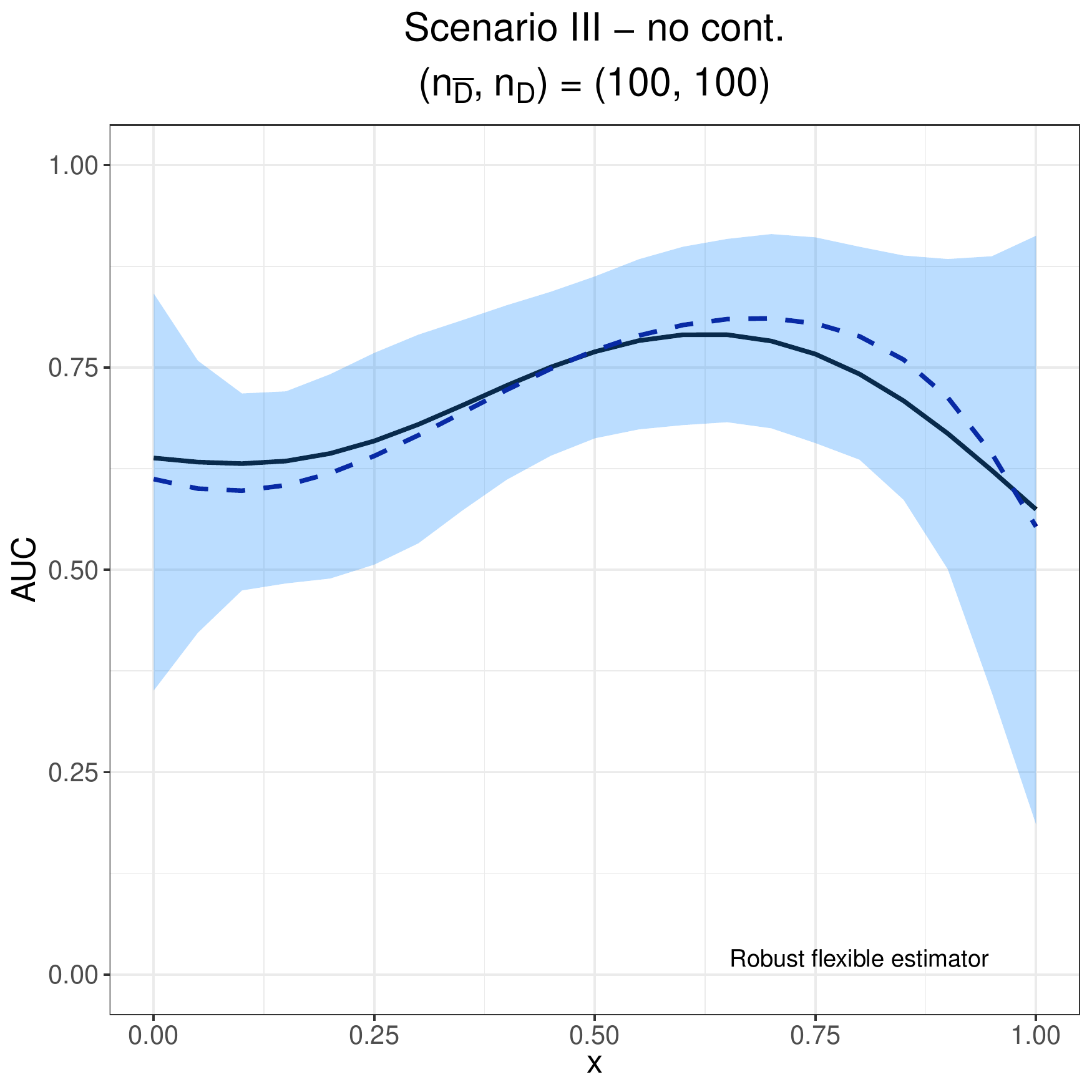}
			\includegraphics[width = 4.65cm]{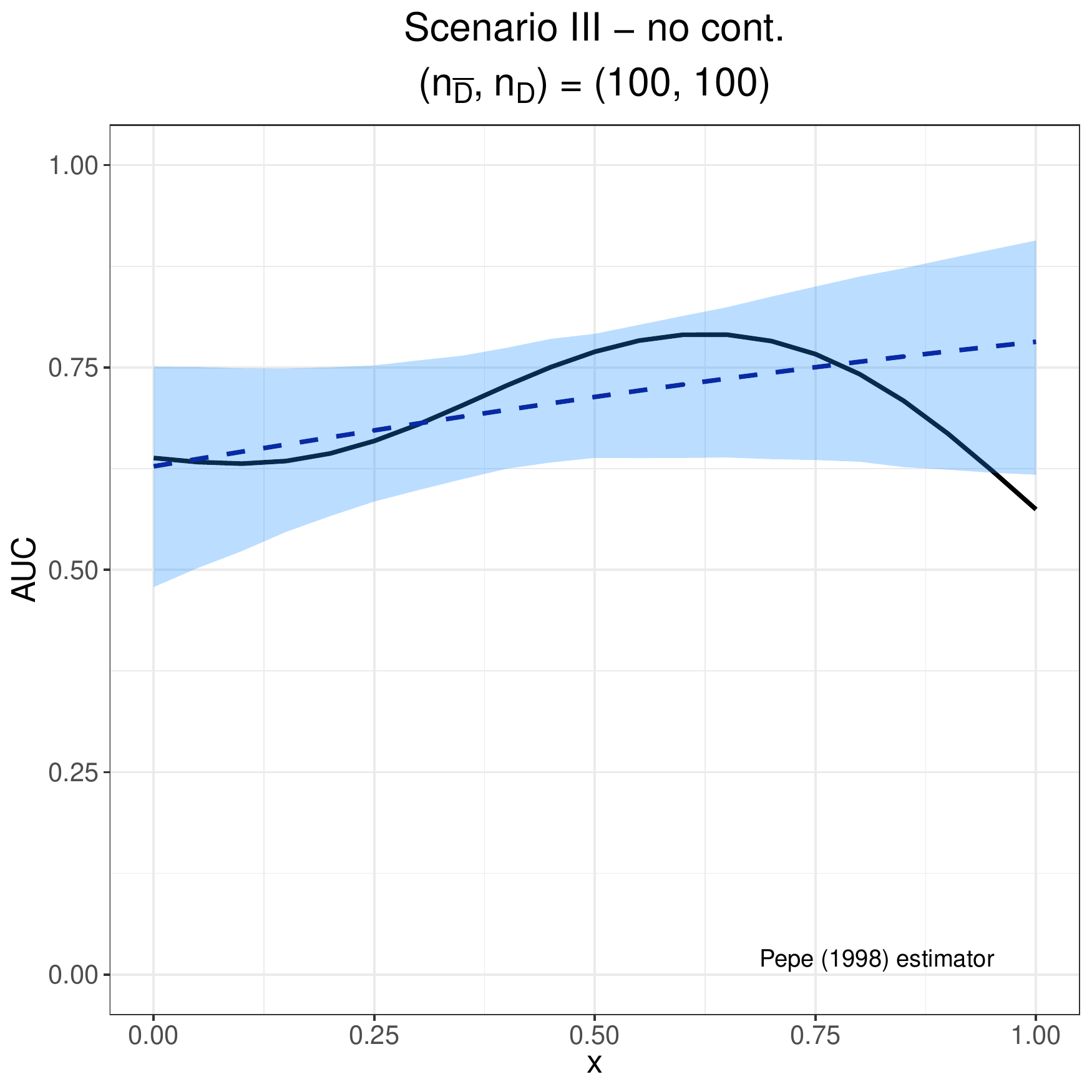}
			\includegraphics[width = 4.65cm]{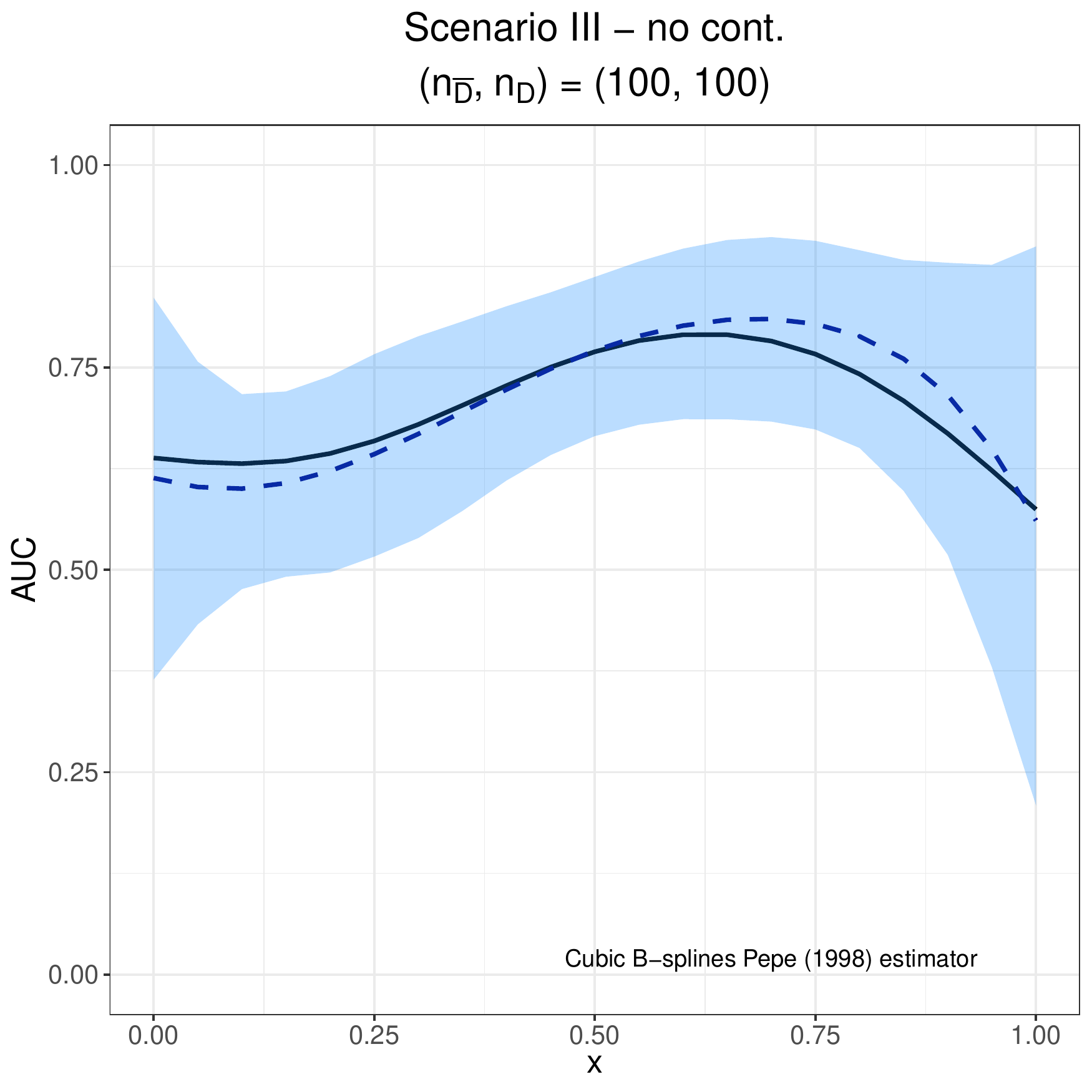}
			\includegraphics[width = 4.65cm]{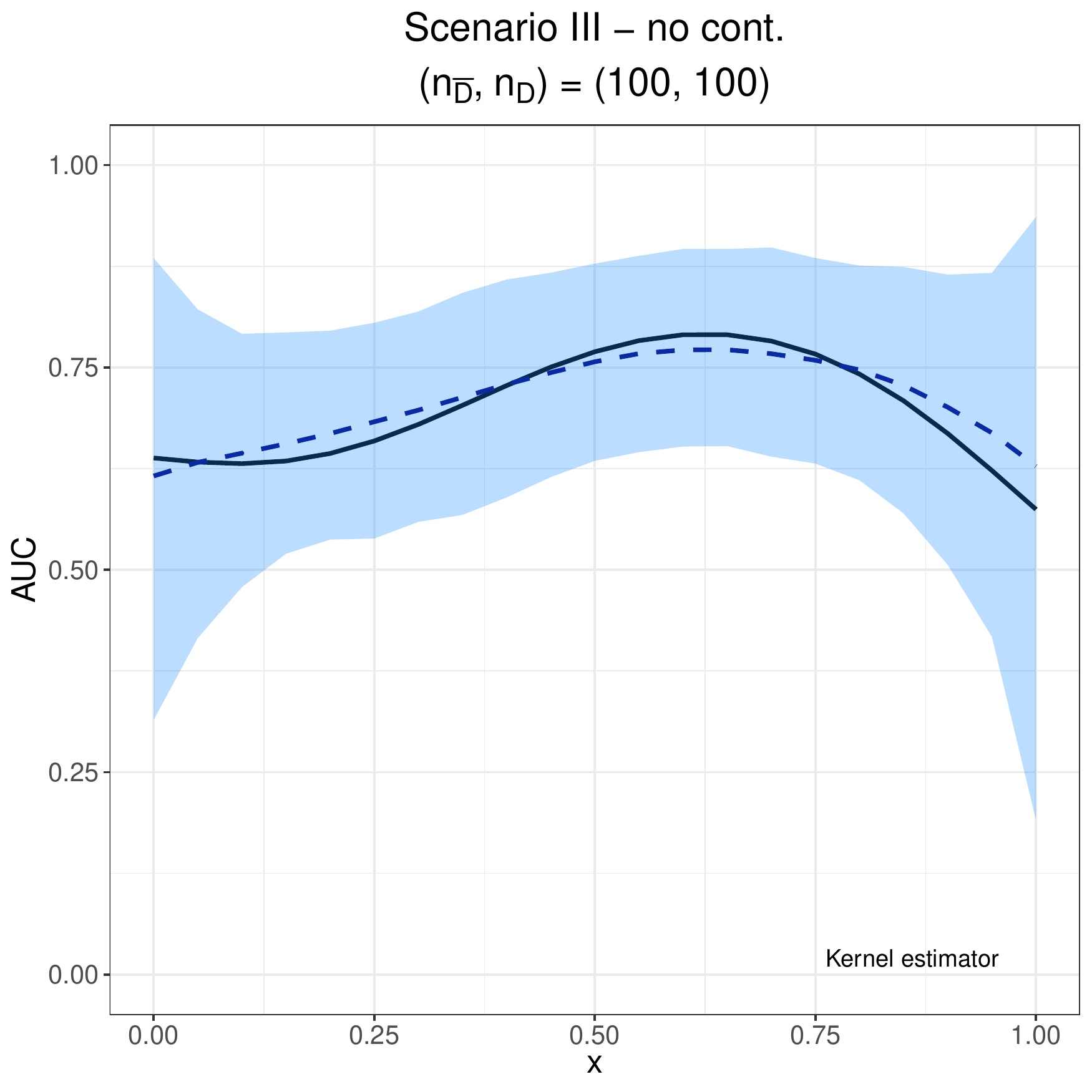}
		}
		\vspace{0.3cm}
		\subfigure{
			\includegraphics[width = 4.65cm]{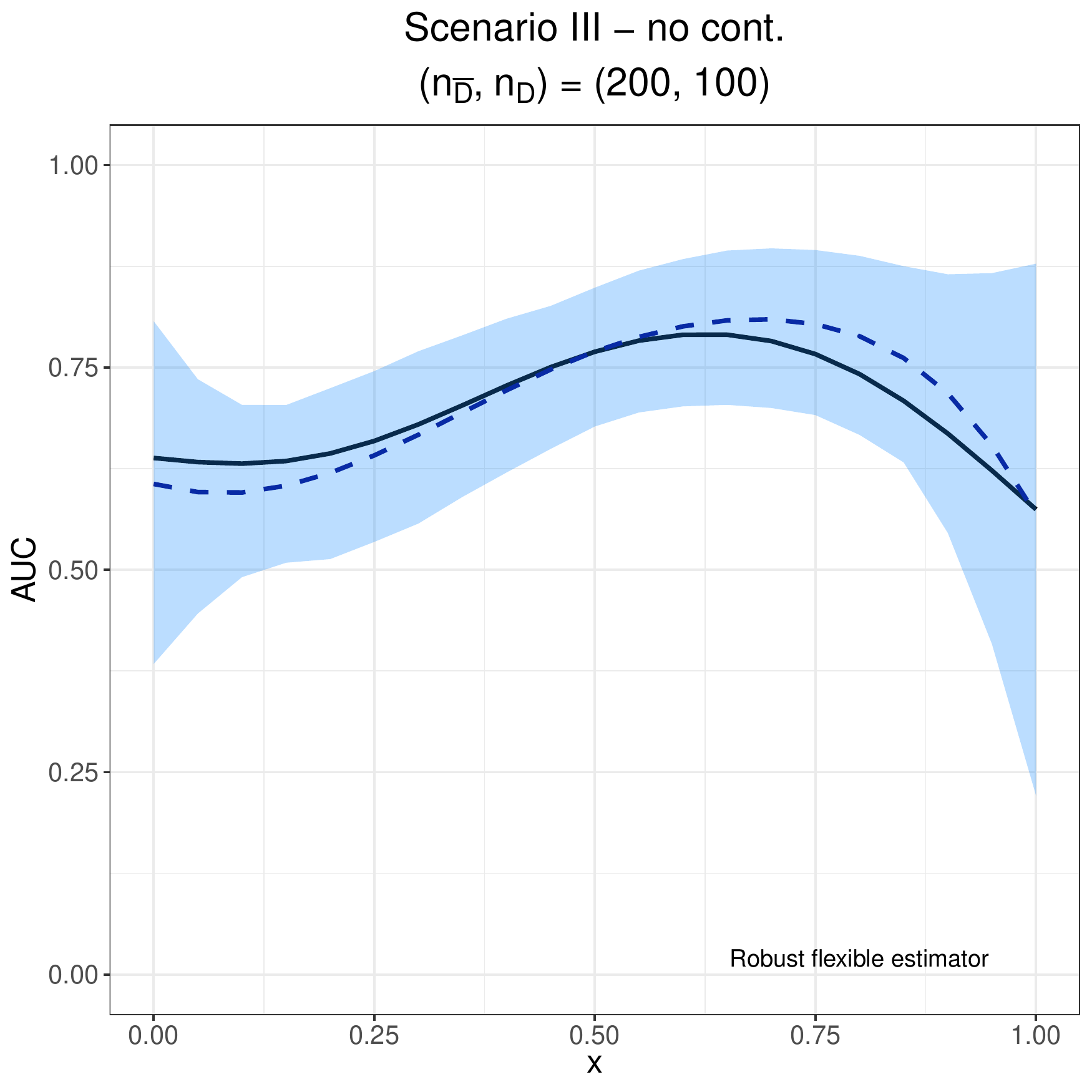}
			\includegraphics[width = 4.65cm]{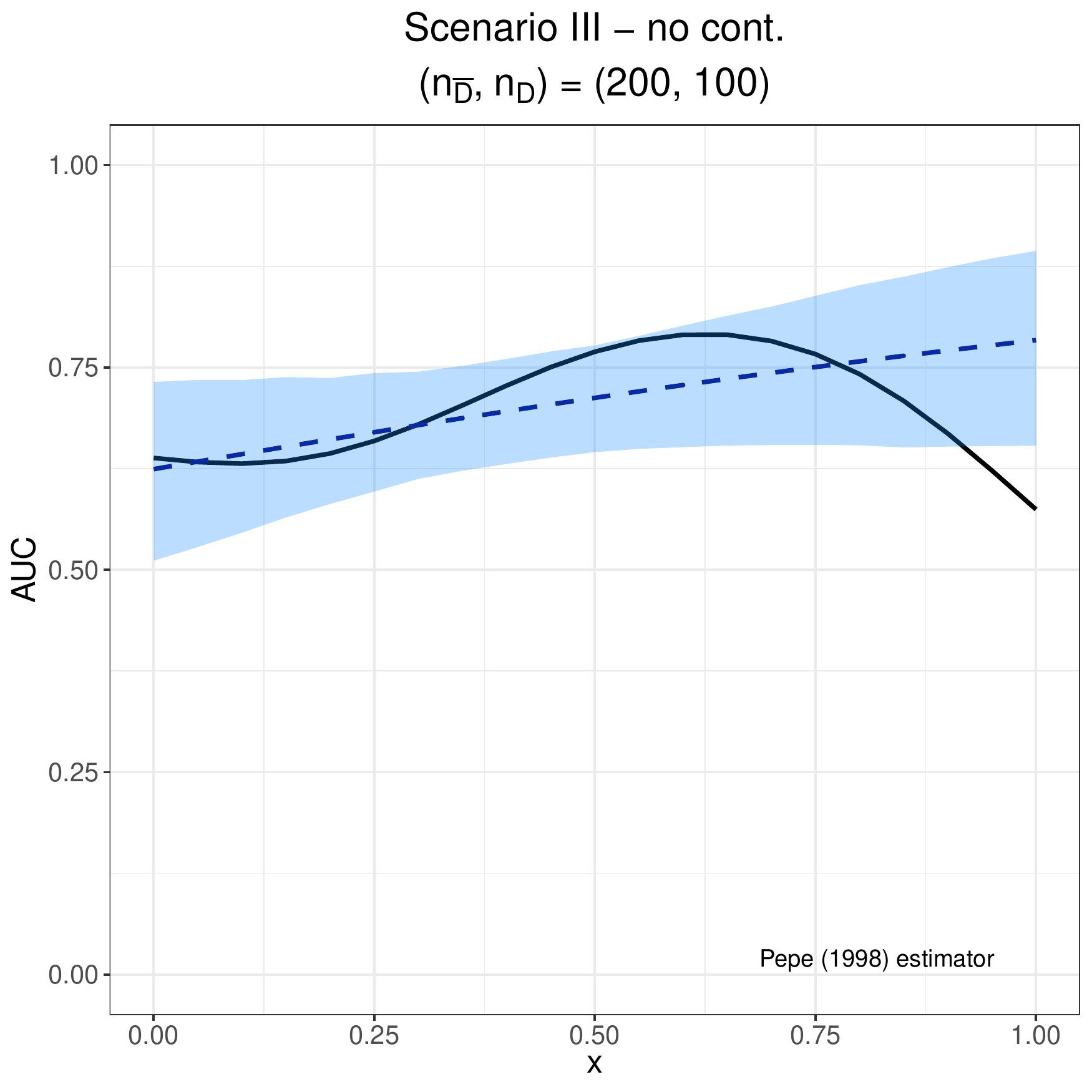}
			\includegraphics[width = 4.65cm]{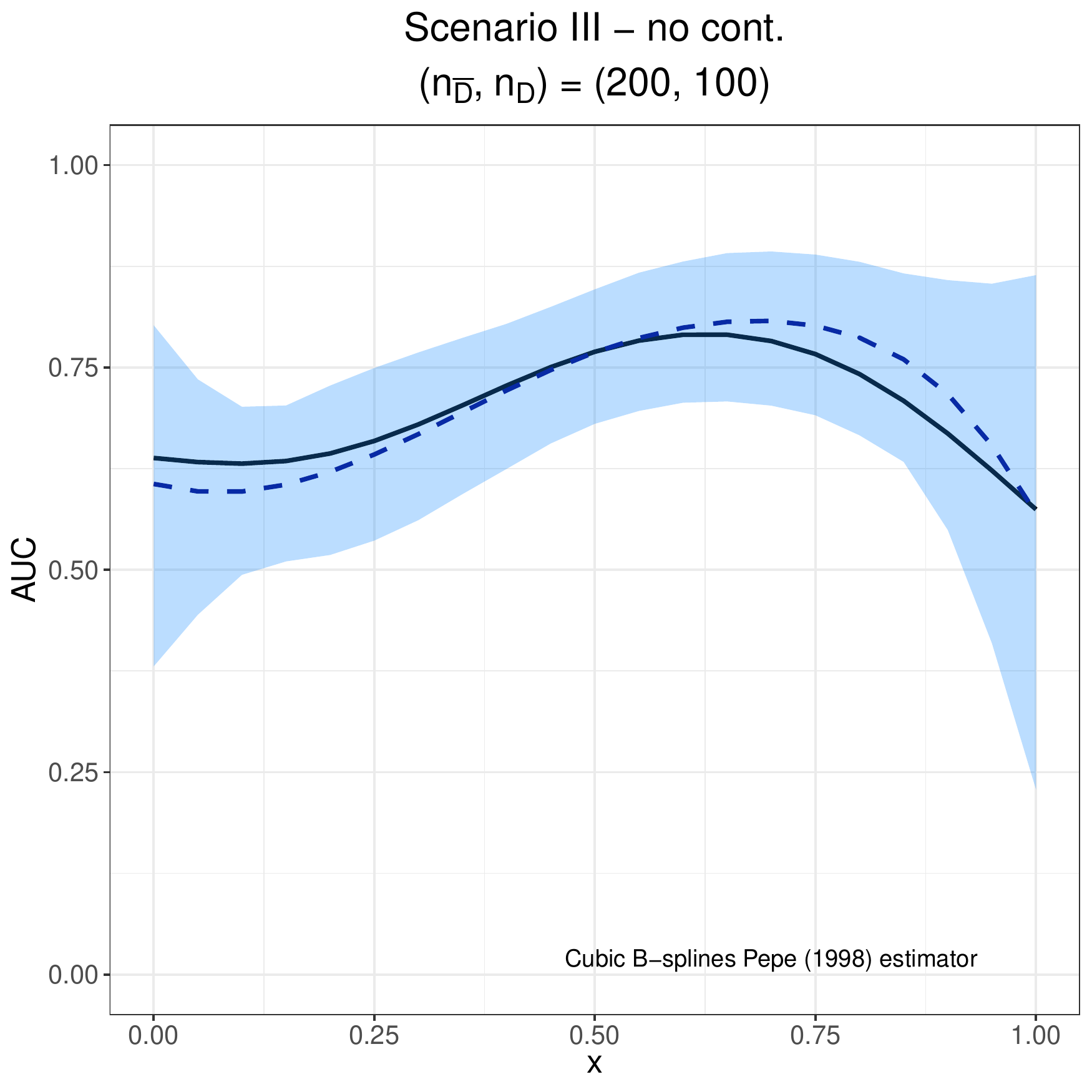}
			\includegraphics[width = 4.65cm]{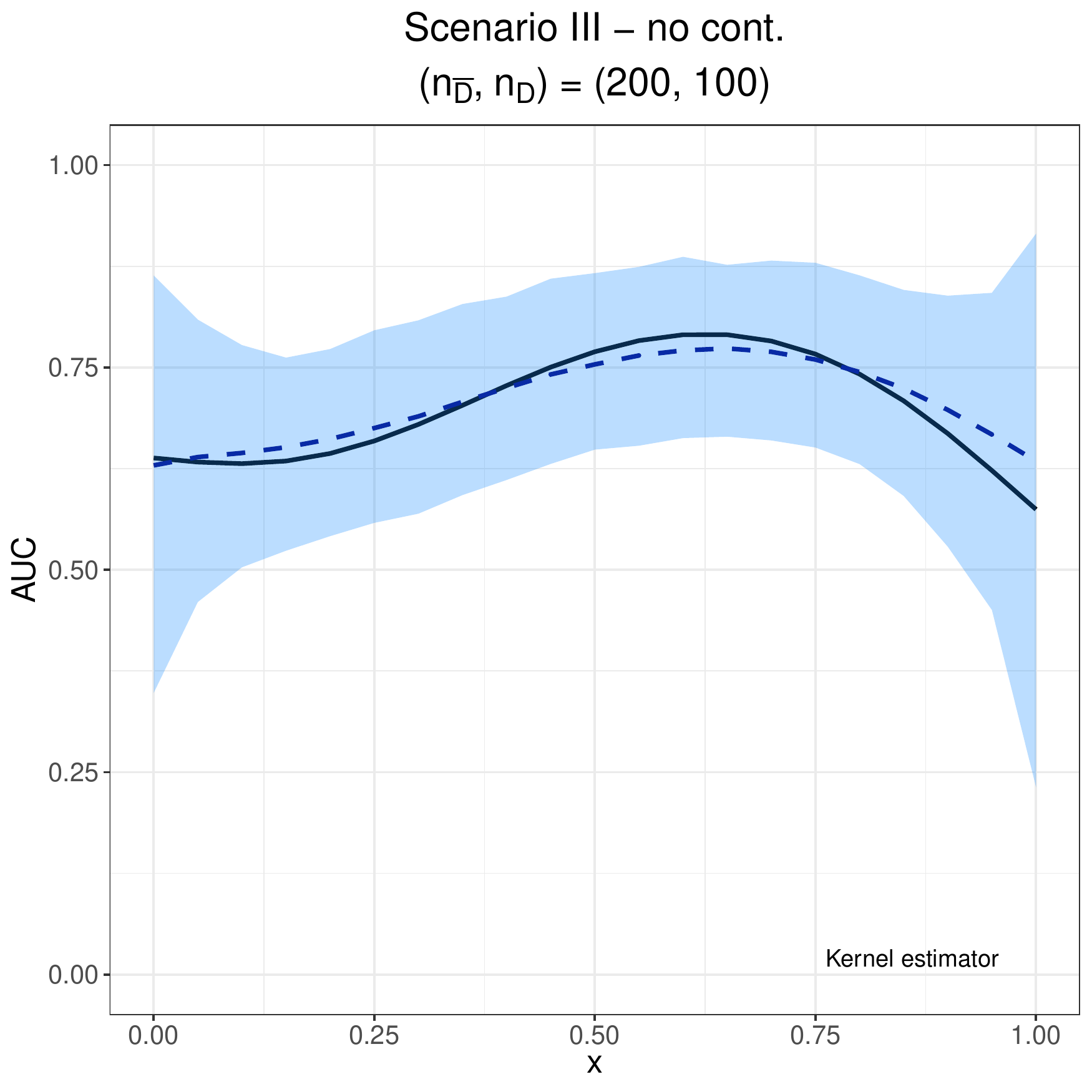}
		}
		\vspace{0.3cm}
		\subfigure{
			\includegraphics[width = 4.65cm]{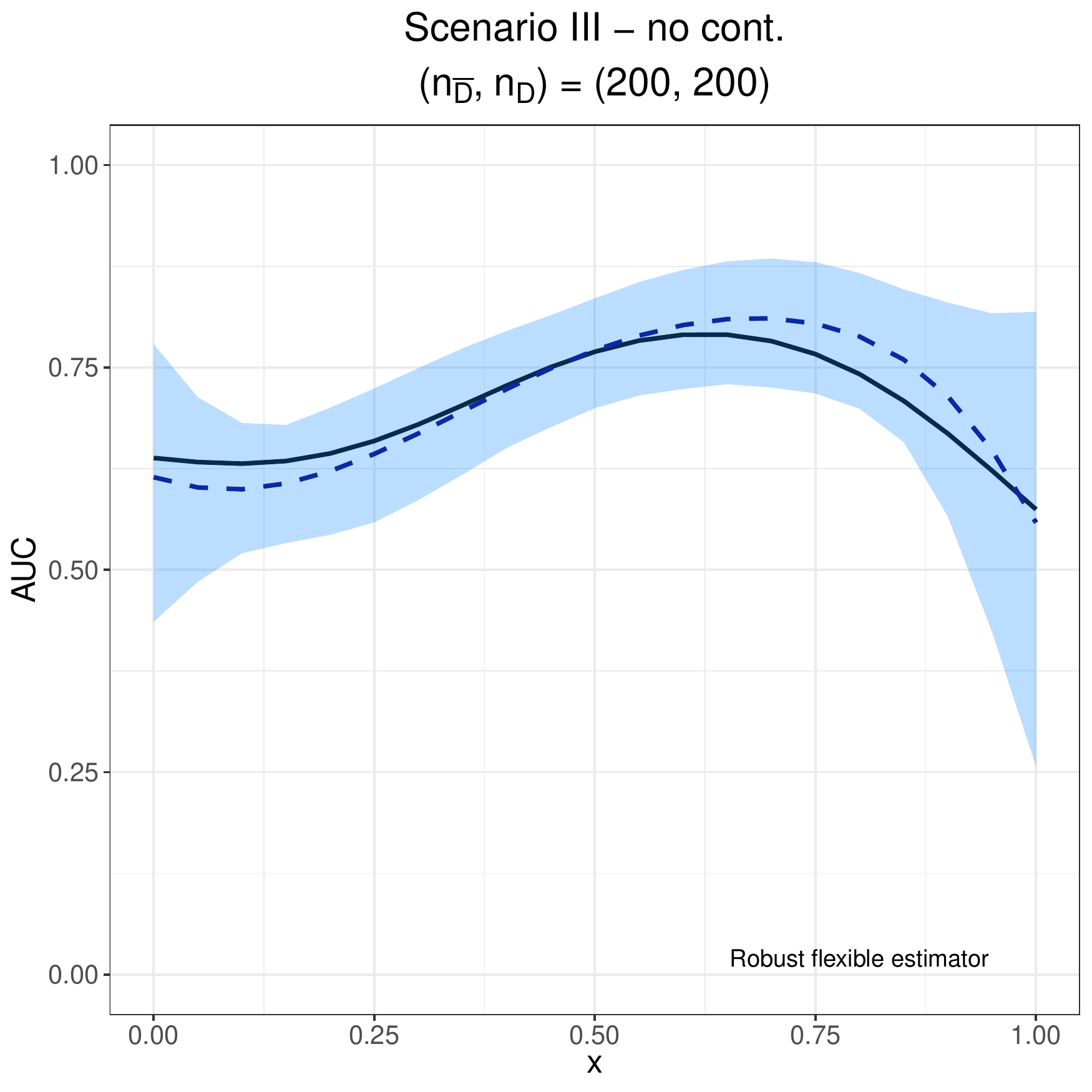}
			\includegraphics[width = 4.65cm]{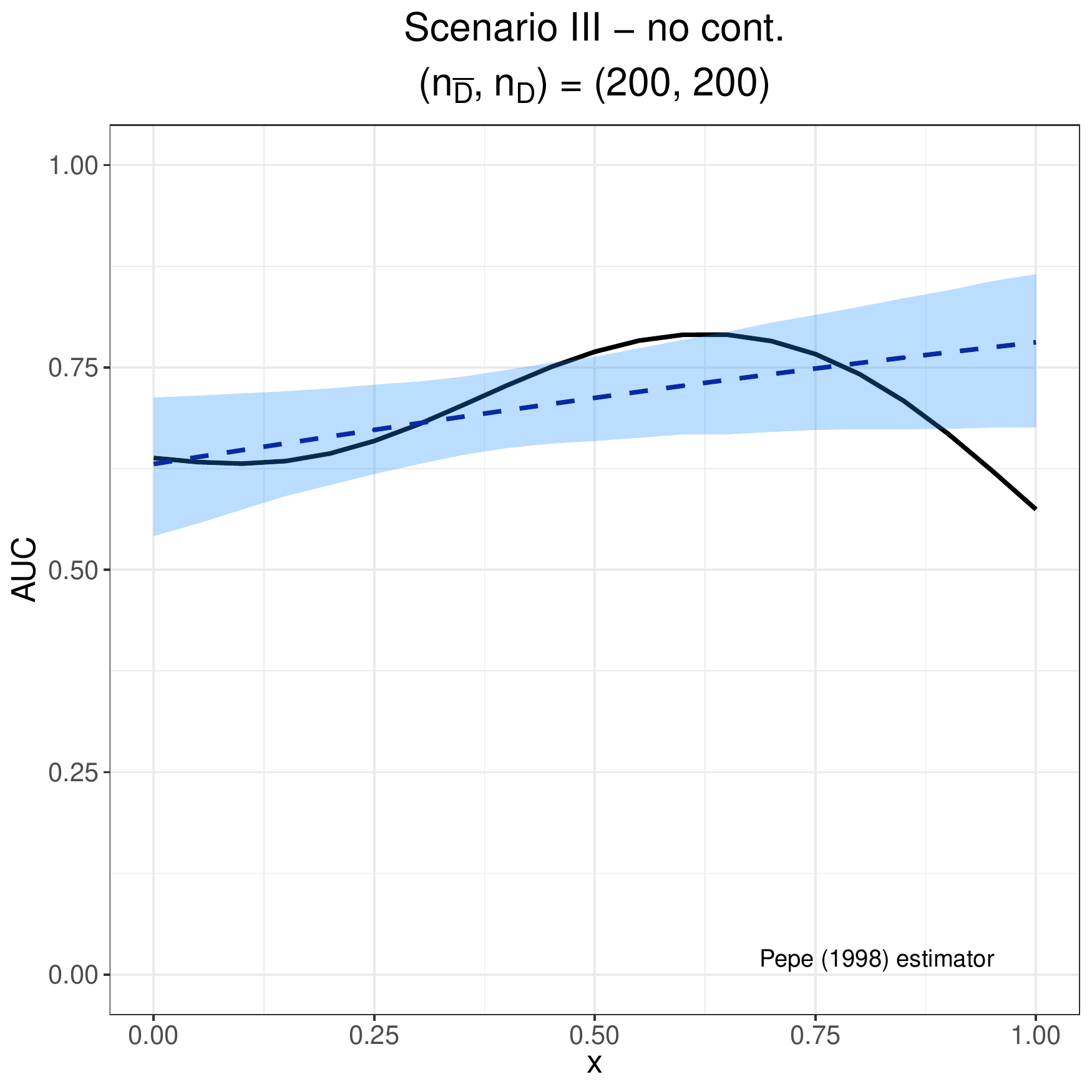}
			\includegraphics[width = 4.65cm]{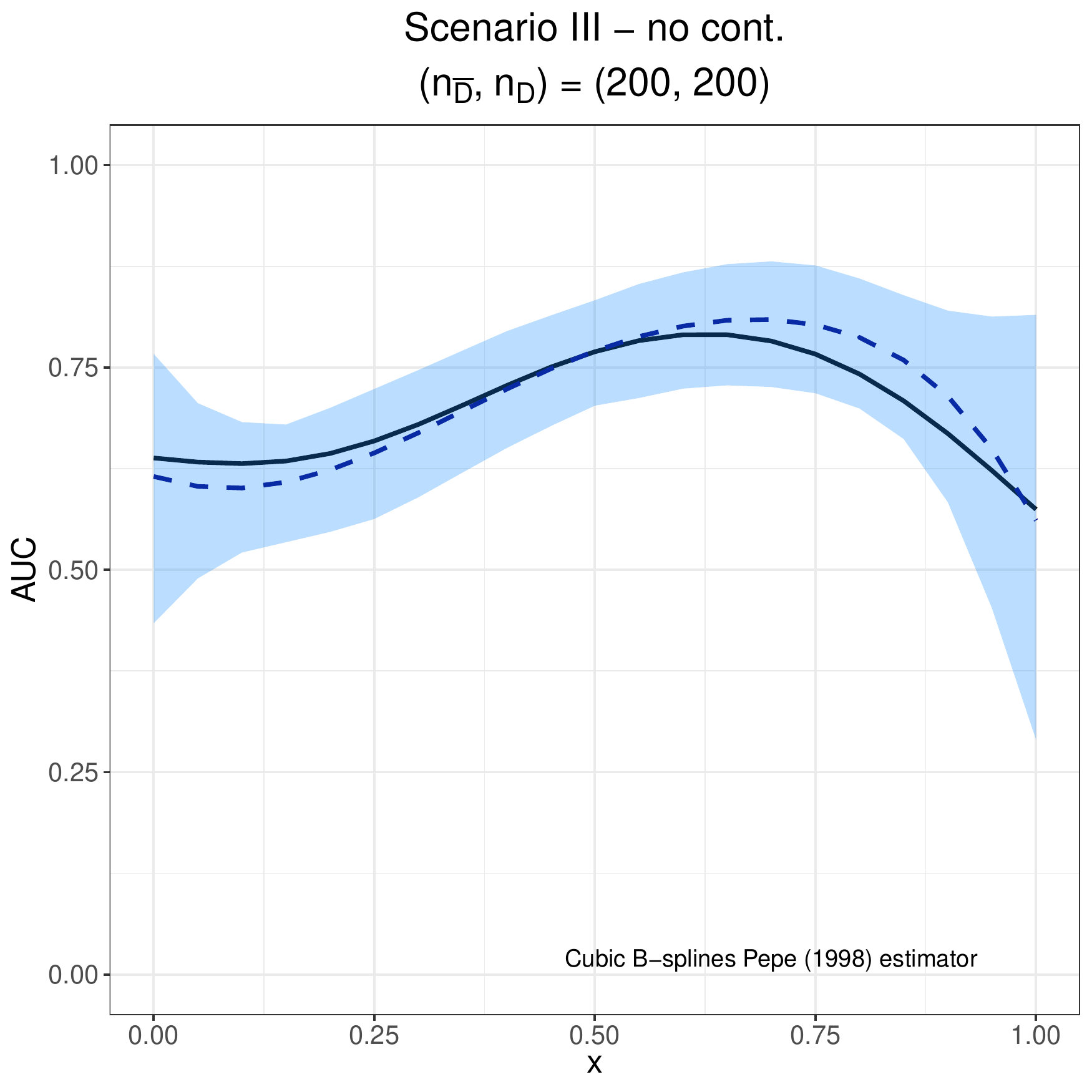}
			\includegraphics[width = 4.65cm]{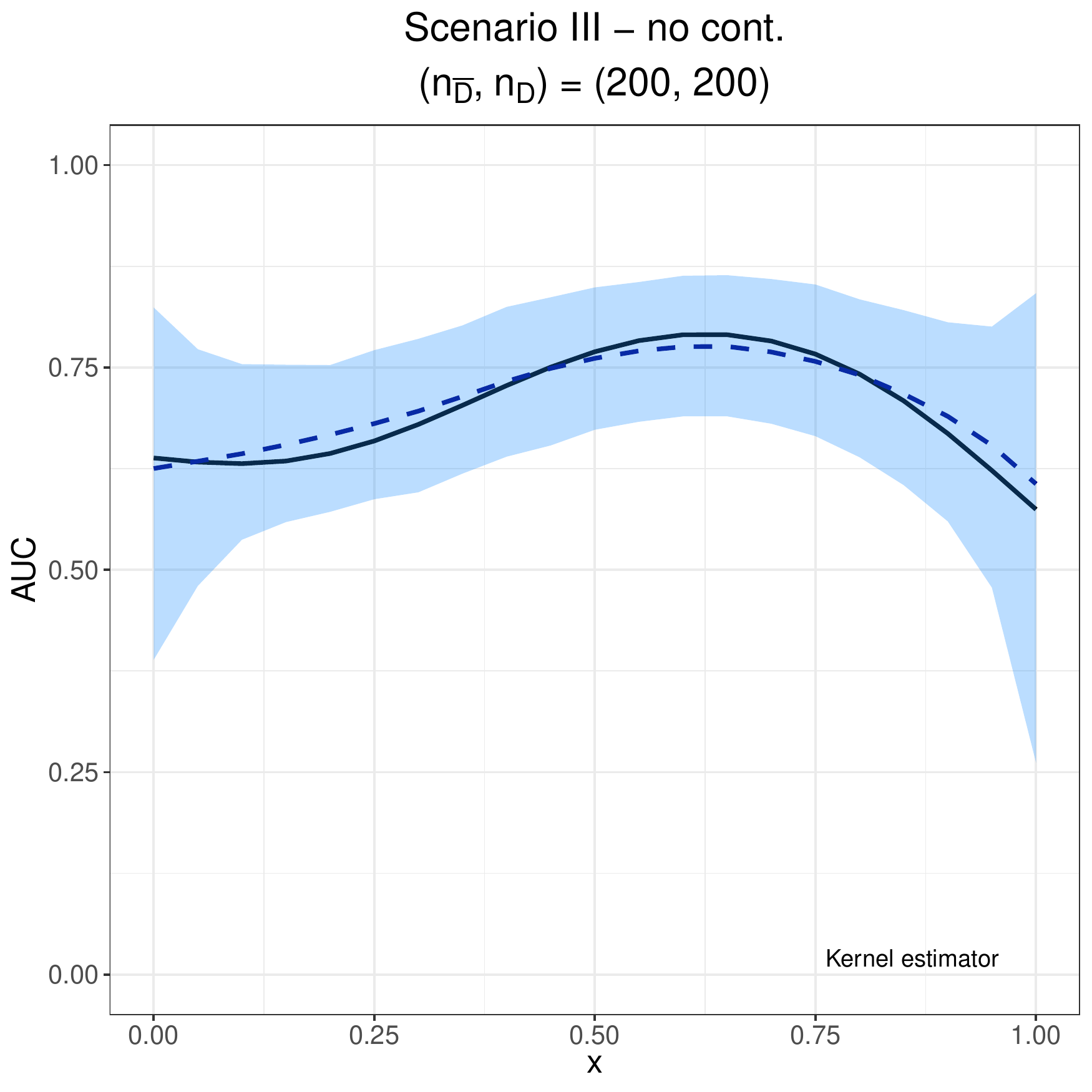}
		}
	\end{center}
	\caption{\footnotesize{Scenario III. True covariate-specific AUC (solid line) versus the mean of the Monte Carlo estimates (dashed line) along with the $2.5\%$ and $97.5\%$ simulation quantiles (shaded area) for the case of no contamination. The first row displays the results for $(n_{\bar{D}}, n_D)=(100,100)$, the second row for $(n_{\bar{D}}, n_D)=(200,100)$, and the third row for $(n_{\bar{D}}, n_D)=(200,200)$. The first column corresponds to our flexible and robust estimator, the second column to the estimator proposed by Pepe (1998), the third one to the cubic B-splines extension of Pepe (1998), and the fourth column to the kernel estimator.}}
\end{figure}

\begin{figure}[H]
	\begin{center}
		\subfigure{
			\includegraphics[width = 4.65cm]{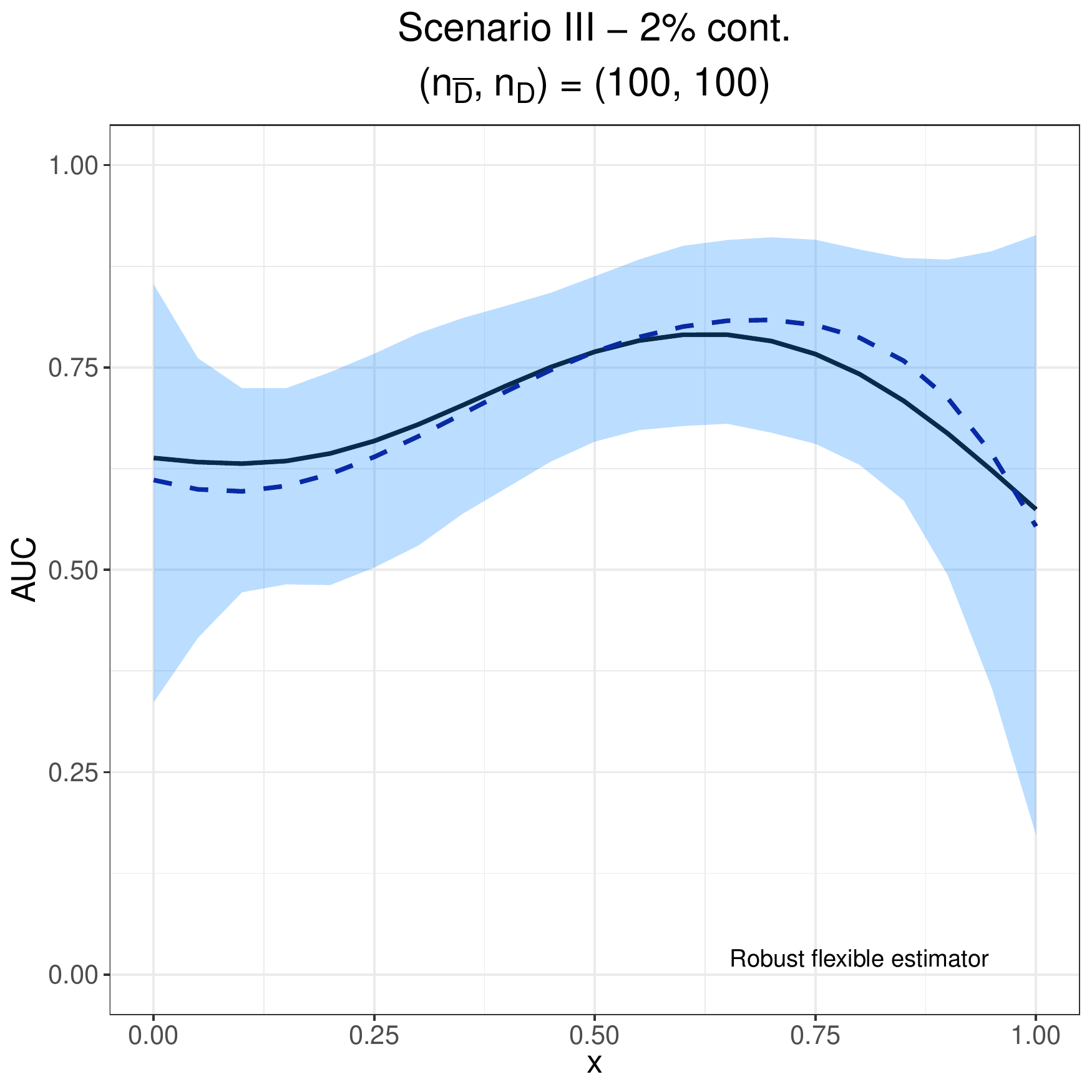}
			\includegraphics[width = 4.65cm]{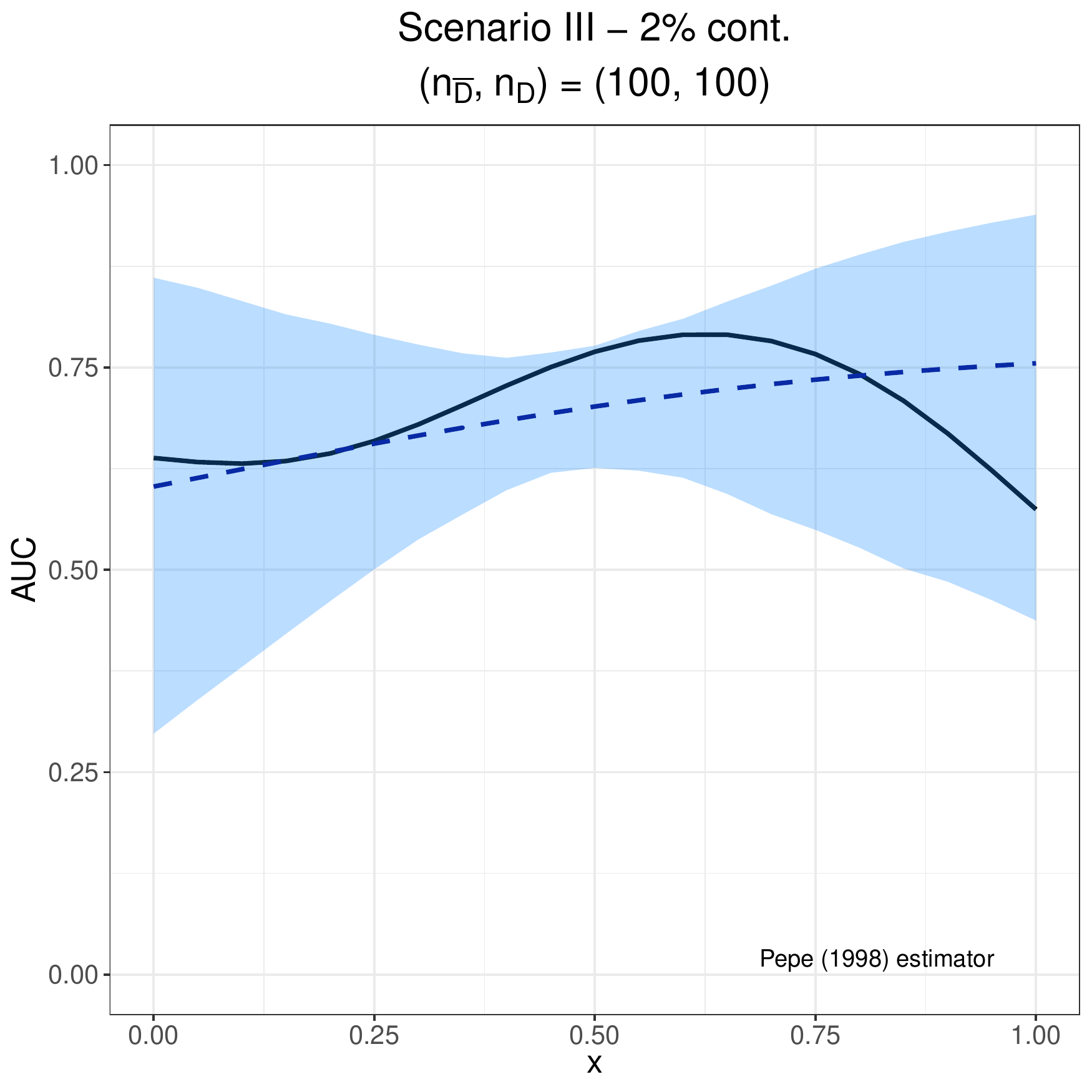}
			\includegraphics[width = 4.65cm]{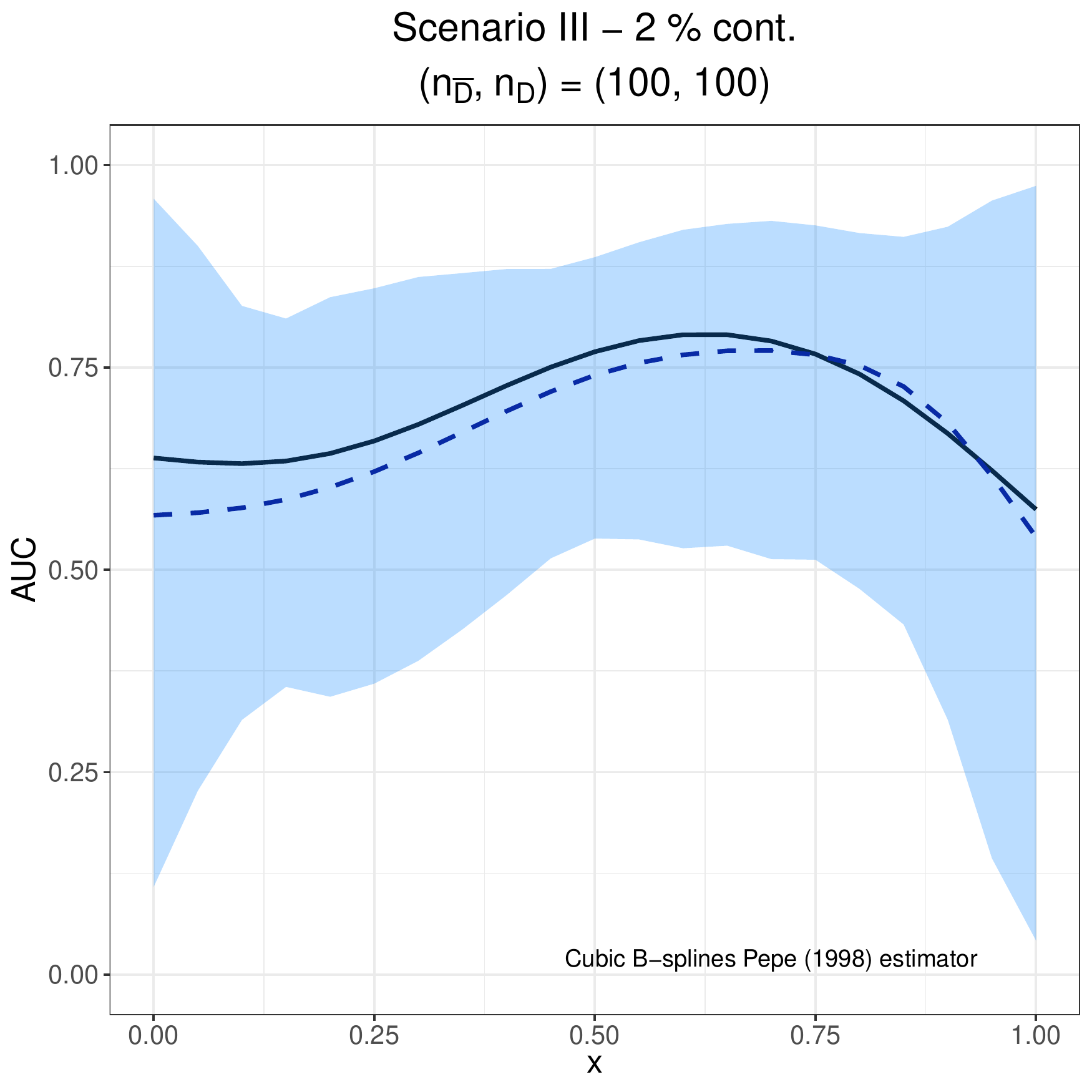}
			\includegraphics[width = 4.65cm]{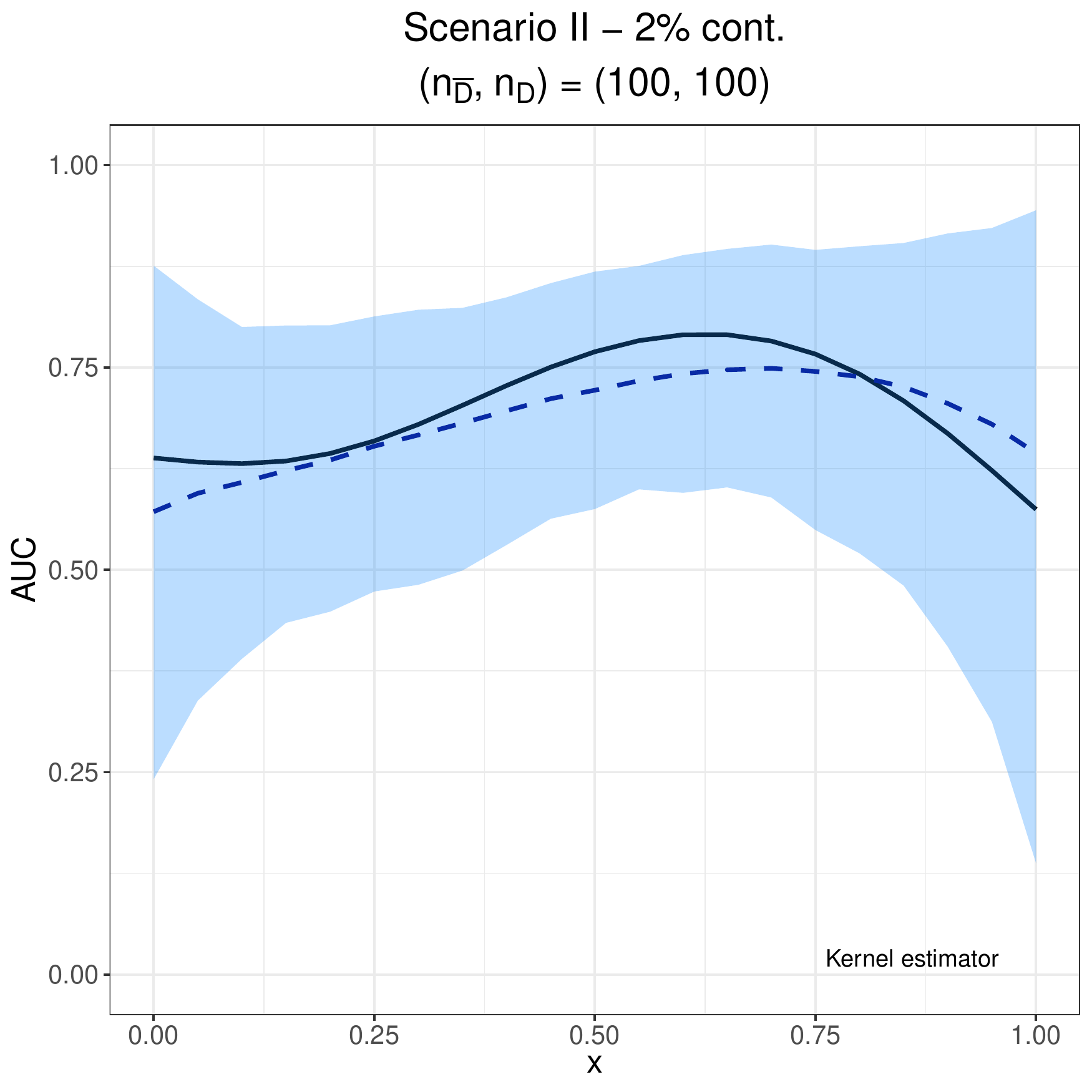}
		}
		\vspace{0.3cm}
		\subfigure{
			\includegraphics[width = 4.65cm]{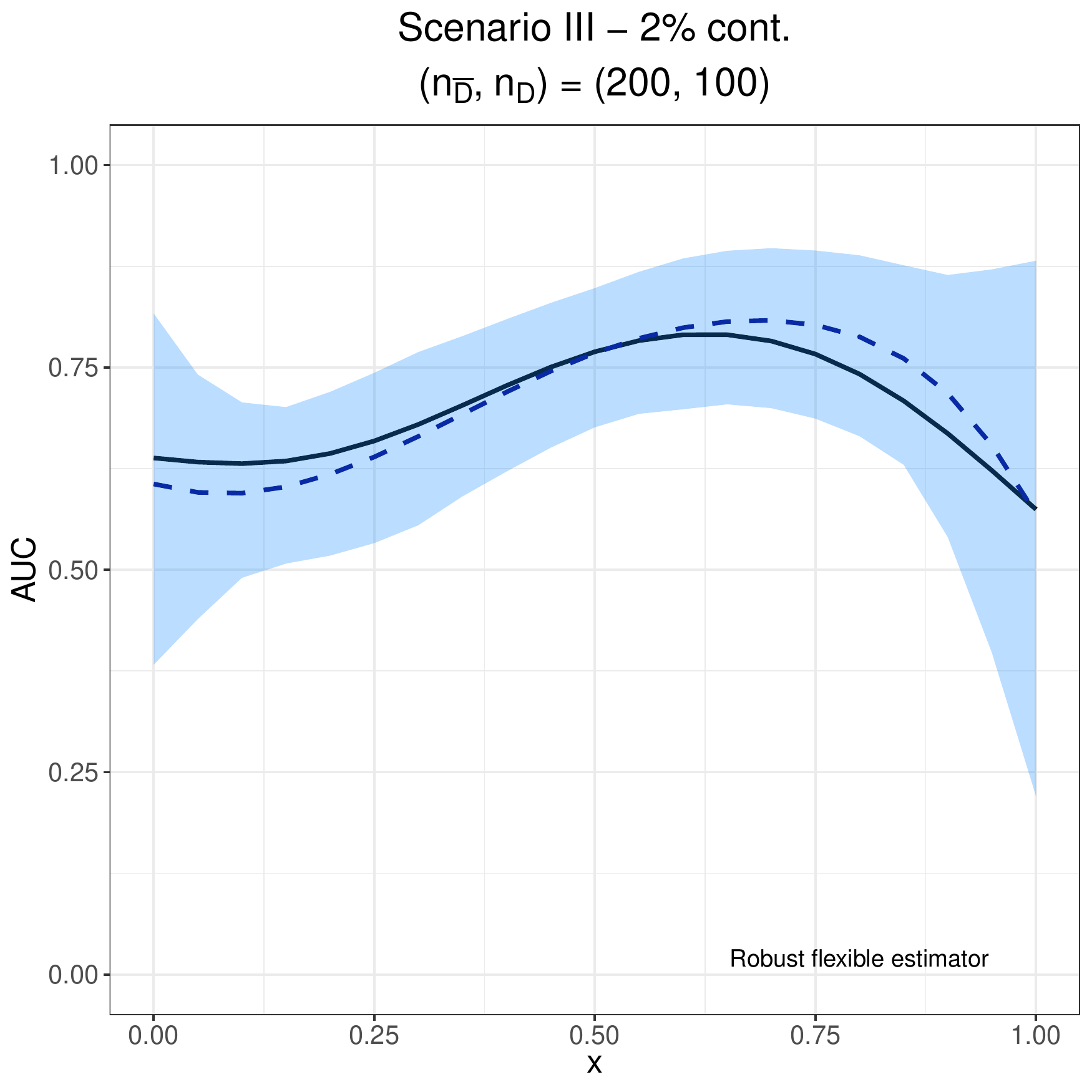}
			\includegraphics[width = 4.65cm]{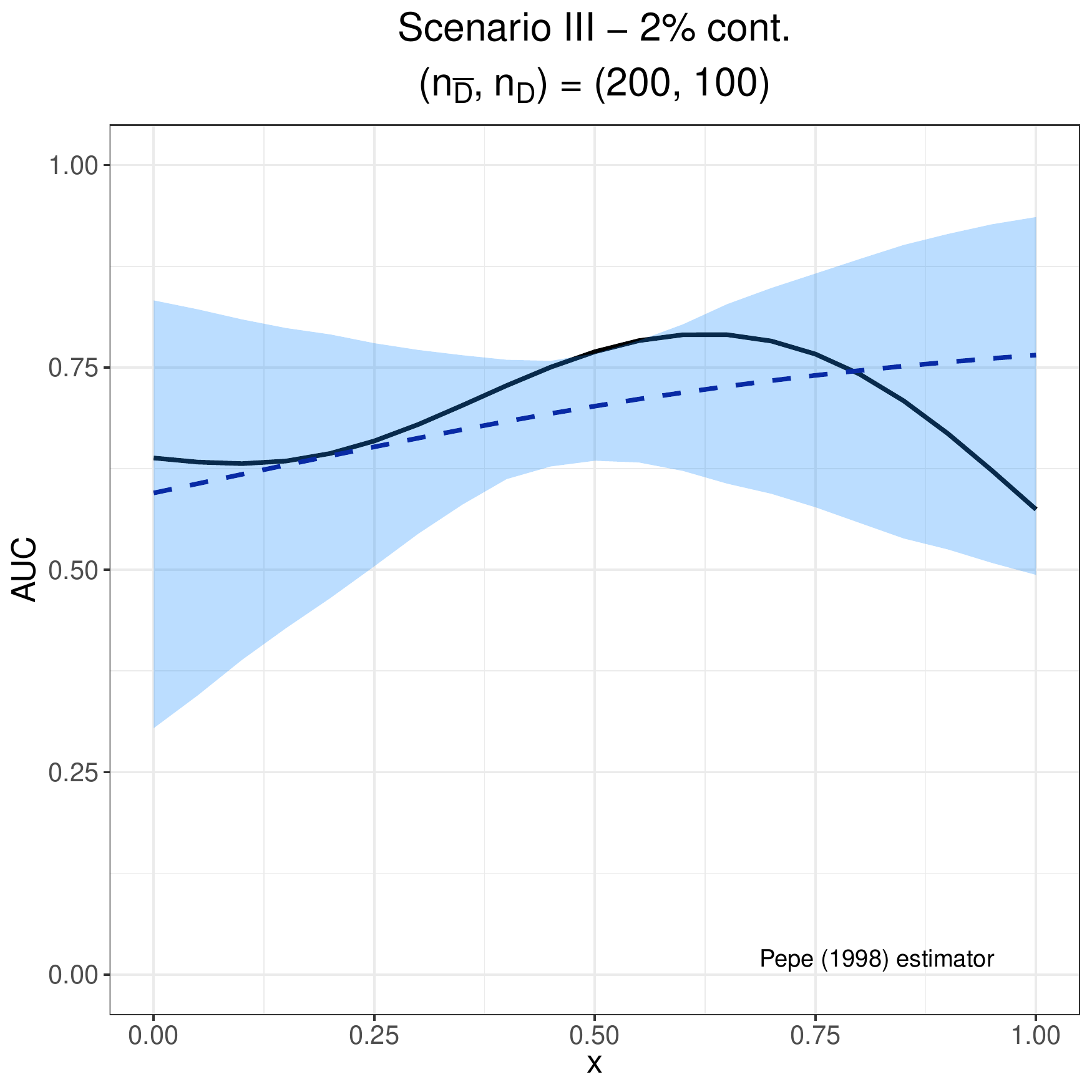}
			\includegraphics[width = 4.65cm]{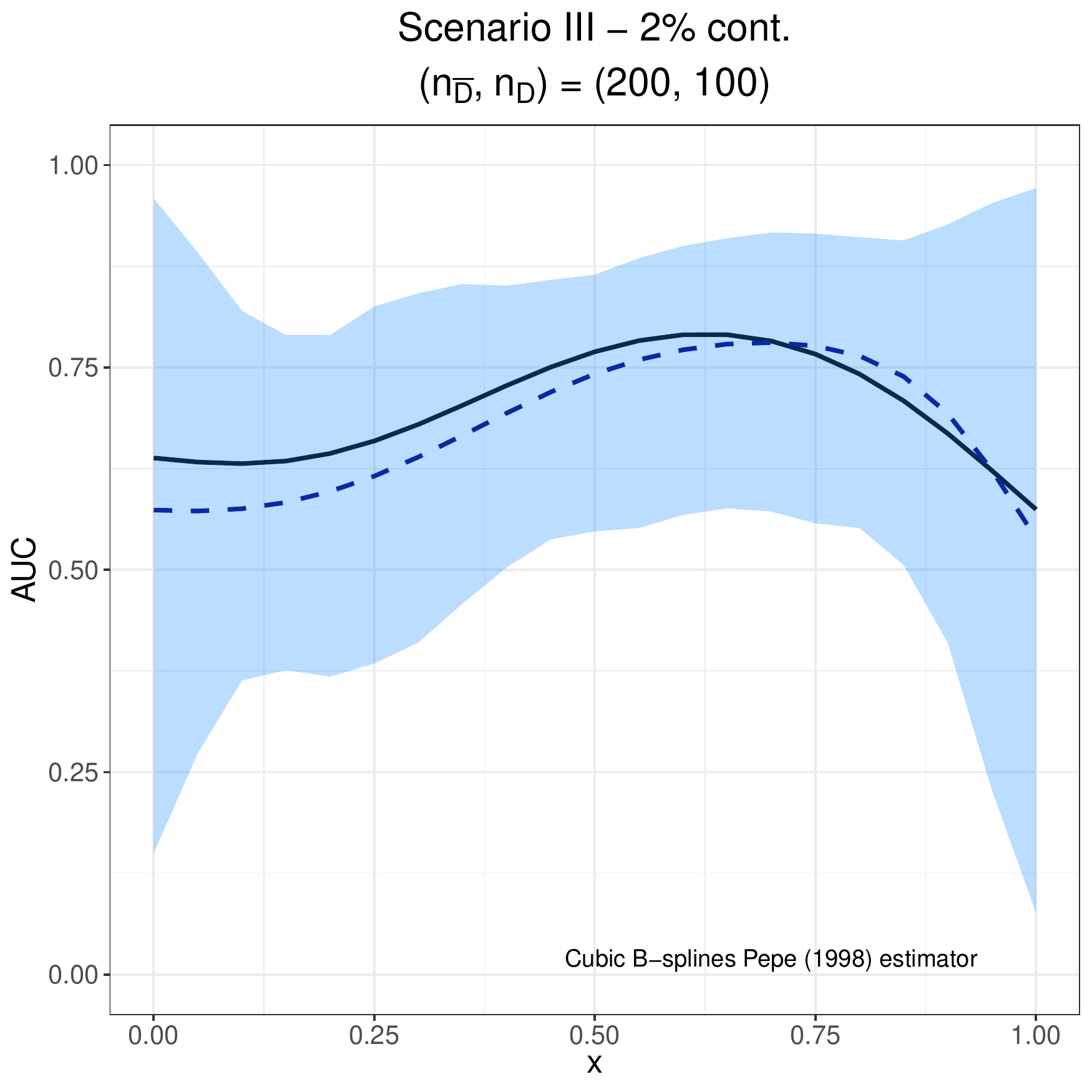}
			\includegraphics[width = 4.65cm]{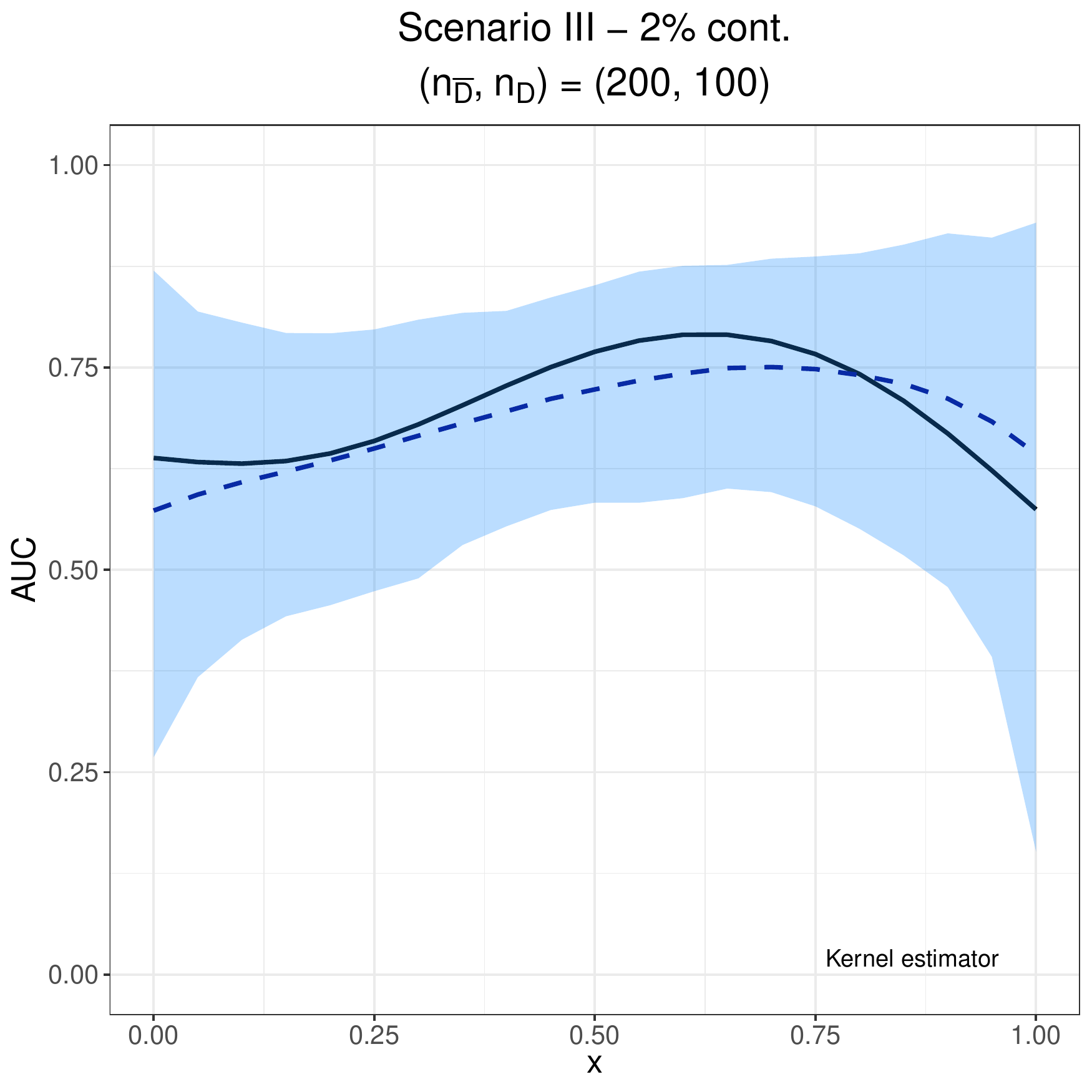}
		}
		\vspace{0.3cm}
		\subfigure{
			\includegraphics[width = 4.65cm]{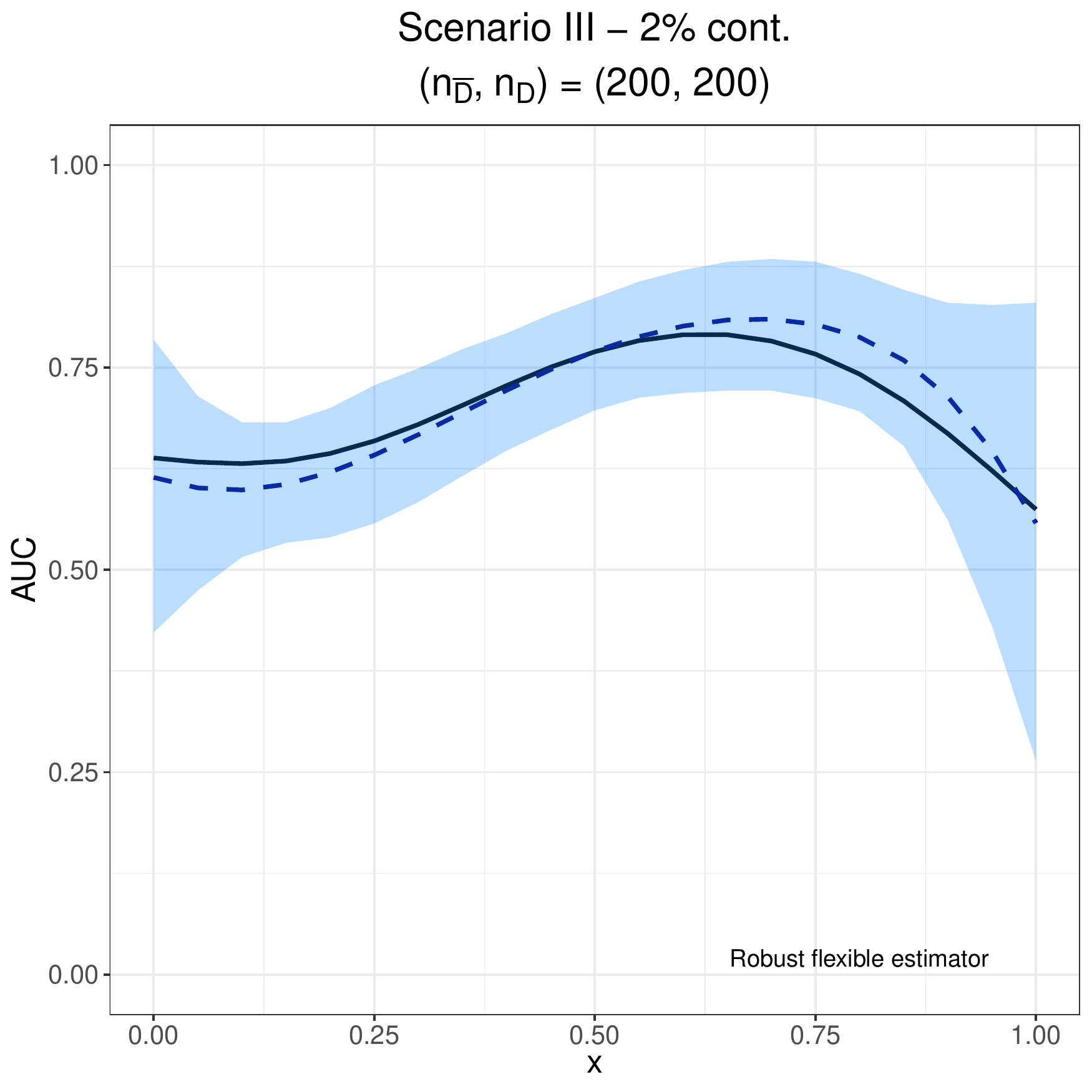}
			\includegraphics[width = 4.65cm]{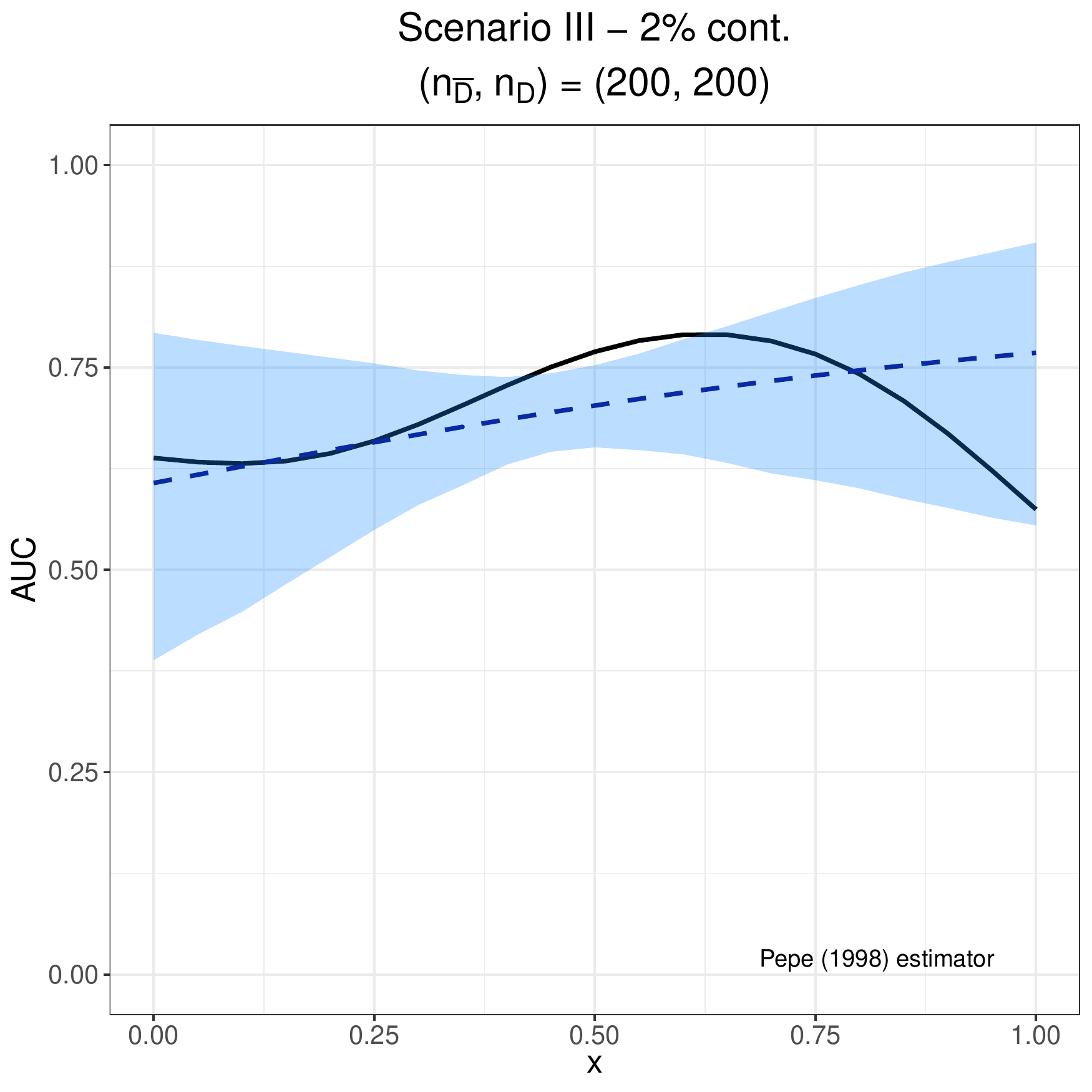}
			\includegraphics[width = 4.65cm]{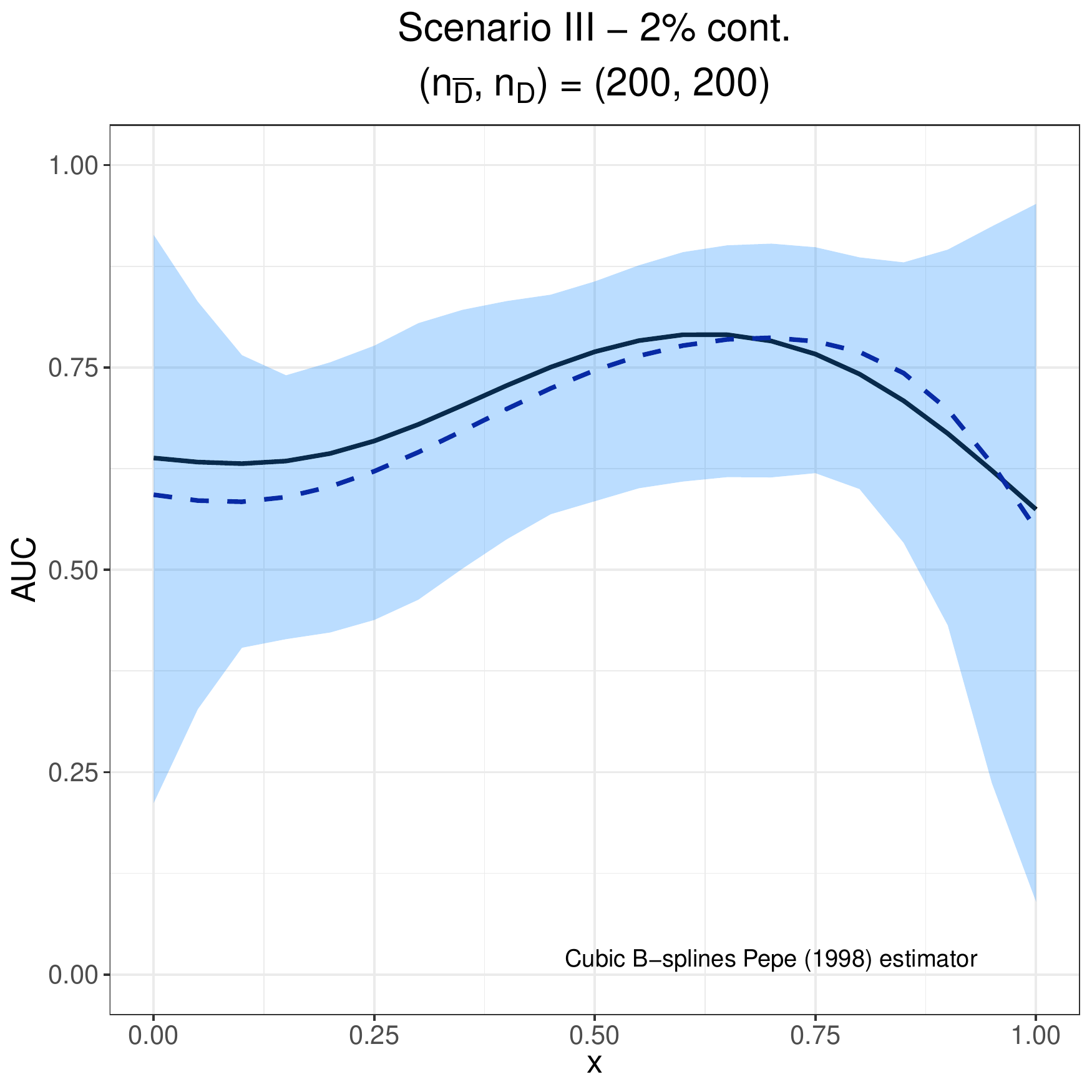}
			\includegraphics[width = 4.65cm]{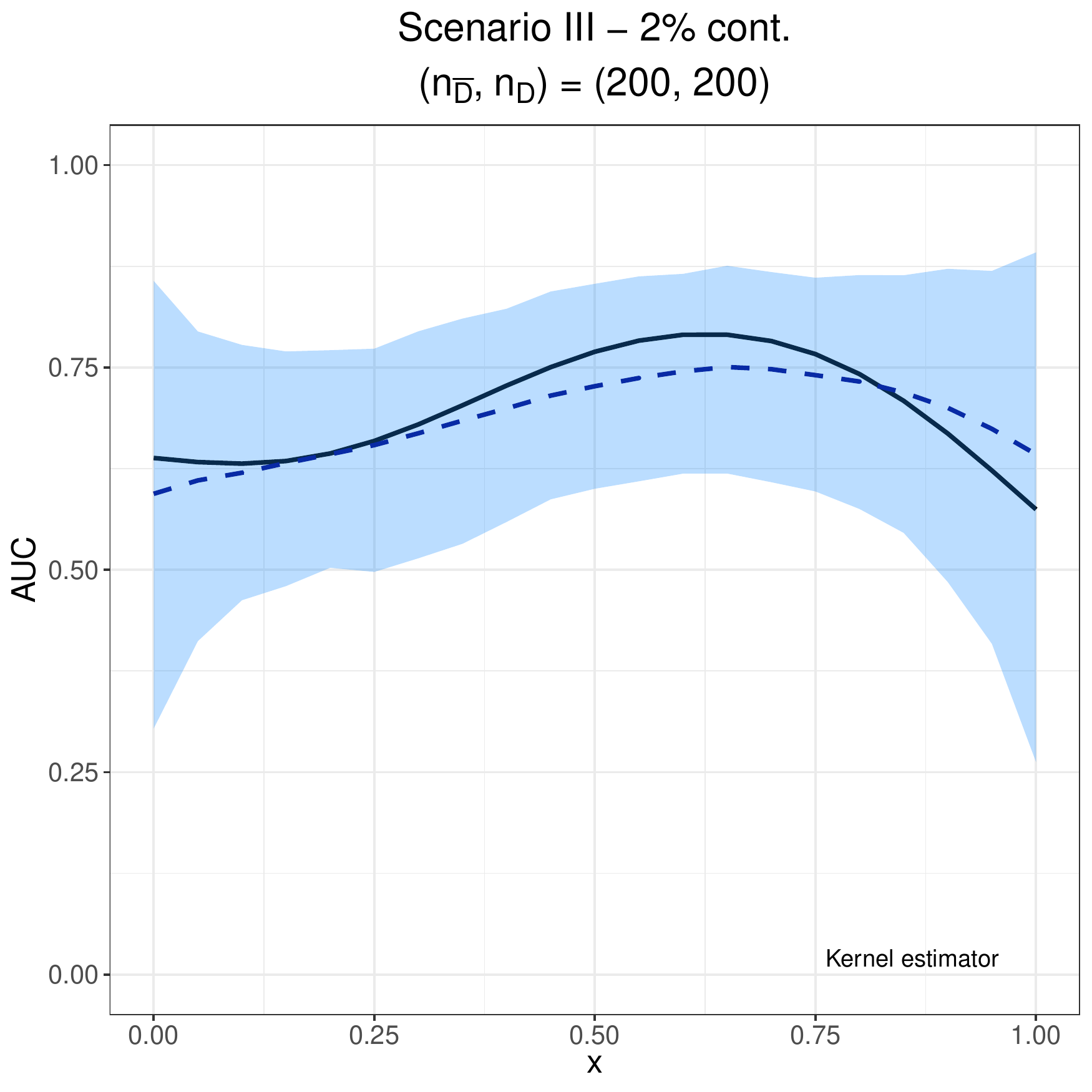}
		}
	\end{center}
	\caption{\footnotesize{Scenario III. True covariate-specific AUC (solid line) versus the mean of the Monte Carlo estimates (dashed line) along with the $2.5\%$ and $97.5\%$ simulation quantiles (shaded area) for the case of $2\%$ of contamination. The first row displays the results for $(n_{\bar{D}}, n_D)=(100,100)$, the second row for $(n_{\bar{D}}, n_D)=(200,100)$, and the third row for $(n_{\bar{D}}, n_D)=(200,200)$. The first column corresponds to our flexible and robust estimator, the second column to the estimator proposed by Pepe (1998), the third one to the cubic B-splines extension of Pepe (1998), and the fourth column to the kernel estimator.}}
\end{figure}

\begin{figure}[H]
	\begin{center}
		\subfigure{
			\includegraphics[width = 4.65cm]{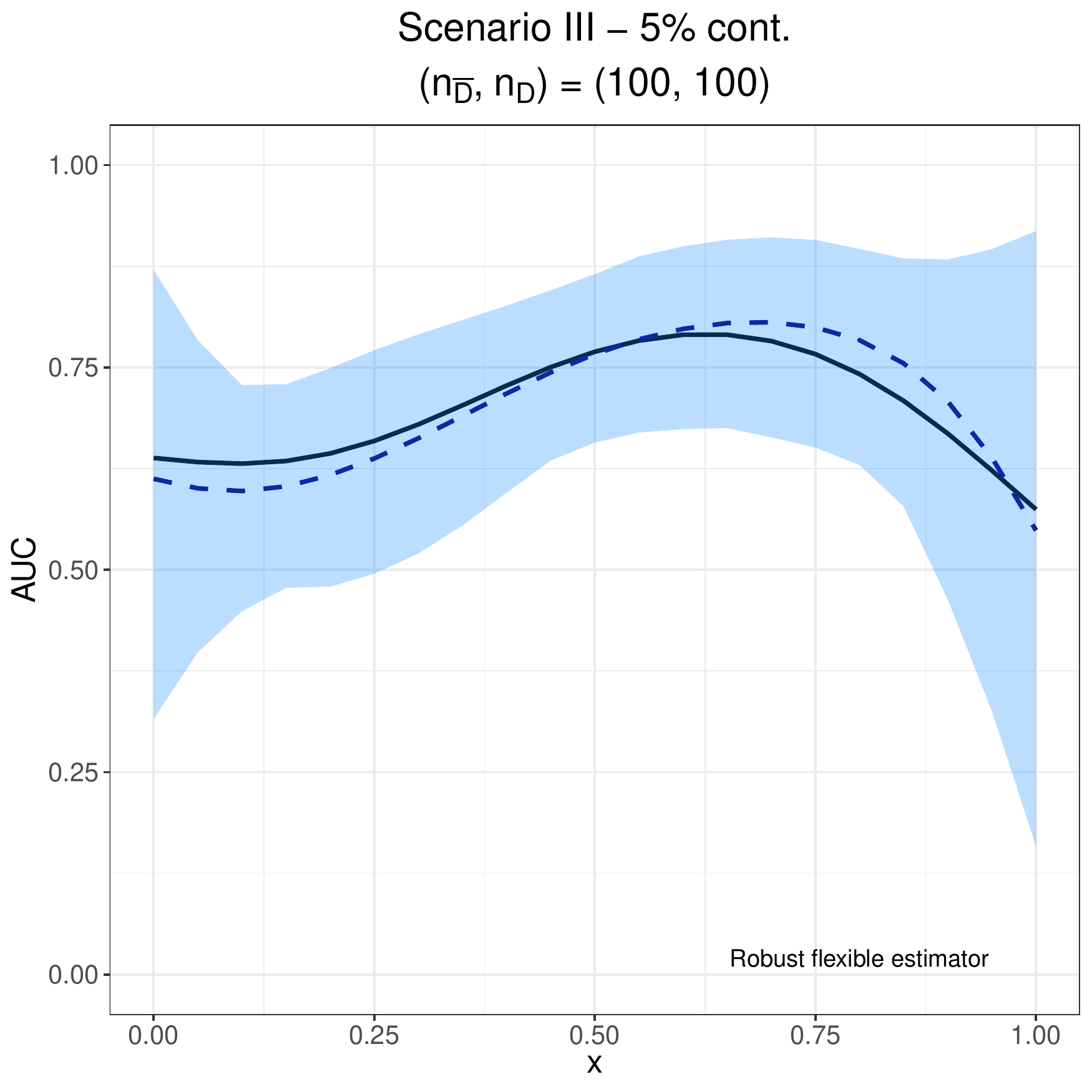}
			\includegraphics[width = 4.65cm]{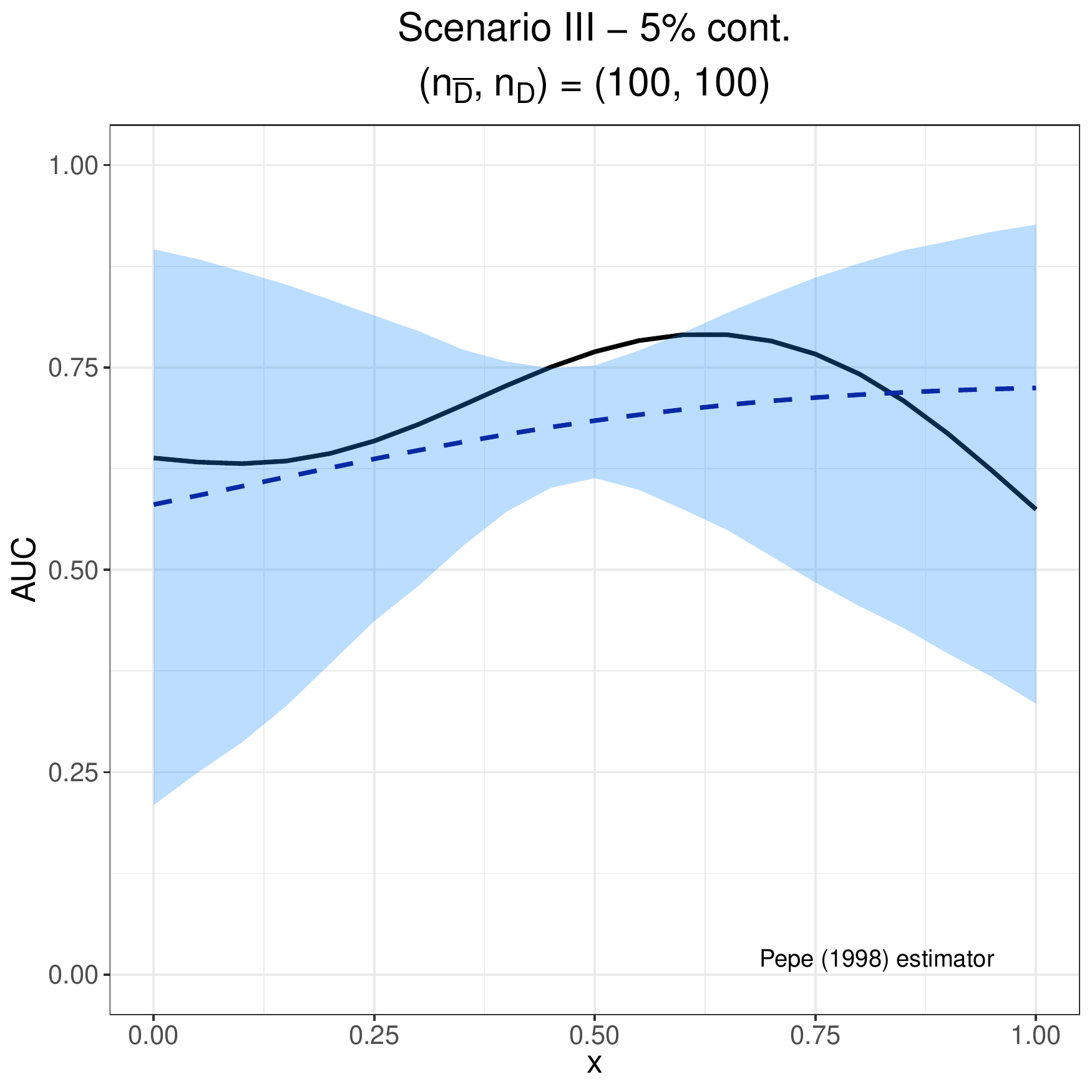}
			\includegraphics[width = 4.65cm]{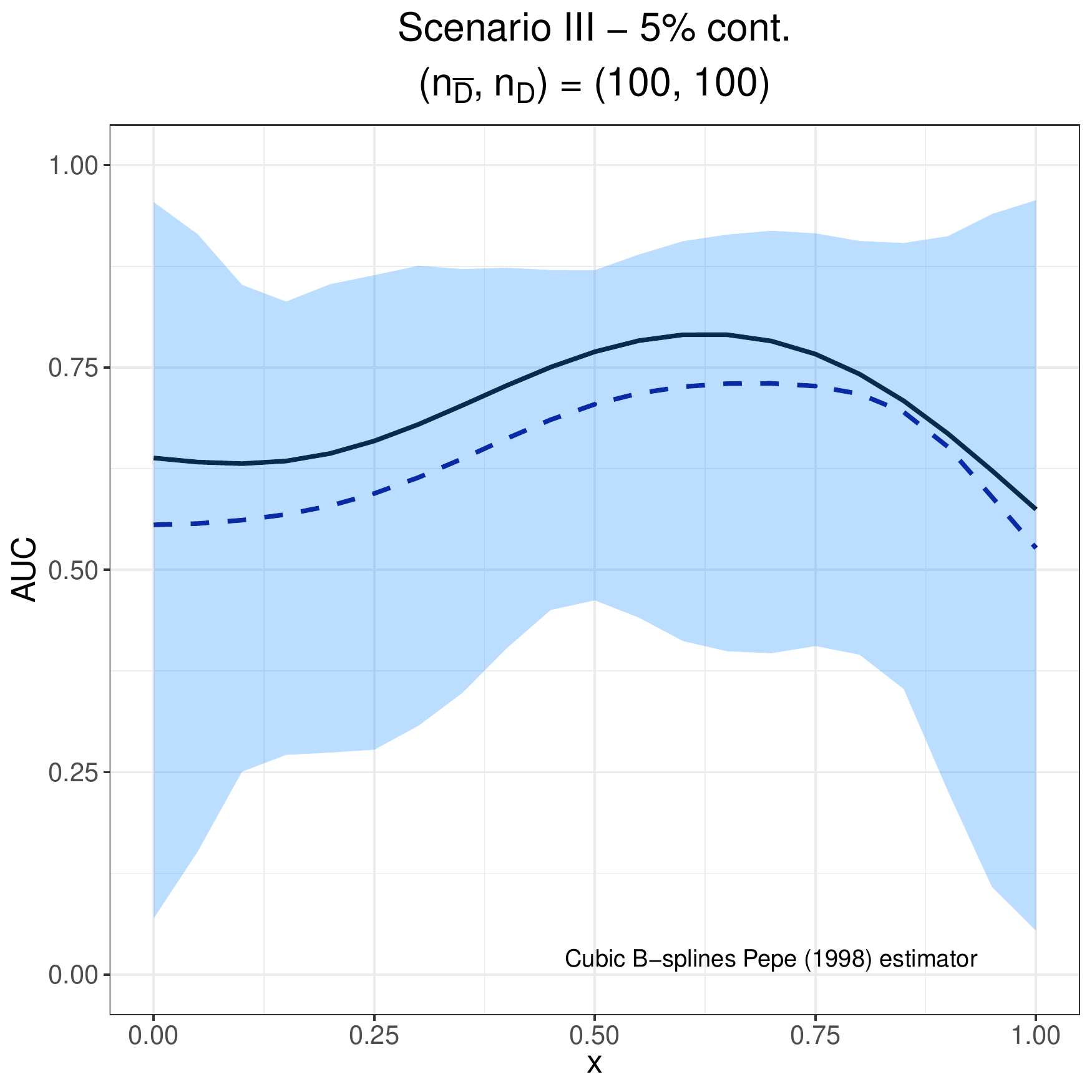}
			\includegraphics[width = 4.65cm]{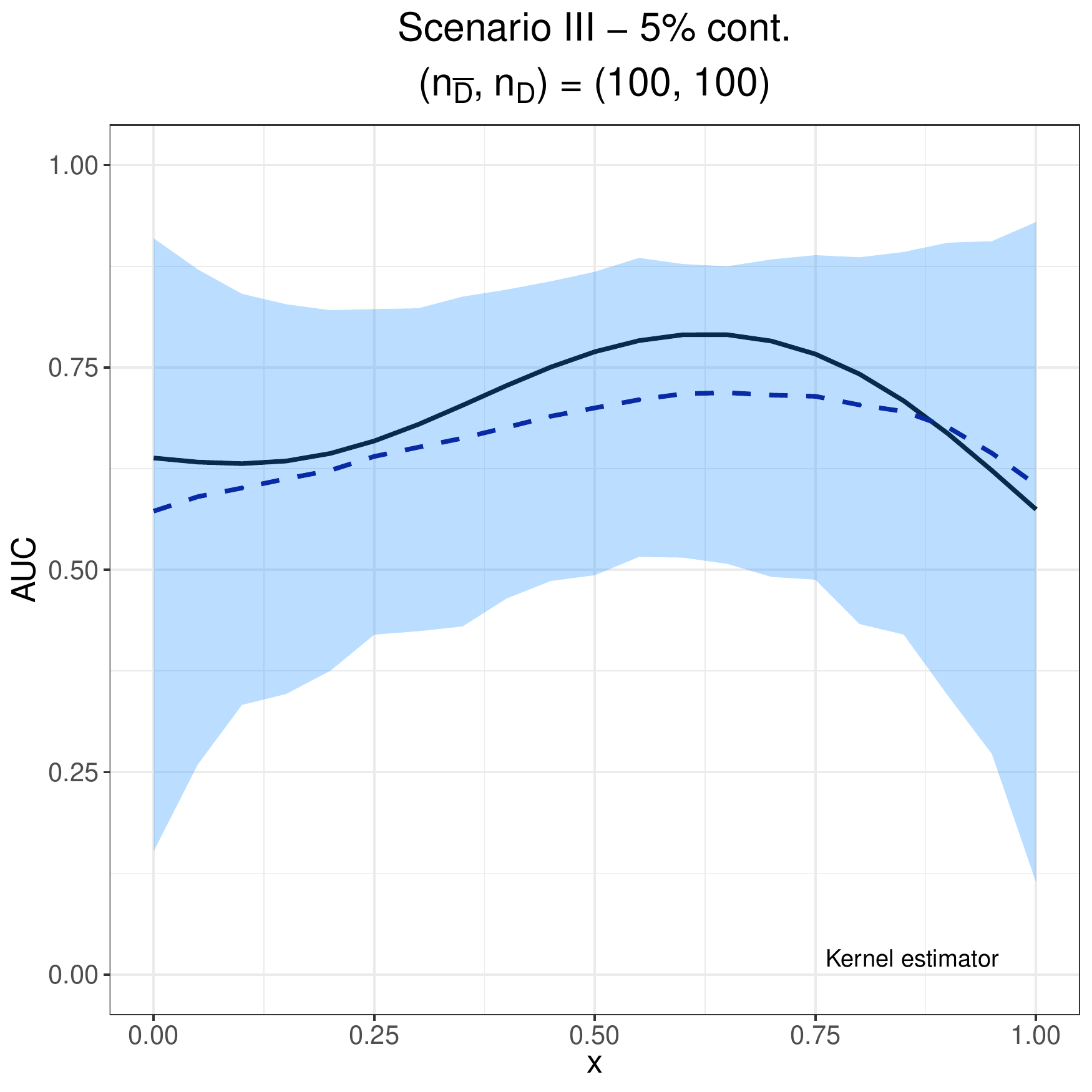}
		}
		\vspace{0.3cm}
		\subfigure{
			\includegraphics[width = 4.65cm]{aucrcont5sampsize2sc3.pdf}
			\includegraphics[width = 4.65cm]{aucspcont5sampsize2sc3.pdf}
			\includegraphics[width = 4.65cm]{aucspbscont5sampsize2sc3.pdf}
			\includegraphics[width = 4.65cm]{auckercont5sampsize2sc3.pdf}
		}
		\vspace{0.3cm}
		\subfigure{
			\includegraphics[width = 4.65cm]{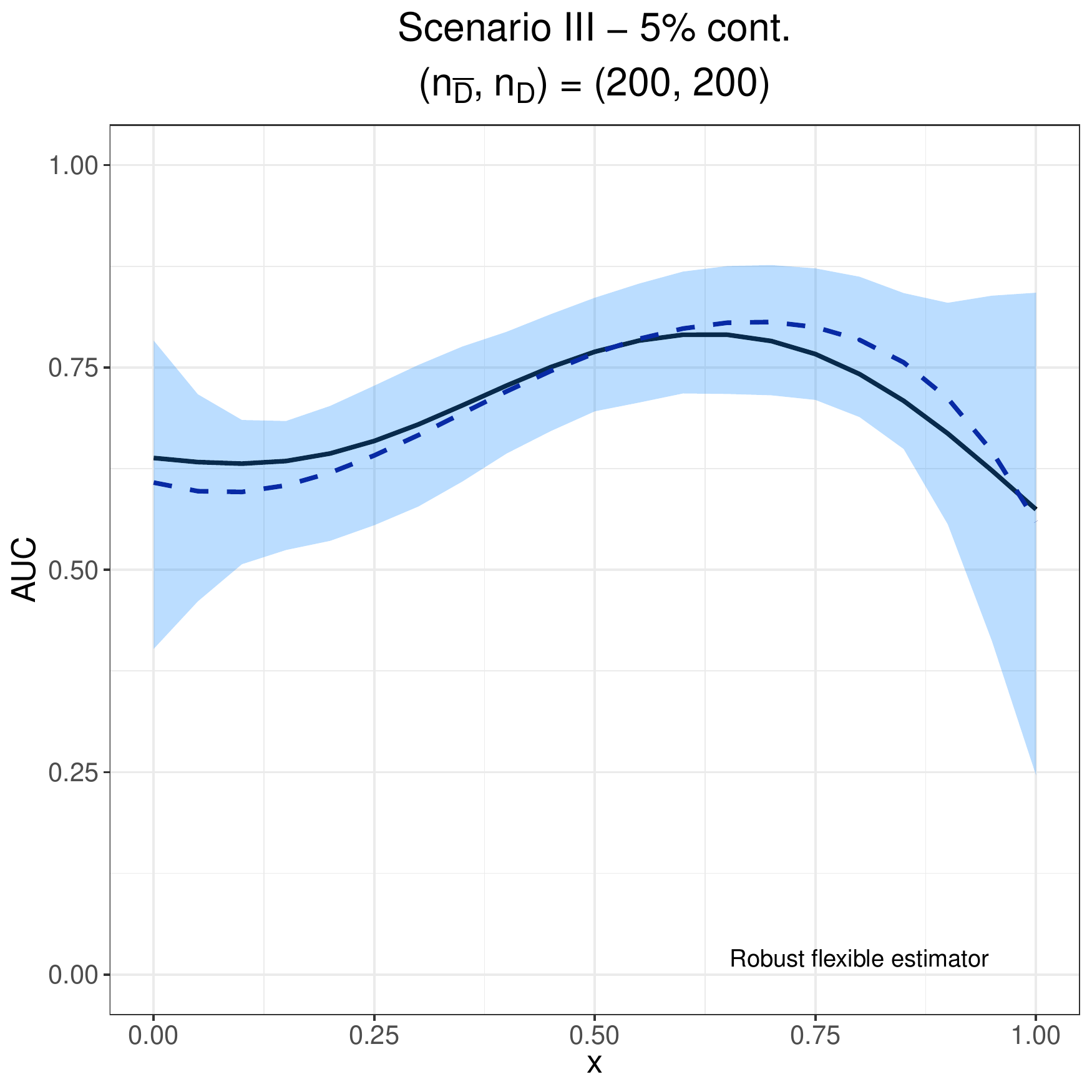}
			\includegraphics[width = 4.65cm]{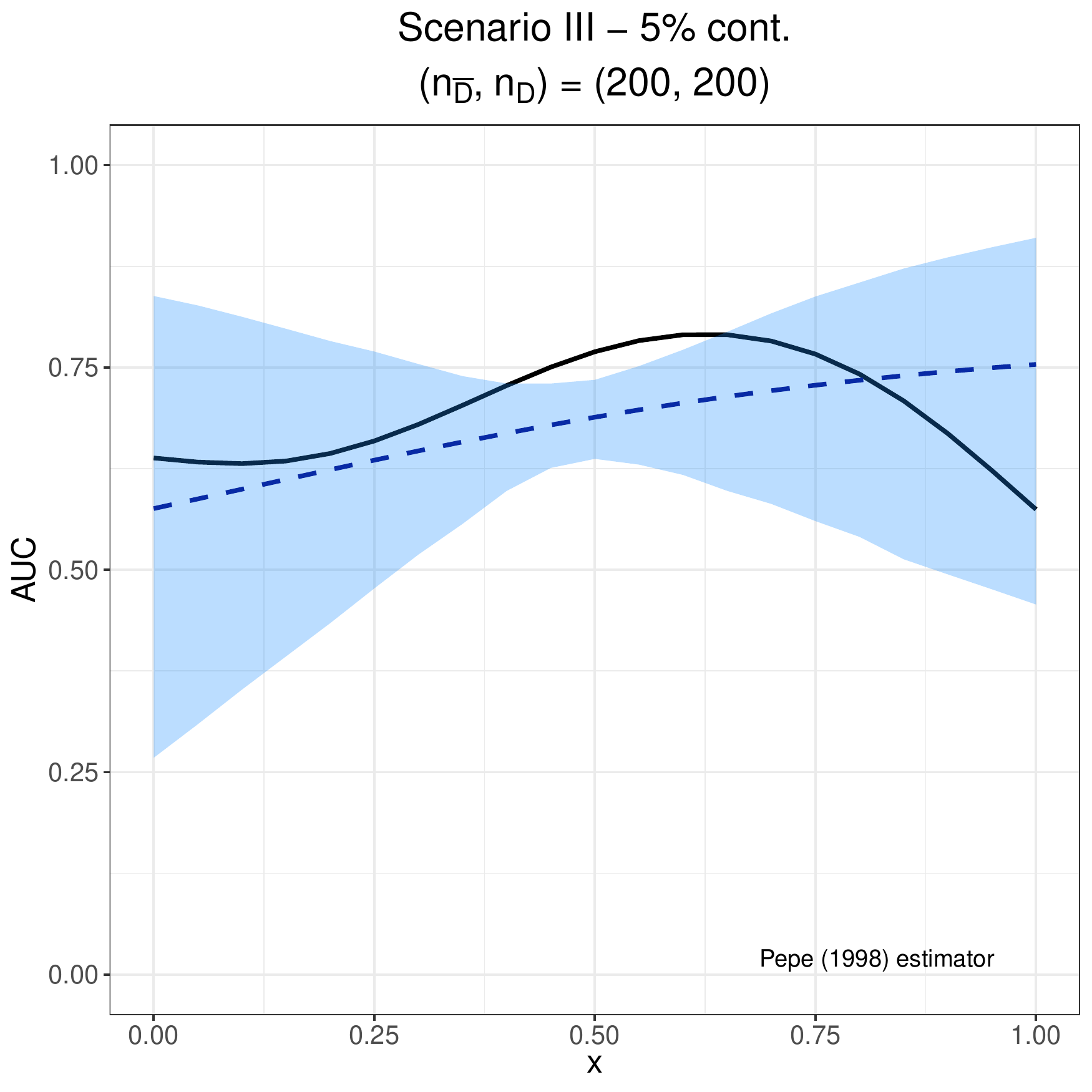}
			\includegraphics[width = 4.65cm]{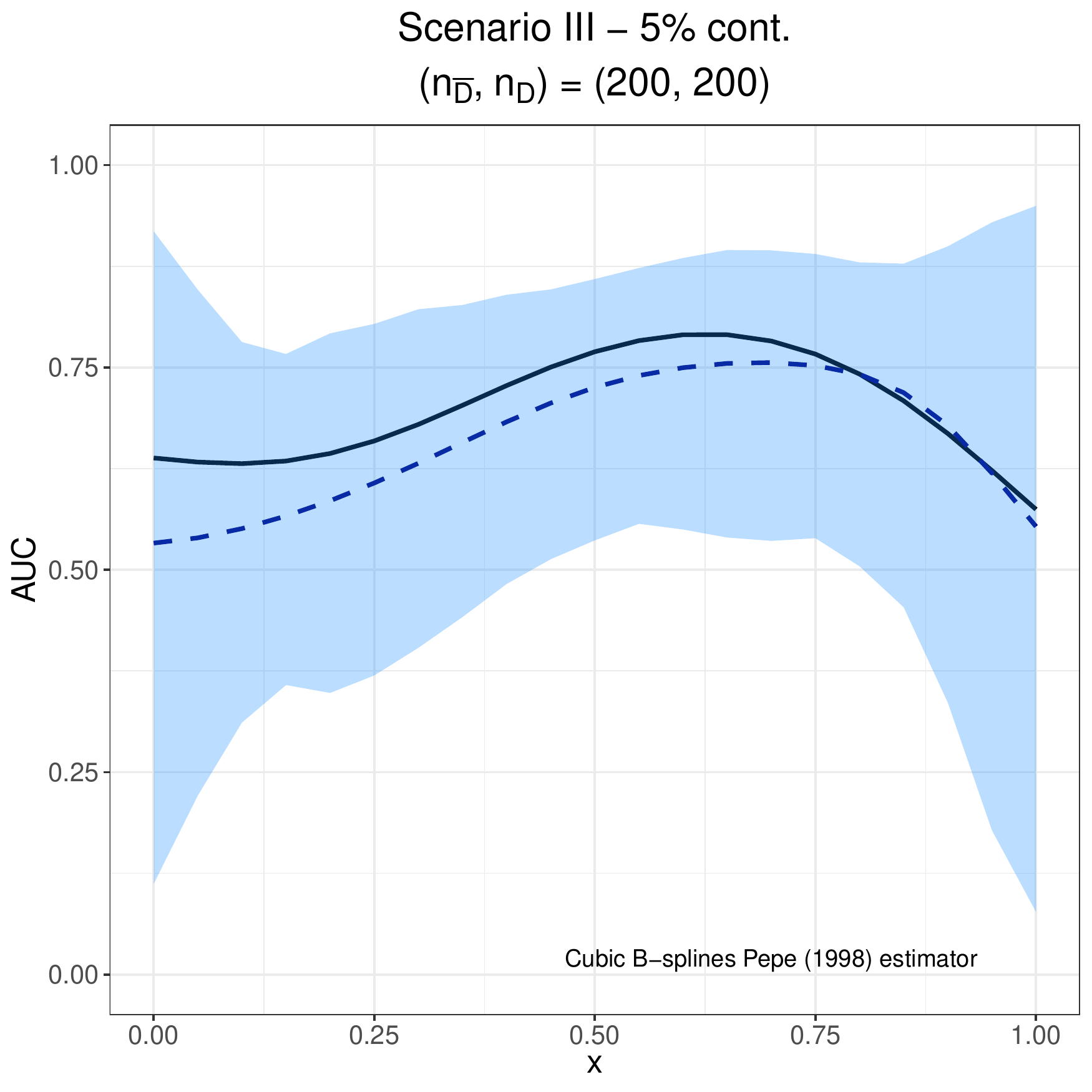}
			\includegraphics[width = 4.65cm]{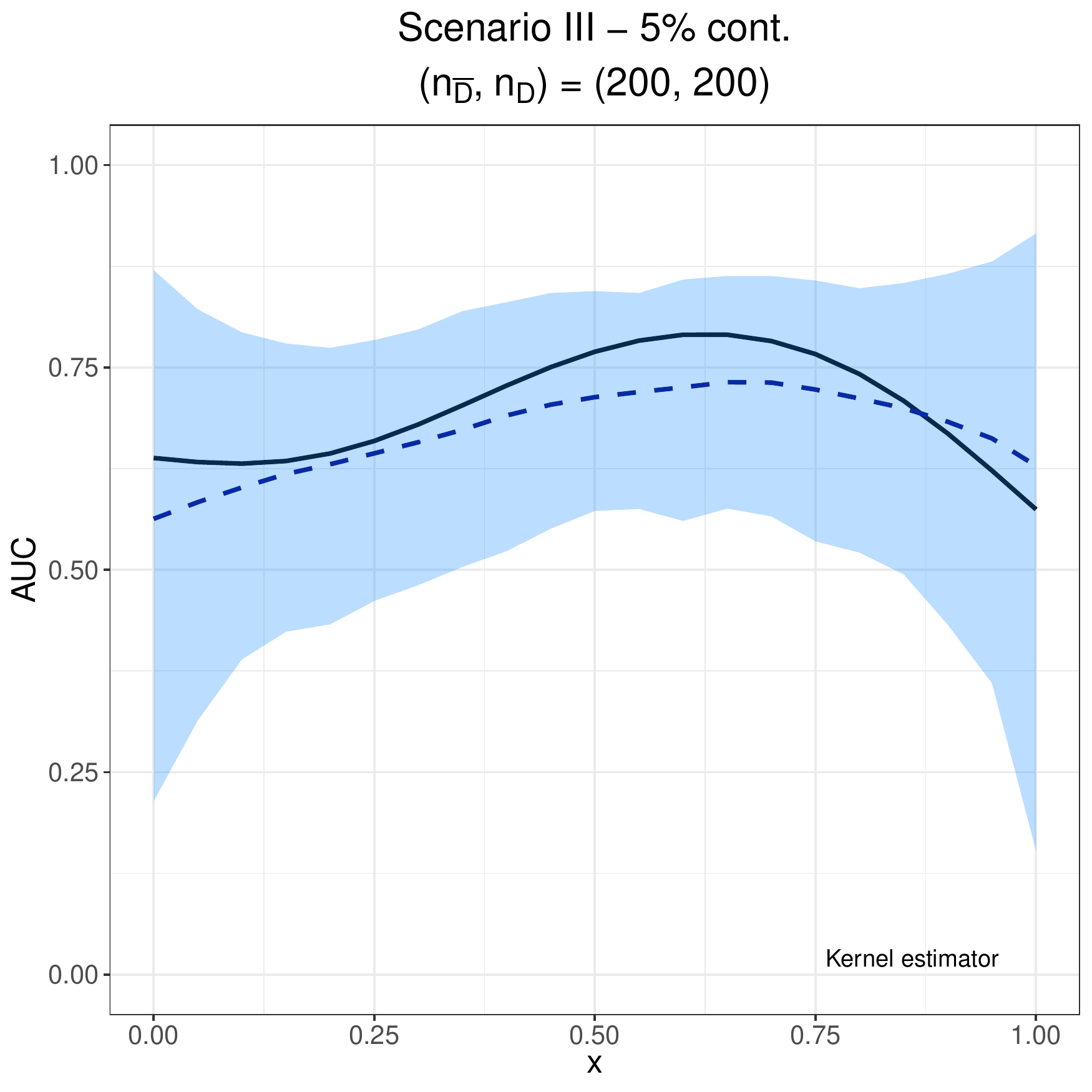}
		}
	\end{center}
	\caption{\footnotesize{Scenario III. True covariate-specific AUC (solid line) versus the mean of the Monte Carlo estimates (dashed line) along with the $2.5\%$ and $97.5\%$ simulation quantiles (shaded area) for the case of $5\%$ of contamination. The first row displays the results for $(n_{\bar{D}}, n_D)=(100,100)$, the second row for $(n_{\bar{D}}, n_D)=(200,100)$, and the third row for $(n_{\bar{D}}, n_D)=(200,200)$. The first column corresponds to our flexible and robust estimator, the second column to the estimator proposed by Pepe (1998), the third one to the cubic B-splines extension of Pepe (1998), and the fourth column to the kernel estimator.}}
\end{figure}

\begin{figure}[H]
	\begin{center}
		\subfigure{
			\includegraphics[width = 4.65cm]{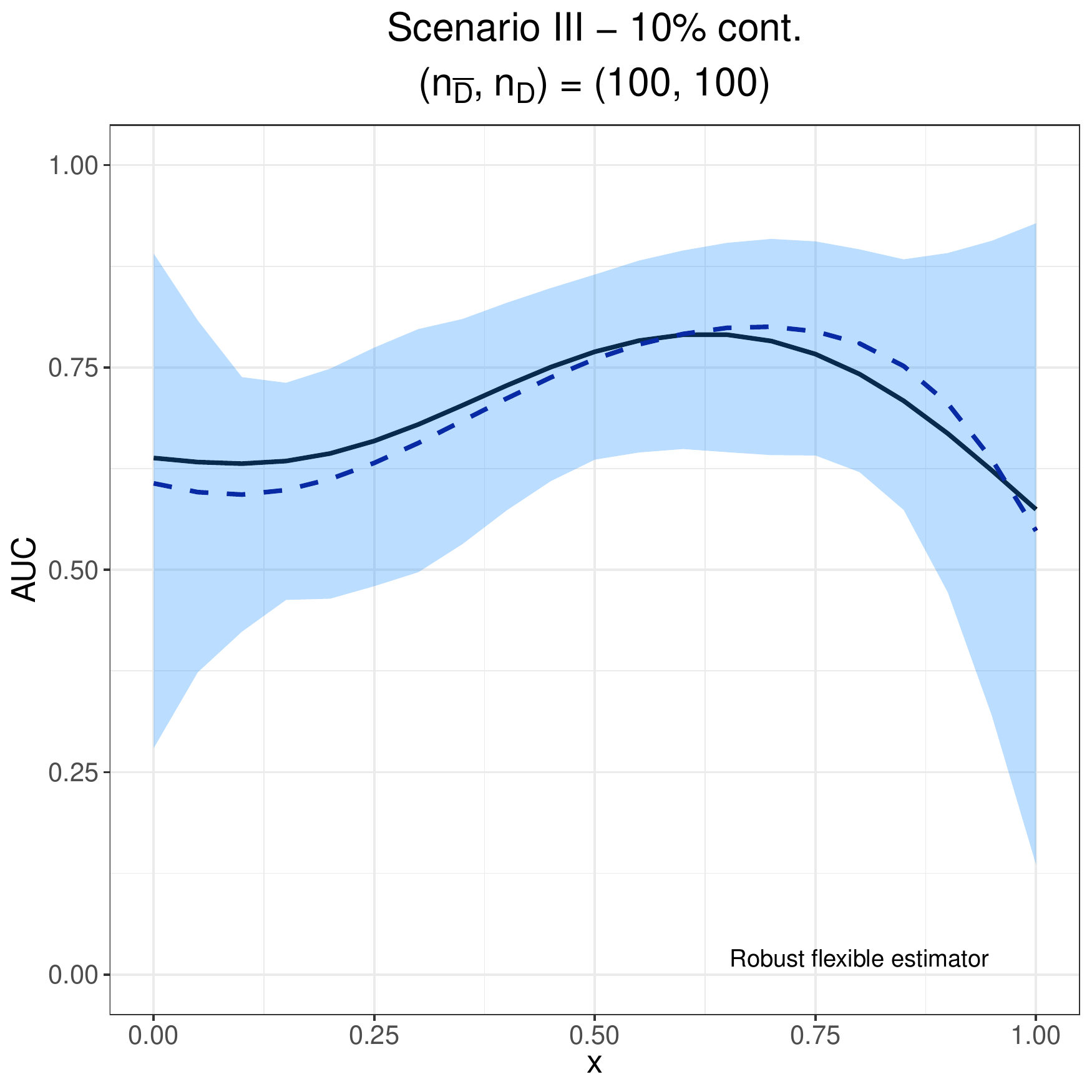}
			\includegraphics[width = 4.65cm]{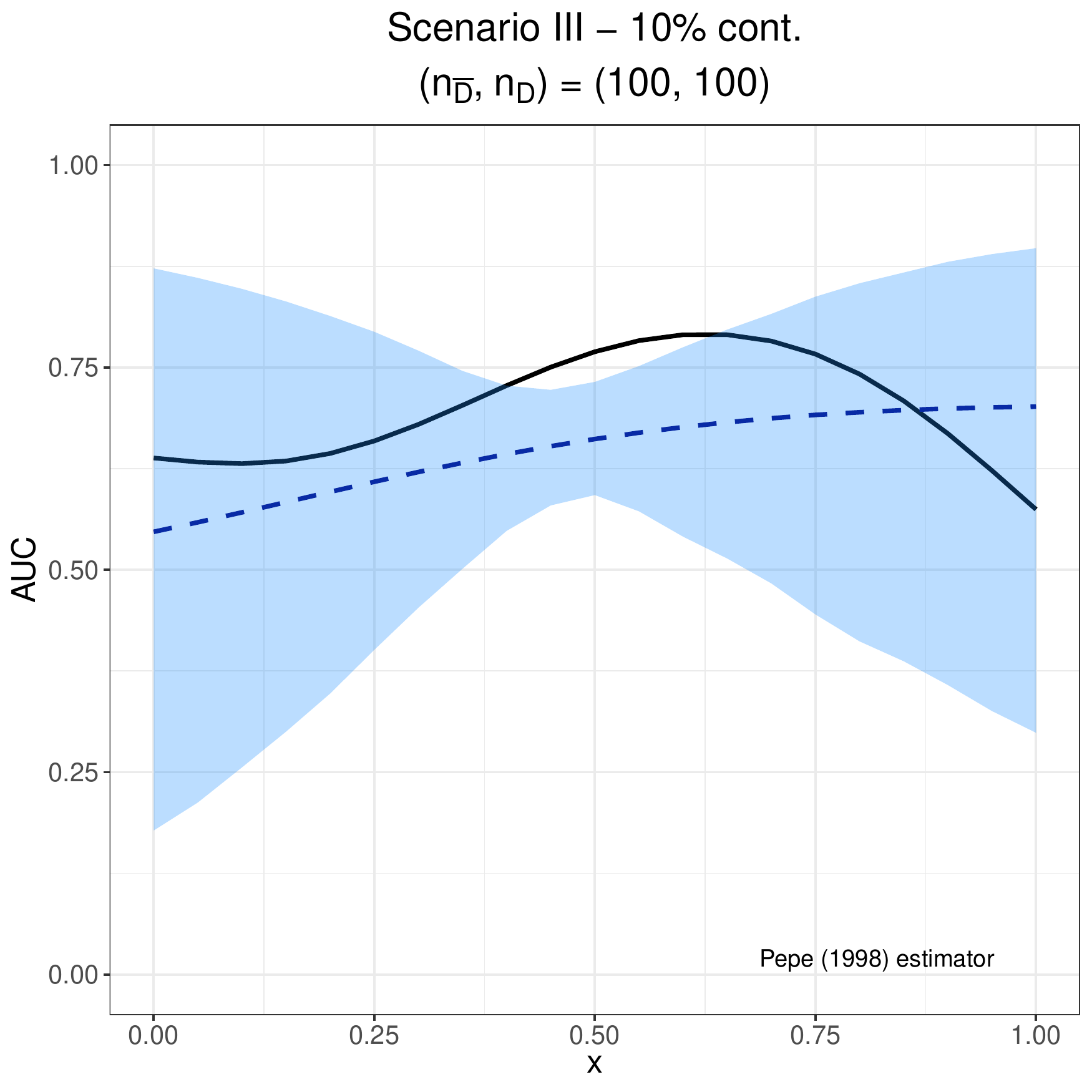}
			\includegraphics[width = 4.65cm]{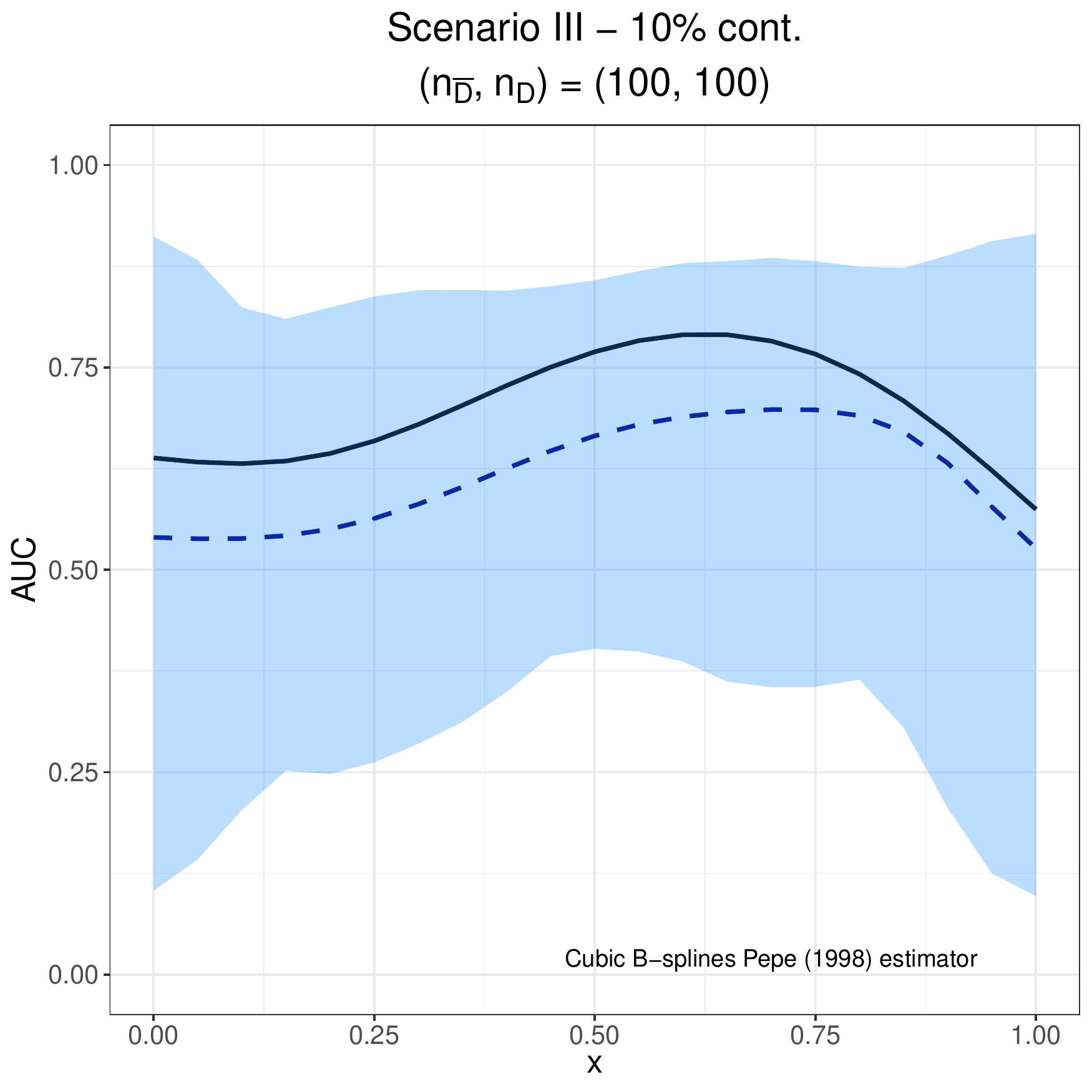}
			\includegraphics[width = 4.65cm]{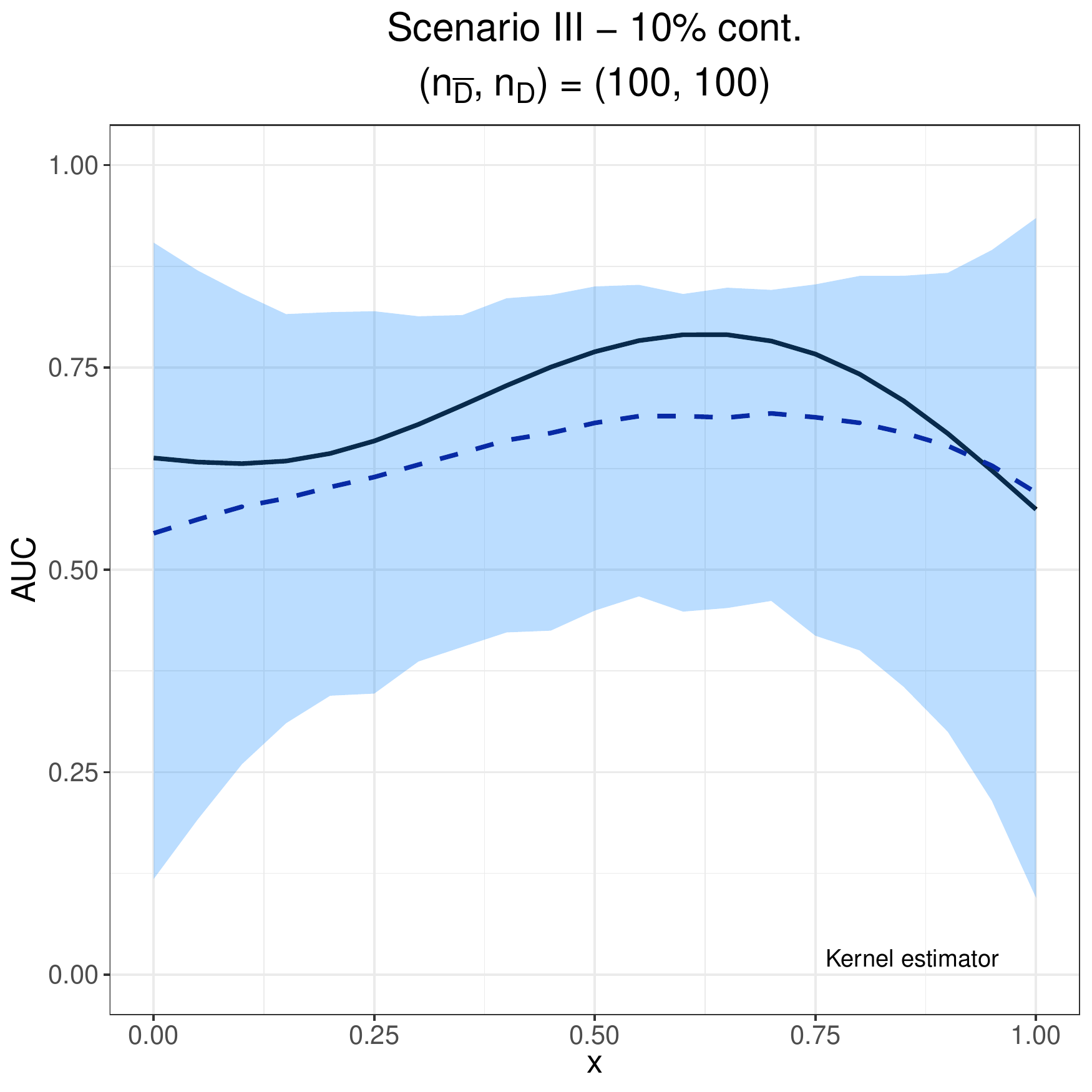}
		}
		\vspace{0.3cm}
		\subfigure{
			\includegraphics[width = 4.65cm]{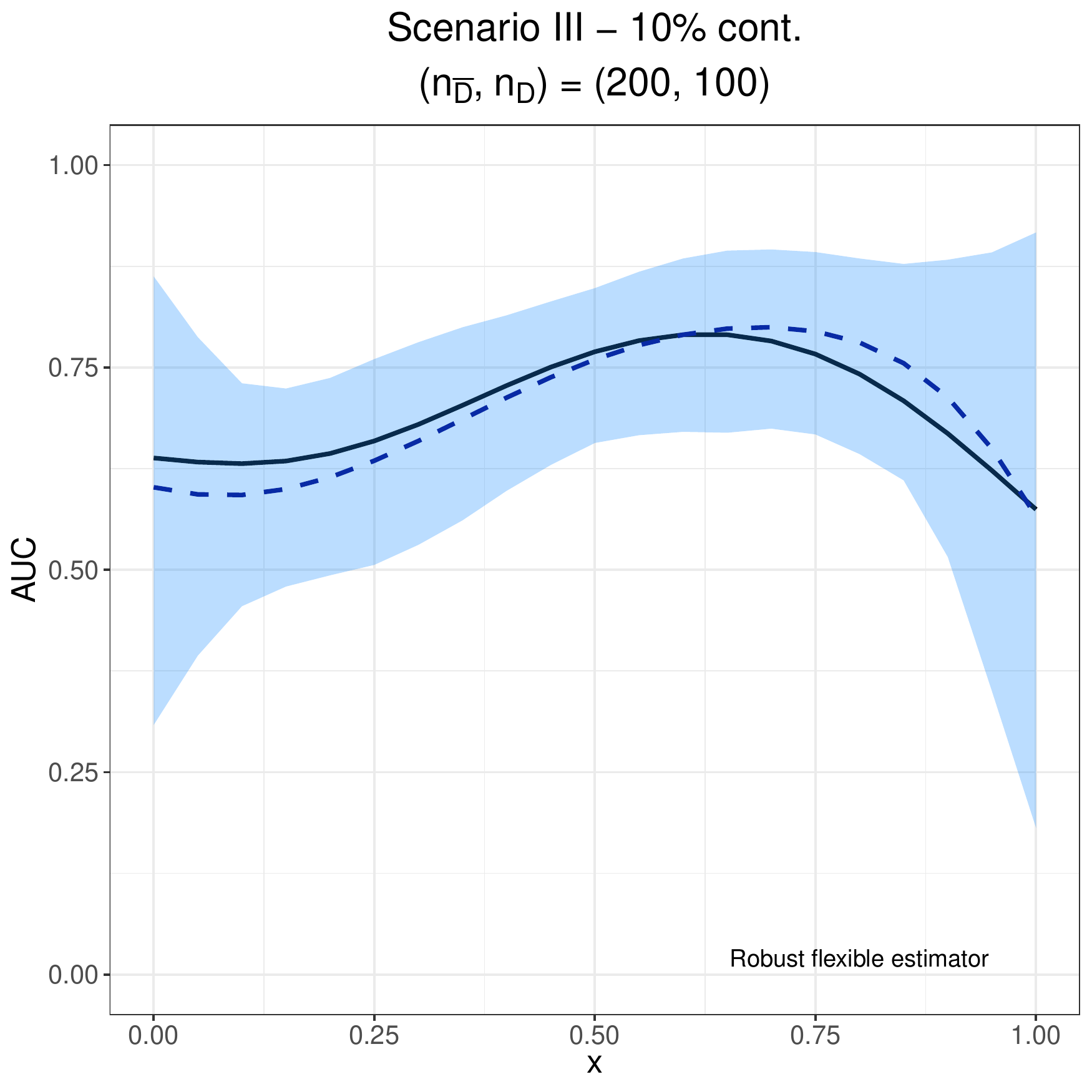}
			\includegraphics[width = 4.65cm]{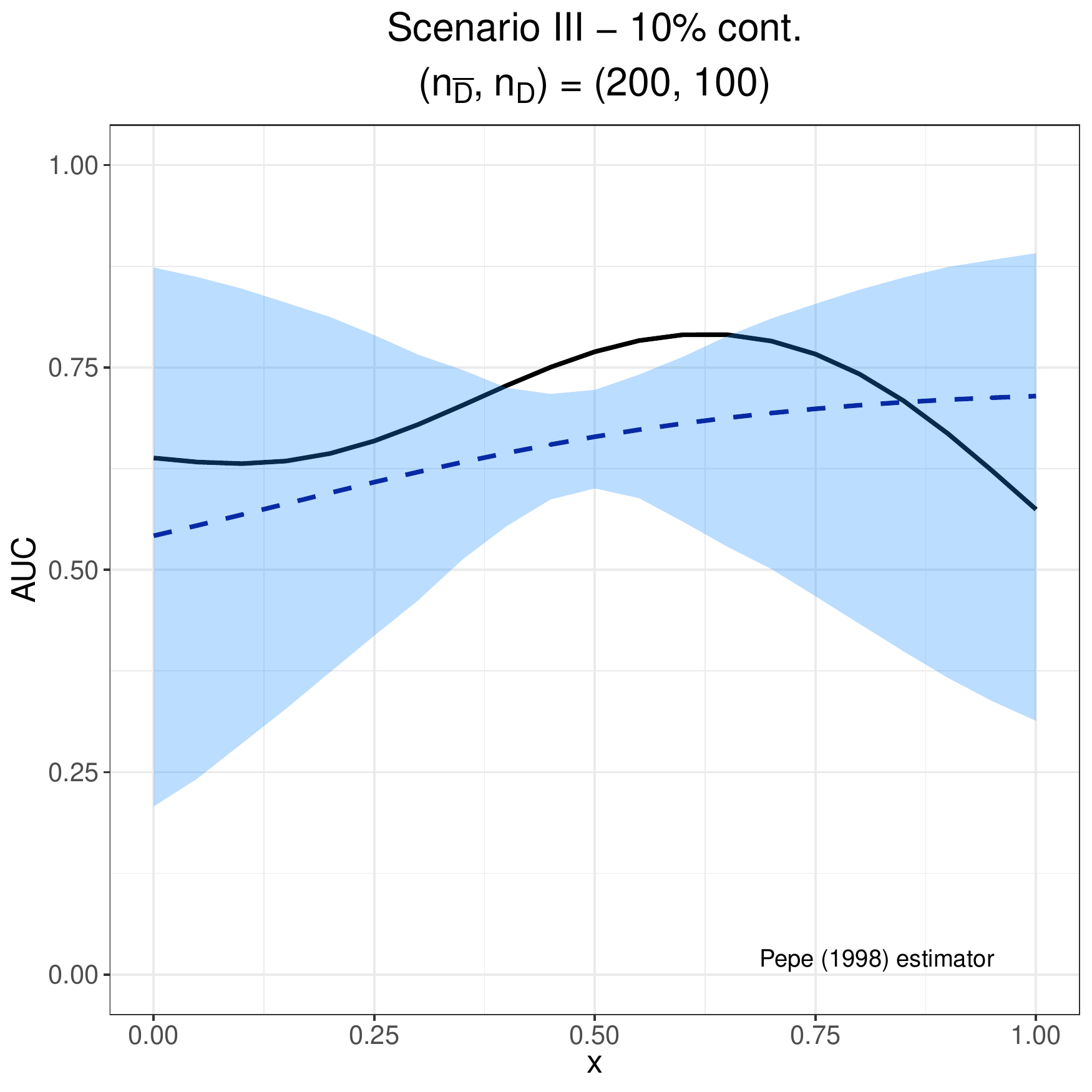}
			\includegraphics[width = 4.65cm]{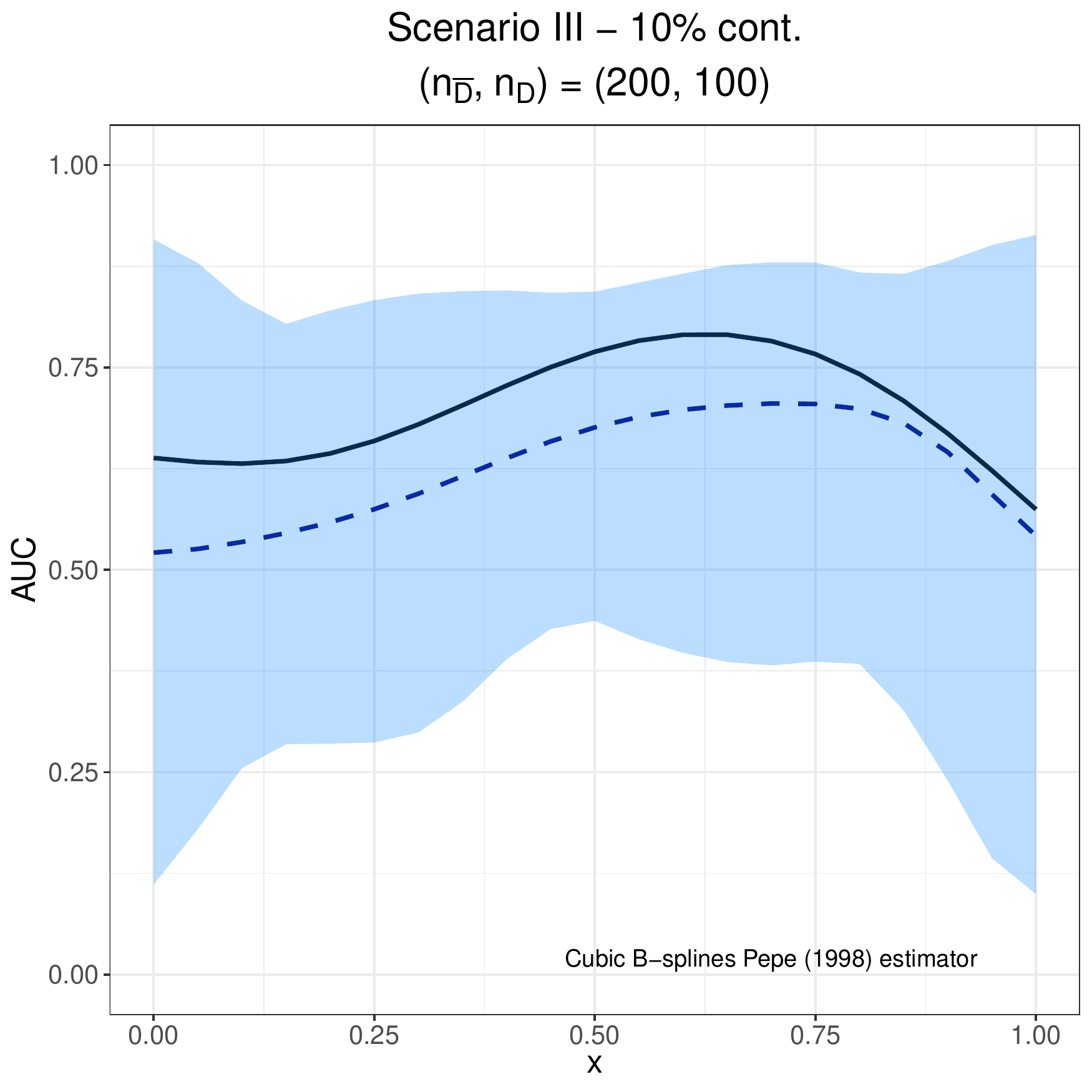}
			\includegraphics[width = 4.65cm]{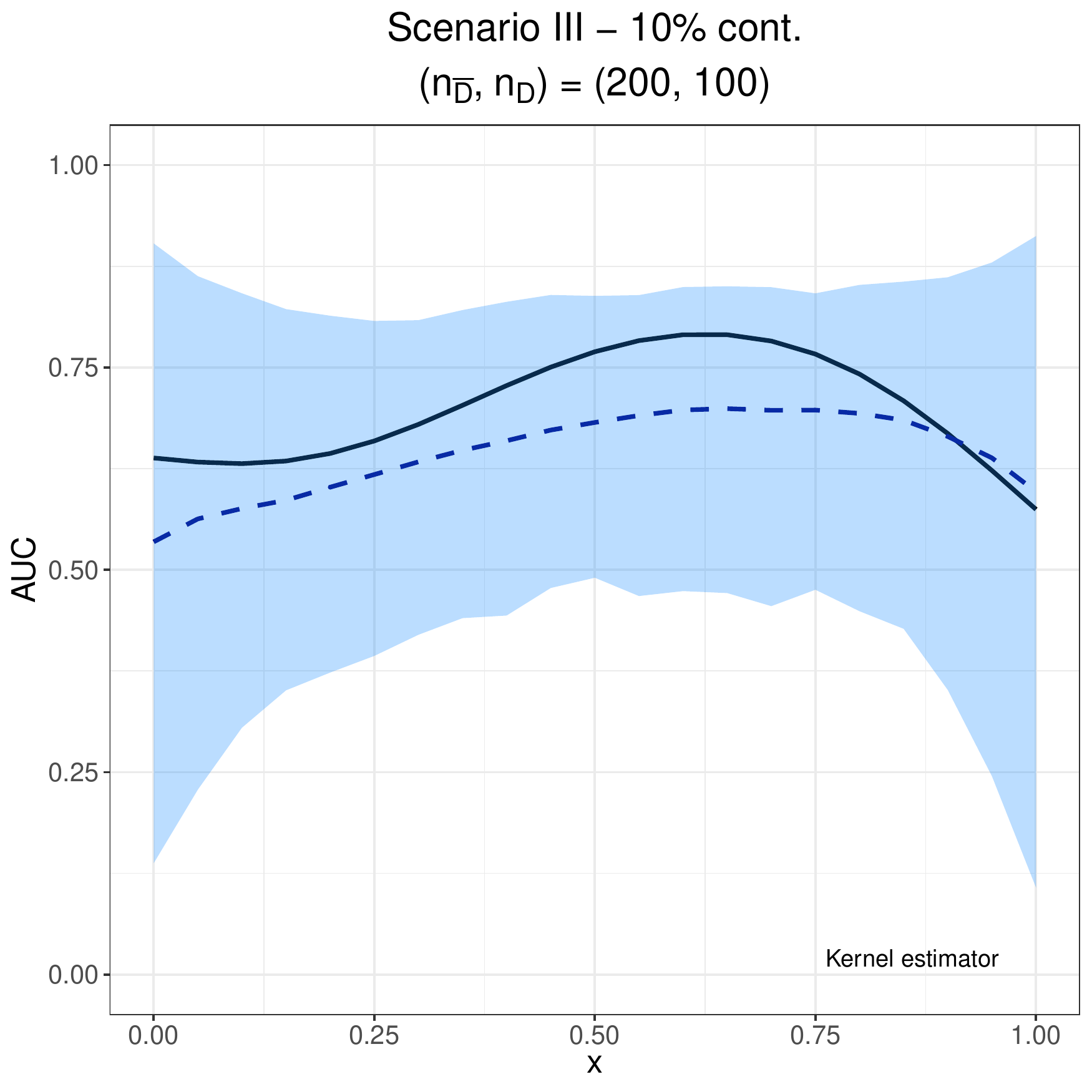}
		}
		\vspace{0.3cm}
		\subfigure{
			\includegraphics[width = 4.65cm]{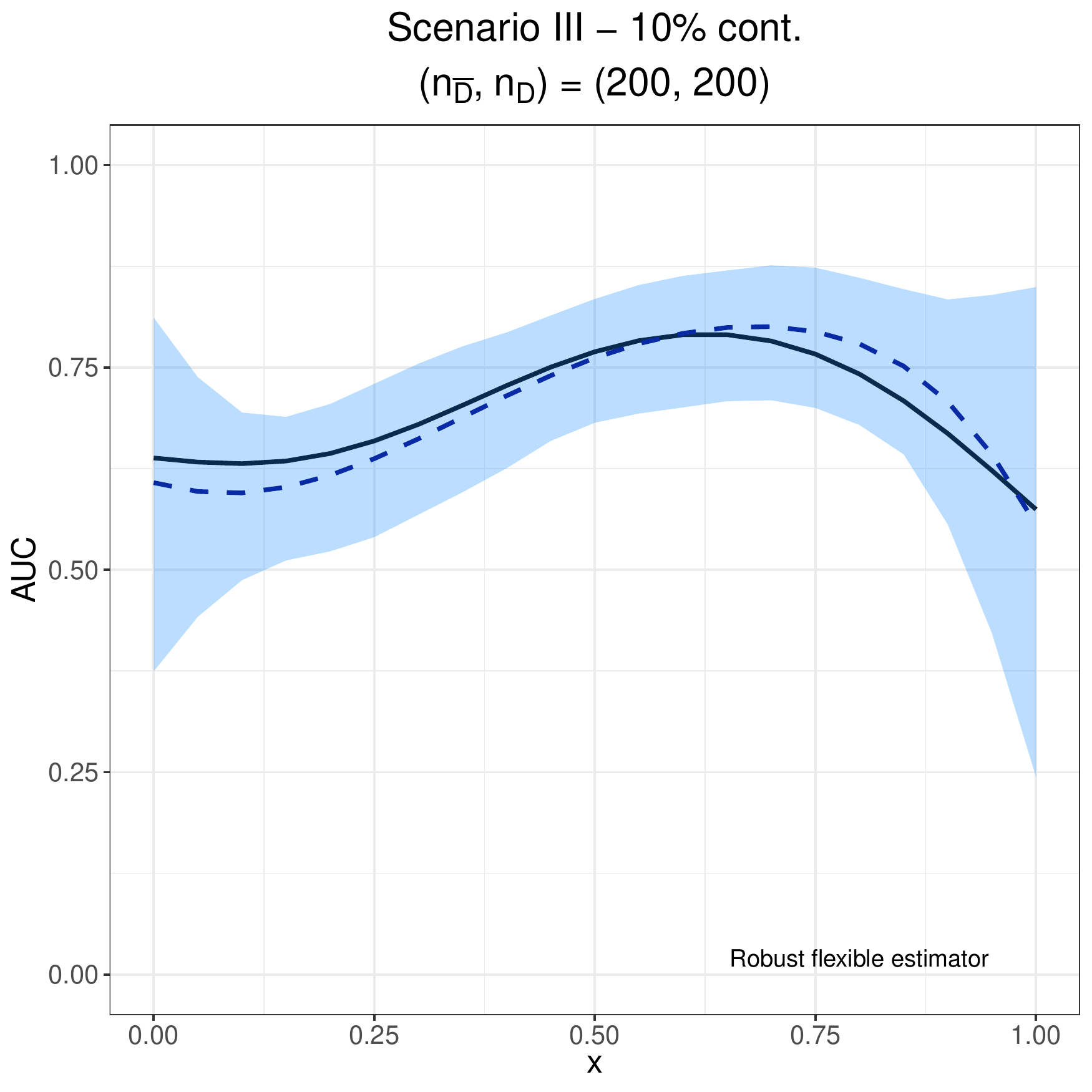}
			\includegraphics[width = 4.65cm]{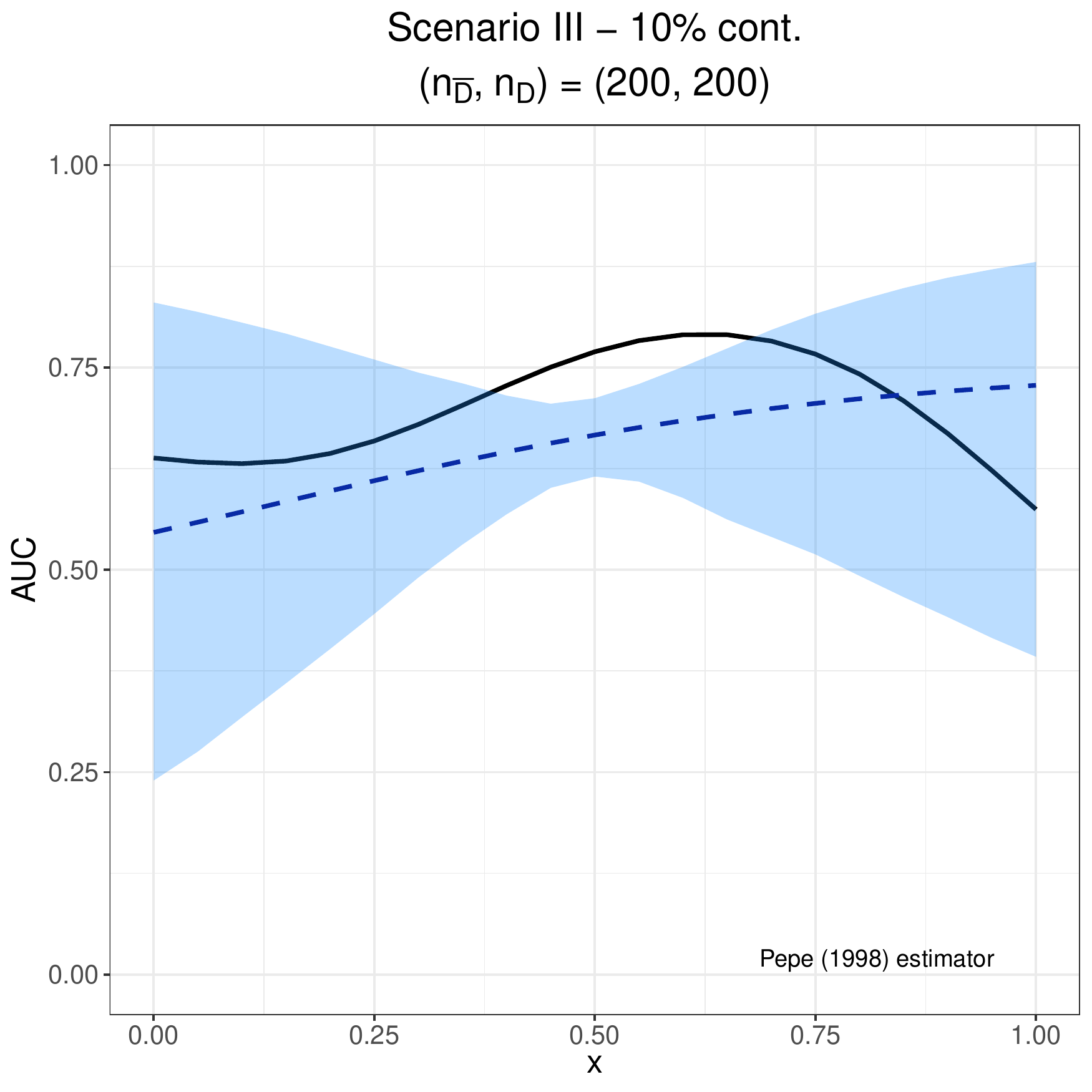}
			\includegraphics[width = 4.65cm]{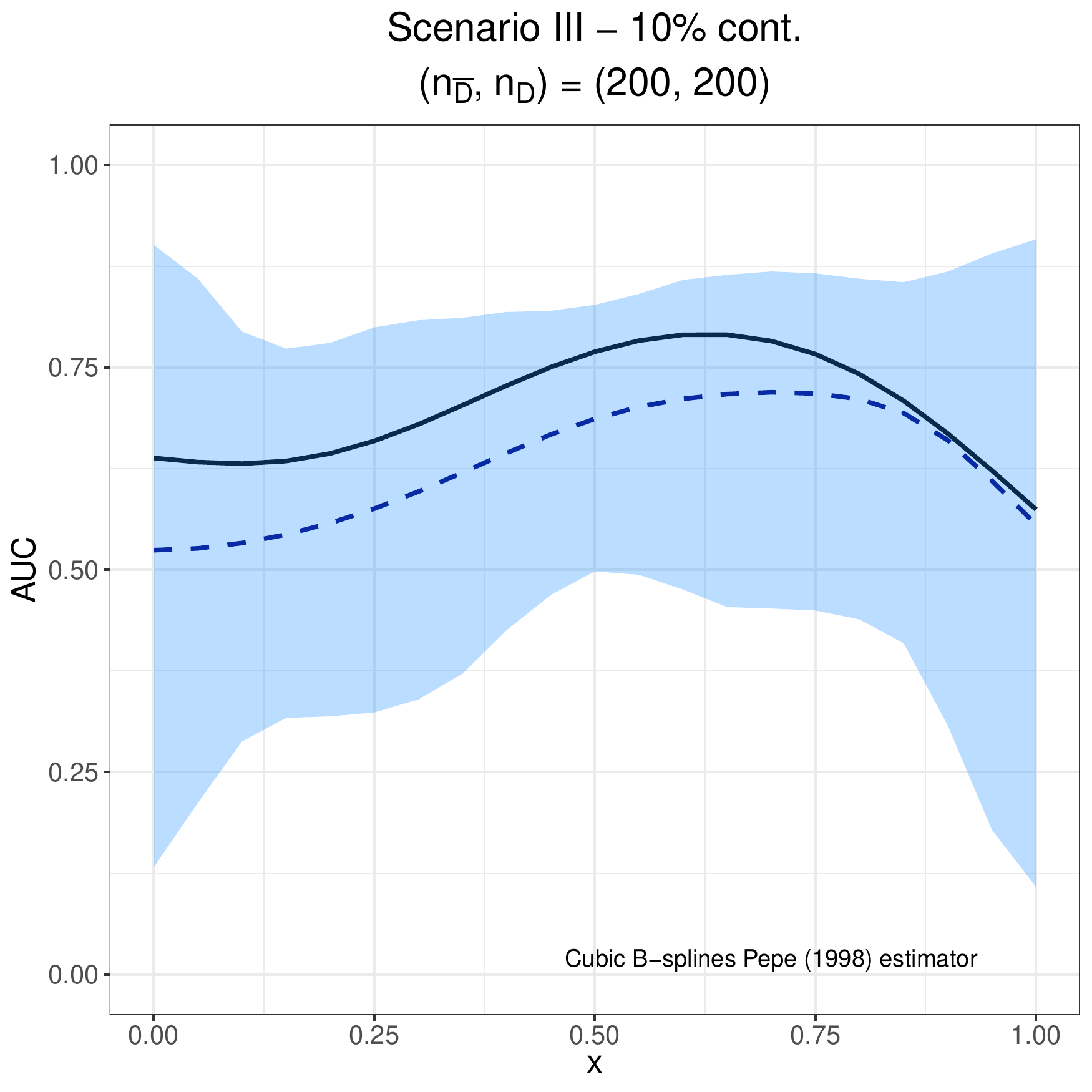}
			\includegraphics[width = 4.65cm]{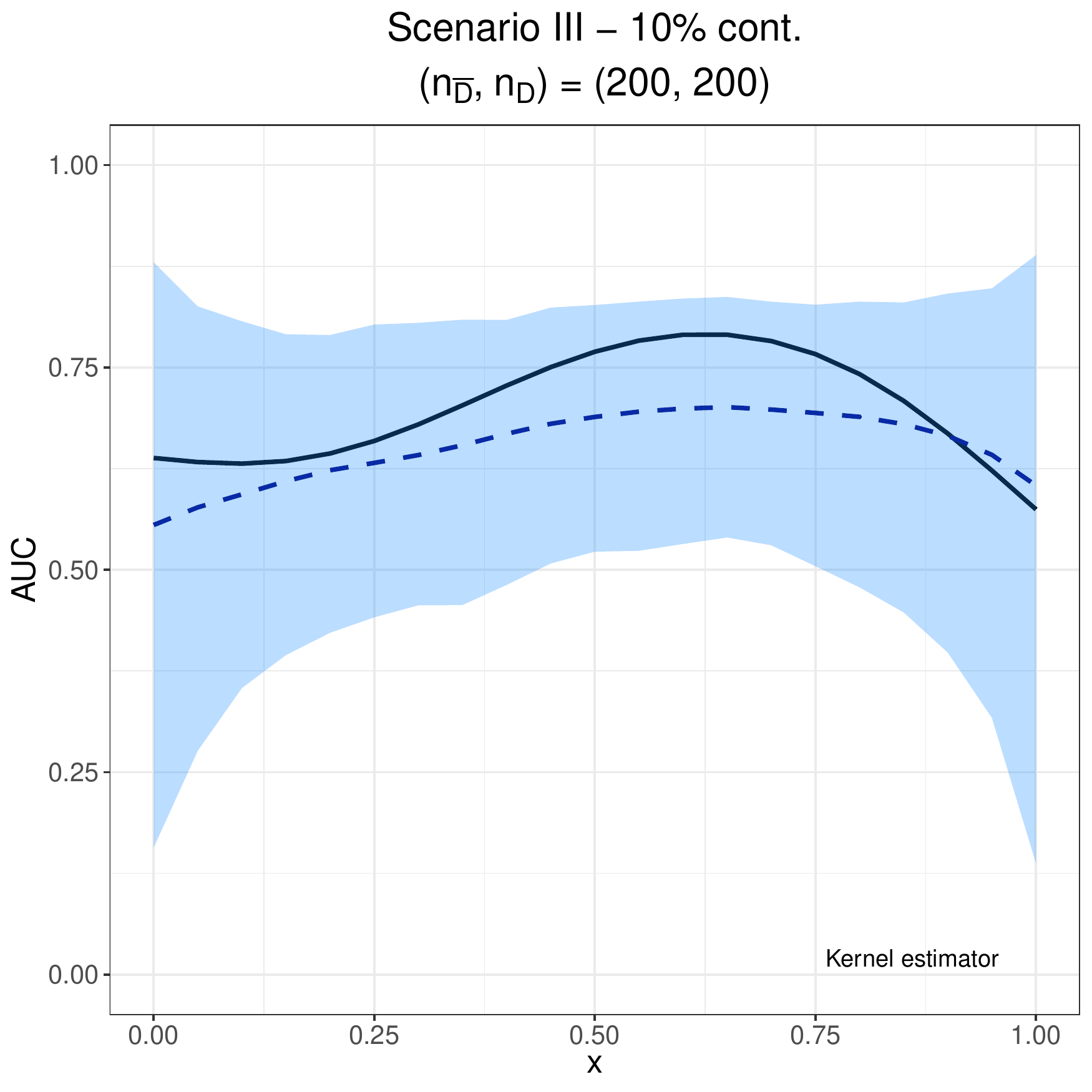}
		}
	\end{center}
	\caption{\footnotesize{Scenario III. True covariate-specific AUC (solid line) versus the mean of the Monte Carlo estimates (dashed line) along with the $2.5\%$ and $97.5\%$ simulation quantiles (shaded area) for the case of $10\%$ of contamination. The first row displays the results for $(n_{\bar{D}}, n_D)=(100,100)$, the second row for $(n_{\bar{D}}, n_D)=(200,100)$, and the third row for $(n_{\bar{D}}, n_D)=(200,200)$. The first column corresponds to our flexible and robust estimator, the second column to the estimator proposed by Pepe (1998), the third one to the cubic B-splines extension of Pepe (1998), and the fourth column to the kernel estimator.}}
\end{figure}

\begin{figure}[H]
	\begin{center}
		\subfigure{
			\includegraphics[width = 3.35cm]{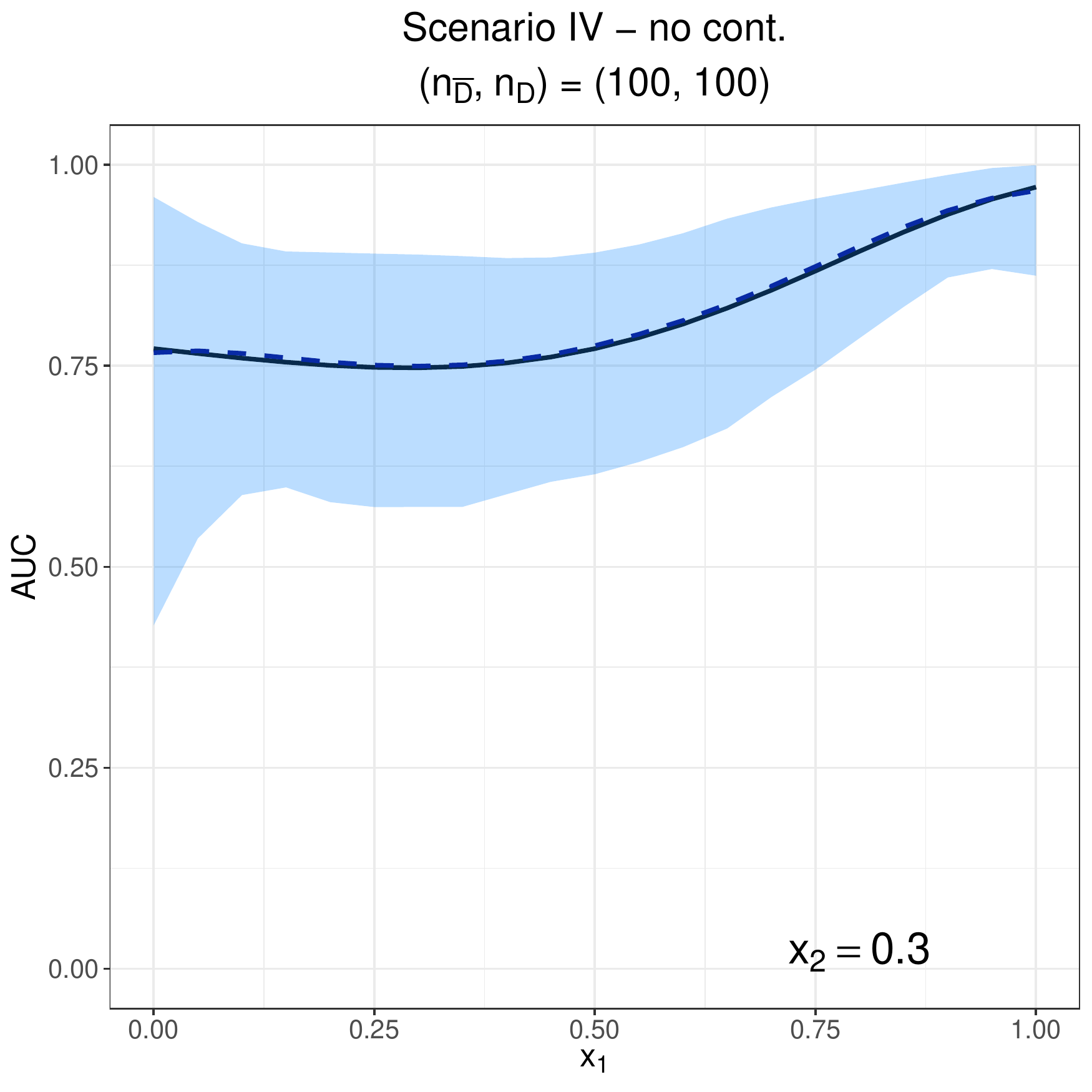}
			\includegraphics[width = 3.35cm]{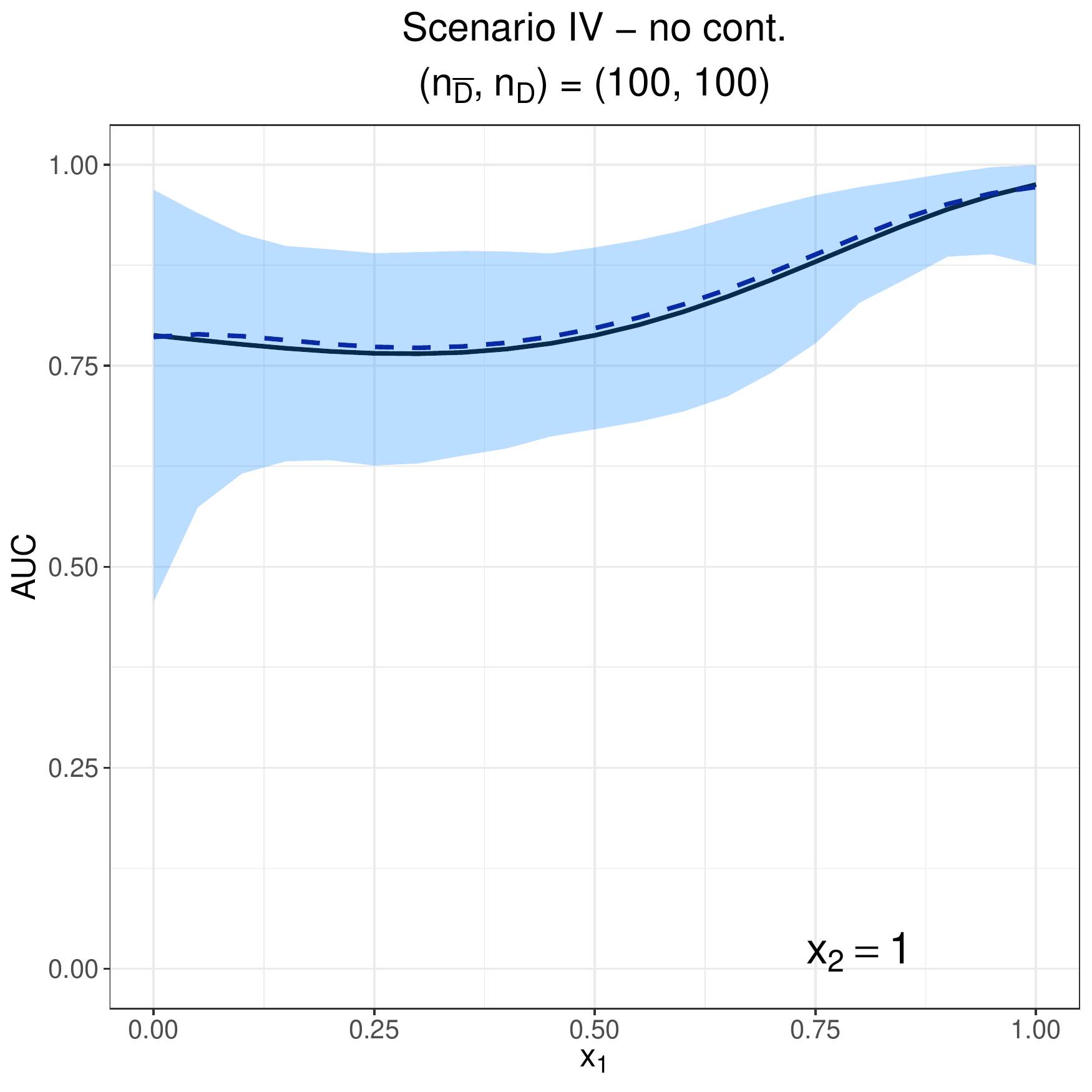}
			\includegraphics[width = 3.35cm]{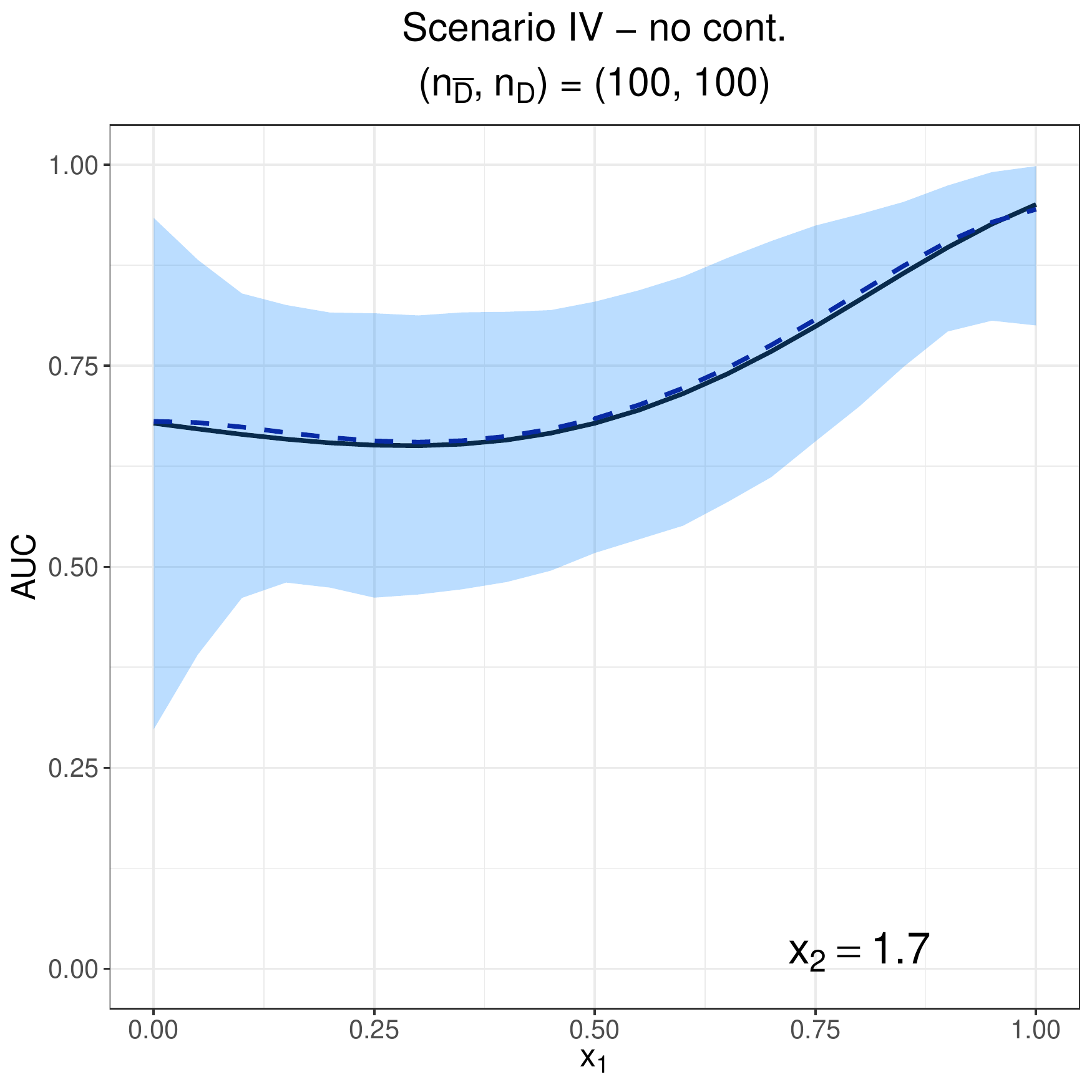}
		}
		\subfigure{
			\includegraphics[width = 3.35cm]{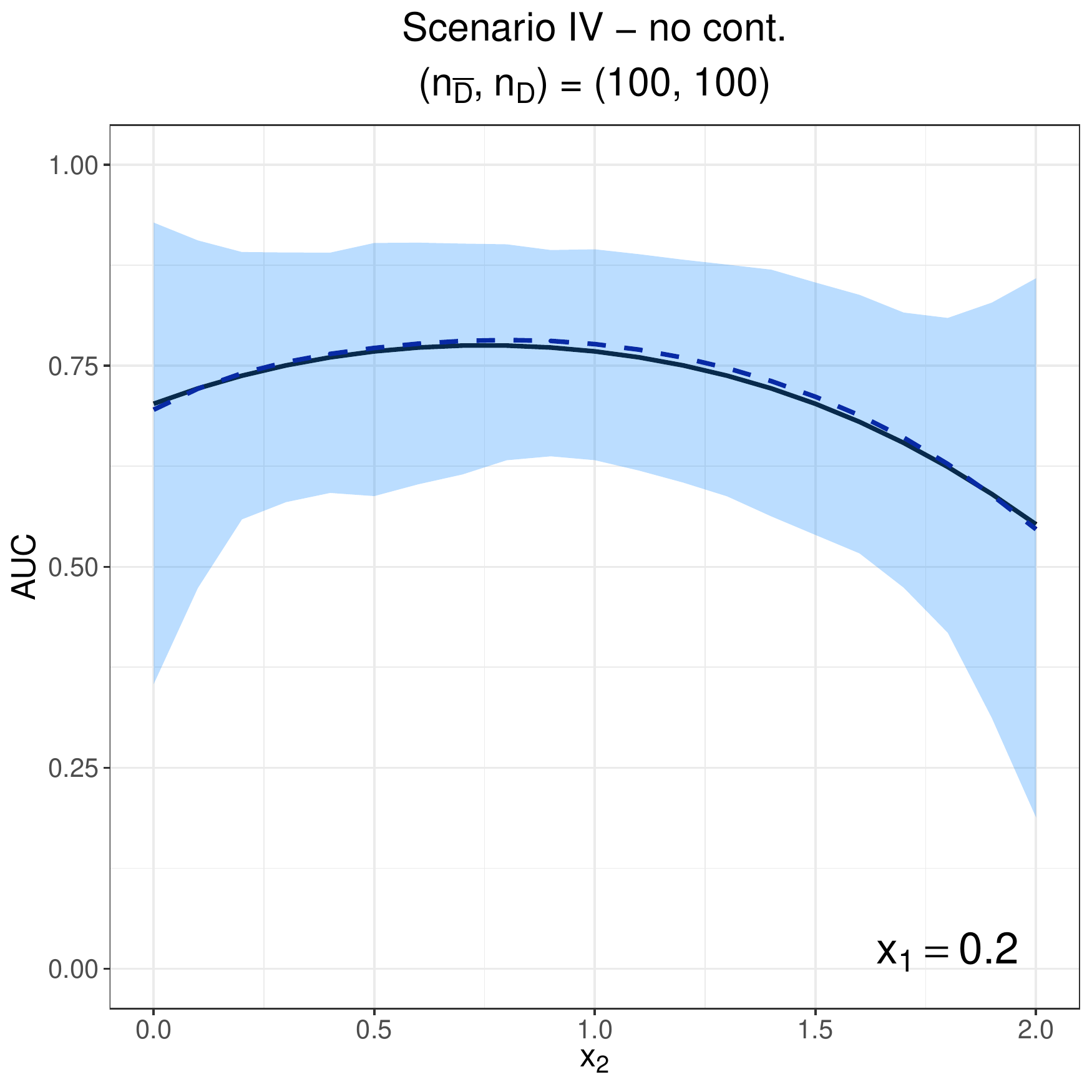}
			\includegraphics[width = 3.35cm]{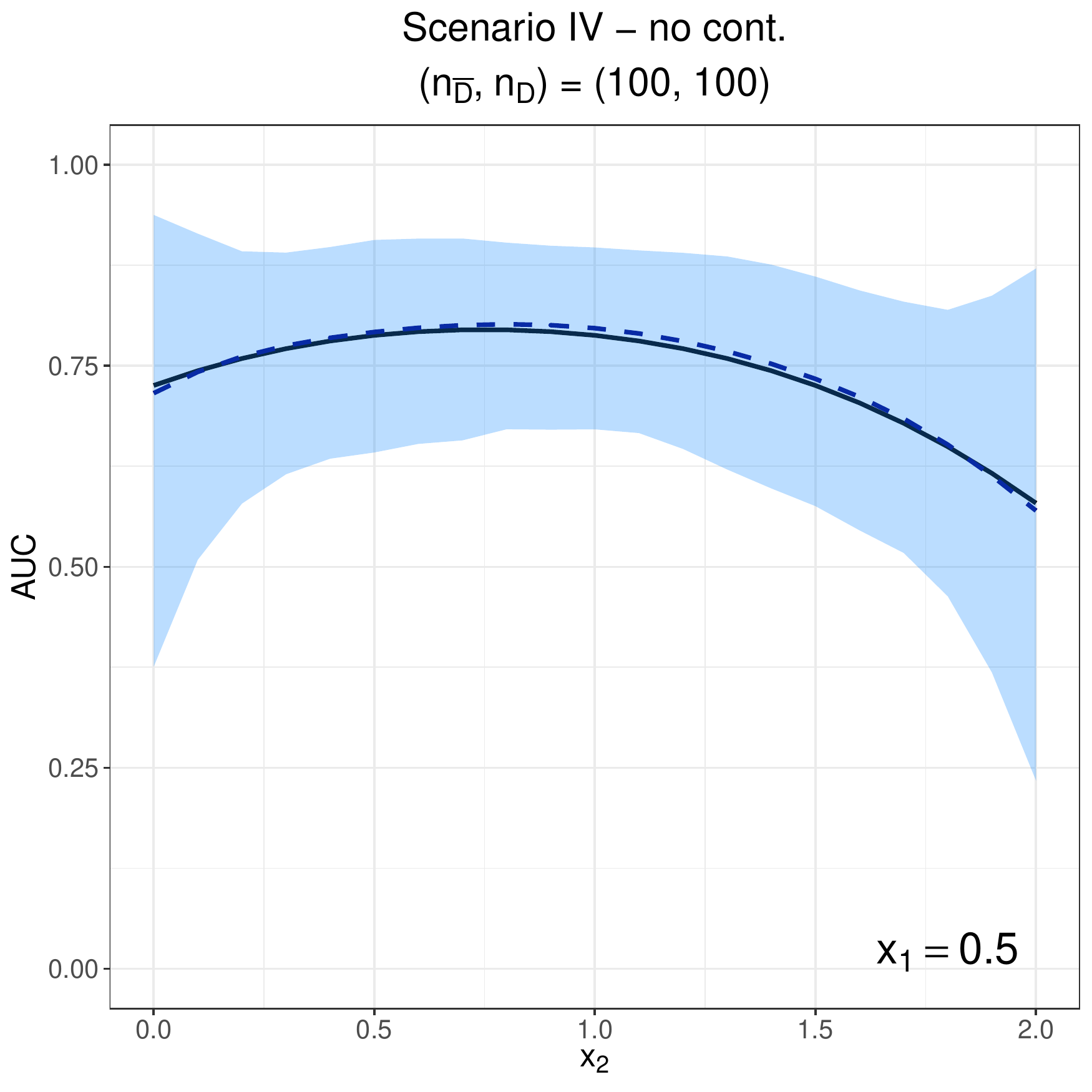}
			\includegraphics[width = 3.35cm]{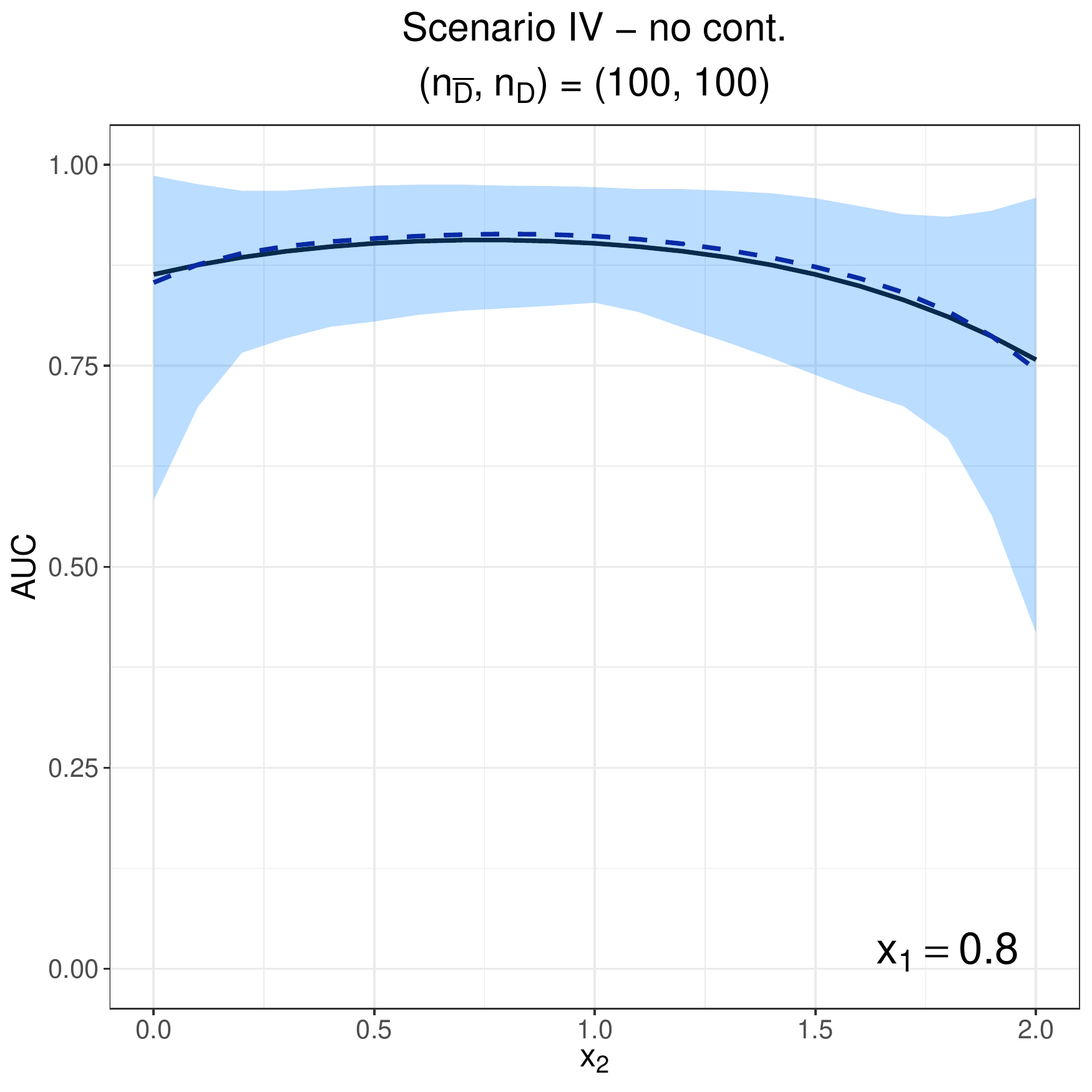}
		}
		\vspace{0.1cm}
		\subfigure{
			\includegraphics[width = 3.35cm]{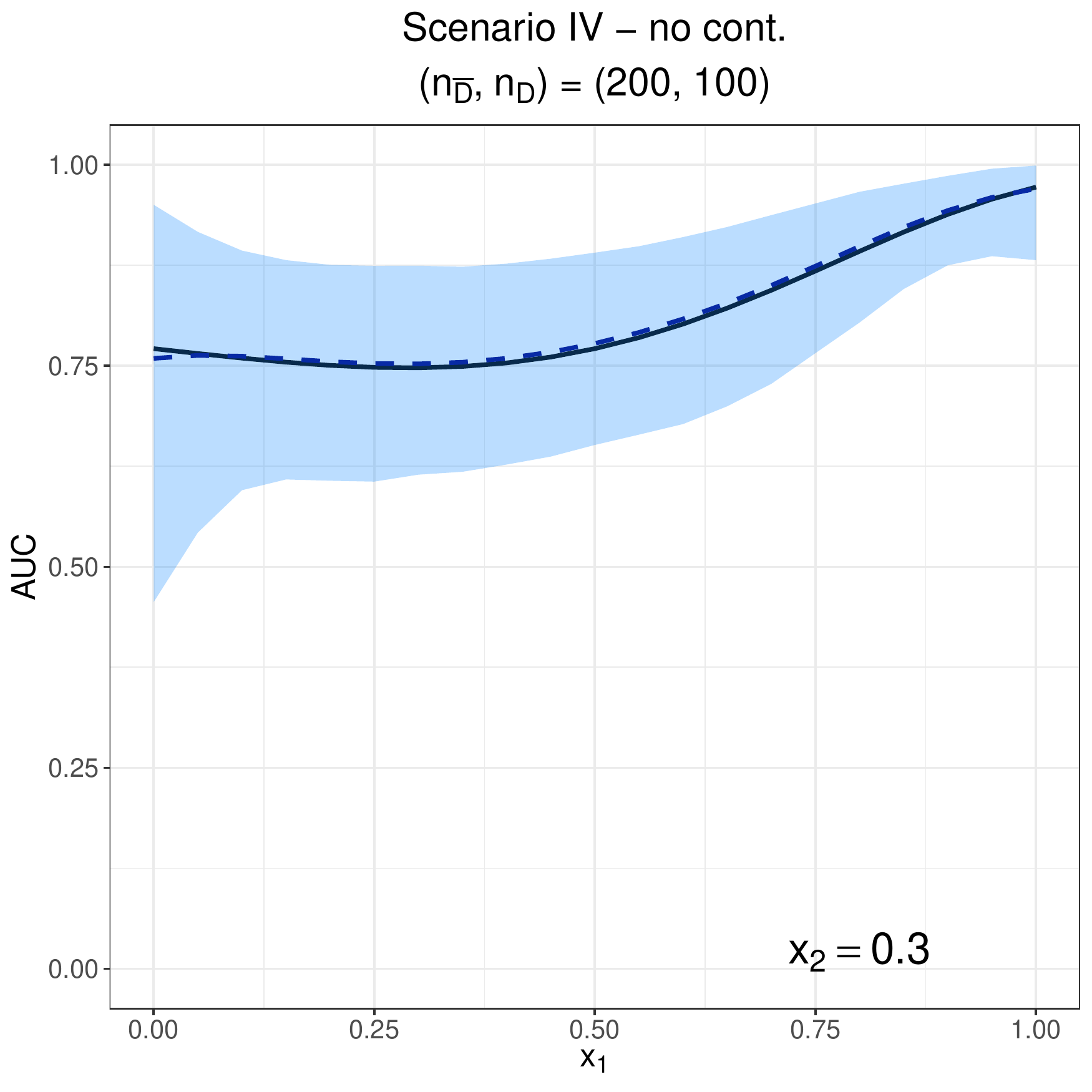}
			\includegraphics[width = 3.35cm]{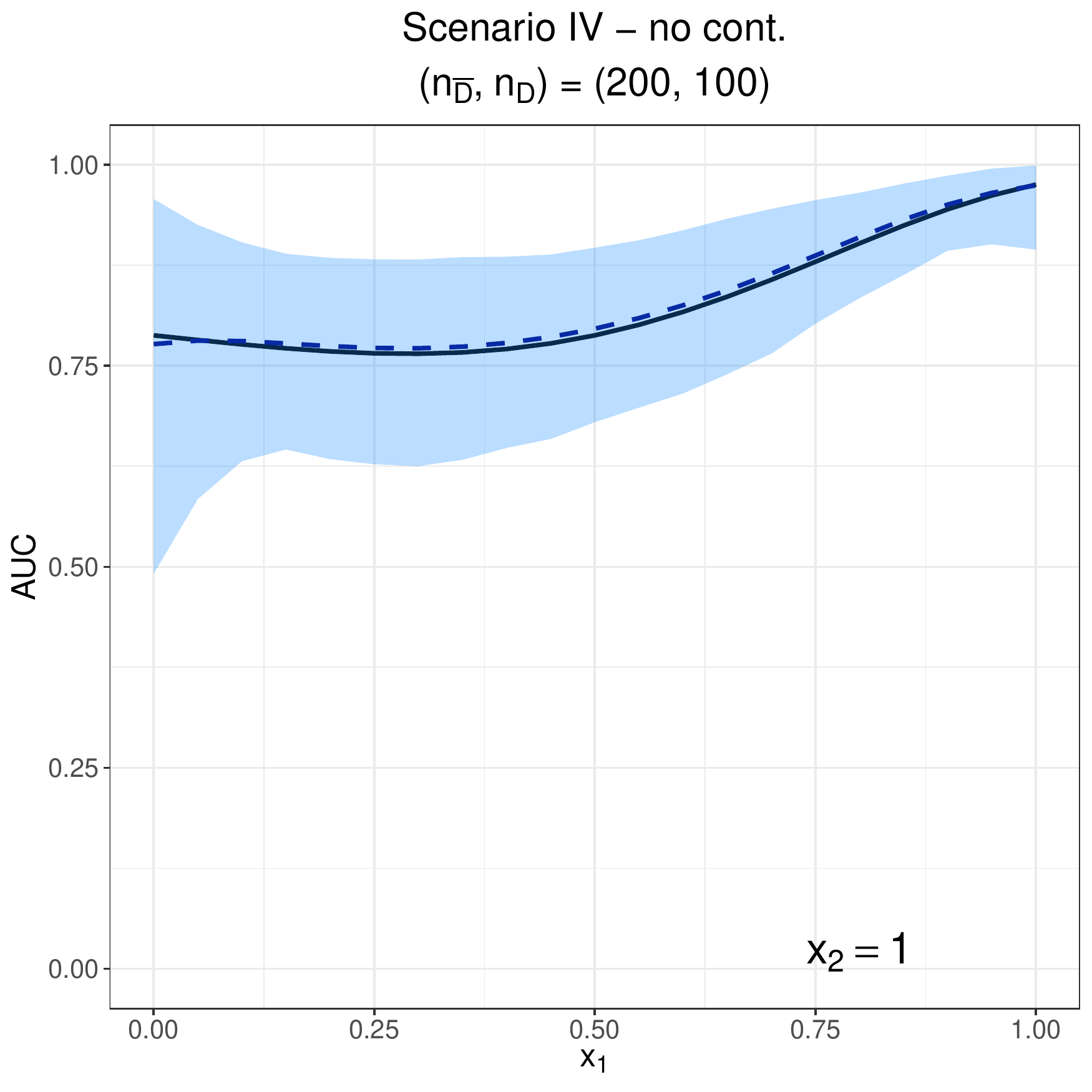}
			\includegraphics[width = 3.35cm]{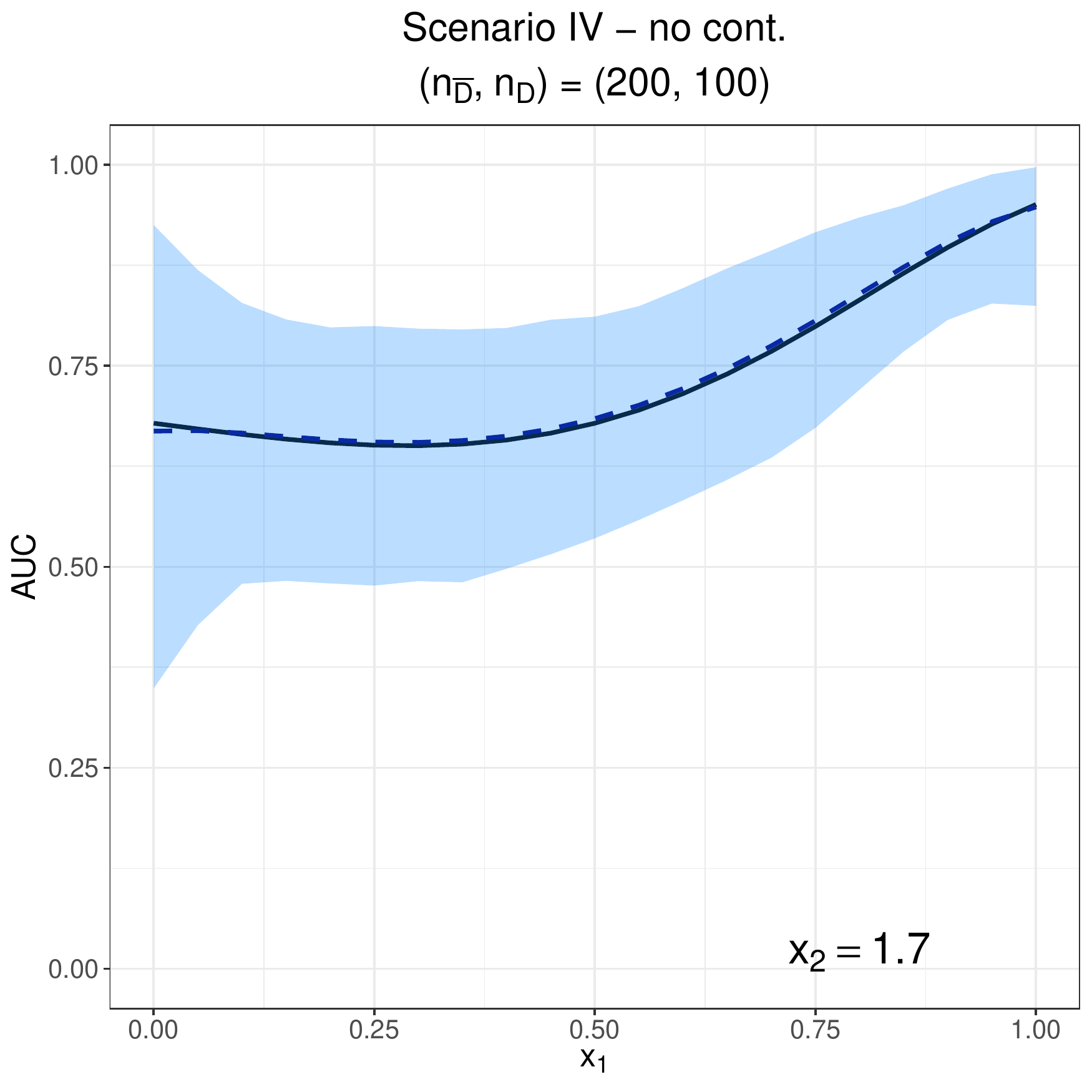}
		}
		\subfigure{
			\includegraphics[width = 3.35cm]{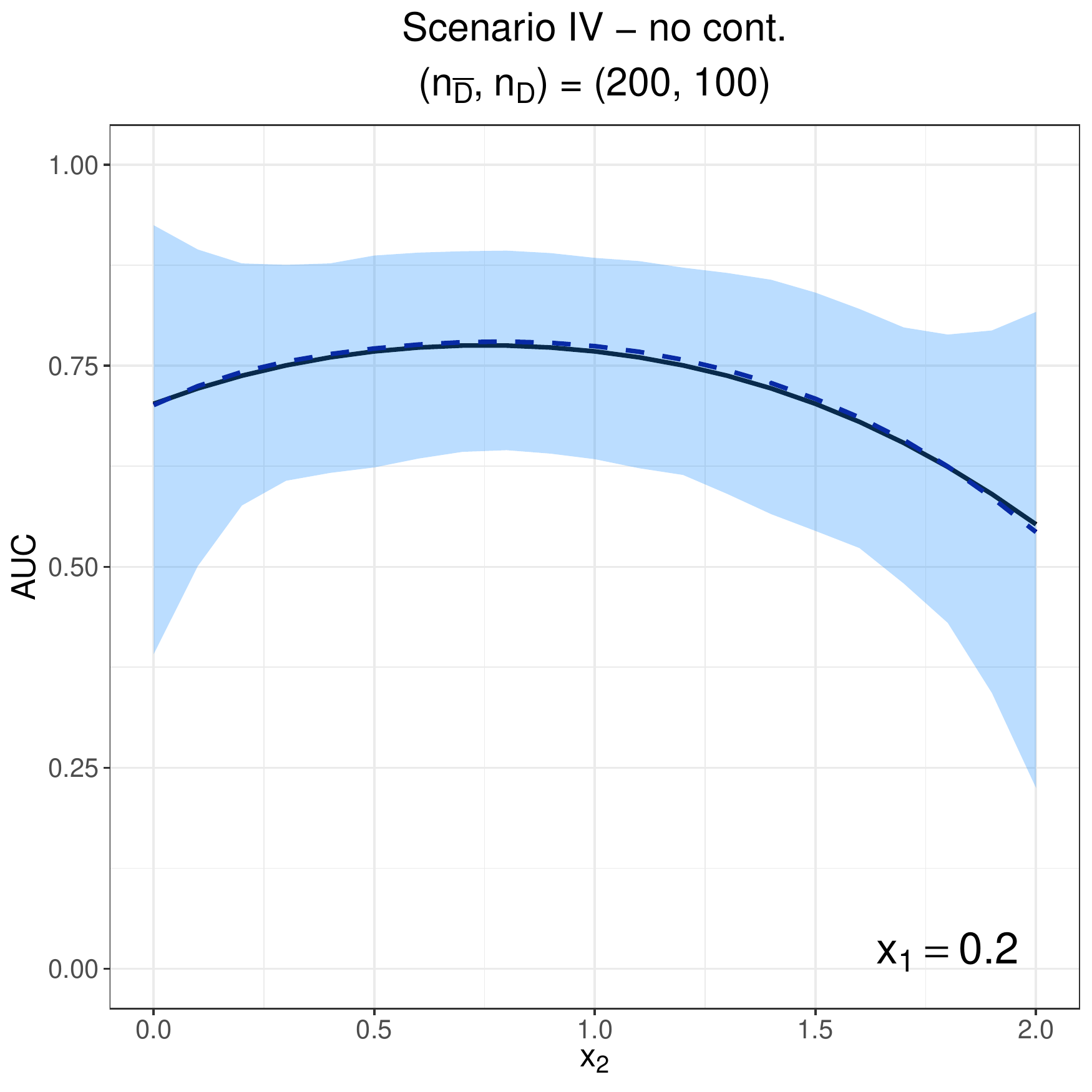}
			\includegraphics[width = 3.35cm]{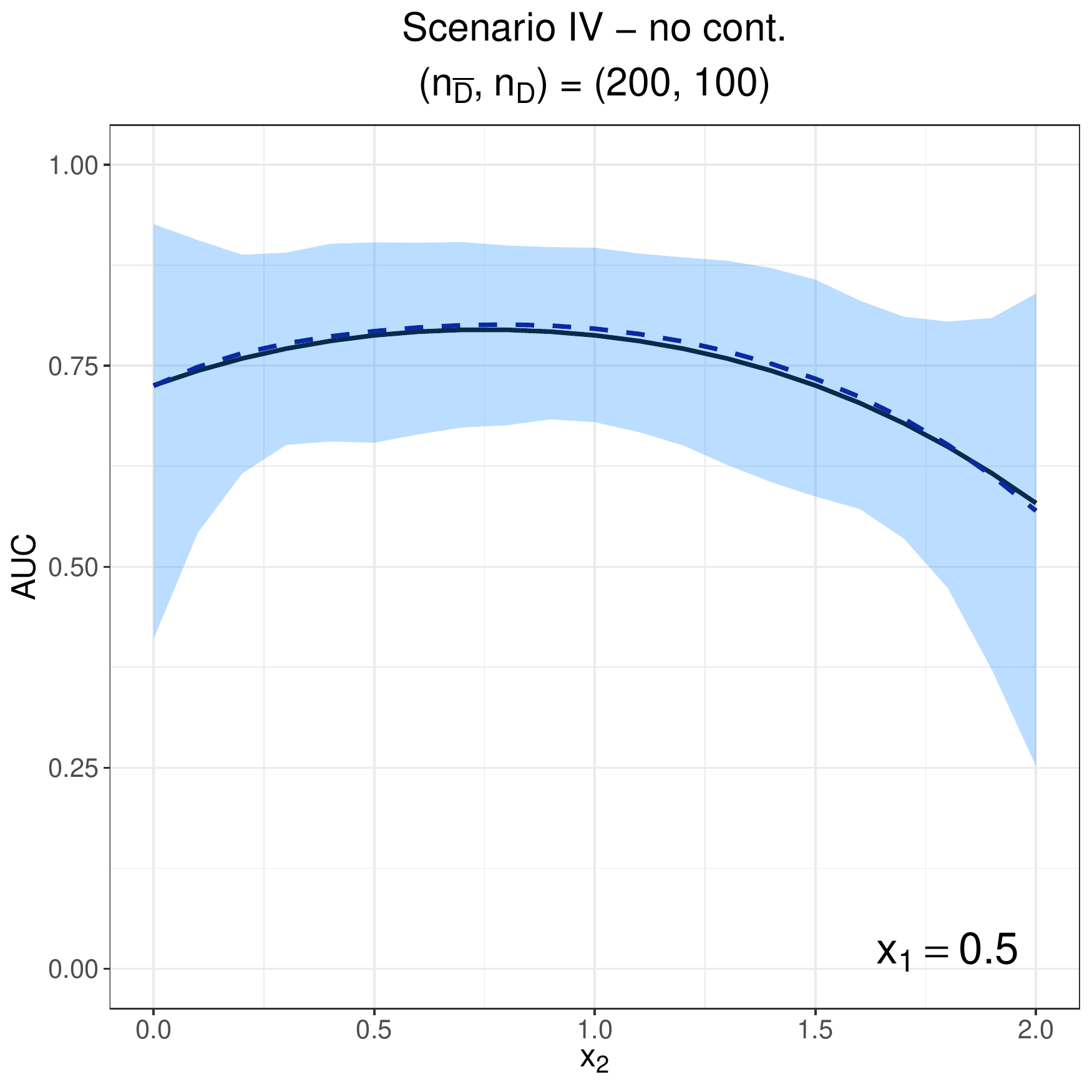}
			\includegraphics[width = 3.35cm]{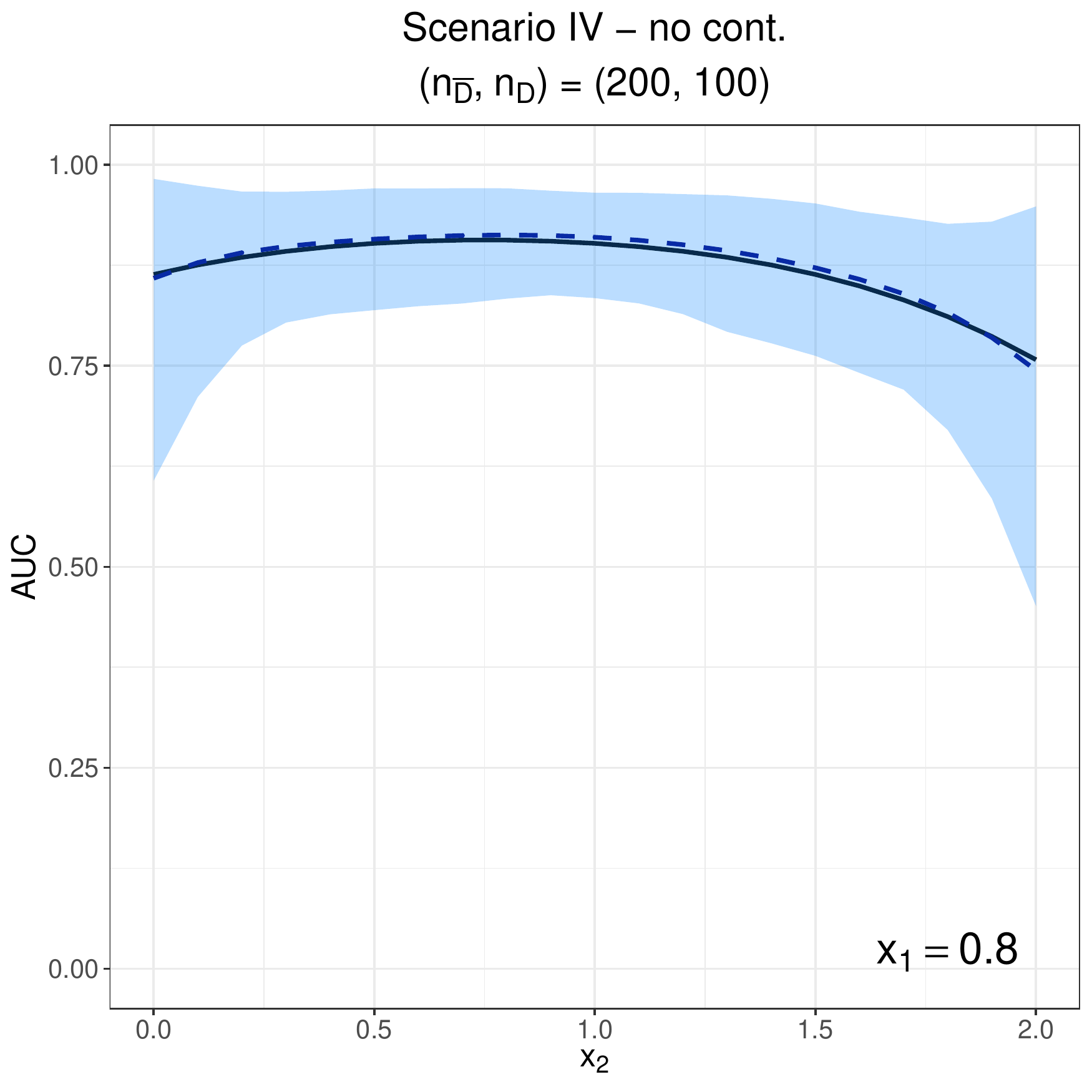}
		}
		\vspace{0.1cm}
		\subfigure{
			\includegraphics[width = 3.35cm]{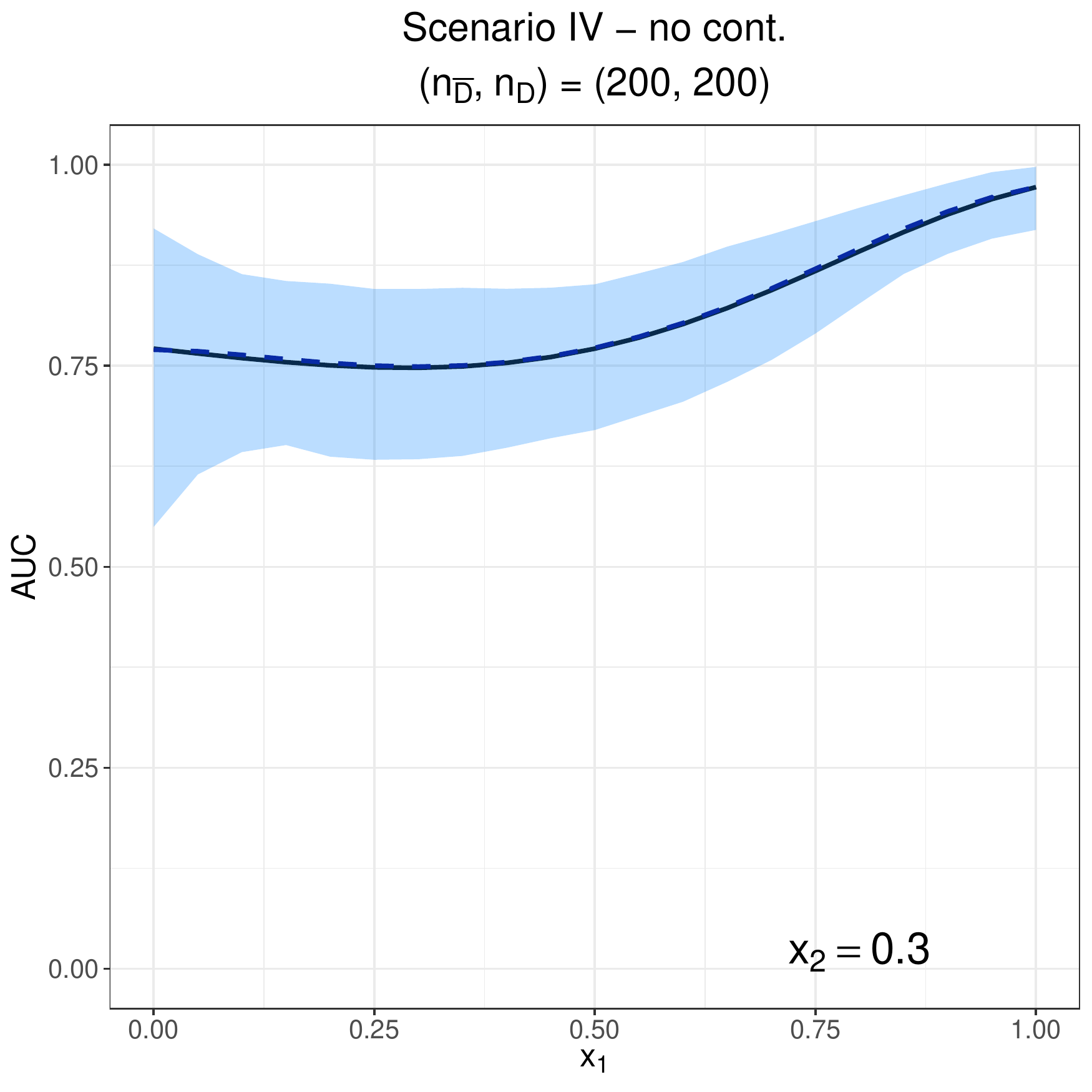}
			\includegraphics[width = 3.35cm]{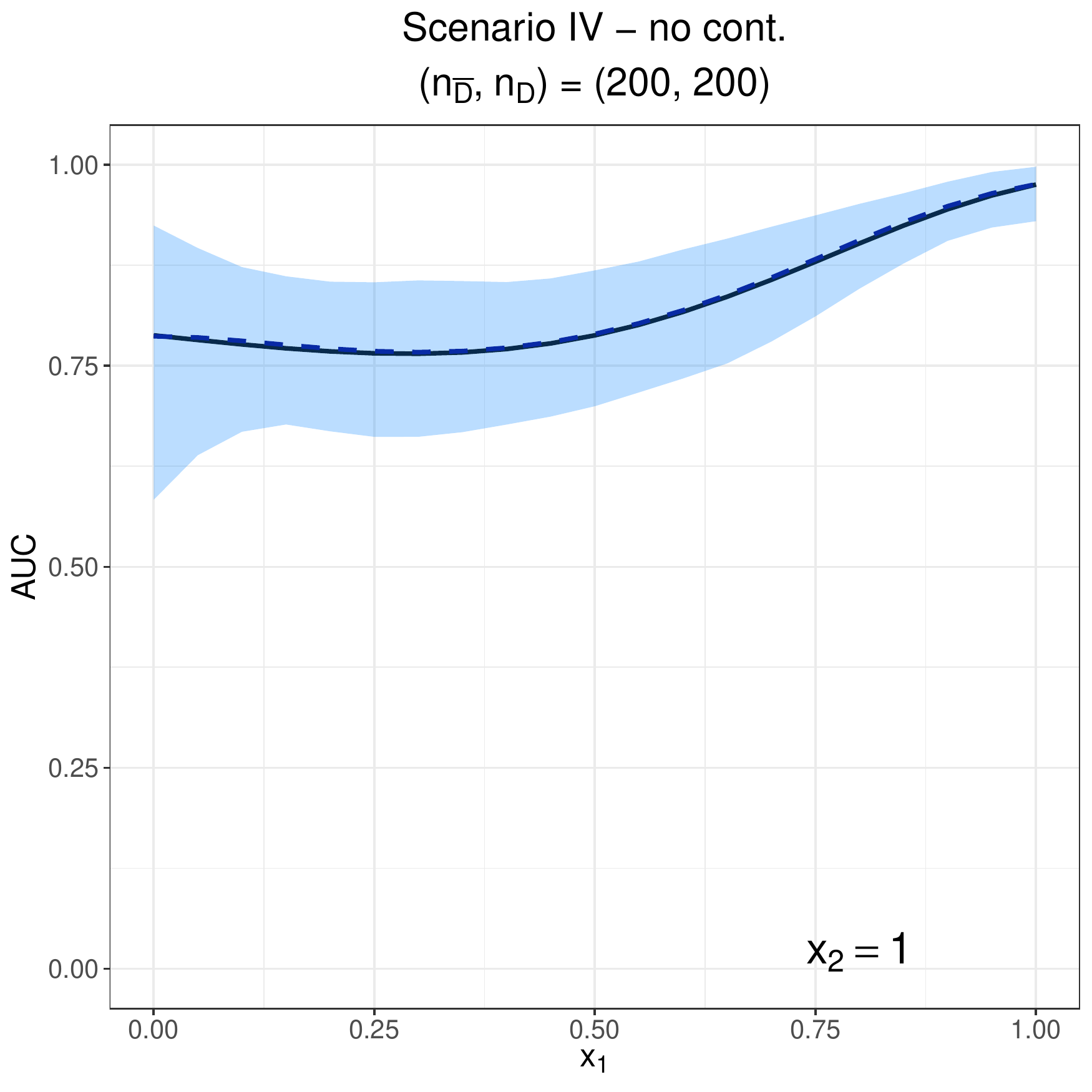}
			\includegraphics[width = 3.35cm]{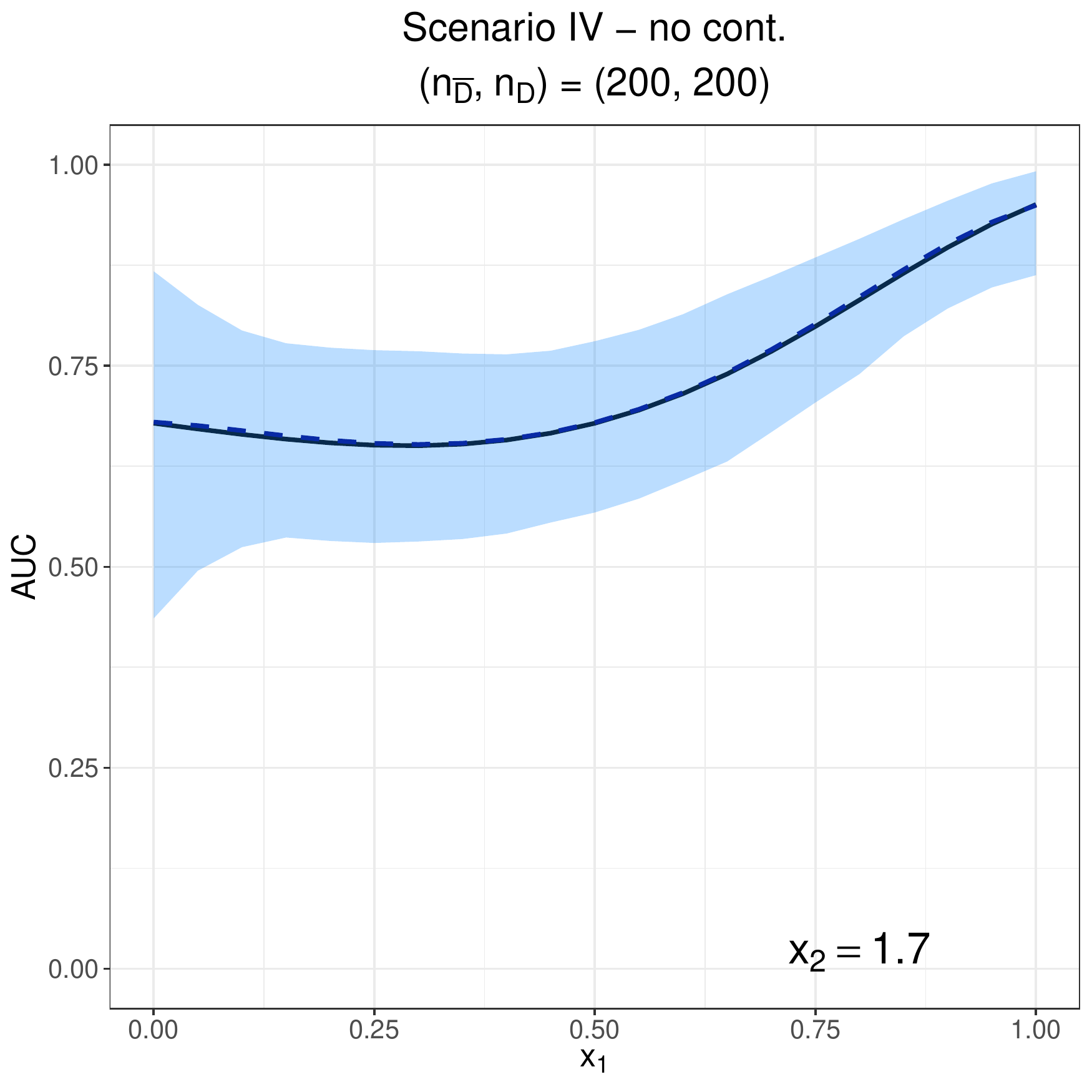}
		}
		\subfigure{
			\includegraphics[width = 3.35cm]{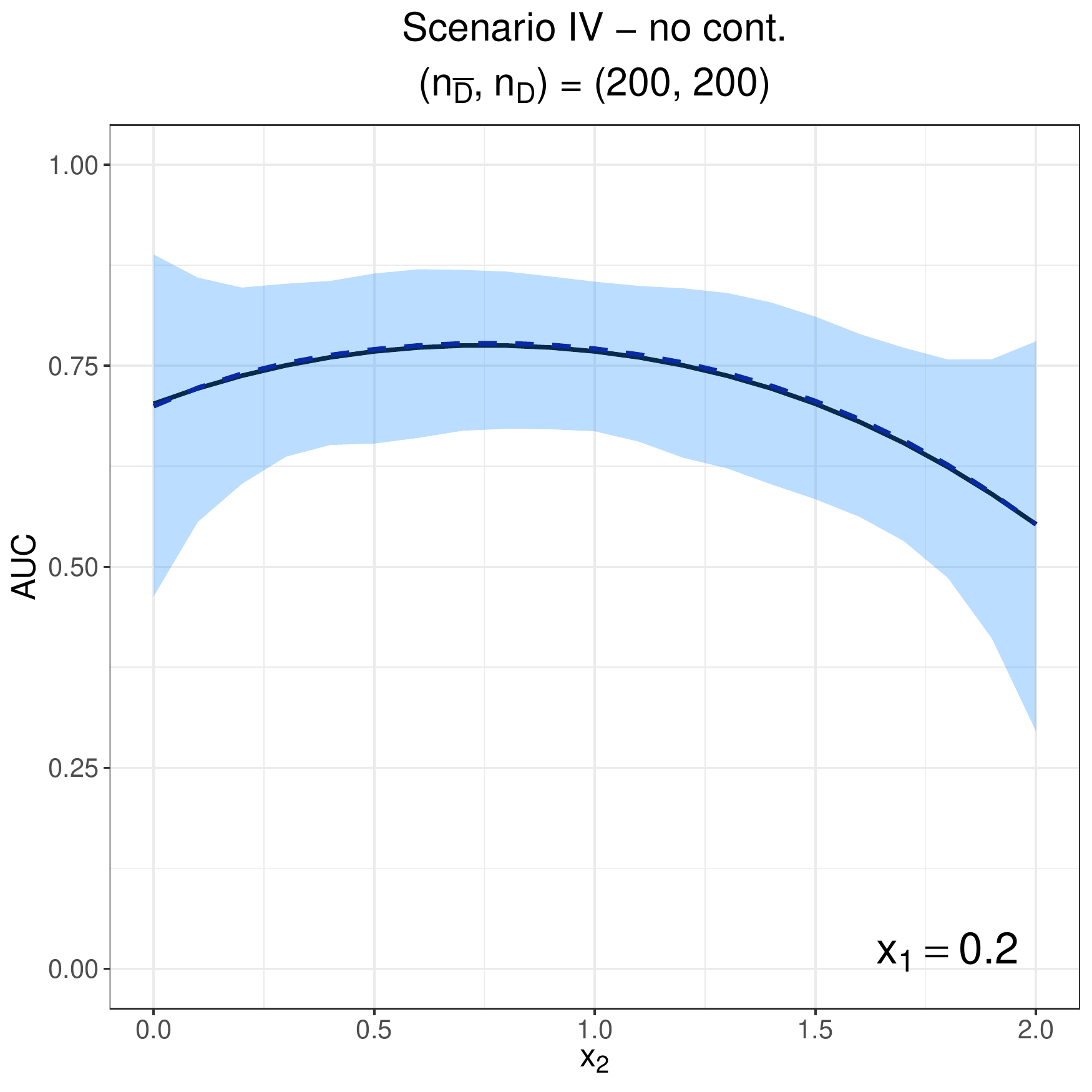}
			\includegraphics[width = 3.35cm]{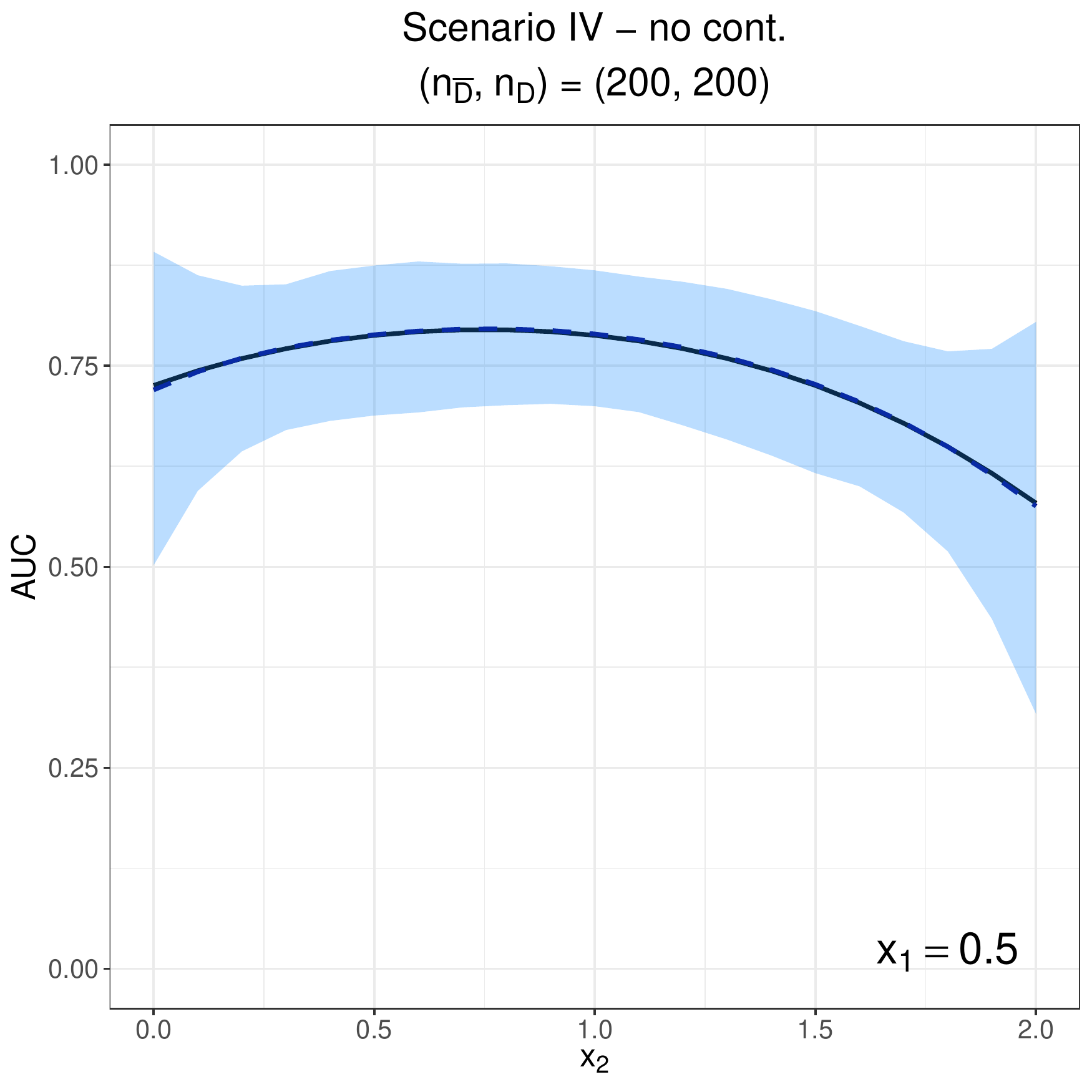}
			\includegraphics[width = 3.35cm]{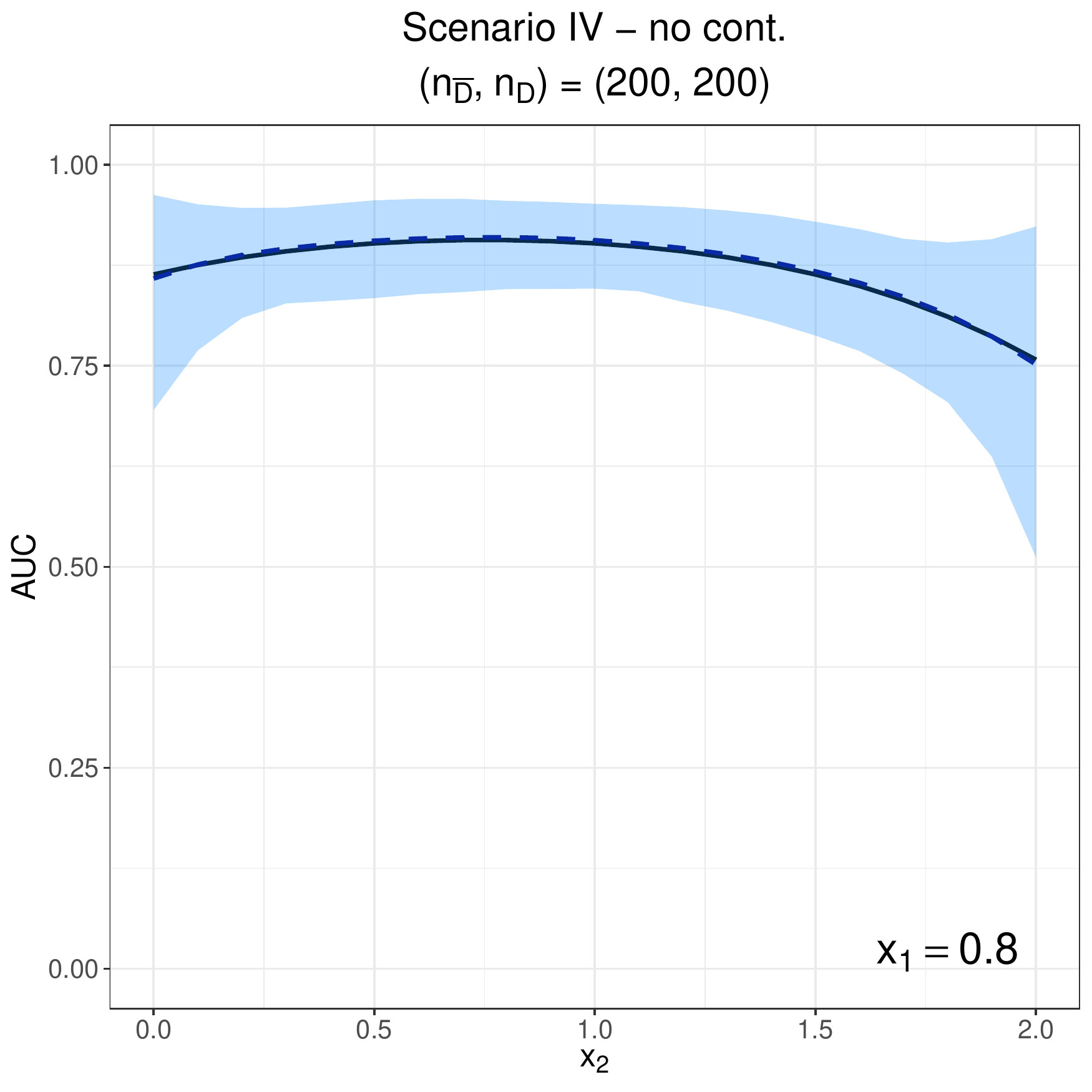}
		}
	\end{center}
	\vspace{-0.3cm}
	\caption{\tiny{Scenario IV. Multiple profiles of the true covariate-specific AUC (solid line) versus the mean of the Monte Carlo estimates (dashed line) along with the $2.5\%$ and $97.5\%$ simulation quantiles (shaded area) for the case of no contamination. Rows 1 and 2 displays the results for $(n_{\bar{D}}, n_D)=(100,100)$, rows 3 and 4 for $(n_{\bar{D}}, n_D)=(200,100)$, and rows 5 and 6 for $(n_{\bar{D}}, n_D)=(200,200)$.}}
\end{figure}

\begin{figure}[H]
	\begin{center}
		\subfigure{
			\includegraphics[width = 3.35cm]{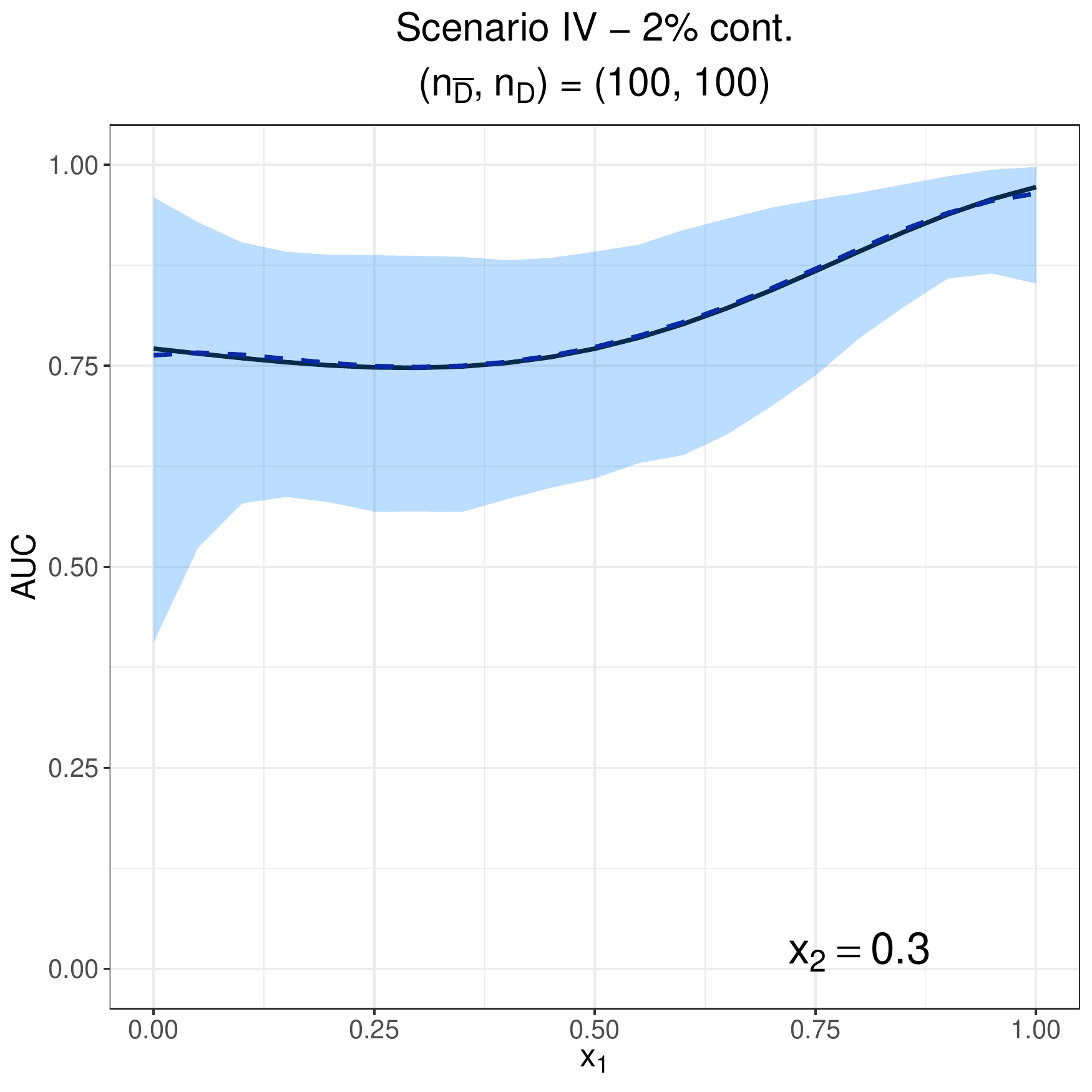}
			\includegraphics[width = 3.35cm]{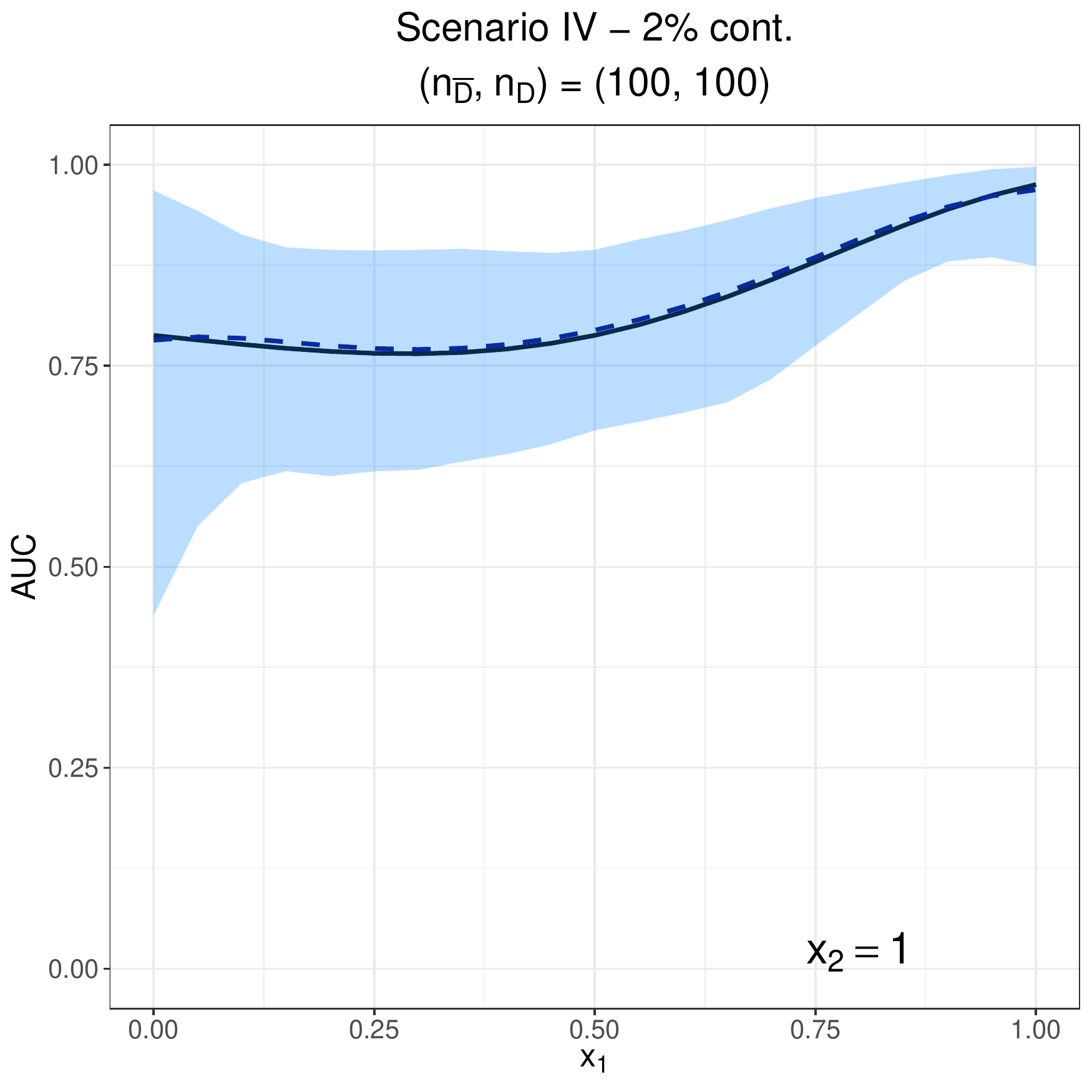}
			\includegraphics[width = 3.35cm]{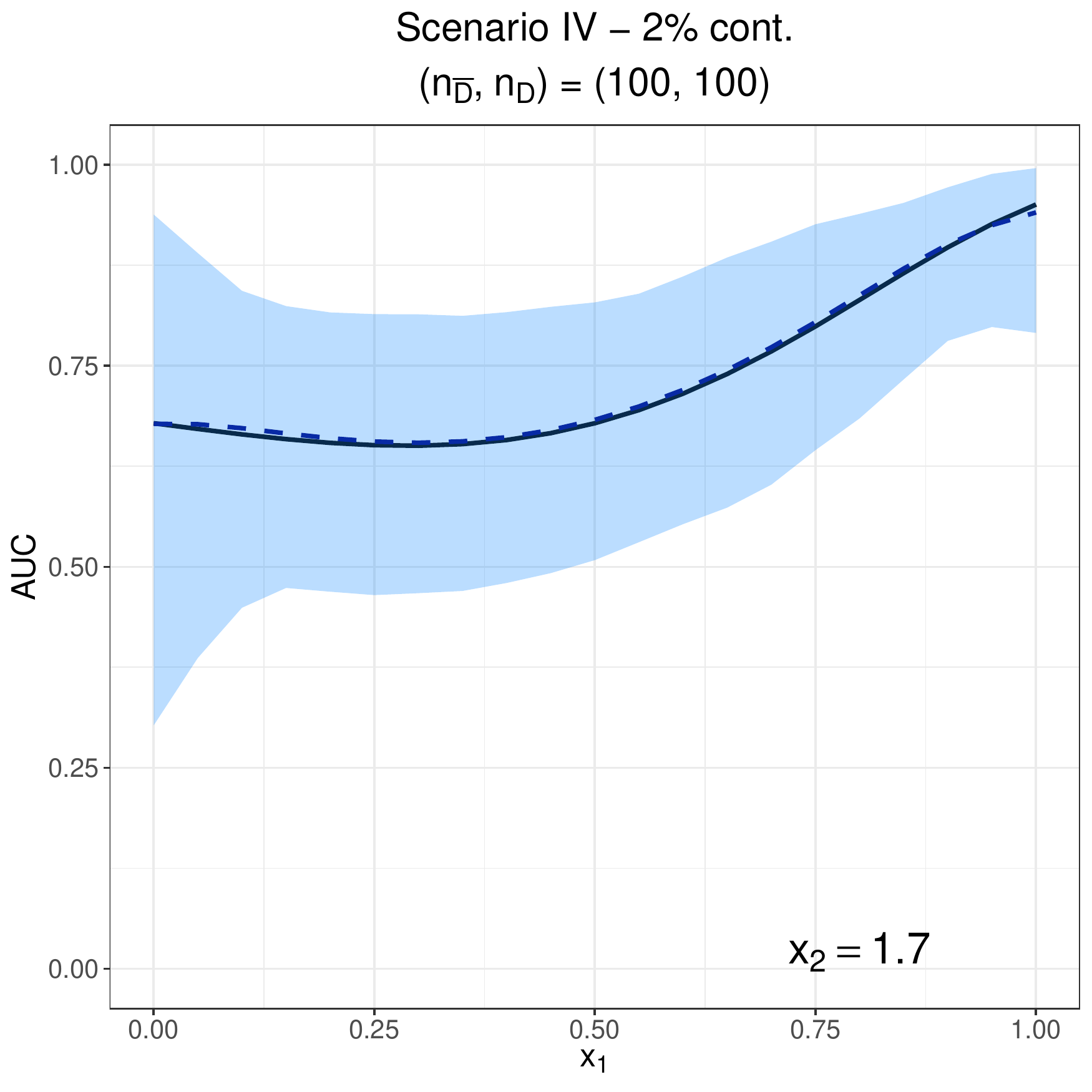}
		}
		\subfigure{
			\includegraphics[width = 3.35cm]{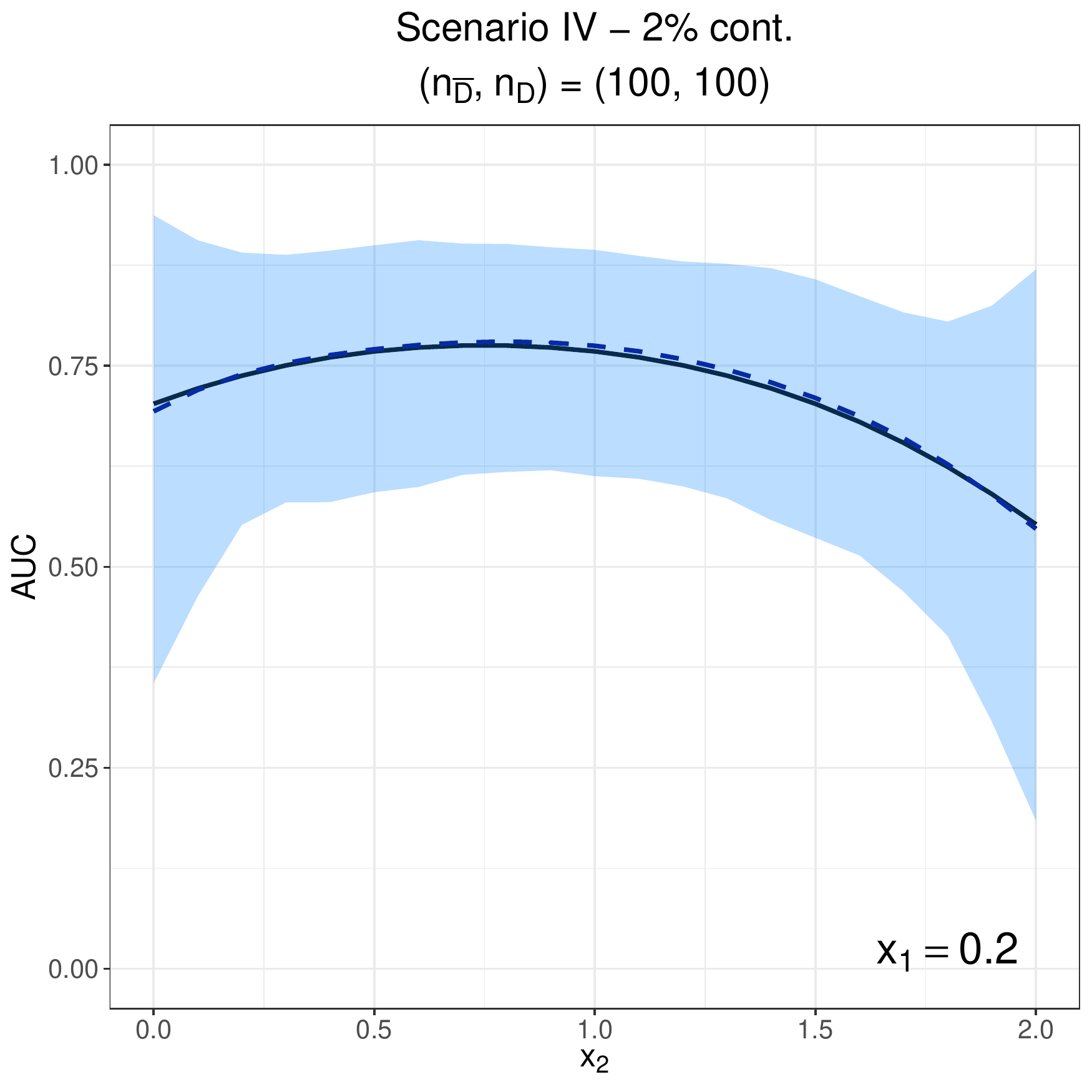}
			\includegraphics[width = 3.35cm]{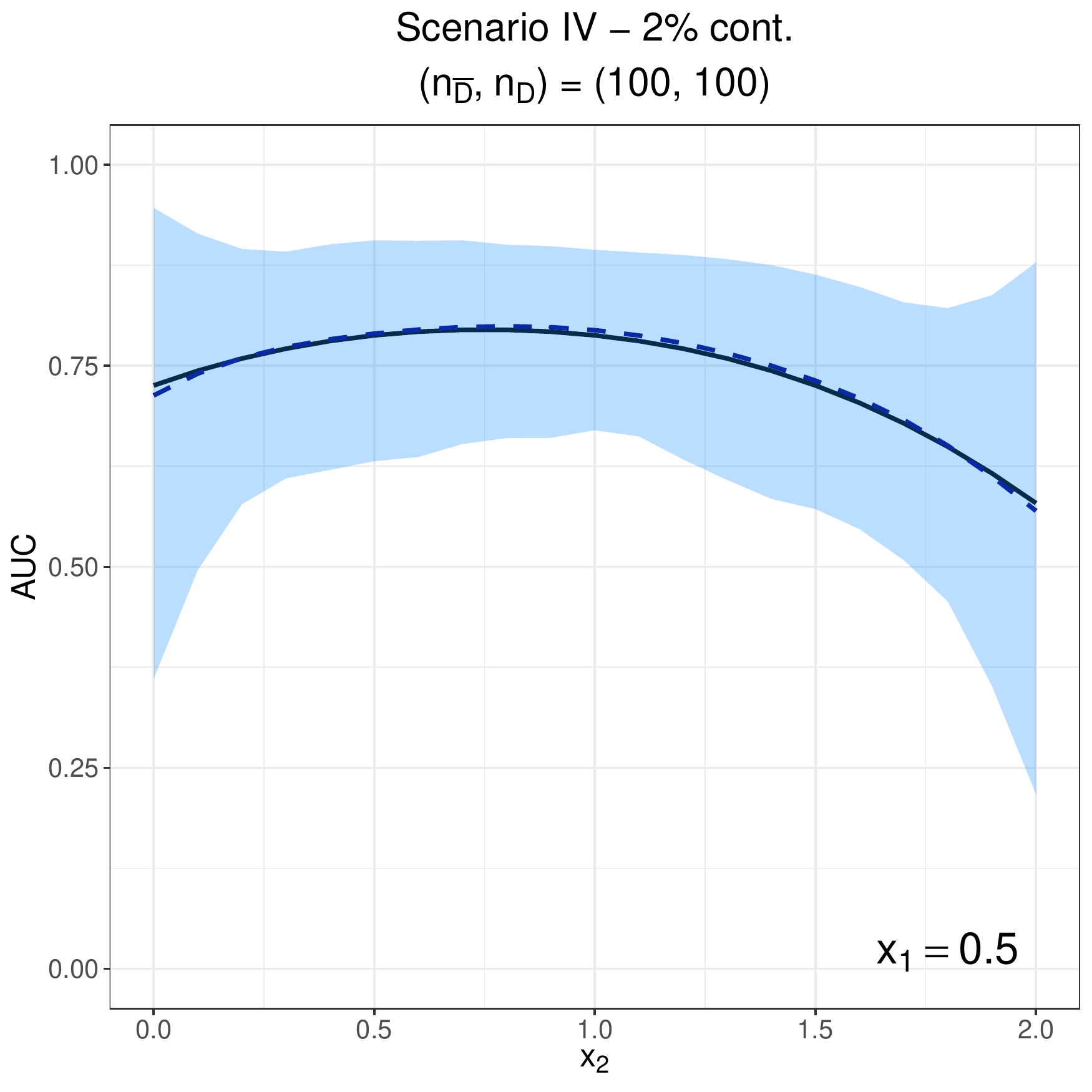}
			\includegraphics[width = 3.35cm]{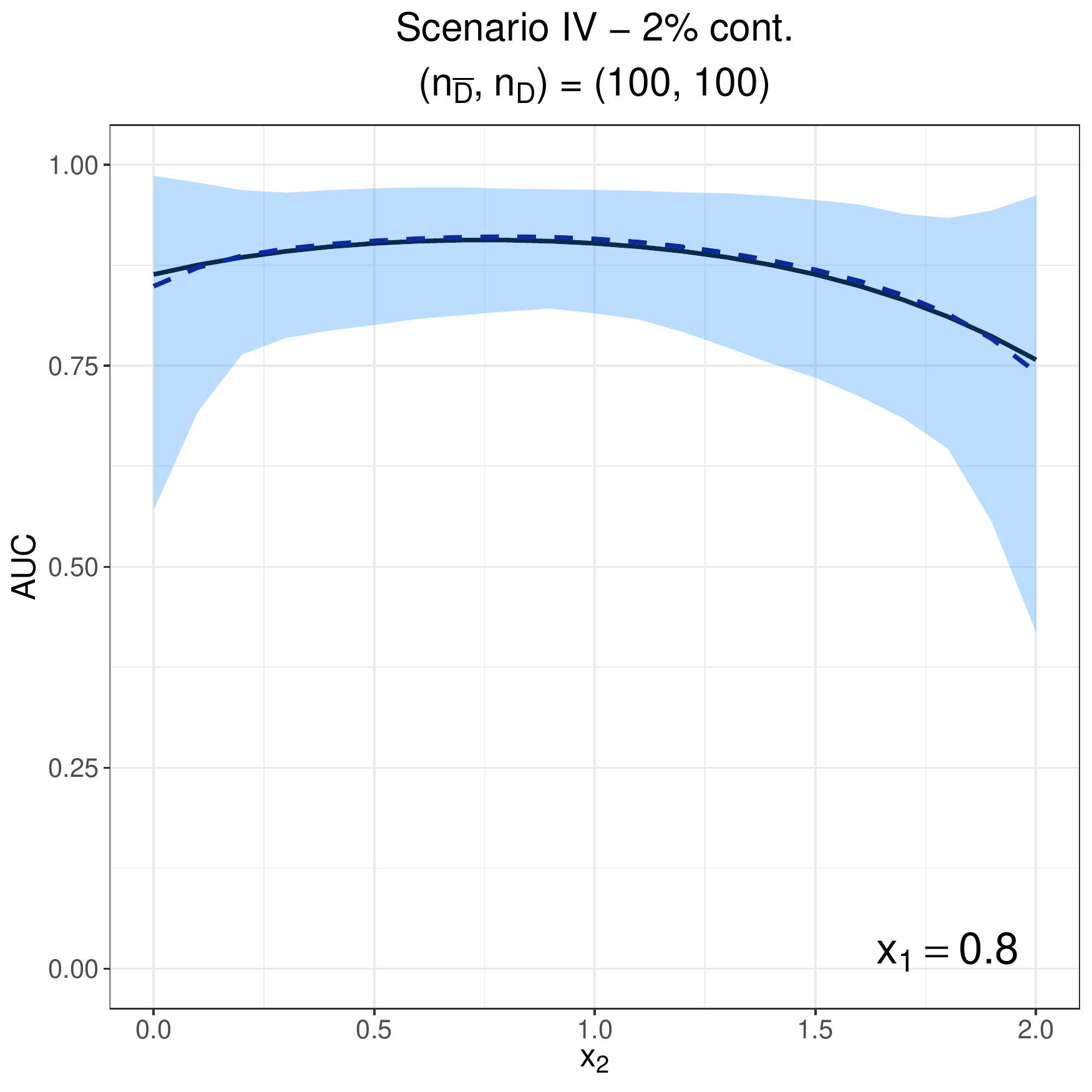}
		}
		\vspace{0.1cm}
		\subfigure{
			\includegraphics[width = 3.35cm]{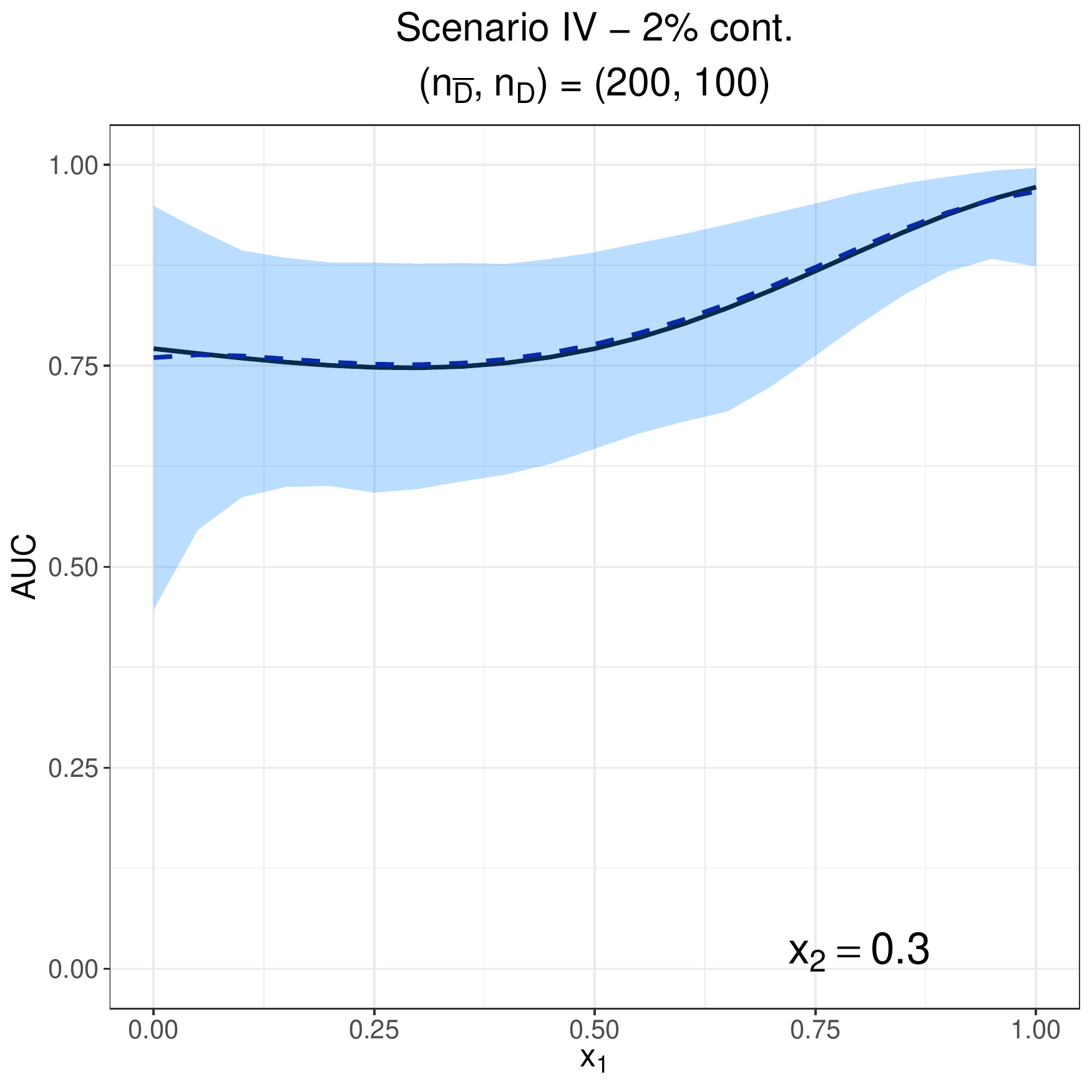}
			\includegraphics[width = 3.35cm]{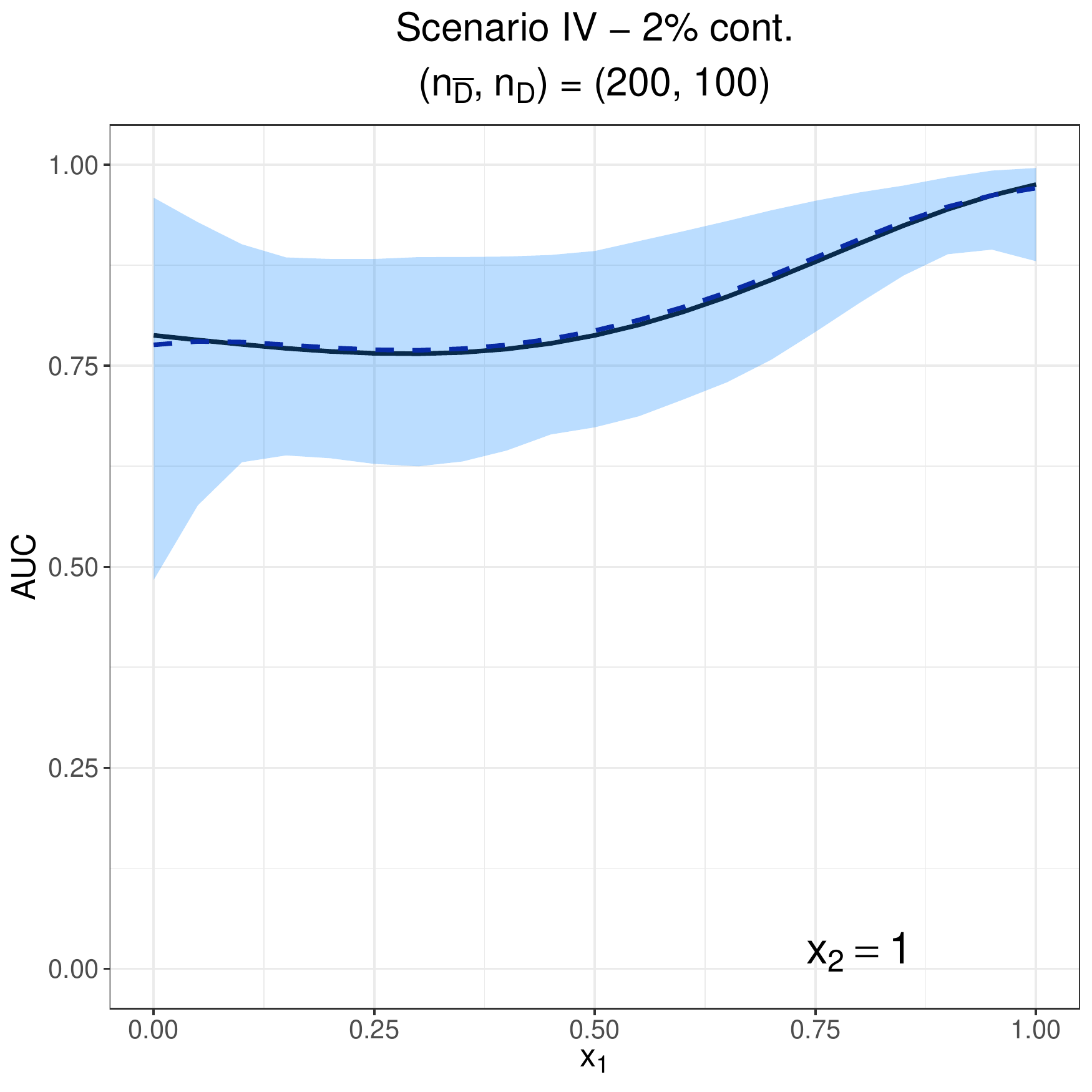}
			\includegraphics[width = 3.35cm]{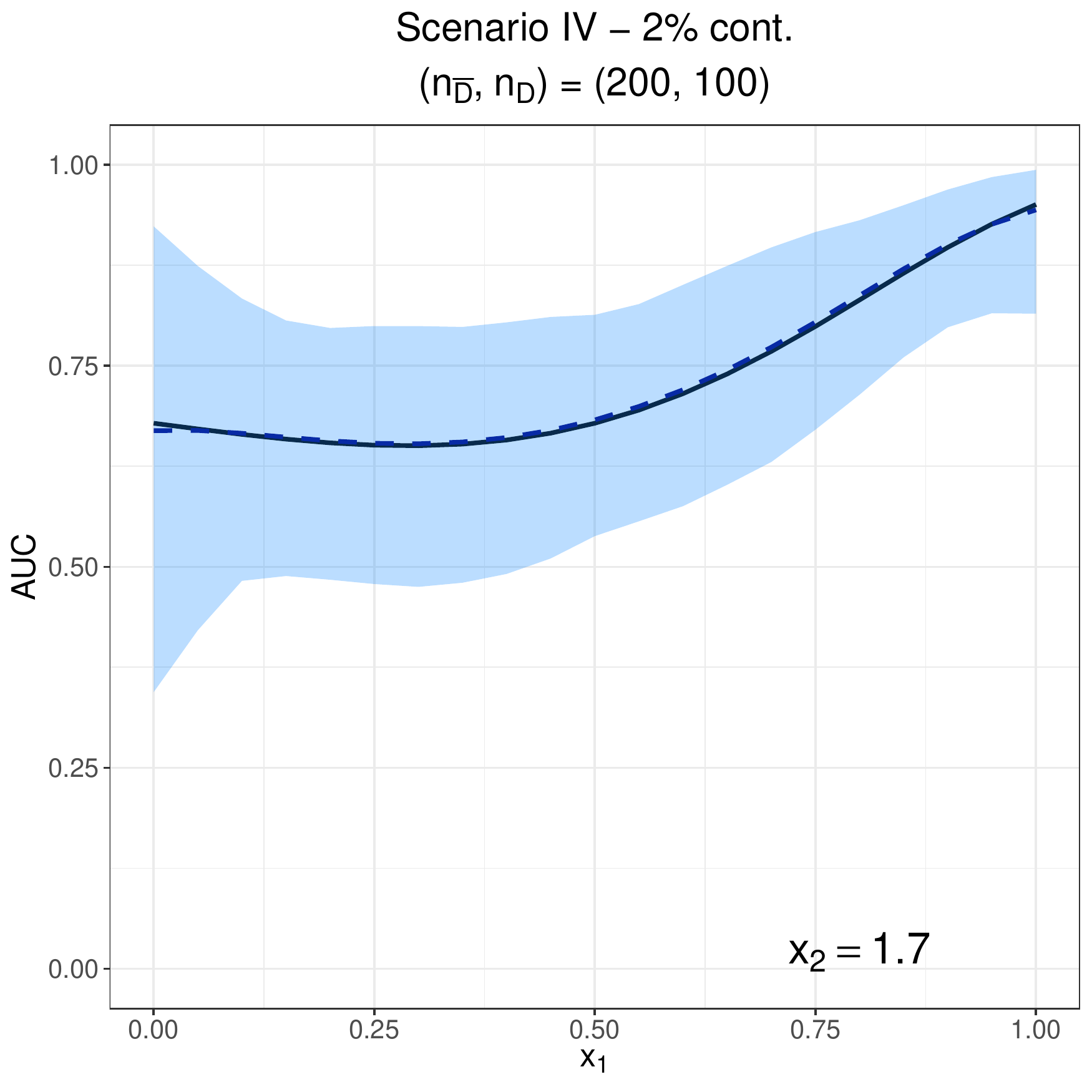}
		}
		\subfigure{
			\includegraphics[width = 3.35cm]{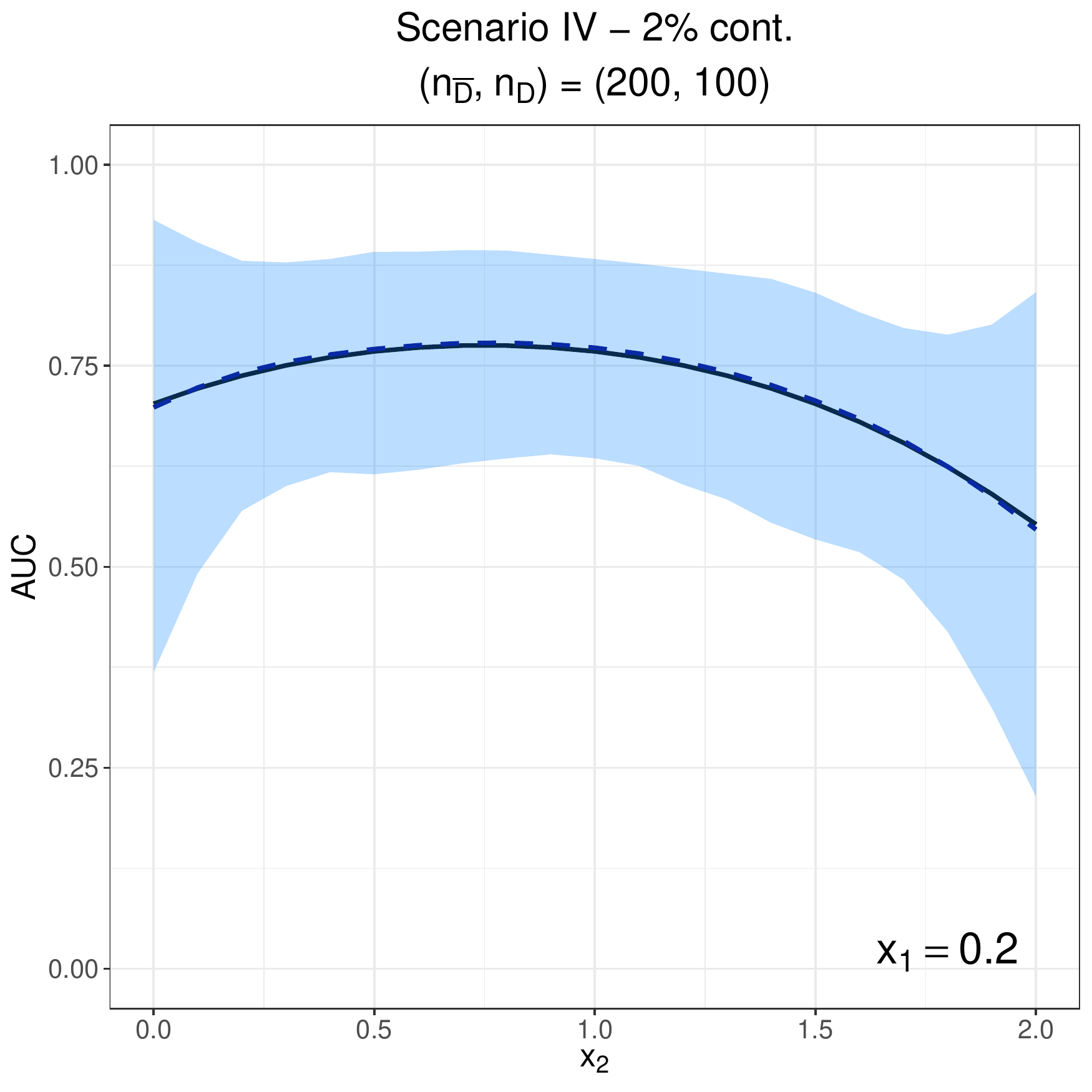}
			\includegraphics[width = 3.35cm]{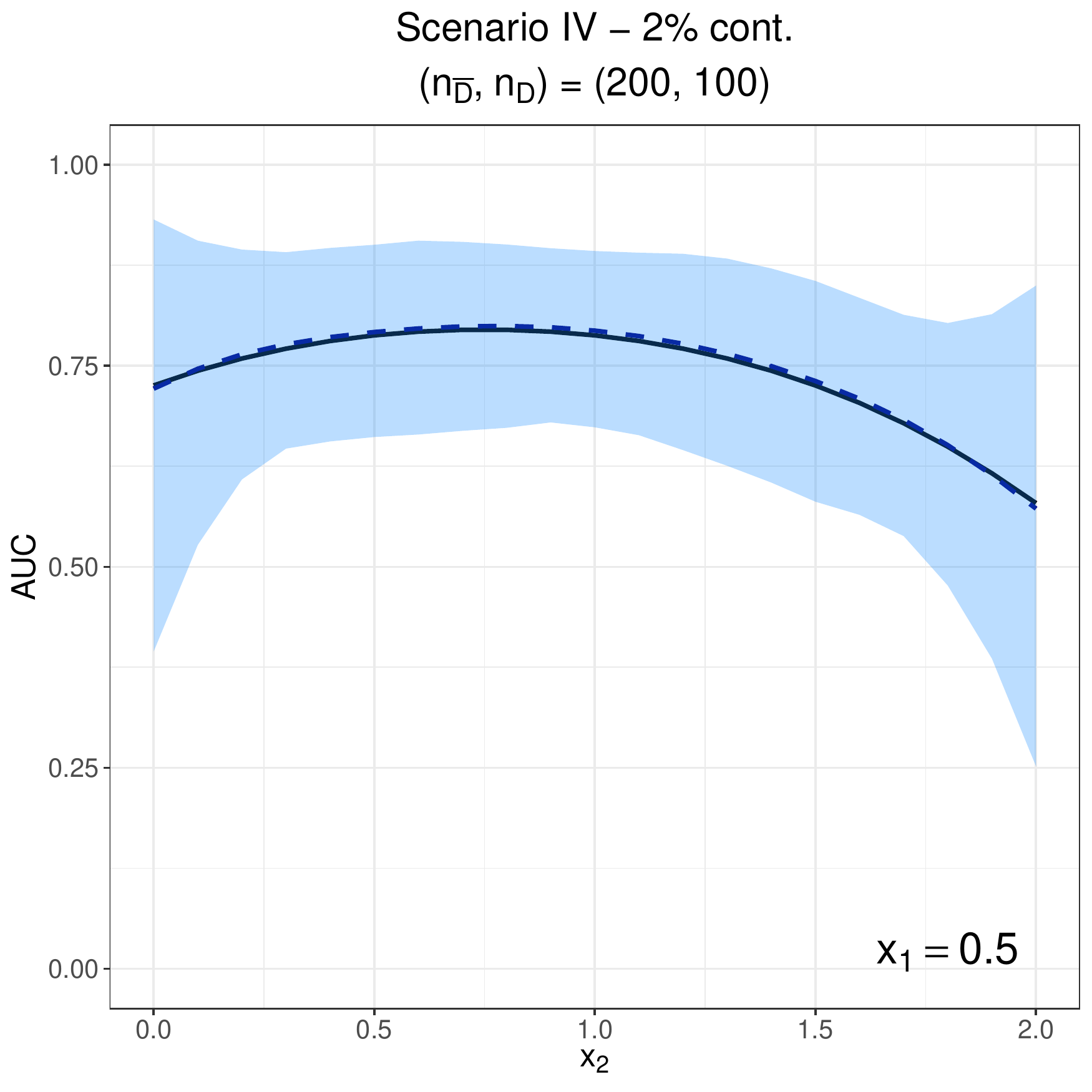}
			\includegraphics[width = 3.35cm]{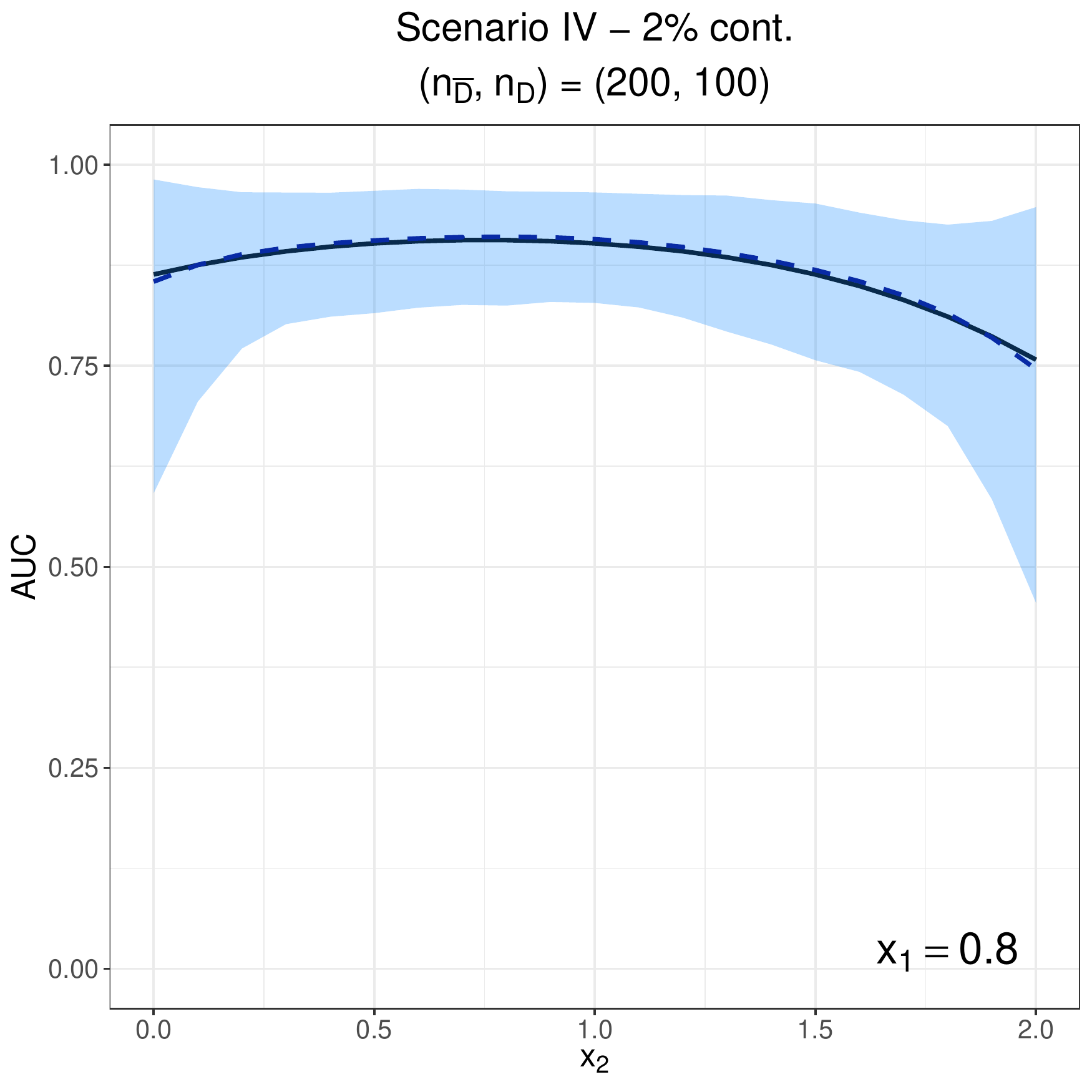}
		}
		\vspace{0.1cm}
		\subfigure{
			\includegraphics[width = 3.35cm]{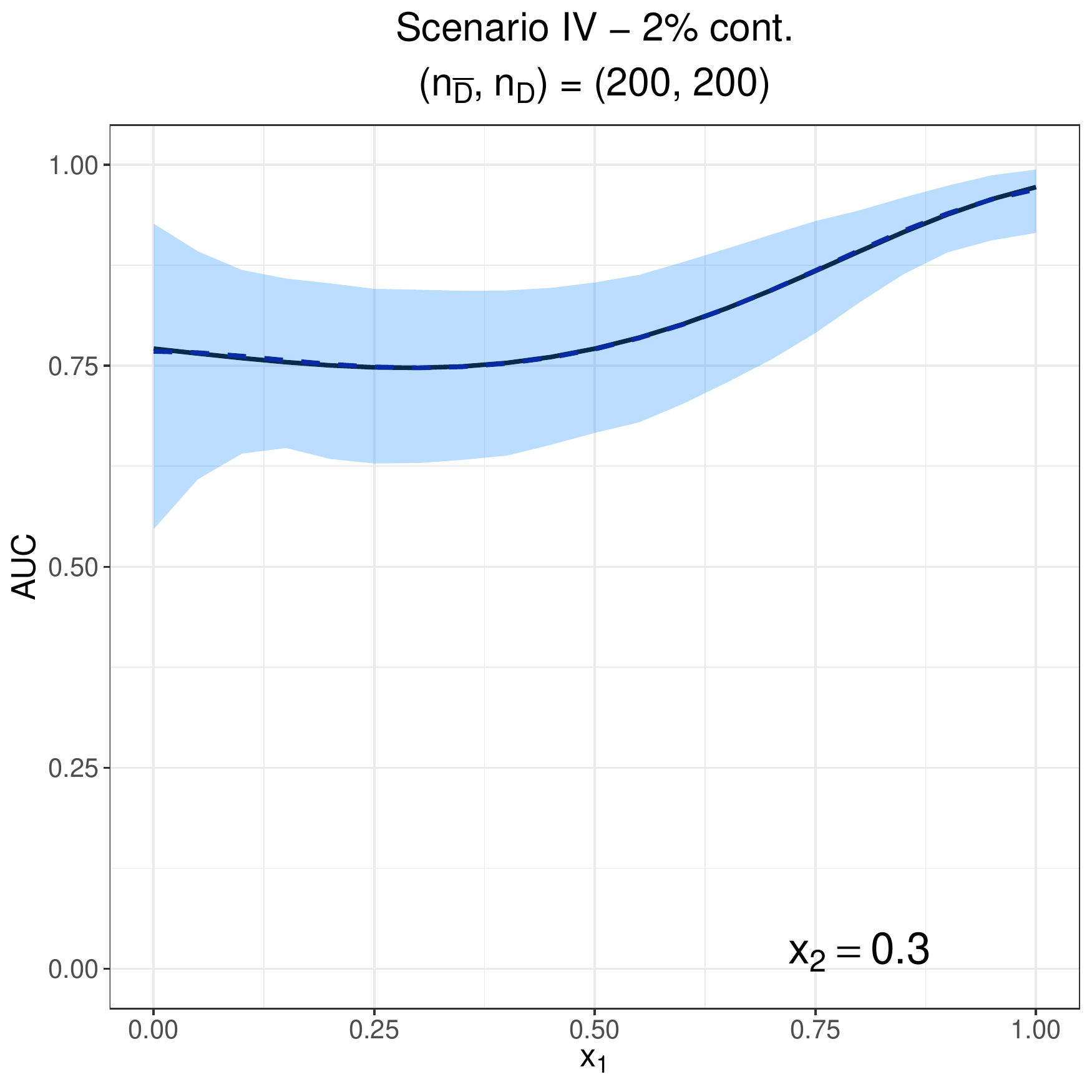}
			\includegraphics[width = 3.35cm]{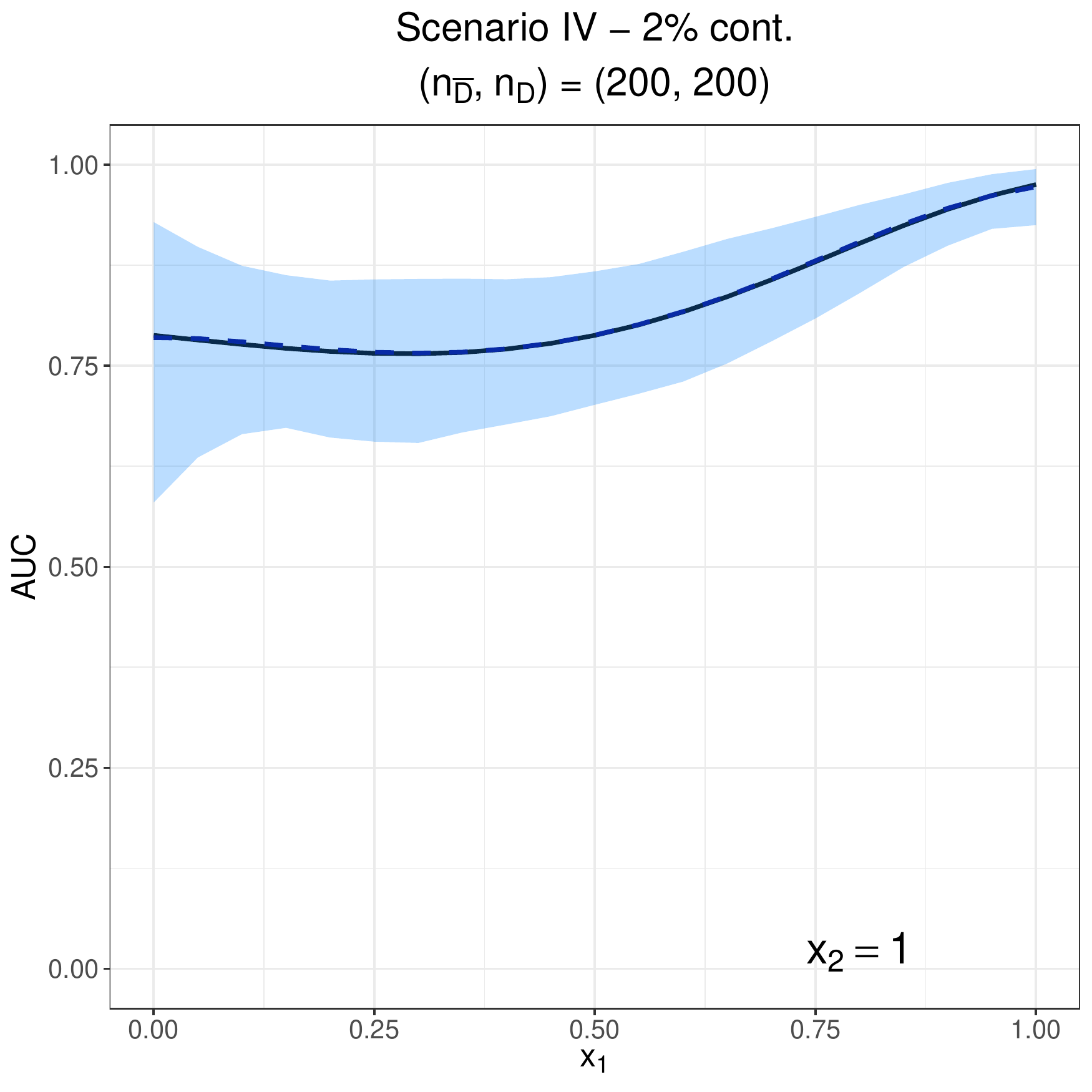}
			\includegraphics[width = 3.35cm]{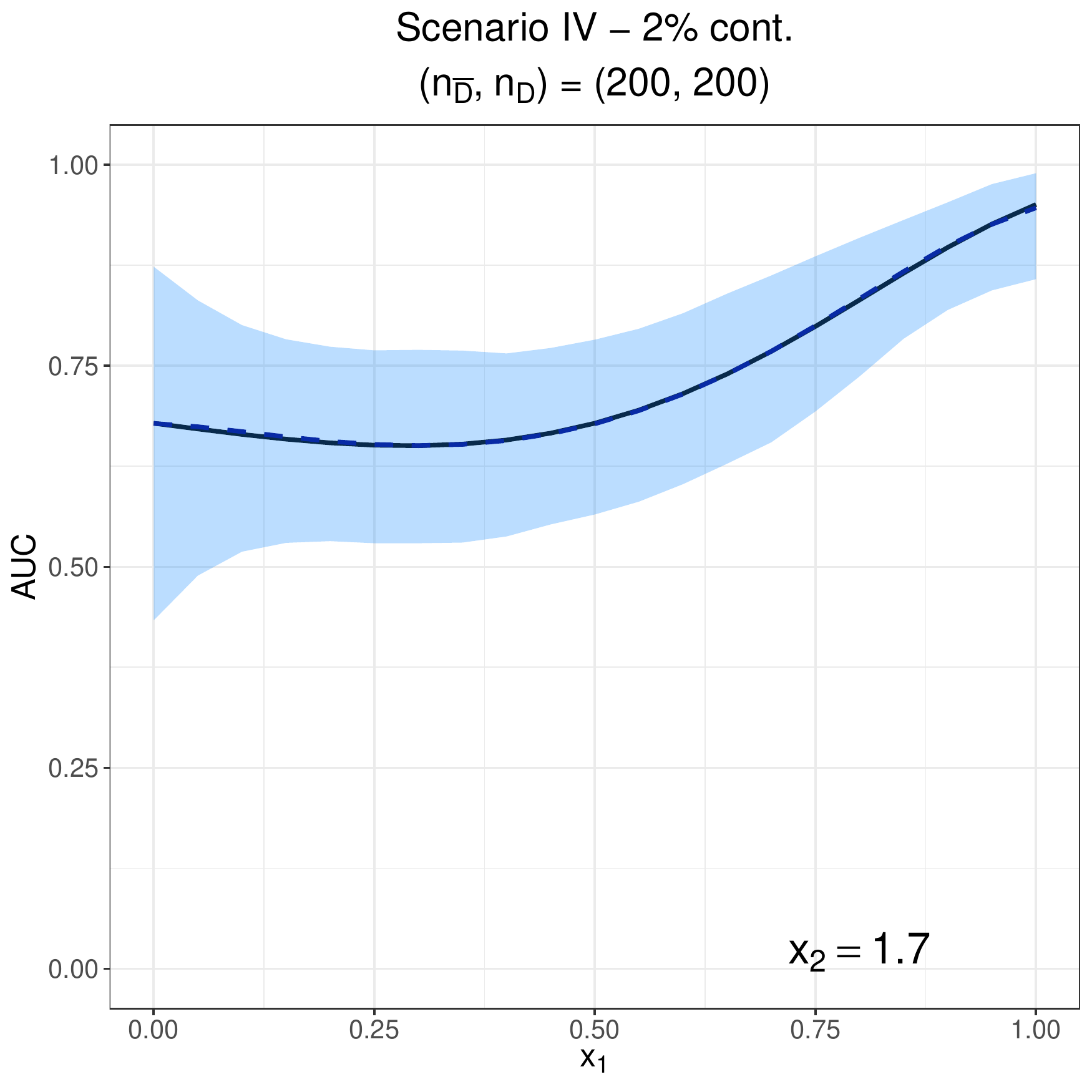}
		}
		\subfigure{
			\includegraphics[width = 3.35cm]{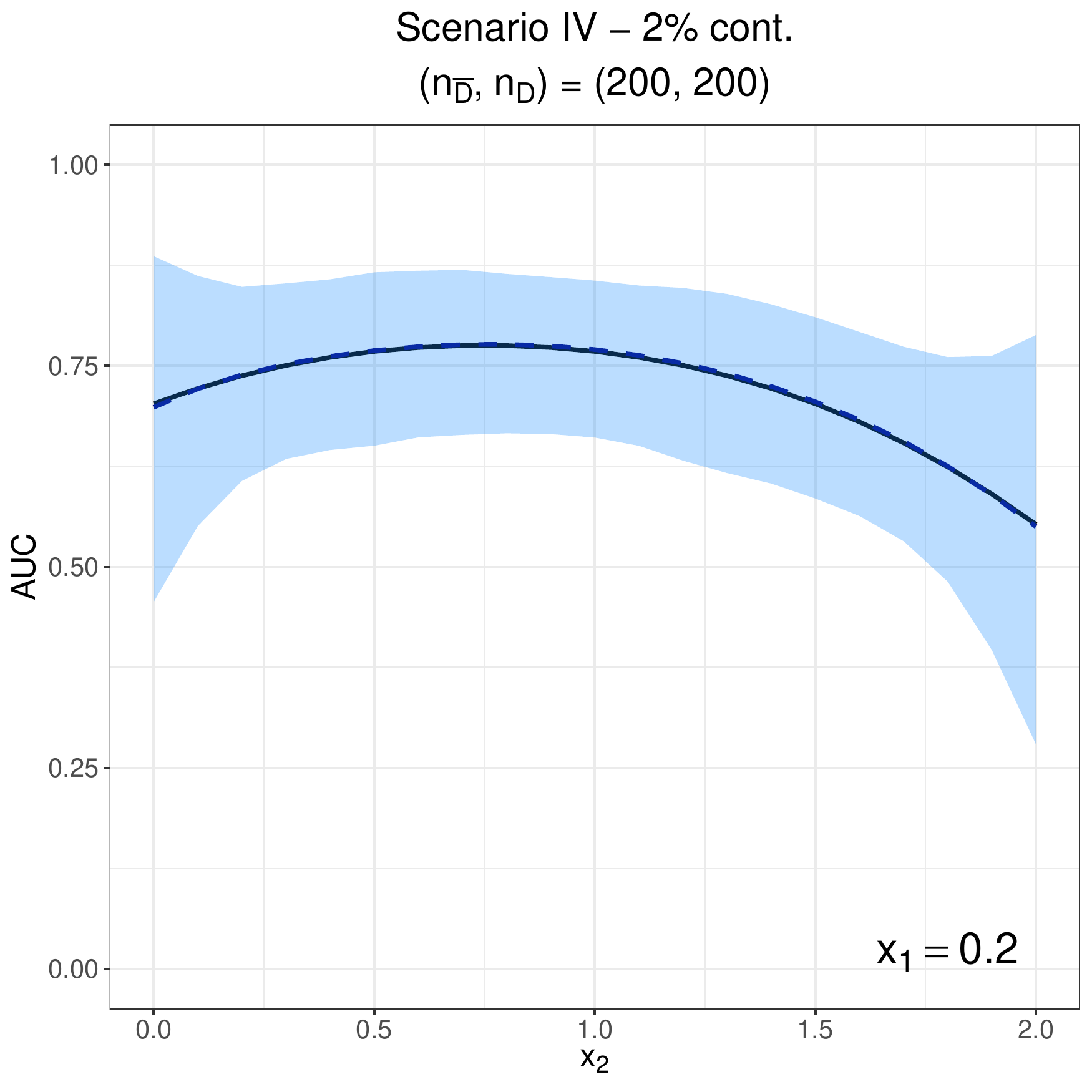}	
			\includegraphics[width = 3.35cm]{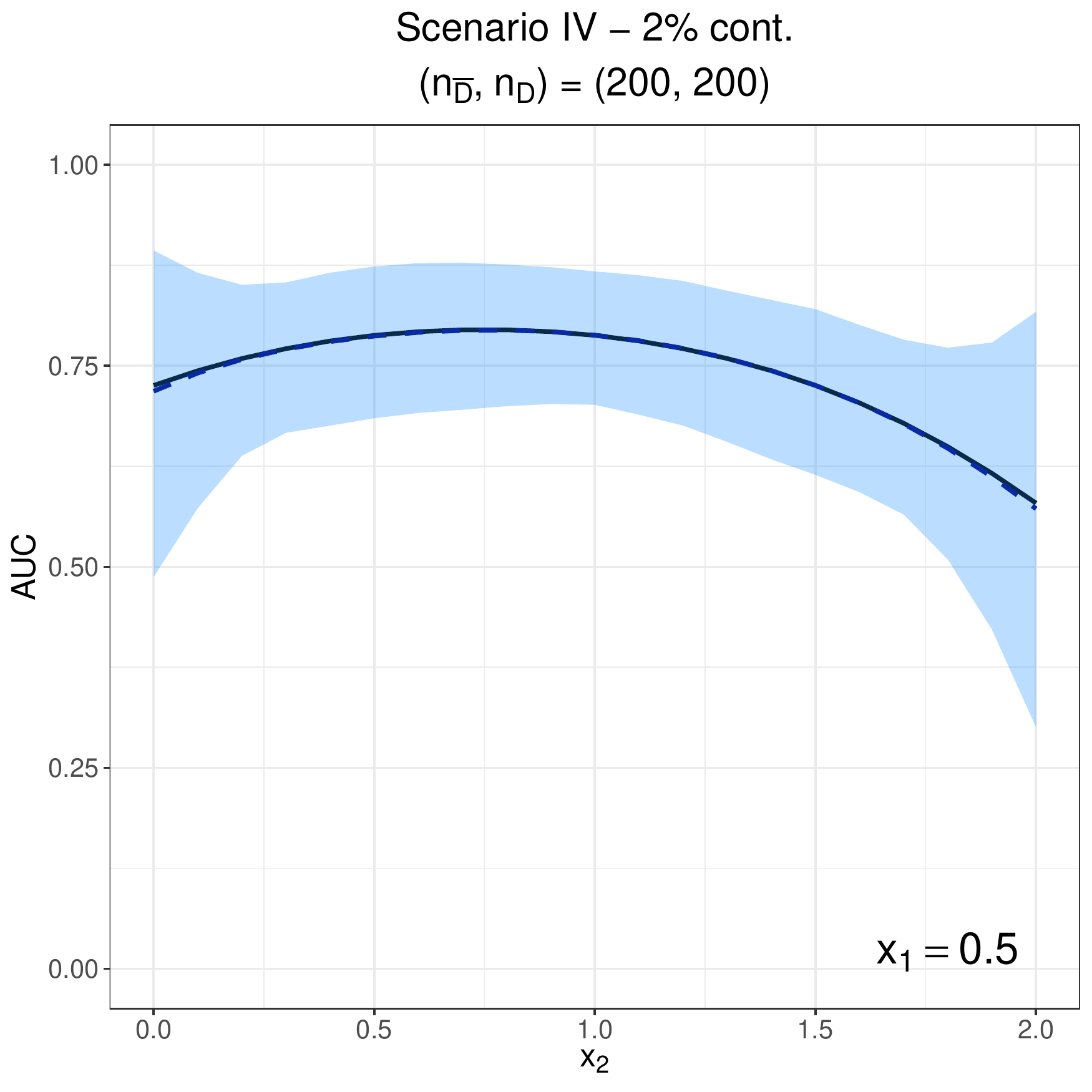}
			\includegraphics[width = 3.35cm]{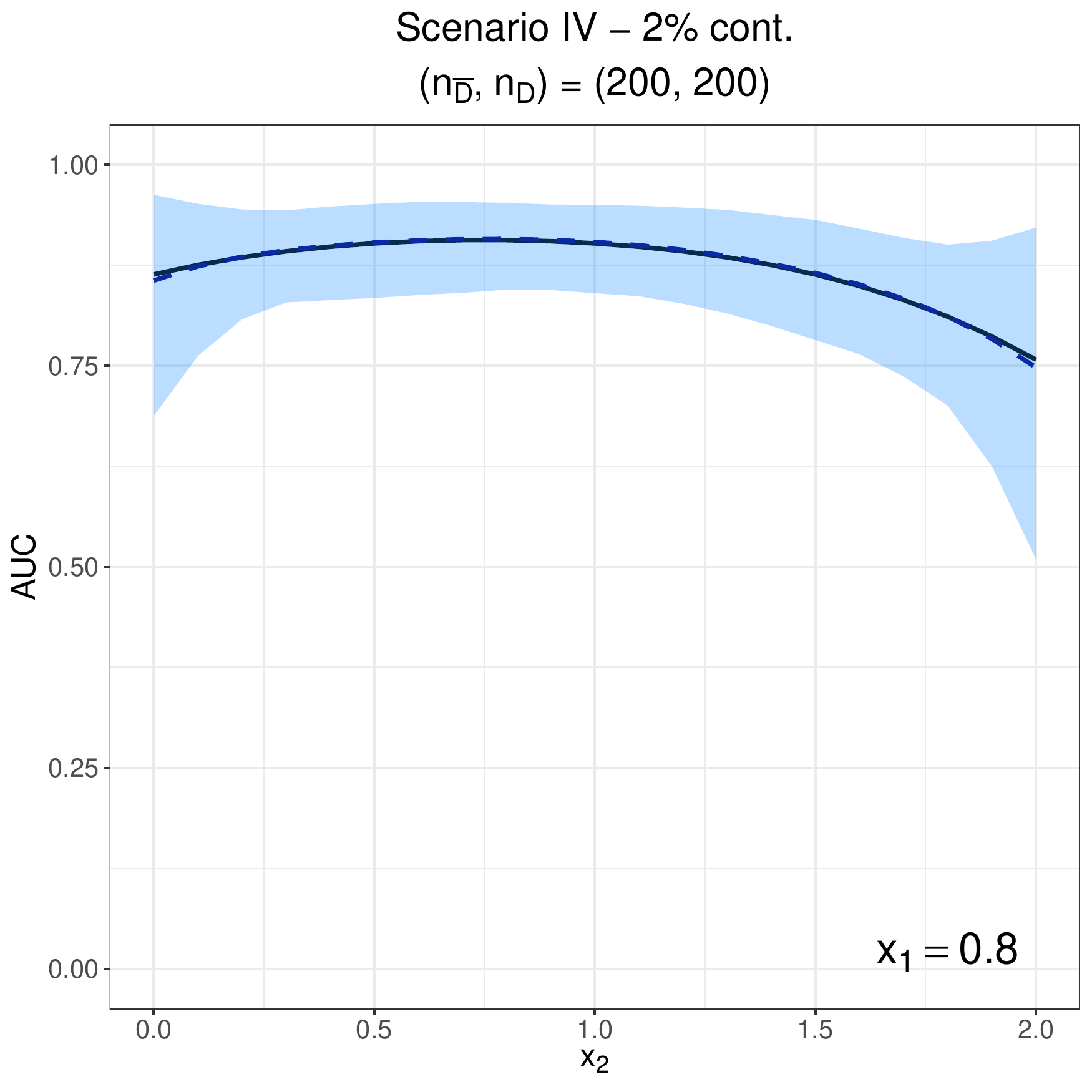}
		}
	\end{center}
	\vspace{-0.3cm}
	\caption{\tiny{Scenario IV. Multiple profiles of the true covariate-specific AUC (solid line) versus the mean of the Monte Carlo estimates (dashed line) along with the $2.5\%$ and $97.5\%$ simulation quantiles (shaded area) for the case of $2\%$ of contamination. Rows 1 and 2 displays the results for $(n_{\bar{D}}, n_D)=(100,100)$, rows 3 and 4 for $(n_{\bar{D}}, n_D)=(200,100)$, and rows 5 and 6 for $(n_{\bar{D}}, n_D)=(200,200)$.}}
\end{figure}

\begin{figure}[H]
	\begin{center}
		\subfigure{
			\includegraphics[width = 3.35cm]{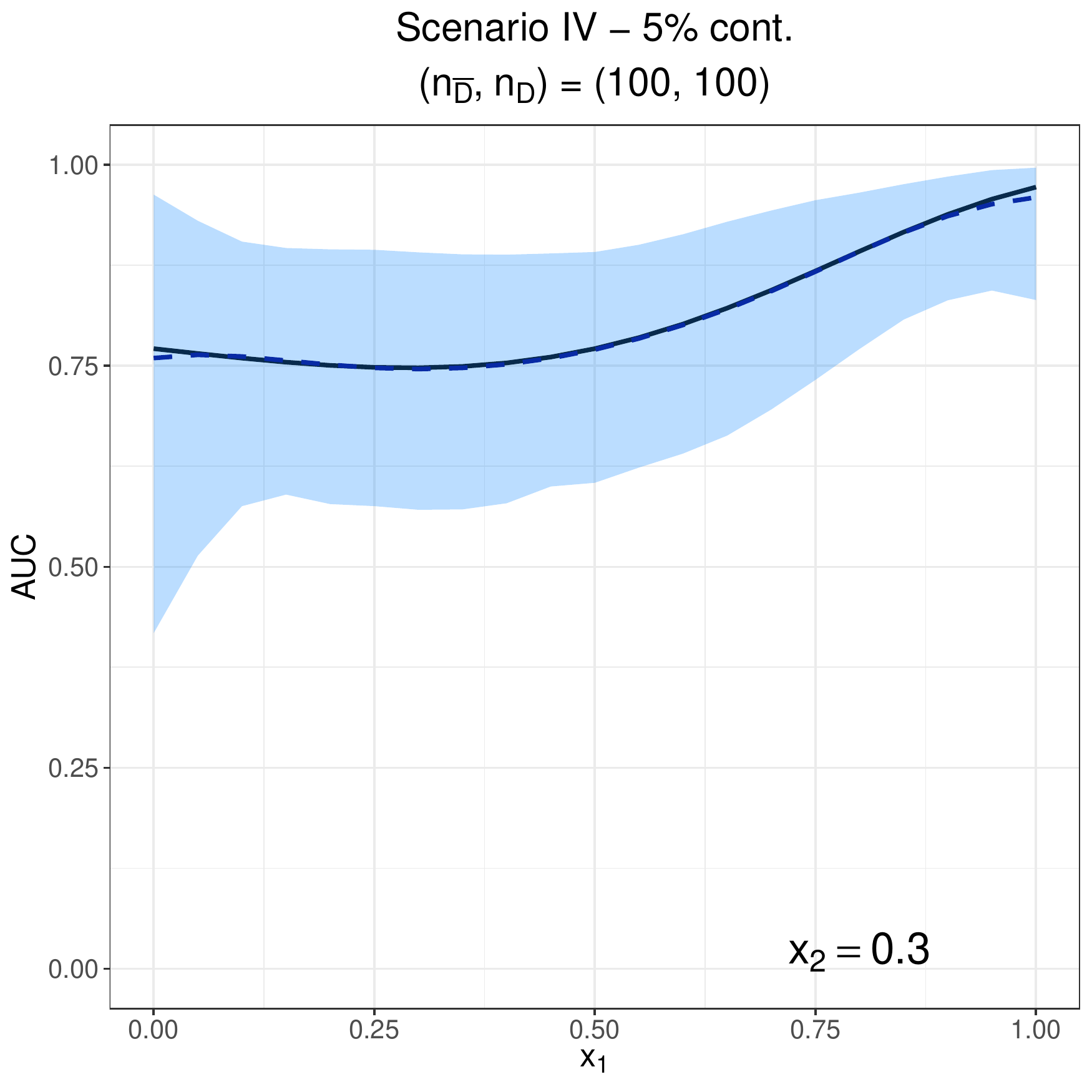}
			\includegraphics[width = 3.35cm]{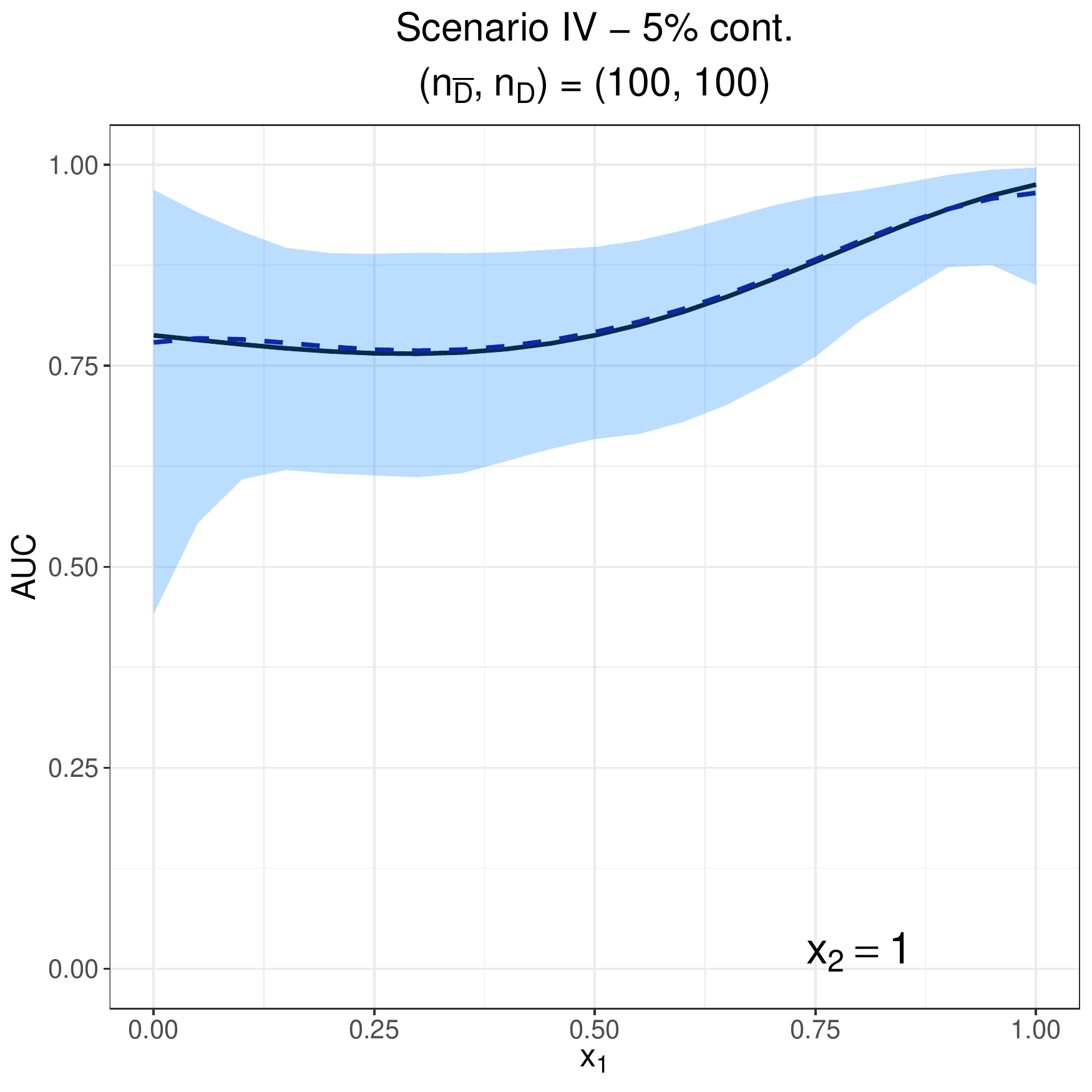}
			\includegraphics[width = 3.35cm]{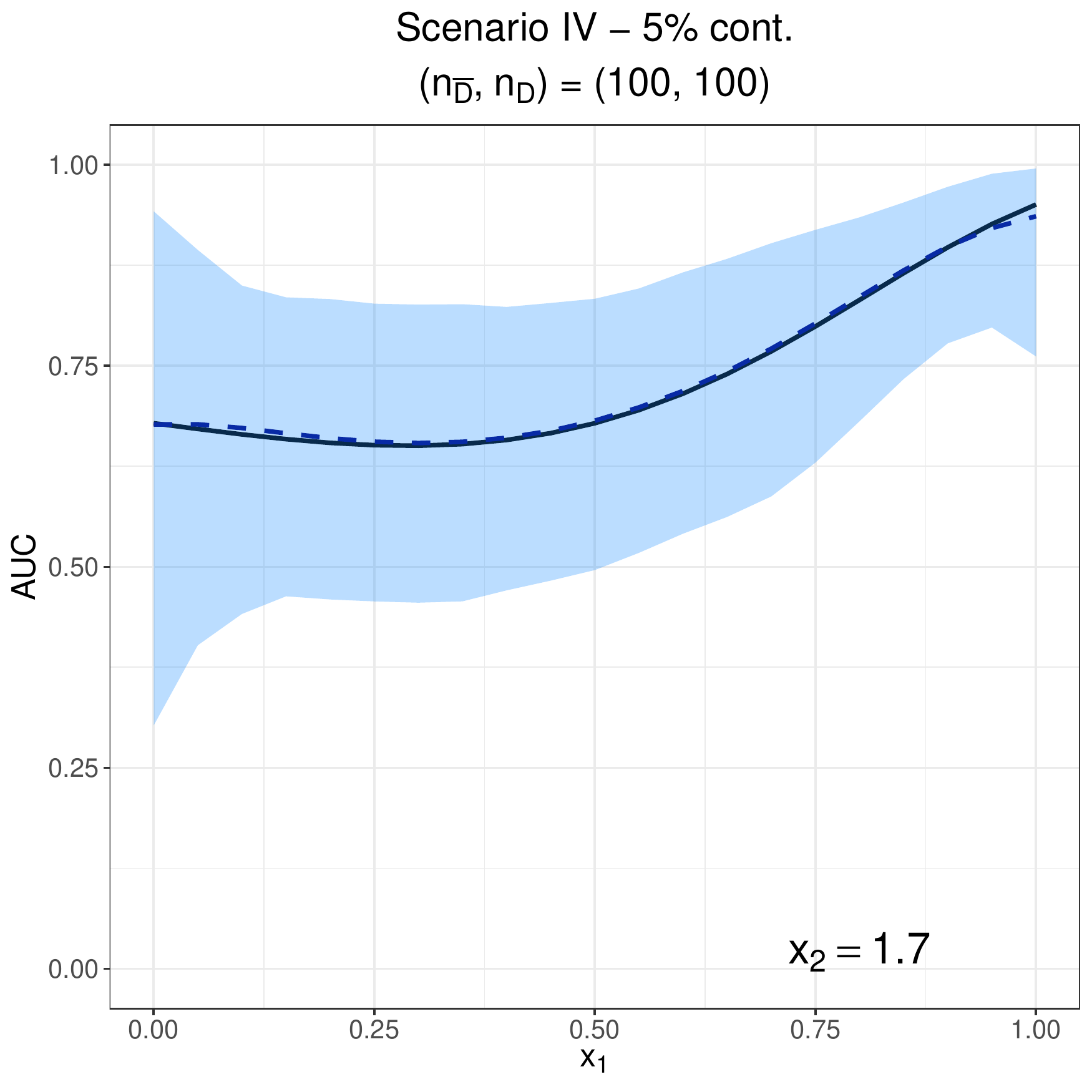}
		}
		\subfigure{
			\includegraphics[width = 3.35cm]{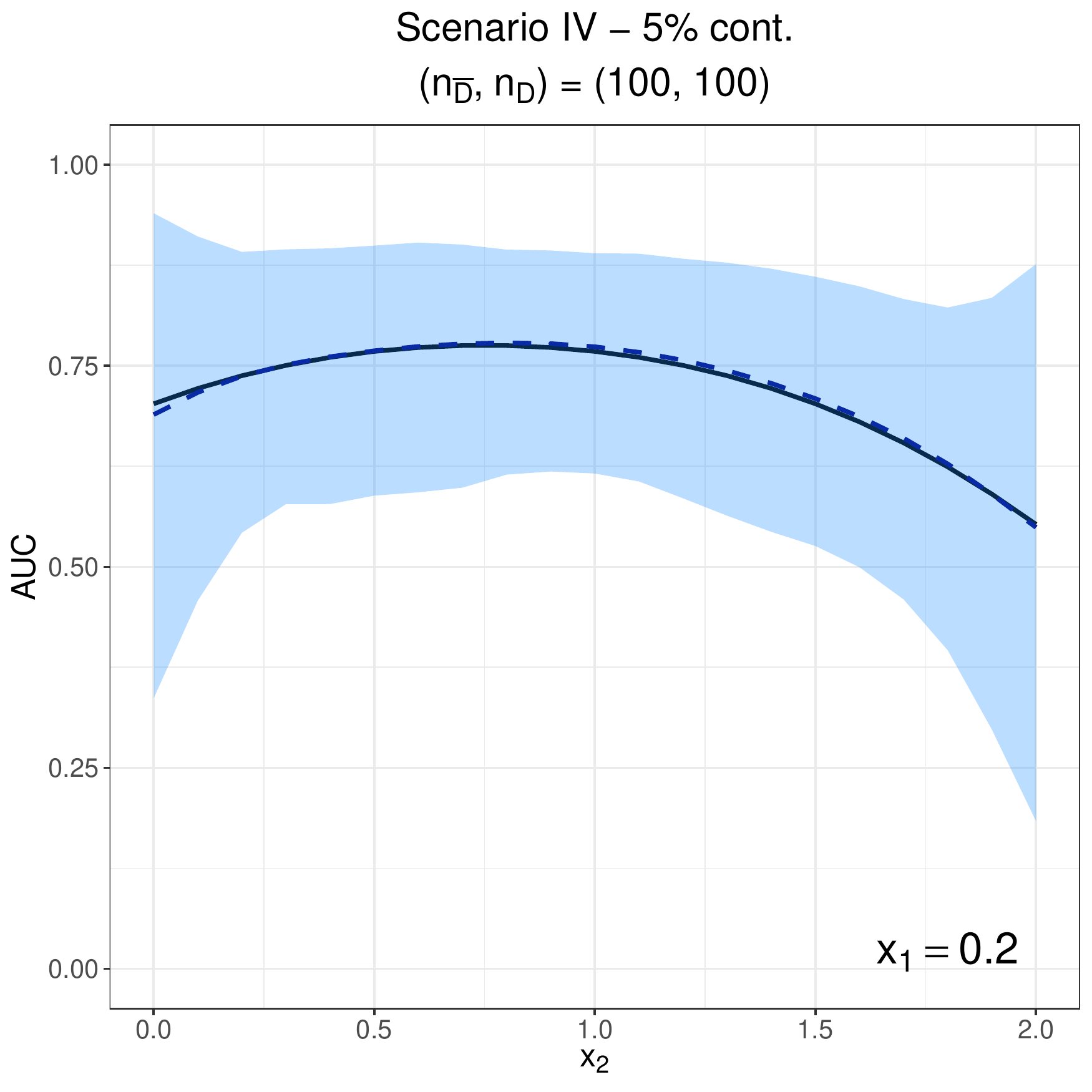}
			\includegraphics[width = 3.35cm]{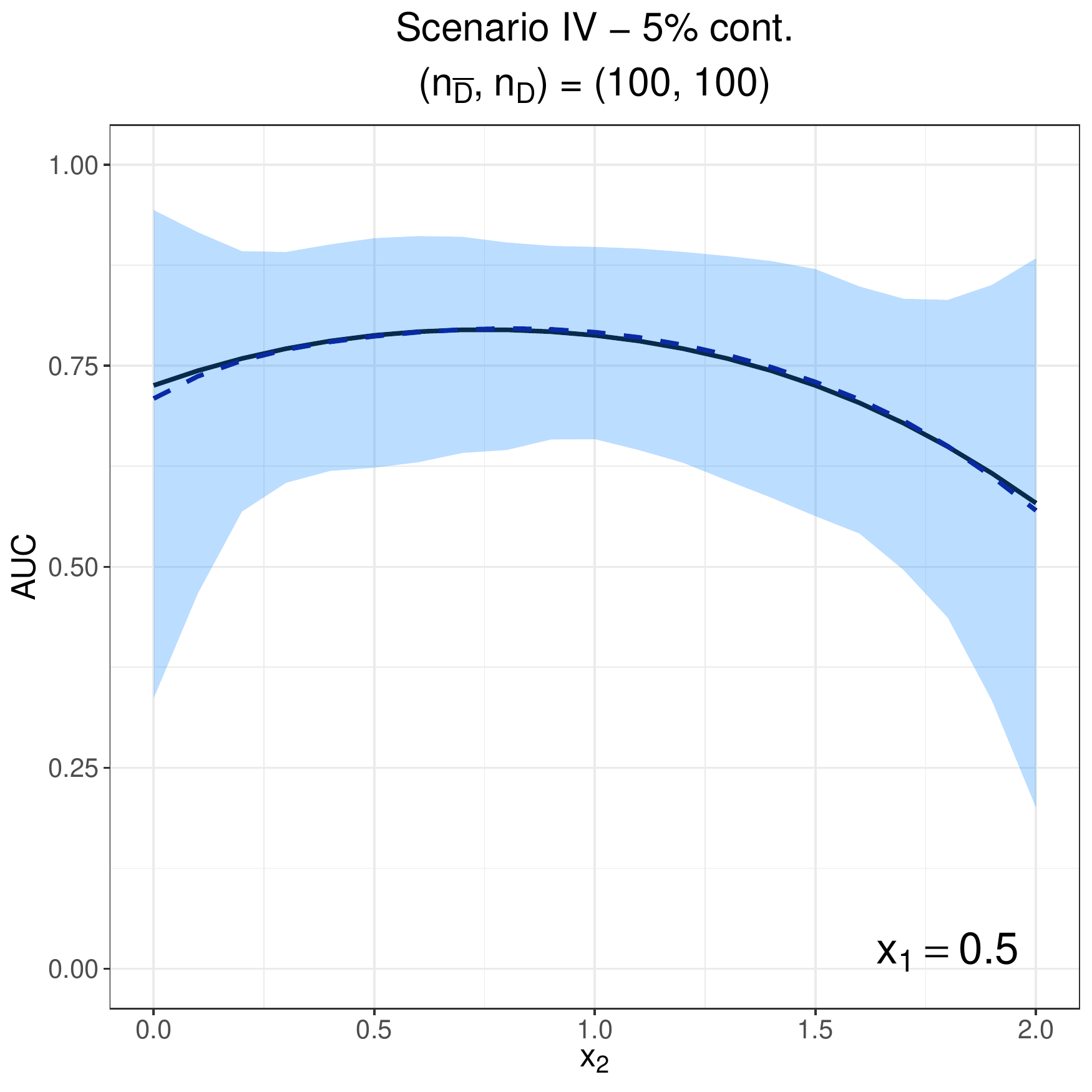}
			\includegraphics[width = 3.35cm]{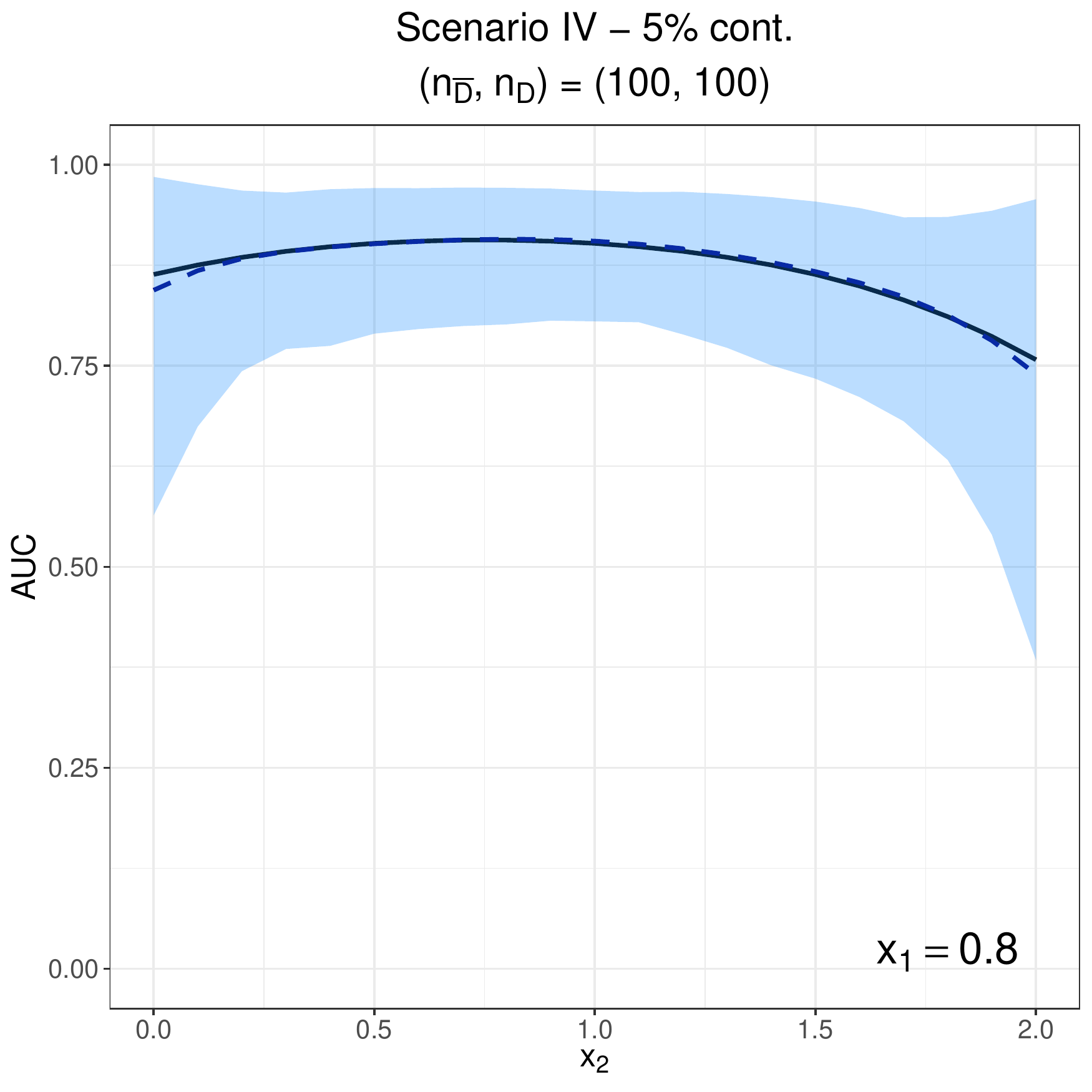}
		}
		\vspace{0.1cm}
		\subfigure{
			\includegraphics[width = 3.35cm]{aucrcont5sampsize2sc4xpred1x2v1.pdf}
			\includegraphics[width = 3.35cm]{aucrcont5sampsize2sc4xpred1x2v2.pdf}
			\includegraphics[width = 3.35cm]{aucrcont5sampsize2sc4xpred1x2v3.pdf}
		}
		\subfigure{
			\includegraphics[width = 3.35cm]{aucrcont5sampsize2sc4xpred2x1v1.pdf}
			\includegraphics[width = 3.35cm]{aucrcont5sampsize2sc4xpred2x1v2.pdf}
			\includegraphics[width = 3.35cm]{aucrcont5sampsize2sc4xpred2x1v3.pdf}
		}
		\vspace{0.1cm}
		\subfigure{
			\includegraphics[width = 3.35cm]{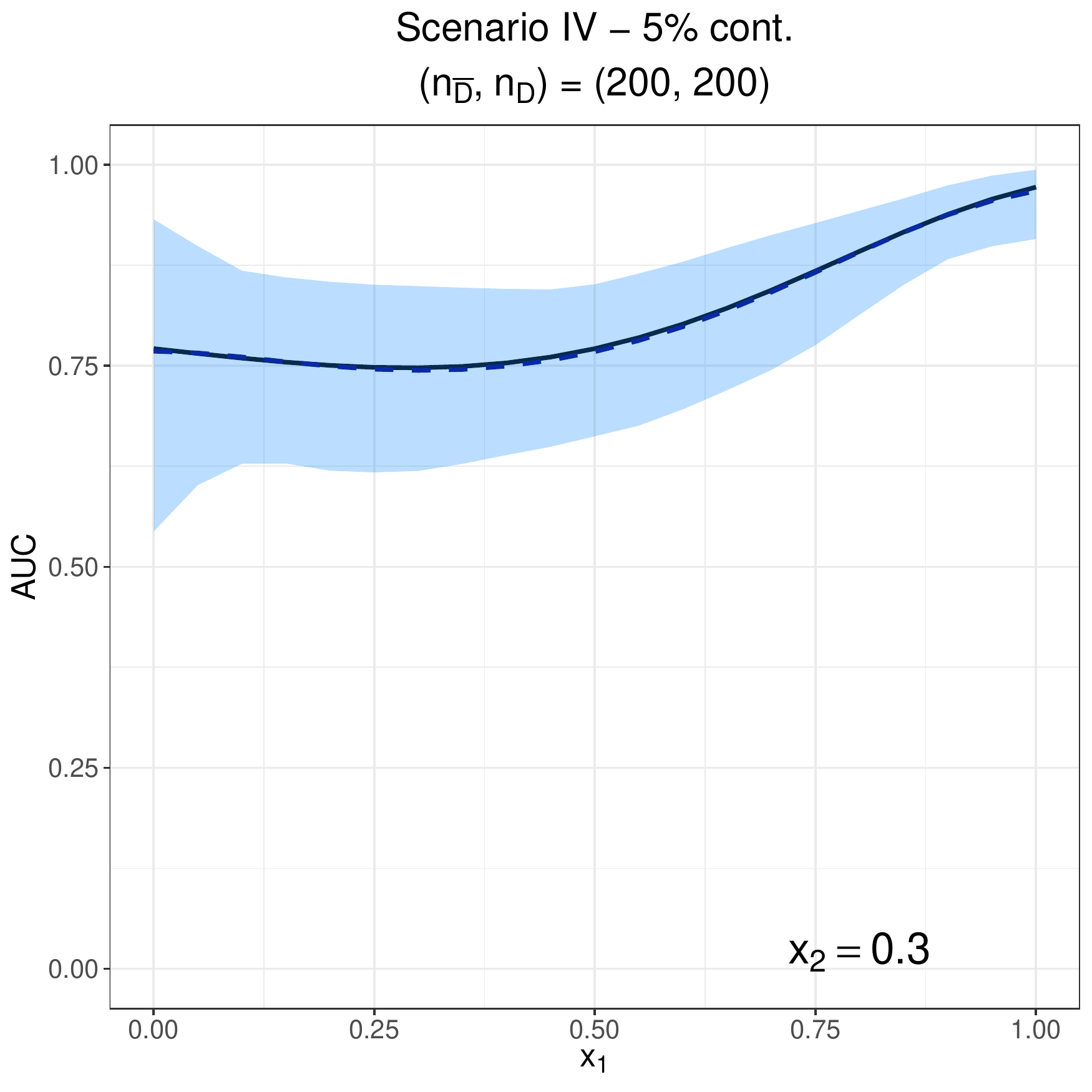}
			\includegraphics[width = 3.35cm]{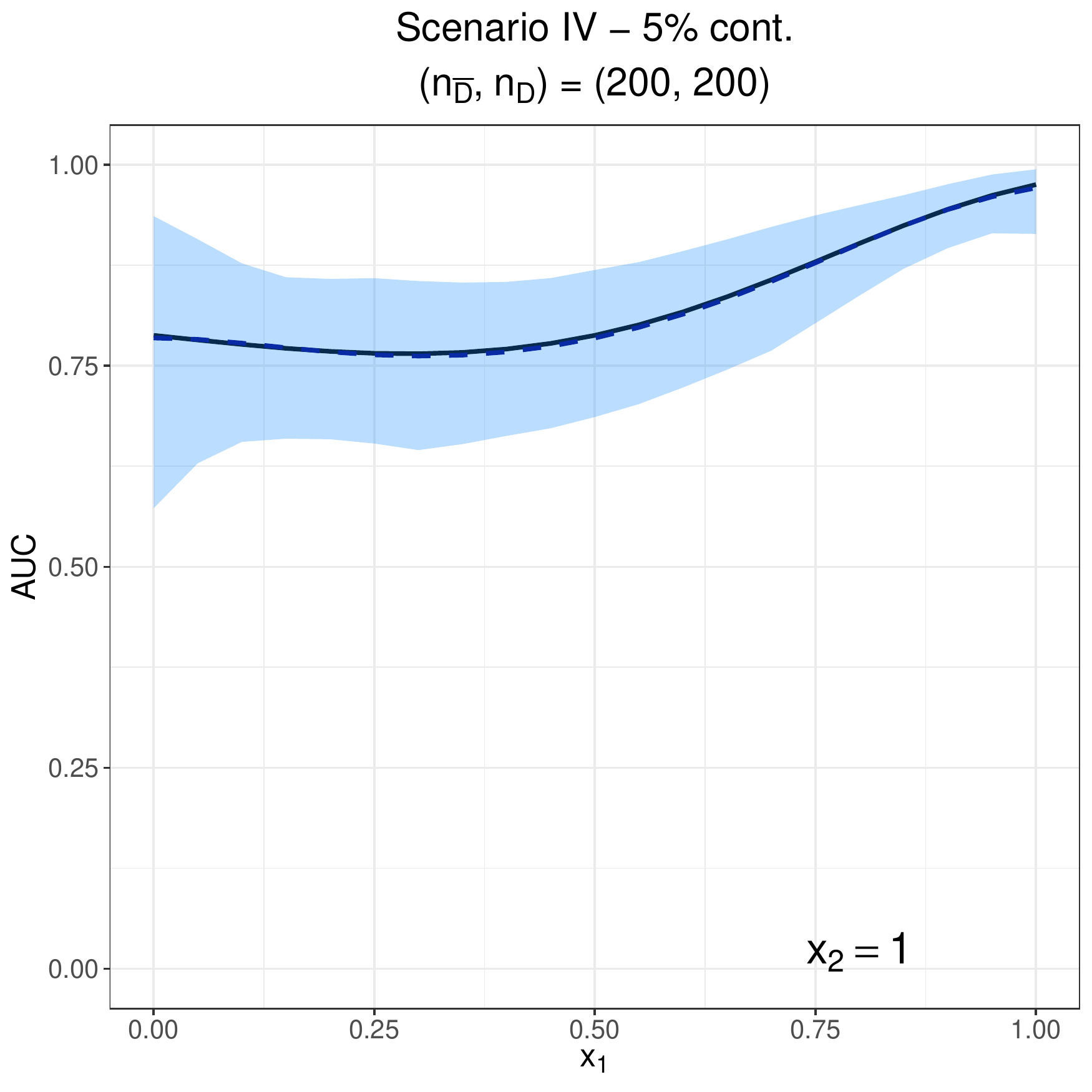}
			\includegraphics[width = 3.35cm]{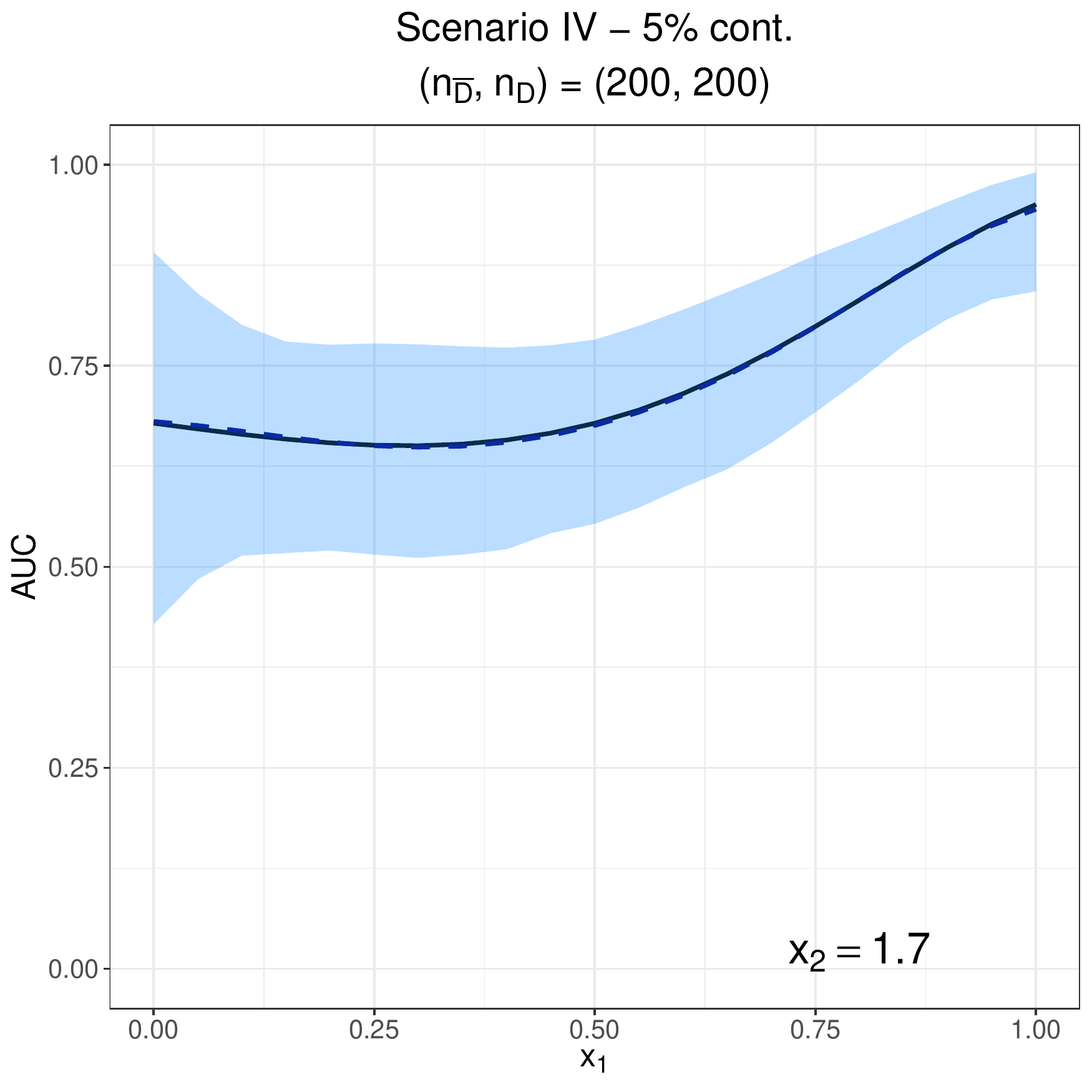}
		}
		\subfigure{
			\includegraphics[width = 3.35cm]{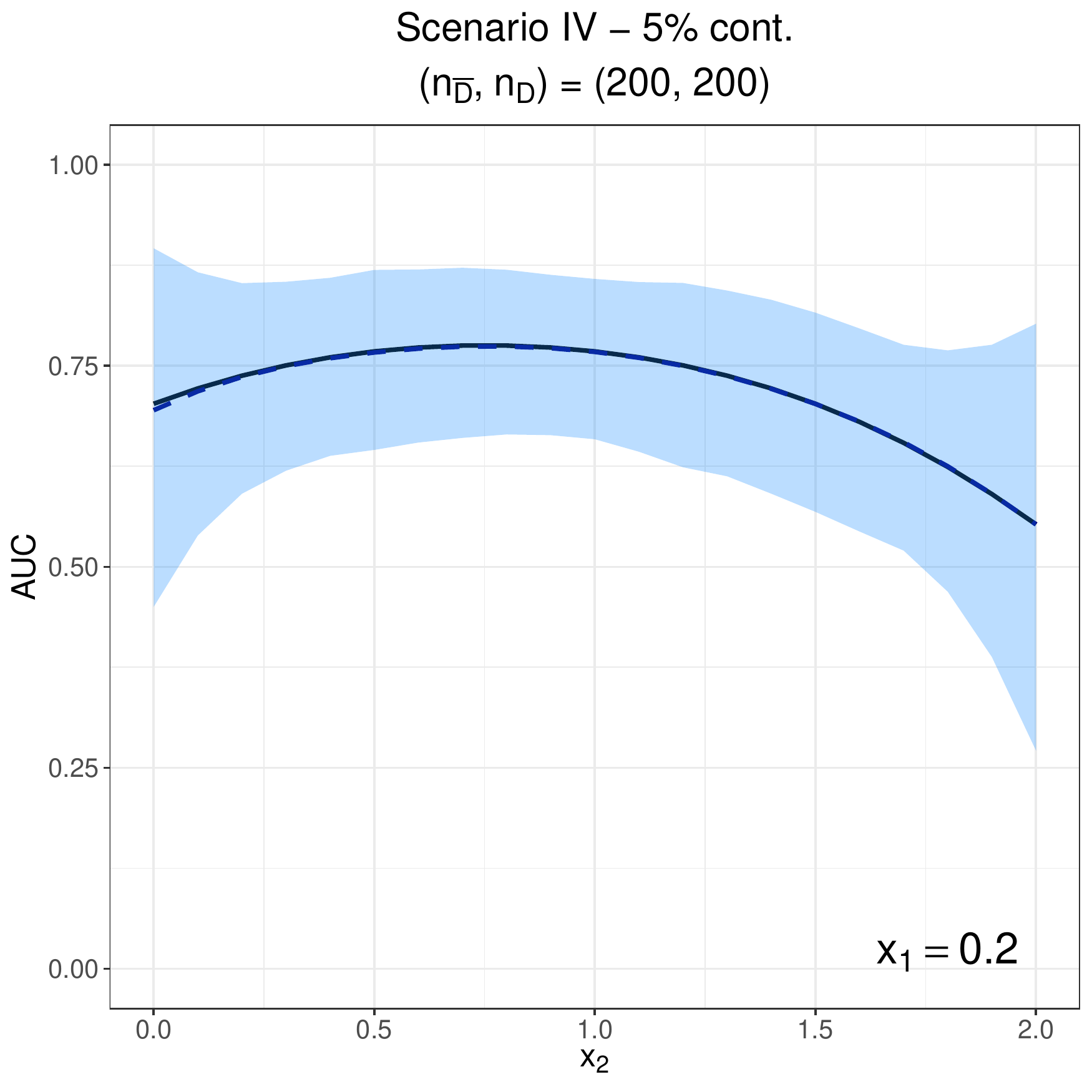}	
			\includegraphics[width = 3.35cm]{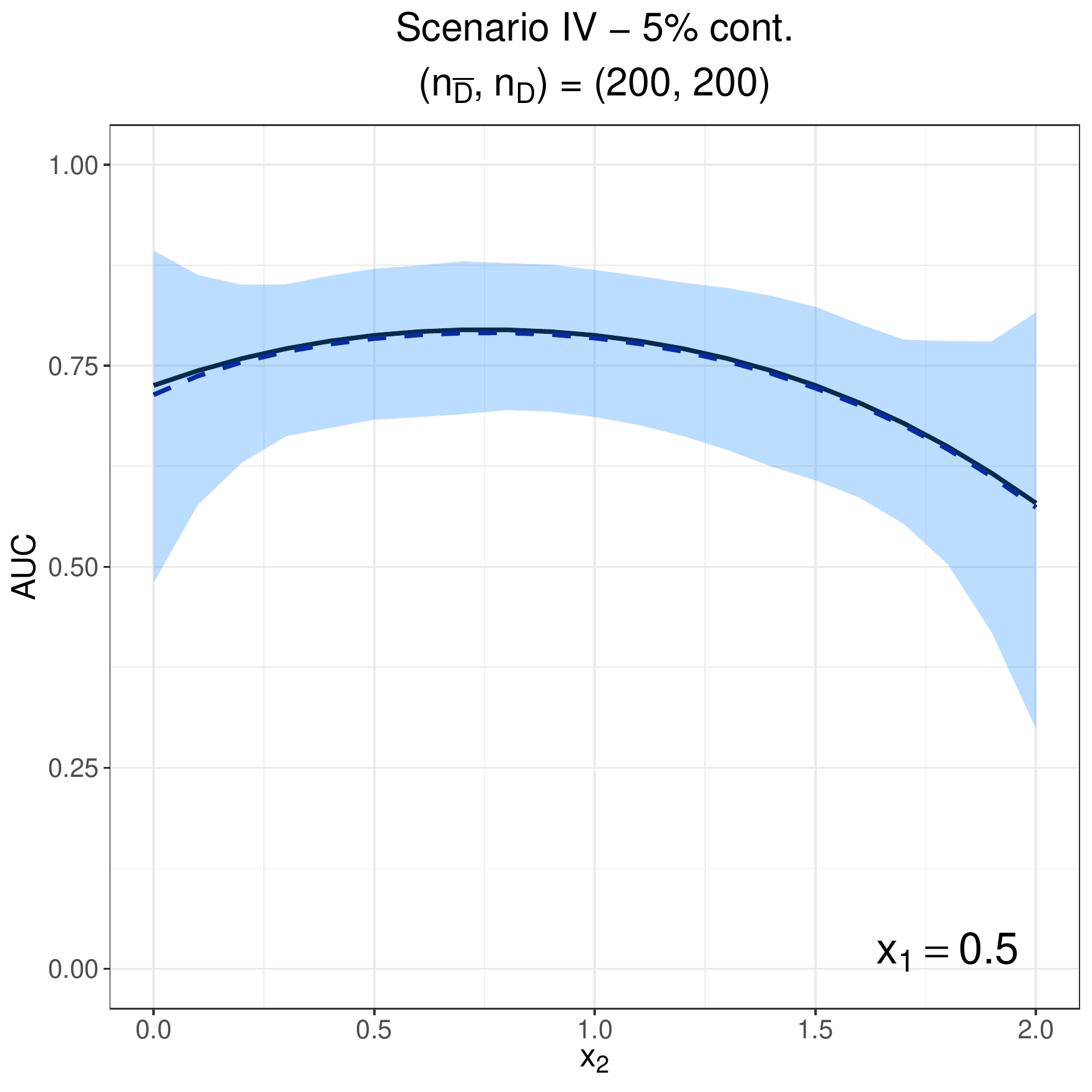}
			\includegraphics[width = 3.35cm]{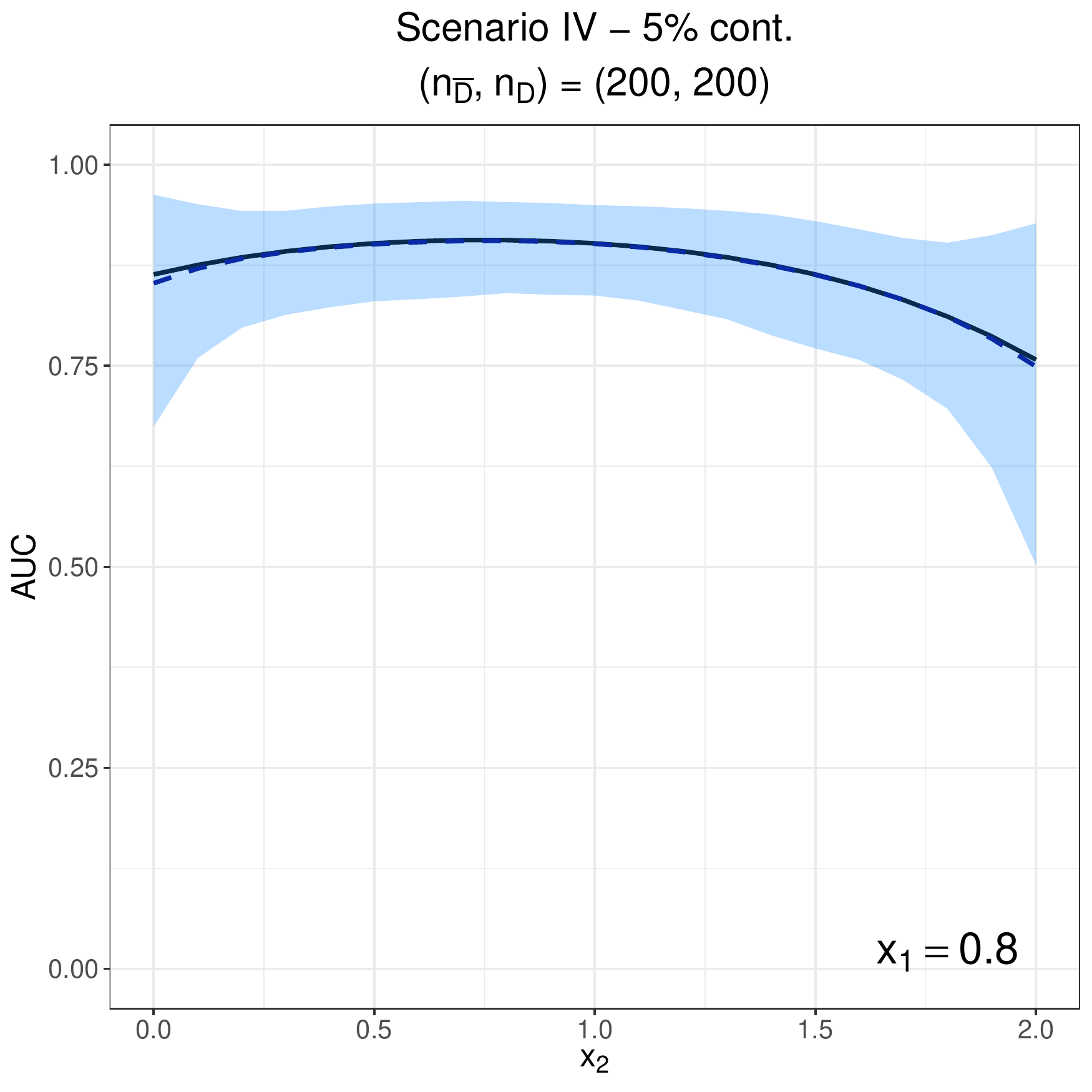}
		}
	\end{center}
	\vspace{-0.3cm}
	\caption{\tiny{Scenario IV. Multiple profiles of the true covariate-specific AUC (solid line) versus the mean of the Monte Carlo estimates (dashed line) along with the $2.5\%$ and $97.5\%$ simulation quantiles (shaded area) for the case of $5\%$ of contamination. Rows 1 and 2 displays the results for $(n_{\bar{D}}, n_D)=(100,100)$, rows 3 and 4 for $(n_{\bar{D}}, n_D)=(200,100)$, and rows 5 and 6 for $(n_{\bar{D}}, n_D)=(200,200)$.}}
\end{figure}

\begin{figure}[H]
	\begin{center}
		\subfigure{
			\includegraphics[width = 3.35cm]{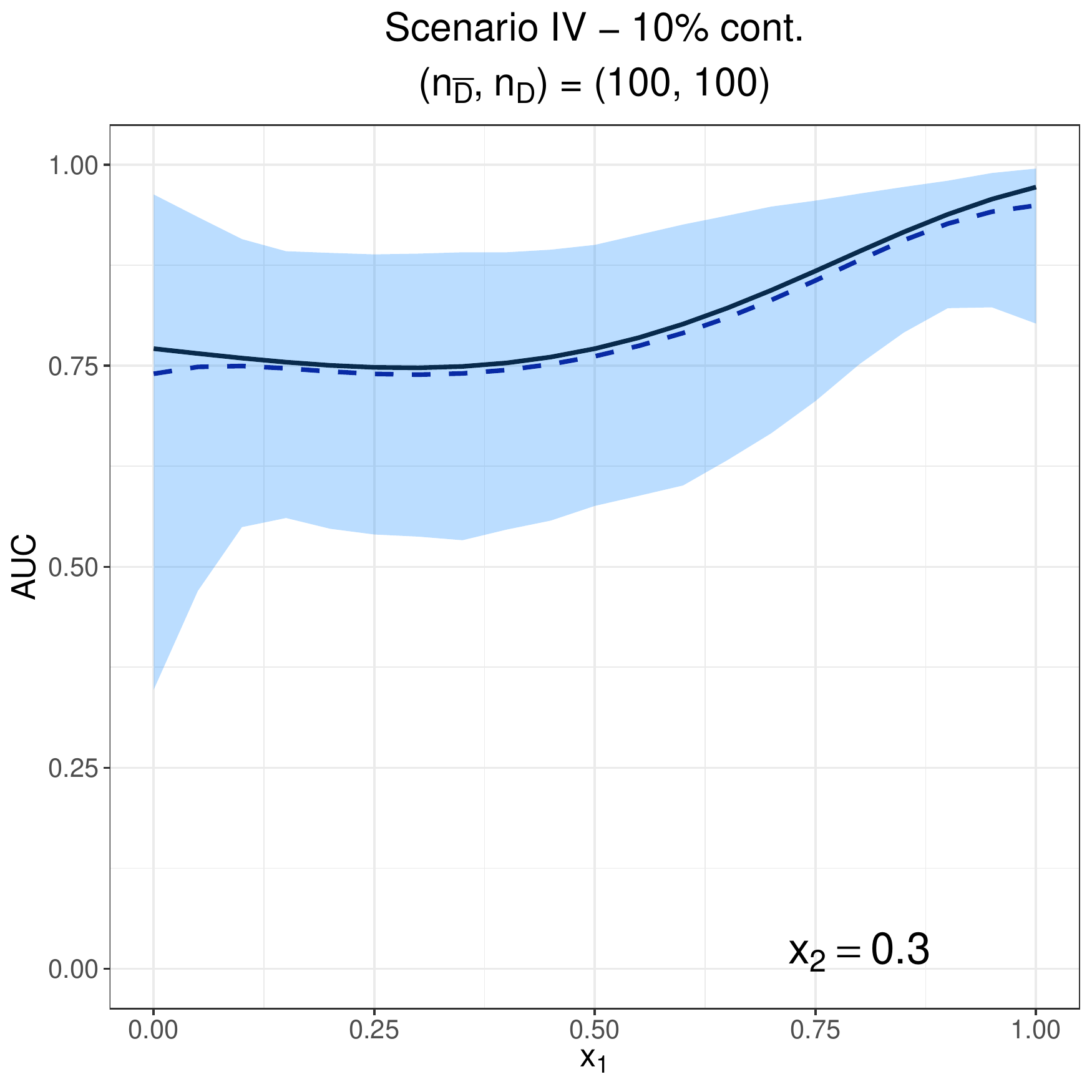}
			\includegraphics[width = 3.35cm]{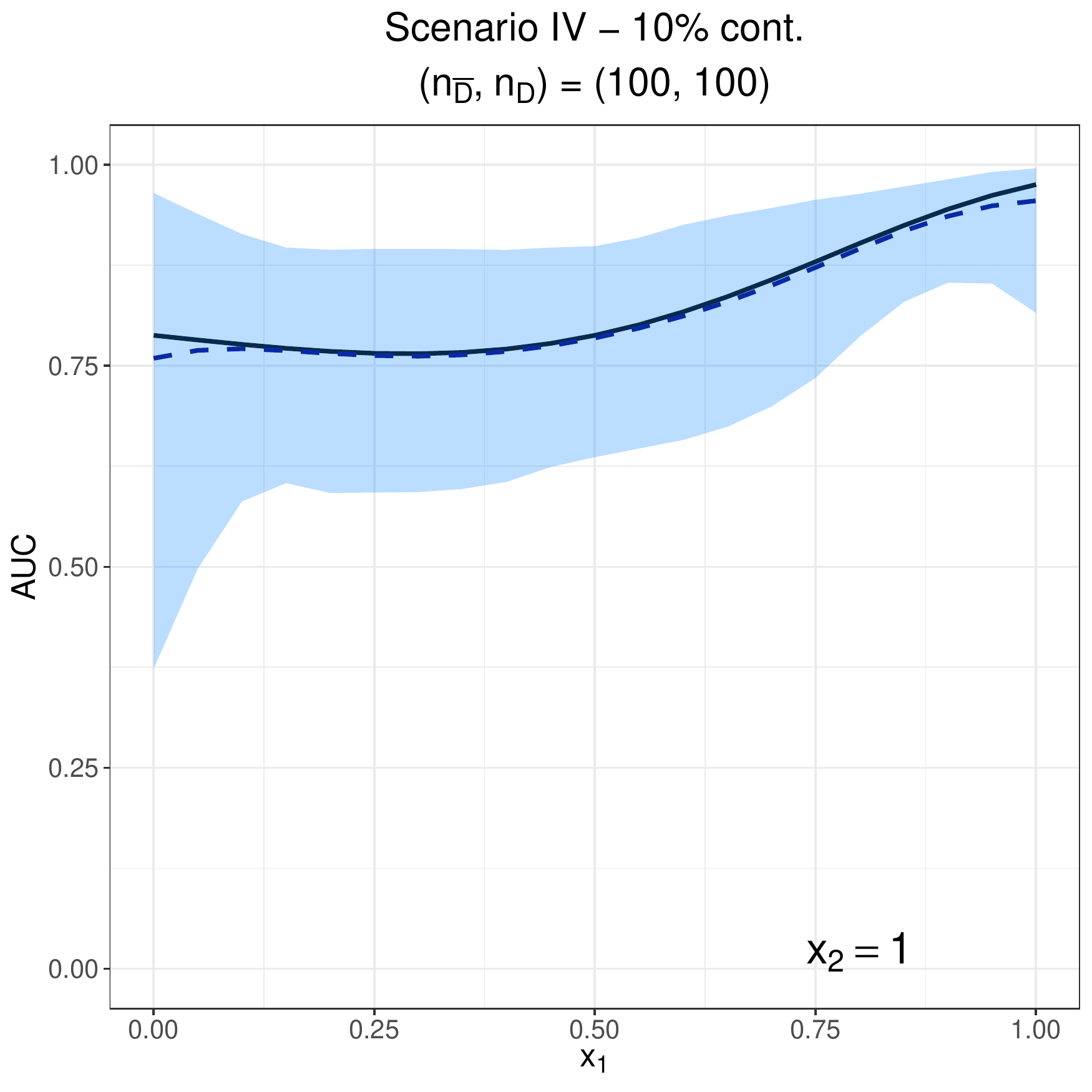}
			\includegraphics[width = 3.35cm]{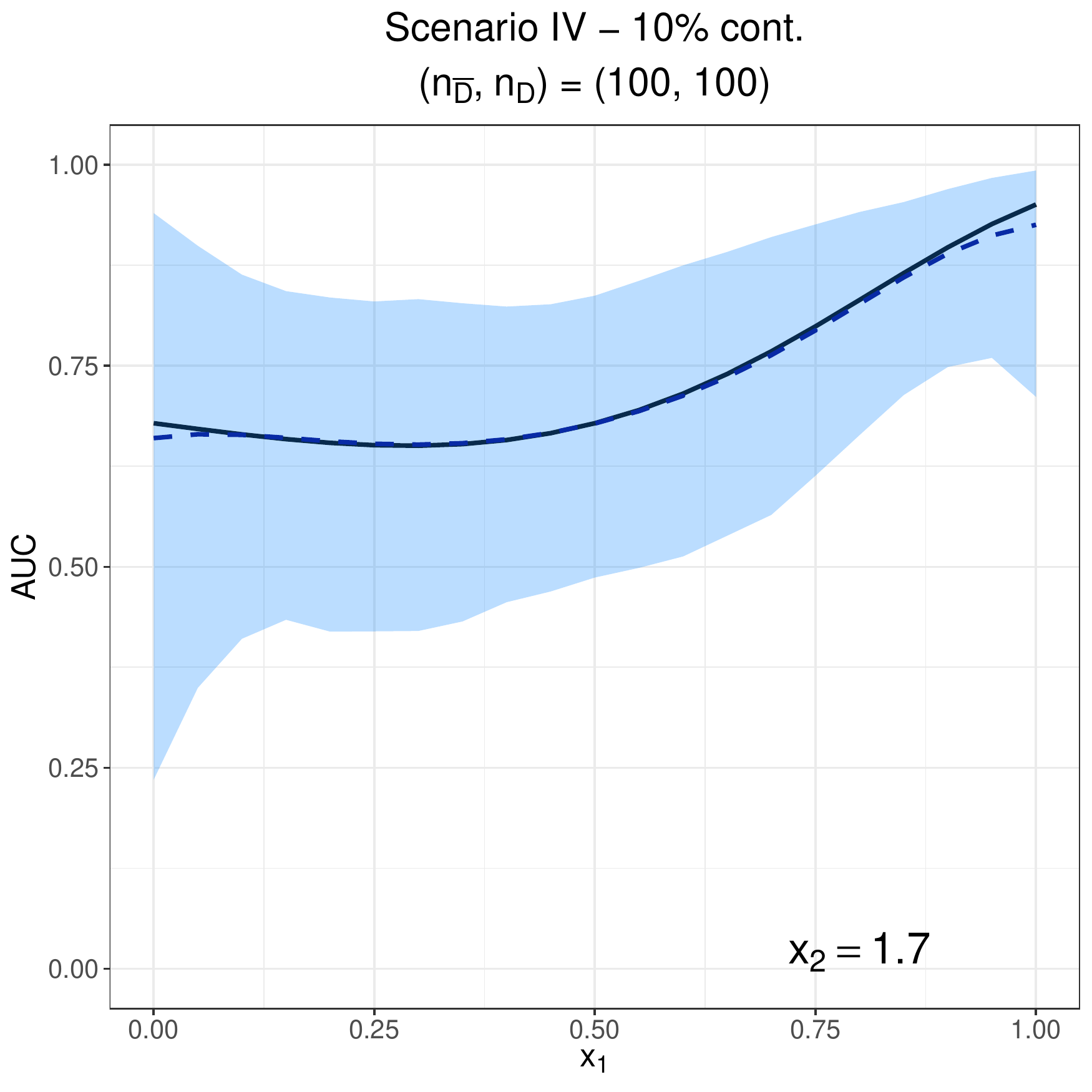}
		}
		\subfigure{
			\includegraphics[width = 3.35cm]{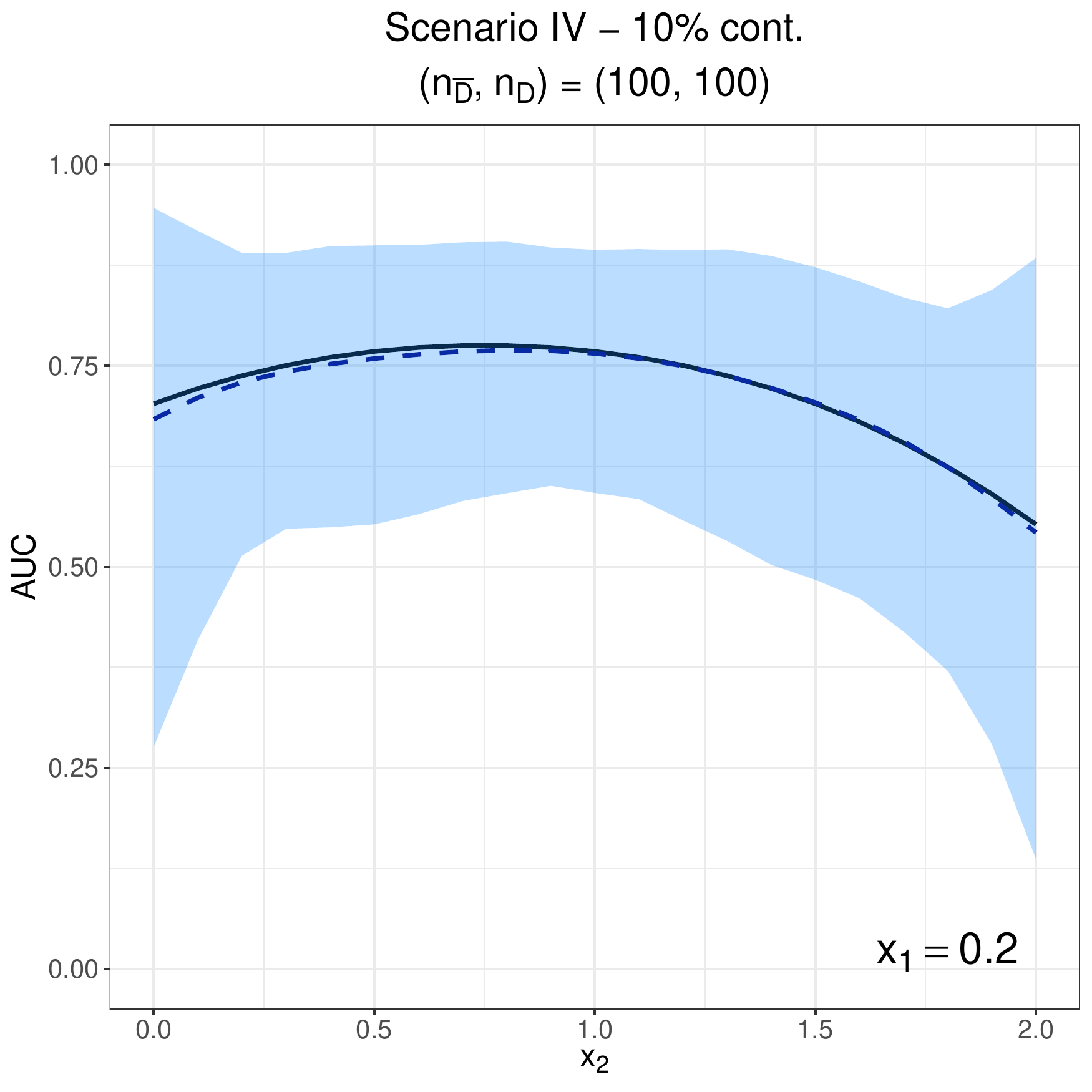}
			\includegraphics[width = 3.35cm]{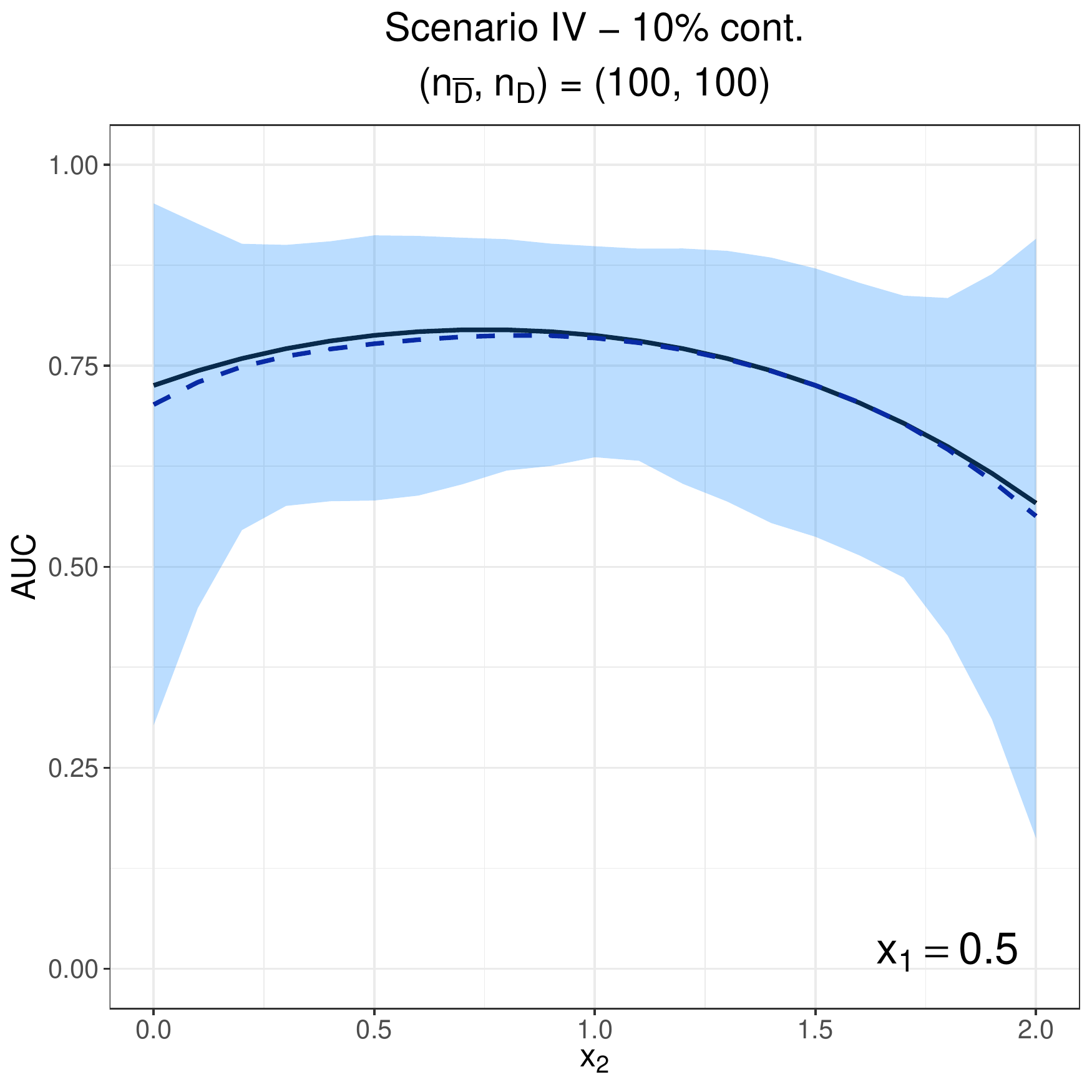}
			\includegraphics[width = 3.35cm]{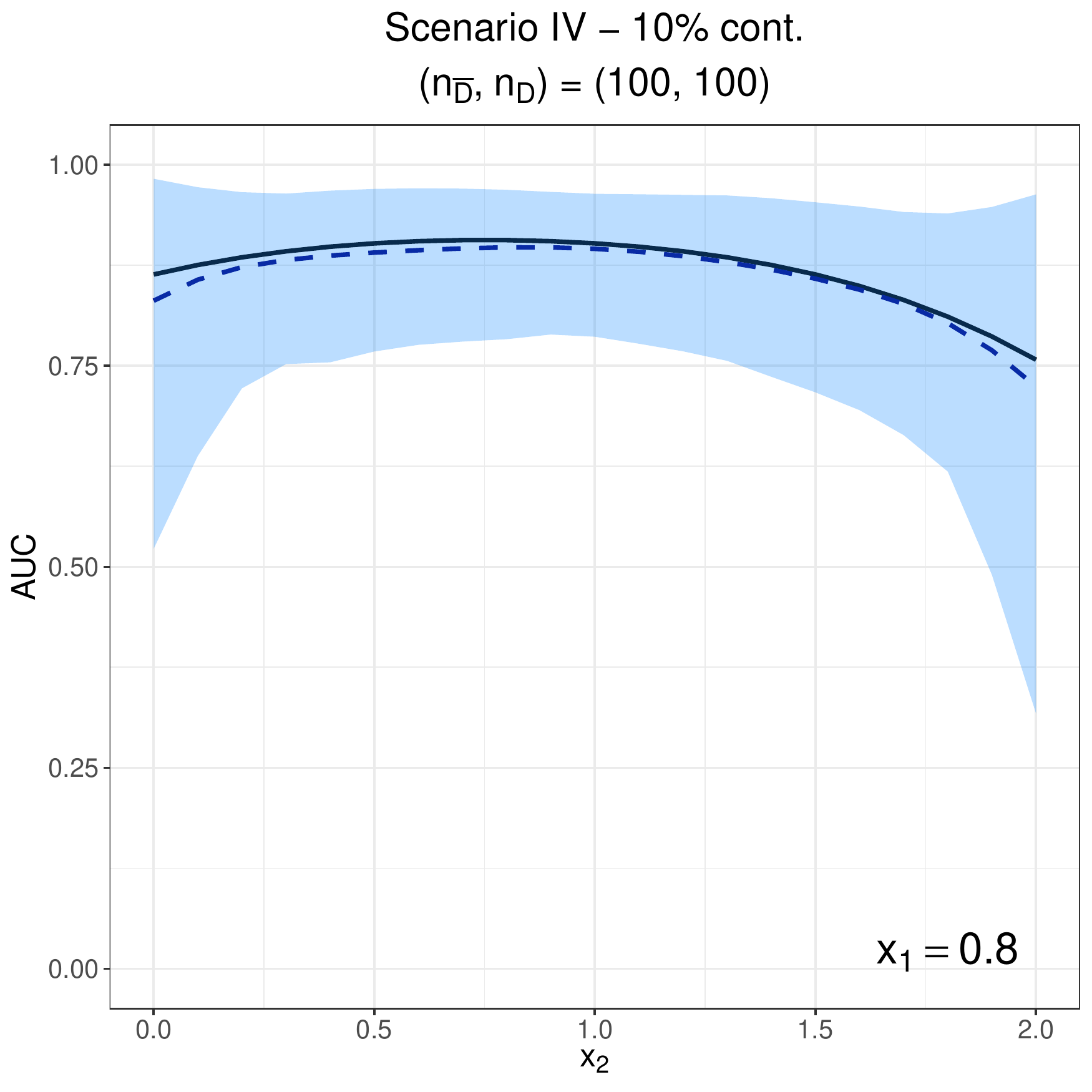}
		}
		\vspace{0.1cm}
		\subfigure{
			\includegraphics[width = 3.35cm]{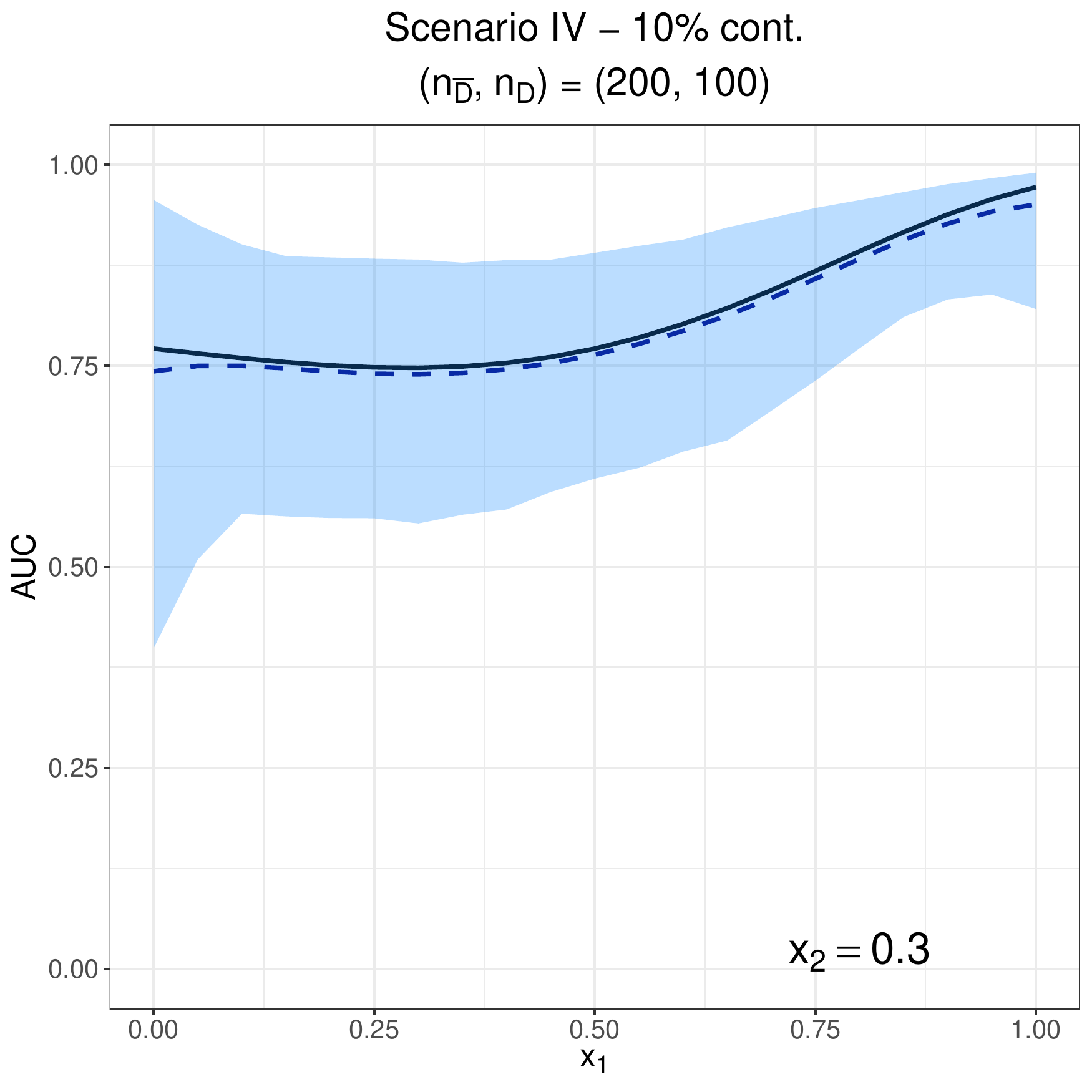}
			\includegraphics[width = 3.35cm]{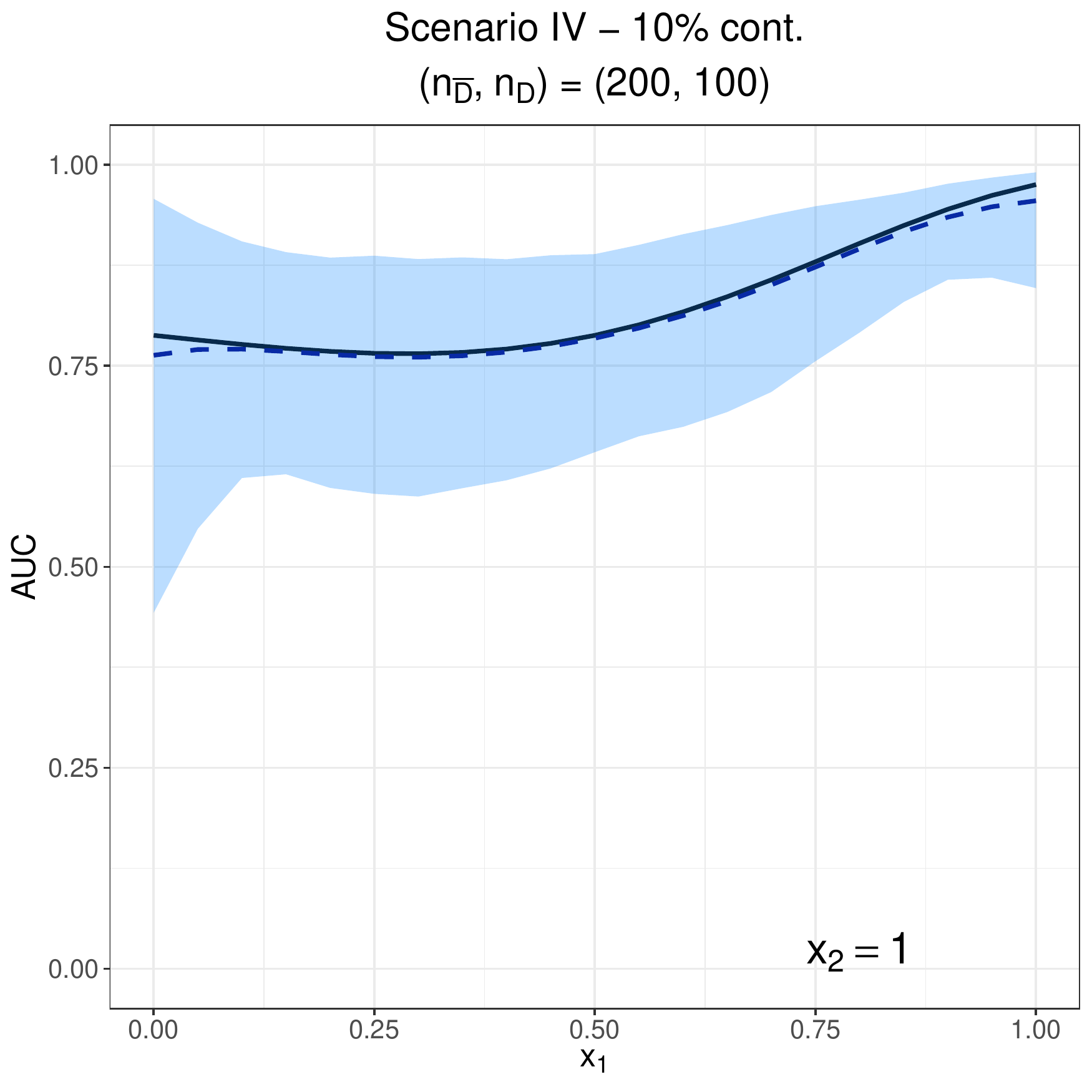}
			\includegraphics[width = 3.35cm]{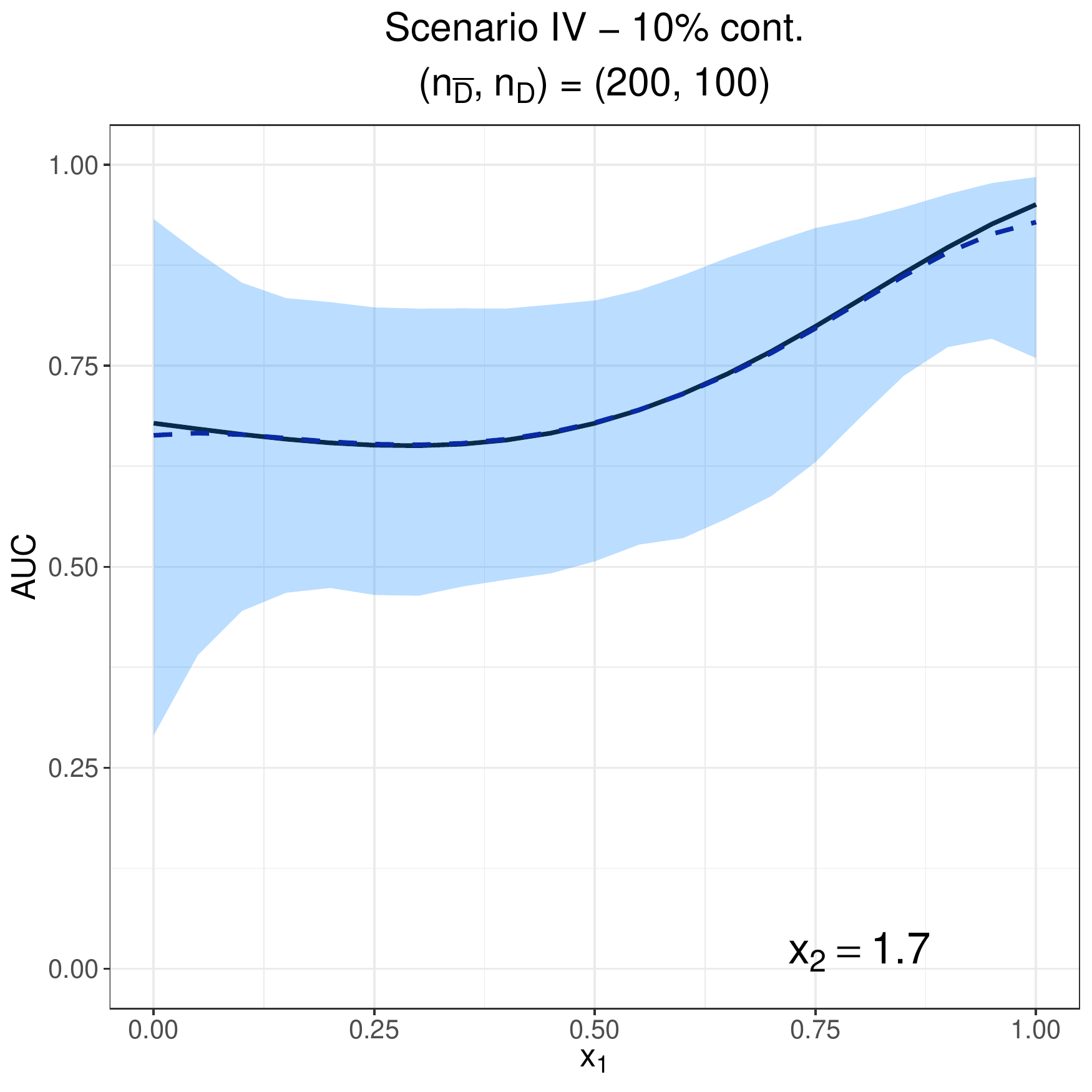}
		}
		\subfigure{
			\includegraphics[width = 3.35cm]{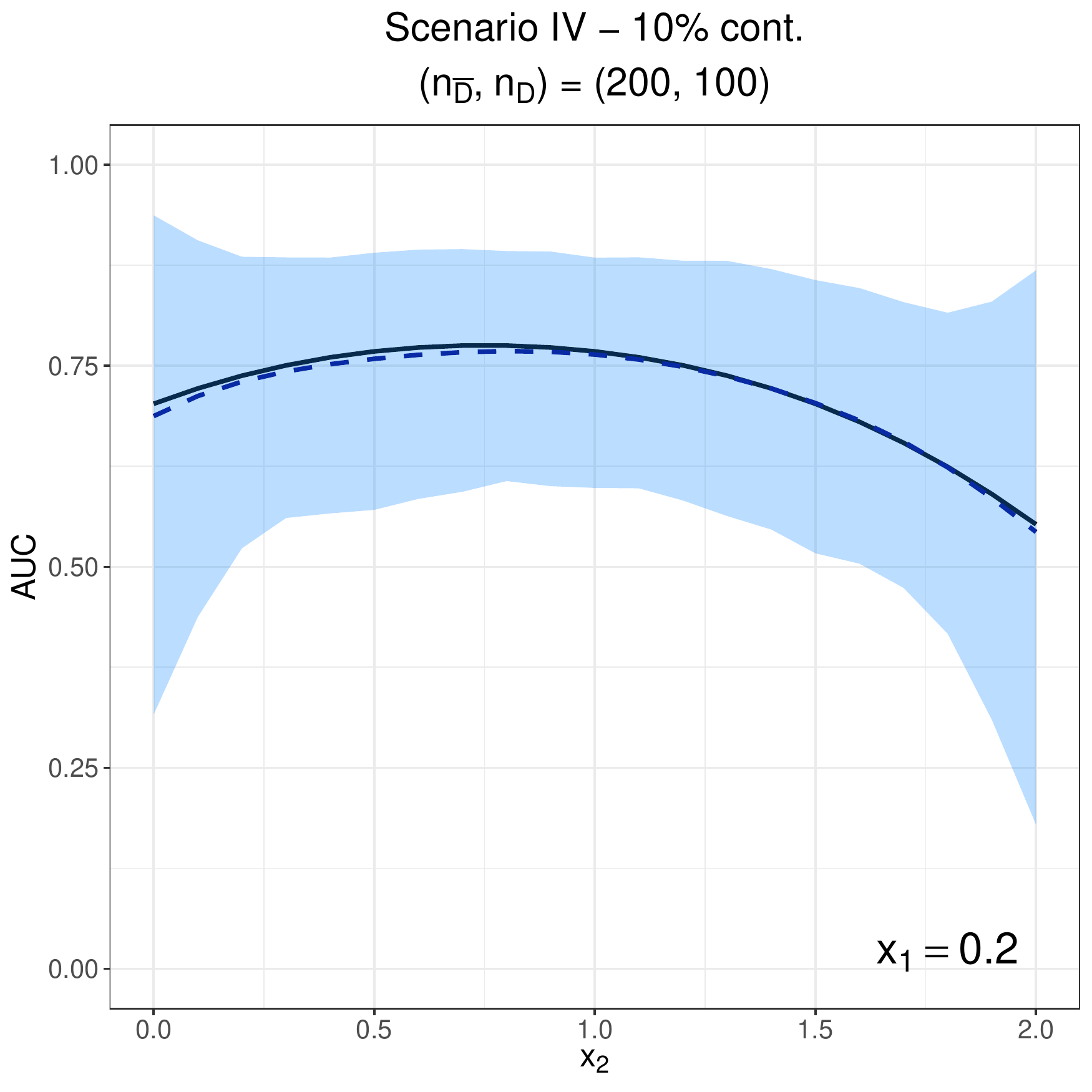}
			\includegraphics[width = 3.35cm]{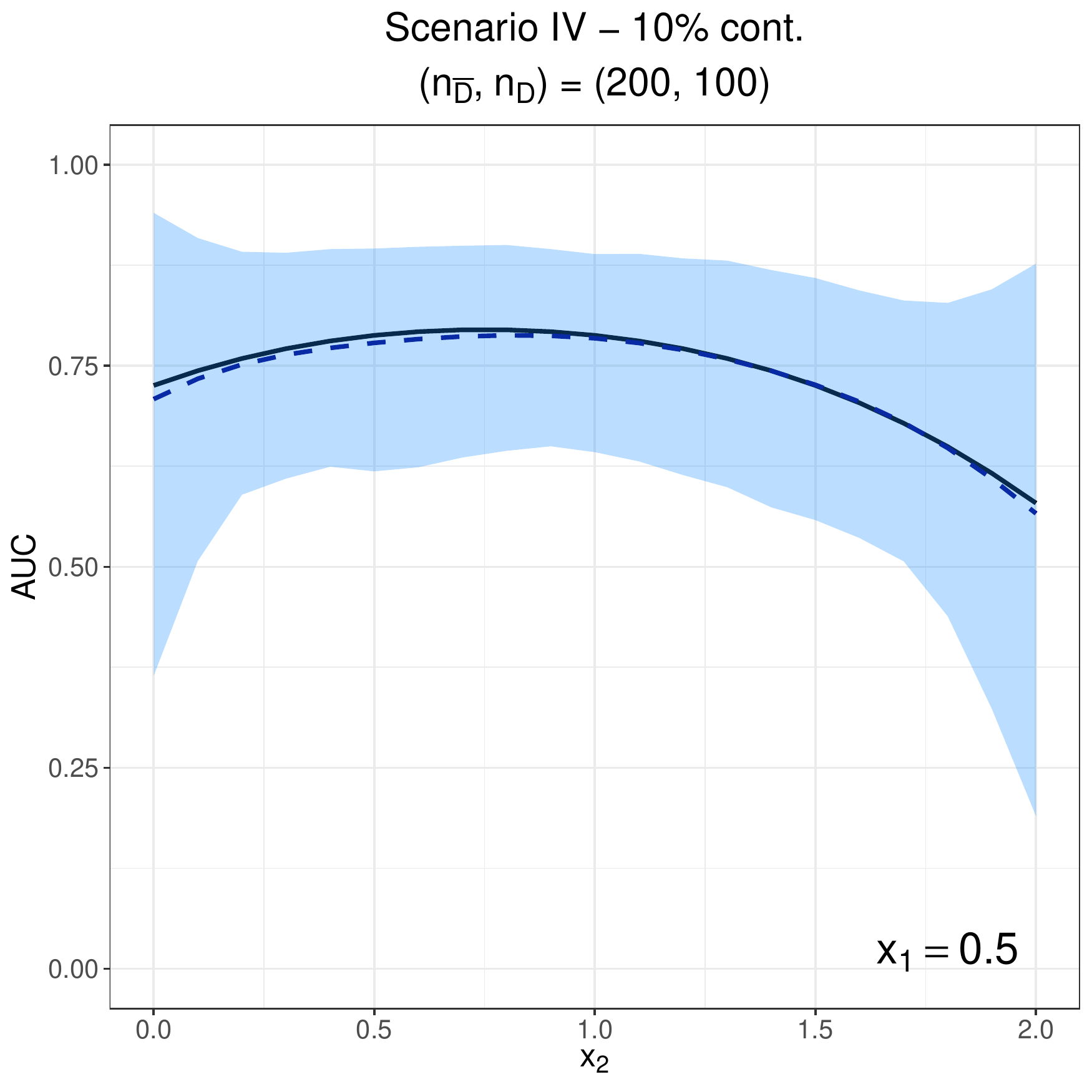}
			\includegraphics[width = 3.35cm]{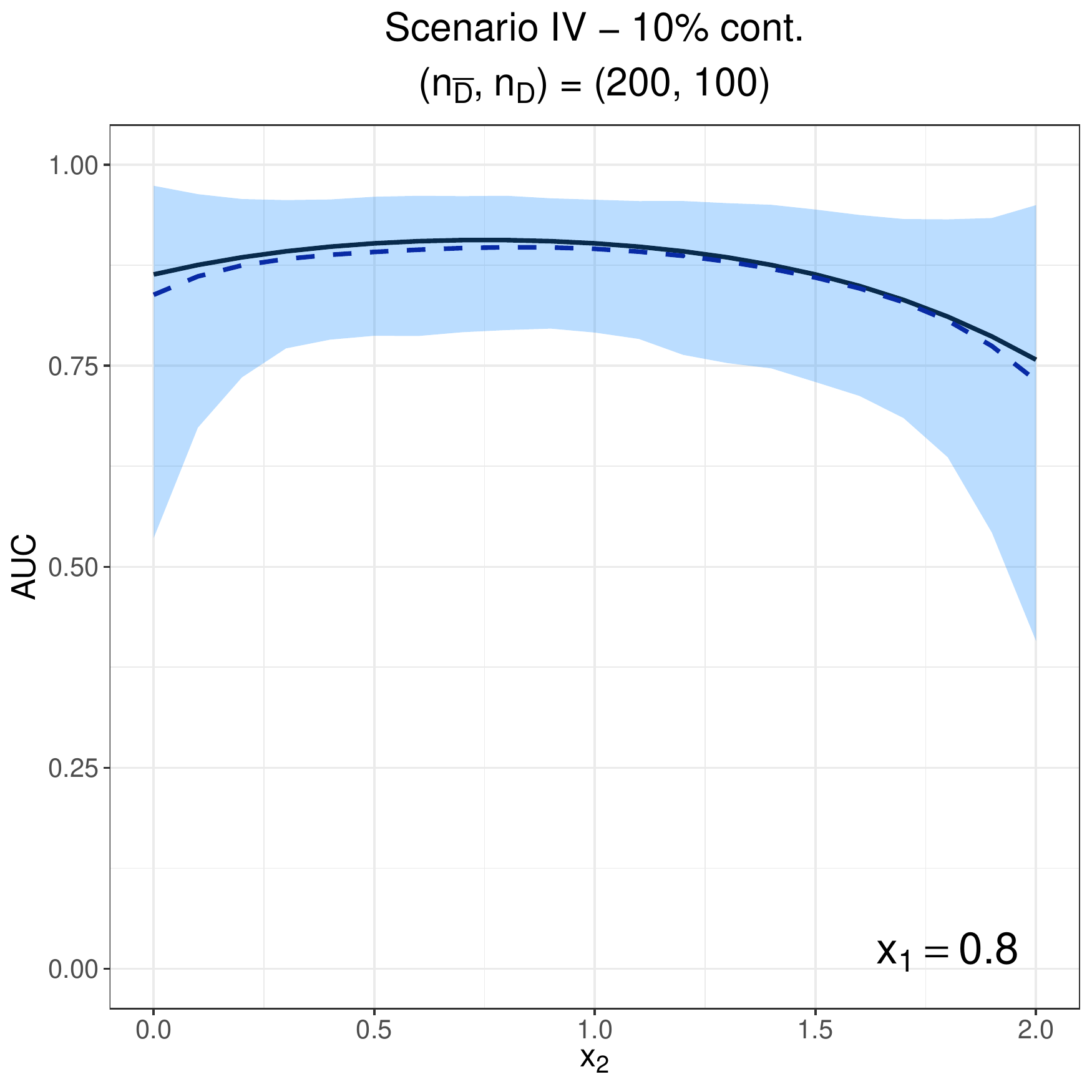}
		}
		\vspace{0.1cm}
		\subfigure{
			\includegraphics[width = 3.35cm]{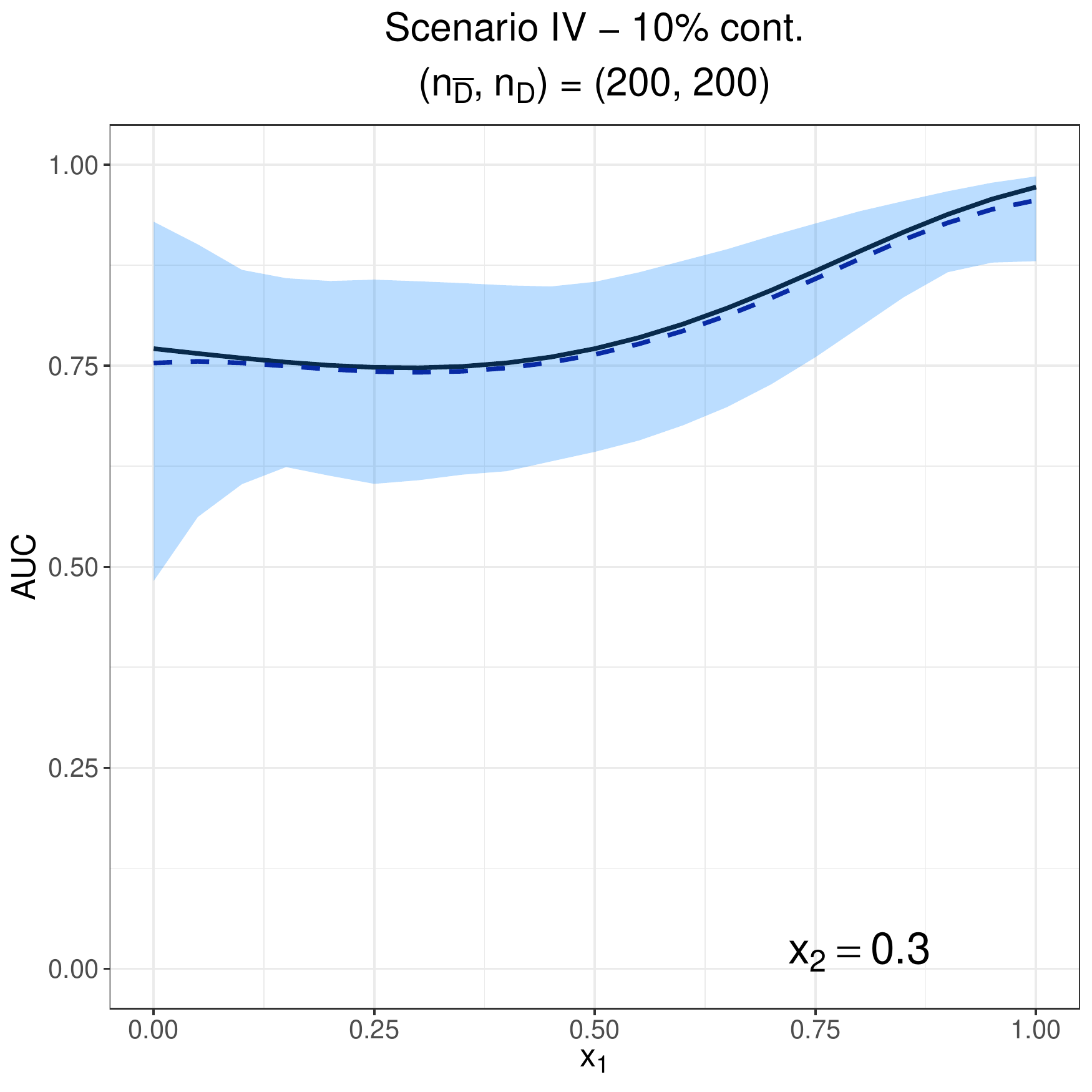}
			\includegraphics[width = 3.35cm]{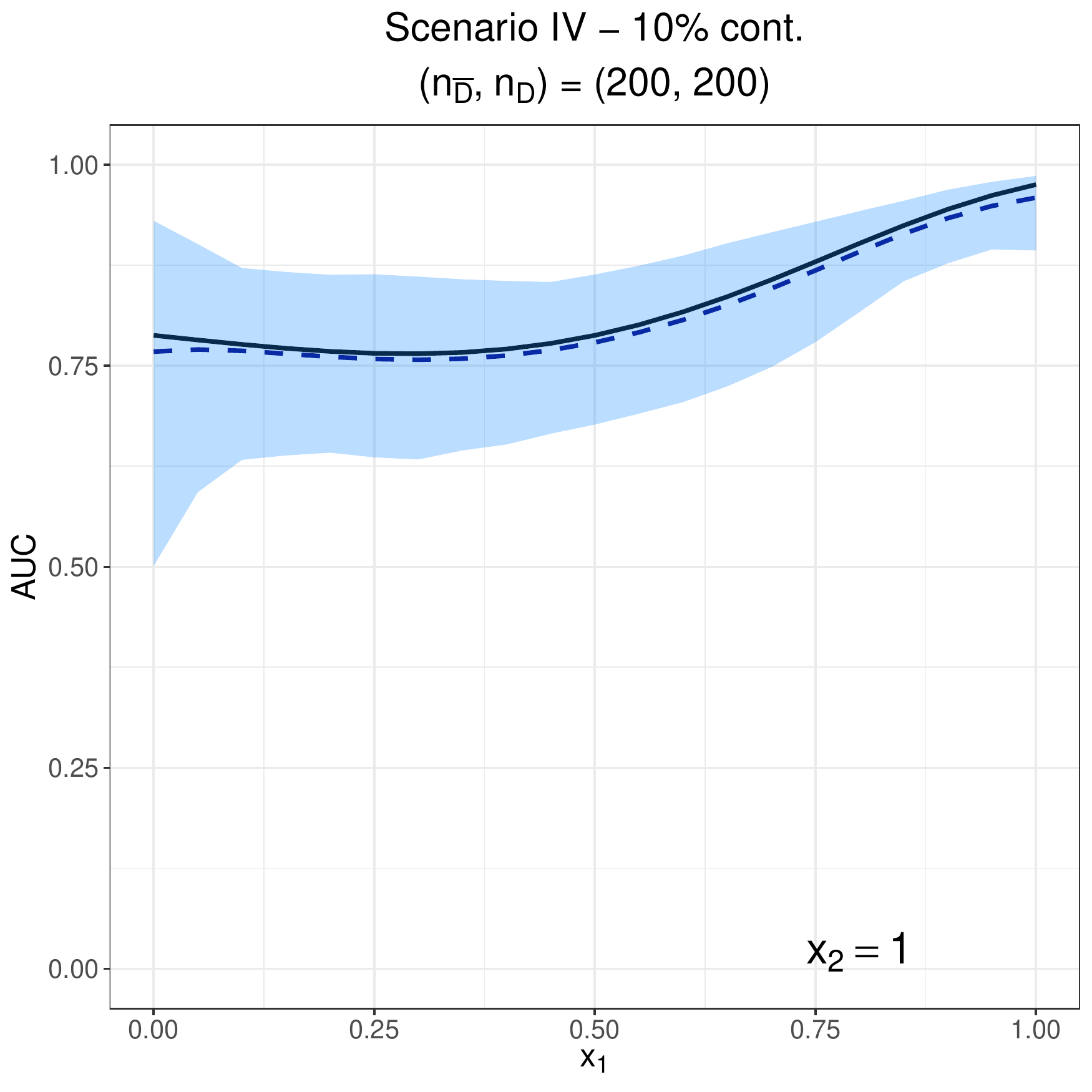}
			\includegraphics[width = 3.35cm]{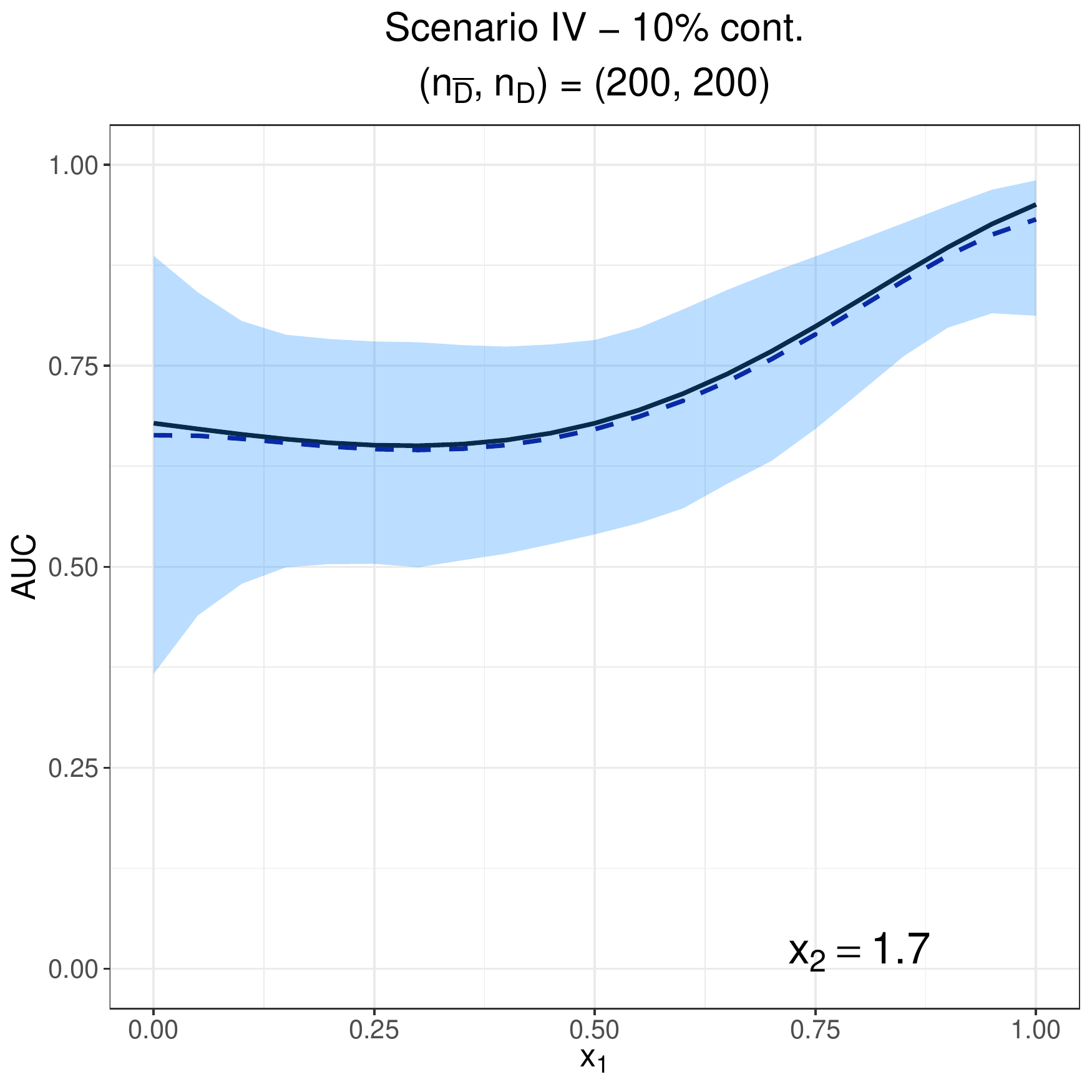}
		}
		\subfigure{
			\includegraphics[width = 3.35cm]{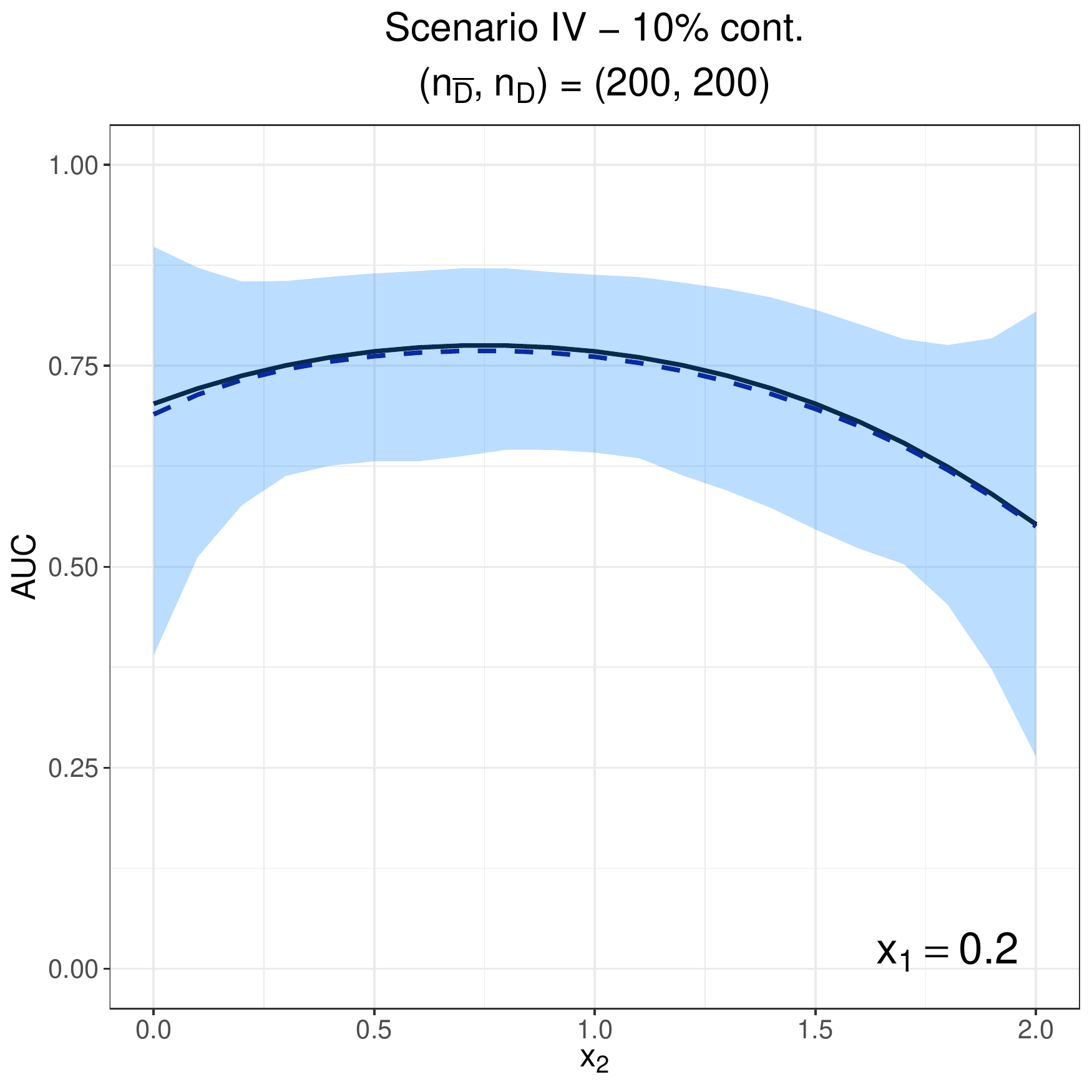}	
			\includegraphics[width = 3.35cm]{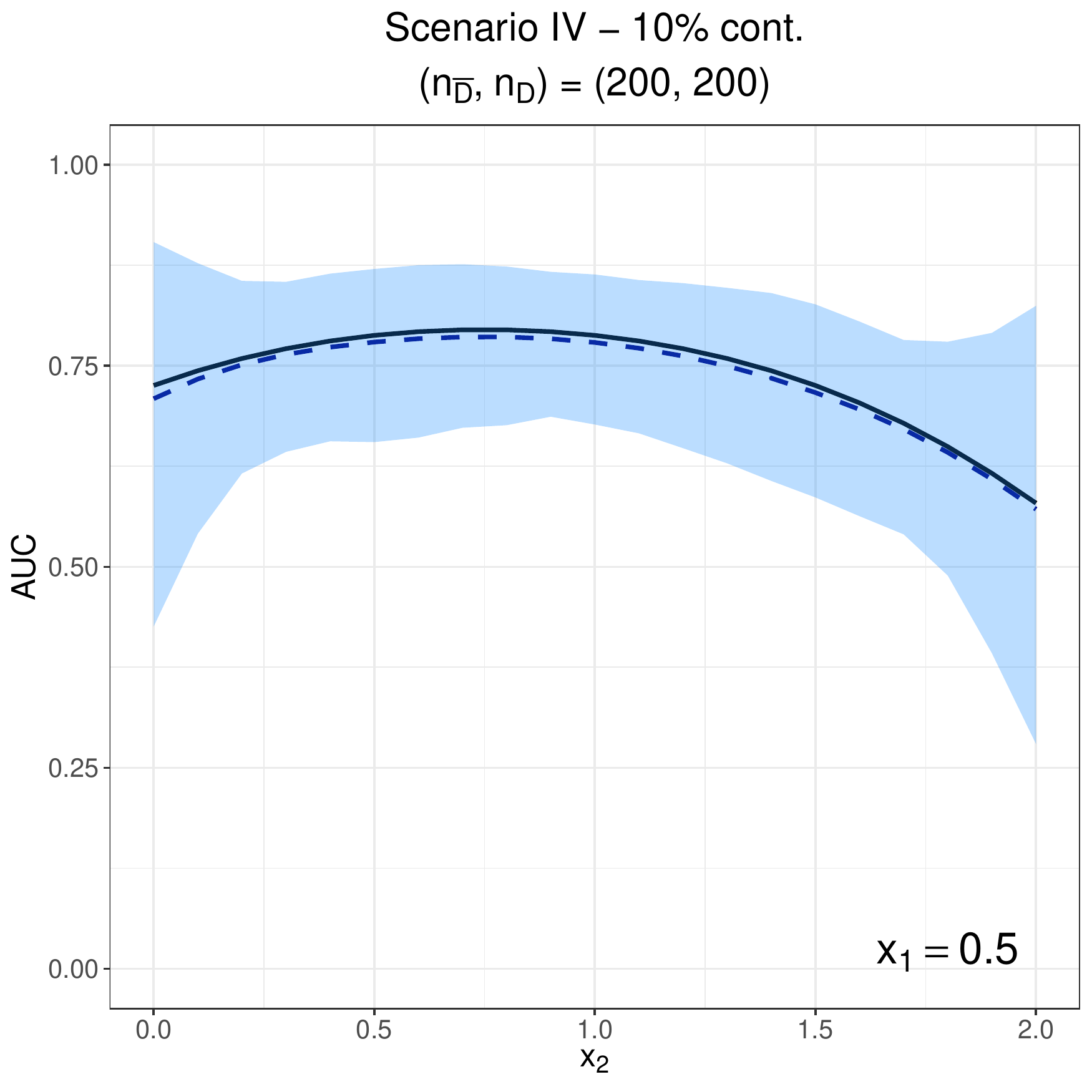}
			\includegraphics[width = 3.35cm]{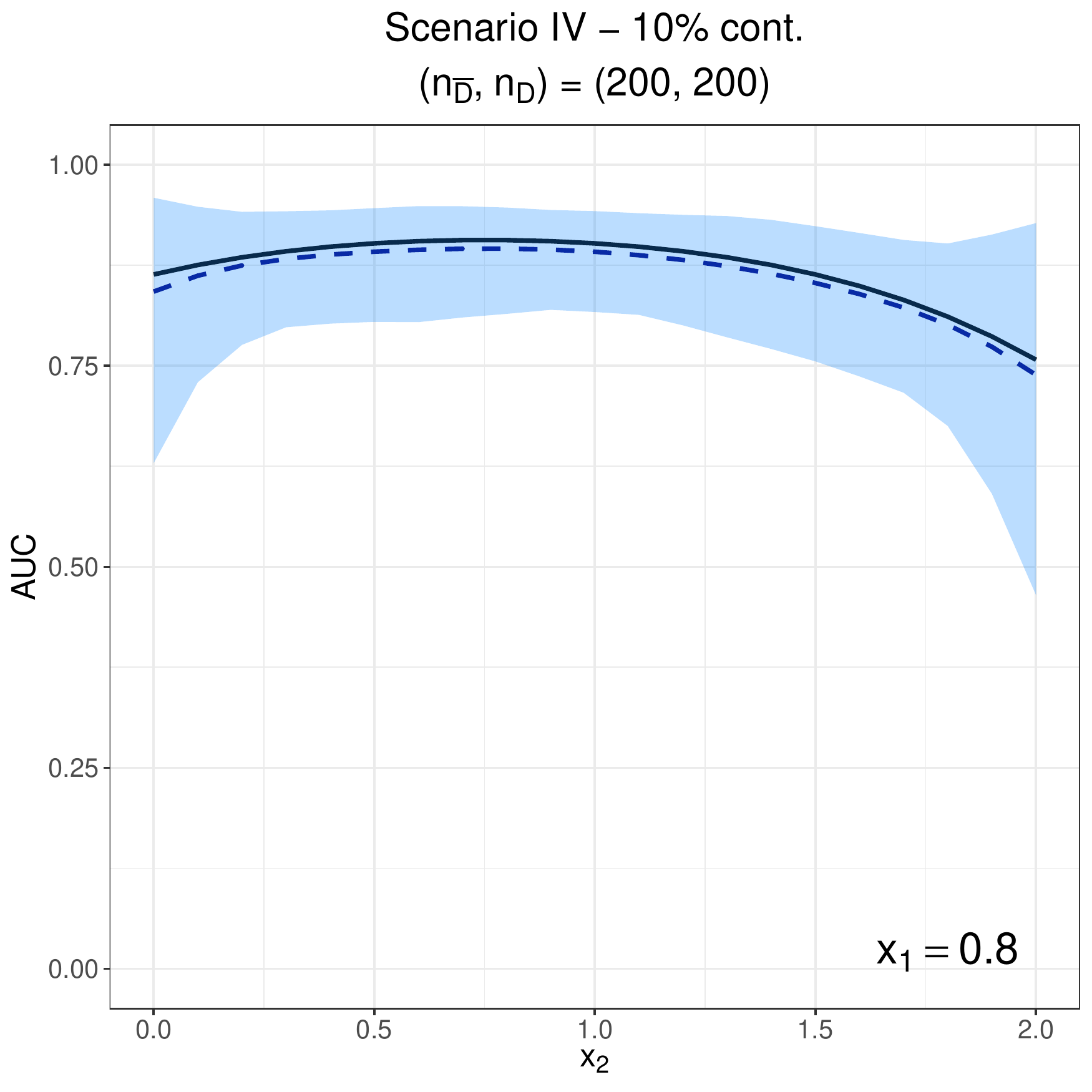}
		}
	\end{center}
	\vspace{-0.3cm}
	\caption{\tiny{Scenario IV. Multiple profiles of the true covariate-specific AUC (solid line) versus the mean of the Monte Carlo estimates (dashed line) along with the $2.5\%$ and $97.5\%$ simulation quantiles (shaded area) for the case of $10\%$ of contamination. Rows 1 and 2 displays the results for $(n_{\bar{D}}, n_D)=(100,100)$, rows 3 and 4 for $(n_{\bar{D}}, n_D)=(200,100)$, and rows 5 and 6 for $(n_{\bar{D}}, n_D)=(200,200)$.}}
\end{figure}

\begin{figure}[H]
	\begin{center}
		\subfigure{
			\includegraphics[width = 5.85cm]{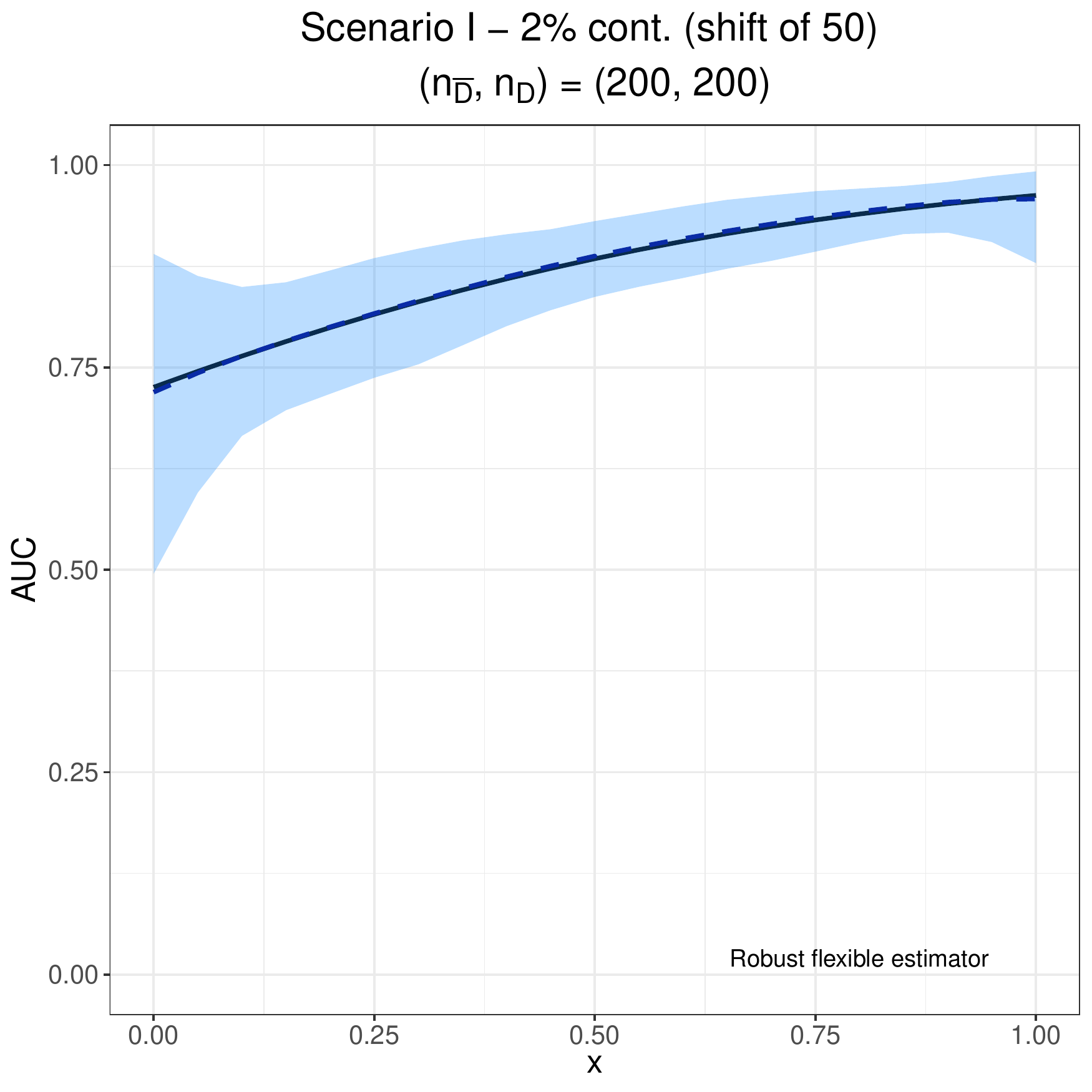}
			\includegraphics[width = 5.85cm]{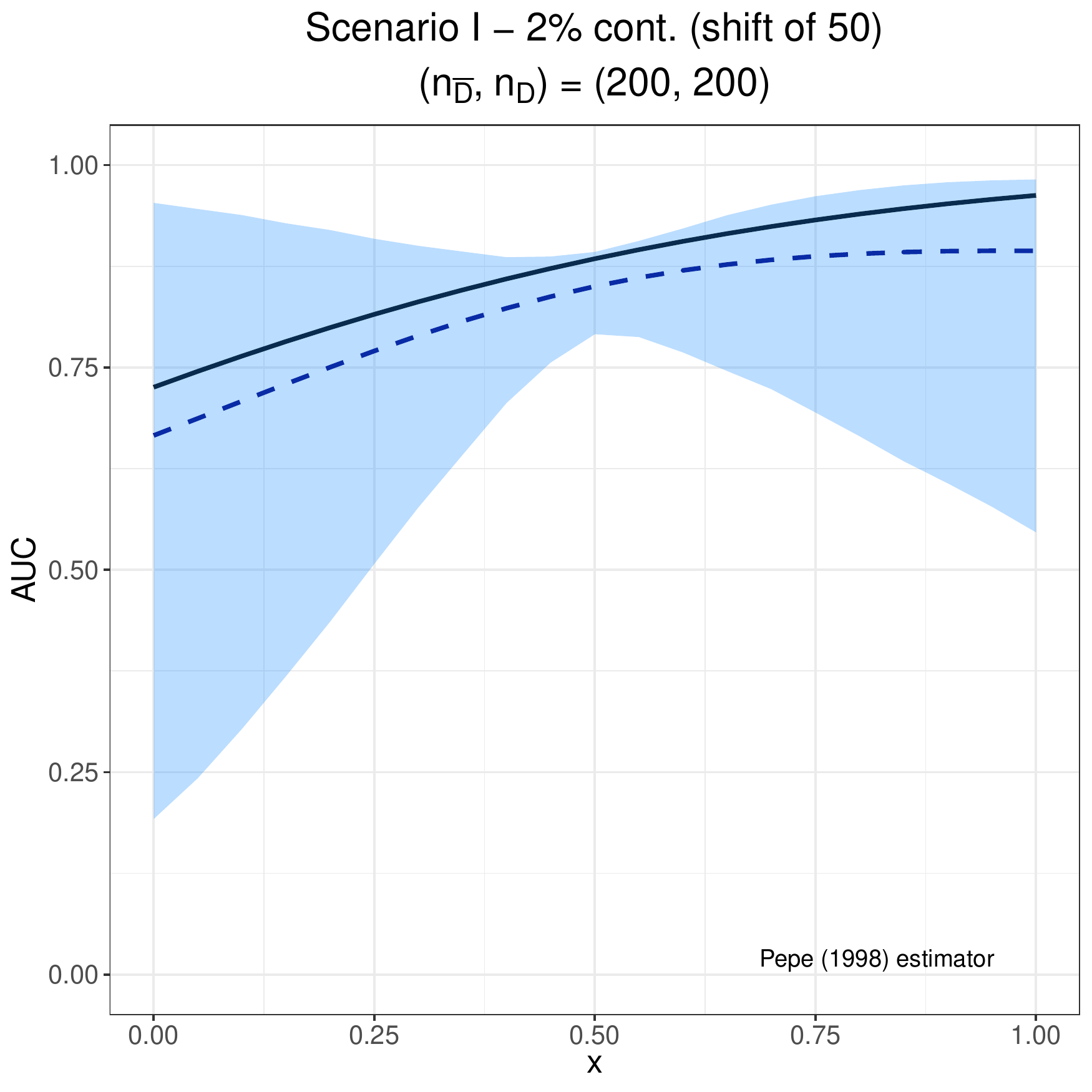}
		}
		\subfigure{
			\includegraphics[width = 5.85cm]{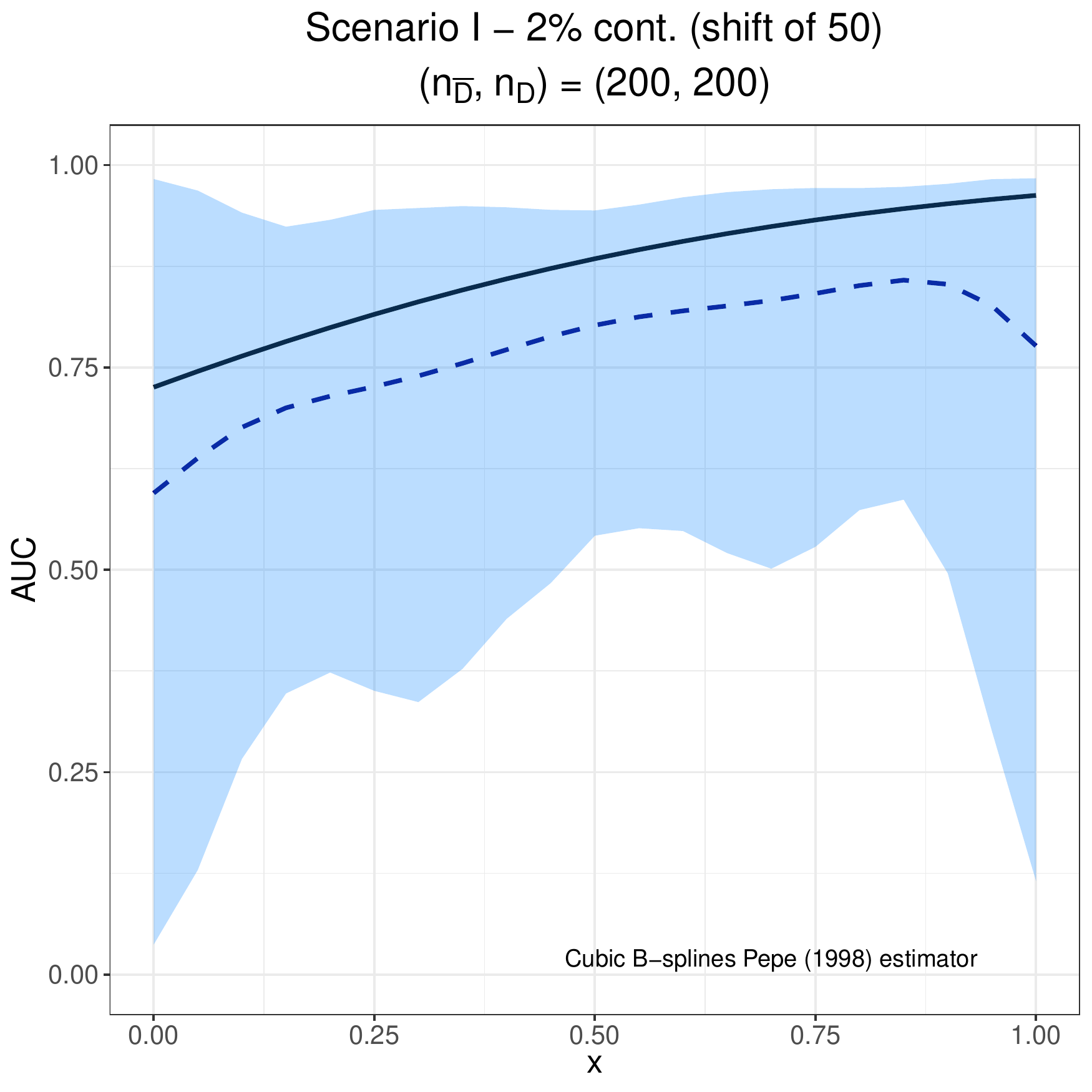}
			\includegraphics[width = 5.85cm]{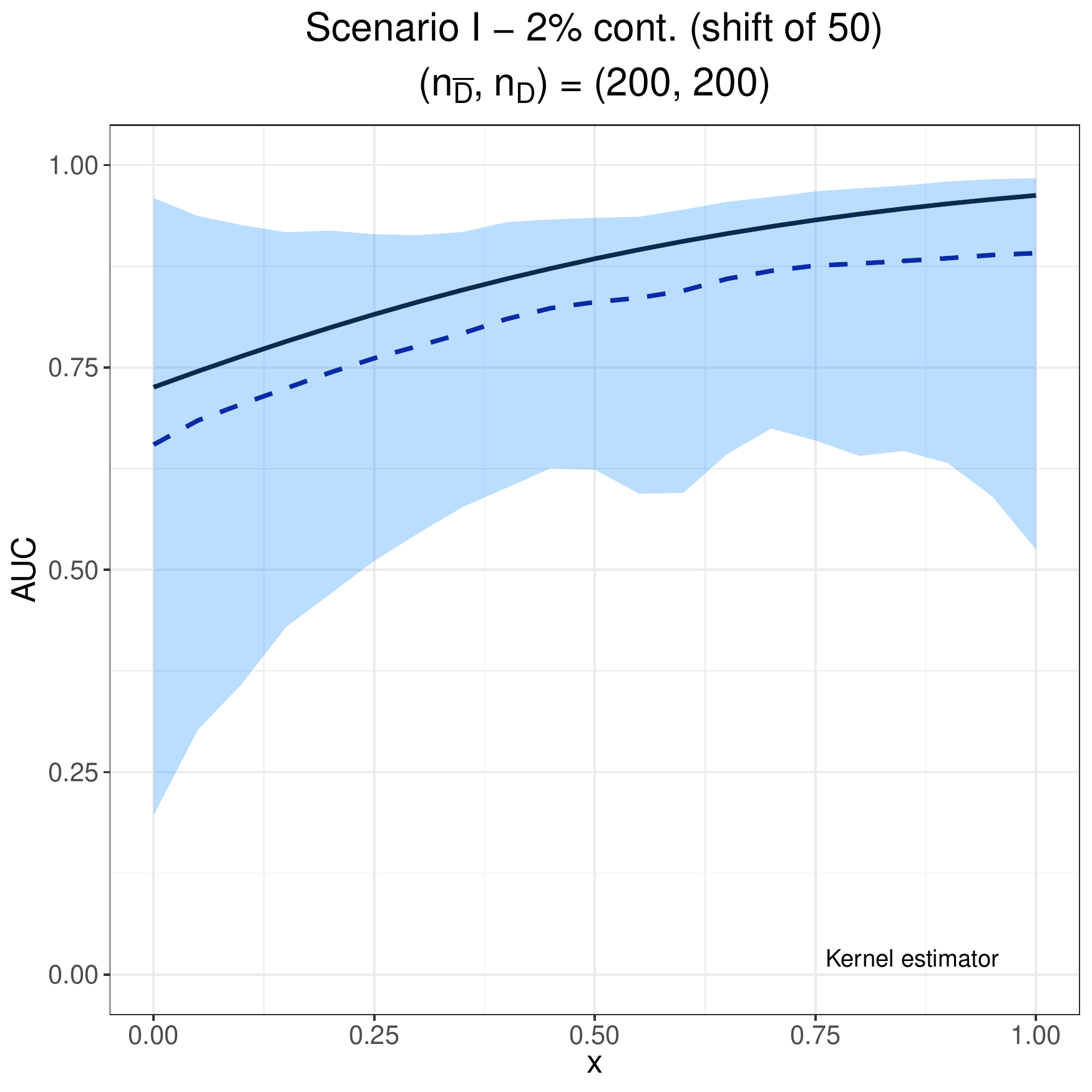}
		}
	\end{center}
	\caption{\footnotesize{Scenario I and $\kappa_{D}=\kappa_{\bar{D}}=50$. True covariate-specific AUC (solid line) versus the mean of the Monte Carlo estimates (dashed line) along with the $2.5\%$ and $97.5\%$ simulation quantiles (shaded area) for the case of $2\%$ contamination and $(n_{\bar{D}}, n_D)=(200,200)$.}}
\end{figure}

\begin{table}[H]
	\begin{center}
		\footnotesize	
		\begin{tabular}{cccc}
			& & \multicolumn{2}{c}{Sample size}\\
			& & \multicolumn{2}{c}{$(n_{\bar{D}},n_{D})$}\\\hline
			Scenario &  & $(100,100)$ & $(200,200)$ \\\hline
			\multirow{2}{*}{No contamination} &	
			$\text{rAIC}_{\bar{D}}(K_{\bar{D}1}=0) < \text{rAIC}_{\bar{D}}(K_{\bar{D}1}=3)$ & 70.0 & 67.0 \\
			&$\text{rAIC}_{D}(K_{D1}=0) < \text{rAIC}_{D}(K_{D1}=3)$ & 71.0 & 69.0 \\ \hline
			\multirow{2}{*}{$2\%$ contamination} &	
			$\text{rAIC}_{\bar{D}}(K_{\bar{D}1}=0) < \text{rAIC}_{\bar{D}}(K_{\bar{D}1}=3)$ & 73.0 & 69.0\\
			& $\text{rAIC}_{D}(K_{D1}=0) < \text{rAIC}_{D}(K_{D1}=3)$  & 72.0 & 70.0\\ \hline
			\multirow{2}{*}{$5\%$ contamination} &	
			$\text{rAIC}_{\bar{D}}(K_{\bar{D}1}=0) < \text{rAIC}_{\bar{D}}(K_{\bar{D}1}=3)$ & 75.0 & 71.0 \\
			& $\text{rAIC}_{D}(K_{D1}=0) < \text{rAIC}_{D}(K_{D1}=3)$ & 73.0 & 71.0\\ \hline
			\multirow{2}{*}{$10\%$ contamination} &	
			$\text{rAIC}_{\bar{D}}(K_{\bar{D}1}=0) < \text{rAIC}_{\bar{D}}(K_{\bar{D}1}=3)$ & 78.0 & 70.0 \\
			& $\text{rAIC}_{D}(K_{D1}=0) < \text{rAIC}_{D}(K_{D1}=3)$ & 75.0 & 74.0\\ \hline
		\end{tabular}
	\end{center}
	\caption{\footnotesize{Scenario I. Percentage  of time (over the $1000$ simulation runs) that the robust AIC favours the robust and flexible model with no interior knots over the same model but with three interior knots.}}
\end{table}

\begin{table}[H]
	\begin{center}
		\footnotesize	
		\begin{tabular}{cccc}
			& & \multicolumn{2}{c}{Sample size}\\
			& & \multicolumn{2}{c}{$(n_{\bar{D}},n_{D})$}\\\hline
			Scenario &  & $(100,100)$ & $(200,200)$ \\\hline
			\multirow{2}{*}{No contamination} &	
			$\text{rAIC}_{\bar{D}}(K_{\bar{D}1}=0) < \text{rAIC}_{\bar{D}}(K_{\bar{D}1}=3)$ & 71.0 & 68.0 \\
			&$\text{rAIC}_{D}(K_{D1}=0) < \text{rAIC}_{D}(K_{D1}=3)$ & 70.0 & 68.0 \\ \hline
			\multirow{2}{*}{$2\%$ contamination} &	
			$\text{rAIC}_{\bar{D}}(K_{\bar{D}1}=0) < \text{rAIC}_{\bar{D}}(K_{\bar{D}1}=3)$ & 73.0 & 69.0\\
			& $\text{rAIC}_{D}(K_{D1}=0) < \text{rAIC}_{D}(K_{D1}=3)$  & 71.0 & 70.0\\ \hline
			\multirow{2}{*}{$5\%$ contamination} &	
			$\text{rAIC}_{\bar{D}}(K_{\bar{D}1}=0) < \text{rAIC}_{\bar{D}}(K_{\bar{D}1}=3)$ & 75.0 & 70.0 \\
			& $\text{rAIC}_{D}(K_{D1}=0) < \text{rAIC}_{D}(K_{D1}=3)$ & 75.0 & 69.0\\ \hline
			\multirow{2}{*}{$10\%$ contamination} &	
			$\text{rAIC}_{\bar{D}}(K_{\bar{D}1}=0) < \text{rAIC}_{\bar{D}}(K_{\bar{D}1}=3)$ & 80.0 & 71.0 \\
			& $\text{rAIC}_{D}(K_{D1}=0) < \text{rAIC}_{D}(K_{D1}=3)$ & 76.0 & 72.0\\ \hline
		\end{tabular}
	\end{center}
	\caption{\footnotesize{Scenario II. Percentage  of time (over the $1000$ simulation runs) that the robust AIC favours the robust and flexible model with no interior knots over the same model but with three interior knots.}}
\end{table}

\begin{table}[H]
	\begin{center}
		\footnotesize	
		\begin{tabular}{cccc}
			& & \multicolumn{2}{c}{Sample size}\\
			& & \multicolumn{2}{c}{$(n_{\bar{D}},n_{D})$}\\\hline
			Scenario &  & $(100,100)$ & $(200,200)$ \\\hline
			\multirow{2}{*}{No contamination} &	
			$\text{rAIC}_{\bar{D}}(K_{\bar{D}1}=0) < \text{rAIC}_{\bar{D}}(K_{\bar{D}1}=3)$ & 72.0 & 68.0 \\
			&$\text{rAIC}_{D}(K_{D1}=0) < \text{rAIC}_{D}(K_{D1}=3)$ & 75.0 & 68.0 \\ \hline
			\multirow{2}{*}{$2\%$ contamination} &	
			$\text{rAIC}_{\bar{D}}(K_{\bar{D}1}=0) < \text{rAIC}_{\bar{D}}(K_{\bar{D}1}=3)$ & 72.0 & 68.0\\
			& $\text{rAIC}_{D}(K_{D1}=0) < \text{rAIC}_{D}(K_{D1}=3)$  & 74.0 & 70.0\\ \hline
			\multirow{2}{*}{$5\%$ contamination} &	
			$\text{rAIC}_{\bar{D}}(K_{\bar{D}1}=0) < \text{rAIC}_{\bar{D}}(K_{\bar{D}1}=3)$ & 75.0 & 71.0 \\
			& $\text{rAIC}_{D}(K_{D1}=0) < \text{rAIC}_{D}(K_{D1}=3)$ & 75.0 & 71.0\\ \hline
			\multirow{2}{*}{$10\%$ contamination} &	
			$\text{rAIC}_{\bar{D}}(K_{\bar{D}1}=0) < \text{rAIC}_{\bar{D}}(K_{\bar{D}1}=3)$ & 82.0 & 75.0 \\
			& $\text{rAIC}_{D}(K_{D1}=0) < \text{rAIC}_{D}(K_{D1}=3)$ & 79.0 & 77.0\\ \hline
		\end{tabular}
	\end{center}
	\caption{\footnotesize{Scenario III. Percentage  of time (over the $1000$ simulation runs) that the robust AIC favours the robust and flexible model with no interior knots over the same model but with three interior knots.}}
\end{table}

\begin{table}[H]
	\begin{center}
		\footnotesize	
		\begin{tabular}{cccc}
			& & \multicolumn{2}{c}{Sample size}\\
			& & \multicolumn{2}{c}{$(n_{\bar{D}},n_{D})$}\\\hline
			Scenario &  & $(100,100)$ & $(200,200)$ \\\hline
			\multirow{2}{*}{No contamination} &	
			$\text{rAIC}_{\bar{D}}((K_{\bar{D}1},K_{\bar{D}2})=(0,0)) < \text{rAIC}_{\bar{D}}((K_{\bar{D}1},K_{\bar{D}2})=(3,3))$ & 82.0 & 78.0 \\
			&$\text{rAIC}_{D}(K_{D1},K_{D2})=(0,0)) < \text{rAIC}_{D}((K_{D1},K_{D2})=(3,3))$ & 83.0 & 78.0 \\ \hline
			\multirow{2}{*}{$2\%$ contamination} &	
			$\text{rAIC}_{\bar{D}}((K_{\bar{D}1},K_{\bar{D}2})=(0,0)) < \text{rAIC}_{\bar{D}}((K_{\bar{D}1},K_{\bar{D}2})=(3,3))$  & 85.0 & 78.0\\
			&$\text{rAIC}_{D}(K_{D1},K_{D2})=(0,0)) < \text{rAIC}_{D}((K_{D1},K_{D2})=(3,3))$  & 84.0 & 78.0\\ \hline
			\multirow{2}{*}{$5\%$ contamination} &	
			$\text{rAIC}_{\bar{D}}((K_{\bar{D}1},K_{\bar{D}2})=(0,0)) < \text{rAIC}_{\bar{D}}((K_{\bar{D}1},K_{\bar{D}2})=(3,3))$   & 88.0 & 82.0 \\
			&$\text{rAIC}_{D}(K_{D1},K_{D2})=(0,0)) < \text{rAIC}_{D}((K_{D1},K_{D2})=(3,3))$ & 87.0 & 81.0\\ \hline
			\multirow{2}{*}{$10\%$ contamination} &	
			$\text{rAIC}_{\bar{D}}((K_{\bar{D}1},K_{\bar{D}2})=(0,0)) < \text{rAIC}_{\bar{D}}((K_{\bar{D}1},K_{\bar{D}2})=(3,3))$   & 90.0 & 84.0 \\
			& $\text{rAIC}_{D}(K_{D1},K_{D2})=(0,0)) < \text{rAIC}_{D}((K_{D1},K_{D2})=(3,3))$ & 91.0 & 86.0\\ \hline
		\end{tabular}
	\end{center}
	\caption{\footnotesize{Scenario IV. Percentage  of time (over the $1000$ simulation runs) that the robust AIC favours the robust and flexible model with no interior knots over the same model but with three interior knots.}}
\end{table}

\begin{figure}[H]
	\begin{center}
		\subfigure{
			\includegraphics[height = 6.15cm]{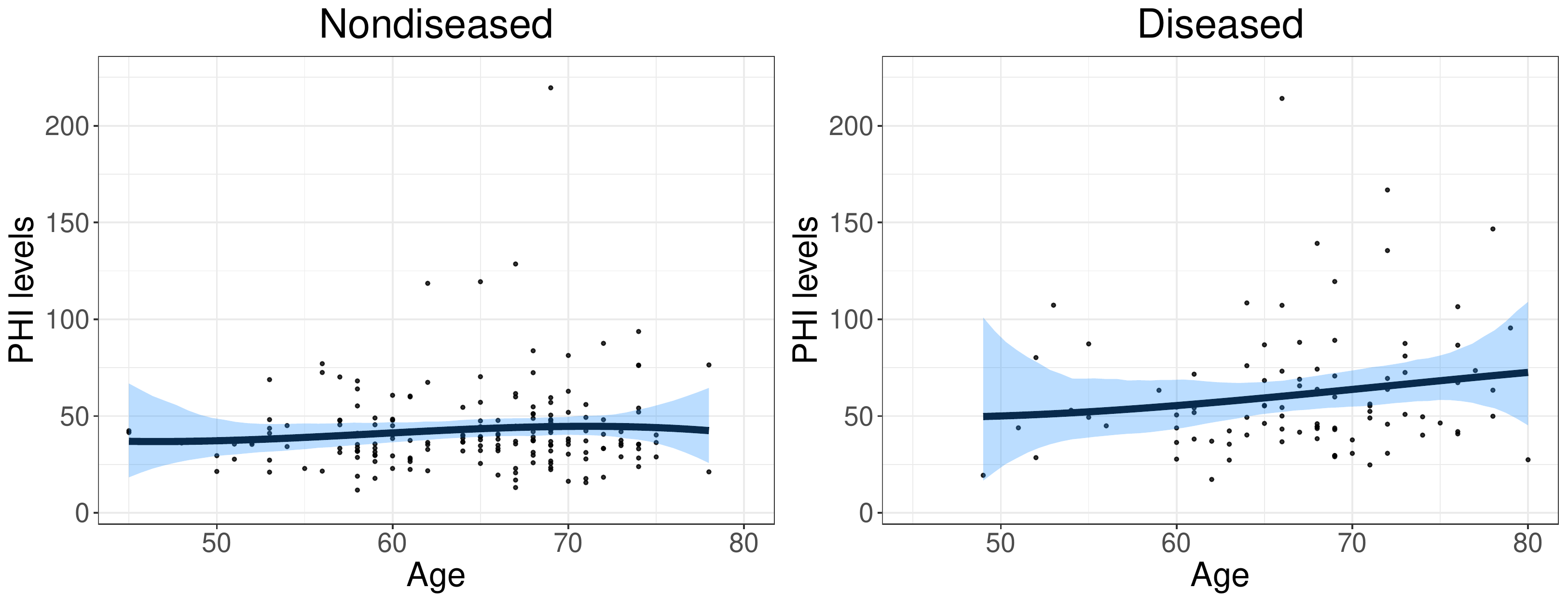} 
		}
		\subfigure{
			\includegraphics[height = 6.15cm]{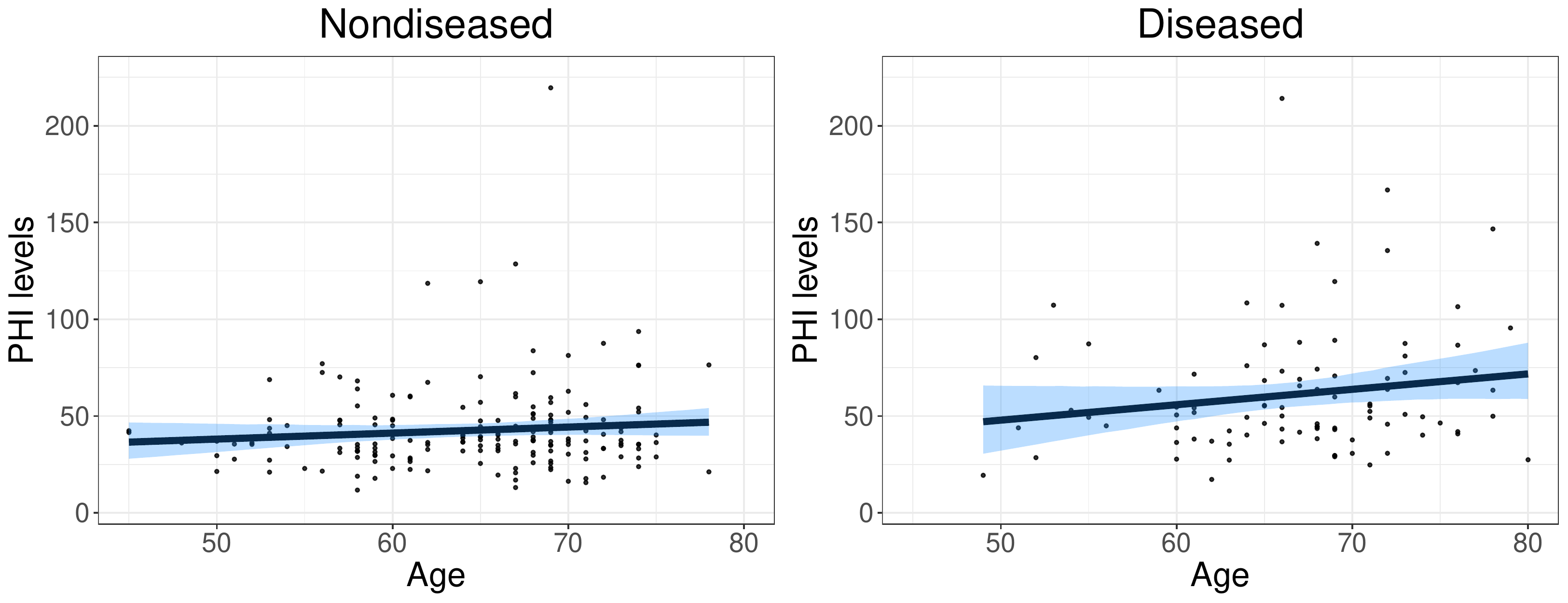} 
		}
		\subfigure{
			\includegraphics[height = 6.15cm]{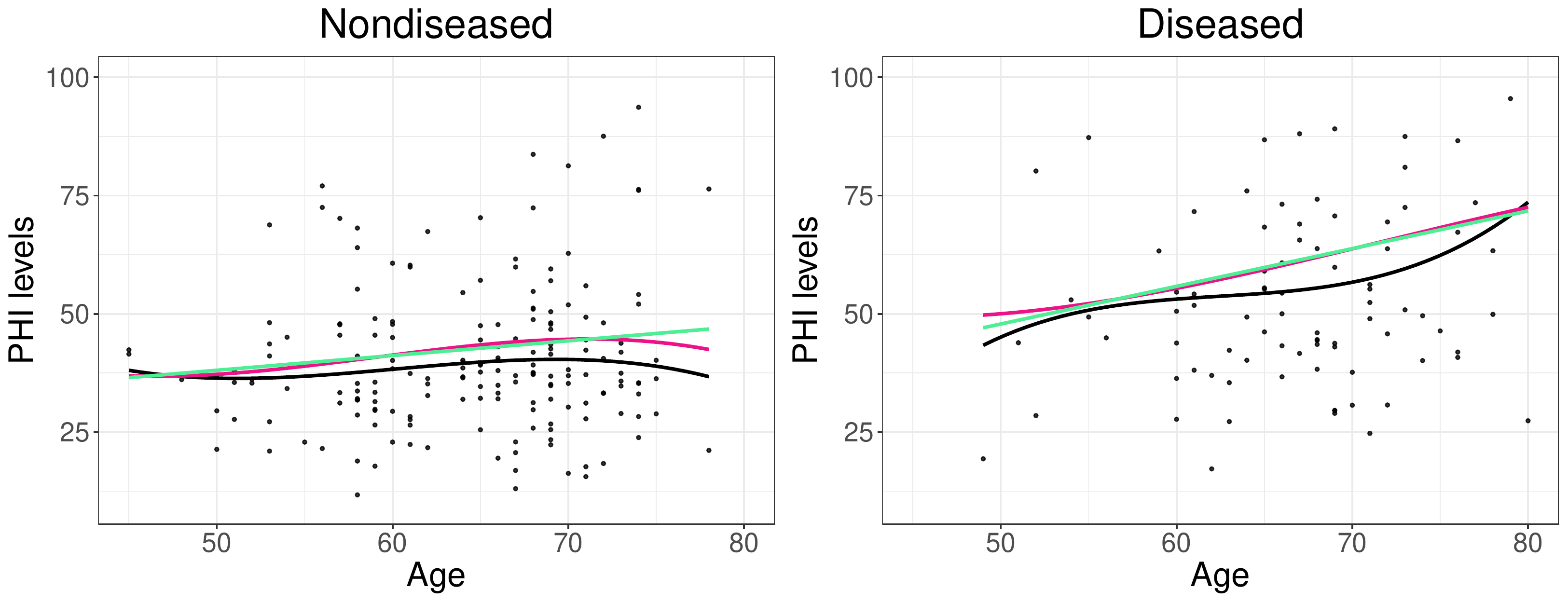} 
		}
	\end{center}
	\vspace{-0.2cm}
	\caption{\scriptsize{First row: Regression functions resulting from fitting a linear model with a cubic B-splines basis expansion (no interior knots) for the mean function. Second row:  Regression functions resulting from fitting a linear model. Third row: Point estimates from the three different fits, where the black line is the point estimate from the robust flexible model, the pink line is the point estimate corresponding to the B-splines linear model and the light green line is the estimate from the linear model. Note that for a better visualization the $y$ axis has been restricted to the range $(10, 100)$. The shaded areas represent the $95\%$ bootstrap confidence bands (based on $1000$ resamples).}}
\end{figure}

\begin{figure}[H]
	\begin{center}
		\subfigure[]{
			\includegraphics[height = 6.5cm]{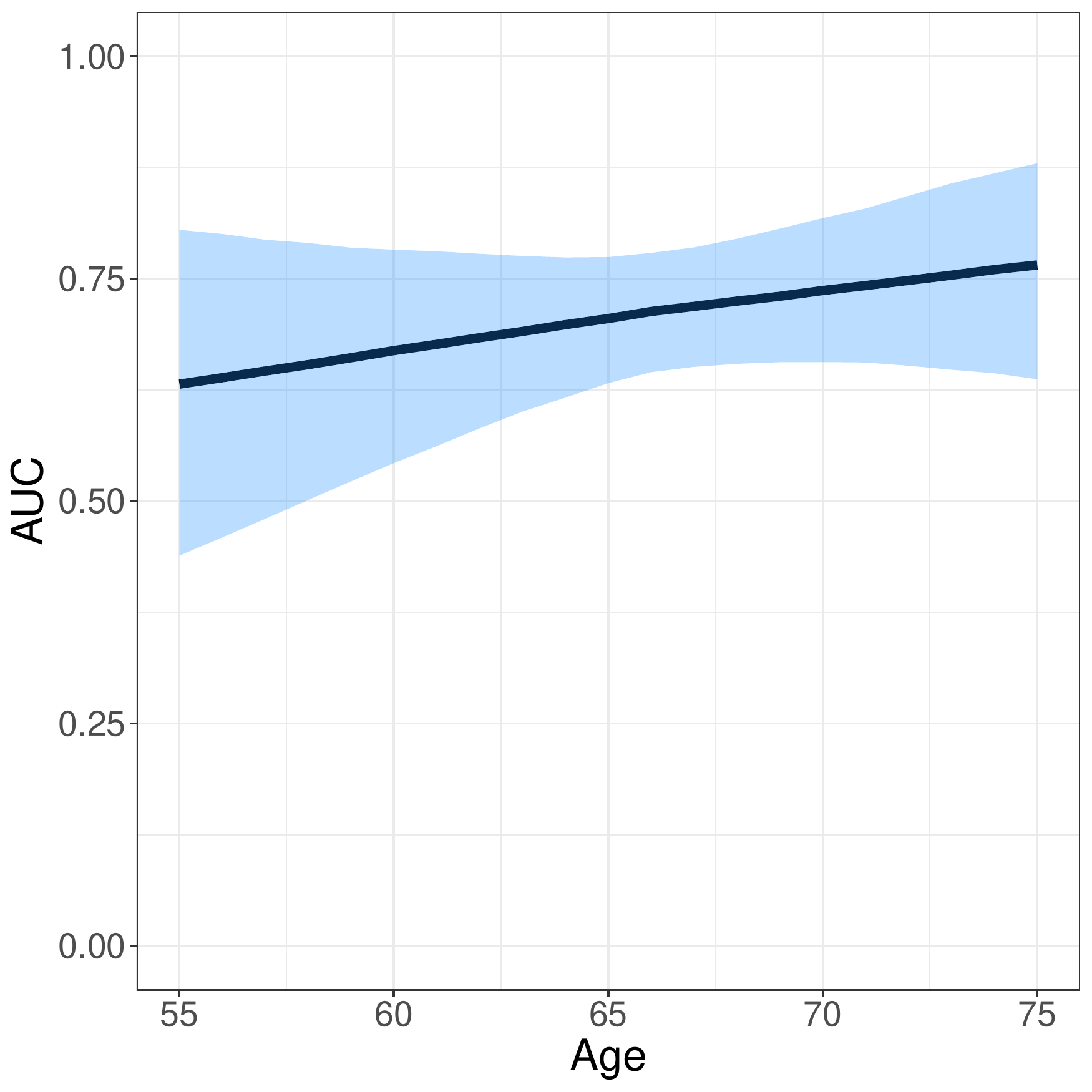} 
		}\hspace{0.5cm}
		\subfigure[]{
			\includegraphics[height = 6.5cm]{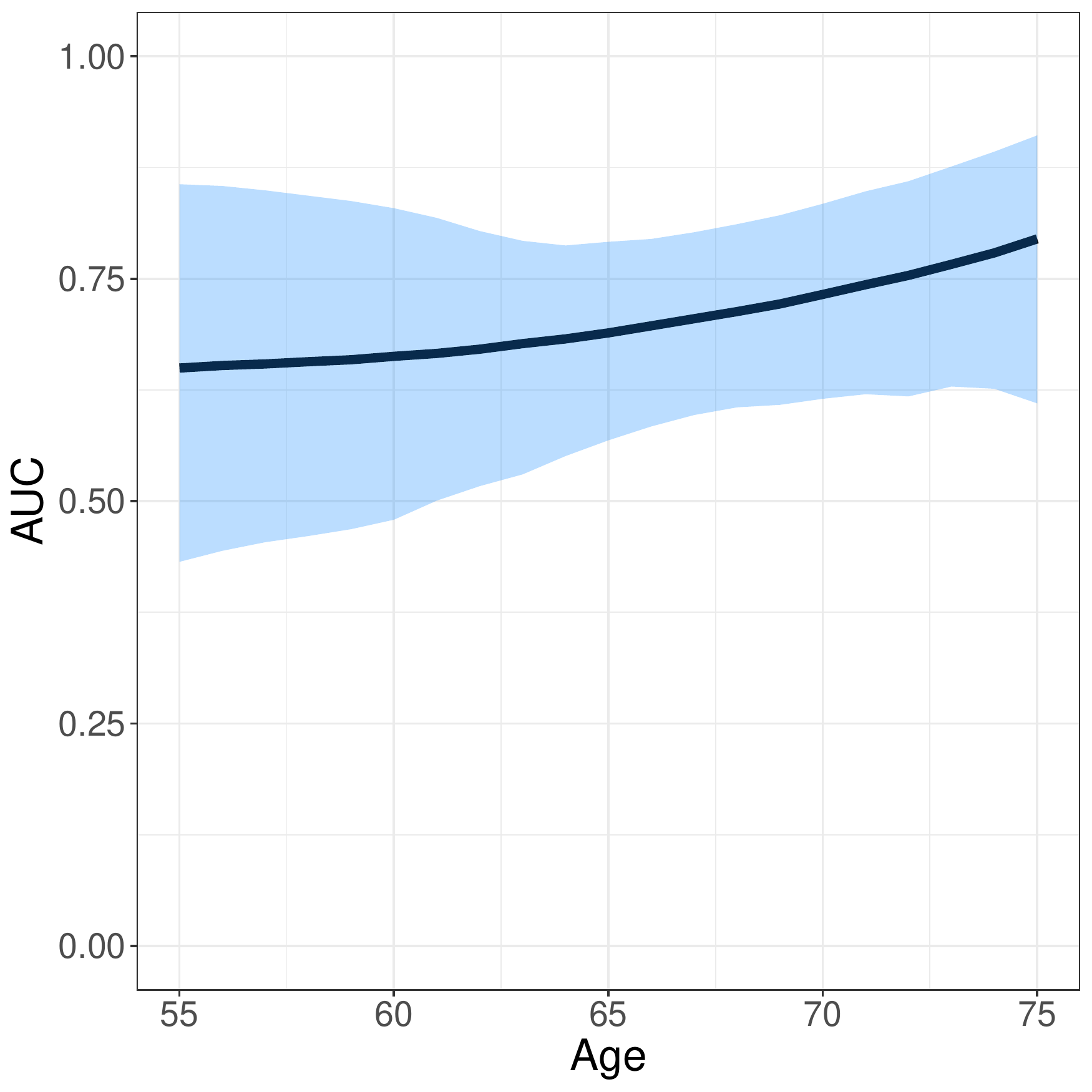} 
		}\\
		\subfigure[]{
			\includegraphics[height = 6.5cm]{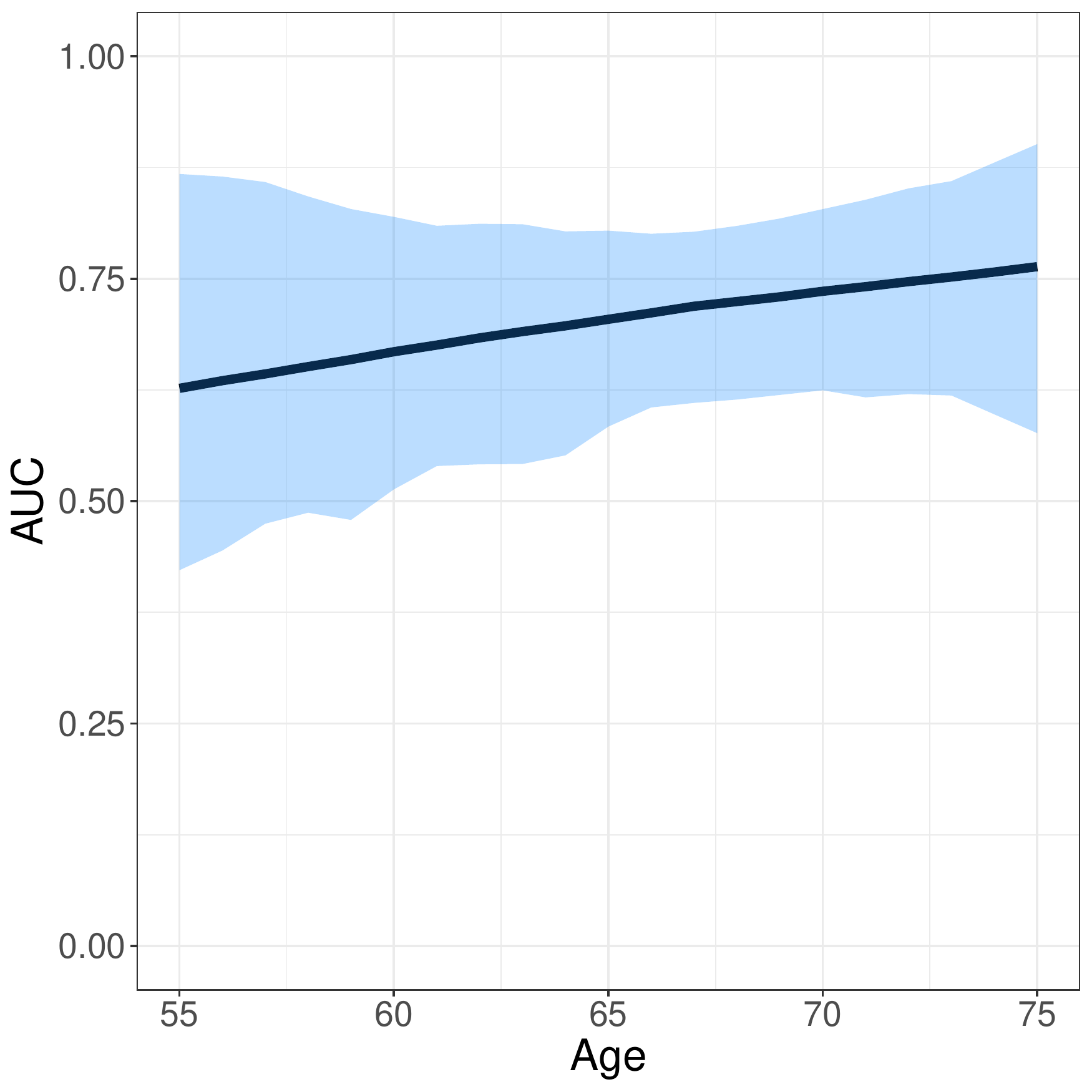} 
		}\hspace{0.5cm}	
		\subfigure[]{
			\includegraphics[height = 6.5cm]{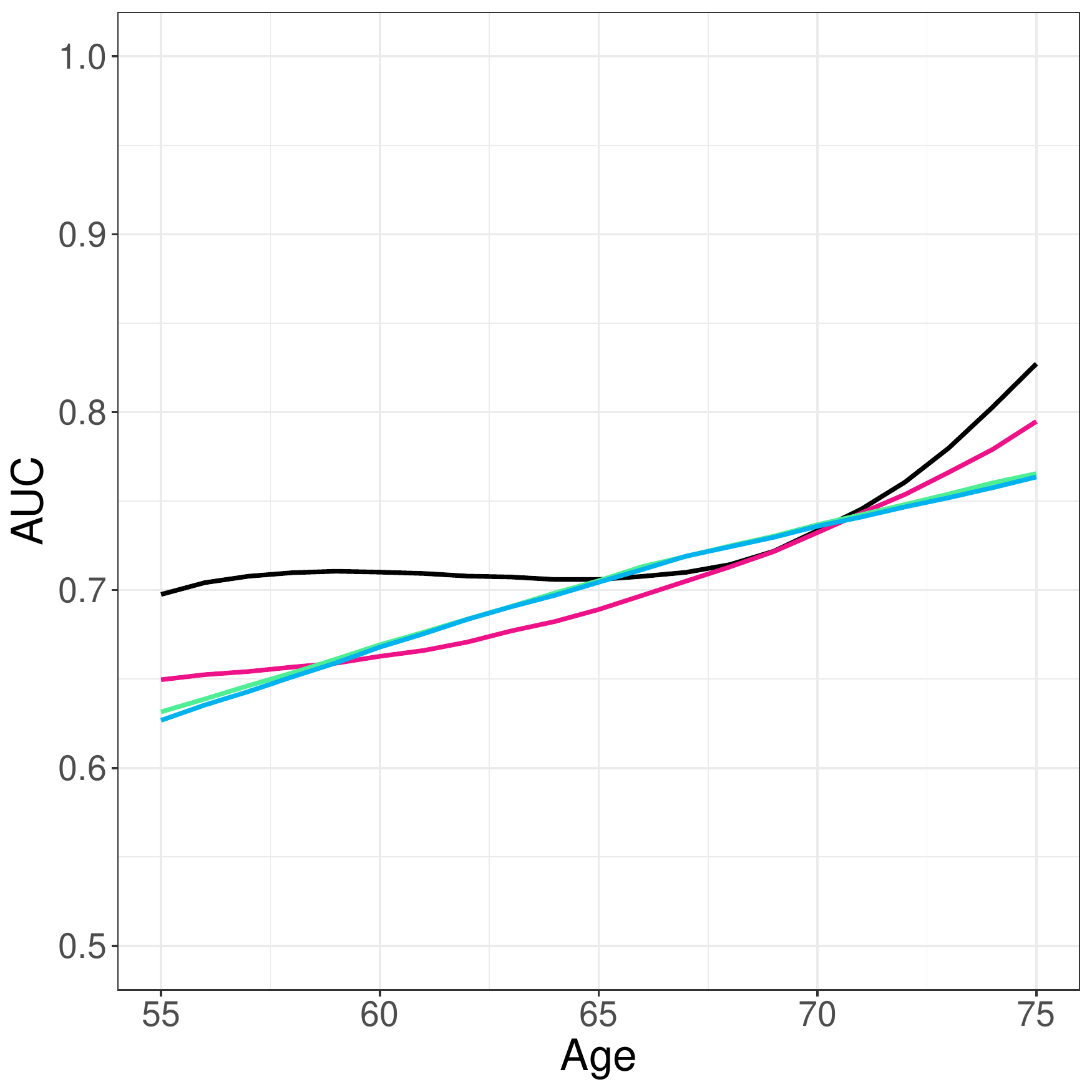} 
		}
	\end{center}
	\vspace{-0.2cm}
	\caption{\footnotesize{Age-specific AUCs. (a) Estimate resulting from fitting a linear model in each group. (b) Estimate resulting from fitting a linear model in each group with a cubic B-splines basis expansion for the mean function. (c) Estimate resulting from fitting the kernel approach in each group.  The shaded areas represent the $95\%$ bootstrap confidence bands (based on $1000$ resamples). (d) Comparison of point estimates from the different approaches. Black line: our approach. Light green: linear model. Pink Line: linear model with B-splines basis expansion for the mean function. Blue line: kernel method. The green and blue line are indistinguishable. }}
\end{figure}

\end{document}